\documentclass[twoside,fleqn]{ActaStyle}
\usepackage{graphicx}
\usepackage[english]{babel}    
\usepackage{times}
\usepackage{amsbsy,amsmath,amssymb}
\usepackage{cite}
\usepackage{enumerate}
\usepackage{lastpage}
\usepackage{epsf}
\usepackage{bm}

\numberwithin{equation}{subsection}

\def\authorheading{M. Hnati\v{c} et al.}
\def\shorttitle{Advanced field-theoretical methods...}
\def\PICS{} 

\newcommand{\beq}{\begin{equation}}
\newcommand{\eeq}{\end{equation}}
\newcommand{\bal}{\begin{align}}
\newcommand{\eal}{\end{align}}
\newcommand{\doo}[2]{\ensuremath{\frac{\partial #1}{ \partial #2}}}
\newcommand{\dee}[2]{\ensuremath{{{\rm d} #1\over {\rm d} #2}}}
\newcommand{\fun}[2]{\ensuremath{{\delta #1\over \delta #2}}}
\newcommand{\fundoo}[2]{\ensuremath{{\delta #1\over \delta #2}}}
\newcommand{\benum}{\begin{enumerate}}
\newcommand{\eenum}{\end{enumerate}}
\newcommand{\bitem}{\begin{itemize}}
\newcommand{\eitem}{\end{itemize}}

\newcommand{\bra}[1]{\ensuremath{\langle\, #1\,\vert}}
\newcommand{\ket}[1]{\ensuremath{\vert\,#1\,\rangle}}
\newcommand{\braket}[2]{\ensuremath{\langle\,#1\,\vert\,#2\,\rangle}}
\newcommand{\matrel}[3]{\ensuremath{\left\langle\,#1\,\left\vert\,#2\,\right\vert\,#3\,\right\rangle}}
\newcommand{\Tr}{\mathrm{Tr\,}}

\newcommand{\fp}[2]{FP$^{\textrm{#1}}_{#2}$}

\def\vx{{\bf x}}

\def\ann{{\hat{a}}}
\def\cre{{\hat{a}^\dagger}}  
\def\Bdag{{B^\dagger}}
\def\tilA{\tilde{A}}
\def\tilphi{\tilde{\varphi}}
\def\Tilphi{\tilde{\phi}}
\def\Lio{{\hat{L}}}
\def\on{\hat{n}}
\def\ophi{\hat{\varphi}}
\def\opi{\hat{\pi}}
\def\orho{\hat{\rho}}
\def\oU{\hat{U}}

\def\vx{{\bf x}}
\def\phi{{\varphi}}
\def\tilphi{\tilde{\varphi}}
\def\eps{\varepsilon}

\def\const{{\rm const\,}}

\def\W{\mathcal{W}}
\def\G{\mathcal{G}}

\def\O{\mathcal{O}}
\def\E{\mathcal{E}}
\def\S{\mathcal{S}}
\def\D{\mathcal{D}}

\def\eRM{{\mathrm e}}
\def\dRM{{\mathrm d}}

\def\mb{{\bm b}}
\def\mB{{\bm B}}
\def\mc{{\bm c}}

\def\mf{{\bm f}}

\def\mh{{\bm h}}

\def\mk{{\bm k}}

\def\mn{{\bm n}}

\def\mp{{\bm p}}

\def\mq{{\bm q}}

\def\mr{{\bm r}}

\def\mv{{\bm v}}

\def\mx{{\bm x}}

\def\my{{\bm y}}

\def\boldnabla{{\bm \nabla}}

\newcommand{\Ord}[1]{\ensuremath{\mathcal{O}( #1  )}}

\allowdisplaybreaks
\allowbreak

\begin{document}

\pagerange{1}{192}   
\title{ADVANCED FIELD-THEORETICAL METHODS IN STOCHASTIC DYNAMICS AND THEORY OF DEVELOPED TURBULENCE}
\author{M.~Hnati\v{c}\footnote{hnatic@saske.sk}$^{1,2,3}$, J.~Honkonen$^4$, 
               T.~Lu\v{c}ivjansk\'y$^{1,5}$}
 {$^1$ Faculty of Sciences, P. J. Safarik University, Park Angelinum 9, 041 54 Ko\v{s}ice, Slovakia \\ 
  $^2$ Institute of Experimental Physics SAS, Watsonova 47, 040 01 Ko\v{s}ice, Slovakia \\ 
  $^3$ BLTP, Joint Institute for Nuclear Research, Dubna, Russia \\
  $^4$ Department of Military Technology, National Defence University, Helsinki, Finland\\
  $^5$ Fakult\"at f\"ur Physik, Universit\"at Duisburg-Essen, D-47048 Duisburg, Germany
  }
\abstract{Selected recent contributions involving fluctuating velocity fields to the rapidly developing domain of stochastic field theory are reviewed.
Functional representations for solutions of stochastic differential equations and
master equations are worked out in detail with an emphasis on multiplicative noise and the inherent ambiguity of the
functional method. Application to stochastic models of isotropic turbulence of multi-parameter expansions in regulators of dimensional
and analytic renormalization is surveyed. Effects of the choice of the renormalization scheme are investigated. Special attention is 
paid to the r\^{o}le and properties of the minimal subtraction scheme. Analysis of the consequences of symmetry breaking of isotropic turbulence with
the use of the renormalization-group method is demonstrated by the effects due to helicity, strong and weak anisotropy.
A careful description is given of the influence of turbulent advection on paradigmatic reaction-diffusion problems.

 }
%
%

\tableofcontents

{\section{Solution of stochastic problems in field theory} \label{sec:stoch_problem}}

In 1928 the relativistic equation of motion for an electron was postulated by P. A. M. Dirac \cite{Dirac1}. This gave rise to tremendous
 development in theoretical physics, which ultimately led to a formulation of relativistic quantum mechanics. From the  historical point of view
 the first example of such a theory is quantum electrodynamics (QED), which describes quantum theory of electromagnetic interactions between
 electrons, positrons and photons. New theoretical methods became available for calculation of 
  differential cross sections for elementary processes in the high energy physics, which offered
 a controlled perturbation calculation in the form of asymptotic series in a small parameter -- the fine-structure constant. However, it turned out
 that integrals, which terms of these series contain, are divergent both in the infrared (IR) and in the ultraviolet (UV) regions
 and mathematical expressions loose their meaning. It was realized that this problem is a property of any quantum field theory, not only quantum electrodynamics.

A method for the removal of ultraviolet divergences was put forward in the 50s by St{\"u}ckelberg and Petermann \cite{stu} giving rise to the
creation of a solid theoretical approach in the form of the renormalization group (RG). Group properties in this approach were found and
mathematically rigorously defined by N. N. Bogolyubov and D. V. Shirkov \cite{bog1,bog2,bog3}.

The methods of perturbative renormalization group, UV renormalization and Feynman diagrammatic technique have become a standard
part of theoretical physics. One of the most important observations originated from RG was
 the discovery of running constants. Roughly speaking, this means that physical quantities such as charge of the electron, its mass etc.
are not constant, but their values depend on the scale at which they are measured. For instance, the fine-structure constant, which
is approximately equal to $1/137$ at energy scales of order $100\mbox{ }\mathrm{ eV}$  (typical scales of atomic and nuclear physics), attains  value $1/128$
at energies of order $100\mbox{ }\mathrm{ GeV}$. This growth leads to a modification of the Coulomb law. QED is an example of abelian gauge theory. Yang and Mills generalized 
the principle of gauge invariance to non-abelian theories \cite{Yang}, which are the main ingredients of models of weak interactions and quantum chromodynamics (QCD).
Gauge theories have played a crucial role in the explanation 
of properties of elementary particles and are the basis of the standard model, which is the most successful attempt of mankind for 
a reductionist description of the universe (without gravity).  

The growth of the fine-structure constant in QED affects results of calculations, which can be experimentally observed, e.g., in the Lamb
shift of the energy levels $2s_{1/2}$ and $2p_{1.2}$ for the hydrogen atom. On the other hand, the coupling constant in QCD exhibits 
completely different  anomalous behavior, which did not correspond to observed in the beginning of 70's. To put it simply, the smaller
distances are probed in the experiment, the weaker is the interaction between particles. Asymptotically the 
coupling constant goes to zero. For this discovery in 1973 \cite{wil,pol}, known as asymptotic freedom, Gross, Wilczek and Politzer, 
were awarded the Nobel Prize in 2004. QCD is the only known (and confirmed) theory of ``string''-like behavior in microworld, which at the larger
scales leads to the confinement of quarks and gluons. A byproduct of all the subsequent discoveries was further development of quantum 
field theory and related mathematical methods.

About the same time, an unrelated branch of physics devoted to the study of continuous phase transitions in classical systems 
experienced many new developments. 
	Let us recall that usually the critical behavior of a system is
	understood as its behavior near the critical point of a continuous phase transition, where a scaling with nontrivial exponents appears.
	For instance, the critical exponent $\beta$
	characterizes the dependence on temperature of the magnetization of a ferromagnet in zero external
	field near critical temperature $T_c$ ($T<T_c$)
	\begin{equation*}
	   M \sim (T_c-T)^{\beta}\,.
	   \label{eq:RG_critM}
	\end{equation*}
	Calculation of the critical exponents, explanation of their universality and
	finding relations between them  is the subject of study of the theory
	of critical behavior. The RG represents a powerful tool for solution of these
	problems. The RG approach was first applied by K.~G.~Wilson
	\cite{WilKog74} (Wilson renormalization group) to static critical phenomena
	and soon after used to study critical dynamics as well
	\cite{Ma74,Ma75}.
Typical
examples are the $\lambda$ transition from normal to superfluid in $\mbox{}^4$He and the transition from paramagnetic to ferromagnetic
phase in uniaxial magnets (Ising model).
The distinguishing physical feature is the presence of  
universal behavior, i.e. independence of the microscopic parameters of the model and importance of only
gross properties such as space dimension, symmetry and nature of the order parameter.
Moreover, in contrast to the typical problems in physics it was clear that the procedure of decoupling of scales is not possible -
a characteristic scale was lacking in the critical system and all length scales must be treated simultaneously. This is the simple
reason why the ordinary perturbation theory was not very useful. 
The lack of a typical scale manifests itself in the powerlike behavior with universal exponents.
  
In order to explain critical behavior
new theoretical approaches were needed.
Kadanoff \cite{Kadanoff66} proposed a block scheme, in which using a specific contraction procedure it was possible to derive a 
macroscopic model from a microscopic one. However,
 the approach led often to uncontrolled approximations and worked only for special cases. An important contribution was made
 by the Soviet school - Landau, Pokrovskii, Patashinkii - who can be regarded as founders of the fluctuation theory of phase
 transitions \cite{Landau_stat_phys,PatPok}. 
 The crucial idea is
 the construction of model Hamiltonians with proper symmetry and physical considerations taken into account. 
 In the beginning of 70s Wilson formulated a renormalization technique based on the integrating out of large-momentum degrees of 
 freedom \cite{wilson1,wilson2}. This approach resulted in
 a conceptual framework, which is still in use. Wilson was awarded the Nobel Prize in 1983.  
Practical analytical tools such as dimensional regularization \cite{Bollini72,Hooft72},  the famous $\eps$ expansion \cite{WilFis} 
and others were found \cite{wilson4}. With them in hand
theoretical physicists have a set of rules which in principle allows
computation of critical exponents (e.g. the Fisher index $\eta$) in a controlled fashion. Values obtained in this way are in a reasonable agreement
 with the experimentally measured values. Needless to say, critical exponents very often assume values, which are different
 from those predicted by the mean-field theory.

Functional formulation of quantum field theory based on the use of the generating functional \cite{Schwinger,Englert,Jona}
made it possible
to forget about quantum nature of the fields and treat them as classical objects. In this sense a quantum-field operator
corresponds to a fluctuating classical field. In the current literature field-theoretic methods has become the common term.
It has to be stressed that in this approach the basic ingredients of the theory are classical fields, Lagrange functions, the corresponding actions
and generating functionals. There is also a change in the underlying geometry - in the case of classical systems coordinates and times are given
in euclidean space rather than in the Minkowski spacetime. Since the models analyzed are invariably non-relativistic, this feature boils down
to the substitution of the Schr\"odinger evolution by the diffusion evolution, which is some cases (e.g. directed percolation) has the interpretation
of time playing a role of a singled out direction. Another important difference is that, contrary to non-relativistic quantum field theory, in classical reaction models the particle number is not conserved.

The first task in solving a classical statistical problem is to find the corresponding action functional. Its construction requires
deep understanding of the physical situation and identification of relevant physical parameters.
The use of methods of quantum field theory has enabled quick solution of problems in statistical physics on the basis of Ginzburg-Landau fluctuation theory. In particular, it has become possible to
calculate critical indices to high orders in perturbation theory for diverse universality classes \cite{Zinn,Vasiliev}.

Later dynamical models were formulated in terms of time-dependent fluctuations of slow variables near equilibrium.
A detailed classification -- which soon became standard -- 
was proposed by Hohenber and Halperin \cite{HH77}. An up-to-date review of dynamic critical phenomena can be found in Ref. \cite{Folk06}.
In this approach dynamics in the vicinity of a critical point are described by Langevin equations, which historically were first used for description of the Brownian motion.

In 1973 Martin, Siggia and Rose \cite{MSR} put forward the idea that in dynamic models an important role is played by a conjugate field and suggested to use it as an additional basic element. Later, Janssen and De Dominicis \cite{Janssen76,Dominicis76} showed how the dynamic models of
the Martin-Siggia-Rose (MSR) approach can be formulated in the language of path integrals. This has the great
advantage of identifying correctly linear terms to construct in an effective manner the perturbation theory. 

In critical dynamics the state of thermodynamic equilibrium must be the stationary solution of the stochastic problem
generated by the Langevin equation. This requirement imposes certain restrictions on the structure of the Langevin equation
and the statistics of the noise (e.g. Onsager relations and fluctuation-dissipation theorems). However, if these restrictions
are lifted, then it is possible to use the Langevin equation to describe stochastic problems with steady states which are not
states of thermodynamic equilibrium. The problem of
fully developed turbulence \cite{Landau_fluid}, stochastic magnetohydrodynamics, transport phenomena in turbulent environment and various advection-diffusion problems \cite{turbo} belong to this class of stochastic problems. There are also situations, in which
a physically reasonable stochastic problem cannot be formulated in terms of the Langevin equation. This is the case, when changes in the relevant variables cannot be treated as continuous functions of time. This happens, for instance, in models of
kinetics of chemical reactions \cite{HHL08,Krapivsky}, ecological models, econophysical models for financial
markets or social systems including combat models of operations research. The stochastic problem must then be formulated with the use of the master equation. The latter allows for a representation very similar to second quantization in quantum mechanics.
Second quantization in quantum mechanics gives rise to non-relativistic quantum field theory, which may be formulated in terms of generating 
functionals. The result of this outflanking is the formulation of the original stochastic problem in field-theoretic terms starting from
the construction of the action functional.
 
We review generalization of the approach based on the Langevin equation with additive noise to the case
with multiplicative noise. We also discuss in detail the choice of the functional representation of the perturbation expansion
for the solution of the Langevin equation. We show that the ambiguity in the Feynman rules is an inherent property of
the functional representation which is present both in case of deterministic and stochastic problems. This ambiguity resembles
the Ito-Stratonovich ambiguity of stochastic differential equations (SDE) with multiplicative noise, for which we give a
detailed account of construction of the field theory as well.

We review the construction of the functional representation for the solution of stochastic problems based on the master equation.
The popular way of construction of this representation with the use of the interpolation approach to the functional integral suffers
from problems in taking into account the initial and final conditions which are often solved by hand-waving arguments. We recall an alternative
way based on the operator formulation of evolution, which is free from ambiguities in dealing with the boundary conditions of evolution.
Although the evolution in these systems is given by first order time derivative, they are similar to relativistic quantum systems in the sense that
the number of particles is not conserved. A typical example is the directed bond percolation process \cite{JanTau04}, where a
continuous creation and annihilation of interacting agents takes place.
Using the method of ''second quantization'' of Doi the entire set of master equations can be cast into a ''Schr\"odinger'' equation
with a given Liouville operator
for a state vector in a Fock space. The large-scale properties of the solution can be then analyzed with the further use of Doi formalism and functional methods, which are presented in this article. 

The final aim of the theory (either in stochastic dynamics or developed turbulence) is to find the time-space dependence of statistical correlations -
mainly those who can be experimentally measured. It turns out that use of quantum field theory methods (RG included) allows to derive a linear
differential equation, which contains stable solutions in the asymptotic region of large macroscopic scales.

The solutions have a form of a product of a power-like term with a nontrivial exponent and scaling function of dimensionless
variables (not determined by the RG). In order to compute critical exponents in the form of asymptotic series one must resort to a certain scheme (we often employ variants of dimensional renormalization). Asymptotic properties of the scaling functions are analyzed by the operator product expansion, which is another theoretical tool developed mainly by Wilson, Wegner and Kadanoff. In the stochastic theory of fully developed turbulence scaling functions may be singular functions of their dimensionless arguments and this can drastically change the critical exponents.
All these features of models turbulent transport are discussed here in a detail on concrete
calculations for models describing diffusion and advection of a passive scalar quantity in a turbulent environment. The results demonstrate intermittent (multifractal) behavior of
statistical correlations of the random fields of concentration of advected particles.

In contrast to critical dynamics, where the expansion parameter in the renormalized theory is -- as a rule -- the deviation of the space dimension
from the upper critical dimension, in the field theory of
fully developed turbulence the basic expansion parameter is the regulator of analytic renormalization, i.e. the deviation of the exponent of
a powerlike correlation function of a random source field from its critical value. There may be several such regulators (as in
stochastic magnetohydrodynamics, for instance) which makes it possible to construct expansions in several regulators. Moreover, at certain
space dimensions additional divergences appear, for which the regulator of dimensional renormalization is usually introduced giving rise to even
more diverse multi-parameter expansions in the renormalized theory. Combining information from different expansions allows to improve significantly
the numerical accuracy of model calculations, as the calculation of the Kolmogorov constant with the use of the double expansion in isotropic 
turbulence demonstrates.

 
A typical approach to stochastic dynamics starts from the analysis of ideal systems - homogeneous in spatial and time, isotropic, incompressible 
(in case of fluids), possessing mirror symmetry etc.
In the present review the corresponding results for fully developed turbulence are summarized. However, real systems almost always exhibit 
some form of anisotropy, compressibility or
violated mirror symmetry. The effect of such deviations from the ideal system on fluctuating random fields has been an object of intensive
research activity, whose arguments and conclusions are described. The results have led to a general conclusion that such effects play a very 
important r\^{o}le. They can drastically change the large-scale behavior predicted by models of ideal systems.

In Sec. \ref{sec:stoch_problem} a detailed account of the two different approaches to the formulation of field-theoretical models of
stochastic problems is presented. 
The first approach based on a transition from a generic stochastic differential equation with additive or multiplicative noise to model
with effective actions generated by the SDE.
The second approach is based on the use of the master equation, which -- with the use of the Doi approach and functional methods -- leads to action
functionals, where
the effect of initial conditions can be explicitly taken into account. In Sec. \ref{sec:RG_theory} the method of RG is briefly described
and the double expansion scheme is presented.

Discussion of stochastic models of developed turbulence and the results of the RG analysis thereof is presented in Sec. \ref{sec:models}.
Sec. \ref{sec:violations} is devoted to an analysis of the effect of violated symmetries on the critical behavior. In particular,
critical amplitudes, values of critical
exponents and the stability of fixed points of the RG that govern the macroscopic (infrared) asymptotic behavior of different statistical mean values 
is of interest.
In Sec. \ref{sec:reactions} results of the analysis of the effects of turbulent flow on certain reaction-diffusion problems are overviewed.

	\subsection{Hydrodynamic kinetic equation}
In the Landau-Khalatnikov approach to kinetic phenomena in phase transitions \cite{Landau54} relaxation to 
equilibrium is described by the kinetic equation
for the time evolution of the order parameter $\varphi$
\begin{equation}
  \label{eq:intro_RelaxCritical}
  \dee{\varphi}{t}=-\,\gamma \doo{F}{\varphi}\,,
\end{equation}
where $F$ is a thermodynamic potential having a minimum at equilibrium. In the microcanonical ensemble it
would be the entropy with the sign minus, in the canonical ensemble the Helmholtz free energy etc. In an inhomogeneous
system the kinetic equation (\ref{eq:intro_RelaxCritical}) is generalized to the form
    	\beq
    	  \label{eq:intro_RelaxCriticalFun}
    	  \doo{\varphi}{t}=-\gamma \fun{F}{\varphi}\,,
    	\eeq
    	where $F$ is a functional of the order parameter. In this case the order parameter may 
    	be a conserved quantity itself, in which case the kinetic coefficient $\gamma$ will be
    	wave-number dependent and vanishing at $k=0$.
    	In the case of the paradigmatic Ginzburg-Landau theory of the ferromagnetic phase transition the
    	thermodynamic potential is of the form
    	\cite{Ginzburg50}
    	\beq
    	    \label{eq:intro_eq:F-GL-Gen}
    	    F(\varphi) = F_0+\int \dRM^3 \mr\, \left[
    	    g(\boldnabla \varphi)^2 + a(T-T_c) \varphi^2 + B\varphi^4
    	    \right]\,. 
    	\eeq
    The generic structure of the kinetic equation (\ref{eq:intro_RelaxCriticalFun}) is basically the same as in 
    hydrodynamic transport equations, in which the right side is not necessarily a (functional) derivative of some
    thermodynamic potential or the like, but a more general functional of the slow parameter $\varphi$, which may, 
    of course, have several components.
    
    The kinetic equation (\ref{eq:intro_RelaxCriticalFun}) provides a deterministic macroscopic description, which 
    does not take into account microscopic fluctuations. To describe effects of fluctuations without resorting to 
    microscopic theory, some kind of randomness may be used in the problem brought about  by the kinetic equation.
    The consistent construction of renormalizable field theories corresponding to nonlinear kinetic equations
    with noise was initially proposed in terms of quantum field theory by Martin, Siggia and Rose \cite{MSR}. It should 
    be noted that in the original paper the nature of the randomness was not specified. The operator approach 
    of Martin, Siggia and Rose (MSR) was soon replaced by the equivalent functional-integral representation for 
    the solution of the Langevin equation, in which a white-noise term is added to the right side of the kinetic
    equation \cite{Dominicis76,Janssen76}. The functional-integral approach avoids any connection to the operator
    formalism of quantum field
    theory. It is ambiguous, however, in a way which makes it difficult to choose the most convenient scheme for 
    calculations and confusing 
    from the point of view of the mathematical ambiguity of stochastic differential equations with multiplicative white noise.
    
    In the MSR approach the starting point is a system of nonlinear equations for field operators. An iterative 
    solution of these equations is the tree-graph solution of a nonlinear differential equation, iterative 
    construction which we shall shortly describe. This solution explicitly depends on random initial conditions 
    or random coefficient functions and implicitly on random boundary conditions through the Green function of 
    the linear problem. Expectation values of products of these tree-graph solutions over sources of randomness
    yield the solution of the stochastic problem.
    
    This iterative solution may be compactly expressed in the functional-differential form of the quantum-field
    perturbation theory, in which the ambiguity of the representation is explicit in the interaction functional. 
    It should be emphasized that the iterative solution is unambiguous, but the form of the functional
    representation, either integral or differential, is not. In this section we shall describe both ways 
    to represent the
    solution of a nonlinear differential equation and stress once more, that for the time being the 
    differential equations are completely
    deterministic. As a specific example we shall use the time-dependent Ginzburg-Landau model.


    The generic kinetic equation for a near-equilibrium system 
    in case of the Ginzburg-Landau free-energy functional \cite{Ginzburg50} gives rise to a nonlinear partial
    differential equation in the form
    \beq
        \label{eq:intro_TDGLEq}
        \partial_t \varphi
        = -\gamma \left[
        -2g\nabla^2 \varphi + 2a(T-T_c) \varphi + 4B\varphi^3
        -h\right],
    \eeq
    where $\partial_t = \partial / \partial t$ is the time derivative, parameter $\gamma$
    sets the time scale.
    Description of dynamic phenomena in the critical region based on the solution
    of this equation is the {\em time-dependent Ginzburg-Landau model} (TDGL model).
    It is widely used especially in the theory of superconductivity, where inhomogeneous solutions are
    important.
    
    Here, however, the TDGL equation (\ref{eq:intro_TDGLEq}) is needed for the description of the effect of small-scale
    fluctuations on the behavior of the system in the critical region. To this end, the external field $h$
    will eventually be regarded as a random noise thus giving rise to a stochastic differential equation (SDE).
    For the construction of perturbative solution of the SDE it is instructive, however, to introduce first
    the iterative solution of the deterministic TDGL equation (\ref{eq:intro_TDGLEq}). With
    the use of this approach a perturbative solution of any polynomially nonlinear differential equation
    may be constructed, provided a linear differential equation may be singled out as the starting point of the iteration.

    In case of the TDGL equation the starting point of the iteration is the linear equation (for simplicity, the symmetric phase
    with vanishing equilibrium value of the field $\varphi$ is considered here)
    \beq
      \label{eq:intro_TDGLEqLin}
     \partial_t \varphi^{(0)}
     = -\gamma\,\left[
      -2g\nabla^2 \varphi^{(0)} + 2a(T-T_c) \varphi^{(0)}
      -h\right]\,.
    \eeq
    Solution of this problem may be expressed in the form
    \beq
      \label{eq:intro_TDGL-0}
      \varphi^{(0)}(t,\mx)=\int \dRM^d\mx'\int \dRM t' \Delta_{12}(t,\mx,t',\mx')\gamma h(t',\mx')\equiv
      \int\!\dRM x' \Delta_{12}(x,x')\gamma h(x')\,,
    \eeq
    where here and henceforth $x=(t,\mx)$, $\dRM x\equiv \dRM^d \mx \dRM t $, $d$ is the dimension of the space and
    $\Delta_{12}$ is the {\em Green function} of the linear differential operator of equation (\ref{eq:intro_TDGLEqLin}), i.e.
    \beq
      \label{eq:intro_GF12}
      \left[
      \partial_t
      -2\gamma g\nabla^2+2a(T-T_c)\right] \Delta_{12}(t,\mx,t',\mx')=\delta_+(t-t')\delta(\mx-\mx')\,.
    \eeq
    Here, $\delta_+$ is the asymmetric $\delta$ function defined by 
    \beq
      \label{eq:intro_delta+}
      \int_{a+0}^b\!f(t')\delta_+(t-t')\,\dRM x'=
      \left\{
      \begin{array}{cc}
        0\,, & t<a \vee t>b\,,\\
        f(t+0)& a\le t<b
      \end{array}
      \right.
    \eeq
    Usually, vanishing boundary conditions are assumed. On time axis it is often technically simpler to consider
    ''initial condition'' $\varphi\to 0$, $t\to -\infty$, in which case the explicit expression for Green
    function in the time-wave-vector representation is
    \beq
      \label{eq:intro_GF12tk}
      \Delta_{12}(t-t',\mk)=\theta(t-t')\exp\left\{-2\gamma\left[g k^2+a(T-T_c)\right](t-t')\right\}\,.
    \eeq
    Should the Cauchy problem at a finite initial time instant $t_0$ be considered, it would be the simplest
    technically to introduce an ''external field'' concentrated at the initial time instant for the initial
    condition $\varphi(0,\mx)=\varphi_0(\mx)$:
    \[
      h(t,\mx)\to h(t,\mx)+\delta_+(t-t_0)\,{\varphi_0(\mx)\over\gamma}\,.
    \]
    It should be noted that the retardedness property of the Green function (propagator) (\ref{eq:intro_GF12tk}) turns
    out to be extremely important in the explicit construction of the perturbation expansion of the solution,
    especially when the small-scale fluctuations will be taken into account.
    
    To construct the iterative solution, it is convenient to cast the TDGL equation (\ref{eq:intro_TDGLEq})
    in the form of an integral equation
    \begin{equation}
      \label{eq:intro_TDGLEqInt}
      \varphi(x)=\int\!\dRM x' \Delta_{12}(x,x')\gamma h(x')
      -4\gamma B\int\!\dRM x' \Delta_{12}(x,x')\varphi^3(x')\,.
    \end{equation}
    The zeroth-order contribution $\varphi^{(0)}$ is obtained by putting the coupling constant equal to zero $B=0$ in
    (\ref{eq:intro_TDGLEqInt}).
    To obtain the first-order contribution, the zeroth-order solution is substituted in the right side
    of (\ref{eq:intro_TDGLEqInt})
    to yield
    \beq
      \label{eq:intro_TDGL-1}
      \varphi^{(1)}(x)=-4\gamma B\int\!\dRM x'
      \Delta_{12}(x,x')\left[\varphi^{(0)}(x')\right]^3\,,
    \eeq
    which then is substituted in the right side of (\ref{eq:intro_TDGLEqInt}) to obtain the second-order contribution etc.
    
    It is convenient to express the iterative solution in a graphical form, the
    two leading terms of which may be depicted as 
    \be
      \label{eq:intro_IterField}
      \varphi=\varphi^{(0)}+\varphi^{(1)}+\ldots
      =
      \raisebox{-0.25ex}{
      \includegraphics[height=0.27truecm]
      {\PICS 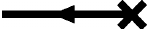}}+ \raisebox{-5.0ex}{
      \includegraphics[height=1.75truecm]{\PICS 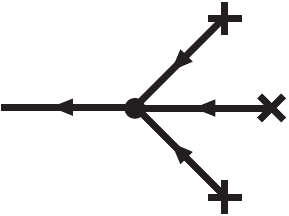}}+\ldots
    \ee
    Here, the directed line
    corresponds to the propagator $\Delta_{12}$, the cross to the external field term field $\gamma h$
    and the full dot to the vertex factor obtained as the third order derivative with respect to $\varphi$ of
    the right side of the kinetic equation (TDGL equation) (\ref{eq:intro_TDGLEq}).
    
    From the iterative solution it is clearly seen that the propagator $\Delta_{12}$ is essentially the
    dynamic (linear) response function of the field $\varphi$:
    \beq
      \label{eq:intro_LinRespFun}
      \chi(x,x')=\fun{\varphi(x)}{h(x')}\biggl\vert_{h=0}=\gamma  \Delta_{12}(x,x')\,.
    \eeq
    Little reflection shows that in the graphical representation of the solution of the kinetic equation
    there are only connected graphs without closed loops of propagators, i.e. the graphical solution of
    the TDGL equation consists of connected {\em tree graphs} only.
    
    
    
    The tree-graph solution of the kinetic equation may be expressed
    in a compact functional form, which is useful to work out the effect of fluctuations. Consider the
    generic kinetic equation
     \beq
        \label{eq:intro_KinEqGen}
        \partial_t\varphi
        =V(\varphi)=-K\varphi+U(\varphi)+f
     \eeq
     in which the right side is not necessarily proportional to a functional derivative. In the last expression
     of (\ref{eq:intro_KinEqGen}) we have singled out the source term for
     calculation of response functions and the first-order term giving rise to the propagator $\Delta_{12}$ 
     \beq
       \label{eq:intro_propagator}
       \left(
       \partial_t \varphi+K\right)\Delta_{12}(t,\mx,t',\mx')=\delta_+(t-t')\delta(\mx-\mx').
     \eeq
    The {\em generating function} of solutions $\varphi[f]$ of the kinetic equation (\ref{eq:intro_KinEqGen}) is defined as
    \beq
      \label{eq:intro_GenFunTree}
      \G(A)=\sum\limits_{n=0}^\infty \!{1\over n!}\left\{\int\! \dRM x
      A(x)\varphi[x,f]\right\}^n\equiv \exp\left\{A\varphi[f]\right\}\,,
    \eeq
    where a {\em universal} notation is introduced, i.e. integrals and sums over parameters of the solution $\varphi[f]$ are implied.
    
    Derivatives of the functional $\G(A)$ (\ref{eq:intro_GenFunTree}) at $A=0$ are simply products of
    the solution $\varphi[f]$. With the use of functional
    $\delta$ function (whose application rules coincide with those of the $\delta$ function in a finite-dimensional Euclidean space)
    the generating function may be expressed as a functional integral \cite{Dominicis78}
    \beq
      \label{eq:intro_GenFunTree2}
      \G(A)=\int\!{\cal D}\varphi\,\delta\left(\varphi-\varphi[f]\right)\, \exp\left(A\varphi\right) \,.
    \eeq
    A change of variables in the $\delta$ function brings about the kinetic equation explicitly:
    \beq
      \label{eq:intro_GenFunTree3}
      \G(A)=\int\!{\cal D}\varphi\,\delta\left[-\partial_t\varphi+V(\varphi)+f\right]\vert\det M\vert
      \, \exp\left(A\varphi\right) \,.
    \eeq
    The functional $U(\varphi)$ contains the nonlinear terms of the functional $V(\varphi)$.
    We stress that the right-side functional $V(\varphi)$
    in the kinetic equation need not be a functional derivative of some other functional, although this is often the case in
    near-equilibrium kinetics.
    
    Representation the functional Jacobi determinant in (\ref{eq:intro_GenFunTree3})
    in the loop expansion of the expression $\displaystyle \exp\left({\rm Tr}\ln M\right)$ yields
    \begin{equation}
      \label{eq:intro_DetMExp}
      \det M =\det\left[\left(\partial_t + K\right)\delta(x-x')-\frac{\delta
      U({\mx},\varphi)}{\delta\varphi(x')}\right]
      =
      \det\left(\partial_t + K\right)\exp\left[-\Delta_{12}(0)U'(\varphi)\right]\,,
    \end{equation}   
    in which the following shorthand notation has been introduced
    \beq
      \label{eq:intro_shorthand12U'}
      \Delta_{12}(0) U'(\varphi)\equiv\int\! \dRM x\int\! \dRM x'
      \Delta_{12}(t,\mx,t,\mx')\fun{U({\mx},\varphi)}{\varphi(x')}\,.
    \eeq
    Recall that the propagator is the retarded Green function of the linear kinetic
    equation, whose value at $t=t'$ is not determined. It is often convenient to use the time-wave-vector representation
    for the diagonal value of the propagator, in which case a wave-vector integral appears
    \[
      \Delta_{12}(t,\mx,t'\mx')\Bigl\vert_{t=t'\atop \mx=\mx'}=\int\!{\dRM^d\mk\over (2\pi)^d}
      \,\Delta_{12}(t-t',\mk)\bigl\vert_{t=t'}\,,
    \]
    for convergence of which an UV cutoff is usually needed. For instance, in case of the TDGL model the diagonal
    value of the propagator may be expressed as
    \[
      \Delta_{12}(t,\mx,t'\mx')\Bigl\vert_{t=t'\atop \mx=\mx'}=
      \theta(t-t')\bigl\vert_{t=t'}\int\!{\dRM^d\mk\over (2\pi)^d}\,{\theta(\Lambda-k)\over
      2\gamma\left[gk^2+a(T-T_c)\right]}\,,
    \]
    where the sharp UV cutoff has been used. Thus, not only are measures of caution to be taken to make the
    diagonal value of the propagator finite, but also value of the step function of the retarded propagator needs to
    be defined at the origin.
    
    With the use of the Fourier-integral representation of the $\delta$ function in (\ref{eq:intro_GenFunTree3}) we arrive at
    the functional integral over two fields (in the Fourier integral for the functional $\delta$ function the imaginary unit
    is usually omitted and the auxiliary field $\tilphi$ taken as imaginary-number quantity, if necessary)
    \begin{align}
      \label{eq:intro_GenFunTree4}
      \G(A) & = \int\!{\cal D}\varphi\,\int\!{\cal D}\tilde{\varphi} \,
      \bigl\vert \det\left(\partial_t+K\right)\bigr\vert \exp\left[-\Delta_{12}(0)U'(\varphi)\right]
      \exp\left\{\tilde{\varphi}\left[-\partial_t\varphi+V(\varphi)+f\right]\right\}\nonumber\\
      & \times \exp\left(A\varphi\right)\,.
    \end{align}
    Henceforth, the field-independent determinant of the differential operator $\det\left(\partial_t+K\right)$ will be included in
    the measure of integration and therefore not expressed explicitly.
    
    In contrast with the usual approach \cite{Dominicis78}, there is no Gaussian integral over the 
    random noise in (\ref{eq:intro_GenFunTree4}), which would render the resulting functional 
    integral convergent as an iterated integral. In fact, the formal functional integral (\ref{eq:intro_GenFunTree4})
    is hardly convergent in any reasonable sense. However, by the standard transformation 
    rules \cite{Vasilev98} it generates a functional-differential
    representation (S-matrix functional) for the generating function (\ref{eq:intro_GenFunTree})
    \begin{equation}
      \label{eq:intro_GenFunTree5}
      \G(A)=\exp\left(\fun{}{\varphi}\Delta_{12}\fun{}{\tilphi}\right)
      \exp\left[\tilphi U(\varphi)-\Delta_{12}(0)U'(\varphi)+\tilphi f +A\varphi\right]\bigr\vert_0
    \end{equation}
    where $|_0$ stands for $|_{\tilphi=\varphi=0}$ and this notation will be employed in what follows.
    Further, the exponential functional differential operator in (\ref{eq:intro_GenFunTree5})
    is the {\em reduction operator}
    \[
      {\cal P}=\exp\left(\fun{}{\varphi}\Delta_{12}\fun{}{\tilphi}\right)
    \]
    spanning propagator lines between {\em vertex factors}
    \beq
      \label{eq:intro_Vfactor}
      V_{n,m}(x_1,\ldots,x_n;y_1,\ldots, y_m)={\delta^{n+m}V(\varphi,\tilphi)
      \over \delta\varphi(x_1)\cdots\delta\varphi(x_n)\delta\tilphi(y_1)\cdots\delta\tilphi(y_n)}
    \eeq
    where, in the present case,
    \[
      V(\varphi,\tilphi)=\tilphi U(\varphi)-\Delta_{12}(0)U'(\varphi)+\tilphi f +A\varphi\,.
    \]
    In the S-matrix functional (\ref{eq:intro_GenFunTree5})) all building blocks are mathematically
    well defined. Problems may, of course, arise with the convergence of the series. Since this 
    representation is obtained by a heuristic functional-integral argument, in what follows it will separately 
    demonstrated that the representation (\ref{eq:intro_GenFunTree5}) yields the iterative solution 
    of the kinetic equation (\ref{eq:intro_KinEqGen}).
    It should be noted that representation (\ref{eq:intro_GenFunTree5}) may be obtained directly --
    without any functional integrals and determinants --
    from the MSR operator equations of motion with the use of standard rules of construction of
    perturbation theory for Green functions \cite{Honkonen05}.
    
    It is a generic property of the functional representation (\ref{eq:intro_GenFunTree5}) that $\ln \G(A)$ 
    consists of the connected graphs of $\G(A)$.
    The interaction term $\tilphi U(\varphi)$ in (\ref{eq:intro_GenFunTree5}) is linear in the auxiliary field $\varphi$, therefore
    only one directed propagator line comes out of the corresponding vertex in the graphical representation, whereas the number
    of incoming lines is equal to the power of the field $\varphi$ in each monomial of $U(\varphi)$ (which is three in the TDGL model). 
    Propagator lines form continuous directed chains, which either form closed loops and vanish due to retardation of the
    propagator (except the single-propagator closed loop $\Delta_{12}(0)$), or go through the graph starting form the
    external field $f$ and ending at the source field $A$.
    
    At the vertex corresponding to $\tilphi U(\varphi)$ directed chains with a start at the external field $f$
    merge to produce a directed chain ending at $A$, whereas at the vertex corresponding to $-\Delta_{12}(0)U'(\varphi)$
    directed incoming chains end without producing any outcoming line. Therefore, connected graphs of $\G(A)$ either
    contain one source vertex corresponding to $A\varphi $ or do not contain such a vertex at all. On the contrary, the 
    number of vertices
    corresponding to external field $\tilphi f$ in a connected graph is limited only by the order of perturbation theory.
    It is not difficult to see by direct inspection that all connected graphs of $\G(A)$ without $A$ actually vanish, because the 
    vertex factors produced by terms $\tilphi V(\varphi)$ and $-\Delta_{12}(0)U'(\varphi)$ cancel each other in them, therefore
    we end up at the result
    \beq
      \label{eq:intro_GenFunTree6}
      \G(A)=\exp\left[A\W_1(f)\right]\,,
    \eeq
    which, of course, coincides with (\ref{eq:intro_GenFunTree}), but is expressed in terms of standard generating functions.
    
    The point of introducing the functional representation
    (\ref{eq:intro_GenFunTree5}) is that the explicit linear exponential of the external field $f$ is convenient in 
    working out generating functions in case of random external field (noise). The basic equation of the generating functional
    $\G(A)$ is the Schwinger equation with respect to the source $A$. It is most 
    straightforwardly obtained from the Gauss law for the functional integral (\ref{eq:intro_GenFunTree4})
    with respect to the auxiliary field $\tilphi$, which yields
    \begin{align}
      0 & =\!\int\!{\cal D}\!\varphi\!\int\!{\cal D}\tilde{\varphi}
      \fun{}{\tilphi}
      \exp\{\tilde{\varphi}\left[-\partial_t\varphi+V(\varphi)+f\right]-\Delta_{12}(0)U'(\varphi)+A\varphi\}
      \nonumber \\
      & =\!\int\!{\cal D}\varphi\!\int\!{\cal D}\tilde{\varphi} 
      [-\partial_t\varphi+V(\varphi)+f]
      \exp\{\tilde{\varphi}\left[-\partial_t\varphi+V(\varphi)+f\right]-
      \Delta_{12}(0)U'(\varphi)+A\varphi\}.\nonumber
    \end{align}
    Pulling out the field variables as derivatives with respect to the source we obtain
    \begin{equation}
      \label{eq:intro_GaussA}
      \G(A)^{-1}\left[-\partial_t\fun{}{A}+V\left(\fun{}{A}\right)+f\right]\G(A)=
      \left[-\partial_t\W_1+V\left(\W_1\right)+f\right]=0\,,
    \end{equation}
    which, of course, coincides with the original kinetic equation for the field $\varphi$.
    
    
    
    The reader already familiar with the functional integral
    would probably like to ask, how the Schwinger equation obtained from the Gauss law with respect to the basic field $\varphi$ fits here.
    The answer is that it actually reproduces the equation for the "response function"
    \beq
      \label{eq:intro_response function}
      \chi_{A\tilde{A}}(t,\mx; t'\mx')=\fundoo{\W_1(t,\mx)}{\tilde{A}(t',\mx')}\,.
    \eeq
    Here, $\tilde{A}=f$ and
    quotation marks are used here, because the physical response function is obtained at vanishing source fields.

    From the original kinetic equation with an additional source term
    \beq
      \label{eq:intro_KinEq}
      \partial_t \varphi
      =V(\varphi)=-K\varphi+U(\varphi)+\tilde{A}
    \eeq
    an equation for $\chi_{A\tilde{A}}$ is obtained by differentiation with respect to $\tilde{A}$ with the result
    \beq
      \label{eq:intro_ChiEq}
      \partial_t\chi_{A\tilde{A}} 
      =V'(\varphi)\chi_{A\tilde{A}}+{ 1}\,.
    \eeq
    The Schwinger equation from the Gauss law with respect to the basic field $\varphi$
    for the functional (\ref{eq:intro_GenFunTree4}) yields (with the account of expression (\ref{eq:intro_GenFunTree6}))
    \beq
      \label{eq:intro_Schwinger2}
      -\Delta_{12}(0)U''(\W_1)+\fundoo{V'(\W_1)}{\tilde{A}}+A\partial_t\chi_{A\tilde{A}}+V'(\W_1)A\chi_{A\tilde{A}}+A=0
    \eeq
    The dependence on the source field $A$ is explicit here ($\chi_{A\tilde{A}}$ does not depend on $A$), therefore
    the independent of $A$ terms and the sum of those proportional to $A$ must vanish independently. Careful tracing of
    the arguments reveals that the condition of vanishing of the coefficient of $A$ in (\ref{eq:intro_Schwinger2}) is actually the 
    equation (\ref{eq:intro_ChiEq}), thereafter it is seen that the sum of two first terms in (\ref{eq:intro_Schwinger2}) 
    vanishes identically
    on the solution $W_1$ of the kinetic equation (\ref{eq:intro_GaussA}).
    
    It should be noted that so far a functional representation (\ref{eq:intro_GenFunTree5}) has
    been constructed for generating functional
    of solutions of a {\em deterministic} nonlinear partial differential equation. There is no randomness here at all
    and the tree-graph solution of the equation is unique.
    It is an intrinsic property of the functional representation that it contains an ambiguity in the form of the diagonal value
    of the propagator $\Delta_{12}(0)$, but this has nothing to do with any stochastic problems, as it is sometimes presented in the
    literature \cite{Munoz89}.

    It is instructive to analyze in a more formal fashion the independence of the generating
    functional (\ref{eq:intro_GenFunTree5}) of the diagonal value of the propagator. To this end, use 
    the following trick \cite{Vasilev98}. Consider a product of
    $N$ functionals $F_n(\varphi_n,\tilphi_n)$ of a set of $N$ fields $\varphi_n$ and $\tilphi_n$ acted upon by
    a reduction operator containing all pairs of fields:
    \beq
      \label{eq:intro_SquareTrick1}
      \exp\left(\sum_{i,j=1}^N\fun{}{\varphi_i}\Delta_{12}\fun{}{\tilphi_j}\right)\prod_{n=1}^NF_n(\varphi_n,\tilphi_n)\,.
    \eeq
    Here, the reduction operator is a ''complete square'' of differential operators, i.e.
    \beq
      \label{eq:intro_CompleteSquare}
      \sum_{i,j=1}^N\fun{}{\varphi_i}\Delta_{12}\fun{}{\tilphi_j}=
      \left(\sum_{i=1}^N\fun{}{\varphi_i}\right)\Delta_{12}\left(\sum_{j=1}^N\fun{}{\tilphi_j}\right)\,.
    \eeq
    With the aid of the
    change of variables to ''center of mass'' $\varphi$ defined as
    \begin{equation}
      \varphi \equiv \frac{\varphi_1+\ldots+\varphi_N}{N}
      \label{eq:intro_center}
    \end{equation}
    and relative
    coordinates $\varphi_i-\varphi_{i+1}$ for $\left\{\varphi_n\right\}_{n=1}^N$ (together with a similar change for
     $\left\{\tilphi_n\right\}_{n=1}^N$) we arrive at the conclusion \cite{Vasilev98}
    \begin{equation}
      \label{eq:intro_SquareTrick}
      \exp\left(\fun{}{\varphi}\Delta_{12}\fun{}{\tilphi}\right)\prod_{n=1}^NF_n(\varphi,\tilphi)
      =\exp\left(\sum_{i,j=1}^N\fun{}{\varphi_i}\Delta_{12}\fun{}{\tilphi_j}\right)\prod_{n=1}^NF_n(\varphi_n,\tilphi_n)
      \biggl\vert_{\varphi_i=\varphi\atop\tilphi_i=\tilphi}\,.
    \end{equation}
    This relation may be used to separate generation of lines attached to the single vertex described by
    any individual functional from spanning
    of lines between different vertices. Extract the diagonal terms in the reduction operator on the 
    right side of (\ref{eq:intro_SquareTrick}) to obtain
    \begin{equation}
      \exp\left(\sum_{i\ne j=1}^N\fun{}{\varphi_i}\Delta_{12}\fun{}{\tilphi_j}\right)\prod_{n=1}^N
      \exp\left(\fun{}{\varphi_n}\Delta_{12}\fun{}{\tilphi_n}\right)F_n(\varphi_n,\tilphi_n)\,.
    \end{equation}
    An effective vertex functional -- {\em the normal form}\index{normal form} of the interaction functional
    (or {\em reduced vertex} functional) -- defined as
    \beq
      \label{eq:intro_NormalF}
      F'(\varphi,\tilphi)\equiv \exp\left(\fun{}{\varphi}\Delta_{12}\fun{}{\tilphi}\right)F(\varphi,\tilphi)
    \eeq
    appears and we see that
    \beq
      \label{eq:intro_DiagVertex}
      \exp\left(\fun{}{\varphi}\Delta_{12}\fun{}{\tilphi}\right)\prod_{n=1}^NF_n(\varphi,\tilphi)=
      \exp\left(\fun{}{\varphi}\Delta'_{12}\fun{}{\tilphi}\right)\prod_{n=1}^NF'_n(\varphi,\tilphi)\,,
    \eeq
    where the reduction operator $\exp\left(\fun{}{\varphi}\Delta'_{12}\fun{}{\tilphi}\right)$ generates lines
    between different functionals (vertices) $F'_n$ only. An alert reader might notice a correspondence with the normal-product form
    of interaction operators in quantum field theory.
    
    It should be emphasized that this procedure is a rearrangement of the perturbation expansion.
    However, if the vertex functional is local, i.e. a one-fold integral of
    a function of fields and their derivatives at a single space-time point, then the statement 
    (\ref{eq:intro_DiagVertex}) is tantamount to
    saying that the diagonal value of the propagator $\Delta'_{12}$ is equal to zero: $\Delta'_{12}(x,x)=0$.
    
    It follows from here, in particular, that the generating function (\ref{eq:intro_GenFunTree5}) may be 
    expressed in the simple form
    \beq
      \label{eq:intro_GenFunTree7}
      \G(A)=\exp\left(\fun{}{\varphi}\Delta'_{12}\fun{}{\tilphi}\right)
      \exp\left[\tilphi U(\varphi)+\tilphi f +A\varphi\right]\bigr\vert_0
    \eeq
    completely independent of the variable $\Delta_{12}(0)$.

    {\subsection{Langevin equation} \label{subsec:langevin}}

    Hydrodynamic kinetic equations are written as mean-field equations for averages of
    quantities, which intrinsically are random processes to some extent. 
    To take fluctuations around the averages into account, a straightforward
    way to proceed is to introduce a source of randomness directly in the
    mean-field equation. Then the quantities solved from the mean-field
    equations become {\em stochastic processes} depending on coordinate variables,
    i.e. {\em stochastic fields}. This approach has been widely used in the description of dynamic
    critical phenomena (see, e.g., \cite{Folk06} for a fairly recent review) as well as in the analysis of fluctuations
    in reaction kinetics and transport phenomena \cite{Tauber05,Vasiliev,Antonov06,Tauber2014}.
    
    \subsubsection{Models A and B of critical dynamics.} Although we intend to concentrate on the field theory of reactions and
    transport phenomena, let us remind the simplest paradigmatic models of critical dynamics.
    
    In the Landau theory of phase
    transitions the dynamics of the order parameter $\varphi$ near
    equilibrium are described time-dependent
    Ginzburg-Landau (TDGL) equation. In the theory of critical phenomena the
    effect fluctuations is crucial. It is customary to take fluctuations into account
    with the use of the Langevin approach, in which the external field term(s) of
    the TDGL equations are random quantities. For instance, in case of wave-number independent
    kinetic coefficient (i.e in case of non-conserved order parameter) the corresponding Langevin equation
    \beq
      \label{eq:intro_TDGL-SDE}
      \partial_t \varphi
      =-\,\gamma \fun{F[\varphi]}{\varphi}+f
      =-\gamma\left[
      -2g\nabla^2 \varphi + 2a(T-T_c) \varphi + 4B\varphi^3
      \right]+\gamma h\,.
    \eeq
    gives rise to the model A of critical dynamics. Here, the random noise field $f=\gamma h$ is
    assumed to have Gaussian distribution law with zero mean and the white-in-time-noise correlation function
    \beq
      \label{eq:intro_correlator-A}
      \langle
      f(t,{\mx})f(t',{\mx}')\rangle=\delta(t-t')D({\mx}-{\mx}')\,,
    \eeq
    where the spatial correlation function is chosen from the condition that the long-time asymptotic state
    is described by the Gibbs ensemble with the functional $F[\varphi]$, i.e. the probability density function
    in this limit $\rho[\varphi]\to \exp\{-F[\varphi]/T\}/{\rm Tr} \,\exp\{-F[\varphi]/T\}$. This means that
    in case of model A the position-dependent part of the correlation function (\ref{eq:intro_correlator-A}) is
    \beq
      \label{eq:intro_D-A}
      D_A({\mx}-{\mx}')=2\gamma T\delta({\mx}-{\mx}')\,.
    \eeq
    In model B the order parameter is a conserved quantity and the kinetic coefficient vanishes at zero 
    wave number: $\gamma \to -\gamma\nabla^2$. Correspondingly the Langevin equation is 
    \beq
      \label{eq:intro_TDGL-SDE-B}
      \partial_t \varphi
      =\gamma\nabla^2\,\left[
      -2g\nabla^2 \varphi + 2a(T-T_c) \varphi + 4B\varphi^3
      \right]+\gamma h
    \eeq
    and the correlation function contains the Laplace operator as well:
    \beq
      \label{eq:intro_D-B}
      D_B({\mx}-{\mx}')=-2\gamma T\nabla^2\delta({\mx}-{\mx}')\,.
    \eeq

    {\subsubsection{Diffusion-limited reactions.} \label{subsubsec:diff_limit}}
    In reaction kinetics and population dynamics
    the simplest kinetic description of the  dynamics of the average particle numbers is
    given by the {\em rate equation}.
    The rate equation is a deterministic differential equation for average particle numbers in
    a homogeneous system, therefore
    it does not take into account boundary conditions,
    spatial inhomogeneities and randomness in the individual reaction
    events. Spatial dependence is often accounted for by a diffusion term, which
    gives rise to models of {\em diffusion-limited reactions} (DLR).
    
    As a simple example, consider the
    coagulation reaction $A+A\to A$. The diffusion-limited rate equation for the
    concentration $\varphi$ of the compound $A$ is
    \begin{equation}
       \partial_t\varphi
       =D\nabla^2\varphi-k\varphi^2\,,
       \label{eq:intro_react}
    \end{equation}          
    where $k$ is the {\em rate constant}\,.
    
    The most straightforward way to take into account various effects of
    randomness is to add a random source and sink term to the rate equation:
    \be
      \label{eq:intro_AAtoAnaive}
      \partial_t\varphi
      =D\nabla^2\varphi-k\varphi^2+f\,.
      \label{eq:intro_react2}
    \ee
    This is a nonlinear Langevin equation for the field $\varphi$. Physically, in the case of concentration
    $\varphi\ge 0$.
    
    There is an important difference between the reaction models and the critical dynamics:
    in the latter, deviations of the fluctuating order parameter from the (usually zero)
    mean may physically be of any sign (or direction). In particular, deviations from the equilibrium
    value are always allowed.
    In the reaction there is often an absorbing steady state, which does not permit fluctuations
    therefrom: once the system arrives at the absorbing state, it stays there forever.
    In particular, if the empty state is an absorbing state of the reaction, the random source
    should be multiplied by a factor vanishing in the limit $\varphi\to 0$ to prevent the system returning
    from the absorbing state by the noise. The simplest choice
    yields
    \begin{equation}
      \partial_t\varphi
      =D\nabla^2\varphi-k\varphi^2+f\varphi
      \label{eq:intro_react3}
    \end{equation}
    instead of (\ref{eq:intro_AAtoAnaive}). This is an equation with a {\em multiplicative noise}.\\
    
    {\subsubsection{Multiplicative noise in diffusion-advection models} \label{subsubsec:multinoise}}
    The issue of multiplicative noise practically does not arise in the field theory of critical phenomena.
    As seen from the previous
    example, it appears as consequence of boundary conditions in reaction problems. In transport
    equations with advection by random field
    it is customary to introduce the coupling of, say, a scalar quantity $\varphi$ to the advecting field $\mv$ through the 
    substantial (or covariant) time derivative
    \begin{equation}
      \partial_t \varphi
      +(\mv\cdot\boldnabla)\varphi\,,
      \label{eq:intro_convective}
    \end{equation}
    which, in case of random advection field $\mv$, gives rise to a multiplicative noise. A stochastic 
    differential equation (SDE) with
    multiplicative noise is not well-defined mathematically. Therefore, below we shall analyze 
    construction of the field theory on the basis
    of a Langevin equation (SDE) in detail to sort out ambiguities appearing within this process.\\

    {\subsubsection{Functional representation.} \label{subsubsec:functional_repre}}
    In a more generic setup \cite{Honkonen11}, standard models of fluctuations in critical dynamics and
    stochastic reaction and transport equations
    are based on nonlinear Langevin equations
    \be
      \label{eq:intro_Langevin}
      \partial_t \varphi
      =V(\varphi)+f\,,
    \ee
    For the random source (''force'') a Gaussian white-in-time
    distribution is assumed:
    \be
      \label{eq:intro_WhiteNoiseCorr}
      \langle f(t,{\mx})f(t',{\mx}')\rangle=
      \delta(t-t')D({\mx},{\mx}')\,,\quad \langle f(t,{\mx})\rangle=0 \,.
    \ee
    The static correlation function $D({\mx},{\mx}')$ is determined through the connection
    to the static equilibrium (fluctuation-dissipation theorem) in case of near-equilibrium fluctuations.
    In case of reaction and transport models far from equilibrium physical properties of a steady state
    are used to establish properties of the random source.
    
    The Langevin equation with white-in-time noise $f$ is mathematically
    inconsistent, because the time integral of the noise $\int f \dRM t$ is
    a Wiener process which is not differentiable anywhere as a function of time.
    The Langevin equation (\ref{eq:intro_Langevin}) is written mathematically correctly in the integral form
    \be
      \label{eq:intro_Langevin-Wiener}
      \dRM\varphi=V(\varphi)\dRM t+\dRM W\,,
    \ee
    where $\dRM W$ is an increment of the Wiener process (see, e.g. \cite{Gardiner}). 
    
    To describe solution of the stochastic problem in terms of
    standard perturbation theory it is convenient to use
    a set of
    correlation functions consisting of a $\delta$ sequence in time,
    i.e.
    \be
      \label{eq:intro_deltaCorrAdditive}
      \langle f(t,{\mx})f(t',{\mx}')\rangle=\overline{D}(t,{\mx};t',{\mx}')
      \to \delta(t-t')D({\mx},{\mx}')
    \ee
    and passing to the
    white-noise limit at a later stage. The point here is that the noise
    may be regarded as a smooth function of time and the differential form
    of the stochastic differential equation (SDE) (\ref{eq:intro_Langevin}) may be used
    literally.
    From the mathematical point of
    view, this treatment gives rise to the solution of the stochastic
    differential equation (\ref{eq:intro_Langevin}) in the Stratonovich sense \cite{Wong65,Sussmann78}.
    Physically, this is often the most natural way to approach the white-noise
    case.
    
    Averaging of the generating functional of solutions of the Langevin equation (\ref{eq:intro_Langevin})
    over the Gaussian noise determined by the correlation function (\ref{eq:intro_correlator-A}) is readily carried out
    with the use of the functional representation (\ref{eq:intro_GenFunTree5}). The result of calculation
    of the Gaussian integral is
    (new notation for the generating functional of solutions of the stochastic differential equation is not introduced)
    \begin{equation}
      \label{eq:intro_GenFunNoise1}
      \G(A)=\exp\left(\fun{}{\varphi}\Delta_{12}\fun{}{\tilphi}\right)      
      \exp\left[\tilphi V(\varphi)-\Delta_{12}(0)U'(\varphi)+{1\over 2}\,\tilphi \overline{D}
      \tilphi +A\varphi\right]\biggr\vert_0\,.
    \end{equation}
    The functional integral accompanying relation (\ref{eq:intro_GenFunNoise1}) obtained with the use of (\ref{eq:intro_GenFunTree4}) 
    \beq
      \label{eq:intro_GenFunNoise2}
      \G(A)=\int\!{\cal D}\varphi\,\int\!{\cal D}\tilde{\varphi} \,
      \exp\left\{-\Delta_{12}(0)U'(\varphi)+{1\over 2}\,\tilphi \overline{D} \tilphi
      +\tilde{\varphi}\left[-\partial_t\varphi+V(\varphi)\right]+A\varphi\right\}\,
    \eeq
    is formally convergent as an iterated integral, in which the auxiliary field
    $\tilphi$ -- which should be considered having {\em imaginary}  values -- is integrated first \cite{Janssen92}.
    It should be noted that with the ''coloured-noise'' correlation function $\overline{D}$ (inter)action functionals in these
    representations of the generating function are not time local.
    
    Functional derivatives of the generating functional averaged over noise do not factorize any more, of course, and
    give correlation functions of the random fields $\varphi$:
    \beq
      \label{eq:intro_GenFun2}
      \G(A)=\sum\limits_{n=0}^\infty\!{1\over n!}\int\!\dRM x_1\cdots\int\!\dRM 
      x_nG_n(x_1,\ldots,x_n)A(x_1)\cdots A(x_n)
    \eeq
    where
    \beq
      \label{eq:intro_CorrFunPhi}
      G_n(x_1,\ldots,x_n)=\langle\varphi(x_1)\cdots\varphi(x_n)\rangle_f=
      {\delta^n \G(A)\over \delta A(x_1)\cdots \delta A(x_n)}\biggr\vert_{A=0}\,.
    \eeq
    With the use of the normal form of the interaction functional the absence of $\Delta_{12}(0)$ in the perturbation
    expansion is made explicit
    \begin{equation}
      \label{eq:intro_GenFunNoise1N}
      \G(A)=\exp\left(\fun{}{\varphi}\Delta'_{12}\fun{}{\tilphi}\right)
      \exp\left[\tilphi V(\varphi)+{1\over 2}\,\tilphi \overline{D} \tilphi +A\varphi\right]\biggr\vert_0\,
    \end{equation}
    The corresponding functional integral is also simplified
    \beq
      \label{eq:intro_GenFunNoise2N}
      \G(A)=\int\!{\cal D}'\varphi\,\int\!{\cal D}'\tilde{\varphi} \,
      \exp\left\{{1\over 2}\,\tilphi \overline{D} \tilphi+\tilde{\varphi}
      \left[-\partial_t\varphi+V(\varphi)\right]+A\varphi\right\}\,.
    \eeq
    The prime in the measure denotes the additional rule $\Delta_{12}(0)=0$ (which is tantamount to 
    amending the definition of the temporal step function as $\theta(0)=0$). Since appearances of 
    integrands in (\ref{eq:intro_GenFunNoise2}) and
    (\ref{eq:intro_GenFunNoise2N}) are different, it is obviously quite essential to bear the differences in the measure in mind,
    if the functional integral is calculated by any other means than perturbation theory.

    The $\delta$ sequence of correlation functions $\overline{D}$ appears in all
    representations (\ref{eq:intro_GenFunNoise1}), (\ref{eq:intro_GenFunNoise2}) and (\ref{eq:intro_GenFunNoise1N}) 
     only in the
    quadratic form with the auxiliary field: $\tilphi\overline{D}\tilphi$ and passing to the white-noise limit 
    in this expression does not cause any problems either in the S-matrix functional or in the
    functional integral. Therefore,
    functional representations for the solution of the Langevin equation 
    (\ref{eq:intro_Langevin}), (\ref{eq:intro_WhiteNoiseCorr}) are obtained
    from (\ref{eq:intro_GenFunNoise1}), (\ref{eq:intro_GenFunNoise2}) and (\ref{eq:intro_GenFunNoise1N}) 
     simply by the
    replacement 
    \begin{equation}
      \label{eq:intro_simpleWhiteLimit}
      \int\!\dRM x\int\!\dRM x'
      \tilphi(x)\overline{D}(x;x')\tilphi(x')      
      \to      
      \int\!\dRM t\int\!\dRM^d\mx\int\!\dRM^d\mx'
      \tilphi(t,\mx){D}({\mx},{\mx}')\tilphi(t,\mx')\,.     
    \end{equation}
    Explicit time-dependence is written here to emphasize that in the white-noise limit the action
    functional is time local.

    Let us construct an iterative solution of the simplest multiplicative
    SDE  of the {\em multiplicative linear white-noise} process
    \be
      \label{eq:intro_LangevinMultiLinear}
      \partial_t\varphi
      =-{K}\varphi +f\varphi\,,
    \ee
    where ${K}$ is a time-independent operator acting on the field $\varphi$ (e.g. ${ K}=
    -\nabla^2 +a$, $a>0$) and $f$ a Gaussian random field with zero mean and the ''coloured''
    correlation function $\overline{D}$ (\ref{eq:intro_deltaCorrAdditive}).
    
    The iterative solution of the SDE (\ref{eq:intro_LangevinMultiLinear}) may expressed as the series
    \be
      \label{eq:intro_solution}
      \varphi=u{\chi}+ \Delta_{12} f u{\chi} +  \Delta_{12} f  \Delta_{12} fu{\chi}+
      \ldots
    \ee
    where ${\chi}$ is the initial condition of the solution
    \[
      u {\chi}=\int\! \dRM^d{\mx}'u(t,{\mx}-{\mx}'){\chi}({\mx}')
    \]
    of the homogeneous equation
    \be
      \label{eq:intro_HomEq}
      \left(
      \partial_t
      +{ K}\right)u{\chi}=0\,.
    \ee
    Here, $u(t,{\mx})$ is the singular solution of the homogeneous equation , i.e. a solution with 
    the initial condition $u(0,{\mx})=\delta({\mx})$, whereas
    $ \Delta_{12}$ in (\ref{eq:intro_solution}) is the (retarded)
    Green function of the same equation, i.e.
    \be
      \left(
      \partial_t + { K}\right) \Delta_{12}(t,{\mx},t',{\mx}') =
      \delta_+(t-t')\delta({\mx}-{\mx}')\,.\nonumber
    \ee
    In (\ref{eq:intro_solution}) a shorthand notation has been used in which all time and space convolution integrals
    in the nonlinear terms are implied.

    The solution (\ref{eq:intro_solution}) may be conveniently expressed graphically
    as a sum of chains of identically
    oriented lines corresponding to retarded propagators $\Delta_{12}$. To this end, it is customary to
    start from the equation with the additive source field $\delta(t)\chi(\mx)$
    \be
      \label{eq:intro_LangevinMultiLinearDelta}
      \partial_t\varphi(t)
      =-{K}\varphi(t) +f(t)\varphi(t)+\delta(t)\chi\,,
    \ee
    instead of the original SDE (\ref{eq:intro_LangevinMultiLinear}).
    
    The graphical representation
    of the solution of (\ref{eq:intro_LangevinMultiLinearDelta}) is of the form
    \be
      \label{eq:intro_GraphSolution}
      \varphi=
      \raisebox{-0.5ex}{
      \includegraphics[height=0.35truecm]{\PICS 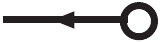}
      }+
      \raisebox{-0.7ex}{
      \includegraphics[height=1.1truecm]{\PICS 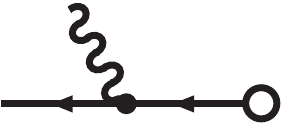}
      } +
      \raisebox{-0.7ex}{
      \includegraphics[height=1.1truecm]{\PICS 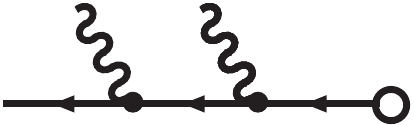}
      }+ \ldots
    \ee
    where the circle stands for the initial condition term $\delta(t){\chi}$,
    the wavy line corresponds to
    the random field $f$ and the full dot represents the vertex factor (equal to one here) brought about by the last term
    of the Langevin equation (\ref{eq:intro_LangevinMultiLinear}). From the iterative construction
    it is seen that to each vertex (crossing of lines and the open circle) of any
    graph a separate time variable (and position vector variable in case of fields) is prescribed and integrated over.
    Solution (\ref{eq:intro_GraphSolution})
    is the {\em tree-graph} representation of the solution of the stochastic differential equation (\ref{eq:intro_LangevinMultiLinearDelta}).

    The perturbative solution of the SDE (\ref{eq:intro_LangevinMultiLinear})
    is given by Wick's theorem for the Gaussian distribution of $f$, which graphically amounts to
    replacing any pair of $f$ by the correlation function $\overline{D}$ depicted by an unoriented line in all
    possible ways and -- in case of zero mean -- discarding all graphs with an odd number of wavy lines.
    For instance, the graphical expression (\ref{eq:intro_GraphSolution}) gives rise to the
    representation
    \be
      \label{eq:intro_GraphSolutionAve}
      \langle\varphi\rangle=
      \raisebox{-0.5ex}{
      \includegraphics[height=0.35truecm]{\PICS intro_emptydot.pdf}
      }+
      \raisebox{-0.5ex}{
      \includegraphics[height=0.525truecm]{\PICS 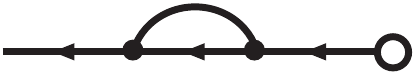}
      }+\ \ldots
    \ee
    Since the free-field differential operator in (\ref{eq:intro_LangevinMultiLinear}) is of first order in time,
    the physically interesting Green function is the retarded propagator. In the limit of white-in-time correlations
    this leads to an enormous truncation of the averaged iterative series (\ref{eq:intro_solution}), because it brings about
    temporal $\delta$ functions contracting the ends of chains of the retarded propagators. Any graph
    containing at least one such closed loop
    of at least two propagators vanishes. Only those terms, in which the correlation function is multiplied by
    a single retarded propagator do not vanish automatically.

    For instance, the one-loop graph in (\ref{eq:intro_GraphSolutionAve}) in case of white-in-time noise
    gives rise to an ambiguity, which is directly related
    to that in the interpretation of the stochastic differential equation (\ref{eq:intro_LangevinMultiLinear}).
    A straightforward substitution of the white-noise correlation function in this graph gives rise to the expression
    \begin{align}
      \label{eq:intro_1multiEx}
      \raisebox{-0.5ex}{
      \includegraphics[height=0.525truecm]{\PICS intro_oneloopdot.pdf}
      }
      & =
      \int\!\dRM t_1\int\!\dRM^d{\mx}_1\int\!\dRM^d {\mx}_2\int\!\dRM^d {\mx}_3\, \Delta_{12}(t,{\mx},t_1,{\mx}_1)\nonumber\\
      &\times
      \Delta_{12}(t_1,{\mx}_1,t_1{\mx}_2)D({\mx}_1,{\mx}_2) 
      \Delta_{12}(t_1,{\mx}_2,0,{\mx}_3){\chi}({\mx}_3)\,,
    \end{align}
    where the value of the propagator at coinciding
    time arguments $ \Delta_{12}(t_1,{\mx}_1,t_1{\mx}_2)=\theta(0)\delta({\mx}_1-{\mx}_2)$ is ambiguous.

    With the use of the $\delta$-sequence of correlation functions (\ref{eq:intro_deltaCorrAdditive}) this ambiguity
    is readily resolved and gives rise to the expression
    \begin{align}
      \label{eq:intro_1multiExStrato}
      \raisebox{-0.5ex}{
      \includegraphics[height=0.525truecm]{\PICS intro_oneloopdot.pdf}
      }
      & = {1\over 2}
      \int\!\dRM t_1\int\!d{\mx}_1\int\!\dRM{\mx}_3\, \Delta_{12}(t,{\mx},t_1,{\mx}_1)
      D({\mx}_1,{\mx}_1) \nonumber\\
      &\times \Delta_{12}(t_1,{\mx}_1,0,{\mx}_3) {\chi}({\mx}_3)\,,
    \end{align}
    with the coefficient ${1\over 2}$ in front of the spatial $\delta$ function.
    As noted before, this
    procedure corresponds to the interpretation of the SDE (\ref{eq:intro_LangevinMultiLinear}) in the Stratonovich
    sense. Formally, this result may be obtained by amending the definition of the propagator
    according to the rule 
    \begin{equation}
      \Delta_{12}(t,{\mx},t,{\mx}')={1\over 2}\delta({\mx}-{\mx}').
      \label{eq:intro_rule}
    \end{equation}   
    Within this choice the graphical
    expression in (\ref{eq:intro_1multiExStrato}) appears excessive, because it hints to twice the number of integrations
    than actually is carried out.
    
    However, the white-noise limit may graphically be depicted
    as replacement of the one-loop graph with the noise correlation function
    by a new vertex factor with the coefficient ${1\over 2}D({\mx},{\mx})\equiv{1\over 2}D(0)$:
    \beq
      \label{eq:intro_VertexShrink}
      \raisebox{-0.5ex}{
      \includegraphics[height=0.525truecm]{\PICS intro_oneloopdot.pdf}
      }\to
      \raisebox{-0.5ex}{
      \includegraphics[height=0.8truecm]{\PICS 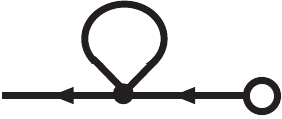}
      }\,.
    \eeq
    
    The iterative solution of (\ref{eq:intro_LangevinMultiLinear}) with the eventual
    $\delta$-function limit in the correlation function of the noise may be obtained also
    as the solution of a slightly different SDE, but with the convention $ \Delta_{12}(0)=0$ for the equal-time value
    of the propagator.
    In this case, passing to the white-noise limit is carried out solely with the use of the
    properties of the propagator and, therefore, the contribution of the graph (\ref{eq:intro_1multiEx}) and the like vanishes.
    To restore this contribution in the solution, a term corresponding to the vertex factor of the
    effective vertex brought about by shrinking to point of the one-loop graph in (\ref{eq:intro_VertexShrink})
    must be added to the SDE. From inspection of (\ref{eq:intro_VertexShrink})
    we see that in this approach the SDE generating the solution constructed above is of the form
    \be
      \label{eq:intro_LangevinMultiLinear2}
      \partial_t\varphi
      =-{K}\varphi+ {1\over 2}D(0)\varphi +f\varphi
    \ee
    and coincides with the SDE in the Ito interpretation for the present linear multiplicative noise problem.
    It should be emphasized that the rules of construction of the solution are different in (\ref{eq:intro_LangevinMultiLinear})
    and (\ref{eq:intro_LangevinMultiLinear2}), but the solution is the same. If the two sets 
    of rules are applied to the same equation,
    different solutions are, of course, produced. It is interesting to note that the choice of interpretation of SDE
    in the iterative solution seems to boil down to the choice of the value of the propagator at coinciding times.
    
    For fixed function $f$ the 
    generating function of solutions of (\ref{eq:intro_LangevinMultiLinear}) may be expressed
    in the form of an S-matrix functional
    \begin{equation}
      \label{eq:intro_GenFunMultiLinear1}
      \G(A)=\exp\left(\fun{}{\varphi}\Delta_{12}\fun{}{\tilphi}\right)
      \exp\left[\tilphi f\varphi-\Delta_{12}(0)f+\tilphi \tilA +A\varphi\right]\Bigr\vert_0\,.
    \end{equation}
    Calculating expected value with respect to random $f$ we obtain (no new notation for $\G(A)$ is introduced,
    field-independent determinant is included in the functional measure)
    \begin{equation}
      \label{eq:intro_GenFunMultiLinear2}
      \G(A)=\exp\left(\fun{}{\varphi}\Delta_{12}\fun{}{\tilphi}\right)
      \exp\left\{{1\over 2}\,\left[\tilphi \varphi-\Delta_{12}(0)\right]
      \overline{D}\left[\tilphi \varphi-\Delta_{12}(0)\right]+\tilphi \tilA +A\varphi\right\}\Bigr
      \vert_0\,.
    \end{equation}
    With the coloured-noise correlation function the interaction functional here is not time local. The
    graphical analysis has revealed that
    technical problems arise from graphs in which the propagator $\Delta_{12}$ is spanned between 
    fields $\varphi$ and $\tilphi$ 
    belonging to the same compound vertex functional 
    ${1\over 2}\,\left[\tilphi \varphi-\Delta_{12}(0)\right]\overline{D}\left[\tilphi \varphi-\Delta_{12}(0)\right]$. 
    This situation is conveniently analyzed with the use of the normal form of the interaction functional, in which
    the effect of these ''diagonal'' terms is expressed explicitly and once for all in the normal form of the interaction
    functional. Calculation yields \cite{Honkonen11}
    \begin{equation}
      \exp\left(\fun{}{\varphi}\Delta_{12}\fun{}{\tilphi}\right)
      {1\over 2}\left[\tilphi \varphi-\Delta_{12}(0)\right]\overline{D}\left[\tilphi \varphi-\Delta_{12}(0)\right]
      ={1\over 2}\,\tilphi \varphi\overline{D}\tilphi \varphi+ \tilphi\Delta_{12}\overline{D}\varphi+{1\over 2}\Delta_{21}
      \overline{D}\Delta_{21}\,.
    \end{equation}
    Here, the white-noise limit in the first term of right side yields the time-local 
    contribution ${1\over 2}\,\tilphi \varphi{D}\tilphi \varphi$
    in the fashion (\ref{eq:intro_simpleWhiteLimit}), whereas the limit of the second term is familiar from the analysis of graphs
    \begin{align}
      \label{eq:intro_multiWhiteLimit}
     \int\! \dRM x\int\!\dRM x'
    \tilphi(x)\Delta_{12}(x,x')
      \overline{D}(x,x')\varphi(x')
      &\to
      {1\over 2}\int\!\dRM t\int\!\dRM^d \mx\, 
      \tilphi(t,\mx){D}({\mx},{\mx})\varphi(t,\mx)\nonumber\\
      & \equiv  {1\over 2}D(0)\tilphi\varphi\,.
    \end{align}
    The explicit expression for the third term ($\Delta_{21}=\Delta_{12}^\top$) contains 
    a closed loop of two retarded propagators and therefore vanishes:
    \begin{equation}
      \label{eq:intro_multiWhiteLimit2}
      \int\!\dRM t\int\!\dRM t'\int\!\dRM^d\mx\int\!\dRM^d\mx'
      \Delta_{12}(t,{\mx};t',{\mx}')\overline{D}(t,{\mx};t',{\mx}')\Delta_{12}(t',{\mx}';t,{\mx})
      =0\,.
    \end{equation}
    Therefore, the white-noise limit of the generating functional (\ref{eq:intro_GenFunMultiLinear2}) is
    \begin{equation}
      \label{eq:intro_GenFunMultiLinear3}
      \G(A)=\exp\left(\fun{}{\varphi}\Delta'_{12}\fun{}{\tilphi}\right)
      \exp\left[{1\over 2}\tilphi\varphi D\tilphi\varphi+{1\over 2}D(0)\tilphi\varphi+\tilphi 
      \tilA +A\varphi\right]\biggr\vert_0\,.
    \end{equation}
    In the corresponding functional integral
    \beq
      \label{eq:intro_GenFunIntMultiLinear}
      \G(A)=\int\!{\cal D}'\varphi\,\int\!{\cal D}'\tilde{\varphi} \,
      \exp\left\{{1\over 2}\tilphi\varphi D\tilphi\varphi+\tilde{\varphi}
      \left[-\partial_t\varphi-K\varphi+{1\over 2}D(0)\varphi\right]+A\varphi\right\}
    \eeq
    the linear in $\tilphi$ term of the dynamic action
    \begin{equation}
      \S'[\varphi,\tilphi]=
      {1\over 2}\tilphi\varphi D\tilphi\varphi+\tilde{\varphi}\left[-\partial_t\varphi-K\varphi+{1\over 2}D(0)\varphi\right]
    \end{equation}
    apart from the deterministic part of the right side of the Langevin equation 
    (\ref{eq:intro_LangevinMultiLinear}) contains an additional
    term corresponding to the Langevin equation in the Ito interpretation (\ref{eq:intro_LangevinMultiLinear2}). It 
    should be borne in mind that
    the generating function has (\ref{eq:intro_GenFunIntMultiLinear}) has been constructed to yield the 
    solution of the SDE in the Stratonovich
    interpretation. However, this solution is most conveniently expressed with the use of the
    normal form of the dynamic action, whose form
    and rules of calculation remind of the SDE in Ito interpretation.

    Consider the Langevin equation with the multiplicative noise of generic form
    \be
      \label{eq:intro_Langevin_multi}
      \partial_t\varphi
      =V(\varphi)+fB(\varphi):=-K\varphi+U(\varphi)+fB(\varphi)\,.
    \ee
    Here, $B(\varphi)$ is a functional of $\varphi$ and
    $U(\varphi)$ is a nonlinear functional
    of $\varphi$. Both functionals are time-local, i.e. depend only on the current times instant of the
    SDE. Again the starting point of the construction of the generating function of solutions is
    the $\delta$ sequence of ''coloured-noise'' correlation functions (\ref{eq:intro_deltaCorrAdditive}).
    
    Generating function of solutions may be expressed in the functional-differential form
    in the same fashion as preceding cases of additive noise and linear multiplicative noise
    with the result
    problem
    \begin{align}
      \label{eq:intro_GenFunMulti1}
      \G(A) &= \exp\left(\fundoo{}{\varphi}\Delta_{12}\fundoo{}{\tilphi}\right)\int {\cal D}f 
      \exp\left(-{1\over 2}f\overline{D}^{-1}f\right) \nonumber \\
      &\times
      \exp\left\{\tilphi\left[ U(\varphi)+fB(\varphi)\right]-\Delta_{12}(0)\left[U'(\varphi)+fB'(\varphi)\right]
      +A\varphi\right\}\biggr\vert_0
      \nonumber \\
      &= \exp\left(\fundoo{}{\varphi}\Delta_{12}\fundoo{}{\tilphi}\right)
      \exp\biggl\{\tilphi U(\varphi)-\Delta_{12}(0)U'(\varphi)
      {1\over 2}\left[\tilphi B-\Delta_{12}(0)B'\right]\overline{D} \nonumber\\
      &\times \left[\tilphi B-\Delta_{12}(0)B'\right] +A\varphi\biggr\}\biggr\vert_0\,.
    \end{align}
    Here, the term corresponding to closed loop of two propagators $\Delta_{12}$ (cf. (\ref{eq:intro_multiWhiteLimit2}))
    has already been omitted.
    
    With the aid of the normal form the diagonal terms of the reduction operator are singled
    out to give rise to representation
    \begin{align}
      \label{eq:intro_GenFunMulti1N}
      \G(A)
      & = \exp\left(\fundoo{}{\varphi}\Delta'_{12}\fundoo{}{\tilphi}\right)
      \exp\biggl\{\tilphi U(\varphi)+{1\over 2}\tilphi B\overline{D}
      \tilphi B +
      \Delta_{12}(1,2)B(2,\varphi)\nonumber \\
      & \times \overline{D}(2,3)\fun{B(3,\varphi)}{\varphi(1)}\tilphi(3)
       +A\varphi\biggr\}\biggr\vert_0\,.
    \end{align}
    Here, contractions of variables are no more obvious and they have been expressed explicitly
    with the use of condensed notation, e.g. $\varphi(1)\equiv\varphi(t_1,\mx_1)$.    
    
    The white-noise limit is obtained by the same argument as in the case of linear multiplicative noise and
    yields the generating function
    in the relatively simple form \cite{Honkonen11}
    \begin{equation}
      \label{eq:intro_GenFunMulti1NWhite}
      \G(A)
      = \exp\left(\fundoo{}{\varphi}\Delta'_{12}\fundoo{}{\tilphi}\right)
      \exp\left[\tilphi U+ \frac{\tilphi B{D}
      \tilphi B}{2}+
      \frac{B(2,\varphi){D}(2,1)}{2}\fun{B(1,\varphi)}{\varphi(2)}\tilphi(1)
      +A\varphi\right]\Biggr\vert_0.
      \nonumber
    \end{equation}
    Here, the dynamic action is already time local
    \beq
    \label{eq:intro_ActionMultiN}
      S'[\varphi,\tilde{\varphi}]={1\over 2}\tilphi B{D}
      \tilphi B+\tilde{\varphi}\left[-
      \partial_t\varphi
      +V+{1\over 2}BDB'\right]
    \eeq
    and the absence of diagonal loops is thus tantamount to definition $\Delta_{12}(0)=0$.
    We see that the linear in $\tilphi$ term corresponds to the deterministic part of 
    the SDE in the Ito form also in the generic case.
    
    Let us remind that the dynamic action (\ref{eq:intro_ActionMultiN}) corresponds to
    the Stratonovich interpretation of
    the SDE (\ref{eq:intro_Langevin_multi}). Henceforth, we shall refer to the additional term -- on top of those
    brought about by the deterministic part of the SDE -- in the dynamic action as the white-noise contraction term.
    Due to the straightforward connection between 
    the white-noise and coloured-noise cases we
    shall stick to the Stratonovich interpretation in the following, if not stated otherwise. However, for 
    completeness of representation we will describe the dynamic action corresponding to the Ito interpretation
    of the stochastic problem posed by the SDE (\ref{eq:intro_Langevin}) in
    the next section.

    \subsection{Fokker-Planck equation}
    
    Account of fluctuations in kinetic problems with the use of Langevin equations is straightforward, because 
    the starting point is the
    hydrodynamic kinetic equation. The solution is given in terms of a generating functional for correlation and response functions
    (Green functions). The disadvantage of this approach lies in its mathematical ambiguity in the case of multiplicative noise. 
    Therefore, it is instructive to infer the generating functional from the mathematically well-defined setup
    of the stochastic problem with the use of evolution equations for probability density functions of
    the relevant random quantities.
    In case of continuous stochastic processes the usual starting point is the Fokker-Planck equation
    and in case of jump processes the
    master equation -- both consequences of the fundamental Chapman-Kolmogorov equation of Markov 
    processes, which is the most important
    class of stochastic processes from the point of view of fluctuation kinetics.
    
    The basic tool to describe stochastic processes are the joint probability density functions (PDF)
    $p\left(\varphi_1,t_1;\varphi_2,t_2;\ldots ;\varphi_n,t_n\right)$, which give the probability to observe
    the values $\varphi_1$, $\varphi_2$,\ldots $\varphi_n$ of the random variable 
    (field) at time instances $t_1$, $t_2$, \ldots, $t_n$
    in a suitable volume element of the space of values of $\{\varphi_i\}_{i=1}^n$.
    For simplicity, only zero-dimensional fields (i.e. functions of time only) will be discussed in these notes. Spatial dependence
    of fields may be taken into account in most cases simply by replacing partial derivatives by functional
    derivatives. Some attention has to be paid to contractions of arguments, though.
    
    The most important stochastic processes in physics are 
    Markov processes, in which the joint probability density function is completely expressed
    in terms of the conditional PDF $p\left(\varphi,t\vert\varphi_0,t_0\right)$
    and the single-event PDF $p\left(\varphi,t\right)$:
    \begin{equation}
     p\left(\varphi_1,t_1;\ldots ;\varphi_n,t_n\right)\\
      =p\left(\varphi_1,t_1\vert\varphi_2,t_2\right)p\left(\varphi_2,t_2\vert\varphi_3,t_3\right)\cdots
      p\left(\varphi_{n-1},t_{n-1}\vert\varphi_n,t_n\right)p\left(\varphi_n,t_n\right)\nonumber
    \end{equation}
    provided $t_1\ge t_2\ge t_3\ge \ldots \ge t_{n-1}\ge t_n$. Both these functions can be found as
    solutions of the Fokker-Planck equation.
    
    Consider evolution generated by the generic Fokker-Planck equation
    \begin{equation}
      \label{eq:intro_FokkerEvo}
      \doo{}{t}p\left(\varphi,t\vert\varphi_0,t_0\right)
      =-\doo{}{\varphi}\left\{\left[-K\varphi+U(\varphi)\right]p\left(\varphi,t\vert\varphi_0,t_0\right)\right\}\\
      +{1\over
      2}\doo{^2}{\varphi^2}\left[b(\varphi)Db(\varphi)p\left(\varphi,t\vert\varphi_0,t_0\right)\right]\,.
    \end{equation}
    The conditional probability density is the fundamental solution of this equation, i.e.
    \[
      p\left(\varphi,t_0\vert\varphi_0,t_0\right)=\delta(\varphi-\varphi_0)
    \]
    and is also properly normalized
    \[
      \int\!\dRM\varphi\,p\left(\varphi,t\vert\varphi_0,t_0\right)=1\,.
    \]
    The probability density function $p(\varphi,t)$ is the solution of the FPE (\ref{eq:intro_FokkerEvo}) as well, but with the
    initial condition $p(\varphi,t_0)=p_0(\varphi)$.
    
    The Fokker-Planck equation (\ref{eq:intro_FokkerEvo}) is similar to the Schr\"odinger
    equation of quantum mechanics. Solution of both
    equations is a function related to probability density function of dynamic variables. This 
    analogy allows to use the
   	the quantum-mechanical operator approach to construct a (formal) solution of the 
   	Fokker-Planck equation \cite{Leschke77,Honkonen12}.
   	
    Therefore, let us introduce -- in analogy with quantum mechanics -- the state vector $\ket{p_t}$ according to
    representation
    \[
      p(\varphi,t):=\braket{\varphi}{p_t}\,.
    \]
    To construct the evolution operator for the state vector, introduce the (nearly) quantum-mechanical
    momentum and coordinate operators by
    \[
      \hat{\pi}f(\varphi)=-\doo{}{\varphi}f(\varphi)\,,\quad \hat{\varphi}f(\varphi)=\varphi f(\varphi)\,.
    \]
    The non-trivial commutation relation is
    \[
      \left[\hat{\varphi},\hat{\pi}\right]=1\,.
    \]
    In these terms, the FPE for the PDF gives rise to the evolution equation for the state vector in the form
    \[
      \doo{}{t}\ket{p_t}=\hat{L}\ket{p_t}\,,
    \]
    where the Liouville operator, according to (\ref{eq:intro_FokkerEvo}), assumes the form
    \be
      \label{eq:intro_LioFPEIto}
      \Lio=
      \opi\left[-K\ophi+U(\ophi)\right]
      +{1\over
      2}\opi^2b(\ophi)Db(\ophi)\,.
    \ee
    In this notation, the conditional PDF may be expressed as the matrix element
    \be
      \label{eq:intro_CondPDFMatrEl}
      p\left(\varphi,t\vert\varphi_0,t_0\right)=\matrel{\varphi}{\eRM^{\Lio(t-t_0)}}{\varphi_0}\,.
    \ee
    Introduce time-dependent operators $\hat{\varphi}(t)$ in the Heisenberg picture
    of imaginary time quantum mechanics (i.e. Euclidean quantum mechanics):
    \be
      \label{eq:intro_LioOpe}
      \hat{\varphi}_H(t)= \eRM^{-\hat{L}t}\,\hat{\varphi}\,\eRM^{\hat{L}t}\,,
    \ee
    and define the time-ordered product (chronological product, $T$ product) of time-dependent operators
    \beq
      \label{eq:intro_TprodLio}
      T\left[\ophi_H(t_1)\cdots\ophi_H(t_n)\right]=\sum\limits_{P(1,\ldots,n)}
      P\left[\theta\left(t_1\ldots t_n\right)\ophi_H(t_1)\cdots\ophi_H(t_n)\right]\,,
    \eeq
    where
    \[
      \theta\left(t_1\ldots t_n\right)\equiv\theta\left(t_1-t_2\right)\theta\left(t_2-t_3\right)\cdots
      \theta\left(t_{n-1}-t_n\right)\,.
    \]
    In definition (\ref{eq:intro_TprodLio}) the sum is taken over all permutations of the labels of
    the time arguments $\left\{t_i\right\}_{i=1}^n$ and the operators in each term are put in the
    order of growing time arguments from the right to the left. Thus, under the $T$-product sign operators
    commute. It should be noted that the definition of the time-ordered product
    should be amended for coinciding time arguments. We shall return to this point later and exclude this case
    for the time being.
    
    Introduce then the $n$-point Green function of the Heisenberg fields (\ref{eq:intro_LioOpe})
    \be
      \label{eq:intro_GnLio}
      G_n(t_1,t_2,\ldots t_n):=\Tr\left\{\orho\, T\left[\ophi_H(t_1)\ophi_H(t_2)\cdots\ophi_H(t_n)\right]\right\}
    \ee
    with the density operator
    \be
      \label{eq:intro_IniDensityOpe}
      \orho:=\int\!\dRM\varphi\ket{p_0}\bra{\varphi}\,.
    \ee
    Choosing, for definiteness, the time sequence $t_1> t_2> t_3> \ldots > t_{n-1}> t_n>t_0$ it is readily seen by
    direct substitution of relations (\ref{eq:intro_CondPDFMatrEl}), (\ref{eq:intro_LioOpe}) and
    (\ref{eq:intro_IniDensityOpe}) in (\ref{eq:intro_GnLio})
    with the aid of the
    normalization conditions of the PDF
    and insertions of the resolution of the identity
    $
      \int\!d\varphi \,\ket{\varphi}\bra{\varphi}=1
    $
    that
    \be
      \label{eq:intro_MomentGF}
      \int\!\dRM\varphi_1\ldots\int\!\dRM\varphi_n\,\varphi_1\cdots\varphi_n
      p\left(\varphi_1,t_1;\ldots ;\varphi_n,t_n\right)=
      G_n(t_1,\ldots t_n)\,,
    \ee
    i.e. the Green function (\ref{eq:intro_GnLio}) is equal to the moment function (\ref{eq:intro_MomentGF}). This 
    relation connects the
    operator approach to evaluation of moments of the random process: moments and correlation functions
    of the random field $\varphi$ may be calculated as derivatives of the {\em generating function} of
    Green functions
    \begin{align}
      \label{eq:intro_G(A)Def}
      \G(A) & = \sum\limits_{n=0}^\infty\! {1\over n!}\,\int\!\dRM t_1\cdots \int\!\dRM t_n\, G_n(t_1,\ldots,t_n)A(t_1)
      \cdots A(t_n)\nonumber \\
      &=
      \Tr\left\{\orho\, T\left[\exp\left(\int\! dt\,A(t)\ophi_H(t)\right)\right]\right\}\,.
    \end{align}
    
    To construct perturbation expansion for the generating function it is convenient to introduce field operators
    in the interaction (or Dirac) representation, in which the time evolution is generated by the free Liouville
    equation
    \be
      \label{eq:intro_LioOpeDirac}
      \hat{\varphi}(t)= \eRM^{-\hat{L}_0t}\,\hat{\varphi}\, \eRM^{\hat{L}_0t}\,,\qquad \Lio_0=-\opi K\ophi\,.
    \ee
    There is some freedom in the choice of the free Liouvillean $L_0$. The expression adopted here on the
    basis of the expression (\ref{eq:intro_LioFPEIto}) is quite convenient for practical calculations.
    
    In the interaction representation the {\em evolution operator} $\oU$ is of the form
    \beq
      \label{eq:intro_EvoOpFP}
      \oU(t,t')= \eRM^{-\Lio_0t}\, \eRM^{\Lio(t-t')}\,\eRM^{\Lio t'}\,.
    \eeq
    It is a fundamental theorem of quantum field theory that the time-ordered product of Heisenberg 
    operators (\ref{eq:intro_LioOpe})
    may be expressed in terms of Dirac operators (\ref{eq:intro_LioOpeDirac}) as
    \begin{equation}
      \label{eq:intro_HtoD}
      T\exp\left(\int_{t_i}^{t_f}\!{\ophi_H(t) A(t)\dRM t}\right)
      =
      \oU(0,t_f)
      T \exp\left[\int_{t_i}^{t_f}
      \!\!\!\Lio_I(t)\,\dRM t+\int_{t_i}^{t_f}\ophi(t)A(t)\,\dRM t
      \right]
      \oU(t_i,0)\,.
    \end{equation}
    
    In quantum field theory calculation of various expectation values of operators in the interaction
    representation is most conveniently carried out, when
    the operators are expressed in the form of a {\em normal product}. In case of quantum-mechanical momentum and position
    operators in the normal product of an operator monomial all momentum operators stand to the left of all position
    operators. When an arbitrary operator is cast in the form of a normal product, the result is a linear combination
    of normal products of operator monomials, whose coefficients are determined by the commutation rules of operators,
    e.g.
    \[
      \ophi\opi^2=\opi^2\ophi+2\opi\,.
    \]
    Wick's theorems state results of representation of operator products in normal form. For the time-ordered
    exponential of generating function (\ref{eq:intro_HtoD}) it follows from Wick's theorems \cite{Vasilev98} that 
    \begin{align}
      \label{eq:intro_WickG(A)}
    T \exp\left\{\int_{t_i}^{t_f}
     \!\!\! [\Lio_I(t) + \ophi(t)A(t)]\, \dRM t
     \right\}    
       & = N\biggl\{ \exp\left(\fundoo{}{\varphi}\Delta_{12}
      \fundoo{}{\pi}\right)\exp\biggl[\int_{t_i}^{t_f}
      \!\!\!L_I(\varphi(t),\pi(t))\,\dRM t\nonumber\\
      &+\int_{t_i}^{t_f}\varphi(t)A(t)\,\dRM t
      \biggl]\biggr\}\biggr\vert_{\varphi=\ophi\atop \pi=\opi}\,.
    \end{align}
    On the right side $L_I(\varphi,\pi)$ is the interaction functional, in which the operators
    $\ophi$ and $\opi$ in $\Lio_I=\Lio-\Lio_0$ are replaced by functions $\varphi$ and $\pi$.
    Then
    the {\em reduction operator}
    \[
      {\cal P}= \exp\left(\fundoo{}{\varphi}\Delta_{12}
      \fundoo{}{\pi}\right)
    \]
    replaces pairs of functions $\varphi$ and $\pi$ by the {\em propagator}
    $$
      \Delta_{12}(t,t')=T\left[\ophi(t)\opi(t')\right]-N\left[\ophi(t)\opi(t')\right]
    $$
    after which all remaining functions $\varphi$ and $\pi$ are replaced by corresponding operators
    in the normal order, because by definition
    \[
       N\left\{P\left[\opi(t_1)\cdots \opi(t_m)\ophi(t'_1)\cdots \ophi(t'_n)\right]\right\}=\opi(t_1)\cdots \opi(t_m)\ophi(t'_1)\cdots \ophi(t'_n)\,,
    \]
    where $P$ stands for any ordering of operators $\opi$ and $\ophi$.
    
    Thus, we arrive at the representation
    \begin{align}
      \label{eq:intro_G(A)FP2}
      \G(A)
      & =\Tr\Biggl(\oU(t_i,0)\,\orho\,\oU(0,t_f)
      \, N\Biggl\{
      \exp\left(\fun{}{\varphi}\Delta_{12}\fun{}{\pi}\right)     
      \exp\biggl[
      \int_{t_i}^{t_f}\!\!\!L_I(\varphi(t),\pi(t))\,dt \nonumber \\
      & +\int_{t_i}^{t_f}\varphi(t)A(t)\,dt
      \biggl]
      \Biggr\}
      \Biggr\vert_{\varphi=\ophi\atop \pi=\opi}
      \Biggr)\,.
    \end{align}
    for the generating function.  At this point the ambiguity in the definition of the $T$ product should be fixed,
    because it affects the explicit form of the interaction functional. The point is that contributions to
    perturbation expansion produced by the action of the reduction operator on the interaction functional $L_I(\varphi,\pi)$
    give rise to a {\em reduced} interaction functional (or to the {\em normal form} of the interaction functional)
    \[
      L'_I(\varphi,\pi)=\exp\left(\fundoo{}{\varphi}\Delta_{12}
      \fundoo{}{\pi}\right)L_I(\varphi,\pi)\,,
    \]
    which, in general, is different from $L_I(\varphi,\pi)$. The form of the interaction functional
    generated by the Fokker-Planck is preserved, if the $T$ product at coinciding time arguments is 
    defined as the $N$ product, in which case, in particular, $\Delta_{12}(t,t)=0$ and the normal form
    of the time-local interaction functional coincides with the original interaction functional.

    In (\ref{eq:intro_G(A)FP2}) under the sign of normal product $N$ there stands 
    some {\em operator functional} $F[\ophi,\opi]$, i.e.
    a Taylor series of operators $\ophi$ and $\opi$. To calculate the expectation value (\ref{eq:intro_G(A)FP2})
    of an arbitrary operator functional, it is sufficient to calculate the expectation value of a linear exponential,
    because
    \begin{align}
      \label{eq:intro_ExpectLinExp}
      \Tr\left\{ N\left[F(\ophi,\opi )\right]\oU(t_i,0)\,\orho\,\oU(0,t_f)\right\}
      & =F\left(\fundoo{}{A},\fundoo{}{B}\right)
      \Tr\biggl\{\oU(t_i,0)\,\orho\,\oU(0,t_f) \nonumber\\
      &\times N\,\exp\left({\ophi A+\opi B}\right)\biggl\}\Bigl\vert_{A=B=0}\,.
    \end{align}
    
    Writing out explicitly the trace in the coordinate representation, the density operator
    $
      \orho=\int\!\dRM \varphi\ket{p_0}\bra{\varphi}
    $, 
    and an identity resolution
    $
      \int\!\dRM \zeta \,\ket{\zeta}\bra{\zeta}=1
    $
    to produce matrix elements of evolution operators we arrive at the representation
    \begin{align}
      \label{eq:intro_CoordAveLinExp}
      \Tr &\left\{\oU(t_i,0)\,\orho\,\oU(0,t_f) N\,\,\exp\left({\ophi A+\opi B }\right)\right\}\nonumber\\
      &=\int \dRM\chi\,\bra{\chi}\oU(t_i,0)\int \dRM\varphi\,\ket{p_0}\bra{\varphi}\oU(0,t_f) 
      N\,\exp\left({\ophi A+\opi B }\right)\ket{\chi}\nonumber \\
      &=\int \dRM\chi\int \dRM\varphi\int \dRM\zeta\,\bra{\chi}\oU(t_i,0)\ket{p_0}\bra{\varphi}\oU(0,t_f)\ket{\zeta}\bra{\zeta}
      N\,\exp\left({\ophi A+\opi B }\right)\ket{\chi}\,.
    \end{align}
    With the choice of the free Liouvillean in the form $\Lio_0= -\opi K\ophi$
    the time-dependent operators are
    \be
      \label{eq:intro_DiracCoordMome}
      \opi(t)=\opi\, \eRM^{Kt}\,,\quad \ophi(t)=\ophi \, \eRM^{-Kt}\,.
    \ee
    Calculate first the matrix element of the exponential in the coordinate basis:
    \be
      \label{eq:intro_NA}
      \bra{\zeta}N\,\exp\left( {\ophi A+ \opi B}\right)\ket{\chi}=\bra{\zeta}
      \exp\left({\opi \tilde{B}}\right) \exp\left({\ophi \tilde{A}}\right)\ket{\chi}\,.
    \ee
    where $\tilde{A}=\int\!e^{-Kt}A(t)\, \dRM t$ and $\tilde{B}=\int\!e^{Kt}B(t) \, \dRM t$.
    In the coordinate basis the operator $\ophi$ is the multiplication operator, whereas the exponential of the momentum
    operator is the translation operator. Therefore (\ref{eq:intro_NA}) immediately yields
    \be
      \label{eq:intro_NA2}
      \bra{\zeta}N\,\exp\left( {\ophi A+ \opi B}\right)\ket{\chi}=\delta\left(\zeta -\chi -\tilde{B}\,\right)\exp\left({\chi \tilde{A}}\right)\,.
    \ee
    The simple and sufficient choice for the Cauchy problem is to put $t_i=0$, after which the matrix element
    $\bra{\chi}\oU(t_i,0)\ket{p_0}$ in (\ref{eq:intro_CoordAveLinExp}) reduces to $\braket{\chi}{p_0}=p_0(\chi)$.
    The matrix element of the other evolution operator in (\ref{eq:intro_CoordAveLinExp}) yields the unity due
    to probability conservation:
    \[
       \int \dRM \varphi \bra{\varphi}\oU(0,t_f)\ket{\zeta}=1\,.
    \]
    Thus, the
    average of the linear exponential (\ref{eq:intro_CoordAveLinExp}) is reduced to 
    \begin{align}
      \label{eq:intro_CoordAveLinExp2}
      \Tr & \left\{\oU(t_i,0)\,\orho\,\oU(0,t_f) N\,\exp\left({\ophi A+\opi B }\right)\right\} \nonumber\\
      &=\int d\chi\int d\zeta\,\delta\left(\zeta -\chi -\tilde{B}\,\right)\exp\left({\chi \tilde{A}}\right)p_0(\chi)
      =\int d\chi\,\exp\left({\chi \tilde{A}}\right)p_0(\chi)\,.
    \end{align}
    Therefore, the expectation value of the normal product of any operator function(al) $F\left[\ophi(t),\opi(t)\right]$ is
    (here, $t_i=0$ at the outset)\cite{Honkonen13}
    \begin{align}
      \label{eq:intro_CoordAve}
      &\int \dRM\chi\int \dRM\varphi\,\braket{\chi}{p_0}\bra{\varphi}\,\oU(0,t_f) N\left[F(\ophi(t),\opi(t) )\right]\ket{\chi}
       \nonumber\\
       &= F\left[\fundoo{}{A(t)},\fundoo{}{B(t)}\right]      
      \int\!\dRM\chi\int\!\dRM\varphi\,\braket{\chi}{p_0}\bra{\varphi}\,\oU(0,t_f) N\,\exp\left({\ophi A+ \opi B}
      \right)\ket{\chi}\bigr\vert_{A=B=0} \nonumber \\
      &=\int\!\dRM\chi F\left[\fundoo{}{A(t)},0\right]p_0(\chi)\exp\left({\chi \tilde{A}}\right)\bigr\vert_{A=0}
      =\int\!\dRM\chi F\left[\varphi_\chi(t),0\right]p_0(\chi)\,,
    \end{align}
    where $\varphi_\chi$ is the solution of the free-field equation
    \[
      \left(
      \partial_t + K\right)\varphi_\chi(t)=0
    \]
    with the initial condition $\varphi_\chi(0)=\chi$. It should be borne in mind that functional 
    variables here may be fields, in which case
    $K$ usually is a second-order differential operator in space.
    
    Introducing the Liouville operator $\Lio=\Lio_0+\Lio_I$ and corresponding functionals 
    explicitly, we obtain the generating function of Green functions of
    the Cauchy problem of the Fokker-Planck equation in the form
    \begin{align}
      \label{eq:intro_GenericGFokker}
      \G(A) & = \int\! \dRM \chi\int\! \dRM \varphi\,p_0(\chi)\matrel{\varphi}{ T\exp\left( {\ophi_H A}\right)}{\chi}
      \nonumber\\
      & = \int\! \dRM\chi\,p_0(\chi)\Biggl\{\exp\left(\fundoo{}{\varphi}\Delta
      \fundoo{}{\pi}\right)
      \exp\left[\int_{0}^{t_f}
      [ L_I(\varphi(t),\pi(t)) + \varphi(t)A(t)]\dRM t\right]
      \Biggr\}_{\varphi=\varphi_\chi\atop \pi=0\ \ }
      .
    \end{align}
    It should be noted that
    $\ophi_H$ here refers to the position operator of the Fokker-Planck equation, not to the
    generic field operator and that the propagator
    \beq
      \label{eq:intro_propagatorFPE}
      \Delta(t,t')=T\left[\ophi(t)\opi(t')\right]-N\left[\ophi(t)\opi(t')\right]=\theta(t-t') \eRM^{-K(t-t')}\,,
    \eeq
    is also written for fields of this particular problem.
    
    Representation in the form of a functional integral may be introduced by the
    trick using a formal functional integral for the reduction operator:
    \beq
      \label{eq:intro_FunIntReduction}
      \exp\left(\,\fundoo{}{\varphi}\Delta_{12}
      \fundoo{}{\pi}\right)= \int\!{\cal D}\phi\,\int\!{\cal D}\tilde{\phi}\,\exp\left[\Tilphi\left(-
      \partial_t-K\right)\phi+
      \phi\fundoo{}{\varphi}+ \Tilphi\fundoo{}{\pi}\right]\,.
    \eeq
    Here, the determinant of the differential operator has been included in the integration measure and for integration
    variables a notation has been introduced which will be used in the analysis of the Langevin equation.
    
    Functional shift operators on the right side of (\ref{eq:intro_FunIntReduction}) generate argument
    changes 
    $\varphi\to \varphi+\phi$ and $\pi\to \pi+\Tilphi$ in the interaction functional in
    (\ref{eq:intro_GenericGFokker}), after which the
    substitutions $\varphi=\varphi_\chi$ and $\pi=0$ may be carried out leading to expression
    \beq
    \label{eq:intro_GFokkerFunInt}
      \G(A)=\int\! \dRM \chi\,p_0(\chi)\int\!{\cal D}\phi\,\int\!{\cal D}\tilde{\phi}\,
      \exp\left\{ \S[\phi+\varphi_\chi,\Tilphi]+
      \int_{0}^{t_f}\negthickspace\left[\phi(t)+\varphi_\chi(t)\right]A(t) \dRM t \right\}\,,
    \eeq
    where the {\em De Dominicis-Janssen dynamic action} $\S[\phi,\Tilphi]$ is
    \beq
      \label{eq:intro_DDJ-FP}
      \S[\varphi,\tilphi]={1\over
      2}\tilphi^2 B(\varphi)DB(\varphi)+\tilphi\left[-
      \partial_t
      \varphi-K\varphi+U(\varphi)\right]\,.
    \eeq
    The form of the dynamic action is unambiguously determined by the appearance of the Fokker-Planck
    equation. It has to be kept in mind that functional representations 
    (\ref{eq:intro_GenericGFokker}) and (\ref{eq:intro_GFokkerFunInt})
    have been derived with the use of the convention that the $T$ product at coinciding time arguments is
    defined as the $N$ product. In particular, this means that the propagator vanishes at
    coinciding time arguments $\Delta_{12}(t,t)=0$.
    
    The explicit shift of the field $\phi+\varphi_\chi$ in the functional integral (\ref{eq:intro_GFokkerFunInt})
    may be absorbed in the definition of the space of integration. The point is that formal convergence of
    the Gaussian integral (\ref{eq:intro_FunIntReduction}) requires that solutions of the homogeneous equation
    $(\partial_t+K)\varphi_0=0$ must be excluded from the space of integration of the Gaussian integral.
    The space of integration may be constructed as $\phi(t)=\int_0^t\!\dRM t'\Delta(t-t')\eta(t')$, where
    $\eta(t)$ is a function vanishing at all borders of the spacetime. Then, by definition, $\phi(0)=0$
    and the combination $\varphi(t)=\phi(t)+\varphi_\chi(t)$ is the solution of the inhomogeneous
    equation $(\partial_t+K)\varphi(t)=\eta(t)$ satisfying the initial condition $\varphi(0)=\chi$.
    Defining the space of integration as a manifold consisting of such functions we may write
    \beq
      \label{eq:intro_GFokkerFunInt2}
      \G(A)=\int\! \dRM\chi\,p_0(\chi)\int\!{\cal D}_\chi\varphi\,\int\!{\cal D}\tilde{\varphi}\,
      \exp\left\{\S[\varphi,\tilphi]+
      \int_{0}^{t_f}\negthickspace\varphi(t)A(t) \dRM t\right\}\,,
    \eeq
    where the subscript $\chi$ in the measure reminds on the dependence on the initial condition for $\varphi$.
    
    The dynamic action (\ref{eq:intro_DDJ-FP}) obtained for the functional representation of the solution of the Fokker-Planck
    equation (\ref{eq:intro_FokkerEvo}) is different from the dynamic action (\ref{eq:intro_ActionMultiN}) obtained for the Langevin
    equation (\ref{eq:intro_Langevin}). This difference is connected with different interpretations of the stochastic
    differential equation (\ref{eq:intro_Langevin}). Dynamic action (\ref{eq:intro_ActionMultiN})
    corresponds to the Stratonovich interpretation
    of the SDE, whereas dynamic action (\ref{eq:intro_DDJ-FP}) corresponds to the Ito interpretation of the same SDE.
    Which interpretation is chosen as the basis of the field theory is largely a matter of 
    model construction. In those cases, however, 
    when the SDE is considered as white-noise limit of a sequence of coloured noises, the Stratonovich interpretation must be used.

    {\subsection{Master equation} \label{subsec:master}}
    
    Markov processes described in terms of the Fokker-Planck equation
    have continuous sample paths. Not all interesting stochastic
    processes belong to this category. A wide class of such processes
    describe changes in occupation numbers (e.g. individuals of some
    population, molecules in chemical reaction) which cannot be
    naturally described by continuous paths. This kind of processes are
    described by {\em master equations} -- a special case of
    (differential) Kolmogorov equations.
    
    The generic form of a master equation
    written for the conditional probability density\\ $p\left(\varphi,t\vert\varphi_0,t_0\right)$ of
    a Markov process is
    \begin{equation}
      \label{eq:intro_GenMaster}
      \doo{}{t}p\left(\varphi,t\vert\varphi_0,t_0\right)
      =\int\!\dRM \chi\left[W(\varphi\vert \chi,t)p\left(\chi,t\vert\varphi_0,t_0\right)-
      W(\chi\vert\varphi ,t)p\left(\varphi,t\vert\varphi_0,t_0\right)\right]\,,
    \end{equation}
    where
    $W(\varphi\vert \chi,t)$
    is the {\em transition probability} per unit time, whose formal definition from
    the differential Kolmogorov equation is (for all $\varepsilon>0$)
    \[
      W(\varphi\vert \chi,t)=\lim_{\Delta t\to 0}{p\left(\varphi,t+\Delta t\vert\chi,t\right)\over \Delta t}\,,
    \]
    uniformly in $\varphi$, $\chi$ and $t$ for all $\vert \varphi-\chi\vert \ge\varepsilon$.
    
    We shall be using the master equation for discrete variables, in this case the discontinuous
    character of the paths of the {\em jump processes} described by the master equation is especially transparent:
    \be
      \label{eq:intro_NMaster}
      \doo{}{t}p\left(n,t\vert m,t_0\right)
      =\sum_l\left[W(n\vert l,t)p\left(l,t\vert m,t_0\right)-
      W(l\vert n ,t)p\left(n,t\vert m,t_0\right)\right]\,.
    \ee
    This set of master equations shall also be cast in the form of an evolution equation of the type of
    Schr\"odinger equation, but the representation is not as straightforward as in the case of the Fokker-Planck equation.
    
    
    In the occupation-number basis and in the stationary field operators
    there is no explicit Planck constant. The ideas of the Fock-space
    representation and creation/annihilation operators may therefore as
    well be used in classical problems with a variable number of
    particles or some other entities.
    
    Various processes in biology and chemistry
    are described in terms of variable numbers of some agents or
    representatives of species ("particle numbers" or "occupation
    numbers"). In many cases changes in the particle number are caused
    by interactions between colliding particles, i.e. {\em reactions}.
    
    Description of a particular reaction may usually be given by the
    {\em reaction equation}. For instance, for a two species process
    with the {\em rate constants} $k_+$ and $k_-$ the reaction equation
    is
    \begin{equation}
      \label{eq:intro_reaction} s_AA+s_BB\ {{\scriptstyle k_{+}\atop
      \scriptstyle\rightleftharpoons}\atop \scriptstyle k_{-}}\
      r_AA+r_BB\,,
    \end{equation}
    where $A$ and $B$ denote the two species and $s_A$, $s_B$, $r_A$ and
    $r_B$ are coefficients (usually integers) describing in which
    proportions the agents react.
    
    The simplest kinetic description of the  dynamics of the average
    particle numbers is given by the {\em rate equation}. For the binary
    reaction (\ref{eq:intro_reaction}), e.g.
    \begin{equation}
      \dee{c_A}{t}=k_+(r_A-s_A)c_A^{s_A}c_B^{s_B}+k_-(s_A-r_A)c_A^{r_A}c_B^{r_B}\,,
    \end{equation}
    where $c_X$ is the concentration of the species $X$.
    
    The rate equation is a deterministic differential equation for
    average particle numbers in a homogeneous system, therefore it does
    not take into account boundary conditions, spatial inhomogeneities
    and randomness in the individual reaction events.

    To keep things simple, consider a system with just one variable. A
    classic example is the {\em Verhulst} ({\em logistic}$\,$) {\em
    model}\index{logistic model}\index{Verhulst model} of population
    growth. The rate equation for the particle number $n$ (the number of
    individuals in the population) may be written as
    \begin{equation}
      \label{eq:intro_Verhulst}
      \dee{n}{t}=-\beta n+ \lambda n-\gamma n^2\,,
    \end{equation}
    where $\beta$ is the death rate, $\lambda$ the birth rate and
    $\gamma$ the damping coefficient necessary to bring about a
    saturation for the population.
    
    A complete (microscopic) description of a stochastic problem of
    random particle number is given by the master equations written
    for the probabilities $P(t,n)$ to find exactly $n$ particles in the
    system at the time instant $t$.
    Transition rates (transition probabilities per unit time) of master equations are related to
    coefficients of the rate equation, but the correspondence is not
    unambiguous: given rate equations may generate different master
    equations, therefore a verbal description of the process is often
    necessary.

    Regarding birth and death as reactions 
    \begin{equation*}
      A\ {{\scriptstyle \lambda\atop
      \scriptstyle\leftrightharpoons}\atop \scriptstyle\gamma}
      2A\,,\qquad      
      A \xrightarrow{\beta} \varnothing
    \end{equation*}
    master equations for the {\em stochastic Verhulst model} may be
    written as (this is slightly different from the original formulation by Feller \cite{Feller39})
    \begin{align}
      \label{eq:intro_VerhulstMaster} \dee{P(t,N)}{t}&=\lambda(N-1)P(t,N-1)
      -\left(\beta
      N+\gamma N^2\right)P(t,N)\,,\nonumber\\
      \dee{P(t,n)}{t}&=[\beta
      (n+1)+\gamma(n+1)^2]P(t,n+1)+\lambda(n-1)P(t,n-1)\nonumber\\
      & -\left(\beta
      n+\lambda n+ \gamma n^2\right)P(t,n)\,,\quad 0<n<N\,,\\
      \dee{P(t,0)}{t}&=(\beta+\gamma)P(t,1) \,.\nonumber
    \end{align}
    Here, boundary conditions for the particle number are made explicit
    in the equations for the probabilities of the boundary values of the
    particle number. Usually, the empty state ($n=0$) is an {\em
    absorbing state}\index{absorbing state} (once the system occurs in
    an absorbing state, it will stay there forever) and the state with
    the maximum sustainable population ($n=N$) is a {\em reflecting
    state}\index{reflecting state} (there are no available states beyond a
    reflecting state, but the system does not get stuck to that state)
    as here.
    
    
    Using master equations (\ref{eq:intro_VerhulstMaster}) a coupled set of
    evolution equations for the moments of the particle number $\langle
    n^m\rangle$ may be written. The evolution equation for the average
    particle number $\langle n\rangle$ coincides with the rate equation
    (\ref{eq:intro_Verhulst}), when all correlations are neglected, i.e. moments
    replaced by corresponding powers of the average particle number
    $\langle n^m\rangle\to\langle n\rangle^m$. This, however, is not
    always the case, but only {\em if} the transition rates vanish with
    $n$.

    The set of master equations for the probabilities $P(t,n)$ may be cast into a single
    kinetic equation by the ''second quantization'' of Doi \cite{Doi76a,Doi76b}. Recall that the conditional probabilities
    $P(t,n\vert t_0,m)$ obey the same set of equations, but with a different initial condition:
    \begin{equation}
      P(t_0,n\vert t_0,m)=\delta_{nm}. 
    \end{equation}    
    Let us first
    construct a Fock space spanned by the annihilation and
    creation operators $\ann$, $\cre$ 
    \beq
      \label{eq:intro_adefs}
      [\,\ann,\cre]=1\,,\quad [\,\ann,\ann]=[\,\cre,\cre]=0
    \eeq
    and the vacuum vector $\ket{0}$
    \beq
      \label{eq:intro_vacuum}
      \ann\ket{0}=0
    \eeq
    so that the basis vectors are
    \beq
      \label{eq:intro_basis}
      \ket{n}=\left(\cre\right)^n\ket{0}\,.
    \eeq
    From these relations it follows that
    \beq
      \label{eq:intro_adefs2}
      \ann\ket{n}=n\ket{n-1}\,,\quad
      \cre\ket{n}=\ket{n+1}\,,\quad
      \braket{n}{m}=n!\delta_{nm}\,.
    \eeq
    Note that the action of creation and annihilation operators on
    the basis vector as well as the normalization of the latter are different
    from those used in quantum mechanics.
    Connection between Fock space and occupation numbers is given by the
    occupation number operator $\on=\cre\ann$, which obeys
    \beq
      \label{eq:intro_n-operator}
      \on\ket{n}=\cre\ann\ket{n}=n\ket{n}\,.
    \eeq

    To cast the set of master equations into a single kinetic equation, define the state vector
    \begin{equation}
      \label{eq:intro_statevector} \ket{P_t}=\sum\limits_{n=0}^\infty
      P(t,n)\ket{n}\,.
    \end{equation}
    The probability of occupation number $n$ may by extracted from the state vector
    through the following scalar product
    \[
      P(t,n)={1\over n!}\,\braket{n}{P_t}\,.
    \]
    The set of master equations
    yields for the state vector (\ref{eq:intro_statevector})
    a single kinetic equation without any explicit dependence
    on the occupation number:
    \begin{equation}
      \label{eq:intro_L} \dee{\ket{P_t}}{t}=\Lio(\cre,\,\ann)\ket{P_t}\,,
    \end{equation}
    where the Liouville operator $\Lio(\cre,\,\ann)$ is constructed from
    the set of master equations according to rules:
    \begin{align*}
      nP(t,n)\ket{n}&=\cre\ann P(t,n)\ket{n}\,,\\
      {n}P(t,n){\ket{n-1}}&=
      {\ann} P(t,n){\ket{n}}\,,\\
      nP(t,n){\ket{n+1}}&= {\cre}\cre\ann P(t,n){\ket{n}}\,.
    \end{align*}
    For instance, the Liouville operator for the stochastic Verhulst
    model is
    \begin{equation}
      \label{eq:intro_VerhulstLio} \Lio(\cre,\ann)= \beta(I-\cre)\ann
      +\gamma(I-\cre)\ann\cre\ann +\lambda(\cre-I)\cre\ann\,.
    \end{equation}
    The formal solution of (\ref{eq:intro_L}) is
    \begin{equation}
      \label{eq:intro_Lsol} \ket{P_t} = \eRM^{t\Lio(\cre,\,\ann)}\ket{P_0}\,,
    \end{equation}
    where the initial state vector is determined by the initial
    condition for the probabilities:
    $$
      \ket{P_0}=\sum_{n=0}^\infty
      P(0,n)\ket{n}\,.
    $$
    In principle, the relation (\ref{eq:intro_Lsol}) may be used
    for a compact representation of the solution of the set of master
    equations, particularly if it is possible to
    calculate the matrix exponential in a closed form.

    To represent expectation
    values of occupation-number dependent quantities in the operator
    formalism use the projection vector $\bra{P}$:
    \begin{equation}
      \label{eq:intro_Pvec} \bra{P}=\sum\limits_{n=0}^\infty{1\over
      n!}\,\bra{n}=\sum\limits_{n=0}^\infty{1\over n!}\,\bra{0}\ann^n
      =\bra{0}
      \eRM^\ann \,.
    \end{equation}
    The projection vector is a left eigenvector of the creation operator
    with the eigenvalue equal to unity
    \begin{equation}
      \label{eq:intro_Peigen} \bra{P}\,\cre=\bra{P}\,.
    \end{equation}
    From here it follows that
    \begin{equation}
      \label{eq:intro_Pbasis} \braket{P}{n}=1\,,\quad \forall\  n\,.
    \end{equation}
    Basis states are eigenstates of the number operator $\on=\cre\ann$,
    therefore from relation (\ref{eq:intro_Pbasis}) it follows
    \begin{equation}
      \label{eq:intro_PQn} \bra{P}(\cre\ann)^m\ket{n}=n^m\braket{P}{n}=n^m\,.
    \end{equation}

    Equation (\ref{eq:intro_L}) gives rise to the Heisenberg evolution of
    operators
    \beq
      \label{eq:intro_HOpME}
      \cre_H(t)= \eRM^{-\hat{L}t}\cre \eRM^{\hat{L}t}\,,\qquad
      \ann_H(t)= \eRM^{-\hat{L}t}\ann \eRM^{\hat{L}t}\,.
    \eeq
    To construct interaction representation, we need Dirac operators as well:
    \beq
      \label{eq:intro_DOpME}
      \cre(t)= \eRM^{-\hat{L}_0t}\cre \eRM^{\hat{L}_0t}\,,\qquad
    \ann(t)= \eRM^{-\hat{L}_0t}\ann \eRM^{\hat{L}_0t}\,.
    \eeq
    It is convenient to choose the free Liouville operator in the form
    \beq
      \label{eq:intro_freeL}
      \Lio_0=-(\cre-I) K\ann\,,
    \eeq
    because it has the property of ''probability conservation''
    \beq
      \label{eq:intro_ProbCon0}
      \bra{P}\Lio_0=0\,.
    \eeq
    From the structure of master equations it follows that the Liouville operator
    always has the structure
    \[
      \Lio(\cre,\ann)=(\cre-I)K(\cre,\ann)
    \]
    and therefore obeys the probability conservation 
    \beq
      \label{eq:intro_ProbConL}
      \bra{P}\Lio=0\
    \eeq
    as well.

    The basic relation for calculation of expectation values follows
    from the definition in terms of probabilities, which may be cast in
    the form of a matrix element between the projection vector $\bra{P}$
    and the state vector $\ket{P_t}$:
    \begin{equation}
      \label{eq:intro_Qexp} \langle Q(n)\rangle=\sum\limits_{n=0}^\infty Q(n)P(t,n)
      =\bra{P}Q(\cre\ann)\ket{P_t}\,.
    \end{equation}
    The last equality, again without explicit occupation-number
    dependence, comes from the the relation (\ref{eq:intro_PQn}). Here and henceforth,
    it is assumed that $Q(n)$ is a polynomial function or Taylor series of $n$.
    
    Thus, with the use of the solution (\ref{eq:intro_Lsol}) the expectation value of an
    occupation-number dependent quantity $Q(n)$ may be written as
    \begin{equation}
      \label{eq:intro_QexpL}
      \langle Q(n)\rangle=\bra{P}Q(\cre\ann)\, \eRM^{t\Lio(\cre,\,\ann)}\ket{P_0}\,.
    \end{equation}
    We are dealing with non-commuting operators, therefore the question of operator
    ordering arises here as well. Define the normal product of creation and annihilation
    operators as a product, in which all creation operators stand to the left of all 
    annihilation operators. Define further the normal form of the operator $Q(\cre\ann)$ as
    \beq
      \label{eq:intro_NformQ}
      Q(\cre\ann)=N\left[Q_N(\cre,\ann)\right]\,.
    \eeq
    Then we see that the expectation value of $Q(n)$
    is equal to
    \begin{equation}
      \label{eq:intro_QNexpL}
      \langle Q(n)\rangle=\bra{P}Q_N(1,\ann)\, \eRM^{t\Lio(\cre,\,\ann)}\ket{P_0}\,.
    \end{equation}
    Introducing identity resolutions $1=\exp(\Lio t)\exp(-\Lio t)$
    between operators $\ann$ and using probability conservation to write
    \[
      \bra{P}=\bra{P}\exp(-\Lio t)
    \] 
    we obtain
    \begin{equation}
      \label{eq:intro_QNexpH}
      \langle Q(n)\rangle=\bra{P}Q_N(1,\ann_H(t))\,\ket{P_0}\,.
    \end{equation}
    Here, the right side is a linear combination of equal-time {\em Green functions}
    of the annihilation operators.

    Consider the Green function of
    creation and annihilation operators $\cre_H(t)$ and $\ann_H(t)$:
    \begin{equation}
      \label{eq:intro_GnLioMasterLinear}
      G_{m,n}(t_1,\ldots
      t_m;t_1',\ldots t_n')\\
      =\Tr\left\{\orho\,T\left[\cre_H(t_1)\cdots\cre_H(t_m)\ann_H(t'_1) \cdots\ann_H(t'_n)\right]\right\}\,,
    \end{equation}
    with the density operator
    \beq
      \label{eq:intro_IniDensityOpeMaster}
      \orho=\ket{P_0}\bra{P}\,.
    \eeq
    From definitions it follows that
    the conditional probability density function for the master equation
    may be written as (the factorial in front of the matrix element is
    due to the unusual normalization of the basis states)
    \beq
      \label{eq:intro_CondPDFMatrElMaster}
      P\left(n,t\vert n_0,t_0\right)={1\over
      n!}\,\matrel{n}{\eRM^{\Lio(t-t_0)}}{n_0}\,.
    \eeq
    Choosing, for
    definiteness, the time sequence $t_1> t_2> t_3> \ldots > t_{n-1}>
    t_n>t_0$ it is readily seen by direct substitution of relations
    (\ref{eq:intro_CondPDFMatrElMaster}) and (\ref{eq:intro_IniDensityOpeMaster}) in
    (\ref{eq:intro_GnLioMasterLinear}) with the aid of the normalization conditions of
    the PDF and insertions of the resolution of the identity $
    \displaystyle \sum_n {1\over n!}\,\ket{n}\bra{n}=1 $ that
    \begin{equation}
      \label{eq:intro_MomentGFMaster}
      \sum_{n_1}\ldots\,\sum_{n_m}\,{n_1}\cdots n_m
      P\left(n_1,t_1;\ldots ;n_m,t_m\right)
      =G_{m,m}(t_1,,\ldots t_m;t_1,\ldots t_m)\,,
    \end{equation}
    i.e. the Green function (\ref{eq:intro_GnLioMasterLinear}) is equal to
    the moment function (\ref{eq:intro_MomentGFMaster}). This relation connects
    the operator approach to evaluation of moments of the random process
    described by a master equation.
    
    Reduction to the interaction representation and Wick's theorem for the normal product
    are -- up to notation -- the same as in the case of the Fokker-Planck equation. Calculation
    of the expectation value of the linear exponential is, however, different.
    
    In case of master equation the form of the evolution operator is just the same as in the case of
    Fokker-Planck equation, but the density operator is different. Operators $\cre$ and $\ann$ specific for this
    problem shall be used in what follows.
    \begin{align}
      \label{eq:intro_DoiAveLinExp}
      \Tr &\left\{\oU(t_i,0)\,\orho\,\oU(0,t_f) N\,\,\exp\left({\ann \Bdag + \cre B }\right)\right\} \nonumber \\
      & =\sum\limits_n{1\over
      n!}\bra{n}\oU(t_i,0)\ket{P_0}\bra{P}\oU(0,t_f) N\,\exp\left({\ann
      \Bdag + \cre B  }\right)\ket{n}\nonumber \\
      & =\bra{P}\oU(0,t_f) N\,\exp\left({\ann \Bdag + \cre B  }\right)\oU(t_i,0)\ket{ {P_0}}\,.
    \end{align}
    Again, to keep things simple, choose $t_i\to 0$, which yields
    $$
      \oU(t_i,0)\ket{ {P_0}}\to \ket{ {P_0}}\,.
    $$
    Choose the free Liouville operator in the form
    \beq
      \label{eq:intro_freeLioDoi} \Lio_0= -\left(\cre-I\right) K \ann
    \eeq
    for
    which the ''conservation of probability'' holds:
    \[
      \bra{P} \eRM^{\hat{L}_0t}=\bra{P}
    \]
    because the projection vector is the left eigenstate of the creation
    operator
    \[
      \bra{P}\cre=\bra{P}\,.
    \]
    The evolution operator $\oU(0,t_f)$ is a product of operator exponentials of $\Lio_0$ and $\Lio$. The
    ''conservation of probability'' holds also for the latter by the very construction, thus
    \[
      \bra{P}\,\Lio_0=0\,,\qquad \bra{P}\,\Lio=0\,.
    \]
    Therefore, we obtain
    \begin{align}
      \label{eq:intro_eq:middle}
      \bra{P}\oU(0,t_f) N\,\exp\left({\ann \Bdag + \cre B  }\right)\ket{ {P_0}} &= 
      \bra{P} N\,\exp\left({\ann \Bdag + \cre B  }\right)\ket{ {P_0}} \nonumber\\
      &=\bra{P} \,\exp\left({\cre B  }\right)\exp\left({\ann \Bdag }\right)\ket{ {P_0}}\,.
    \end{align}
    The free Liouvillean (\ref{eq:intro_freeLioDoi}) gives rise to time-dependent
    operators
    \beq
      \label{eq:intro_DiracDoiMome}
      \cre(t)=\cre\,
      \eRM^{Kt}+\left(1-\eRM^{Kt}\right)\,,\quad \ann(t)=\ann \, \eRM^{-Kt}\,.
    \eeq
    Therefore, writing the time integrals in (\ref{eq:intro_eq:middle}) explicitly, we arrive at
    representation
    \begin{equation}
      \label{eq:intro_2nd step}
      \bra{P} \,\eRM^{\cre B } \,\eRM^{\ann \Bdag }\ket{P_0}
      =\exp\left[{\int_0^{t_f}\!\left(1-\eRM^{Kt}\right)B(t)\, \dRM t}\right]
      \bra{P}\, \eRM^{\cre \tilde{B}  } \, \eRM^{\ann \tilde{ \Bdag }}\ket{P_0}\,,
    \end{equation}
    where
    $$
      \tilde{B}=\int\!\eRM^{Kt}B(t)\dRM t\,,\qquad
      \tilde{\Bdag}=\int\! \eRM^{-Kt}\Bdag (t)\dRM t\,.
    $$
    Pull now the operator
    exponential $\exp \ann$ from $\bra{P}$ to the right by the rule
    \begin{equation}
       \label{eq:intro_transfer} \left(\exp\ann\right)\cre=(\cre + I)\exp \ann\,,\nonumber
    \end{equation}
    which boils down to the shift $\cre\to \cre+I$ in operators, through
    which the coherent-state exponential is pulled:
    \begin{equation}
      \label{eq:intro_3rd step}
      \bra{P}\,\exp\bigl({\cre \tilde{B}  }\bigr)\,\exp\bigl({\ann \tilde{
      \Bdag }}\bigr)\ket{n} =\bra{0}\,\exp\bigl({\left(\cre+I\right)
      \tilde{B}  }\bigr)\,\exp\left({\ann \tilde{
      \Bdag}}\right)\left(\cre+I\right)^n\ket{0}
    \end{equation}
    Combining (\ref{eq:intro_2nd step}) and
    (\ref{eq:intro_3rd step}) we obtain
    \begin{equation}
      \label{eq:intro_mat el final}
      \matrel{P}{\exp\left({\cre B }\right)\, \exp\left({\ann
      \Bdag}\right)}{ {P_0}}
      =P(0,n)\exp\left[{\int_0^{t_f}
      B(t)\dRM t}\right]\,\left[\int_0^{t_f}\! \eRM^{-Kt} \Bdag(t) \dRM t+1\right]^n\,.
    \end{equation}
    Thus,
    the expectation value of the linear exponential (\ref{eq:intro_DoiAveLinExp})
    is \cite{Honkonen13}
    \begin{align}
      \label{eq:intro_DoiAveLinExResult}
      \matrel{P}{\oU(0,t_f) N\,\exp\left({\ann
      \Bdag + \cre B }\right)}{ {P_0}}
      &=\sum\limits_nP(0,n)\exp\left[{\int
      B(t)dt}\right]\nonumber\\
      &\times \left[\int_0^\infty\!\eRM^{-Kt} \Bdag (t) \dRM t+1\right]^n\,.
    \end{align}
    The result for the expectation value of the normal form of an
    operator functional $F\left[\cre(t),\ann(t) \right]$ may be written
    in a compact form with the use of the identity
    \[
      x^n={n!\over 2\pi i}\,\oint_C {\eRM^{xz}\over z^{n+1}}\,\dRM z\,,
    \]
    where $C$ is a closed Jordan contour encircling the origin of the
    complex $z$ plane. We obtain
    \begin{align}
      \label{eq:intro_DoiAve}
      \matrel{P}{ N\,F\left[\cre(t),\ann(t) \right]}{P_0}
      & =
      F\left[\fundoo{}{B(t)},\fundoo{}{\Bdag(t)} \right]\matrel{P}{ N\,
      \eRM^{\ann \Bdag + \cre B}} {P_0}
      \biggr\vert_{B=\Bdag=0}\nonumber \\
      & =F\left[\fundoo{}{B(t)},\fundoo{}{\Bdag(t)} \right]\,\exp\left({\int B(t)\dRM t
       }\right) 
   \end{align}       
   \begin{align}       
       \sum\limits_n &P(0,n){n!\over 2\pi i}
       \,\oint_C {\exp\left[{z\left(\int_0^{t_f}\!e^{-Kt} \Bdag(t)dt+1 \right)}\right]\over z^{n+1}}\,\dRM z
       \biggr\vert_{B=\Bdag=0}
       \nonumber \\
       & =\,\sum\limits_nP(0,n){n!\over 2\pi i}\,\oint_C {\eRM^{z}\over
       z^{n+1}}\,F\left[1,nz \right]\,\dRM z \nonumber\\
       & ={1\over 2\pi i}\,\oint_C
       \eRM^{z}\widetilde{G}(z)\,F\left[1,nz \right]\,\dRM z\,,
    \end{align}
    where $nz=\eRM^{-Kt}z$. In (\ref{eq:intro_DoiAve}) the shorthand
    notation
    \[
      \widetilde{G}(z)=\sum\limits_n{P(0,n)n!\over z^{n+1}}
    \]
    has been used on the right side.
    
    If the initial PDF is that of the {\em Poisson distribution}, which is the assumption usually made \cite{Tauber05},
    \beq
      \label{eq:intro_Poisson}
      P(0,n)={n_0^n \eRM^{-n_0}\over n!}\,,
    \eeq
    then the expression for the expected value of linear exponential is tremendously simplified, since
     \begin{align}
       &\matrel{P}{ \oU(0,t_f) N\,\exp\biggl({\ann
      \Bdag + \cre B }\biggl)}{P_0} \nonumber \\
       &=\sum\limits_nP(0,n)\exp\biggl[{\int
       B(t)\, \dRM t}\biggl]\,\biggl[\int_0^\infty\!\eRM^{-Kt}\Bdag(t)\dRM t+1\biggl]^n
       \nonumber\\
       &=
       \exp\biggl[\int
      B(t)\dRM t+n_0\int_0^\infty\!\eRM^{-Kt} \Bdag(t)\dRM t\biggl]
       \,.
     \end{align}
    Therefore, we arrive at the representation
    \beq
      \label{eq:intro_DoiAvePoisson}
      \matrel{P}{\oU(0,t_f) N\,F(\cre,\ann)}{P_0}=F(1,\eRM^{-Kt}n_0)\,.
    \eeq

    Introducing again the Liouville operator
    $\Lio=\Lio_0+\Lio_I$ and corresponding functionals explicitly, we
    obtain the generating function of Green functions of the Cauchy
    problem of the master equation equation in the form:
     \begin{align}
       \label{eq:intro_GenericGMaster}
       \G(A) & = \bra{P}\oU(0,t_f) T\exp\biggl({\ann_H A^\dagger +\cre_HA}\biggl)\ket{P_0}
       \nonumber\\
       &={1\over 2\pi i}\,\oint_C \eRM^{z}\widetilde{G}(z)\,\Biggl[ \exp\biggl(\,\fundoo{}{a}\Delta
       \fundoo{}{a^\dagger}\biggl)
       \exp\biggl\{\,\int_{0}^{t_f}
        L_I(a(t),a^\dagger(t))\,\dRM t\nonumber\\
        &+\int_{0}^{t_f} \biggl[a(t)A^\dagger(t)+a^\dagger(t)A(t)\biggl]\,\dRM t\biggl\}
       \Biggr]_{a=nz\atop a^\dagger=1}
       \,.
     \end{align}
    In case of initial Poisson distribution the generating function is
    \begin{align}
      \label{eq:intro_PoissonGMaster}
      \G(A)
      &= \exp\left(\,\fundoo{}{a}\Delta
      \fundoo{}{a^\dagger}\right)\nonumber\\
      &\times
      \exp\left\{\,\int_{0}^{t_f}
      \negthickspace L_I(a(t),a^\dagger(t))\,\dRM t+\int_{0}^{t_f}\negthickspace\left[a(t)A^\dagger(t) 
      + a^\dagger(t)A(t)\right]
      \,\dRM t\right\}
      \Biggr\vert_{a=\eRM^{-Kt}n_0\atop a^\dagger=1\phantom{\eRM^{-Kt}n}}
      \,.
    \end{align}
    
    The functional-differential representation may again be transformed to a functional integral by
    a trick similar to that used in the Fokker-Planck case with the result
    \begin{multline}
      \label{eq:intro_GMasterFunInt}
      \G(A)=
      \int\!{\cal D}\varphi\,\int\!{\cal D}\tilde{\varphi}\,
      \exp\left\{ \S[\varphi,\varphi^\dagger]+
      \int_{0}^{t_f}\negthickspace\left[\varphi(t)A^\dagger(t)+\varphi^\dagger(t)A(t)\right]\right\}\,,
    \end{multline}
    where the {\em Peliti dynamic action} \cite{Peliti85} is
    \beq
      \label{eq:intro_PD-A}
      \S[\varphi,\varphi^\dagger]=\varphi^\dagger \left(-
      \partial_t
      \varphi-K\varphi\right)+L_I(\varphi^\dagger + 1,\varphi + \eRM^{-Kt}n_0)\,.
    \eeq
    The explicit shift of the field argument by the solution of the homogeneous free-field equation $\varphi+e^{-Kt}n_0$
    is the same as in the Fokker-Planck case. The shift $\varphi^\dagger + 1$ may be carried out
    explicitly and gives rise to the
    ''shifted'' action discussed in the literature. 
    
    For the Verhulst model the Peliti action after this procedure is
    \beq
      \label{eq:intro_Peliti-Verhulst}
      \S[\varphi,\varphi^\dagger]=\varphi^\dagger \left[-
      \partial_t \varphi+(\lambda-\beta-\gamma)\varphi\right]
      -\gamma \varphi^\dagger \varphi^2 + \lambda {\varphi^\dagger}^2\varphi-\gamma {\varphi^\dagger}^2\varphi^2\,.
    \eeq
    The structure of the dynamic action here is similar to that of the Fokker-Planck problem.
    We have seen that a very similar expression arises in the case, when the account of
    randomness is introduced by constructing a Langevin equation instead of the master equation.
    Therefore, functional representation leads also to a possibility make conclusions on this basis
    about connections between different
    approaches to stochastic problems.\\

    \subsection{Random sources and sinks in the master equation}
    \label{eq:intro_sec:sourcesink}
    
    Apart from reactions and transport of particles effect of random sources and sinks
    may sometimes be of interest, e.g., to maintain a steady state in the system.
    In most cases this is carried out by including an additive noise
    term in the Langevin equation of the stochastic process. When the 
    analysis is based on the master equation, this is not quite
    appropriate and a more elaborated approach is called for \cite{Hnatic11}.
    Unfortunately, there is no unique way to introduce
    random sources in the master equation corresponding to the random
    noise of the mean-field (Langevin) description. We use the simplest
    choice, described in detail, e.g., in \cite{Kampen}, which is
    tantamount to amending the problem by reactions $A\to X$ and $Y\to
    A$, where $X$ and $Y$ stand for particle baths of the sink and the
    source, respectively. In a homogeneous system these reactions give
    rise to the master equations
    \begin{equation}
      \dee{P(t,n)}{t}=\mu_+V\left[P(t,n-1)-P(t,n)\right]
      +\mu_-\left[(n+1)P(t,n+1)-nP(t,n)\right]+\ldots
      \label{eq:intro_MasterSource}
    \end{equation}
    where $P(t,n)$ is the probability to find $n$ particles at the
    time instant $t$ in the system. The ellipsis in (\ref{eq:intro_MasterSource})
    stands for the terms describing the annihilation reaction, diffusion
    and advection in the system. In (\ref{eq:intro_MasterSource}) $\mu_+$ and $\mu_-$ are the
    reaction constants of the creation and annihilation reactions, respectively. The transition rate
    has been chosen proportional to the particle number $n$, which seems the quite natural and
    also preserves the empty state as an absorbing state. In the transition rate for creation process
    $V$ is the volume of the (for the time being) homogeneous system and will be important in passing to
    the continuum limit of the inhomogeneous system. We recall that the master equation (\ref{eq:intro_MasterSource})
    gives rise to the reaction-rate equation
    \beq
      \label{eq:intro_RateSource}
      \dee{\langle n\rangle}{t}=\mu_+V-\mu_-\langle n\rangle+\ldots
    \eeq
    Master equations (\ref{eq:intro_MasterSource}) give rise to the following terms in
    the Liouville operator
    \beq
      \label{eq:intro_LSource}
      \Lio_g(\cre,\ann)= \mu_+V\left(\cre-I\right)+\mu_-\left(I-\cre\right)\ann\,,
    \eeq
    where $I$ is the identity operator. The corresponding contribution to the
    dynamic action is
    \begin{equation}
      \label{eq:intro_action1}
      \S_g=\int\limits_0^\infty\!\! \dRM t\,\left[\mu_+ Va^+(t)-\mu_-a^+ (t)a(t)\right]\ldots
    \end{equation}
    Only the generic time-derivative term and terms brought about by the random source model are expressed
    here explicitly, while the ellipsis stands for terms corresponding to other reactions and initial conditions.
    
    Let the transition rates $\mu_\pm$ be random functions uncorrelated in time with a probability distribution given
    in terms of the moments $\langle\mu_\pm^n\rangle=E_{\pm\,,n}$. At this point we also generalize the treatment to the
    case of a spatially inhomogeneous system and introduce a lattice subscript as the spatial argument, e.g.
    $a(t)\to a_i(t)$. In this case the volume $V$ becomes the volume element attached to the lattice site.
    For simplicity, we replace the time integral with the integral sum $\int_0^\infty\! \dRM t\to \sum_\alpha\Delta t$
    and assume that the transition rates at each time instant and lattice site $\mu_{\pm\\,\alpha,i}$
    are independent random variables. Then the average 
    over the distribution of random sources reduces to the calculation of the expectation value
    \beq
      \label{eq:intro_g-average}
      \prod_{\alpha,i}
      \langle \eRM^{\mu_{+\,,\alpha,i}V\tilde{a} _{\alpha,i}\Delta t-
      \mu_{-\,,\alpha,i}\tilde{a} _{\alpha,i}a_{\alpha,i}\Delta t}\rangle\,.
    \eeq
    For each particular time instant and lattice this yields (we assume that the moments of $\mu_\pm$ are the same for
    all $\alpha$ and $i$ and omit labels for brevity) this gives rise to the usual cumulant expansion
    \begin{align}      
      \langle \eRM^{\mu b\Delta t}\rangle & =
      1+b\Delta t E_1+{1\over 2}E_2(b\Delta t)^2+{1\over 6}E_3(b\Delta t)^3+\cdots \nonumber \\
      & = \eRM^{b\Delta t E_1+{1\over 2}\left(E_2-E_1^2\right)(b\Delta t)^2
      +{1\over 6}\left(E_3-3E_1E_2+E_1^3\right)(b\Delta t)^3+\cdots}
      \label{eq:intro_expansion-average}
    \end{align}
    Here, $b$ stands for either $V\tilde{a} $ or $-\tilde{a} a$. Thus, for instance the average over $\mu_+$ assumes the form
    \begin{align}
      \nonumber
      \prod_{\alpha,i}
      \langle \eRM^{\mu_{+\,,\alpha,i}V\tilde{a} _{\alpha,i}\Delta t}\rangle
      & =  \eRM^{\sum_\alpha\sum_i\left[\Delta t E_{+1}V\tilde{a} _{\alpha,i}
      +{1\over 2}\left(E_{+2}-E_{+1}^2\right)(V\tilde{a} _{\alpha,i}\Delta t)^2\right]} \nonumber \\
      &\times \eRM^{\sum_\alpha\sum_i\left[{1\over 6}\left(E_{+3}-3E_{+1}E_{+2}+E_{+1}^3\right)
      (V\tilde{a} _{\alpha,i}\Delta t)^3+\cdots\right]}\,.
      \label{eq:intro_+average}
    \end{align}
    In the continuum limit the function $\tilde{a} _{\alpha,i}$ is replaced
    by the field $\varphi^\dagger(t,\mx)$, whereas in the limit $V\to 0$ the expression
    $a_{\alpha,i}/V$ gives rise to the field
    $\varphi(t,\mx)$. The sum over $\alpha$ together with $\Delta t$ gives rise to the time integral and the
    sum over $i$ together with the volume element gives rise to the spatial integral $\sum_iV\to \int \dRM^d\mx$.
    In the first term of the exponential in (\ref{eq:intro_+average}) we thus obtain
    \[
      \sum_\alpha\sum_i\Delta t E_{+1}V\tilde{a} _{\alpha,i}\to E_{+1}\int \dRM t\int \dRM^d \mx\,\varphi^\dagger(t,\mx)\,.
    \]
    In the cumulants of second and higher order the continuum limit is not so obvious.
    We assume the simplest nontrivial distribution for $\mu_\pm$, in which only the variance term has
    a finite limit, when $\Delta t\to 0$ and $V\to 0$, whereas the contributions of higher-order cumulants
    vanish, for instance
    \begin{align}
      \label{eq:intro_limit variance}
      \left(E_{+2}-E_{+1}^2\right)V\Delta t \to \sigma_+\,,\quad &\Delta t\to 0\,,\ V\to 0\,, \\
      \label{eq:intro_limit cumulants}
      \left(E_{+3}-3E_{+1}E_{+2}+E_{+1}^3\right)(V\Delta t)^2\to 0\,,\quad &\Delta t\to 0\,,\ V\to 0\,.
    \end{align}
    Therefore, the contribution of the average over $\mu_+$ to the effective dynamic action assumes the form
    \beq
      \label{eq:intro_+}
      \S_+= \int\! \dRM t\int\! \dRM^d \mx\,
      \left\{E_{+1}\varphi^\dagger(t,\mx)+{1\over 2}\sigma_+\left[\varphi^\dagger(t,\mx)\right]^2\right\}\,.
    \eeq
    For the average over $\mu_-$ a similar argument yields
    \beq
      \label{eq:intro_-}
      \S_-= \int\! \dRM t\int\! \dRM^d\mx\,\left\{-E_{-1}\varphi^\dagger(t,\mx)\varphi(t,\mx)
      +{1\over 2}\sigma_-\left[\varphi^\dagger(t,\mx)\varphi(t,\mx)\right]^2\right\}\,.
    \eeq
    These contributions to the effective dynamic action may, of course, be generated by suitably chosen normal
    distributions of $\mu_\pm$.
    
    This way of introduction of random sources and sinks has the annoying feature that it does not conserve the
    number of particles in the system. For a comparison with the treatment of this problem in the Langevin approach the
    random sources and sinks should be introduced in such a way that the particle number is conserved.
    
    The simplest way to effect this is to add to the random source a term proportional to the particle number, i.e.
    use the ''reaction constant'' $\mu_+V+\mu_{1+}n$ instead of $\mu_+V$ in the master equation. The source terms on
    the right-hand side of the master equation (\ref{eq:intro_MasterSource}) in this case the assume the form
    \begin{equation}
      \nonumber
      \dee{P(t,n)}{t}=\mu_+V\left[P(t,n-1)-P(t,n)\right]
      +\mu_{1+}\left[(n-1)P(t,n-1)-nP(t,n)\right]\ldots
      \label{eq:intro_MasterSource2}
    \end{equation}
    The new part of the master equation corresponds to a branching process \cite{Kampen}.
    
    The added term gives rise to the following contribution to the Liouville operator
    \beq
      \label{eq:intro_LSource2}
      \Lio_{g_2}(\cre,\ann)= \mu_{1+}\left(\cre-I\right)\cre\ann\,.
    \eeq
    Performing the steps described above we arrive at the contribution to the dynamic action in
    the form
    \beq
      \label{eq:intro_1+}
      \S_{1+}= \int\! \dRM t\int\! \dRM^d \mx\,\left\{E_{1+1}\varphi^\dagger\left(\varphi^\dagger + 1\right)\varphi
      +{1\over 2}\sigma_{1+}{\varphi^\dagger}^2\left(\varphi^\dagger + 1\right)^2\varphi^2\right\}\,.
    \eeq
    It is easy to see now that if we exclude the plain source (i.e. put $E_{+1}=\sigma_+=0$) and choose
    $E_{1+1}=E_{-1}$, then the empty state remains absorbing and the ''mass term'' $\propto \varphi^+\varphi$
    disappears in the dynamic action and we arrive at the dynamic action of random sources and sinks
    \begin{equation}
      \nonumber
      \S_{gc}=\int\! \dRM t\int\! \dRM^d\mx\,\biggl\{E_{1+1}{\varphi^\dagger}^2\varphi+
      {1\over 2}\sigma_-\left(\varphi^\dagger\varphi\right)^2
      +{1\over 2}\sigma_{1+}{\varphi^\dagger}^2\left(\varphi^\dagger + 1\right)^2\varphi^2\biggl\}
      \label{eq:intro_gc}
    \end{equation}
    which conserves the average number of particles.\\
{\section{Renormalization group in field theory} \label{sec:RG_theory}}

\subsection{Renormalization procedure and renormalization group}

Usually, the Feynman graphs of Green functions, which are the graphical
representation of some integrals, contain divergences in the range
of large and small scales (wave vectors.) Therefore it is
necessary  to find an effective procedure to eliminate these
divergences  step by step in each order of concrete perturbation scheme.
Below we will demonstrate  renormalization  methods in the framework 
of the stochastic model of developed turbulence and related applications.

The method of renormalization group (RG) has been proposed in the
framework of the quantum field theory in the fifties of the previous century
 \cite{stu,sir,gell,bog1,bog2,bog3}. From the practical point of view RG method
 represents an effective way to determine non-trivial asymptotic
 behavior of Green functions in the range of large (ultraviolet)
or small (infrared) wave vectors (scales). The asymptotic behavior is
non-trivial if in a given order of a perturbative calculation
the divergences in a certain range of wave vectors appear (e.g.  so called large
logarithms) which compensate the smallness of coupling constant
 $g$. In such a case one needs to sum  all  terms of perturbation series.
 This summation can be carried out by means of RG approach.
Technically,  one obtains  linear partial differential
RG equations for the Green functions. The coefficient functions
(RG-functions) in the differential operator (see below) are calculated
at a given order of the perturbation scheme. However, the solution of the RG
equation represents the sum of the infinite series.
For example, if the RG-functions are calculated at the lowest
non-trivial order of the perturbation theory and after corresponding
RG-equation is solved, obtained result is a sum of leading
logarithms of all the perturbation series. Moreover if the RG-functions
will be  calculated with an improved precision the solution of the RG
equation will include corrections to the leading logarithms.

A simple criterion how to determine the true asymptotic range exist in
the framework of RG. One of the RG-functions is the $\beta$-function,
which is a coefficient at the operation  $\partial_g$ in the RG equation.
The $\beta$-function is calculated perturbatively as infinite series
of powers of the coupling constant $g$ and for relativistic models has
form: $\beta (g)=\beta_2 g^2+ \beta_3 g^3 + ...$. Non-trivial
asymptotic behavior is governed by  {\it RG fixed points}
 $g^*$, which are roots of  $\beta$-functions (solutions
of equation $\beta (g)=0$). A fixed point can be IR or
 UV stable  depending on behavior of
the $\beta$-function  in th vicinity  $g^*$.
If the coefficient  $\beta_2>0$ then $g^*$ is an IR stable fixed point. In the opposite case it is
UV stable  fixed point.
In the time when the RG technique  appeared
no physical models with non-trivial UV  asymptotic behavior were known.
Moreover, as non-trivial IR behavior is possible only for massless models, which also
were not known in this time, RG method remained  unused up to the seventies of the previous century.

In 1977  D.~Forster, D.~R.~Nelson and M.~J.~Stephen applied the RG method to
calculate  the correlations of velocity field
\cite{FNS77} governed by stochastic Navier-Stokes equation with
external random forcing. Later it was shown by C.~De Dominicis and P.~C.~Martin \cite{Dominicis79} that in the 
range 
of small wave numbers
the correlations of the velocity field manifest a scaling behavior with the celebrated  Kolmogorov exponents.
The basic idea of application of RG in the theory of developed turbulence
consists in elimination of the direct influence of the modes with high frequencies and wave
numbers on observed quantities. Effectively their influence is included
to  some effective variables, e.g. to the turbulent viscosity.

Numerous versions of RG methods exist. Constricted all they are equivalent
but technically can be quite different.
The most formalized is RG developed in the framework of the quantum field theory (field-theoretic RG), which has been used in
papers \cite{Adzhemyan83,Dominicis76}.

The field theoretic  RG  is based on non-trivial techniques of UV  renormalization.
The basic procedure consists of calculation of the RG-functions in the framework of a prescribed scheme of
regularization \cite{Zinn}.
To find and analyze all possible UV divergences in concrete field-theoretic models a counting of 
canonical scaling dimensions of fields and
parameters of the model is used. Let us briefly remind the essence of such a power counting which is closely connected with the 
existence of a scale invariance in the model.

{\subsection{Extended homogeneity}\label{eq:RG_jednatridva}}

This conception is introduced for the formulation of the hypothesis of similarity
(critical scaling, scale invariance) and it is useful for classification of all UV divergences
in field-theoretical models: critical statics and dynamics, developed turbulence and so on
\cite{Zinn,Vasiliev,Tauber2014}.
We give its definition and basic properties.

A function $F(e)$ which depends on variables $e\equiv \{e_1 \ldots e_N\}$ is termed
extended homogeneous (or simply speaking dimensional) if for
a set of numbers  $\Delta$ and arbitrary $\lambda>0$ the following equation is valid:
\begin{equation}
  F(\lambda^{\Delta_1} e_1,...,\lambda^{\Delta_N}
  e_N)=\lambda^{\Delta_W} F(e_1,...,e_N)\,,
  \label{eq:RG_eh}
\end{equation}
or shortly
\begin{equation}
  F(e_{\lambda})=\lambda^{\Delta_F} F(e),
  \label{eq:RG_zovs}
\end{equation}
where $e_{\lambda}=\lambda^{\Delta_i} e_i$ and $i=1,...,N$. The parameters
$\Delta_i\equiv\Delta_{e_i}\equiv\Delta[e_i]$  are (canonical or critical or anomalous) scaling dimensions 
(exponents) of corresponding
variables $e_i$ and $\Delta_F \equiv \Delta[F]$
represents dimension (exponent)  of  function $F$. If $\Delta_F=0$ then
the function $F$ is scale invariant (dimensionless).
A dimensional function $F(e)$ depending on one variable $e$ is proportional
to the power $|e|^{\beta}$ with the exponent $\beta=\Delta_F/\Delta_e$.
A function   $F$ depending on $N$ variables  can be expressed in the form of product of a
power function and
{\it scaling function} $R$ which is a function of $N-1$
dimensionless combinations of its arguments, e.g.
\begin{equation}
  F(e_1,\dots,e_N)=|e_N|^{\beta}
  R\left(\frac{e_1}{|e_N|^{\beta_1}},\dots,\frac{e_{N-1}}{|e_N|^{\beta_{N-1}}}
  \right)\,,
  \label{eq:RG_homog}
\end{equation}
where $\beta=\Delta_F/\Delta_n$ and $\beta_i=\Delta_i/\Delta_n$
$i=1,2,\dots n-1$. If one differentiates the  equation (\ref{eq:RG_zovs}) with respect to
$\lambda$ and afterwards puts $\lambda=1$, then will obtain Euler
differential equation, which represents another equivalent formulation of the extended homogeneity (\ref{eq:RG_zovs}):
\begin{equation}
  \left[\sum_e \Delta_e e \partial_e -\Delta_F\right] F(e)=0.
  \label{eq:RG_eh1}
\end{equation}

The substitution  $\lambda \rightarrow \lambda^a$ in the equation (\ref{eq:RG_zovs}) is
equivalent to multiplying of all exponents  $\Delta$ by
parameter $a$, therefore the exponent of one of the variables can be fixed.
Usually the dimension of the wave number $k$ is selected to be the unity (or, equivalently, dimension
of the space coordinate is equal to $-1$):
\begin{equation}
  \Delta[k]=1\,.
\end{equation}\label{eq:RG_hyb}
This definition is standard and we use it everywhere.
We note that in dynamical models (\ref{eq:intro_DDJ-FP}) also the dimension of
frequency can be fixed (see below).\\

{\subsection{UV-renormalization. RG equations} \label{subsec:UV}}

Let us summarize main ideas of the
quantum-field theory of renormalization and RG technique; a
detailed account can be found in monographs \cite{Amit,Collins,Zinn,Vasiliev,Tauber2014}.

We will be mainly concerned with  models whose diagrams are calculated without
UV-cut-off $\Lambda$
and UV-divergences manifest themselves  as poles in a certain
dimensionless  ``parameter of deviation from logarithmic theory $\varepsilon$''.
Historically, this term appeared in connection with infinite summation of main logarithms
(see aforementioned discussion)),
which is necessary to make in the case when the bare coupling constant $g_0$ (or constants) is canonically
dimensionless ($\Delta_g\equiv d_g=0.$)
The procedure of multiplicative renormalization removing
UV-divergences (in the given case, poles in in a parameter $\varepsilon$) consists in the
following: the original action $\S(\phi, e_0)$ is declared to be unrenormalized;
its parameters $e_0$ (letter $e_0$ stands for the whole set of parameters)
are the bare parameters, and they are considered to be some functions of the new renormalized parameters $e$, whereas a
new renormalized action is assumed to be the functional $\S_R[\varphi]=\S[\varphi Z_{\varphi}]$ with
certain (also to be  determined) renormalization constants
of fields $Z_{\varphi}$ (one per each independent component of the field).
In unrenormalized full Green functions $G_N= \langle \varphi\dots \varphi\rangle$ the functional averaging
$\langle \dots \rangle$ is performed with the "weight" $\exp \S[\varphi]$; while in renormalized functions ,
$G_N^R$ with the "weight" $\exp \S_R[\phi]$. The connection between the functionals $\S[\varphi]$ and $\S_R[\varphi]$
leads to the relation  between the corresponding Green
functions $G_N^R=Z_{\varphi}^{-N}G_N$, where by definition  $G_N=G_N(e_0,\varepsilon \dots )$(ellipsis denotes other arguments like
coordinates or wave numbers), and, by convention, the quantities $G_N^R$ and $Z_{\varphi}$
are expressed in terms of the parameters $e$. The correspondence $e_0\leftrightarrow e$ within
perturbation theory is assumed to be one-to-one, therefore
either of the sets $e_0, e$ can be taken as the independent variables.

For translation invariant theories it is convenient to deal not with the full
Green functions $G_N$, but either with their connected parts $W_N$ (their generating functional being
$\W(A)=\ln \G(A))$ or with 1-irreducible functions $\Gamma_N$ (also called one particle irreducible functions), which
generating functional is defined by the functional
Legendre transform \cite{Vasilev98}
\begin{equation}
  \label{eq:RG_transform} {\Gamma}(\alpha)=W(A)-A\alpha, \quad 
  \alpha=\frac{\delta \W(A)}{\delta A(x)}\,.
\end{equation}

These  unrenormalized and renormalized Green functions
satisfy relation:
\begin{eqnarray}
  W_N^R(e,\varepsilon,\dots )&=&Z_{\varphi}^{-N}(e,\varepsilon)
  W_N(e_0(e,\varepsilon),\varepsilon,\dots )\,,\\
  \Gamma_N^R(e,\varepsilon,\dots )&=&Z_{\varphi}^{N}(e,\varepsilon)
 \Gamma_N(e_0(e,\varepsilon),\varepsilon,\dots )\,,-
\end{eqnarray}
where the functions $e_0(e,\varepsilon),$ $Z_{\varphi}^{N}(e,\varepsilon)$ can be chosen arbitrarily, which implies an
arbitrary choice of normalization of the fields and parameters $e$ at
given $e_0$. 
In the present text we also interchangeably use the following notation for the connected Green functions
\begin{equation}
  W_{\varphi_1 \ldots \varphi_N} \equiv \langle \varphi_1 \ldots \varphi_N \rangle_\text{conn}. 
  \label{eq:1PI_writing}
\end{equation}
and for the one particle irreducible (1PI) Green functions
\begin{equation}
  \Gamma_{\varphi_1 \ldots \varphi_N} \equiv \langle \varphi_1 \ldots \varphi_N \rangle_\text{1-ir}. 
  \label{eq:1PI_writing}
\end{equation}
In the present section it is superfluous, because for simplicity we are assuming one field theory. However,
later on we will discuss theories with more than one field and then it becomes necessary to have more general notation at hand.

The basic statement of the theory of renormalization is
that for the  multiplicatively renormalizable models
these functions can be chosen  to provide UV-finiteness of Green functions as $\varepsilon \rightarrow 0.$
With
this choice, all UV-divergences (poles in $\varepsilon$) contained in the functions $e_0(e,\varepsilon),$ $Z_{\varphi}^{N}(e,\varepsilon)$ are
absent in
renormalized Green functions $W_N^R(e,\varepsilon)$.
We note that the UV-finiteness in this sense of any one set of
Green functions (full, connected, 1-irreducible)
automatically leads to the UV-finiteness of any other.
The RG equations are written for the renormalized functions $W_N^R$ which differ
from the original unrenormalized functions $W_N$ only by normalization,
and therefore, can  be used equally well to analyze  the critical
scaling. Let us demonstrate an elementary derivation of the RG equations
\cite{Amit,Vasiliev}. The requirement of elimination of divergences does not
uniquely determine the functions  $e_0(e, \varepsilon)$ and
$Z_{\varphi}(e, \varepsilon)$.  An
arbitrariness  remains which allows  to introduce in these functions (and via them also into  $W_N^R$)
an additional dimensional
parameter -  scale setting parameter (renormalization mass) $\mu$:
\begin{equation}
  W_N^R(e,\mu,\varepsilon,\dots )=Z_{\varphi}^{-N}(e,\mu,\varepsilon)
  W_N(e_0(e,\mu,\varepsilon),\varepsilon,\dots )\,.
\end{equation}
A change of $\mu$ at fixed $e_0$ leads to a change of $e, Z_{\varphi}$ and $W_R$
for unchanged $W_N(e_0,\varepsilon,\dots )$.
We use $\tilde{{\cal D}}_{\mu}$ to denote the differential operator $\mu
\partial_{\mu}$ for fixed $e_0$ and operate on both sides of the equation
$Z_{\phi}W_N^R=W_N$ with it. This gives the basic RG differential equation :
\begin{equation}
  \left[\mu \partial_{\mu}+\sum_e \tilde{{\cal D}}_{\mu} e\partial_e+N
  \gamma_{\varphi}
  \right]W_N^R(e,\mu,\varepsilon,\dots )=0,\,\, \gamma_{\varphi}\equiv \tilde{{\cal D}}_{\mu} \ln Z_{\varphi}
  \label{eq:RG_rgrovnica}
\end{equation}
where the operator $\tilde{{\cal D}}_{\mu}$  is expressed in the variables
$\mu, e.$
The coefficients  $\tilde{{\cal D}}_{\mu} e$ and
$\gamma_{\varphi}$
are
called the RG functions and are calculated in terms of
various renormalization constants $Z$. Coupling constants (charges) $g$ are those parameters $e$, on
which the renormalization constants $Z=Z(g)$
depend. Logarithmic derivatives of charges in (\ref{eq:RG_rgrovnica}) are $\beta$ functions
\beq
\label{beta-def}
\beta_g=\tilde{{\cal D}}_{\mu} g\,.
\eeq
All the RG-functions are
UV-finite, i.e. have no poles in $\varepsilon$, which is a consequence of the
functions $W_N^R$ being UV-finite in (\ref{eq:RG_rgrovnica}).

The general theory of renormalization \cite{Collins,Vasiliev} distinguishes
unrenormalized $\S$, renormalized $\S_R$, and base
$\S_B$
actions; the last is  obtained from $\S$ by replacement of all the
bare parameters $e_0$ by their renormalized counterparts $e$.
The UV-divergences are removed by adding to the base  action
$\S_B$  all necessary counterterms $\Delta \S$ which are determined by the
known rules (see below). If the renormalized action thus obtained
$\S_R=\S_B+\Delta \S$
 can be reproduced by the above procedure of redefinition of
fields and parameters in the original  unrenormalized action $\S$, the
model is multiplicatively renormalizable. Therefore, the first step
in the RG analysis of any model is to explicitly determine all
counter-terms required for the removal of UV-divergences and to verify
its multiplicative renormalizability.

The form of the required counterterms is determined by the analysis of canonical
 (engineering) dimensions of the 1PI Green functions of model with the action $\S_B$, which
satisfy the equation of extended homogeneity (\ref{eq:RG_eh}) (or, equivalently, (\ref{eq:RG_eh1}))
with definite canonical exponents $\Delta_e\equiv d_e, d_{\mu}=1, d_{\phi}$
of parameters $e_0, e, \mu$ and fields $\phi$ respectively.
Dynamical models (\ref{eq:intro_DDJ-FP}), in
contrast to static ones, are two-scale models, i.e. two independent canonical
wave-number (momentum)  $d^Q_k$ and frequency  $d^Q_{\omega}$ dimensions can be  assigned
to every quantity $Q$ (fields and parameters in the action).
They are easily determined from  the natural normalization conditions
\begin{equation}
  d_k^k=-d_x^k=1,\qquad d_{\omega}^{\omega}=-d_t^{\omega}=1 
  \label{eq:RG_normal}
\end{equation}
and from
the requirement that every term $T$ of the actions $S, S_B$
is dimensionless $(d^T_k=d^T_{\omega}=0).$
After that summarized full canonical  (engineering) dimension $d^Q$ can be determined by means of $d^Q_k$ and $ d^Q_{\omega}.$
Formally one can write
\begin{equation}
   d_F = d^Q_k + d_\omega d^Q_\omega,\
   \label{eq:RG_total_dim}
\end{equation}
where a value of $d_\omega$ depends on the particular model \cite{Vasiliev}. For example,
for stochastic developed turbulence and relaxational models A and B we have $d_\omega = 2$, whereas
for models D and H $d_\omega=4$.

 Of course,  the existence of the two aforementioned   wave-number and frequency scale invariance
can be expressed by means of  two differential equations similar to the Eq. (\ref{eq:RG_eh1}) with corresponding
exponents $d^e_k, d^e_{\omega}$ of parameters $e.$

In the scheme of renormalization of
dynamical models (\ref{eq:intro_DDJ-FP}) the full dimension $d^Q$ plays
 the same role as  the conventional (momentum)
dimension does in static problems. Canonical dimensions of an arbitrary
1PI Green function $\Gamma$ with $n_{\phi}$ (multiple index) fields $\varphi', \varphi$
for a $d+1$-dimensional problem are given by the relations
\begin{equation}
  d_{\Gamma}^k=d-\sum_{\phi}n_{\phi}d_{\phi}^k\,,\quad
  d_{\Gamma}^{\omega}=d-\sum_{\phi}n_{\phi}d_{\phi}^{\omega}\,,\quad
  d_{\Gamma}=d+2-\sum_{\phi}n_{\phi}d_{\phi}
  \label{eq:RG_diver}
\end{equation}
with summation over all the fields $\phi$ entering into given function
$\Gamma$. In a logarithmic theory, which corresponds to  $\varepsilon=0$ when bare
coupling constant(s) $g$ of concrete model is (are) dimensionless ($d^g=d^g_k=d^g_{\omega}=0$),
full canonical dimension of $\Gamma_N$ is equal to  a
formal index of UV-divergence $\delta.$  The UV-divergences which must be removed by
suitable  counterterms are allowable only in those functions $\Gamma_N$ for
which index $\delta$ is  nonnegative and integer \cite{Vasiliev}. All counterterms are polynomial functions of 
wave vector $\mk$ and frequency $\omega.$

For models considered in the present work the analysis of divergences should be based on the following auxiliary
considerations:

\begin{enumerate}[a)]
  \item For any dynamic model (\ref{eq:intro_DDJ-FP}) all 1PI Green
        functions containing only the original fields $\varphi$ are proportional to the closed circles of step functions
        hence they  vanish, and  thus do not generate counterterms.
  \item If for some reason several external momenta or frequencies occur as
	an overall factor in all the graphs of a particular Green function,
	the real degree of divergence $\delta^{\prime}$ is less than $\delta\equiv d_{\Gamma}(\varepsilon=0)$
	by the corresponding number of units.
  \item Sometimes the divergences formally allowed dimensionally
	are absent due to symmetry requirements, for instance, the
	Galilean invariance of the fully developed turbulence \cite{turbo} restricts
	the form of possible counterterms.  
  \item Nonlocal terms of the model are not renormalized. 
\end{enumerate}

These general considerations and formula (\ref{eq:RG_diver}) permit us to
determine all 
superficially divergent functions  and to explicitly obtain the
corresponding counter-terms for any concrete dynamic model.

{\subsection{Solution of RG equations. Invariant variables. RG-rep\-re\-sen\-ta\-tions
of correlation functions.} \label{sec:RGsolution}}

In this section we demonstrate general mathematical methods how to find  solutions
of RG equations of type (\ref{eq:RG_rgrovnica}) which are typical of models under consideration.
Consider linear differential  equation which is typical of RG approach:
\begin{equation}
  L F(u)=\gamma(u)\,,\quad L=-s\partial_s + \sum_{i=2}^n Q_i(u)
  \partial_{e_i}\,,
  \label{eq:RG_prdel}
\end{equation}
where $s=u_1$ is the scaling parameter, $e_i=u_i, i=2,\dots ,n$, $Q_i$ are given functions of 
the parameters $u$ and $F$ is a sought function.
The general solution of this inhomogeneous  equation is the sum of its particular solution and
 a solution of the homogeneous equation.
The latter is an arbitrary function of the full set of independent first integrals, which represent arbitrary solutions of the homogeneous
equation. The number of independent  first integrals is equal to the number of
parameters $e$. It is convenient to choose  first integrals $\bar
e_i=\bar e_i(s,e)$, which are defined as follows:
\begin{equation}
  L \bar e_i(s,e)=0\,,\quad \bar e_i(s,e)|_{s=1}=e_i\,.
  \label{eq:RG_in}
\end{equation}
These quantities are usually called  {\it invariant (running)
variables (charges)}.

The differential operator in the RG-equation (\ref{eq:RG_rgrovnica}) belongs to an important type of operators ${\cal D_{RG}}$ defined by the equation
\begin{equation}
  {\cal D_{RG}}\Upsilon(s,g,a)=0 \, ,\quad
  {\cal D_{RG}}\equiv \left[-s \partial_s+\beta(g) \partial_g -\sum_a \gamma_a(g)
  a\partial_a + \gamma(g) \right]\,,
  \label{eq:RG_priklad}
\end{equation}
where $g$ is the charge which defines $\beta$-function, $a$ are other parameters ($e=g,a$) and
the functions $\beta(g)$, $\gamma_a(g)$ and $\gamma(g)$ are
independent of  $s$. It is possible to show, that in this special case
the invariant charge $\bar g = \bar g(s,g)$ is independent of the parameters $a$ and satisfies
the differential equation known as the Gell-Mann-Low  equation:
\begin{equation}
  s \partial_s \bar g = \beta(\bar
  g)\,, \quad \bar g|_{s=1}=g\,.
  \label{eq:RG_gellmann}
\end{equation}
This equation is easily integrable:
\begin{equation}
  \ln s= \int_g^{\bar g} \frac{\dRM x}{\beta(x)}\,.
  \label{eq:RG_logar}
\end{equation}
The last expression implicitly defines $\bar g=\bar g(s,g)$ as a function of the 
scale parameter $s$ and the charge $g$.
For models with $n$ charges $g_i$ $(i=1,2,\dots n)$ one obtains a set of $n$ equations
\begin{equation}  
  s \frac{\dRM \bar g_i}{\dRM s}=\beta_{g_i}(\bar g)\,,\quad 
  \bar g \equiv ({\bar g_1}, {\bar g_2},\dots  {\bar g_n}) \, ,
  \label{eq:RG_RNG}
\end{equation}
where $\bar g_i$ is a set of invariant charges with initial values equal to $g_i$ $(\bar g_i|_{s=1}=g_i).$
A straightforward  integration (at least numerically) of these equations  gives a way  to find their fixed points.
Instead,  very often one solves the set of equations
\begin{equation}
  \beta_{g_i^*}=0
  \label{eq:RG_betanula}
\end{equation}
 which defines so-called fixed points $g^*\equiv (g_{1}^*, \dots  g_{n}^*$) of the model. To determine the
 type of a fixed point one calculates the matrix
 $\Omega\equiv \Omega_{ik}$ of the first derivatives of $\beta$ functions at a given fixed point $g^*$:
\begin{equation}
  \Omega_{ik}=\frac{\partial \beta_{g_i}(g)}{\partial g_k} \biggl|_{*} \,,\quad i, k = 1,2,\dots n,\quad {\ldots} \biggl|_* \equiv {\ldots}  \biggl|_{g=g^*}.
  \label{eq:RG_Omega}
\end{equation}
If the matrix $\Omega$ is positive (negative) definite, then  the fixed point is IR (UV) stable.
Technically one needs to determine the eigenvalues $\{ {\lambda}_j \},j=1,\ldots,n$ of
the $\Omega$ matrix.  A given fixed
point is infrared (or ultraviolet) if  all real parts of the eigenvalues
are positive (or negative). In other words
 for ${\overline g}(s)$ close to $g^{\ast}$ we obtain a system
 of linearized equations
\begin{equation}
  \left( I\,\frac{\mbox{d}}{\mbox{d} \ln s}  - \Omega\, \right) \,({\overline g} - g^{\ast})\,= 0,
\end{equation}
 where $I$ is the unit matrix of the size $n$. Solutions of this system
 behave like ${\overline g}=g^{\ast}+{\cal O}( s^{{\lambda}_j}) $,
 when $s\rightarrow 0$. 

The other invariant variables  $\bar a$ satisfy equations
\begin{equation}
  s\partial_s \bar a= -\bar a \gamma_a(\bar g)\,,\quad \bar
  a|_{s=1}=a\,.
\end{equation}
In  models with one charge (coupling constant) $g$ these equations are easily integrable and 
the solution can be written \cite{Amit,Zinn} the form
 \begin{equation}
   \bar a=\bar a(s,g,a)= a\, \exp {\left[-\int_g^{\bar g} \dRM x
   \frac{\gamma_a(x)}{\beta(x)}\right]}\,.
   \label{eq:RG_acka}
 \end{equation}
Finally, the general solution of the equation (\ref{eq:RG_priklad}) has the form
\begin{equation}
  \Upsilon(s,g,a)=\Upsilon(1,\bar g, \bar a)\, \exp{\left[\int_g^{\bar g} \dRM x
  \frac{\gamma(x)}{\beta(x)} \right]}\,,
  \label{eq:RG_or}
\end{equation}
where $\Upsilon(1,\bar g, \bar a)$ is an arbitrary (scaling) function of the first integrals.\\

{\subsection{Dimensional renormalization and the scheme of minimal subtractions} \label{subsec:dim_regul} }

Parameters $e$, on which the renormalization constant $Z_\varphi(g,\varepsilon)$ depends, are 
{\em coupling constants} or {\em charges} of the model. It is customary to choose renormalized charges
dimensionless, therefore in concrete models renormalized and unrenormalized 
values of charges $g$ obey relations of the type
\beq
\label{e-connection}
g_0=\mu^{2\varepsilon}gZ_g(g,\varepsilon)\,,
\eeq
where the renormalization constant $Z_g$ is dimensionless and the canonical dimension of the unrenormalized coupling
constant in this example is $2\varepsilon$.
RG functions corresponding to charges are $\beta$ functions (single charge is assumed here for simplicity). From
(\ref{e-connection}) it follows that
\beq
\label{def-beta}
\beta_g=\tilde{{\cal D}}_{\mu} g=g\left[-2\varepsilon -\gamma_g(g,\varepsilon)\right]\,.
\eeq
According to the
main statement of renormalization theory, renormalization constants $Z$ can be chosen so
as to eliminate all the UV divergences in the Green functions, in this case,
poles in $\varepsilon$. This is the main requirement on the functions $Z$, but it does not
determine them uniquely. The remaining arbitrariness is fixed by imposing
some auxiliary conditions. This is referred to as the choice of subtraction
scheme. Various schemes are used in which the
renormalized Green functions differ only by a UV-finite renormalization
and from the viewpoint of physics are equivalent. Therefore, the
scheme is chosen on the basis of convenience.

The most convenient scheme for analytic calculations is the minimal subtraction
(MS) scheme proposed in \cite{Hooft73}, in  which all the constants $Z$
have the following form in perturbation theory:
\beq
\label{Z-MS}
Z^{\rm MS}(g,\varepsilon)=1+\sum_{n=1}^\infty g^n\sum_{k=1}^n\varepsilon^{-k}c_{n,k}\,.
\eeq 
In dimensional renormalization the contribution to the coefficient of $g^n$ in (\ref{Z-MS}) may be expressed as a Laurent
series in $\varepsilon$. In the MS scheme only the singular part of the Laurent expansion of each coefficient is retained.
In any other renormalization scheme the renormalization constant is of the form
\beq
\label{Z-dimensional}
Z(g,\varepsilon)=1+\sum_{n=1}^\infty g^n\sum_{k=-n}^\infty\varepsilon^{k}c_{n, -k}\,,
\eeq
where the regular part of each coefficient $\sum_{k=0}^\infty\varepsilon^{k}c_{n, -k}$ is, by and large, an arbitrary regular function
of $\varepsilon$ at the origin.

It should be emphasized that even in the MS scheme the contribution of a graph to the renormalization constant is not determined
solely by the singular part of the Laurent expansion of the graph itself.
Calculations in perturbation theory are usually carried out order by order in the number of loops. For example,
  in the $\varphi^4-$theory the singular part of a one-loop
graph is a constant (i.e. independent of external wave vectors and frequencies), which is taken as the contribution of the
graph to the renormalization constant at one-loop order. When a superficially divergent one-irreducible graph contains two 
or more loops, its singular part is a function of external wave vectors and frequencies. To extract its contribution to the
renormalization constant, renormalization
of various subgraphs must be taken into account according to the rules of the consistent renormalization procedure
($R$ operation) \cite{Collins}.
In the course of this operation the Laurent series of each one-loop subgraph up to the constant term is included in
the calculation of the
contribution of a two-loop graph to the renormalization constant in the MS scheme. For consistent calculation of the
contribution of a superficially
divergent three-loop graph to the renormalization constant the Laurent expansion of each one-loop subgraph is needed
to the linear order
in $\varepsilon$ etc. Therefore, even in the MS scheme the Laurent series in $\varepsilon$ of each superficially 
divergent graph must
eventually be calculated. The difference is, so to speak, in the order of appearance of the terms of this series.

In the theory of critical phenomena and stochastic dynamics there are several models including long-range correlations or interactions
described -- in the Fourier space -- by fractional powers of wave numbers of the type $k^{-2a}$. When the exponent $a$ is treated
as a parameter of the model, the model may become logarithmic at some critical value $a_c$. In such a case the difference
$\varepsilon=a_c-a$ may be used as a regulator of the model and we arrive at a particular case of analytic
renormalization \cite{Speer69}
with a single regulator. In this particular case all relations about the analytic behavior of graphs, renormalization
constants and RG functions are
the same as in dimensional renormalization and henceforth the terminology of dimensional renormalization will be used 
both in the case
of genuine dimensional renormalization and in the case of analytic renormalization with a single regulator.\\

\subsection{Scheme dependence of critical exponents in dimensional renormalization}

It is one of the basic properties and the main reason of success of the RG approach to asymptotic analysis that values of relevant
physical quantities are independent of the finite renormalization, i.e. independent of the renormalization
scheme (see, e.g. \cite{Amit}).
The basic scheme-independent quantities are the anomalous dimensions (i.e. values of the coefficient
functions $\gamma$ at the fixed point)
and the eigenvalues of the Jacobi matrix of the set of $\beta$ functions at the fixed point. This, however, is a
global statement which does not
take into account the approximation method used in the calculation of these quantities. In particular, in dimensional
renormalization
of perturbation theory the RG functions are calculated as power series in coupling constants at fixed $\varepsilon$. The 
leading term in the
$\beta$ function in this expansion is proportional to $\varepsilon$, whereas coefficient of high-order terms have -- as a rule --
a finite limit, when $\varepsilon\to 0$. Therefore, power counting of RG-functions in the coupling constant and power counting
in $\varepsilon$ of fixed-point values thereof do not coincide. As a consequence, at any finite order of perturbation theory the
condition of the scheme independence is not fulfilled and anomalous dimensions as well as fixed-point eigenvalues of the Jacobi matrix
exhibit heavy dependence of the renormalization scheme, as will be shown below. Calculations within the dimensional renormalization
have been customarily carried out with the use of the MS (minimal subtractions) scheme, therefore the issue of scheme dependence of
critical exponents practically has not appeared. 

Consider a dimensionally renormalized model with a single dimensionless renormalized charge $g$ and a single coefficient function $\gamma_a$ related
to renormalization of a field, temperature (mass) or transport coefficient.
In the dimensional renormalization the generic form of the RG functions is
\begin{align}
\label{beta}
\beta(g,\varepsilon)&=\tilde{{\cal D}}_{\mu} g=g\left[-\varepsilon -\gamma_g(g,\varepsilon)\right]\,,\\
\label{rg_gamma}
\gamma_\phi(g,\varepsilon)&=\tilde{{\cal D}}_{\mu} \ln Z_\phi(g,\varepsilon)\,.
\end{align}
In perturbation theory the $\gamma$ functions of the RG are series expansions in the renormalized charge by 
construction, whose coefficient 
are regular functions of the parameter $\varepsilon$ at the origin, i.e.
\begin{align}
\label{beta-exp}
\gamma_(g,\varepsilon)&=\sum\limits_{n=1}^\infty a_{g\,n}(\varepsilon)g^n\,,\\
\label{gamma-exp}
\gamma_\phi(g,\varepsilon)&=\sum\limits_{n=1}^\infty a_{\phi\,n}(\varepsilon)g^n\,.
\end{align}
The usual argument leading to scheme independence of an anomalous dimension (i.e. the value of a $\gamma$
function at a fixed point of the RG)
goes as follows \cite{Amit}. In two different schemes the renormalized charges $g$ and $g'$ are connected by 
a relation $g'=G(g)$ in the form
of series expansion in $g$. Renormalization constants giving rise to $\gamma$ functions are connected -- due
to the group property -- by the
multiplicative relation
\beq
\label{Zconnection}
Z'_i(g',\varepsilon)=F_i(g,\varepsilon)Z_i(g,\varepsilon)\,,\qquad i=g,\phi\,,
\eeq
where the function $F(g,\varepsilon)$ is also expressed as a series in $g$ with regular in $\varepsilon$ coefficients.
Therefore,
\beq
\label{gamma-connection}
\gamma'_i(g',\varepsilon)=\gamma_i(g,\varepsilon)+\beta(g){\partial\over \partial g}\ln F_i(g,\varepsilon)\,,\qquad i=g,\phi\,.
\eeq
At a fixed point $g^*$ of the RG $\beta(g^*)=0$ and $\beta'(G(g^*))=0$. Therefore, the second term on the right side of (\ref{gamma-connection})
vanishes rendering the anomalous dimensions equal in the two renormalization schemes.

This is a global argument assuming that all functions in relation (\ref{gamma-connection}) are known completely. This 
is not the case, however,
in perturbation theory. Renormalization constants and the RG functions are calculated order by order as power 
series in the charge $g$.
Typically, expansions of the coefficient functions start with a linear term. In that case the linear term on 
the right side of (\ref{gamma-connection})
is produced by the function $\gamma(g,\varepsilon)$ and the term $g\varepsilon$ multiplied by the coefficient of the linear
term of ${\partial_g}\ln F(g,\varepsilon)$. The second contribution to the $\beta$ function (\ref{beta}) is $\O(g^2)$, should not
be included in the linear contribution to the right side (\ref{gamma-connection}) and the vanishing at the fixed point
factor is lost! Obviously, the same property holds at every finite order of perturbation theory and we arrive at the conclusion that
in the perturbative dimensional renormalization the value of the anomalous dimension at a non-trivial ($g^* \ne 0$) fixed point
heavily depends on the renormalization scheme!

To analyze this scheme dependence in more detail, consider renormalization constants and RG functions calculated to
some finite order $N$ of perturbation theory. In this case relation (\ref{gamma-connection}) gives rise to
\beq
\label{gamma-connection-N}
\sum\limits_{n=1}^N a'_{i\,n}(\varepsilon){g'}^n=\sum\limits_{n=1}^N a_{i\,n}(\varepsilon)g^n+\sum\limits_{m=1}^N
\left[g\varepsilon-g\sum\limits_{n=1}^m a_{g\,n}(\varepsilon)g^n\right]
\sum\limits_{n=1}^{N-m-1} f_{i\,n}(\varepsilon)g^n\,,
\eeq
where terms to the order $g^N$ are taken into account on the right side.
In (\ref{gamma-connection-N}) $f_{i\,n}(\varepsilon)$ are the coefficients of the perturbation expansion
of $\ln F(g,\varepsilon)$. They are regular functions of $\varepsilon$ by definition. The
perturbative non-trivial ($g^*\ne 0$) fixed point is sought in the form of an $\varepsilon$ expansion
\beq
\label{ustar}
g^*=\sum_{n=1}g^*_{n}\varepsilon^n
\eeq
for the solution of the equation $\varepsilon +\gamma_g(g,\varepsilon)=0$.
By direct substitution it is seen that at such a fixed point
\beq
\label{beta-rest}
g^* \varepsilon-g^*\sum\limits_{n=1}^m a_{g\,n}(\varepsilon){g^*}^n=\O(\varepsilon^{m+2})\,.
\eeq
At the fixed point (\ref{ustar}) we therefore see that the second term on the right side of (\ref{gamma-connection-N})
is of the order $\O(\varepsilon^{N+1})$ and the statement of the scheme independence is actually that the $\varepsilon$
expansion of a critical dimension is scheme independent up to the order given by a consistent perturbative calculation.
The $\varepsilon$ expansion is given by the MS scheme, therefore in any other renormalization scheme anomalous dimensions
contain $\varepsilon$ dependent contributions which are not controlled by the perturbation theory beyond the
order of calculation of the renormalization constants.

In case of the stability exponent the renormalization invariance is based on the connection between $\beta$ functions
\beq
\label{beta-connection}
\beta'(g')=\dee{G(g)}{g}\beta(g)
\eeq
from which it follows that
\beq
\label{omega-connection}
\dee{\beta'(g')}{ g'}=\dee{\beta(g)}{g}+\left[\dee{G(g)}{g}\right]^{-1}\dee{^2G(g)}{g^2}\beta(g)\,.
\eeq
At a fixed point the rightmost term on the right side of (\ref{omega-connection}) vanishes, if complete 
functions are known and the value of the
derivative of the $\beta$ function at the fixed point (the critical exponent $\omega$) is the same in both
renormalization schemes. Order by order
in perturbation theory this is again not true. Again, the $\beta$ function of the additional term on the right
side of (\ref{omega-connection})
at a fixed point in the $\varepsilon$ expansion produces an excess
factor $\O(\varepsilon)$ and the exponent $\omega$ is renormalization invariant at the leading
order of the $\varepsilon$ expansion, which is easily verified by direct calculation with the use of representations
(\ref{beta-exp}) and (\ref{gamma-exp}). It should also
be recalled that in the usual single-charge case (i.e. when the leading term in $\gamma_g$ is linear in $g$) all 
anomalous dimensions
and the stability exponent $\omega$ are expressed in regular at the origin power expansions in $\varepsilon$. This is not always the case
in multi-charge problems.

Critical dimensions and stability conditions of asymptotic patterns should be independent of the renormalization 
scheme and the different values obtained for them signal that approximations used for their calculation are different
as well. In dimensional renormalization differences in renormalization schemes
show in the $\varepsilon$ dependence of various quantities in the renormalized model. It is customary to carry out 
calculations in the form
of $\varepsilon$ expansions. Practical evaluation is usually performed loop by loop in graphs of perturbation
theory with each consecutive loop improving
the accuracy of results of the $\varepsilon$ expansions by some fixed order in $\varepsilon$. In the prevailing MS scheme of the
field-theoretic approach
the contribution of each superficially divergent graph to a renormalization constant at a given order of the loop expansion
is restricted to the singular part of its Laurent expansion
in $\varepsilon$ which in practice means that within this scheme the calculation of critical exponents and other 
relevant quantities yields 
directly the $\varepsilon$ expansions thereof without any excessive $\varepsilon$ on top of that. If any other 
renormalization scheme is used,
then finite renormalization means, in principle, that to the singular part of the Laurent expansion of the graph an 
arbitrary regular function of
$\varepsilon$ is added. This finite renormalization introduces $\varepsilon$ dependence in coefficients of the
perturbation expansions of RG functions which destroys the connection between the order of the $\varepsilon$ expansion 
and the loop expansion.
This in turn means that values of critical exponents and other relevant quantities calculated in an arbitrary
dimensional renormalization
scheme at each finite order in loops contain terms of the $\varepsilon$ expansion, whose coefficients are subject 
to corrections in
the subsequent orders of the loop expansion. Reliable in this sense information is provided by expanding everything in the results of
the arbitrary renormalization scheme in $\varepsilon$ and retaining only terms whose coefficients are consistently given by the current
order of loop expansion. This returns us to the calculation in the MS scheme. It is quite possible that evaluation of a power series
in some other way than simply adding terms order by order provides advantages in numerical accuracy. However, 
to make such a conclusion, the
functions expanded in the power series should be known in some other, preferably closed form. In perturbation 
expansion of a field theory
this kind of information is not available directly and usually it is quite difficult to obtain, e.g. by instanton 
analysis. The bottom line
here is that although different renormalization schemes provide different information about critical exponents 
and the like, to make conclusions
going beyond the strict $\varepsilon$ expansion some additional arguments are required.

{\subsection{Composite operators and operator product expansion}  \label{subsec:OPE}}

In this section we recall the basic information about renormalization and critical exponents (dimensions) of
composite operators.
A composite operator $F$ is any 
monomial of the basic fields of model and their derivatives.
In models we are interested they are constructed from the velocity field $\mv,$  scalar field $\theta $
or magnetic field $\mb$ at the single space-time point $ x \equiv (t, \mx)$. Examples are $\mv^n,$ $\mb^n,$
$\theta^n,$ $\partial_t \mv^n,$ $\mv\Delta \mv$, $(\boldnabla \theta\cdot \boldnabla \theta)^n$ and so on.
The term is borrowed from the quantum field theory, where fields and anything constructed from them are actually
operators in a Hilbert
space. In statistical field theory they are just random fluctuating quantities.

Study of composite operators and their renormalization is important at least 
for two reasons. First, their critical dimensions and correlation functions can be measured
experimentally and for some operators such data are available \cite{Antonia,Anselmet}. 
In the developed turbulence
the mean of the energy dissipation proportional to the statistical average of 
the composite operator $\mv\Delta \mv$
enters the equation of energy balance, hence playing a crucial role in redistribution of the energy of the turbulent
motion and its dissipation.
Moreover, strong statistical fluctuations of the operator of energy dissipation seem to be responsible for 
deviations from Kolmogorov exponents and lead to the intermittency (multifractality) of 
the turbulent processes \cite{Frisch}.
Second, the general solution of the RG equation (\ref{eq:RG_or})
 contains an unknown
scaling function depending on dimensionless effective variables (coupling constants, viscosity etc.). This 
function can be calculated
in the framework of usual perturbation scheme in an expansion parameter but as was already mentioned above, in 
certain asymptotic ranges
of scales this calculation fails.
In theory of turbulence dependence of correlation functions on outer (integral) scale $L$ is of interest.
In particular, in turbulence one is interested in the scaling function $R(1,g^*, kL)$ (compare with Eq. (\ref{eq:RG_homog}))
in  the inertial interval  $kL\gg  1$.
In the theory of critical phenomena, the asymptotic form of  scaling functions for  $kL\gg  1$
(formally $L\rightarrow \infty$)
is studied using the well known the Wilson operator product expansion (OPE), see e.g.
 \cite{Collins}, \cite{Zinn}; the analog of $L$ is there the correlation length $r_{c}$.
It has turned out that this technique can be used  also in the theory of
turbulence  and simplified (toy) models associated with the genuine 
 turbulence, see e.g. \cite{Vasiliev,Adzhemyan96,turbo,Adzhemyan89,FGV01}.

The generating functional of the correlation functions of the field $\phi$ with one insertion of the composite operator $F(\phi)$
has the form (compare with the generating functional (\ref{eq:intro_GenFunNoise2N}) for the usual
correlation functions of $\phi$)
\begin{equation}
  \label{eq:RG_xx9} 
  \G(A,F)= \int \D\varphi \: F(\varphi)\exp{\left[\S [\varphi] +
  A\varphi\right]}\,,
\end{equation}
where all normalization factors are included in the functional measure.
Since the arguments of the fields in $F$ coincide, correlation functions with
these operators contain additional UV divergences, which are removed
by an additional renormalization procedure, see
e.g. \cite{Collins,Vasiliev,Zinn}. For the renorma\-li\-zed correlation functions
 the standard RG equations are obtained, which describe IR scaling with
definite critical dimensions $\Delta_{F}\equiv\Delta[F]$  of
certain ``basis'' operators  $F$. Owing to the renormalization,
$\Delta[F]$ does not coincide in general with the naive sum
of critical dimensions of the fields and derivatives entering
into $F$.
Detailed description of the renormalization of composite
operators for the stochastic Navier--Stokes equation is given
in the review paper \cite{Adzhemyan96},
below we confine ourselves to only the necessary information.

In general, composite operators are mixed in renormalization,
i.e., an UV finite renormalized operator $F^{R}$ (the correlation functions with one insertion of $F_R$ 
don't possess the UV divergences)  has the form
$F^{R}=F+$ counterterms, where the contribution of the
counterterms is a linear combination of $F$ itself and,
possibly, other unrenormalized operators which ``admix''
to $F$ in renormalization. 
Let $F\equiv\{F_{\alpha}\}$ be a closed set, all of whose
monomials mix only with each other in renormalization.
The renormalization matrix $Z_{F}\equiv\{Z_{\alpha\beta}\}$
and the  matrix of anomalous dimensions
$\gamma_{F}\equiv\{\gamma_{\alpha\beta}\}$
for this set are given by
\begin{equation}
  F_{\alpha }=\sum _{\beta} Z_{\alpha\beta}
  F_{\beta }^{R},\qquad
  \gamma _F=Z_{F}^{-1}\tilde{\mathcal{D}}_{\mu }Z_{F},
  \label{eq:RG_2.2}
\end{equation}
and the corresponding matrix of critical dimensions
$\Delta_{F}\equiv\{\Delta_{\alpha\beta}\}$ is given by
\begin{equation}
  \Delta[F]\equiv\Delta_{F} = d_{F}^{k}+ \Delta_{\omega}
  d_{F}^{\omega}+\gamma_{F}^{*},
  \label{eq:RG_32B}
\end{equation}
in which
$d_{F}^{k}$, $d_{F}^{\omega}$, and $d_{F}$ are understood as the
diagonal matrices of canonical dimensions of the operators under consideration
(with the diagonal elements equal to sums of corresponding
dimensions of all fields, their derivatives and renormalized parameters constituting $F$) and
$\gamma^{*}_F\equiv\gamma_F (g^{*})$ is the matrix (\ref{eq:RG_2.2}) at
the fixed point.

Critical dimensions of the set $F\equiv\{F_{\alpha}\}$ are
given by the eigenvalues of the matrix $\Delta_{F}$. The ``basis''
operators that possess definite critical dimensions have the form
\begin{equation}
  {\bar F}^{R}_{\alpha}=\sum_{\beta}
  U_{\alpha \beta}F^{R}_{\beta} \,
  \label{eq:RG_2.5}
\end{equation}
where the matrix $ U_{F} =  \{U_{\alpha \beta} \}$
is such that $ \Delta'_{F}= U_{F} \Delta_{F} U_{F}^{-1}$
is diagonal.

In general, counterterms to a given operator $F$ are
determined by all possible 1PI Green functions
with one operator $F$ and arbitrary number of primary fields $\varphi$,
\begin{equation}
  \Gamma_{N;F}=
  \langle F(t,\mx) \phi(t,\mx_{1})\dots\phi(t,\mx_{N})\rangle.
  \label{eq:RG_}
\end{equation}
The total canonical dimension (formal index of divergence)
for such functions is given by
\begin{equation}
  d_\Gamma = d_{F} - N_{\Phi}d_{\Phi},
  \label{eq:RG_index}
\end{equation}
with the summation over all types of fields entering into
the function. For 
the UV divergent diagrams,
$d_\Gamma$ is a nonnegative integer (cf. \ref{eq:RG_diver}).

According to the OPE, the single-time product
$F_{1}(t,\mx_{1})F_{2}(t,\mx_{2})$
of two renormalized operators at
$\mx\equiv (\mx_{1} + \mx_{2} )/2 = {const}$, and
$\mr\equiv \mx_{1} - \mx_{2}\to 0$
has the representation
\begin{equation}
  F_{1}(t,\mx_{1})F_{2}(t,\mx_{2})=\sum_{\alpha}A_{\alpha} (\mr)
  \bar F^{R}_{\alpha}(t,{\mx}) ,
  \label{eq:RG_2.44}
\end{equation}
in which the functions $A_{\alpha}$  are the Wilson coefficients
regular in $L$ and  $\bar F^{R}_{\alpha}$ are all possible
renormalized local composite operators of the type (\ref{eq:RG_2.5})
allowed by symmetry, with definite critical
dimensions $\Delta_{\bar F^R_{\alpha}}$.

The renormalized correlator $\langle F_{1}(t,\mx_{1})F_{2}(t,\mx_{2})
\rangle$ is obtained by averaging (\ref{eq:RG_2.44}) with the weight
$\exp S_{R}$, the quantities  
$\langle\bar F^{R}_{\alpha}\rangle \propto L^{-d_{\alpha}}  f_{\alpha} (g,L\mu)$
involving dimensionless (scaling) functions $ f_{\alpha} (g,L\mu)$
appear on the right hand side. Their asymptotic behavior
for $L\mu\to0$ is found from the corresponding RG equations (see \cite{Ant99} for the case of Kraichnan model) and
has the form
\begin{equation}
  \langle\bar F^{R}_{\alpha}\rangle \propto  L^{\Delta_{\bar F^R_{\alpha}}}.
  \label{eq:RG_2.45}
\end{equation}

 From the operator product expansion (\ref{eq:RG_2.44}) we therefore
\begin{equation}
  \langle F_{1}(t,\mx_{1})F_{2}(t,\mx_{2}) \rangle = \sum_{\bar F^R} C_{\bar F^R} (r/L)^{\Delta_{\bar F^R}},\quad r/L\to 0,
  \label{eq:RG_A6}
\end{equation}
Here    $C_{\bar F^R}$ generated by the Wilson coefficients $A_{\alpha}$ in (\ref{eq:RG_2.44}) are 
regular in $L$, the summation
is implied over all possible composite basic renormalized operators $\bar F^R$
allowed by the symmetry of the left-hand side, and
$\Delta_{\bar F^R}$ are their critical dimensions.
The leading contributions for  $r/L\to 0$
are those with the smallest dimension $\Delta_{\bar F^R}.$
In the theory of critical phenomena all the nontrivial composite
operators have positive critical dimensions $\Delta_{\bar F^R}>0$ for small 
$\varepsilon$ and
the leading term in (\ref{eq:RG_A6}) is determined by the simplest
operator $\bar F^R=1$ with $\Delta_{\bar F^R}=0$, i.e., the function $R(r/L)$
is finite as $L\equiv r_{c}\to 0$, see \cite{Zinn}.
However, as has been observed in \cite{Adzhemyan89} in the model of developed turbulence
 composite operators with {\it negative} critical dimensions exist and are
responsible for possible singular behavior of the scaling functions like 
 N-point correlation functions $W_N=\langle \varphi \ldots \varphi \rangle$ as $r/L \to  0$.
We shall term the operators with $\Delta_{\bar F^R}<0$,
if they exist, dangerous \cite{Adzhemyan96},
as they correspond to contributions to (\ref{eq:RG_A6})
which diverge for $r/L\to0$. The scaling functions (\ref{eq:RG_A6}) decomposed in
dangerous operators exhibit  {\it anomalous scaling behavior}
which is a manifestation of a nontrivial multifractal (intermittent) nature
of the statistical fluctuations of the random fields under consideration and globally
all the physical system.

Dangerous composite operators in the stochastic model of turbulence 
occur only for
finite values of the RG expansion parameter $\varepsilon$, and within
the $\varepsilon$ expansion it is impossible to decide whether or not
a given operator is dangerous, provided its critical dimension
is not found exactly using the Schwinger-type functional equations
or the Galilean symmetry, see \cite{Adzhemyan96}, \cite{Adzhemyan88}.
Moreover, dangerous operators enter into the operator product
expansion  in the form of infinite families with the
spectrum of critical dimensions unbounded from below, and
the analysis of the large $L$ behavior implies the summation
of their contributions.

In view of the difficulties encountered by the RG approach to
the  model of developed turbulence 
it is reasonable to apply the formalism to simpler
models, which exhibit some of the features of genuine turbulent
flows, but are easier to study.
Much attention has been attracted by a simple model
of the passive advection of a scalar quantity by a Gaussian
velocity field, introduced by Obukhov \cite{Obu49} and Kraichnan
\cite{Kraichnan94}.
Of special interest are structure functions $S_N$, which for the scalar field $\theta$ can be 
defined as follows
\begin{equation}
  S_{N} (r) \equiv \big\langle [ \theta (t,{\mx}+{\mr}) -
  \theta (t,{\mx})]^{N} \big\rangle, \quad r\equiv |{\mr}|, 
  \label{eq:def_structure}
\end{equation}
where homogeneity and isotropy has been assumed.
It turns out, that the structure functions of
the scalar field in aforementioned Obukhov-Kraichan model exhibit anomalous scaling behavior
 and the corresponding anomalous
exponents can be calculated explicitly within  an expansion in
certain small parameter, see \cite{FGV01,Antonov06} and reference therein.\\

{\section{Stochastic models of developed turbulence} \label{sec:models}}

{\subsection{Stochastic version of Navier-Stokes equation} \label{subsec:stochNS}}
Systems with large number of degrees of freedom
display similar behavior in certain asymptotic regimes
independently of numerous microscopic details of the system.
In the theory of strongly developed turbulence
this {\it universality} is connected with long-distance asymptotics
of velocity correlation functions.
The main indication of the universality in
the turbulence  comes from the celebrated
Kolmogorov scaling theory \cite{Kolmog41}
describing the large-scale behavior of
velocity structure functions.

During the last decade much attention has been paid to the inertial
range of fully developed turbulence, which contains wave numbers
larger then those that pump the energy into the system and smaller
enough then those that are related to the dissipation processes
\cite{Monin,HuPhWi91}. Foundations of theory of the inertial range
turbulence were laid in the well known
Kolmogorov--Obukhov (KO) phenomenological theory
(see, e.g., \cite{Monin,McComb,Frisch}). One of the main
problems in the modern theory of fully developed turbulence is to
verify the validity of the basic principles of  KO theory and their
consequences within the framework of a
microscopic model. Recent experimental and theoretical studies
indicate  possible deviations from the celebrated Kolmogorov scaling
exponents. The scaling behavior  of the velocity fluctuations with
exponents, which values are different from Kolmogorov ones, is
called as anomalous and usually  is associated with
phenomenon of intermittency. Roughly speaking, intermittency means that statistical
properties (for example, correlation or structure functions of the
turbulent velocity field) are dominated by rare spatiotemporal
configurations, in which the regions with strong turbulent activity
have exotic (fractal) geometry and are embedded into the vast
regions with regular (laminar) flow. In the turbulence such
phenomenon is believed to be related to the strong fluctuations of
the energy flux which, therefore leads to deviations from the
predictions of the aforementioned KO theory. Such deviations,
referred to as ``anomalous'' or ``non-dimensional'' scaling,
manifest themselves in singular (arguably power-like) dependence of
correlation or structure functions on the distances and the integral
(external) turbulence scale $L$. The corresponding exponents are
certain nontrivial and nonlinear functions of the order of the
correlation function, the phenomenon referred to as
``multiscaling''.

Although great progress
in the understanding of intermittency and anomalous scaling in
turbulence has been achieved as a result of intensive studies, their
investigation in fully developed turbulence still remains a
major theoretical problem.

Although the theoretical description of the fluid turbulence on the
basis of the "first principles", i.e., on the stochastic
Navier-Stokes (NS) equation \cite{Monin} remains essentially an
open problem, considerable progress has been achieved in
understanding simplified model systems that share some important
properties with the real problem: shell models \cite{Dyn},
stochastic Burgers equation \cite{Burgulence} and passive advection
by random ``synthetic'' velocity fields \cite{FGV01}.

A crucial role in these studies is played by models of advected
passive scalar field \cite{Obu49}.
A simple model of a
passive scalar quantity advected by a random Gaussian velocity
field, white in time and self-similar in space (the latter property
mimics some features of a real turbulent velocity ensemble), the
so-called Kraichnan's rapid-change model \cite{Kra68}, is an
example. The interest to these models is based on two important
facts: first, as were shown by both natural and numerical
experimental investigations, the deviations from the predictions of
the classical Kolmogorov-Obukhov phenomenological theory
\cite{Monin,OrszagBook,Frisch,McComb} is even more strongly
displayed for a passively advected scalar field than for the
velocity field itself (see, e.g.,
\cite{Antonia84,Sreenivasan91,HolSig94,Pumir96,Pumir97,Pumir98,Pumir94,TonWar94,Elperin1,Elperin2,Elperin3} and
references cited therein), and second, the problem of passive
advection is much more easier to be consider from theoretical point
of view. There, for the first time, the anomalous scaling was
established on the basis of a microscopic model \cite{Kraichnan94},
and corresponding anomalous exponents was calculated within
controlled approximations 
\cite{Chertkov95b,Chertkov96,Gawedzki95,Bernard96,Shraiman96,Shraiman97,Pumir96,Pumir97,Pumir98} (see also reviews
\cite{FGV01,Antonov06} and references therein).

The greatest stimulation to study the simple models of passive
advection not only of scalar fields but also of vector fields (e.g.,
weak magnetic field) is related to the fact that even simplified
models with given Gaussian statistics of so-called "synthetic"
velocity field describes a lot of features of anomalous behavior of
genuine turbulent transport of some quantities (as heat or mass)
observed in experiments, see, e.g.,
\,\cite{Kra68,HolSig94,Pumir96,Pumir97,Pumir98,Pumir94,TonWar94,Elperin1,Elperin2,Elperin3,
Chertkov95b,Chertkov96,Gawedzki95,Bernard96,Shraiman96,Shraiman97,Majda90,Majda92,Majda93,Majda94,ZhaGli92,Kraichnan94,
Kraichnan95}.

An effective method for investigation of a self-similar scaling
behavior is the renormalization group (RG) technique
\cite{Zinn,Vasiliev,Collins}. It was widely used in the theory
of critical phenomena to explain the origin of the critical scaling
and also to calculate corresponding universal quantities (e.g.,
critical dimensions). This method can be also directly used in the
theory of turbulence
\cite{Vasiliev,Dominicis79,Adzhemyan83,Adzhemyan96,turbo}, as well as in
related models like a simpler stochastic problem of a passive scalar
advected by prescribed stochastic flow. In these investigations, the diagram technique
of Wyld \cite{Wyld} for the stochastically forced Navier-Stokes
equation with powerlike correlation function of the random force $\mf$.
The exponent of this correlation function gives rise to an expansion
parameter similar to that of the famous $\varepsilon=d_c-d$ expansion
in the theory of critical phenomena. In the wave-vector space the spectrum
of force correlations of the form
\beq
\label{NS-eps}
\langle f_i(\mk)f_j(-\mk)\rangle\propto k^{4-d-2\varepsilon}
\eeq
allows to obtain a regular expansion of
scaling exponents and amplitude coefficients in the small
parameter $\varepsilon$. 
In what follows the
conventional ("quantum field theory" or field-theoretic) RG will be use which is
based on the standard renormalization procedure, i.e., on the
elimination of the UV divergences in the logarithmic model.

In work \cite{AAV98} the field theoretic RG and
operator-product expansion (OPE) were used in the systematic
investigation of the rapid-change model. It was shown that within
the field theoretic approach the anomalous scaling is related to the
very existence of so-called "dangerous" composite operators with
negative critical dimensions in OPE (see, e.g.,
\cite{Vasiliev,turbo} for details). In the subsequent papers
\cite{AdAnBaKaVa01} the anomalous exponents of the model were
calculated within the $\varepsilon$ expansion to order
$\eps^3$ (three-loop approximation). Here $\eps$ is a
parameter which describes a given equal-time pair correlation
function of the velocity field (see subsequent section). 

Afterwards, various generalized descendants of the Kraichnan model,
namely, models with inclusion of large and small scale anisotropy
\cite{AAHN00}, compressibility \cite{AdzAnt98} and finite
correlation time of the velocity field \cite{Ant99,Ant00}
were studied by the field theoretic approach. Moreover, advection of
a passive vector field by the Gaussian self-similar velocity field
(with and without large and small scale anisotropy, pressure,
compressibility, and finite correlation time) has been also
investigated and all possible asymptotic scaling regimes and
cross-over among them have been classified \cite{ALM01,AHMG00,AAR01,AHHJR04,HJMS02}. General
conclusion is: the anomalous scaling, which is the most important
feature of the Kraichnan rapid change model, remains valid for all
generalized models.

Let us describe briefly the solution of the problem in the framework
of the field theoretic approach. It can be divided into two main
stages. On the first stage the multiplicative renormalizability of
the corresponding field theoretic model is demonstrated and the
differential RG equations for its correlation functions are
obtained (See Sec. \ref{subsec:UV}). The asymptotic behavior of the latter on their ultraviolet
argument $(r/\ell)$ for $r\gg\ell$ and any fixed $(r/L)$ is given by
infrared  stable fixed points of those equations. Here $\ell$ and
$L$ are an inner (ultraviolet) and an outer (infrared) scales. It
involves some {}``scaling functions'' of the infrared argument
$(r/L)$, whose form is not determined by the RG equations. On the
second stage, their behavior at $r\ll L$ is found from the OPE (See Sec. \ref{subsec:OPE})
within the framework of the general solution of the RG equations.
There, the crucial role is played by the critical dimensions of
various composite operators, which give rise to an infinite family
of independent aforementioned scaling exponents (and hence to
multiscaling). Of course, these both stages (and thus the phenomenon
of multiscaling) have long been known in the RG theory of critical
behavior.

In Ref.\,\cite{Ant99} the problem of a passive scalar advected
by the Gaussian self-similar velocity field with finite correlation
time \cite{Shraiman94,Shraiman96} was studied by the field theoretic RG method.
There, the systematic study of the possible scaling regimes and
anomalous behavior was present at one-loop level. The two-loop
corrections to the anomalous exponents were obtained in
 work \,\cite{AdAnHo02}. It was shown that the anomalous exponents are
non-universal as a result of their dependence on a dimensionless
parameter, the ratio of the velocity correlation time, and  turnover
time of a scalar field.

The Navier-Stokes equations conserve kinetic energy and helicity in
inviscid limit. Presence of two quadratic invariants leads to the
possibility of appearance of double cascade. It means that cascades
of energy and helicity take place in different ranges of wave
numbers analogously to the two-dimensional turbulence and/or the
helicity cascade  appears concurrently to the energy one in the
direction of small scales \cite{Brissaud73,Moiseev1996}.
Particularly,  helicity cascade is closely connected with the
existence of exact relation between triple and double correlations
of velocity known as ``2/15'' law analogously to the ``4/5''
Kolmogorov law \cite{Chkhet96}. According to \cite{Brissaud73}
aforementioned scenarios of turbulent cascades  differ from each other by
spectral scaling. Theoretical arguments given by Kraichnan
\cite{Kraich73}  and results of numerical calculations  of
Navier-Stokes equations \cite{Andre77,Orszag1997,Chen2003} support
the scenario of concurrent cascades. The appearance of helicity in
turbulent system leads to constraint of non-linear cascade to the
small scales. This phenomenon was firstly demonstrated by Kraichnan
\cite{Kraich73} within the modeling problem of statistically
equilibrium spectra and later in numerical experiments.

{\subsection{Double expansion and the ray scheme} \label{subsec:double_intro}}

RG calculations with two (or even more) small parameters which may serve as regulators in dimensional or analytic renormalization
have been widely used in the analysis of static critical phenomena \cite{Weinrib83,HonNal89,Blavatska01},
dynamic critical phenomena \cite{Antonov06c,Antonov06b,Antonov09,Antonov11,Antonov12b}, diffusion in random environment
\cite{Gevorkian87,HonKar88,Honkonen96b,Goncharenko10}, interface growth \cite{Antonov15} and in stochastic 
hydrodynamics \cite{Fournier82,Adzhemyan85,Ronis87,Hnatich90,HonNal96,Antonov97,Hnatich99,HHJ01,Gladyshev12}.
Critical exponents and other relevant quantities may be expressed in a double expansion in these parameters.
The two parameters may both be regulators of analytic renormalization or one of them is the regulator of dimensional renormalization.
In the following, this pair of parameters will be denoted $\varepsilon$ and $\Delta$. 
 
Analytic renormalization would be a natural renormalization scheme to use to construct a double expansion in
the two regulators, since it yields the RG  functions as analytic functions of the two parameters at the 
origin. The genuine analytic renormalization
involves rather tedious calculations \cite{Zavyalov79}. Moreover,
in analytic renormalization there is no analog of the MS scheme to simplify practical calculations.
Therefore, it is invariably assumed (implicitly or explicitly) that both parameters are of the same 
order of magnitude. This is made explicit by putting
them proportional to each other in the ray scheme \cite{AHKV03,AHKV05}): $\Delta=\zeta\varepsilon$, where
$\zeta$ is fixed and finite. This
assumption effectively restores the dimensional renormalization with a single small parameter and the MS scheme may be used. 

In any case, critical exponents turn out to be scheme dependent in the same sense as in the dimensional
renormalization. Typically there are at least 
two charges in these models and therefore a rather generic case of two charges and a single anomalous 
dimension $\gamma$ (corresponding to a field renormalization)
will be analyzed here. It should be
emphasized that we are considering coupling constants which serve as expansion parameters of the perturbation 
theory. When there are several coupling constants, it is customary to classify the order of perturbation 
theory by the number of loops. In multi-charge problems there
are coupling constants, which should be calculated in closed form at each such order of perturbation 
theory (e.g. ratios of coefficients of viscosity, diffusion and thermal conductivity). We do not
discuss such coupling constants here.

Two different structures of $\beta$ functions are met. 
In stochastic hydrodynamics two (or more) random sources with different powerlike
falloff of correlation functions are often introduced \cite{Fournier82,Adzhemyan85,Hnatich90,Antonov97,Hnatich99}: always 
random force for the stochastic momentum equation (Navier-Stokes equation) and the random source for either the stochastic 
diffusion or heat conduction equation (the passive scalar problem) or for Faraday's law (magnetohydrodynamics). Similar 
constructions have been used in critical dynamics \cite{Antonov06,Antonov06b,Antonov09,Antonov11,Antonov12b} and the
interface growth problem \cite{Antonov15}. Thus, two analytic regulators are used:
deviations of exponents of these powerlike correlation functions from their critical values. The regulators are invariably
put explicitly proportional to each other and renormalization is treated in the framework of the usual dimensional renormalization.
Nevertheless, a double expansion in the regulators is implied, if not always worked out explicitly.
In models of this type the structure of the $\beta$ functions is similar to the single-charge case, i.e.
the renormalized coupling constant is a common factor in the expression for the corresponding $\beta$ function
(for brevity, parameters $\varepsilon$ and $\Delta$ are omitted in the list of arguments):
\begin{align}
\label{beta1}
\beta_1(g_1,g_2)&=\mu{\partial\over\partial\mu}\Bigl\vert_0 g_1 = g_1\left[-\varepsilon -\gamma_1(g_1,g_2)\right]\,,\\
\label{beta2}
\beta_2(g_1,g_2)&=\mu{\partial\over\partial\mu}\Bigl\vert_0 g_2 = g_2\left[-\Delta -\gamma_2(g_1,g_2)\right]\,,\\
\label{gamma}
\gamma_\phi(g_1,g_2)&=\mu{\partial\over\partial\mu}\Bigl\vert_0 \ln Z_\phi(g_1,g_2)
\end{align}
and the coefficient functions $\gamma_1$, $\gamma_2$ and $\gamma_\phi$ are regular expansions 
in powers of $g_1$ and $g_2$, whose coefficients
depend on the regulators $\varepsilon$ and $\Delta$. We shall refer to this situation as the regular multi-charge case.

Connections between renormalization constants and the corresponding RG functions in different schemes in this case are
\begin{align}
\label{Zconnection2}
Z'_i(g'_1,g'_2)&=F_i(g_1,g_2)Z_i(g_1,g_2)\,,\qquad i=1,2,\phi, \\
\label{gamma-connection2}
\gamma'_i(g'_1,g'_2)&=\gamma_i(g_1,g_2)+\sum\limits_{j=1}^2\beta_j(g_1,g_2){\partial\over \partial g_j}\ln F_i(g_1,g_2)\,.
\end{align}
When the problem is treated in the framework of analytic renormalization, coefficients of the
perturbation expansion of the RG functions are regular functions of the two parameters $\varepsilon$
and $\Delta$ at the origin by construction of the renormalization scheme. The argument about the scheme 
dependence of the anomalous dimensions then goes in analogy with the dimensional renormalization argument above.
Due to the analytic properties of the RG functions the perturbative non-trivial fixed point may be found 
in the form of a double expansion in $\varepsilon$ and $\Delta$. Here and henceforth only fixed points with
both non-vanishing charges ($g^*_{1}\ne 0$, $g^*_{2}\ne 0$) will be considered,
if not stated otherwise.

Regularity of the fixed points and RG functions imply that anomalous dimensions are obtained
in the form of regular expansions in $\varepsilon$ and $\Delta$. Little reflection shows that the second 
term in relation (\ref{gamma-connection2})
at a fixed point gives rise to a contribution which is of higher order by $\O(\varepsilon)$ or $\O(\Delta)$ 
in comparison with the double
expansion of the anomalous dimension in the two renormalization schemes.

However, in practical calculations instead of the analytic renormalization the ray scheme is used, in which the
regulators are proportional to each other and the renormalization is carried out as in dimensional renormalization 
with an additional finite and fixed parameter $\zeta=\Delta/\varepsilon$. At one-loop order the $\gamma$'s are
linear functions of the charges vanishing at the origin with coefficients which are regular functions of 
$\varepsilon$ and $\Delta$ in the ray scheme as well. This is because the UV divergences show in the form of
meromorphic functions
with simple poles $A(\varepsilon,\Delta)/\varepsilon$, $B(\varepsilon,\Delta)/\Delta$ 
and $C(\varepsilon,\Delta)/(\varepsilon+\Delta)$, where $A$, $B$ and $C$ are regular functions of $\varepsilon$ 
and $\Delta$ which determine the RG functions at the one-loop order. At higher orders, however, subtraction
of divergent subgraphs in the renormalization gives rise to expressions containing products of terms of the type 
$\left[\varepsilon D(\varepsilon,\Delta)+\Delta E(\varepsilon,\Delta)\right]/(m\varepsilon+n\Delta)$, where -- in
analytic renormalization -- integers $m$ and $n$ are the numbers of correlations functions with exponents $\varepsilon$
and $\Delta$, respectively, in a (sub)graph and
$D$ and $E$ are regular functions of  $\varepsilon$ and $\Delta$ at the origin. In case of combined dimensional and 
analytic renormalization $m$ is the number of loops in the (sub)graph. It should be noted that in the ray scheme such 
expressions are finite
quantities, but in analytic renormalization they are singular functions which should not appear in the RG 
functions. In the ray scheme the common power of
$\varepsilon$ is extracted by the rule $\Delta=\zeta\varepsilon$, which gives rise to meromorphic functions of $\zeta$ as 
coefficients of the
perturbative expansion. From the point of view of dimensional renormalization these meromorphic functions produce
contributions to finite
renormalization. Thus, in the ray scheme the coefficient functions of the perturbation expansion of RG functions
are not analytic in the
variable $\zeta=\Delta/\varepsilon$, i.e. they are not analytic functions of the regulators $\varepsilon$ and $\Delta$.

At the leading one-loop order in both schemes the equations for the fixed point with both $g^*_{1}\ne 0$ and
$g^*_{2}\ne 0$ are linear equations for the fixed point values of the charges, whose solution is a unique linear
function of $\varepsilon$ and $\Delta$. In the analytic renormalization all $\gamma$'s are power series
in charges with analytic in  $\varepsilon$ and $\Delta$ coefficients. Therefore, it is immediately seen that
in analytic renormalization all anomalous dimensions are regular functions of $\varepsilon$ and $\Delta$.
Notwithstanding the non-analytic coefficient functions, the result for the
anomalous dimensions is the same up to the order in  $\varepsilon$ and $\Delta$ guaranteed by the
loop expansion within the ray scheme. In the ray scheme minimal subtractions may be used leading to much simpler
calculation of graphs.

The other possibility is dimensional regularization amended by analytic regularization (only one
analytic regulator will be considered here,
although several have been introduced).
In this case either in propagators or interactions the wave-number dependence contains the combination
$a+bk^{-2\alpha}$, in which $\alpha>0$
(in propagators this combination is usually multiplied by the factor $k^2$). For small $\alpha$ a 
non-trivial problem of renormalization
of field operators with this structure arises 
\cite{Weinrib83,HonNal89,HonNal96,Blavatska01,Antonov06,Antonov06b,Antonov09,Goncharenko10,Antonov11,Antonov12b,Antonov15}, since
in the limit $\alpha\to 0$ the terms in $a+bk^{-2\alpha}$ become indistinguishable and it is not clear, which of them 
should be renormalized. The problem is solved by the prescription of the counter terms to renormalization of the local
(analytic in $k^2$) contribution \cite{HonNal96,AHKV05}. The basic idea is that renormalization produces only local
counterterms. Construction of renormalization constants is carried out in the regularized model, in which
the local and non-local term are clearly distinguishable ($\alpha>0$ although small) and the counterterms 
have the structure of the local term
and thus contribute to the renormalization of that term only. If the original model did not contain the local
term at the outset (which is often
the case when models with long-range effects are constructed), then it is usually brought about by the 
renormalization procedure \cite{Weinrib83}.
In the field-theoretic approach such ''generation terms'' are to
the original model to make it multiplicatively renormalizable, which is very convenient from the technical point of view. 

In many cases the analytic properties of RG functions in problems with combined dimensional and 
analytic regulators are analogous to those of the case with two analytic regulators.
A different situation takes place, for instance, in critical systems with quenched disorder
\cite{Weinrib83,HonNal89,Blavatska01,Goncharenko10} and in stochastic hydrodynamics with competing
long-range and short-range correlations \cite{HonNal96}.
The interplay of long-range and short-range correlations is accompanied by the appearance of generation terms.
Generation terms are contributions to renormalization of a charge produced by other charges only. 
Generation terms produce contributions to renormalization constant of the corresponding charges in which the 
charge corresponding to the generation term stands in the denominator of a polynomial functions of other charges. 
This introduces significant changes to conclusions obtained from connections between renormalization constants and
charges in different schemes. 
First, contrary to the regular multi-charge case the fixed-point values of charges in the analytic renormalization 
are not regular functions of the regulators (although the RG functions are).
Therefore, critical exponents may not be regular functions of regulators either.
Another feature of this class of models is that
the very number of the fixed points becomes scheme dependent. This may be seen in the example of stochastic
hydrodynamics near two dimensions, in which one-loop calculations in four different schemes are available 
\cite{HonNal96,HHJ01,AHKV05,AHH10}. In the MS scheme in the ray approach
the one-loop solution for the two charges  $g^*_{1}\ne 0$ and $g^*_{2}\ne 0$ is obtained from a system of 
equations which is essentially linear
and the solution is unique \cite{HonNal96}, whereas in the other schemes the one-loop equation for charges 
is quadratic \cite{HHJ01,AHKV05} with two different solutions corresponding to different choices of the sign
of the quadratic root in the solution. 
In most cases only the stable
fixed point with a regular expansion in regulators has been discussed, however, with modifications taking into 
account the additional solution 
\cite{HHJ01,Antonov06}. 
The explicit root solutions have been used
in a random walk problem \cite{Goncharenko10}.

In the multi-charge case the stability exponents are eigenvalues of the Jacobi matrix of the set of $\beta$ functions. 
From connections
between charges  
\beq
\label{charge-connection}
g'_i=G_i(g_1,g_2)\,,\qquad i=1,2\,,
\eeq
it follows that
\beq
\label{omega-connection2}
\doo{\beta'_i}{ g'_j}=[J(G)]^{-1}_{ni}\doo{\beta_{m}(g)}{g_n}J(G)_{jm}+[J(G)]^{-1}_{ni}\doo{^2G_j(g)}{g_n g_m}\beta_m(g)\, ,
\eeq
where 
\[
J(G)_{ij}=\doo{G_i}{g_j}
\]
and summation over repeated indices is implied. The matrix transformation of the Jacobi matrix on the right
side leaves eigenvalues intact,
but the $\beta$ functions in the second term produce again non-vanishing terms of higher order
than the matrix transformation of the first term in the double expansion of the Jacobi matrix and thus to
its eigenvalues. Eigenvalues are solutions of algebraic equations containing fractional powers, therefore at
the outset there is no expectation of regularity of the stability exponents, but the very equations are scheme 
independent only up to the consistently calculated order of
the double expansion. The scheme dependence of perturbation expansion has been used in field-theoretic RG approach
to critical dynamics 
to catch qualitative features absent at the leading order of a double expansion \cite{HJJS01,Gladyshev12,Antonov15}
as well as to improve numerical results \cite{AHKV03,AHKV05,AHH10}. It appears that in the momentum-shell RG approach 
this is not considered an issue at all. However, conclusions made on the basis of scheme-dependent behavior
should be corroborated by independent arguments to be reliable.\\

{\subsection{Randomly stirred fluid near two dimensions} \label{subsec:NS-double}}

Let us now analyze the large-distance long-time behavior
of randomly stirred fluid with powerlike correlation
(\ref{NS-eps}) of
the random force near two dimensions. This problem may not be directly related to the
problem of two-dimensional turbulence due to significant physical differences
between turbulence in two and three dimensions. Nevertheless, this analysis allows to
infer useful information about the behavior of the perturbation expansion in the
most generic case which may then be used to improve numerical accuracy of calculation 
of the amplitudes -- which, in general, are scheme dependent -- of the powerlike asymptotics of correlation and response functions.
In the stochastic model of fully developed turbulence a double expansion in dimensional and analytic regulators may be constructed
using the two-dimensional system as the formal starting
point. Two is the critical dimension of the model of randomly stirred fluid \cite{FNS77}, for which an expansion in the
parameter $2\Delta=d - 2$, the deviation of the space dimensionality
from the critical value, has been constructed in full analogy with the theory of the critical phenomena. Near
two dimensions both deviation parameters $\varepsilon$ and
$\Delta $ are small, and a double expansion
may be established. The renormalization is carried out at
the critical values of the parameters, which
leads to the following problem. The long-range correlation function of the random force is a powerlike function
of the momentum $\propto k^{4-d-2\varepsilon}$ and thus, in general, a singular function of
the momentum at the origin. Renormalization gives rise to
regular in the momentum terms only, therefore singular
terms are not renormalized. When $d = 2$ and $\varepsilon=0$, however,
the correlation function becomes a regular function of the
momentum $\propto k^2$. It is not obvious, how the model should
be renormalized in this case. This problem was originally discussed for this model in
\cite{Ronis87}, but due to an incorrect renormalization procedure with false 
 conjectures about the asymptotic behavior of the forced
Navier-Stokes equation. Similar inconsistencies have occurred in the renormalization of magnetohydrodynamics
with long-range correlations of the random force \cite{Fournier82,Adzhemyan85} and
in the renormalization-group approach to two-dimensional turbulence \cite{Olla91,Olla94}.

We consider the stochastic Navier - Stokes equation for
the homogeneous flow of incompressible fluid
\beq
  \nabla_t v_i =\nu_0 \boldnabla^2 v_i 
  - \partial_i p + f_i^v,
  \quad \nabla_t \equiv \partial_t + 
		        \mv\cdot\boldnabla,
  \label{eq:double_NS}
\eeq
 where $\mv(t,\mx)$ is the transverse velocity field, $\nu_0$ is the
 kinematic viscosity. Here and henceforth 
 because of the future use of renormalization group
we distinguish between
unrenormalized (with the subscript ``0'') quantities and renormalized terms
(without the subscript ``0'').
 The renormalized fields will be denoted by the subscript $R$.
 Further,
 $P_{ij}$, is the transverse projection 
  operator, in  the momentum space given as follows
\begin{equation}
  P_{ij} = \delta_{ij} - k_ik_j/k^2.
  \label{eq:double_P}
\end{equation}  
  Further, $p$ is the pressure and $f_i$ is
the random force. Here, and henceforth summation over
repeated indices is implied. As usual \cite{Dominicis79,Adzhemyan83,YO86,YO86PRL}, the random
force is assumed to have a gaussian distribution with zero
mean and the correlation function in the momentum
space of the form
\beq
  \langle f_i^v(t,\mx) f_j^v (t',\mx')\rangle \equiv D_{ij}(t,\mx;t'\mx') =
  \frac{\delta(t-t')}{(2\pi)^d}
  \int \dRM^d \mk\, P_{ij}(\mk)d_f(k) \eRM^{i\mk\cdot(\mx-\mx')},  
  \label{eq:double_random_force_real}
\eeq

\beq
  d_f(k) = D_0 k^{4-d-2\varepsilon},
  \label{eq:double_kernel}
\eeq

\beq
  \langle f_i^v(t,\mk) f_j^v (t',\mk')\rangle = d_f(k) P_{ij}(\mk) \delta(\mk+\mk')
  \delta(t-t').
  \label{eq:double_random_force_fourier}
\eeq
Here, $\varepsilon$ is an arbitrary parameter, the "physical" value
of which is determined by the condition that the parameter $g_0 \nu_0^3$ has the dimension of the energy injection rate
\cite{Dominicis79,Adzhemyan83}. Moreover
as \,$\varepsilon\rightarrow 2,\,$ the amplitude $D_0{\sim} (2-\varepsilon)$\cite{turbo}, from which it follows that
$d_f(k){\sim} \delta(\mk)$, which corresponds to the energy injection by infinitely large eddies.

The connection between $D_{0}$ and $\overline{\E}$ is determined by an exact relation expressing
$\overline{\E}$ in terms of the function $d_f(k)$ in the correlation function (\ref{eq:double_random_force_real})
\begin{equation}
\overline{\E} = \frac{(d-1)}{2 (2\pi)^{d}} \,\int\! \dRM^d {\mk} \, d_f(k).
\label{bal}
\end{equation}
Substituting here function (\ref{eq:double_kernel}) and introducing the UV cutoff  $k\leq \Lambda =
(\overline{\E}/\nu_0^3)^{1/4}$ (the inverse dissipation length), we obtain the following
connection between the parameters $\overline{\E}$ and $D_{0}$
\begin{eqnarray}
D_{0} = \frac{4(2-\eps)\,\Lambda^{2\eps-4}} {\overline
S_{d}(d-1)}\,\, \overline{\E} \,. \label{2.74}
\end{eqnarray}
Idealized injection by infinitely large eddies corresponds to
$d_{f}(k) \propto \delta({\mk})$. More precisely, according to Eq.
(\ref{bal})
\begin{equation}
d_f(k)= {2 (2\pi)^{d}\, \E\, \delta({\mk}) \over d-1}. \label{2.75}
\end{equation}
In view of the relation
\begin{equation*}
\delta({\mk}) = \lim_{\eps\to 2}\,(2\pi)^{-d}\int\! 
\dRM^d {\mx}(\Lambda x)^{2\varepsilon-4}\exp(i{\mk}\cdot{\mx})\\
=S_{d}^{-1}
k^{-d} \lim_{\eps\to 2} \left[ (4-2\eps) (k/\Lambda)^{4-2\eps}
\right],
\end{equation*}
the powerlike injection with
$d_f=D_{0}k^{4-d-2\varepsilon}$ and the amplitude
$D_{0}$ from Eq. (\ref{2.74}) in the limit $\varepsilon\rightarrow 2$ from the
the region $0<\varepsilon<2$ gives rise to the $\delta$ sequence
(\ref{2.75}).

The stochastic problem (\ref{eq:double_NS}) and (\ref{eq:double_random_force_real})
may be cast \cite{Dominicis79,Adzhemyan83} to
a field theory with
the generating functional
\beq
  \G(A) = \int \D v \int \D v' \mbox{ }\eRM^{\S + A v + \tilde{A}v'}
  \label{eq:double_NS_G}
\eeq
where $v'$ is an auxiliary field, $A,\tilde{A}$ are the source fields,
and the action
\begin{align}
  \S[\mv,\mv'] &= \frac{1}{2} \int \dRM t\int\dRM^d\mx \int \dRM t'\int\dRM^d\mx'\mbox{ }v_i'(t,\mx)D_{ij}(t,\mx;t',\mx')
  v_j(t',\mx') \nonumber \\
  &  + \int \dRM t\int\dRM^d\mx \mbox{ }v_i'(t,\mx)\left[-\partial_t v_i(t,\mx)
  +\nu_0 \boldnabla^2 v_i(t,\mx) - v_j(t,\mx) \partial_j v_i(t,\mx)
  \right]\,.
  \label{eq:double_NS_action}
\end{align} 
The canonical scaling dimensions of fields and variables are summarized in Tab. \ref{tab:canon}.
\begin{table}[h!]
\centering
\begin{tabular}{| c | c | c | c | c |}
  \hline\noalign{\smallskip}
  $Q$ & $v$ & ${v'}$ & $\nu_0$ & $g_0$ 
 \\  \noalign{\smallskip}\hline\noalign{\smallskip}
  $d_Q^k$ & $-1$ & $d+1$ & $-2$ & $2\varepsilon$ 
  \\  \noalign{\smallskip}\hline\noalign{\smallskip}
  $d^\omega_Q$ & $1$ & $-1$ & $1$ & $0$ 
  \\  \noalign{\smallskip}\hline\noalign{\smallskip}
  $d_Q$ & $1$ & $d-1$ & $0$ & $2$  
  \\ \noalign{\smallskip}\hline    
\end{tabular}
  \caption{Canonical dimensions of the bare fields and bare parameters 
	  for the model (\ref{eq:double_NS_action}).  }
  \label{tab:canon}
\end{table}
The model (\ref{eq:double_NS_action}) gives rise to the
standard Feynman diagrammatic technique; the bare propagators
(lines in the diagrams) in the time-wave-vector $(t,\mk)$
representation are
\begin{align}
  \bigl\langle v_{i}(t)  v_{j}'(t') \bigr\rangle _0 & =
  \theta(t-t') \exp \left\{ - \nu_0 k^{2} (t-t') \right\} \, P_{ij}({\mk})
  =\raisebox{-0.05cm}{\includegraphics[width=1.5cm]{\PICS 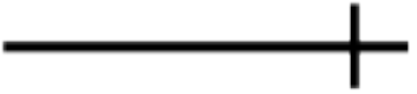}},
  \nonumber \\
  \bigl\langle v_{i}(t) v_{j}(t')
  \bigr\rangle _0
  &= \frac{d_{f}(k)}{2\nu_0 k^{2}}\, \exp \left\{ -
  \nu_0 k^{2} |t-t'| \right\} \, P_{ij}({\mk})
  =\raisebox{0.05cm}{\includegraphics[width=1.5cm]{\PICS 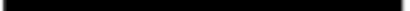}},
  \nonumber \\
  \bigl\langle v_{i}'(t) v_{j}'(t')
  \bigr\rangle _0
  & = 0, 
  \label{eq:double_linesV}
\end{align}
with $d_{f}(k)$ from (\ref{eq:double_kernel}) and the Heaviside step function
$\theta(\dots)$. The interaction in (\ref{eq:double_NS_action}) corresponds to
the three-point vertex 
$-v'(v\partial)v=v'_{i}V_{ijs}v_{j}v_{s}/2$
with the vertex factor
\begin{equation}
  V_{ijs} = {\rm i} (k_{j}\delta_{is}+k_{s}\delta_{ij})
  =\raisebox{-0.40cm}{\includegraphics[height=1.0cm,width=1.0cm]{\PICS 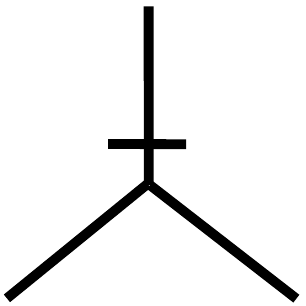}},
  \label{eq:double_vertexV}
\end{equation}
where ${\mk}$ is the wave-vector argument of the field $v'$. The
coupling constant  $g_0\equiv D_0/\nu_0^3$ is the expansion
parameter of the perturbation theory.

The model (\ref{eq:double_NS_action}) is logarithmic, i.e. $d_{g_0} = 0$, when $\varepsilon=0$
independently of the dimension of the space $d$. Due to the
Galilei invariance of the action (\ref{eq:double_NS_action}), the one-particle irreducible (1PI) 
Green function $\Gamma_{v' v v}$, which is superficially
divergent by power counting, is actually convergent \cite{FNS77,Dominicis79,Adzhemyan83}.
Therefore, above two dimensions only the graphs of the
1PI Green function $\Gamma_{v' v}$, yield divergent contributions to
the renormalization of the model, which leads to the
renormalization of the parameter $\nu$.
At $d = 2$, however,
a new set of graphs corresponding to the 1PI Green
function $\Gamma_{v'v'}$ becomes divergent, and they must be taken
into account in the renormalization of the model.

In the stochastic Navier-Stokes problem \cite{Adzhemyan83} critical dimensions are expressed
through $\gamma_\nu$, whose value at the fixed point is determined to all
orders in perturbation theory by the fixed-point equation. Thus, critical dimensions
do not depend on $d$ and thus for them the problem of singularities
in the limit $d\rightarrow 2$ is not relevant.
There are, however, other important physical quantities such as the skewness
factor, Kolmogorov constant, critical dimensions of various composite operators
to which this problem persists. It is important that for these quantities the
problem of anomalous scaling is absent, which cannot be treated in the framework
of the model with massless injection (\ref{eq:double_kernel}) lacking a dimensional parameter
to account for the external scale of turbulence.

For such quantities, the solutions contain full series of the form
\begin{equation}
R(\eps,d)=\sum_{k=0}^\infty R_k(d)\eps^k ,  \label{Reps}
\end{equation}
and the coefficients
$R_k(d)$ in the limit $d\rightarrow 2$
reveal singular behavior of the type
$\sim(d-2)^{-k}\sim\Delta^{-k}$
($2\Delta\equiv d-2$)
giving rise to the growth of the relative part of the correction terms at
$d\rightarrow 2$. The effect of these is fairly discernable also at the
real value $d=3$, hence the natural desire to sum up contributions of the form
$(\varepsilon/\Delta)^k$ at all orders of the
$\varepsilon$ expansion (\ref{Reps}). This may be done with the aid of the
double ($\varepsilon$, $\Delta$) expansion \cite{Ronis87,HonNal96}. 

In a fixed dimension $d>2$ the  value $\varepsilon=2$
corresponds to the ''real problem''. Calculations in the
framework of the
$\varepsilon$ expansion have a rigorous meaning only in the vicinity of
$\varepsilon=0$, whereas continuation of the results to the ''real'' value
$\varepsilon=2$ is always understood as an extrapolation. In the scheme
applicable for $d>2$ this extrapolation corresponds to the continuation
along the vertical ray from the point
$(d,\,\varepsilon=0)$ to the point  $(d,\,\varepsilon=2)$
in the ($d$, $\varepsilon$) plane. The same final point may be reached along
a ray from any starting point
$(d_0\neq d,\,\varepsilon=0)$ at which the model is logarithmic as well.
The extrapolation along the ray starting from the origin
$(d_0=2,\,\varepsilon=0)$ is, however, singled out, because at $d=2$ in model (\ref{eq:double_NS_action})
an additional UV divergence (absent at $d>2$) occurs in the 1PI function
$\Gamma_{v'v'}$. On such a ray we put
\begin{eqnarray}
d=2+2\Delta,\qquad\Delta/\varepsilon=\zeta=const.
 \label{dz}
\end{eqnarray}
The parameters $\varepsilon$ and  $\Delta$ are considered small of the same order
and their ratio $\Delta/\varepsilon=\zeta$  a fixed constant
[$\zeta=1/4$ in the extrapolation to the point
$(d=3,\,\varepsilon=2)$].

Extraction of contributions of the order
$\varepsilon^m$ with
$\Delta/\varepsilon=const$ corresponds to the account of all contributions
of the form
$\varepsilon^m(\varepsilon/\Delta)^n$ with any $n=0,\,1,\,2...$
and $m+n=k$ in Eq.
(\ref{Reps}). Thus the use of the ($\varepsilon$, $\Delta$) expansion
in such a form is directly related to the problem of the account of
the singularities at $\Delta\rightarrow 0$ pointed out in the discussion of
relation (\ref{Reps}).

\begin{figure}[h!]
\centering
\includegraphics[width=9cm]{\PICS 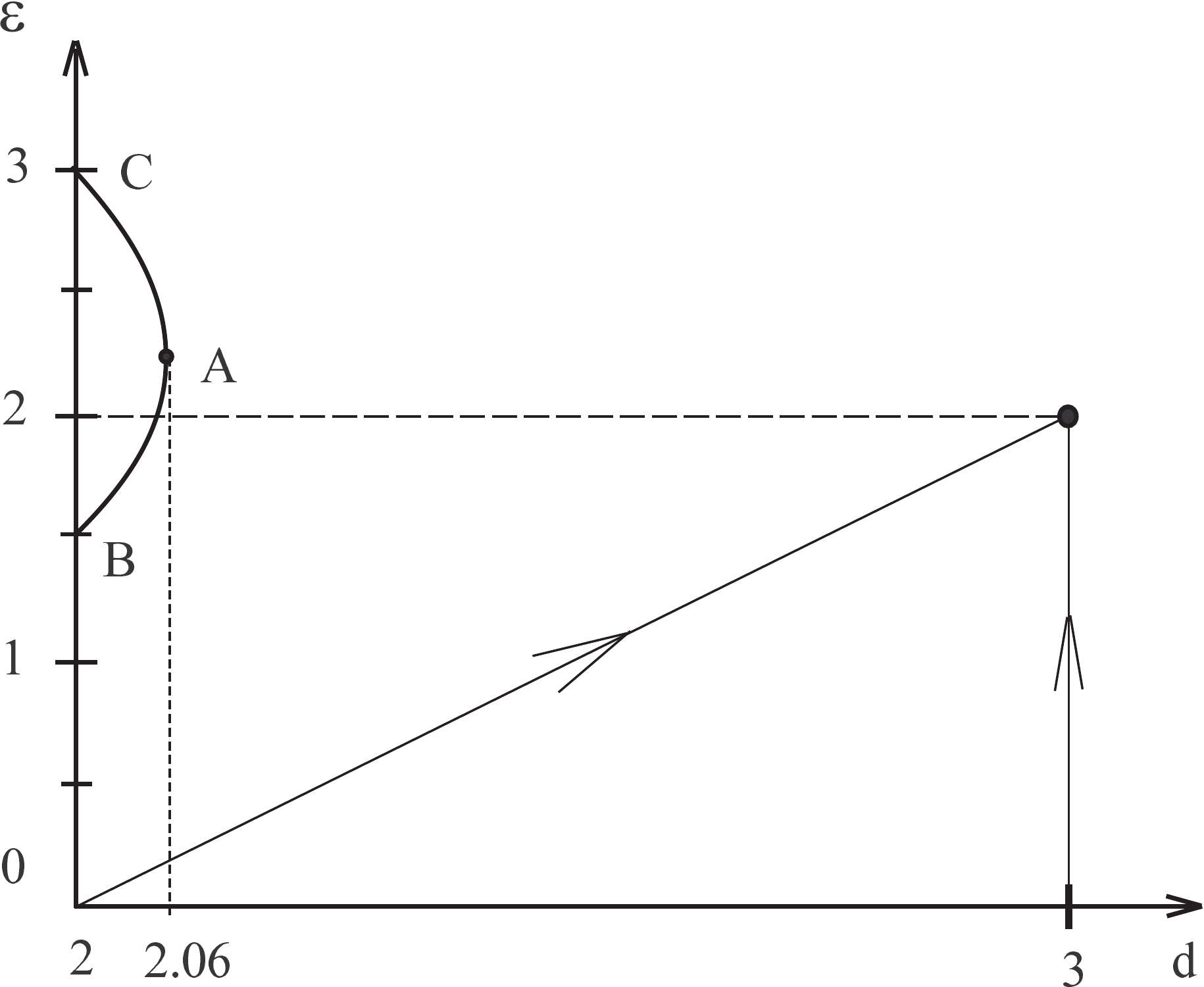}
\caption{\label{frish2}The borderline BAC between the regions of parameter space
$d$, $\eps$ corresponding to direct (to the right from the curve BAC) and inverse
(to the left) energy cascades.}
\end{figure}

It is worth emphasizing that the very process of extrapolation along a ray
from the starting point
$(d=2,\,\varepsilon=0)$ is inapplicable to description of two-dimensional
turbulence in which the physics is totally different from the three-dimensional
problem due to the appearance of the inverse energy cascade \cite{Frisch}.
In Fig. \ref{frish2} we have plotted the borderline curve BAC between the
direct (normal) and inverse energy cascades obtained in Ref. \cite{FNS77}.
The starting point of the extrapolation for the two-dimensional case
$(d=2,\,\varepsilon=0)$ lies in the region of the direct cascade, whereas the final
point $(d=2,\,\varepsilon=2)$ in the region of the inverse cascade. Thus the ray
connecting these points intersects the borderline -- the curve BAC -- so that the
extrapolation becomes impossible. However, the ray connecting the starting point
$(d=2,\,\varepsilon=0)$ and a final point like
$(d=3,\,\varepsilon=2)$ lies completely in the region of the direct cascade, therefore on
such a ray the problem of the change of the cascade pattern does not arise. The rightmost
point of the region of the inverse cascade (point A on Fig. \ref{frish2}) has the coordinate
$d_A\simeq 2.06$ \cite{FF78}. In the preceding discussion of the extrapolation along
the vertical ray from the point
$(d,\,\varepsilon=0)$ to the point $(d,\,\varepsilon=2)$ at $d>2$,
it should have been noted that the condition is not simply $d>2$, but $d>d_A=2.06$.
 From the practical point of view this is irrelevant, because we are interested in the
space dimension $d=3$.

The idea of the double
($\varepsilon$,$\Delta$) expansion together with the extrapolation
along the ray
$\Delta\sim\varepsilon$ of relation (\ref{dz}) in the context of the present problem was first
put forward in Ref. \cite{Ronis87}. The UV divergences are present not only
in the 1PI function
$\Gamma_{v'v}$ but also in
$\Gamma_{v'v'}$ and appear in the form of poles in the parameters
$\varepsilon$ and $\Delta$ and linear combinations thereof, or, equivalently,
as poles in
$\varepsilon$ with the fixed ratio
$\Delta/\varepsilon\equiv\zeta=const$. To remove the additional divergences
from the graphs of the 1PI function
$\Gamma_{v'v'}$ renormalization of the amplitude
$D_0$ in the nonlocal correlation function of the random force
(\ref{eq:double_random_force_real}) and (\ref{eq:double_kernel}) was used in Ref. \cite{Ronis87}.
The renormalization scheme of Ref. \cite{Ronis87} is not internally consistent, however. This
is not obvious in the one-loop approximation, but becomes
apparent already in the two-loop approximation \cite{AHKV05}.

In the
($\eps$ ,$\Delta$) scheme
(\ref{dz}) the multiplicative renormalization \cite{Ronis87}
of the amplitude $D_0$ in Eq. (\ref{eq:double_kernel}) is not acceptable. The reason
is that the counterterm with structure (\ref{eq:double_kernel})
is nonlocal $\sim k^{4-d-2\eps}=k^{2-2\Delta-2\eps}$ on rays (\ref{dz}).

Guided by the general theory of the UV renormalization, the authors of Ref. \cite{HonNal96}
put forward another scheme, in which a local counterterm  $\sim k^2$ instead of the nonlocal
one $ \sim k^{2-2\Delta-2\eps}$ is used to absorb singularities from the graphs of
the 1PI function $\Gamma_{v'v'}$.
This corresponds to addition of the term
$\sim v'\boldnabla^2 v'$ 
to the action functional. In functional (\ref{eq:double_NS_action})
with the correlation function $D$ from Eqs. (\ref{eq:double_random_force_real}) and (\ref{eq:double_kernel}) there is no such term,
so that upon the addition of the term
$\sim v'\boldnabla^2 v'$ 
the renormalization ceases to be multiplicative.
This would be unessential, if our only goal was the elimination of divergences from
Green's functions which is quite possible by a non-multiplicative renormalization. For the
use of the standard technique of the RG multiplicative renormalization is, however, necessary.
This is why the authors of Ref. \cite{HonNal96} proposed to consider a two-charge model in which
to function (\ref{eq:double_kernel})
$\sim k^{4-d-2\eps}=k^{2-2\Delta-2\eps}$ the term
$\sim k^2$ is added at the outset with an independent coefficient:
\begin{equation}
    d_f(k)=D_{10} k^{2-2\Delta-2\eps}+D_{20} k^2
    = g_{10}  \nu_0^{3}\,k^{2-2\Delta-2\eps}+g_{20} \nu_0^{3}\, k^2\,.
\label{nakach2}
\end{equation}
Here, the amplitude
$D_0$ of Eq. (\ref{eq:double_kernel}) is denoted by $D_{10}$. The parameters  $g_{10}$ and $g_{20}$
introduced in Eq. (\ref{nakach2}) play the role of two independent bare charges.

We investigate the model in $d = 2 + 2\Delta$ dimensions
regarding $\varepsilon$ and $\Delta$ as small parameters of
a regular expansion. The renormalized action is
\beq
  \S_R = \frac{1}{2} g_1\nu^3 \mu^{2\varepsilon} v' D v' +
  \frac{1}{2} Z_{D_2} g_2\nu^3 \mu^{-2\Delta} \partial_i v'_j \partial_i v'_j
  + \mv'\cdot[-\partial_t \mv + Z_\nu \nu\nabla^2 \mv - (\mv\cdot\boldnabla)\mv],
  \label{eq:double_NS_action_ren}
\eeq
where $\mu$ is the scaling-setting parameter (or renormalization mass). Note that in
this work we always interpret $\mu$ in this way. Further, the 
renormalized parameters $\nu$ and $g$ are defined by $\nu_0 = \nu Z_\nu$ and
$g_{10} = \mu^{2\varepsilon} g_1Z_{g_1}$, $g_{20} = \mu^{-2\Delta} g_2Z_{g_2}$.
As usual, the renormalized coupling constants $g$ are chosen to be both spatially and temporally
dimensionless. The non-local term of the action (\ref{eq:double_NS_action_ren}) is not
renormalized, therefore the renormalization constants
$Z_{g_1}$ and $Z_\nu$, are related as
\beq
   Z_{g_1} = Z_\nu^{-3}.
   \label{eq:double_NS_relation}
\eeq
We have introduced a new parameter $g_2$
with the canonical scaling dimensions $d_{g_{20}}^k = 2 - d=-2\Delta$,
 $d_{g_{20}}^\omega = 0$, $d_{g_{20}} = -2\Delta$ to account for the new divergences at
two dimensions. This is necessary, because the counterterms from the graphs are always local in space and time,
and therefore cannot be taken into account by renormalization of the parameters of the initial action, in which the
$v'v'$ term has a non-local kernel.
This amendment is crucial to the correct renormalization of the model.
This subtle point has caused confusion on several
other occasions \cite{Fournier82,Adzhemyan85,Olla91,Olla94}. The anomalous asymptotic behavior
in the present problem arises from the singularities of
the perturbation expansion at small momenta and frequencies. The region of large momenta is assumed to have
a physical ultraviolet cutoff parameter $\Lambda$ inversely proportional to the typical microscopic length of the problem.
Here, the cutoff is most conveniently implemented in the
correlation function of the random force, which therefore
is a rapidly decaying function at large momenta. For
instance, the substitution
\beq
  k^{4-d-2\varepsilon} \rightarrow k^{4-d-2\varepsilon} \eRM^{-k^2/\Lambda^2}
\eeq
would suffice. It is not difficult to see by power counting that there are no singularities in the infrared limit, when
$\varepsilon < 0$ and $\Delta > 0$ and the ultraviolet-regularized perturbation expansion may be used as it stands. 
If $\varepsilon \ge 0$ or $\Delta \le 0$
small-momentum singularities do occur. At present, there
is no way to treat these infrared singularities of the 
perturbation expansion in a consistent manner, except for the
logarithmic case $\varepsilon = \Delta = 0$, in which the analysis can be
entirely transferred to the analysis of ultraviolet divergences. 
By the scale transformation $\mk \rightarrow s\mk$, $\omega \rightarrow s^2 \omega$ of all the
momenta and frequencies the analysis of the behavior of the perturbation expansion at small momenta and
frequencies, i.e. in the limit $s \rightarrow 0$, may be transferred to the
large momentum limit of the momentum integrals of the perturbation expansion, since the cutoff becomes $\Lambda/s$ and
there is no other $s$-dependence left in these integrals.
However, the coupling constants scale as $g_{10} \rightarrow g_{10} s^{-2\varepsilon}, g_{20} \rightarrow g_{20} s^{2\Delta}$,
 and the effective expansion parameters $g_{10} s^{-2\varepsilon}$, 
 $g_{20} s^{2\Delta}$ become large in the limit $s \rightarrow 0$, when ($\varepsilon > 0$ or $\Delta < 0$. 
 In general, the effect of these singular terms cannot be estimated in any consistent manner. However, in the 
 logarithmic case $\varepsilon = \Delta = 0$, and the coupling constants remain fixed. Then, by the 
 standard procedure of the field-theoretic renormalization \cite{Collins,HonNal89,Adzhemyan88,Antonia84} the 
 ultraviolet singularities
of the integrals may be absorbed in a redefinition of the parameters of the model.
The renormalization relevant to the anomalous
asymptotic behavior must therefore be carried out at the
critical values of the parameters $\varepsilon=0$, $d = 2$, which
correspond to the logarithmic model. The model cannot
be consistently renormalized at any other values of $\varepsilon>0$
and $d < 2$. This is a somewhat dangerous point, since the inconsistencies do not show in any way in the one-loop
calculations, which are usually performed, although they become obvious
already in two-loop order \cite{AHKV05}. When the model is renormalized at $d = 2$, the two terms 
$\propto v' v'$ of the action (\ref{eq:double_NS_action_ren}) are indistinguishable: both are of the form
$const \times\int \dRM\omega \int \dRM^d\mk \mbox{ } v'_i (\omega,\mk) k^2 v'_i(-\omega,-\mk)$ 
and the question arises, which one of them should be renormalized? 
To answer this question, let us look at the renormalization
procedure at $d > 2, \varepsilon<0$. In this case the terms of the
ultraviolet-regularized perturbation expansion do not
contain any infrared divergences and there is no anomalous asymptotic behavior brought about by the higher
order terms. We may, of course, carry out renormalization
of the model also in this case in the usual way, which
means that we perform the first steps of the gradient
expansion of the perturbation expansion, and add all the
higher order perturbative contributions to the coefficients
of the corresponding terms of the action (\ref{eq:double_NS_action_ren}). The renormalization constants
obtained are finite and do not give
rise to any anomalous asymptotic behavior of the Green
functions. The important feature here is that the gradient
expansion produces only terms, the Fourier transforms of
which are polynomial functions of the momenta, at least if
the large-momentum cutoff is chosen wisely to be smooth
enough. 

Therefore, the non-local term $v' k^{4-d-2\varepsilon}v'$
is not renormalized at all in this case, when it is not a polynomial of the momentum and thus it is clearly
distinguishable from the local term $v'_i k^2 v'_i$. When we go to the limit
$\varepsilon\to 0$, $d\rightarrow 2$, the renormalization constants become singular reflecting the
infrared divergences of the integrals discussed above. Therefore, it is reasonable to prescribe all
the contributions $\propto v'v'$ produced by the renormalization to the local term 
 $\partial_iv'\partial_iv' $ also in the limit $\varepsilon\to 0$, $d \rightarrow 2$. Another
 argument to support this renormalization prescription is obtained, when dimensional and analytic
regularization are used. When the renormalization constants are calculated with $\varepsilon$ and $\Delta$ as 
regulators, the long-range term stands in the non-analytic form $v' k^{2-2\varepsilon-2\Delta} v'$, therefore
 the polynomial in $k^2$ counterterms do not renormalize it, but the short-range term $v' k^2v'$. It
 should be noted that this local term is
always produced by the renormalization, but it is irrelevant for $d > 2$. To keep the model multiplicatively 
 renormalizable, the local term with an independent coupling
constant should be added to the original action at the outset.\\
{\subsubsection{Renormalization-group equations and fixed points} \label{subsubsec:double_renorm}}

From the connection between the renormalized and unrenormalized generating functionals 
$\G_R(g_1, g_2, \nu, \mu ) = \G(g_{10}, g_{20}, \nu_0)$
(there are enough parameters available, no field renormalization is needed here) we obtain the basic RG equation
\begin{equation}
  \D_{RG} \W_R \equiv [\D_\mu + \beta_1\partial_{g_1}+ \beta_2\partial_{g_2} - \gamma_\nu \D_\nu]\W_R = 0
  \label{eq:double_RG_eq}
\end{equation}
for the generating functional $\W_R = \ln \G_R$ of the connected
renormalized Green functions.

To illustrate the idea of asymptotic analysis, we consider the equal-time velocity-velocity 
correlation function   $G_{ij}(\mx-\mx')=\langle v_i(t,\mx)v_j(t,\mx')\rangle$.
It is convenient to express
the Fourier transform of the correlation function
\begin{align}
  \langle v_i(t,{\mx}) v_j (t,{\bm x'})\rangle \equiv
  G_{ij}({\mr})\,,\qquad {\mr}\equiv {\mx}-{\bm x'}
  \label{eq:double_Gr}
\end{align}
in the form
\begin{equation}
  G_{ij}({\mp})= P_{ij}({\mp})G(p),
  \label{eq:double_Gp}
\end{equation}
where $ P_{ij}({\mp})$ is the transverse projection operator and $p\equiv |\mp|$.
By dimensional arguments the scalar function $G(p)$ can be expressed as
 \begin{equation}
  G(p) =  \nu^2p^{-d+2} R(s,g), \quad s \equiv 
  p/\mu
  , \quad g\equiv
  (g_1,g_2),
  \label{eq:double_dimG}
\end{equation}
where $R$ is a scaling function of dimensionless arguments. Introduce a set of invariant
parameters $\bar e (s) = (\bar \nu(s),
\bar g_1(s), \bar g_2(s))$ corresponding to the set of renormalized parameters $e=(\nu,g_1,g_2)$ as solutions
fixed bare parameters $e_0$. In terms of invariant parameters the correlation function assumes the form
\begin{equation}
  G(p) =  \nu^2p^{2-d} R(s,g)=\bar\nu^2 p^{2-d}R(1,\bar g).
  \label{eq:double_dimG1}
\end{equation}
Equation (\ref{eq:double_dimG1})
is valid because both sides of it satisfy the RG equation and
coincide at $s=1$ owing to the normalization of the invariant
parameters. The right-hand side of (\ref{eq:double_dimG1}) depends on $s$ through the invariant parameters $ \bar e(s,e)$.
They have simple asymptotic behavior as 
 $s\rightarrow 0$, which is governed by the infrared-stable fixed
point: the invariant charges $\bar g$ tend to the fixed-point values
$g^{*}=\Ord{\varepsilon}$ and the invariant coefficient of viscosity $\bar \nu$ exhibits
simple power-law behavior. To determine the latter it is
convenient to express the invariant parameters $\bar e = (\bar \nu,\bar g_1, \bar g_2)$ in terms of
the bare variables $e_0=(\nu_0, g_{10},g_{20})$ and the wave number  $p$. Due to definition the bare
variables $e_0$ also satisfy the RG equation
by relations
\begin{equation}
  \nu_0=\bar \nu Z_{\nu}(\bar g), \qquad g_{10}=\bar g_1
  p ^{2\varepsilon}Z_{g_1}(\bar g), \quad g_{20}=\bar g_2 p^{-2\Delta}Z_{g_2}(\bar g).
  \label{eq:double_119}
\end{equation}
Relations (\ref{eq:double_119}) are valid because
both sides of them satisfy the RG equation, and because relations
(\ref{eq:double_119}) at 
$s\equiv p/\mu = 1$ coincide with their counterparts in (\ref{eq:double_ZZ}) owing
to the normalization conditions. Using the
connection $Z_gZ_\nu^3=1$ between the renormalization constants defined in (\ref{eq:double_ZZ}), and
eliminating these constants from the first two expressions in (\ref{eq:double_119})
we find
$g_{10}\nu_0^3=D_{10}={\bar g_1}p^{2\varepsilon}\,{\bar\nu}^{\,3}$ \,, from which it follows that
\begin{equation}
  \bar \nu=(D_{10}p^{-2\varepsilon}/\, {\bar g}_{1})^{1/3}\,.
  \label{eq:double_4}
\end{equation}
In the limit $\bar
g_1 \rightarrow g^*_{1}$  the sought asymptotic behavior of the invariant coefficient of viscosity  as $s\rightarrow 0$
thus assumes the form
\begin{equation}
  \bar\nu \to \bar \nu^*=( D_{10} / g^*_{1})^{1/3}p^{-2\varepsilon/3}, \quad \quad s\rightarrow 0.  
\end{equation}
Substituting this result into (\ref{eq:double_dimG1}) we obtain the relation
 \begin{equation}
  G(p) \simeq (D_{10} / g^*_{1} )^{2/3} p^{2-d-4\varepsilon/3}R(1,
  g^{*}), \quad \quad s\rightarrow 0
  \label{eq:double_asymp}
\end{equation}
describing the large-scale asymptotic behavior of the pair correlation function.

For the physical values of the parameters $\Delta = 1/2, \varepsilon= 2$,
chosen from the condition that the dimensional parameters of the model are viscosity and energy injection rate,
 the scaling behavior of the equal-time correlation function 
 $G$ in the three-dimensional
space corresponds to the Kolmogorov scaling $G (p) \sim p^{-11/3} $ \cite{Dominicis79,Adzhemyan83}.
 The scaling form (\ref{eq:double_asymp}) yields the
large-scale asymptotic behavior of the original correlation function, if the fixed point is infrared 
stable, i.e. if $\overline{g}_1\rightarrow g^*_{1}$, $\overline{g}_2\rightarrow g^*_{2}$, when $p \rightarrow 0$.

One-loop calculation in the ray scheme yields the following expressions
for the $\beta$ functions:
\begin{equation}
   \beta_1=g_1'\left(2\Delta-3g_2'-3g_1'\right) ,\quad \beta_2=g_2'\left(2\varepsilon-\frac{{g_1'}^2}{g_2'}+g_1'+2g_2'\right),
   \label{eq:double_beta}
\end{equation}
where $g_2'= g_2/(32\pi)$, $g_1' = g_1/(32\pi)$. Three fixed points are
determined by the system of equations $ \beta_1(g^*_{1},g^*_{2}) =\beta_2(g^*_{1},g^*_{2}) = 0$.

 From relation (\ref{eq:double_dimG1}) near a fixed point it follows
that the fixed point is infrared stable, when the matrix
$\Omega_{ij}$ defined in Eq. (\ref{eq:RG_Omega}) is positively definite at the fixed point. The
trivial fixed point: ${g'_1}^* = {g'_2}^* = 0$ is infrared stable only if
$\Delta > 0$ , $\varepsilon < 0$. For the nontrivial fixed point ${g'_1}^*= 0$, ${g'_2}^*= -\Delta$ the region
of stability is determined by the
inequalities $\Delta < 0, 2\varepsilon < - 3\Delta$. This fixed point corresponds
to the fixed point for the model A of Forster, Nelson and
Stephen \cite{FNS77} and yields the asymptotic behavior of the
solution of the Navier-Stokes equation with short-range
correlated random force. Physically, this corresponds to the effect of thermal fluctuations.

The anomalous asymptotic behavior of the long-range model above two dimensions is
governed by the third fixed point
\beq
\label{LR-FP}
{g'_1}^* = \frac{2}{9}\,\frac{\varepsilon(3\Delta + 2\varepsilon)}{\Delta + \varepsilon}\,,\qquad {g'_2}^* = \frac{2}{9}\,
\frac{\varepsilon^2}{\Delta + \varepsilon}\,,
\eeq
at which the $\Omega$ matrix is
\begin{equation}
   \Omega = \begin{pmatrix}
  \frac{2(2\varepsilon^2+3\varepsilon\Delta)}{3(\varepsilon+\Delta)} & \frac{2(2\varepsilon^2+3\varepsilon\Delta)}{3(\varepsilon+\Delta)}  \\
  -\frac{2(2\varepsilon^2+3\varepsilon\Delta)}{3(\varepsilon+\Delta)} & \frac{2(3\Delta^2+4\varepsilon\Delta+2\varepsilon^2)}{3(\varepsilon+\Delta)}
 \end{pmatrix}
   \label{eq:double_matrix}
\end{equation}
with the eigenvalues
\begin{equation}
   \Omega_{1,2} = \frac{1}{3}[3\Delta+4\varepsilon \pm \sqrt{9\Delta^2-12\Delta\varepsilon-8\varepsilon^2}]
   \label{eq:double_eigenvalues}
\end{equation}
It should be noted that the fixed point ${g'_1}^*\ne 0$, ${g'_2}^*\ne 0$ is unique due to the degeneracy of the $\beta$ functions in the ray scheme.

The inequalities $\Delta + (2/3)\varepsilon > 0, \varepsilon > 0$ determine the basin
of attraction of the fixed point (\ref{LR-FP}), in which the present
results may be compared with those of the RG analysis
above two dimensions \cite{Dominicis79,Adzhemyan83}. The anomalous dimension $\gamma_\nu^*$ is a continuous
function of the parameters $\Delta$ and $\varepsilon$ at the borderlines of the
basins of attraction on the $(\Delta, \varepsilon)$-plane.\\
{\subsection{Improved $\varepsilon$ expansion in the RG analysis of turbulence} \label{subsubsec:double_tur}}
In the description of developed turbulence in the
framework of the stochastic Navier-Stokes equation representation of the problem in the form
of an effective field-theoretic model opens new possibilities for
understanding of both the Kolmogorov scaling and the deviation therefrom as well as
for calculation of principal physical quantities.
The RG approach with its various perturbation
schemes (e.g. the famous $\varepsilon$ expansion \cite{WilKog74}),
with effective analytical and numerical algorithms for evaluation of quantities involved forms a robust
method, which allows to put into practice this scenario.

A specific feature of the renormalization-group approach and the
$\varepsilon$ expansion in the theory of developed turbulence is that
the formal small parameter $\varepsilon$ is not
connected with the space dimension and it is determined only by the noise correlator of random forcing in the stochastic
Navier-Stokes equation \cite{Dominicis79,Adzhemyan83}.
Its physical value $\varepsilon=2$ is not small \cite{Adzhemyan96,turbo}, hence reasonable doubts arise
about the effectiveness of such an expansion.
For some paramount physical quantities like the critical dimensions of the velocity field and the
coefficient of viscosity the $\varepsilon$
expansion terminates at the first term due to the Galilei invariance of the theory \cite{Dominicis79,Adzhemyan83}. Therefore,
exact values are predicted for these quantities. However, there are other physically important quantities, viz. the skewness
factor, the Kolmogorov constant and critical dimensions of various composite operators, for which the $\varepsilon$ series
do not terminate \cite{Adzhemyan03a,AHKV03,AHKV05}, therefore the question about the effectiveness of the expansion remains open.

Consider a quantity $A$ calculated at the fixed point of the RG in the renormalized field theory of developed turbulence. In $d$
dimensions it is a function of the parameters $\varepsilon$ and $d$: $A=A(\varepsilon,d)$.
In practice, calculations are often carried out in the $\varepsilon$ expansion, whose coefficients
for the quantity $A(\varepsilon,d)$
depend on the space dimension $d$
\begin{equation}
  A(\varepsilon,d)=\sum_{k=0}^\infty A_k(d)\varepsilon^k .  
  \label{eq:double_Aeps}
\end{equation}
Analysis shows that these coefficients  $A_k(d)$ have singularities at small dimension  $d \leq 2$. The singularity at $d=2$ --
the nearest to the physical value
$d=3$ -- gives rise to new divergences as $d\rightarrow 2$ not eliminated by the renormalization of the
$d$-dimensional theory \cite{AHKV05,Ronis87,HonNal96}. These divergences manifest themselves in the form of poles
in the parameter $(d-2)\equiv 2 \Delta$ in
the coefficients of the $\varepsilon$ expansion  $A_k(d)$, which therefore may be expressed as Laurent series of the form
\begin{equation}
  A_k(d) = \sum_{l=0}^\infty a_{kl}\, \Delta^{l-k}.
  \label{eq:double_Qq}
\end{equation}
A two-loop calculation of the Kolmogorov constant and skewness factor at various values of space
dimension  $d$ carried out in \cite{Adzhemyan03a} has shown that at $d=3$ the relative part of the two-loop contribution is
just large -- it is of order 100\% in comparison with the one-loop contribution. The two-loop contribution, however, rapidly
decreases as  $d$ increases, and at
$d=5$ it gives only 30 \%, and  at $d\rightarrow\infty$
decreases to 10 \%. On the contrary, when the space dimension decreases from $d=3$ to
$d=2$ rapid growth of the two-loop correction term is observed. This growth is due to diagrams which contain 
singularities at  $d = 2$.
Analysis has shown that it is just these diagrams which form the main part of two-loop contribution at $d=3$. Therefore, the nearest 
singularity strongly
manifests itself at the realistic value  $d=3$ and allows to improve the $\varepsilon$ expansion by 
means of summation of
singular contributions in all orders of this expansion \cite{AHKV03,AHKV05}.

\begin{figure}[h]
\begin{center}
  \includegraphics[width=106mm]{\PICS 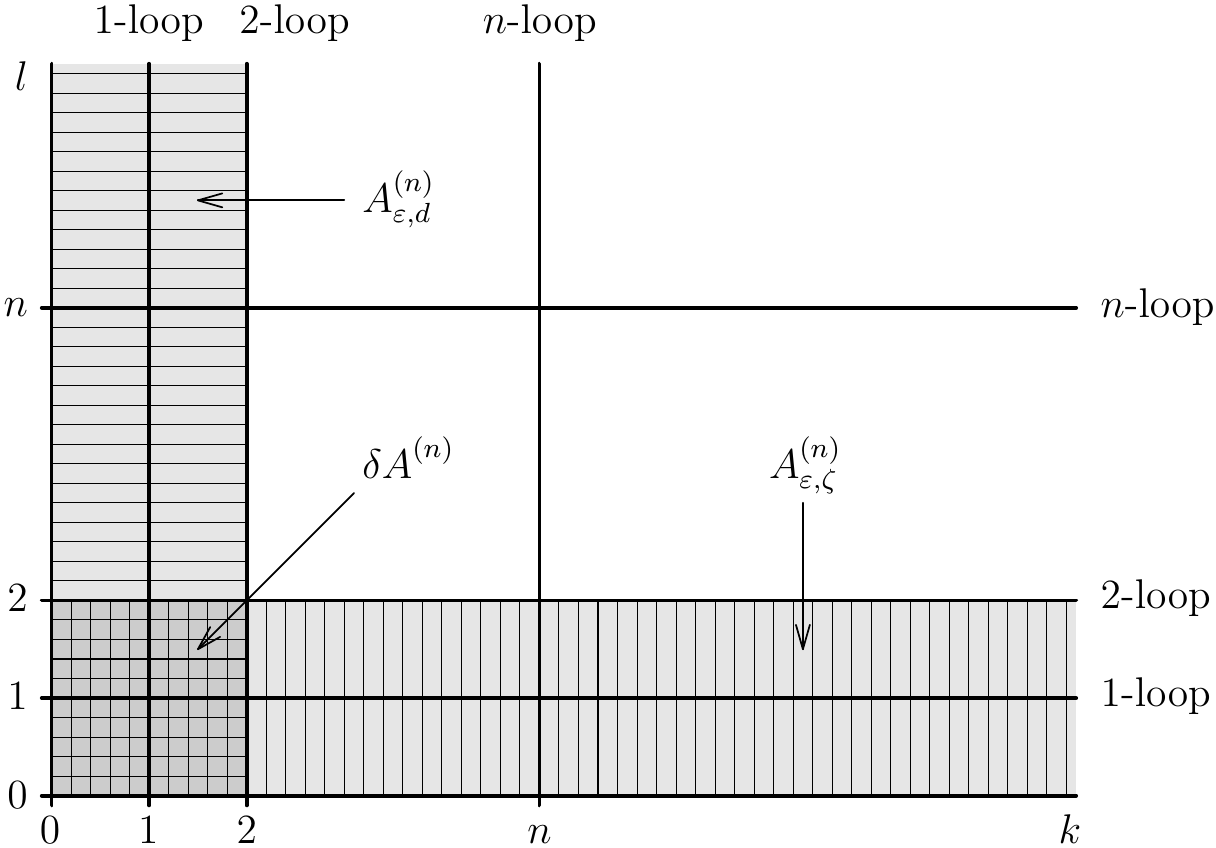}
  \end{center}
\caption{\label{fig:double_qed1}
Illustration of the subsequences of the double sum (\ref{eq:double_Aed})
summed in the calculation in the usual $\varepsilon$ expansion (\ref{eq:double_Aeps})
and in the alternative expansion (\ref{eq:double_Aeps1})
Terms in the double sum (\ref{eq:double_Aed})
taken into account in the approximations $A^{(n)}_{\varepsilon,\,d}$  (\ref{eq:double_Qen})
and $A^{(n)}_{\varepsilon,\, \zeta}$  (\ref{eq:double_Qdn}) at the two-loop order
correspond to the shaded
horizontal and vertical stripes, respectively. The correction term
$\delta A^{(n)}$ corresponds
to the double sum in (\ref{eq:double_Qeffn}) over the double-shaded square.}
\end{figure}

Divergences in $\Delta $ may be absorbed to suitable additional counterterms, which gives rise to a
different renormalized field theory
(the physical unrenormalized field theory is, of course, the same in both cases).
Henceforth, we therefore consider the theory with two formal small parameters  $\varepsilon$ and $\Delta$, which
satisfy the relation
$\zeta\equiv \Delta/\varepsilon = const$,
and then construct a new
$\varepsilon$ expansion proposed by Honkonen and Nalimov \cite{HonNal96}, an alternative to the expansion (\ref{eq:double_Aeps})
\begin{equation}
  A(\varepsilon,\zeta)=\sum_{k=0}^\infty b_k(\zeta)\varepsilon^k \,,\qquad
  \zeta\equiv \Delta/\varepsilon\,.
  \label{eq:double_Aeps1}
\end{equation}
Relation between the two expansions (\ref{eq:double_Aeps}) and (\ref{eq:double_Aeps1}) of the quantity $A$ becomes evident
upon substitution of the Laurent series (\ref{eq:double_Qq}) into the $\varepsilon$ expansion
(\ref{eq:double_Aeps}), which gives rise to the
double series
\begin{equation}
  A(\varepsilon, d)=\sum_{k=0}^\infty \sum_{l=0}^\infty a_{kl} \,
  \frac{\Delta^l}{\zeta^k}. 
  \label{eq:double_Aed}
\end{equation}
To clarify the connection between representations (\ref{eq:double_Aeps}), (\ref{eq:double_Aeps1}) and
(\ref{eq:double_Aed}), consider
Figure \ref{fig:double_qed1}. The point ($k,l$) corresponds to a term with the coefficient $a_{kl}$ in the double
sum (\ref{eq:double_Aed}).
Thus, all points of the first quadrant in the $k,l$ plane correspond to the full double sum (\ref{eq:double_Aed}).
In the usual $\varepsilon$ expansion (\ref{eq:double_Aeps}) the index of summation labels columns in the graphical representation 
of the double sum
(\ref{eq:double_Aed}). In the alternative expansion (\ref{eq:double_Aeps1}) the index of summation labels rows of
points in Figure \ref{fig:double_qed1}
(see \cite{AHKV05} and also \cite{AHH10}). Therefore,
the coefficients $b_k$ are expressed as
\begin{equation}
   b_k(\zeta)=\sum_{l=0}^\infty
   a_{lk} \,\zeta^{k-l}. 
   \label{eq:double_b}
\end{equation}
An $n$-loop calculation in the usual scheme leads to the approximate expression
\begin{equation}
   A^{(n)}_{\varepsilon,\,d} \equiv \sum_{k=0}^{n-1} A_k(d)\varepsilon^k
  \label{eq:double_Qen}
\end{equation}
for the quantity sought,
which corresponds to the inclusion of all terms of the double sum in the first $n$ vertical 
bands shown in Figure \ref{fig:double_qed1}.
Calculation in the alternative scheme gives rise to the approximation
\begin{equation}
  A^{(n)}_{\varepsilon,\, \zeta} \equiv \sum_{k=0}^{n-1} b_k(\zeta)\varepsilon^k ,
  \label{eq:double_Qdn}
\end{equation}
which includes all terms in $n$ horizontal bands.

The $\varepsilon$ expansion may be improved  
by the use of the complementary
information about the quantity $A$ contained in the finite sums (\ref{eq:double_Qen}) and (\ref{eq:double_Qdn}). This is
carried out by means of the approximation
\begin{equation}
  A^{(n)}_{eff} = A^{(n)}_{\varepsilon,\,d} + A^{(n)}_{\varepsilon,\,\zeta} -
  \sum_{k=0}^{n-1} \sum_{l=0}^{n-1} \zeta^{-k} a_{kl} \, \Delta^l\,,
  \label{eq:double_Qeffn}
\end{equation}
which includes all terms in $n$ vertical and horizontal bands simultaneously
("$n$ region") and also eliminates the double counting in the overlap region of these bands.
These calculations have led to notable improvement of the agreement of calculated values of the Kolmogorov constant
and skewness factor with their experimental values \cite{AHKV03,AHKV05} in comparison with the results of the
usual $\varepsilon$ expansion \cite{Adzhemyan03a}. We recall that calculations in \cite{Adzhemyan03a,AHKV03,AHKV05}
were carried out in two-loop order.

The MS ray scheme was used in \cite{AHKV03,AHKV05}. It turns out that even better numerical performance may be obtained
with the use of the scheme with a normalization point, in which the renormalization constants are normalized
by prescribing to suitable renormalized correlation functions finite values at given wave numbers. 
An $n$-loop calculation in such a scheme
({\it without} expansion in  $\varepsilon$) guarantees true reproduction of terms from the
$n$ region and leads to good agreement with experiment already in the one-loop approximation.
 From the point of view of renormalization of the $d$-dimensional theory this approach corresponds to the inclusion in the action
(see below) of an infrared
irrelevant at $d>2$ operator, which becomes relevant, when $d\rightarrow 2$.\\

{\subsubsection{Renormalization with inclusion of divergences at $d = 2$} \label{subsubsec:AddRen}}

We shall not dwell on the detailed analysis of the rather cumbersome results of the two-loop calculation
of the amended double expansion in the MS scheme \cite{AHKV03,AHKV05}. Instead, we present main steps 
of this analysis in the case of an NP scheme following \cite{AHH10}. The only difference between
the two cases is in the renormalization schemes, the rest of the analysis is completely analogous in
both cases.

We start the analysis from the stochastic Navier-Stokes model in $d$ dimensions
(\ref{eq:double_NS_action}), in which, however,
to the energy pumping kernel (\ref{eq:double_kernel}) we prefer
the more realistic function
\begin{equation}
d_f(k)=D_0\,k^{4-d-2\varepsilon}\, h(m/k), \qquad  h(0)=1,
\label{eq:double_kernel_h}
\end{equation}
where $m=1/L$ is the reciprocal to the integral turbulence scale $L$
and $h(m/k)$ is some well-behaved function that provides the
infrared regularization. Its specific form is inessential and  we
shall always use the sharp cutoff
\begin{equation}
h(m/k)=\theta(k-m), \label{1.99}\,.
\end{equation}
Choice (\ref{1.99})
is the most convenient from the point of view of calculation of graphs of perturbation theory.

Model
(\ref{eq:double_NS_action}) is logarithmic (i.e. the coupling constant $g_0$ is
dimensionless) at $\varepsilon=0$, and the ultraviolet (UV) divergences
appear in the form of the poles in $\varepsilon$ in the correlation functions
of the fields $\{v,v'\}$. The standard analysis of
canonical dimensions supplemented by arguments of Galilei invariance shows
that for $d>2$ superficial UV divergences in the model
(\ref{eq:double_NS_action}) are present only in the one-irreducible function
$\langle v'v\rangle_{{\rm 1-ir}}$ (notation introduced in (\ref{eq:1PI_writing}) is employed)
 and the compensating counterterm enters in the form
$v'\partial^{2}v$.
In the special case $d=2$ a new UV divergence appears in the
one-irreducible function $\langle v'v'\rangle_{{\rm 1-ir}}$.

The inclusion of the counterterm  $v'\boldnabla^{2}v$ in the
action (\ref{eq:double_NS_action}) is tantamount to multiplicative
renormalization of the parameters $ \nu_0$ and $g_0$ :
\begin{equation}
  \nu_0=\nu Z_{\nu},\qquad D_{0} = g_{0}\nu_0^{3} = g\mu^{2\varepsilon}
  \nu^{3}, \qquad g_{0}=g\mu^{2\varepsilon}Z_{g}, \qquad Z_{g}=Z_{\nu}^{-3}.
  \label{eq:double_18}
\end{equation}
The renormalization constant of viscosity $Z_{\nu}$ is the only independent renormalization constant in (\ref{eq:double_18}).
Relation between the renormalization constants of the coupling constant and coefficient of viscosity shown in (\ref{eq:double_18}) 
 follows from the fact that
the amplitude of the correlator of the random force $D_{0}$ is not renormalized [no renormalization constant is needed for the term
$v' v'$  in action (\ref{eq:double_NS_action})]. The quantity
$\mu$ in (\ref{eq:double_18}) is the scale setting parameter,
$\nu$ is the renormalized viscosity and $g$ is the dimensionless renormalized coupling constant (charge).

An RG analysis of the model  (\ref{eq:double_NS_action}) shows the presence of an
IR-stable fixed point $g^* {\sim} \varepsilon$ at small $\varepsilon>0$, which
governs the infrared scaling. The parameters of this scaling  --
critical dimensions and universal scaling functions -- are
calculated in the form of series in $\varepsilon$. Due to
the Galilei invariance of the theory the series for critical
dimensions of fields $\Delta_v\,, \,\Delta_{v'}$ and the
frequency $\Delta_\omega$ are terminated at the first order
\begin{equation}
  \Delta_v=1-2\varepsilon/3,\qquad
  \Delta_{v'}=d-1+2\varepsilon/3, \qquad
  \Delta_\omega=2-2\varepsilon/3.
  \label{eq:double_razm}
\end{equation}
These formulas are exact without  higher order corrections with respect to
$\varepsilon$. They follow from relation (\ref{eq:double_18}) between the
renormalization constants $Z_g$ and $Z_\nu$, which, in turn, is a consequence of the
absence of renormalization of the non-local term in
(\ref{eq:double_NS_action}). At the realistic value
$\varepsilon=2$ quantities (\ref{eq:double_razm}) assume the Kolmogorov values
\beq
  \Delta_v=-1/3,\qquad \Delta_\omega=2/3.
  \label{eq:double_kolm}
\eeq
 Critical dimensions (\ref{eq:double_razm})
 are free from singularities at $d\rightarrow 2$ referred to in section \ref{subsec:double_intro}.
 However, other physical quantities like the
 skewness factor, the Kolmogorov constant and critical dimensions of various composite operators
 strongly depend on the space dimension. Contrary to (\ref{eq:double_razm}), these quantities
 are expressed in the form of infinite series of types (\ref{eq:double_Aeps}), (\ref{eq:double_Qq}), (\ref{eq:double_Aeps1}).
 The following sections will be devoted to the analysis of the renormalization of the model at the exceptional space dimension
 $d=2$ and to the construction of a renormalization scheme which provides for improved
 $\varepsilon$ expansions of the aforementioned quantities.\\

As it was already emphasized, in model (\ref{eq:double_NS_action})
additional UV divergences appear in the one-irreducible function
 $ \bigl\langle v'v'\bigr\rangle_{{\rm 1-ir}} $
as $d\rightarrow 2$. In particular, these divergences manifest themselves
in the form of singularities in $\Delta = (d-2)/2$ in coefficients
of the usual $\varepsilon$-expansion (\ref{eq:double_Aeps}). They can be eliminated
by addition of a counterterm of the form $v'\boldnabla^{2}v'$ to the
action. This counterterm is local and quite different from the
nonlocal contribution $v 'Dv '/2$ to action
(\ref{eq:double_NS_action}).  To restore the (technically convenient) multiplicative renormalization of the
model, the authors of \cite{HonNal96} have proposed to pass
to the two-charge model (i.e. with two coupling constants) in which to the function
(\ref{eq:double_kernel}) ${\sim} k^{4-d-2\varepsilon}=k^{2-2\Delta-2\varepsilon}$ the
term ${\sim} k^2$ with an independent coefficient is added:
\begin{equation}
    d_f(k)=D_{10} k^{2-2\Delta-2\varepsilon}+D_{20} k^2,\quad  D_{i0}=g_{i0} \nu_0^{3},
    \;\;i=1,2\,.
  \label{eq:double_kernel2}
\end{equation}
Here, the amplitude $D_0$ from (\ref{eq:double_kernel}) is now denoted by $D_{10}$. The parameters  $g_{10}$ and 
$g_{20}$ introduced in (\ref{eq:double_kernel2})
play the r\^{o}le of two independent bare charges.
The noise correlation function chosen for the velocity field reflects the detailed intrinsic
statistical definition of forcing,
 whose consequences are thoroughly
 discussed in Ref. \cite{Horvath95}.
 Correlation functions of noise with long-range and short-range terms include
 two principal --- low and high wave
 number scale --- kinetic forcings separated
 by a transition region at the vicinity
 of the characteristic wavenumber of order
 ${\cal O}([g_{v10}/g_{v20}]^{\frac{1}{5}}).$
 In  the  language of classical
 hydrodynamics the forcing contribution $\propto k^2$
 corresponds to the appearance of large eddies  convected
 by small and active eddies.
 
As before, the unrenormalized action is taken in the form (\ref{eq:double_NS_action}), but now instead of the injection function
(\ref{eq:double_kernel}) the function (\ref{eq:double_kernel2}) is used in function (\ref{eq:double_kernel_h}); thus 
the unrenormalized action
of the extended model is
\beq
  \S= \frac{1}{2}v'(D_{1 0}k^{2-2\Delta-2\varepsilon}+D_{2 0}k^{2} )v ' + \mv' \cdot[-\partial
  _t \mv +\nu _0\boldnabla^{2} \mv -(\mv\cdot\boldnabla )\mv ] \,. 
  \label{eq:double_NS_action3}
\eeq
Relations between renormalized and bare parameters are expressed by the formulas
\begin{align}
  &D_{10} = g_{10}\nu_0^{3} =
  g_1\mu^{ 2\varepsilon} \nu^{3},  &D_{20}& = g_{20}\nu_0^{3} =
  g_2\mu^{-2\Delta} \nu^{3}{Z}_{D_2}, &\nu_0&=\nu {Z}_{\nu}, \nonumber \\
  &g_{10}=g_1\mu^{2\varepsilon}{Z}_{g_1}, &g_{20}&=g_2\mu^{-2\Delta}{Z}_{g_2}, 
  \label{eq:double_ZZ}
\end{align}
with two independent renormalization constants for the coefficient of viscosity
${\nu_0}$ and for the amplitude ${D_{20}}$. The amplitude ${D_{10}}$ of the non-local term of the correlator of the random force
is not renormalized.  Therefore two additional relations follow
\beq 
   {Z}_{g_1}{Z}_{\nu}^{3}=1, \qquad { Z}_{g_2}{Z}_{\nu}^{3}={Z}_{D_2}.
   \label{eq:double_ZZ2}
\eeq
The independent
renormalization constants $Z_{\nu}$ and $Z_{D_2}$  are found from the condition that the one-irreducible functions
$\bigl\langle v'v\bigr\rangle_{{\rm 1-ir}}\big|_
{\omega=0}$ and $ \bigl\langle v'v'\bigr\rangle_{{\rm
1-ir}}\big|_ {\omega=0}$ are UV finite
(are free from poles with respect to  $\varepsilon$ at
$\Delta/\varepsilon =const$). In \cite{HonNal96} the renormalization constants
are calculated in the minimal subtraction scheme (MS)
[only poles with respect to $\varepsilon$, $\Delta=\O(\varepsilon)$ and their
linear combinations are subtracted]. To reach our goal a different
subtraction scheme with a normalization point (NP) is more suitable.
To this end, introduce normalized scalar one-irreducible functions as
\beq
  \Gamma_{v'v}=
  \frac{\bigl\langle v'_{i}v_{i}\bigr\rangle_{{\rm  1-ir}}
  \big|_{ \omega=0}} {\nu p^2(1-d)}, \quad
  \Gamma_{v'v'}=
  \frac{\bigl\langle v'_{i}v'_{i}\bigr\rangle_{{\rm 1-ir}}\big|_{ \omega=0}} {\nu^3
  p^2(d-1)}-g_1(\mu/p)^{2\Delta+2\varepsilon}-g_2 
  \label{eq:double_Rat}
\eeq
and determine the renormalization constants from the conditions
\beq
  \Gamma_{v'v}\Big|_{{p=0,\atop\mu=m}}=1\,,\qquad
  \Gamma_{v'v'}\Big|_{ p=0,\atop\mu=m}=0 \,.
  \label{eq:double_norm}
\eeq
Note that these normalization conditions are different from those used in \cite{AHKV03,AHKV05}.

Application of the RG approach to the renormalized model
specified by the condition (\ref{eq:double_norm}) leads to the same
$\varepsilon, \Delta$ expansions in the ray scheme as in \cite{HonNal96}.
Instead, to exploit the scheme dependence of perturbation theory, we propose to use
the renormalization scheme
(\ref{eq:double_norm}) {\em without} the $\varepsilon, \,
\Delta$ expansion.
Such a scheme reproduces correctly the leading terms of expansion in both regimes
$\varepsilon \rightarrow 0\,,
\Delta=const$ and $\varepsilon {\sim} \Delta \rightarrow 0$ simultaneously.
For all that, in the first case, the additional term
$v'\boldnabla^{2}v'$ is an infrared
irrelevant operator in the action and the renormalization constant
$Z_{D_2}$ describes the renormalization of this additional operator.

One-loop calculation of these renormalization constants
yields
\begin{equation}
  Z_\nu=1+\frac{d-1}{4(d+2)}\Big(-\frac{u_1}{2\varepsilon}+\frac{u_2}{2\Delta}\Big)\,,
  \label{eq:double_Znu}
\end{equation}

\begin{equation}
  Z_{D_2}=1+\frac{d^2-2}{4d(d+2)}\Big( -\frac{u_1^2}{
  2(2\varepsilon+\Delta)u_2}
  -\frac{u_1}{\varepsilon}+\frac{u_2}{2\Delta}\Big)\,,
  \label{eq:double_ZD}
\end{equation}
where instead of $g_1$ and $g_2$ more convenient charges $u_1$ and $u_2$ are used:
\beq
  u_1\equiv \bar S_d g_1\,, \quad  u_2\equiv \bar S_d g_2\,,
  \label{eq:double_ui}
\eeq
where
\begin{equation}
  \quad\bar S_d \equiv S_d /
  (2\pi)^{d},
  \quad S_d \equiv 2\pi^{d/2}/\Gamma (d/2)
  \label{eq:double_gfactors}
\end{equation}
are often encountered geometrical factors.
Here, $ S_d $
is the surface area of the unit sphere in $d$-dimensional space
and $\Gamma$ is Euler's Gamma function. From (\ref{eq:double_ZZ}), (\ref{eq:double_Znu}) and
(\ref{eq:double_ZD}) the renormalization constants of the charges $u_1$ and $u_2$ are determined as
\beq
  Z_{u_1}=1+\frac{3(d-1)}{4(d+2)}\Big(\frac{u_1}{2\varepsilon}-\frac{u_2}{2\Delta}\Big)\,,
\eeq
\beq
  Z_{u_2}=1+\frac{d^2-2}{4d(d+2)}\Big( -\frac{u_1^2}{
  2(2\varepsilon+\Delta)\,u_2}
  -\frac{u_1}{\varepsilon}+\frac{u_2}{2\Delta}\Big)+
  \frac{3(d-1)}{4(d+2)}\Big(\frac{u_1}{2\varepsilon}-\frac{u_2}{2\Delta}\Big)\,.
\eeq
The
corresponding RG functions are then found straightforwardly
\begin{equation}
  \gamma_i=(\beta_1 \partial_{u_1}+\beta_2
  \partial_{u_2})\ln Z_{u_i}, \qquad  i=1,\,2\,, 
  \label{eq:double_gamma}
\end{equation}
\begin{equation}
  \beta_1=-u_1(2\varepsilon+\gamma_1)\,, \qquad
  \beta_2=-u_2(-2\Delta+\gamma_2)\,.
  \label{eq:double_beta2}
\end{equation}
In the linear approximation for $\gamma$'s with respect to charges $u_1$ and $u_2$ it is enough to use $\beta_1\simeq-2u_1
\varepsilon $\,, $\beta_2\simeq 2 u_2 \Delta $\,, because  $\gamma_1=\Ord{u}$ and $\gamma_2=\Ord{u}$
in (\ref{eq:double_gamma}). We find
\begin{equation}
  \gamma_1=-\frac{3(d-1)}{4(d+2)}(u_1+u_2)\,, 
  \label{eq:double_gamma1}
\end{equation}
\begin{equation}
  \gamma_2=\frac{(d^2-2)(u_1+u_2)^2}{4d(d+2)\,u_2}-\frac{3(d-1)}{4(d+2)}(u_1+u_2)
  \label{eq:double_gamma2}.
\end{equation}
With the use of (\ref{eq:double_beta2}), (\ref{eq:double_gamma1}) and (\ref{eq:double_gamma2}), the
coordinates of the nontrivial fixed
point $u^*_{1}>0$, $u^*_{2}>0$ are found as the solution of
the equations
$\beta_1(u^*)=0$, $\beta_2(u^*)=0$ in the form
\begin{equation}
   u^*_{1}+u^*_{2}=\frac{8\varepsilon(d+2)}{3(d-1)} \,,
   \label{eq:double_u*}
\end{equation}
\begin{equation}
  u^*_{2}=\frac{\varepsilon^2}{\varepsilon+\Delta}\,\frac{8(d^2-2)(d+2)}{9d(d-1)^2}\,.
  \label{eq:double_u2*}
\end{equation}
Stability of the fixed point is determined by the sign of the real part of the eigenvalues $\Omega_\pm$ of the matrix $\Omega$:
\begin{align}
  \Omega_\pm=\Delta+\frac{2\varepsilon(2d^2-3d+2)}{3d(d-1)} \pm
  \sqrt{\Delta^2-\frac{4(d^2-2)}{3d(d-1)}\,\varepsilon\,\Delta-\frac{4(d^2-2)(2d^2-3d+2)}{9d^2(d-1)^2}\,\varepsilon^2}\,.
  \label{eq:double_omega}
\end{align}
We are interested in the region $\varepsilon > 0,\, \Delta > 0$, in which the
eigenvalues $\Omega_\pm$ are positive. Therefore, the fixed point (\ref{eq:double_u*}), (\ref{eq:double_u2*}) is infrared stable.
In the regime  $\varepsilon {\sim} \Delta \rightarrow 0$ with the use of (\ref{eq:double_omega})
we obtain
\begin{equation}
  \Omega_\pm=\Delta+\frac{4\varepsilon}{3}\pm
  \sqrt{\Delta^2-\frac{4}{3}\varepsilon\,\Delta-\frac{8}{9}\,\varepsilon^2}+\Ord{\varepsilon^2}
  \label{eq:double_omega2}
\end{equation}
which is consistent with the result of \cite{HonNal96}. At
$\varepsilon \rightarrow 0, \, \Delta=const$ it follows that
\begin{equation}
  \Omega_{-}=2\varepsilon+\frac{2(d^2-2)}{3d(d-1)}\cdot
  \frac{\varepsilon^2}{\Delta} +\Ord{\varepsilon^3}\,, 
  \label{eq:double_omega-d1}
\end{equation}
\begin{equation}
  \Omega_{+}=2\Delta+\frac{2(d^2-3d+4)}{3d(d-1)}\varepsilon-\frac{2(d^2-2)}{3d(d-1)}\cdot
  \frac{\varepsilon^2}{\Delta} +\Ord{\varepsilon^3}\,. 
  \label{eq:double_omega-d2}
\end{equation}
The quantity $\Omega_{-}$ plays the r\^{o}le of the correction index $\omega$ in the framework of the
prevailing $\varepsilon$ expansion in the theory of developed turbulence whereas  $\Omega_{+}$
determines the critical exponent of the infrared-irrelevant composite operator
$v'\partial^{2}v'$.

The terms ${\sim} \varepsilon$ in (\ref{eq:double_omega-d1}) and
(\ref{eq:double_omega-d2}) are reliable: for $\Omega_{-}$  relation
(\ref{eq:double_omega-d1} reproduces the known one-loop expression
\cite{Dominicis79} for the exponent $\omega$; for $\Omega_{+}$ we
have checked the result by a direct calculation of the critical
dimension of the composite operator $v'\partial^{2}v'$
in the usual $\varepsilon$ expansion.
Moreover,
the terms ${\sim} \varepsilon^2/\Delta$ yield the true singular part
with respect to $\Delta$ in the coefficients of $\varepsilon^2$: for
$\Omega_{-}$ it was confirmed by the pioneering two-loop calculation
\cite{Adzhemyan03a}, for $\Omega_{+}$ we have checked it by
calculation of the critical dimension of the composite operator
$v'\partial^{2}v'$ in two-loop approximation.
Expression (\ref{eq:double_omega}), in fact,  correctly
reproduces main singular terms of the form $\varepsilon(\varepsilon /
\Delta)^k$ and all leading terms of the $\varepsilon$ expansion,
i.e. the first terms of the corresponding Laurent series (\ref{eq:double_Qq}).

Calculation of graphs with increasing number of loops in our renormalization scheme
guarantees that results become more precise step by step
in the sense that the number of true terms of the $\varepsilon$ expansion and the number of singular in $\Delta$ contributions
is increased: an
$n$-loop calculation correctly reproduces first $n$ of all coefficients $A_k(d)$
in (\ref{eq:double_Aeps}), while simultaneously of the coefficients $A_k(d)$ with $k>n$ the first
$n$ terms of their Laurent series (\ref{eq:double_Qq}) will be reproduced correctly.\\

{\subsubsection{Calculation of the Kolmogorov constant through the skewness factor} \label{subsubsec:double_CK&S}}

Even in the field-theoretic RG approach several ways have been proposed 
\cite{Adzhemyan89,Honkonen98,Hnatich99,HHJ01,Adzhemyan03a,AHKV05,AHH10}
to calculate the (non-universal) amplitude factor
-- the Kolmogorov constant -- in Kolmogorov's 5/3 law for the turbulent energy spectrum
\beq
\label{C1def}
E(k)\sim C_K'\overline{\cal E}^{2/3}k^{-5/3}\,,
\eeq
where $\overline{\cal E}$ is the average energy injection rate per unit mass. The notation 
in (\ref{C1def}) follows that of \cite{Adzhemyan03a}.

Different approximations for the connection of model parameters and the average energy injection rate lead have resulted
in different values for the Kolmogorov constant. At present the most reliable approach appears to be that based on the
connection between the Kolmogorov constant and the skewness factor  \cite{Adzhemyan03a}. For the 
latter consistent expansions both in
$\varepsilon$ and $\varepsilon$, $\Delta$ may be constructed and actually have been calculated in two-loop approximation.
As in case of critical exponents, scheme-dependence or perturbative calculation may be used to improve the
performance of regulator expansions. The renormalization scheme presented in preceding sections following \cite{AHH10}
allows to obtain best match with experimental data up to date.

The Kolmogorov constant is not determined uniquely in the
$\varepsilon$ expansion in the model with power-law injection
(\ref{eq:double_kernel}) (for details, see \cite{AHKV05}). On the other hand
physical quantities independent of the amplitude $D_{10}$ (\ref{eq:double_kernel}) (universal quantities) are
determined unambiguously
in the framework of the $\varepsilon$ expansion. The skewness factor
\begin{equation}
  {\cal S} \equiv S_{3}/S_{2}^{3/2}, \label{eq:double_sviaz_2}
\end{equation}
is an example of such a quantity.
In (\ref{eq:double_sviaz_2}) $S_n$ are structure functions defined by relations
\begin{equation}
  S_{n} (r) \equiv \big\langle [ v_{r} (t,{\mx}+{\mr}) -
  v_{r} (t,{\mx})]^{n} \big\rangle, \qquad v_{r}\equiv
  (\mv \cdot \mr)/r, \quad r\equiv |{\mr}|. 
  \label{eq:double_struc}
\end{equation}
According to the Kolmogorov theory, the second-order structure function  $S_{2}(r)$ in the inertial range
is of the form
\begin{equation}
 S_{2}(r) = C_K  \E^{2/3}r^{2/3},
 \label{eq:double_CK}
\end{equation}
where $\E$ is the average energy dissipation rate per unit mass (in the
steady state it coincides with the mean energy injection rate $\overline{\mathcal{E}}$, see Eq. (\ref{C1def})) and $C_K$ is the
Kolmogorov constant, the value of which is not determined in the
framework of the phenomenological approach. Although there is strong
experimental evidence that the Kolmogorov scaling $S_n(r) {\sim}
r^{n/3}$ does not hold in the inertial range for the structure
functions of order $n\geq 4$, for the second-order structure
function $S_2(r)$ the experimental situation about anomalous
scaling [i.e., deviation of the power of $r$ from the Kolmogorov
value $2 / 3$ in (\ref{eq:double_asymp})] in the inertial range is still
controversial and in any case this deviation is small
\cite{Barenblatt99,Benzi99}. Therefore, we shall use the Kolmogorov
asymptotic expression (\ref{eq:double_CK}) for the second-order structure
function $S_2(r)$ in the following analysis.

The amplitude of the third-order structure function  $S_{3}(r)$ is determined in the Kolmogorov theory exactly
\cite{Monin,Frisch}:
\begin{equation}
   S_{3}(r) = -\frac{12}{d(d+2)}\,\E\, r. \label{eq:double_S}
\end{equation}
All these expressions together with (\ref{eq:double_sviaz_2}), (\ref{eq:double_CK}) allow to connect the Kolmogorov constant
with the skewness factor:
\begin{equation}
  C_K=\Big[-\frac{12}{d(d+2)\cal S}\Big]^{2/3}. 
  \label{eq:double_sviaz}
\end{equation}
Among the three quantities  $S_{2}(r)$,  $S_{3}(r)$ and $\cal S$ only the last one
has a unique well-defined $\varepsilon$ expansion. Thus, relation
(\ref{eq:double_sviaz}) (valid only at the physical value
$\varepsilon=2$) may be used to determine $C_K$ by means of the calculated value
${\cal S} (\varepsilon=2)$.

To find the RG representation of the skewness factor (\ref{eq:double_sviaz_2})
the RG representations of the functions $S_{2}(r)$ and $S_{3}(r)$ have to be determined.
The function $S_{2}(r)$ is connected with the Fourier transform of the pair correlation function  $G(p)$
by relation
\begin{align}
  S_{2}(r) = 2\int \frac{\dRM^d{\mk}}{(2\pi)^{d}} \,G(k)\,
  \left[1-({\mk}\cdot{\mr})^{2}/(kr)^{2}\right] \left\{1- \exp
  \left[{\rm i} ({\mk}\cdot{\mr})\right]\right\}. 
  \label{eq:double_atas}
\end{align}
Therefore, the RG representation of $S_{2}(r)$ can be specified with the aid of the
RG representation (\ref{eq:double_asymp}). A similar RG representation can be
written for the function $S_{3}(r)$. It is, however, more convenient
to use the exact result analogous to expression (\ref{eq:double_S})
\begin{equation}
  S_3(r)= -\frac{3(d-1) \, \Gamma(2-\varepsilon) \, (r/2)^{2\varepsilon-3}D_{10}}
  {(4\pi)^{d/2} \, \Gamma(d/2+\varepsilon)}\,. 
  \label{eq:double_S3}
\end{equation}
This relation clearly demonstrates that the amplitude of the
structure function expressed  through $D_{10}$ has a singularity at
$\varepsilon\rightarrow 2$. In this case  the singularity is in
the form ${\sim} (2-\varepsilon)^{-1}$. After the substitution of the
amplitude $D_{10}{\sim} (2-\varepsilon)$ into (\ref{eq:double_S3}) the
singularity on the right-hand side of
(\ref{eq:double_S3}) is canceled by the node of $D_{10}$. This leads to a finite expression for
$S_{3}(r)$ at $\varepsilon=2$ which coincides with (\ref{eq:double_S}).

Relations (\ref{eq:double_asymp}), (\ref{eq:double_atas}) and (\ref{eq:double_S3}) could be used as the 
basis for the construction of the
$\varepsilon$ expansion of the skewness factor
(\ref{eq:double_sviaz_2}), but on this way there is an additional complication.
The point is that
the behavior  $S_{2}(r){\sim} r^{2-2\varepsilon/3}$,
which is determined by power counting from
(\ref{eq:double_atas}) and (\ref{eq:double_asymp}), is valid only at
$\varepsilon>3/2$, because at  $\varepsilon<3/2$ the integral
(\ref{eq:double_atas}) diverges as $k\rightarrow \infty$ [it means that in this case the leading
contribution to $S_{2}(r)$ is given by the term
$\big\langle v_{r}^{2} (t,{\mx}) \big\rangle$ independent of $r$].
The derivative
$r\partial_r S_{2}(r)$, however, is free from this flaw and according to (\ref{eq:double_atas}) it
assumes the form
\begin{equation}
  r\partial_r S_{2}(r) = 2 \int \frac{\dRM^d{\mk}}{(2\pi)^{d}}\,
  G(k)\, \left[1-({\mk}\cdot{\mr})^{2}/(kr)^{2}\right]\, ({\mk}\cdot{\mr}) \,\sin ({\mk}\cdot{\mr}).
  \label{eq:double_atas1}
\end{equation}
The integral in (\ref{eq:double_atas1}) converges at all values $0<\varepsilon <2$.
On the other hand, at the physical value $\varepsilon=2$ the amplitudes in
$S_{2}(r)$ and  $r\partial_rS_{2}(r)$ differ from each other only by the factor
2/3, therefore the $\varepsilon$ expansion has been constructed for
the following analogue of the skewness factor
\cite{Adzhemyan03a,AHKV03,slovac}
\begin{equation}
  Q(\varepsilon)\equiv\,{  r\partial_r S_{2}(r) \over |S_{3}(r)|^{2/3}}= {
  r\partial_r S_{2}(r) \over (-S_{3}(r))^{2/3}}. 
  \label{eq:double_Q}
\end{equation}
The Kolmogorov constant and the skewness factor are expressed through
the value
$Q(\varepsilon=2)$ according to  (\ref{eq:double_sviaz_2}), (\ref{eq:double_CK}) and
(\ref{eq:double_S}) by relations:
\begin{equation}
  C_{K}= \left[3Q(2)/2\right]\left[12/d(d+2)\right]^{2/3}, \quad
  {\cal S} = - \left[3Q(2)/2 \right]^{-3/2}. 
  \label{eq:double_trud}
\end{equation}
Substituting   expressions (\ref{eq:double_atas1}), (\ref{eq:double_asymp}) and
(\ref{eq:double_S3}) into (\ref{eq:double_Q}) we obtain
\begin{equation}
  Q(\varepsilon)= \left[4(d-1)/9/u_{1*}^2 \right]^{1/3}\, A(\varepsilon)\,
  R(1,u_{1*},u_{2*}),
  \label{eq:double_atas2}
\end{equation}
where
\begin{equation}
  A(\varepsilon)= \frac{\Gamma(2- 2\varepsilon /3) \Gamma^{1/3}(d/2)
  \Gamma^{2/3}(d/2+\varepsilon) } {\Gamma(d/2+2\varepsilon/3)
  \Gamma^{2/3}(2-\varepsilon)}= 1 + \O(\varepsilon^2). 
  \label{eq:double_amp}
\end{equation}
Consecutive loop calculations of the coordinates of the fixed point $u^*_{1},u^*_{2}$ and the scaling function
$R(1,u^*_{1},u^*_{2})$ in (\ref{eq:double_atas2}) in the framework of the scheme used increase
the number of true terms of the usual $\varepsilon$ expansion in these quantities. In all higher 
coefficients of the $\varepsilon$ expansion
the number of true terms of their Laurent series (\ref{eq:double_Qq}) increases as well. The quantity $A(\varepsilon)$ is 
not singular at $d=2$ therefore
it is sufficient to use the corresponding part of its $\varepsilon$-expansion.

At the leading order the scaling function
$R(1,u^*_{1},u^*_{2})$ in (\ref{eq:double_atas2}) is
\begin{align}
  R(1,u^*_{1},u^*_{2}) \approx \frac{u^*_{1}+u^*_{2}}{2} . 
  \label{eq:double_Q12}
\end{align}
Putting $A(\varepsilon)\approx 1$ we obtain in one-loop approximation
\begin{align}
  Q(\varepsilon)= \left[4(d-1)/9/(u^*_{1})^2 \right]^{1/3}\cdot
  \frac{u^*_{1}+u^*_{2}}{2}, 
  \label{eq:double_Q1}
\end{align}
where for the coordinates $u^*_{1}, u^*_{2}$ values (\ref{eq:double_u*}) and (\ref{eq:double_u2*})
are implied.
Calculating $Q$ in (\ref{eq:double_Q1}) at $d=3$ and $\varepsilon = 2$ we obtain $Q(2)\approx 1.461\,.$
Further, using this value for calculation of the Kolmogorov constant and skewness factor (\ref{eq:double_trud})
we arrive at the values $C_K \approx 1.889$ and ${\cal S}\approx -0.308\,.$
The values $C_K \approx 2.01$ and ${\cal S}\approx -0.28$ are considered the most reliable experimental
values of these quantities
\cite{Sreenivasan95}. Therefore, the scheme which we have suggested for calculations by means of improvement 
of the $\varepsilon$ expansion
provides quite reasonable agreement with the experiment. It should be recalled that the prevailing $\varepsilon$ 
expansion at one-loop
order gives the values
$C_K \approx 1.47$ and ${\cal S}\approx - 0.45\,.$

The version of the RG approach used in the present paper bears certain resemblance
with the well known RG method in the real space, which is
widely used in the theory of critical phenomena
(it is also called the $"g$ expansion"). In the theory of phase transitions
the parameter $\varepsilon$ has the meaning of deviation from the critical
space dimension  (e.g, $\varepsilon=4-d $ for the $\varphi^4$-model). In the framework of the $g$ expansion
renormalization constants -- adopted from the logarithmic theory -- are calculated in the form of power series in the coupling
constant $g$ directly at $\varepsilon=1$, i.e. at real value of the space dimension  $d=3$. The use of term
"RG in real space" is just explained by this feature. In the theory of turbulence
the parameter  $\varepsilon$ has different meaning and it is more pragmatic to use the term
 $g$ expansion.
In the framework of the $g$ expansion calculations are notably simplified because it is
much easier to calculate finite integrals in three dimensions than to calculate integrals with singularities at
$d \rightarrow 4$. This is the reason why
the use of the $g$ expansion in the theory of critical phenomena has
allowed to achieve better accuracy in perturbative calculations than in the usual $\varepsilon$ expansion.
However, we must keep in mind a drawback of the $g$ expansion.
While the results of the $\varepsilon$ expansion are unambiguous and independent of the renormalization scheme,
the $g$ expansion approximates higher terms of the
$\varepsilon$ expansion, and the result of such an approximation depends, in particular, on the choice of the form of the
 IR regularization. In the theory of critical phenomena a natural regularization
present in the model at $T-T_c\neq 0$ is usually used, but it does not eliminate the problem of ambiguity.

Here, an approach akin to the $g$ expansion has been used to achieve a different aim.
The choice of the renormalization constants from natural normalization conditions for the
response and correlation functions together with the additional renormalization
of random forcing allowed to include singular in  $d-2$ contributions to the coefficients
in all orders of the $\varepsilon$ expansion. This summation led to a remarkable improvement
of agreement of the theoretical prediction with the experimental value of the Kolmogorov constant already in 
the one-loop approximation.\\

{\subsection{Advection of passive scalar} \label{subsec:scalar2D_advection}}
One of the most studied models related to the turbulence is an advection of passive
scalar quantity \cite{turbo,FGV01}. The main object of study are statistical properties
of a field, which is coupled to the velocity field $\mv$ through advective term (See Sec. \ref{subsubsec:multinoise}).
From general point of view there are two main approaches depending on the way $\mv$ is generated:
\begin{enumerate}[a)]
  \item field $\mv$ is governed by the stochastic Navier-Stokes equation.
  \item field $\mv$ is considered as an random variable with prescribed properties.
\end{enumerate}

In what follows, we apply the first approach, whereas the latter will be analyzed in Sec. \ref{sec:violations}.
{\subsubsection{Functional formulation  of the passive scalar problem near two dimensions} \label{subsubsec:scalar2D_passive_model}}
Here, we study the problem of the advection of the passive
scalar using a random velocity field generated by the stochastically forced
Navier-Stokes equation~\cite{Wyld}, which has been widely used to produce stochastic
velocity field with the Kolmogorov scaling behavior
obtained by the use of the field-theoretic renormalization
group~\cite{Dominicis79,Adzhemyan83}.
The passive scalar problem has already been treated within
the RG
approach of the randomly forced Navier-Stokes equation
for both the local~\cite{FNS77} and
long-range~\cite{Adzhemyan84} correlations of the random force,
but without random pumping of the passive scalar, due to which
the behavior of the correlation functions of the passive
scalar was not addressed at all.

 The velocity field in the functional-integral approach is generated by
 the dynamic action of the randomly forced Navier-Stokes equation (\ref{eq:double_NS_action3}),
 to which a contribution corresponding to the advection of the passive scalar is added.
 The statistical model of the advection of the passive scalar
 characterized by the concentration field $\theta(x)$ in
 the turbulent environment (see e.g.\cite{Adzhemyan84},\cite{Hnatich90})
 is given by the stochastic differential equation
\begin{align}
   \label{eq:scalar2D_pasivo}
   \partial_t \theta +({\mv} \cdot {\boldnabla}) \theta
   - \nu_0 u_0\boldnabla^2 \theta = f^\theta\,,
\end{align}
where $u_0$ is the inverse Prandtl number. Physically it 
represents the ratio between diffusion and viscosity in a liquid.
 The random source field
 $f^\theta$ is assumed to be Gaussian with zero mean and the correlation function'
  \begin{align}
   \langle\,f^\theta(x_1) f^\theta(x_2)
   \,\rangle   & \equiv D^\theta(x_1-x_2) \nonumber\\
   & = \frac{u_0^3 \nu_0^3}{d-1}
   \delta(t_1-t_2)\int
     \frac{{\mbox d}^d {\mk}}{(2\pi)^d}     
     \eRM^{i{\mk} \cdot {\mx}}
     \left[
      g_{\theta 10}\, k^{ 2- 2\Delta-2a \eps}
     + g_{\theta 20}k^2\right].
    \label{eq:scalar2D_corel3}
 \end{align}
 Since near two dimensions a local term is generated to the correlation function of the scalar noise (describing thermal noise),
 we have taken the correlation function with both the long-range and local term at the outset.
 
 A detailed analysis \cite{Adzhemyan84}
 of the renormalization of
 the passive scalar model
 has shown that there are
 superficial divergences
 in the graphs corresponding to the 1PI Green functions
$\Gamma_{v'v}$, and $\Gamma_{\theta'\theta} $
in the renormalization scheme of Refs.~\cite{Dominicis79,Adzhemyan83}
applicable for space dimensions $d>2$. We will refer to the
approach of~\cite{Dominicis79,Adzhemyan83} as the standard scheme.
The 1PI Green functions
$\Gamma_{v'vvv}$, and $ \Gamma_{\theta'\theta v }$,
which could, by standard power counting,
give rise to the renormalization of the nonlinear terms in the Navier-Stokes
and advection-diffusion equation, are actually finite due to the Galilei invariance
of the stochastic equations with temporally white noise.

The stochastic problem
of the passive scalar (\ref{eq:scalar2D_pasivo}), (\ref{eq:scalar2D_corel3}) gives rise to the field-theoretic action
\begin{align}
  \S_\text{PS}[\theta,\theta',\mv] & = 
  \frac{1}{2}
  \theta' D^\theta \theta'+
   \theta'\,\left[- \partial_t \theta +  u_0 \nu_0 \boldnabla^2 \theta - ({\mv}\cdot {\boldnabla}) \theta\,\right] \,.
  \label{eq:scalar2D_passive_action}
\end{align}
As explained above, this action containing
 non-analytic terms (proportional to the coupling constants
 $g_{v10}$ and $g_{\theta 10}$) requires also the analytic terms
 (proportional to $g_{v20}$ and $g_{\theta 20}$) in order
to be multiplicatively renormalizable.
 All dimensional constants
 $g_{v10}$, $g_{v20}$, $g_{\theta 10}$ and $g_{\theta 20}$, which
 control the amount of randomly
 injected energy and mass 
 play the role of the expansion parameters of the perturbation theory.

 For convenience of further calculations the factors
 $\nu_0^3$ and $\nu_0^3 u_0^3  $ including the
 "bare" (molecular) viscosity $\nu_0$ and the
 "bare" (molecular or microscopic ) diffusion coefficient
 $\nu_0u_0$ have been extracted.

{\subsubsection{Calculation of the fixed points of the renormalization group} \label{subsubsec:passive_RG}}

\begin{figure}
   \begin{center}
   \includegraphics[width=9cm]{\PICS 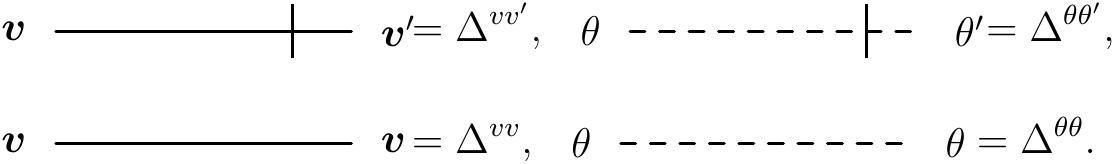}
   \caption{Diagrammatic representation of the bare
   propagators. The time flows from right to left.}
   \label{fig:scalar_prop}
   \end{center}
\end{figure}

\begin{figure}
   \begin{center}
   \includegraphics[width=9cm]{\PICS 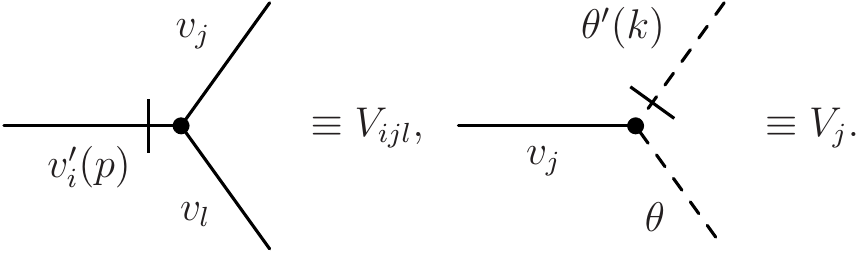}
   \caption{Diagrammatic representation of the interaction vertices 
   describing an interactions between velocity components (left) and advection interaction (right).}
   \label{fig:scalar_vertex}
   \end{center}
\end{figure}

 The model given by the sum of dynamic actions (\ref{eq:double_NS_action3}) and (\ref{eq:scalar2D_passive_action}) 
 is renormalizable by the standard
 power-counting rules for $\Delta=0$ (see (\ref{dz}) for the definition of $\Delta$) and $\eps=0$.
The divergent 1PI Green functions are $\Gamma_{vv'}$, $\Gamma_{\theta\theta'}$ as in the
standard case~\cite{Adzhemyan83} and
also $\Gamma_{v'v'}$, $\Gamma_{\theta'\theta'}$ typical of $d=2$, so that the model is multiplicatively renormalizable.
The standard Feynman diagrammatic expansion can be used in
a straightforward fashion.
The inverse matrix of the quadratic part in the actions determines a form of the bare propagators.
 The propagators are presented in the wave-number-time representation, which is
for the translationally invariant systems the most convenient way for doing explicit calculations.
Graphical representation of propagators
is depicted in Fig.~\ref{fig:scalar_prop} and the corresponding algebraic expressions 
are
\begin{align}
  \Delta^{v v'}_{ij}({\mk},t) & =  \Delta^{v'v}_{ij}(-{\mk},-t) =
  \theta(t)\,P_{ij}({\mk})\,\eRM^{-\nu_0\, k^2 t}, \nonumber  \\
  \Delta^{\theta\theta'}({\mk},t) & =  \Delta^{\theta'\theta}(-{\mk},-t) 
  = \theta(t) \eRM^{ - u_0 \nu_0 \, k^2 t }, \nonumber\\
  \Delta^{vv}_{ij}({\mk},t) &= \frac{1}{2} \nu_0^2\,P_{ij}({\mk}) \left( g_{v10}\,k^{-2\eps-2\Delta}+g_{v20}
  \right) \eRM^{-\nu_0 \, k^2 | t | }, \nonumber\\
  \Delta^{\theta\theta}({\mk},t) & = \frac{1}{2}\,u_0^2\,\nu_0^2  \left(\,g_{\theta 10}k^{-2\,a \eps- 2\Delta} + g_{\theta 20}
  \right) \eRM^{ - u_0 \nu_0\,k^2 | t | }.
  \label{eq:scalar2D_PropY}
\end{align}
In our model there are two different interaction vertices, which are
graphically depicted in Fig.~\ref{fig:scalar_vertex}, where 
we have included also the corresponding vertex factors. The corresponding expressions in momentum representation are
\begin{equation}
   V_{ijl}  =i(p_j\delta_{il}+p_l\delta_{ij})
   \label{eq:scalar2D_vertexNS}
\end{equation}
and
\begin{equation}
   V_j = ik_j.
   \label{eq:scalar2D_vertexADV}
\end{equation}
Note that the momentum is always carried by the slashed field.

\begin{figure}
  \begin{center}
  \includegraphics[width=12cm]{\PICS 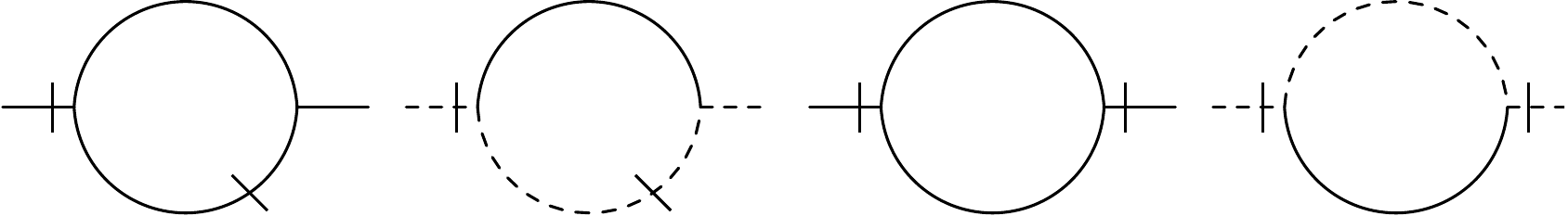}
  \caption{One-loop graphs giving rise to the divergent terms in the perturbative
 expansion of the one-particle irreducible Green functions $\Gamma_{vv'}$,
 $\Gamma_{\theta\theta'}$, $\Gamma_{\theta'\theta'}$, and $\Gamma_{\theta'\theta'}$, respectively. The slashes
 on the lines denote the derivatives appearing in the cubic interaction terms.}
  \label{fig:scalar2D_passive}
 \end{center} 
\end{figure}

Graphical representation of the
one-loop contributions to diverging 1PI functions is depicted in Fig. \ref{fig:scalar2D_passive}.

The divergences show in the form of poles in $\Delta$, $\varepsilon$ and their linear
combinations and the ray scheme is used (which implies that $\Delta$ and $\eps$ are treated as small
parameters of same order).

 Renormali\-zed Green functions are expressed
 in terms of the renormalized pa\-ra\-me\-ters
\begin{align}  
   g_{v1} & =  g_{v10} {\mu}^{-2\eps}\, Z_1^3,  &g_{v2} &=  g_{v20}\,{\mu}^{2\Delta}\, Z_1^3 Z_3^{-1}, &\nu & = \nu_0 Z_1^{-1}, \nonumber\\
   g_{\theta 1} & =    g_{\theta 10} {\mu}^{-2 a \eps- 2\Delta} Z_2^3, &g_{\theta 2} &=  g_{\theta 20} Z_2^3 Z_4^{-1}, &u&=u_0 Z_1 Z_2^{-1} 
  \label{eq:scalar2D_Zkaka1}
\end{align}
 appearing in the renormalized action.
 From connections (\ref{eq:scalar2D_Zkaka1}) it follows that
the $\beta$ functions are
\begin{align}
 \beta_{v1} &= g_{v1} \left( - 2 \eps + 3\gamma_1  \right),  &\beta_{v2}& = g_{v2}
 \left(   2 \Delta + 3\gamma_1- \gamma_3\right), &\beta_u  =  u \left( \gamma_1 - \gamma_2 \right), 
 \nonumber \\
  \beta_{\theta 1} & =  g_{\theta 1}
  \left( -2 a\eps -2\Delta + 3\gamma_2 \right),
  &\beta_{\theta 2}& =  g_{\theta 2} \left( 3\gamma_2- \gamma_4\right).
  \label{eq:scalar2D_beta0}                          
\end{align}

 The function $\gamma_4$ may be expressed in the following form
 \beq
  \label{eq:scalar2D_gamma4}
   \gamma_4=\gamma_4'(g_{v1},g_{v2},u)+\frac{g_{\theta 1}}{g_{\theta 2}}
   \gamma_4''(g_{v1},g_{v2},u).
 \eeq
 As a consequence, the RG functions $\beta_{\theta 1}$ and $\beta_{\theta 2}$ are linear functions
of the coupling constants $g_{\theta 1}$ and $g_{\theta 2}$
\begin{equation}
\label{eq:scalar2D_newbeta}
\beta_{\theta 1}=g_{\theta 1}(-2a\eps-2\Delta+3\gamma_2)\,,\quad
\beta_{\theta 2}=g_{\theta 2}(3\gamma_2-\gamma_4')-g_{c1}\gamma_4''\,.
\end{equation}
Since $\gamma_1$, $\gamma_2$ and $\gamma_3$ functions depend on
$g_{v1}$, $g_{v2}$ and $u$ only, vanishing of the first three $\beta$ functions
in (\ref{eq:scalar2D_beta0}) already yields a closed system of equations for
the fixed point values of $g_{v1}$, $g_{v2}$ and $u$.
 
Thus, fixed points in the model with scalar pumping are determined by the system of equations
\beq  
g_{v1} \left( - 2 \eps + 3\gamma_1  \right) = 0,\quad
g_{v2}  \left(   2 \Delta + 3\,\gamma_1-\gamma_3  \right) = 0,\quad
u \left( \gamma_1 - \gamma_2 \right) = 0,
\label{eq:scalar2D_FP}
\eeq
of the passive scalar model without the random scalar source.

Apart from the Gaussian fixed point $g_{v1}^*=g_{v2}^*=0$, which is stable for
$\Delta>0$, $\eps<0$,
there are two nontrivial fixed
points of the RG: the fixed point corresponding to
short-range correlations of the random
force~\cite{FNS77} with
\beq
 g_{v1}^*=0,\quad g_{v2}^*=-32\pi\Delta,
 \label{eq:scalar2D_FPFNS} 
\eeq
and the inverse Prandtl number
\begin{equation}
  u^{\ast}=\frac{\sqrt{17}-1}{2}\simeq 1.562.
  \label{eq:scalar2D_Pr}
\end{equation}
The region of stability of this short-range fixed point
$2\Delta+3\eps<0$, $\Delta<0$ is determined by the positivity of the
eigenvalues $\Omega_i,i=1,2,3$ of the $\Omega$ matrix
\beq
  \Omega_1=-2\eps-3\Delta,\quad \Omega_2=-2\Delta,\quad
  \Omega_3=-2\Delta\frac{\sqrt{17}}{\sqrt{17}+1}.
  \label{eq:scalar2D_OmegaFNS}
\eeq
The third fixed point is the
{\em kinetic} fixed point, which is the fixed point relevant to the
description of turbulent diffusion.
At the kinetic fixed point
the value of the renormalized inverse Prandtl number $u$ is given by (\ref{eq:scalar2D_Pr})
and the values of the other relevant coupling constants
\begin{equation}
  g_{v1}^{\ast}= \frac{64\,\pi}{9}\,\frac{\eps\,(2\eps+3\Delta)}
  {\eps+\Delta}, \quad
  g_{v2}^{\ast}=\frac{64\pi}{9}\,
  \frac{\eps^2}{\Delta+\eps}\,,
  \label{eq:scalar2D_fix11}
\end{equation}
of the double expansion of the stochastic
Navier-Stokes equation~\cite{HonNal96}.

 The calculation of the $\Omega$ matrix (up to the order
 ${\O}(\eps,\Delta)$)
 at this fixed point yields the eigenvalues
\begin{align}
 \Omega_{1,2} = \frac{1}{3} \left[(3\Delta +4\eps )\pm\sqrt
 {9\Delta^2-12\Delta
 \eps -8\eps ^2}\right],\quad
 \Omega_3  = \frac{4\sqrt{17}}{\sqrt{17}+1}\eps.
 \label{eq:scalar2D_omega1}
\end{align} 
 From these expressions we see that the region of stability
 of the kinetic fixed point is $\eps>0$, $2\eps+3\Delta >0$.\\

{\subsection{Stochastic magnetohydrodynamics $d\ge 2$} \label{subsec:mhd2D_magneto}}

During the past four decades asymptotic analysis of stochastic
transport equations [Navier-Stokes equation,
magnetohydrodynamic (MHD) equations, advection-diffusion equation and the
like] has attracted increasing attention.
Somewhat less effort has been devoted to the asymptotic analysis of
stochastic magnetohydrodynamics since the pioneering work of
Fournier et al~\cite{Fournier82} and Adzhemyan et
al~\cite{Adzhemyan85}. In particular, in these papers the existence of two
different anomalous scaling regimes (kinetic and magnetic) in three dimensions
was established corresponding to two non-trivial infrared-stable
fixed points of the renormalization group. It was also conjectured
that in two dimensions the magnetic scaling regime does not exist
due to instability of the magnetic fixed point. However, in both
papers there were flaws in the renormalization of the model in two
dimensions~\cite{turbo,HonNal96}.
Even more serious shortcomings are present in investigations of
MHD turbulence~\cite{Liang93,Kim99}, in which a
specifically two-dimensional setup has been applied with the use of the stream
function and magnetic potential. Therefore, results obtained for the
two-dimensional case in these papers cannot be considered completely conclusive.

Here, we present a field-theoretic
RG analysis of the stochastically forced
equations of magnetohydrodynamics with the proper account of
additional divergences which arise in two dimensions. This
gives rise to a double expansion in the analytic and dimensional regulators $\varepsilon$ and $\Delta$, respectively.
In the ray scheme of the double
expansion the standard procedure of minimal subtractions was used.
The RG analysis of the large-scale asymptotic
behavior of the model confirms the basic conclusions of the
previous analyses~\cite{Fournier82,Adzhemyan85} that near two dimensions
a scaling regime driven by the velocity fluctuations may exist, but no
magnetically driven scaling regime can occur.

Second, 
a renormalization of the model with a
different choice of finite renormalization is performed in order to find at which
non-integer dimension the magnetic fixed point ceases to be
stable. This borderline dimension was found in Ref.~\cite{Fournier82}
with the use of the momentum-shell RG in a
setup valid in a fixed space dimension $d>2$. In the double expansion
in $\eps$ and $2\Delta=d-2$ this effect cannot be
traced at one-loop order, but higher-order calculation is required.
Therefore, 
also a description based on the RG analysis
according to the "principle of maximum divergences" is included.
This procedure gives rise to
RG functions such that in the limit of small
$\Delta$, $\eps$ they reproduce the results of the two-parameter
expansion,
and in the limit of small $\eps$ (but finite $\Delta$) they
yield the results of the usual $\eps$
expansion~\cite{Fournier82,Adzhemyan85}. These properties are similar to those
of the normalization point schemes in the stochastic Navier-Stokes problem.\\

{\subsubsection{Field theory for stochastic magnetohydrodynamics} \label{subsubsec:mhd2D_mag_model}} 
The model of stochastically forced conducting fluid is
 described by the system of magnetohydrodynamic equations
 for the fluctuating velocity field ${\mv} (t,{\mx})\equiv{{\mv}}(x)$
 of an incompressible conducting fluid
 and the magnetic induction ${\mB}=\sqrt{\rho\mu}{\mb}$
 ($\rho$ is the density and $\mu$ the permeability of the fluid) in the form
 \cite{Fournier82,Adzhemyan85}
\begin{align}
   \partial_t {\mb} +
  ({\mv}\cdot {\boldnabla}){\mb}
  - ({\mb}\cdot {\boldnabla}){\mv}
  -\nu_0 u_0\nabla^2 {{\mb}}  
  &= {\mf}^b\,,\quad  {\boldnabla}\cdot{{\mf}}^b = 0
  \label{eq:mhd2D_magnet}
\end{align}
 In (\ref{eq:mhd2D_magnet}) parameter $1/u_0$ is the
 unrenormalized magnetic Prandtl number.

 The statistics of ${{\mv}}$ and ${{\mb}}$
 are completely determined by the non-linear equations
 (\ref{eq:double_NS}), (\ref{eq:mhd2D_magnet})
 and the probability distribution
of the external large-scale random forces
 ${\mf}^v$, ${\mf}^b$.

 To analyze renormalization near two dimensions
 the SDE (\ref{eq:mhd2D_magnet}) is 
 supplemented by the forcing statistics
\begin{align}
  \langle\, f_i^b(x_1) f_j^b(x_2)
  \,\rangle &=
  u_0^2\,\nu_0^3\,D_{ij}
    \delta(t)\int
    \frac{{\mbox d}^d {\mk}}{(2\pi)^d}
    \,\,
    P_{ij}({\mk})\,
    \eRM^{i{\mk} \cdot {\mx_1-\mx_2}}\, \nonumber \\
    &\times
    \left[
    g_{b10}\, k^{ 2- 2\Delta-2\,a\, \eps}
    + g_{b20}\,k^2\,\right]
 \equiv D_{ij}^b(x_1-x_2)\,, 
  \label{eq:mhd2D_corel2}
 \\
  \langle\, f_i^v(x_1) f_j^b(x_2)\,\rangle &=  0 \,.
  \label{eq:mhd2D_corel}
\end{align}

 We choose uncorrelated kinematic and magnetic driving
 [$ \langle\, {f}_i^v {f}_j^b\,\rangle=0$], because we are considering arbitrary
 space dimension $d\ge 2$ and it is not possible to define a
 non-vanishing correlation function of a vector field and a pseudovector field in
 this case. This can be done separately for integer dimensions of space,
 but, contrary to claims of some
 authors~\cite{Fournier82,Camargo92}, is no obstacle for
 application of the RG~\cite{Adzhemyan85}.
 As usual the prefactors $ u_0\,\nu_0^3$ and
 $ u_0^2\,\nu_0^3$ in (\ref{eq:mhd2D_corel}) have been extracted for the
 dimensional reasons.

We are working in an arbitrary dimension, but the renormalization will be carried out
within the two-dimensional model. In magnetohydrodynamic turbulence, in contrast to fluid
turbulence, there are direct energy cascades in both two and
three dimensions. Therefore it is natural to expect that the
scaling behavior is rather similar in both cases, and
we apply the same forcing spectrum in all space dimensions $d\ge 2$.

The system of the stochastic MHD equations
 (\ref{eq:double_NS}), (\ref{eq:mhd2D_magnet}) and (\ref{eq:mhd2D_corel})
gives rise to the following De Dominicis-Janssen action:
\begin{align}
  \S[\mv,\mv',\mb,\mb'] &=\frac{1}{2} \mb' D^b \mb' + {{\mb}}'\cdot [ - \partial_t {{\mb}} + u_0 \nu_0 \nabla^2 {{\mb}}
   + ({{\mb}}\cdot{\boldnabla}) {{\mv}} - ({{\mv}}\cdot{\boldnabla}) {{\mb}}
   ] \nonumber\\
    &+ \frac{1}{2}\mv' D^v \mv' + {\mv}'\cdot[  -\partial_t {\mv} - ({\mv}\cdot {\boldnabla} ) {{\mv}} +
  \nu_0 \nabla^2 {\mv}+ ({{\mb}}\cdot{\boldnabla}) {{\mb}} ].  
   \label{eq:mhd2D_ACTION}
 \end{align}
 The dimensional constants
 $g_{v10}$, $g_{b10}$, $g_{v20}$, and $g_{b20}$, which
 control  the amount of randomly
 injected energy through (\ref{eq:mhd2D_corel}),
 play the role of expansion parameters of the perturbation theory.\\

{\subsubsection{Double expansion of the model} \label{subsubsec:mhd2D_mag_double}}

The action~(\ref{eq:mhd2D_ACTION}) gives rise to
four three-point interaction vertices defined by the standard rules~\cite{Vasiliev}, and
the following set of propagators
 \begin{align}  
  &\Delta^{v v'}_{ij}({{\mk}},t)  = \Delta^{v'v}_{ij}(-{{\mk}},-t)
  = \theta(t)\,P_{ij}({{\mk}})\,\eRM^{-\nu_0\,k^2 t}\,,
  \\
  &\Delta^{bb'}_{ij}({{\mk}},t) = 
  \Delta^{b'b}_{ij}(-{{\mk}},-t)
  = \theta(t)\,P_{ij}({{\mk}})\,  \eRM^{- u_0 \nu_0 k^2 t }\,,
  \\
  &\Delta^{vv}_{ij}({{\mk}},t) = \frac{1}{2}\,u_0\,\nu_0^2\,P_{ij}({{\mk}})\,
  \eRM^{ -\nu_0 k^2 | t | } \left(\, g_{v10}\,k^{-2\eps-2\delta}+g_{v20} \,\right)\,, \\
  &\Delta^{bb}_{ij}({{\mk}},t) = \frac{1}{2}
  \,u_0\,\nu_0^2\,P_{ij}({{\mk}})\, \eRM^{ -u_0 \nu_0 k^2 | t | } \left(\, g_{b10}\,k^{-2\,a \eps- 2\delta} + g_{b20}
  \,\right) 
  \label{eq:mhd2D_PropY}
\end{align}
in the time-wave-number representation.
With due account of Galilei invariance of the action~(\ref{eq:mhd2D_ACTION}), and
careful analysis of the structure of the perturbation expansion it can be shown \cite{Adzhemyan85}
that for any fixed space dimension $d>2$ only
three one-particle irreducible (1PI) Green functions
$\Gamma_{v'v}$, $\Gamma_{b'b}$ and $ \Gamma_{v'bb}$
with superficial UV divergences are generated by the
action. They give rise to counterterms of the form
already present in the action, which thus is multiplicatively renormalizable by
power counting for space dimensions $d>2$.

We would like to emphasize that the structure of renormalization
should always be analyzed separately and it is not at all obvious
that the nonlinear terms are not renormalized in the solution of
the stochastic MHD equations. In fact, direct calculation shows
that the Lorentz-force term is renormalized. There seems to be a
certain amount of confusion about this point in the 
literature. For instance, in Refs.~\cite{Liang93,Camargo92} the
authors erroneously neglect renormalization of nonlinear terms
as high-order effect. The approach adopted in Ref.~\cite{Liang93}
for two-dimensional MHD turbulence
was criticized by Kim and Young~\cite{Kim99}, who,
alas, in their field-theoretic treatment of the same problem ignore
renormalization of the Lorentz force without any justification.
They also neglect renormalization of the forcing correlations
by effectively considering renormalization of the model at
$d>2$, which does not seem to be appropriate in a setup in which
the strictly two-dimensional quantities, the stream function and magnetic
potential, are used for the description of the problem.

The analysis of the correlation functions of the velocity
field and magnetic induction is essential near two dimensions,
since
in two dimensions additional divergences in the graphs of the 1PI
Green functions $\Gamma_{v'v'}$ and  $\Gamma_{b'b'}$ occur.
The simplest way to include the corresponding local counter terms
${\mv}'\boldnabla^2 {\mv}'$ and ${\mb}'\boldnabla^2 {\mb}'$
in the renormalization is to add corresponding
{\em local} terms  to the force
correlation function at the outset in order to keep the model multiplicatively
renormalizable. This is why
we have used the force correlation functions~(\ref{eq:mhd2D_corel}) and
(\ref{eq:double_random_force_fourier}) with the relation (\ref{nakach2}) with both long-range and short-range correlations
 taken into account.
As a result, the action~(\ref{eq:mhd2D_ACTION}) is multiplicatively
renormalizable and allows for a standard RG asymptotic analysis~\cite{Vasiliev}.

The model is regularized using a combination of analytic and
dimensional regularization with the parameters $\eps$ and
$\Delta$, the latter was introduced in Eq. (\ref{dz}). As a consequence, the UV divergences appear as
poles in the following linear combinations of the regularizing
parameters: $\eps$, $\Delta$, $2\eps+\Delta$, and
$(a+1)\eps+\Delta$. In principle, the exponent of magnetic forcing correlations $2a\eps$
may be treated as the second analytic regulator to construct a triple expansion
of renormalized quantities at a suitable fixed point. However, in the ray
scheme used here all regulators are assumed to be of same order and the
discussion of a triple expansion in this case would be rather formal.

The UV divergences  can be
removed by adding suitable counterterms to the basic action
$S_B$ obtained from the unrenormalized one (\ref{eq:mhd2D_ACTION}) by
the substitution of the renormalized
parameters for the bare ones:
$g_{v10}  \rightarrow \mu^{2\eps}  g_{v1}$,
$g_{v20}  \rightarrow \mu^{-2 \Delta} g_{v2}$,
$g_{b10}  \rightarrow \mu^{2 a \eps} g_{b1}$,
$g_{b20}  \rightarrow \mu^{-2 \Delta} g_{b2}$,
$\nu_0 \rightarrow \nu$,
$u_0 \rightarrow u$,
where $\mu$ is a scale setting parameter having the same canonical
dimension as the wave number.

For the actual calculations the ray scheme with minimal subtractions is convenient. The counter terms for
the basic action
corresponding to the unrenormalized action~(\ref{eq:mhd2D_ACTION}) are
\begin{align} 
  \Delta \S & = 
  \nu \left(Z_1-1\right) {{\mv}}' \boldnabla^2\, {{\mv}}+
   u \nu\,\left(Z_2-1\right) {{\mb}}'\boldnabla^2\, {{\mb}}
   + \frac{1}{2}(1-Z_4)  u {\nu^3} g_{v2} {\mu}^{-2\delta}
  {{\mv}}'{\boldnabla^2} {{\mv}}'\nonumber\\
  & + \frac{1}{2} (1- Z_5 ) u^2 {\nu^3} g_{b2} {\mu}^{-2\Delta}
  {{\mb}}'{\boldnabla^2} {{\mb}}' 
  + (Z_3-1) {{\mv}}'({{\mb}}\cdot{\boldnabla}) {{\mb}} 
  ,
  \label{eq:mhd2D_kontra}
\end{align}
where the renormalization constants $Z_1$, $Z_2,$
$Z_4$, $Z_5$  renor\-mali\-zing the unrenormalized (bare)
para\-me\-ters
$e_0=\{g_{v10},g_{v20},g_{b10},g_{b20},u_0,\nu_0\}$ and the constant  $Z_3$ renormalizing
the fields ${\mb}$, and ${{\mb}}'$, are
chosen to cancel the UV divergences appearing in the Green functions constructed
using the basic action.
Due to the Galilean invariance of the action
the fields ${{\mv}}'$, and ${{\mv}}$ are not renormalized.

 In a multiplicatively renormalizable model, such as~ (\ref{eq:mhd2D_ACTION}),
 the counter terms (\ref{eq:mhd2D_kontra})
 can be chosen in a form containing a finite number of
 terms of the same algebraic structure
 as the terms of the original action (\ref{eq:mhd2D_ACTION}).
 Thus, all UV divergences of the graphs of the perturbation expansion
 may be eliminated
 by a redefinition of the parameters of the original model.

 Renormalized Green functions are expressed
 in terms of the renormalized parameters
\begin{align}
  g_{v1}&= g_{v10} {\mu}^{-2\eps} Z_1^2 Z_2, & g_{v2} &=g_{v20}{ \mu}^{2\Delta} Z_1^2 Z_2 Z_4^{-1}, &\nu& = \nu_0 Z_{1}^{-1},
  \nonumber\\  
  g_{b1} & =  g_{b10} {\mu}^{-2\,a\eps} Z_1 Z_2^{2} Z_3^{-1},
  &g_{b2}& = g_{b20} {\mu}^{2\Delta} Z_1 Z_2^{2} Z_3^{-1} Z_5^{-1}
  , &u&=u_0\, Z_2^{-1} Z_1 \,.
   \label{eq:mhd2D_Zkaka1}
 \end{align}
 The definitions  (\ref{def-beta}) and the relations  (\ref{eq:mhd2D_Zkaka1})
 yield $\beta$ functions of the form
 \begin{align}
   \beta_{g_{v1}} & = g_{v1}(-2\eps+2\gamma_1+\gamma_2), 
   &\beta_{g_{v2}} & = g_{v2}(2\Delta+2\gamma_1+\gamma_2-\gamma_4),\nonumber\\
   \beta_{g_{b1}} & =  g_{b1}(-2a\eps+\gamma_1+2\gamma_2-\gamma_3),
   &\beta_{g_{b2}} & = g_{b2}(2\Delta+\gamma_1+2\gamma_2-\gamma_3-\gamma_5),\nonumber\\
   \beta_u & =  u(\gamma_1-\gamma_2).
   \label{eq:mhd2D_beee1}
\end{align}

Apart from the Gaussian fixed point $g_{v1}^*=g_{v2}^*=g_{b1}^*=g_{b2}^*=0$
with no fluctuation effects on the large-scale asymptotics,
which is IR stable for $\Delta>0$, $\eps<0$, $a>0$,
there are two nontrivial  IR stable fixed
points of the RG with nonnegative $g_{v1}^*$,  $g_{v2}^*$, $g_{b1}^*$,$g_{b2}^*$, and $u^*$.

The thermal fixed point is generated by
short-range correlations of the random
force with
\begin{align}
  g_{v1}^{\ast}   = 0,\quad
  g_{v2}^{\ast}    = -4\pi (1 + \sqrt{17} ){\Delta}, \quad
  g_{b1}^{\ast}    = 0, \quad
  g_{b2}^{\ast}    = 0,
   \label{eq:mhd2D_thermal}
\end{align}
and the inverse magnetic Prandtl number
\begin{equation}
  u^{\ast}=\frac{\sqrt{17}-1}{2}\simeq 1.562\,.
  \label{eq:mhd2D_Pr}
\end{equation}
Physically, the asymptotic behavior described by this fixed point
is brought about by thermal
fluctuations of the velocity field~\cite{FNS77}.
The region of stability of the thermal fixed point (\ref{eq:mhd2D_thermal}), (\ref{eq:mhd2D_Pr}) is
$2\eps+3\Delta<0$, $\Delta<0$ in the $\Delta$, $\eps$
plane. For the magnetic forcing-decay parameter
$a$ the stability region is determined by the inequality
$8a\eps+(13+\sqrt{17})\Delta<0$.

The {\em kinetic} fixed point~\cite{Fournier82} generated by the
forced fluctuations of the velocity field
is given by the universal inverse magnetic Prandtl number (\ref{eq:mhd2D_Pr}),
the parameters
\begin{align}
  g_{v1}^{\ast}=
  \frac{128\pi}{9(\sqrt{17}-1)} \frac{\eps(2\eps+3\Delta)} {\eps+\Delta},\quad
  g_{v2}^{\ast}=\frac{128\pi}{9(\sqrt{17}-1)}\frac{\eps^2}{\Delta+\eps},
  \label{eq:mhd2D_fix11}
\end{align}
and zero couplings of the magnetic forcing
\beq
 g_{b1}^{\ast}=g_{b2}^{\ast}=0\,,
\eeq
and it may be associated with
turbulent advection of the magnetic field.
The values of $g_{v1}^{\ast}$ and $g_{v2}^{\ast}$ in
(\ref{eq:mhd2D_fix11})
correspond to those find previously in Ref.~\cite{HonNal96}.
The region of stability
of the kinetic fixed point in the $\Delta$, $\eps$ plane
is $\eps>0$, $2\eps+3\Delta >0$.
The stability of this fixed point also requires that the parameter
$a< (13+\sqrt{17})/12\approx 1.427$
independent of the ratio $\Delta/\eps$. In spite of the
absence of
renormalization of the forcing correlation, the momentum-shell
approach \cite{Fournier82} yields the same condition.

The system of equations 
for the fixed points in this multi-charge problem is rather
complicated and thus has several (in general complex-number) solutions, which we do note quote
explicitly here, because they are not physically relevant:
apart from the fixed points listed above there are eight IR unstable real-number fixed points
in the physical region (all $g\ge 0$) of the parameter space, and
several unphysical ones.
Among the unstable fixed points are, in particular, all the
possible candidates to {\em magnetic} fixed points,
i.e. fixed points with a non-vanishing magnetic coupling constant. Therefore,
the conclusion made in Refs.~\cite{Fournier82,Adzhemyan85}
(although on inconsistent grounds) that the RG does
not predict any magnetically driven scaling regime at and near two dimensions,
is confirmed in the double-expansion approach.

An analysis of beta functions reveals that at the leading order of the ray scheme
there is no fixed point with both $g_{b1}$ and $g_{b2}$
non-vanishing: at least one of them must be zero \cite{HHJ01}. This, of course,
severely reduces the set of possible magnetic fixed points at the
outset. 

In three dimensions there are stable magnetic fixed points, whose stability
as a function of space dimension may be followed in the usual $\eps$
expansion. Stability of the magnetic fixed point disappears at a borderline
dimension, whose leading-order value is $d_c=(3+\sqrt{649})/10\approx 2.848$~\cite{Fournier82}.
As seen from the results presented above, to follow the crossover between the two regimes from below within the ray scheme of
triple expansion calculation beyond the one-loop order is needed.
\\

{\subsubsection{Renormalization with maximum divergences above two dimensions} \label{subsubsec:mhd2D_mag_max}}

 All the renormalization constants and the RG functions quoted above may be calculated
 also at finite $\Delta$.
 The resulting system of fixed-point
 equations allows for a solution in a form of an
 $\varepsilon$ expansion (with finite $\Delta$) and yields the same
 result as the usual $\varepsilon$ expansion at the leading order.
 However, this approach is not self-consistent in the sense that
 the field theory is not renormalizable at finite $\Delta>0$, but
 only in the form of a simultaneous expansion in the coupling
 constants and $\Delta$~\cite{Vasiliev}.

The aim is now  to maintain the model UV finite for $\Delta>0$ and simultaneously
keep track of the effect of the additional divergences near two dimensions. To
this end we introduce an additional UV cutoff in all propagators, i.e.
instead of the set (\ref{eq:mhd2D_PropY}) we use the propagators
\begin{align}
  \Delta^{v v'\Lambda}_{ij}({{\mk}},t) & =
   \theta(t)\,\theta(\Lambda-k)\,P_{ij}({{\mk}}) \eRM^{-\nu_0\,k^2 t}, \\
   \Delta^{bb'\Lambda}_{ij}({{\mk}},t) & = 
   \theta(t) \theta(\Lambda-k) P_{ij}({{\mk}}) \eRM^{- u_0 \nu_0 k^2 t }, \\
   \Delta^{vv\Lambda}_{ij}({{\mk}},t) & =\frac{1}{2}\,\theta(\Lambda-k)
    u_0\nu_0^2\,P_{ij}({{\mk}}) \eRM^{ -\nu_0 k^2 | t | } \left( g_{v10} k^{-2\eps-2\Delta}+g_{v20}
   \right),\\
   \Delta^{bb\Lambda}_{ij}({{\mk}},t) & =  \frac{1}{2}\theta(\Lambda-k) u_0\nu_0^2 P_{ij}({{\mk}})
   \eRM^{ -u_0 \nu_0 k^2 | t | } \left( g_{b10} k^{-2a \eps- 2\Delta} + g_{b20} \right),
\end{align}
where $\Lambda$ is the cutoff wave number. This change obviously does not affect
the large-scale properties of the model. We would like to
emphasize that the additional cutoff must be introduced uniformly
in all lines in order to maintain the model multiplicatively
renormalizable. An attempt to introduce the cutoff, say, in the
local part of the correlation functions only by the substitution
$k^2\to \theta(\Lambda-k)k^2$ would fail to renormalize the model
multiplicatively, because loop contributions to the complete (dressed)
correlation function would not reproduce such a structure in the wave-vector space.

In contrast with particle field theories we will keep the cutoff
parameter $\Lambda$ fixed, although large compared with the
physically relevant wave-number scale. This introduces an explicit
dependence on $\Lambda$ in the coefficient functions of the RG,
which has to be analyzed separately in the large-scale limit in
the coordinate space. The setup is thus very similar to that of
Polchinski~\cite{Polchinski84}.

The coefficient functions of the RG equations become, in general, functions of
the parameters $\mu$ and $\Lambda$ through the dimensionless ratio
$\mu/\Lambda$. Solution for the scalar coefficients of equal-time velocity
and magnetic induction pair correlation functions is (cf. (\ref{eq:double_Gp})):
\begin{align}
  G^{vv}(sk,\Lambda;g) & = 
  \overline{ \nu}^2k^{-2\Delta}R_v\left(1,{\mu s\over \Lambda};
  \overline{g}\right), \\
  G^{bb}(sk,\Lambda;g) & = 
  \eRM^{\int_1^s\!{\rm d}x\,\gamma_{3}(x)/x}
   \overline{\nu}^2k^{-2\Delta}R_b\left(1,{ \mu s\over \Lambda};
  \overline{g}\right),
\end{align}
where  $\overline{g}$ is now the solution of the Gell-Mann-Low equations:
\beq
  \frac{\dRM\overline{g}(s)}{\dRM\ln s}= \beta_g
  \left[ \overline{g}(s),{\mu s\over \Lambda}\right]\,,
\eeq
with the $\beta$ functions explicitly depending on $s$, the dimensionless wave number.

Above two dimensions an UV renormalization of the model would
require and infinite number of counter terms and in this sense it
is not renormalizable in the limit $\Lambda\to \infty$.
To avoid dealing with these formal complications, we keep the additional
cutoff $\Lambda$ fixed (although large), and choose the
renormalization procedure according to the principle of maximum
divergences~\cite{HonNal89}: the same terms of the action are
renormalized as in the two-dimensional case in the previous section
(\ref{eq:mhd2D_kontra}), but the renormalization constants may
have an explicit dependence on the scale-setting
parameter through the ratio $\mu/\Lambda$.
The two counter terms
\beq
  \int\!\mbox{d} x\,
  \big[\frac{1}{2}\,(1-Z_4)   u {\nu^3} g_{v2}\,{\mu}^{-2\Delta}
  {{\mv}}'{\nabla^2} {{\mv}}' +\frac{1}{2} (1- Z_5 ) u^2 {\nu^3} g_{b2}{\mu}^{-2\Delta}
  {{\mb}}'{\nabla^2} {{\mb}}'
  \big]
\eeq
are superfluous in the sense that in the limit $\mu/\Lambda\to 0$
the contribution to the Green functions of the graphs containing
the coupling constants $g_{v2}$ and  $g_{b2}$ remains finite
provided $2\Delta=d-2$ is fixed and finite and the other
counter terms are properly chosen. This is
guaranteed by Polchinski's theorem~\cite{Polchinski84}. We retain
these counter terms in order to have a possibility to pass to the
limit $\Delta\to 0$ in the RG equations.

The presence of the explicit cutoff implies some technical difficulties
in the calculation of the renormalization constants in the
traditional field-theoretic approach, which arise because
we are dealing with massless vector fields. It turns out that
the coefficient functions of the
RG equation are simplest in a
renormalization procedure, which is similar to the
momentum-shell renormalization of Wilson~\cite{WilKog74}.
If we were calculating over the whole wave-vector space without an
explicit UV cutoff, there would not be much difference between the
effort required in both approaches. The presence of the UV cutoff
makes calculations with non-vanishing external wave vectors rather
tedious.

Let us remind that the choice of a renormalization procedure
basically is the choice of the rule according to which the counter-term
contributions are extracted from the perturbation expansion
of the Green functions of the model.
The usual field-theoretic prescription goes as
follows~\cite{Collins}:
Consider a 1PI graph $\gamma$, let $R(\gamma)$ be the renormalized value
of the graph, and let $\overline{R}(\gamma)$ the value of the graph with
subtracted counter terms of all the subgraphs, then
\begin{equation}
  \label{eq:mhd2D_R}
  R(\gamma)=\overline{R}(\gamma)- T\overline{R}(\gamma)
\end{equation}
where the operator $T$ extracts the counter-term contribution
from $\overline{R}(\gamma)$. Usually $\overline{R}(\gamma)=\gamma$ on
one-loop graphs, and the renormalization scheme is specified by
the action of $\overline{R}(\gamma)$ on multi-loop graphs together
with the definition of the operator $T$. The counter terms may
then be constructed recursively with the use of (\ref{eq:mhd2D_R}) and the
definitions of $T$ and $\overline{R}(\gamma)$. There is rather large
freedom in the choice of the counter-term operator, but to arrive
at Green functions finite in the limit $\Lambda\to \infty$
in two dimensions -- which we want to have a connection with the
double expansion -- the operation $\overline{R}$ must be chosen
properly.

Here, we have used a renormalization procedure, in which the
operation $\overline{R}(\gamma)$ is standard~\cite{Collins}, and the
subtraction operator $T$ is defined as follows:
let $F_\gamma(\omega,{\mk},\Lambda)$ be the function of external
frequencies and wave-vectors (which also depends on the cutoff parameter $\Lambda$)
corresponding to the expression $\overline{R}(\gamma)$ (this is not a
1PI graph, in general).
The subtraction operator $T$ extracts the same set of terms
of the Maclaurin-expansion in the external wave-vectors,
which generate the counter terms (\ref{eq:mhd2D_kontra}), from the {\em
difference}
$F_\gamma(\omega,{\mk},\Lambda)-F_\gamma(\omega,{\mk},\mu)$. These coefficients of the Maclaurin-expansion are
calculated at vanishing external frequencies and wave-vectors.
It should be noted that the coefficients of this Maclaurin
expansion of the function $F_\gamma(\omega,{\mk},\Lambda)$
itself may not exist in the limit $\omega\to 0$ in this "massless" model, but the difference
$F_\gamma(\omega,{\mk},\Lambda)-F_\gamma(\omega,{\mk},\mu)$ allows for a Maclaurin
expansion finite in the limit $\omega\to 0$ to the order required
for the renormalization. The counter-term operator $T$ and the combinatorics of the
renormalization procedure for higher-order graphs may then be
constructed in the standard fashion. Although this is actually not needed
in the present one-loop calculation, the very possibility of this
extension is necessary to guarantee that the renormalization renders the model finite
in the limit $\Lambda\to\infty$ in two dimensions.

Effectively, at one-loop order this prescription reduces the region of
integration to the momentum shell $\mu<k<\Lambda$, which leads to
the same calculation as in the momentum-shell renormalization. In
higher orders, however, our renormalization scheme does not coincide with
the momentum-shell renormalization.
The point of the present renormalization procedure is that without some sort of IR cutoff the
subtraction at zero momenta and frequencies is, in general, not
possible in a massless model, whereas a subtraction at vanishing
frequencies and external momenta of the order of $\mu$ leads to
much more complicated calculations due the heavy index structure.

At one-loop accuracy in this scheme the $\gamma$ functions are
\begin{align}
  \gamma_1 & = \frac{1}{2B}\Bigl[\left(d^2-d-2\eps\right)u\,g_{v1}+\left(d^2+d-4+2
  a \eps\right) g_{b1} + \left(d^2-2\right)\left(u\,g_{v2} + g_{b2}\right)\Bigr],
  \nonumber\\
  \gamma_2&=\frac{1}{(1+u)B}\Bigl[\left(d-1\right)\left(d+2\right)\left(g_{v1}+
  g_{v2}\right)+ \left(d+2\right)\left(d-3\right)\left(g_{b1}+g_{b2}\right)\Bigr],
  \nonumber\\
  \gamma_3 & = {2\over B}\left[g_{b1}+g_{b2}-g_{v1}-g_{v2} \right],  \nonumber\\
  \gamma_4 & = \frac{d^2-2}{2 g_{v2}B}\left[u\left(g_{v1}+g_{v2}\right)^2+
  \left(g_{b1}+g_{b2}\right)^2\right], \nonumber\\
  \gamma_5 & = \frac{2(d-2)(d+2)}{g_{b2}(1+u)B} \left(g_{b1}+g_{b2}\right)\left(g_{v1}+g_{v2}\right),
  \label{eq:mhd2D_GGGa2}
\end{align}
where $B=d(d+2)\Gamma(d/2)(4\pi)^{d/2}$. These expressions reveal
an additional advantage of this renormalization scheme: at
one-loop order there is no explicit dependence on $\mu/\Lambda$ in
the coefficient functions of the RG. At one-loop level
a direct comparison with the expressions obtained in the
framework of the Wilson RG is also possible:
the dependence on $g_{v1}$,  $g_{b1}$, and $u$ of the $\beta$ functions
$\beta_{gv1}$, $\beta_{gb1}$ and $\beta_{u}$ corresponding to
(\ref{eq:mhd2D_GGGa2}) coincides with that of their
counterparts of Ref.~\cite{Fournier82} up to notation.

The set of $\beta$ functions generated by (\ref{eq:mhd2D_GGGa2}) allows for
a fixed-point solution in the form of an $\eps$ expansion.
Little reflection shows that the fixed-point equations in this
case have a self-consistent solution with the following
leading-order behavior: $u=\O(1)$,
$g_{v1}=\O(\eps)$, $g_{b1}=\O(\eps)$,
$g_{v2}=\O(\eps^2)$ and $g_{b2}=\O(\eps^2)$. The
actual fixed-point values of $g_{v1}$, $g_{b1}$ and $u$ in the
$\eps$ expansion as well as
the stability regions with respect to $\eps$ are determined by
the same set of equations as in the earlier momentum-shell~\cite{Fournier82}
and field-theoretic~\cite{Adzhemyan85} calculation
above two dimensions. The stability condition with respect to the
dimension of space of these fixed points is, as expected, $d>2$.

It should be noted that the function $\gamma_5$ is finite in the
set (\ref{eq:mhd2D_GGGa2}), whereas in the double-expansion approach it was
equal to zero 
This means that magnetic fixed
points with both $g_{b1}$ and $g_{b2}$ may exist. In
fact, there is one such fixed point stable at high dimensions of
space which gives rise to a magnetically driven scaling regime.
This fixed point may be found in the $\eps$ expansion, and we
have also investigated its stability numerically. Technically
speaking, the appearance of a magnetic fixed point with both
magnetic couplings non-vanishing would be a completely expected thing to
happen in the two-loop approximation, since we have not found any
symmetry reasons or the like to prevent the renormalization of the
magnetic forcing at higher orders. Thus, to investigate this
effect consistently in the $\eps$ expansion would
require a full two-loop renormalization of the model, which is
beyond the scope of the present analysis.

\begin{figure}[!htb]
    \begin{minipage}{0.475\textwidth}
       \mbox{ } 
        \includegraphics[width=6cm]{\PICS 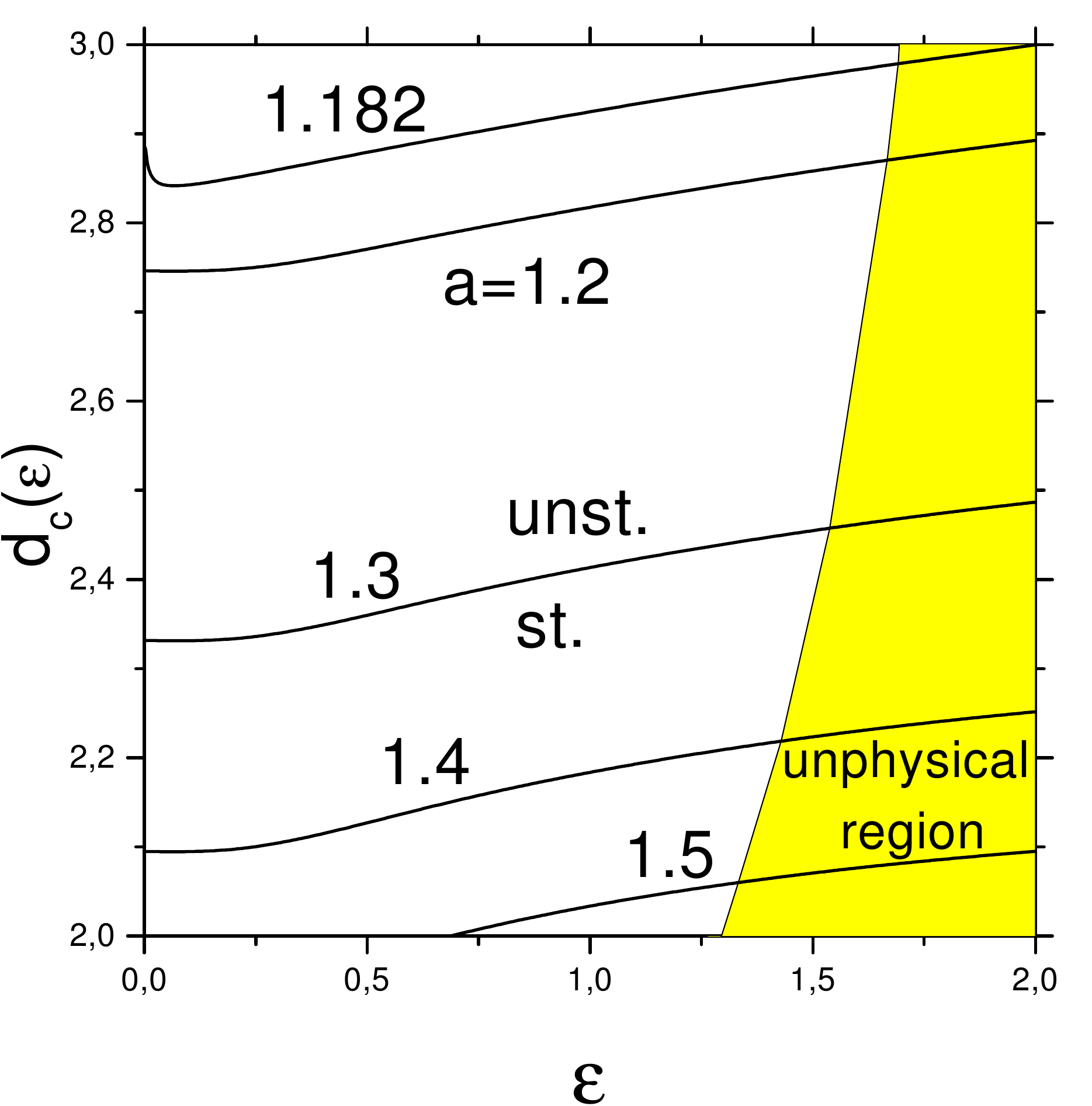}
        \caption{The borderline dimension $d_c$ between the stability regions of the kinetic fixed point
of the RG equations for (\ref{eq:mhd2D_beee1}) and (\ref{eq:mhd2D_GGGa2}) for
magnetic forcing-decay parameter $a$ near the double-expansion threshold $a=1.427$.
This plot reveals the strong dependence of the borderline
dimension on the parameter $a$. The shaded region on the right
corresponds to values $\eps>2/a$, for which the forcing
correlation function in the powerlike form (\ref{eq:double_kernel}) leads to intractable IR
divergences, and a corresponding IR cutoff (magnetic integral
length scale) must be introduced.}
\label{fig:mhd_fig1}
    \end{minipage}%
    \hfill
    \begin{minipage}{0.475\textwidth}
         \mbox{ } 
        \includegraphics[width=6cm]{\PICS 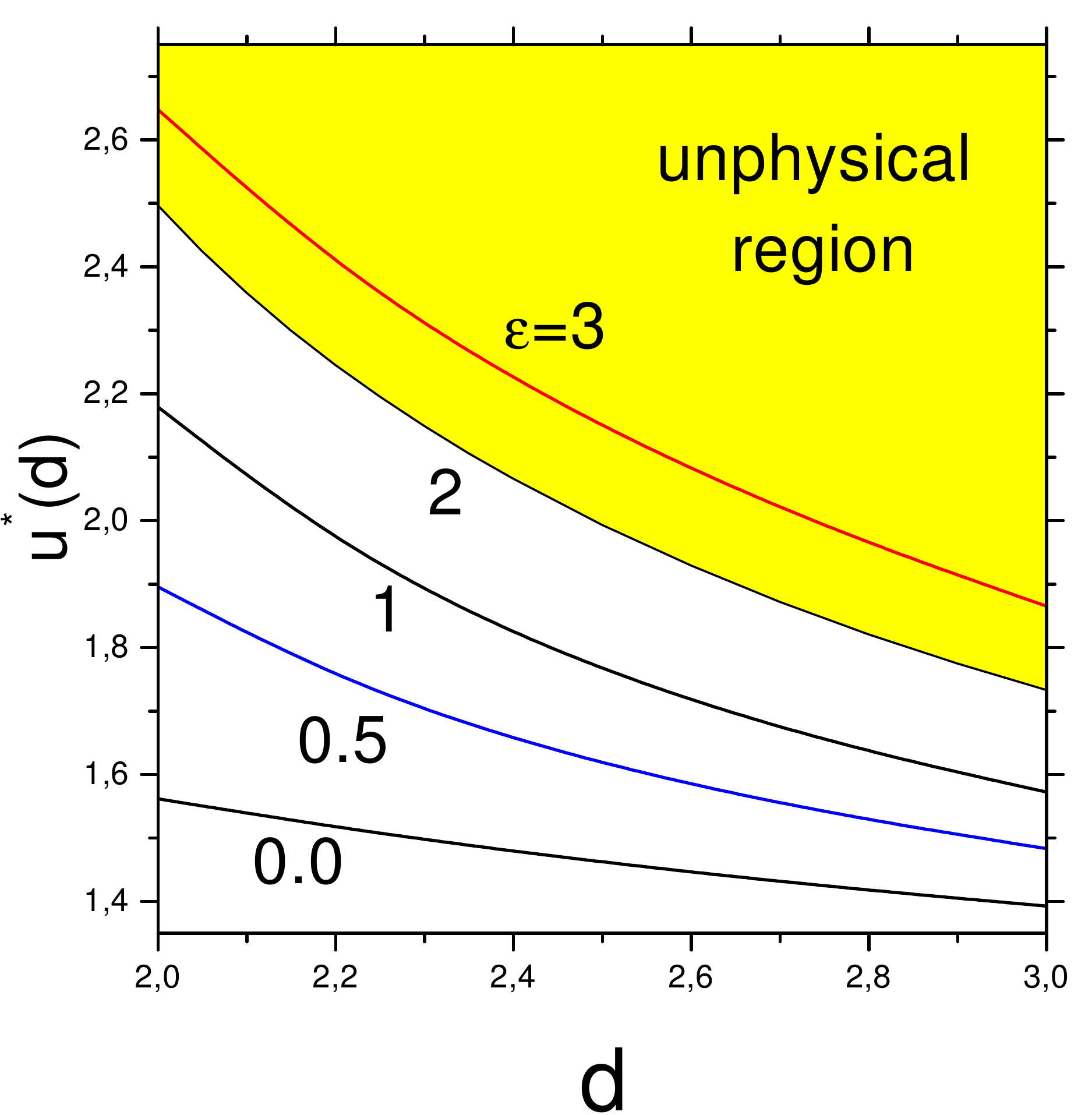}        
        \caption{The fixed-point value of the inverse magnetic Prandtl number $u^*$ as a
function of the space dimension $d$ and the decay parameter
$\eps$. The lowest curve corresponds to the leading order in
the $\eps$ expansion, which is not affected by thermal fluctuations.
The shaded region in the upper part
of the plot
corresponds to values $\eps > 2$, for which an IR cutoff (kinetic integral
length scale) must be introduced in the
correlation function (\ref{eq:double_kernel}).
}
\label{fig:mhd_fig2}
\mbox{ }\\ \mbox{ }\\ 
    \end{minipage}
\end{figure}

\begin{figure}[!htb]
    \begin{minipage}{0.475\textwidth}
  \includegraphics[width=6cm]{\PICS 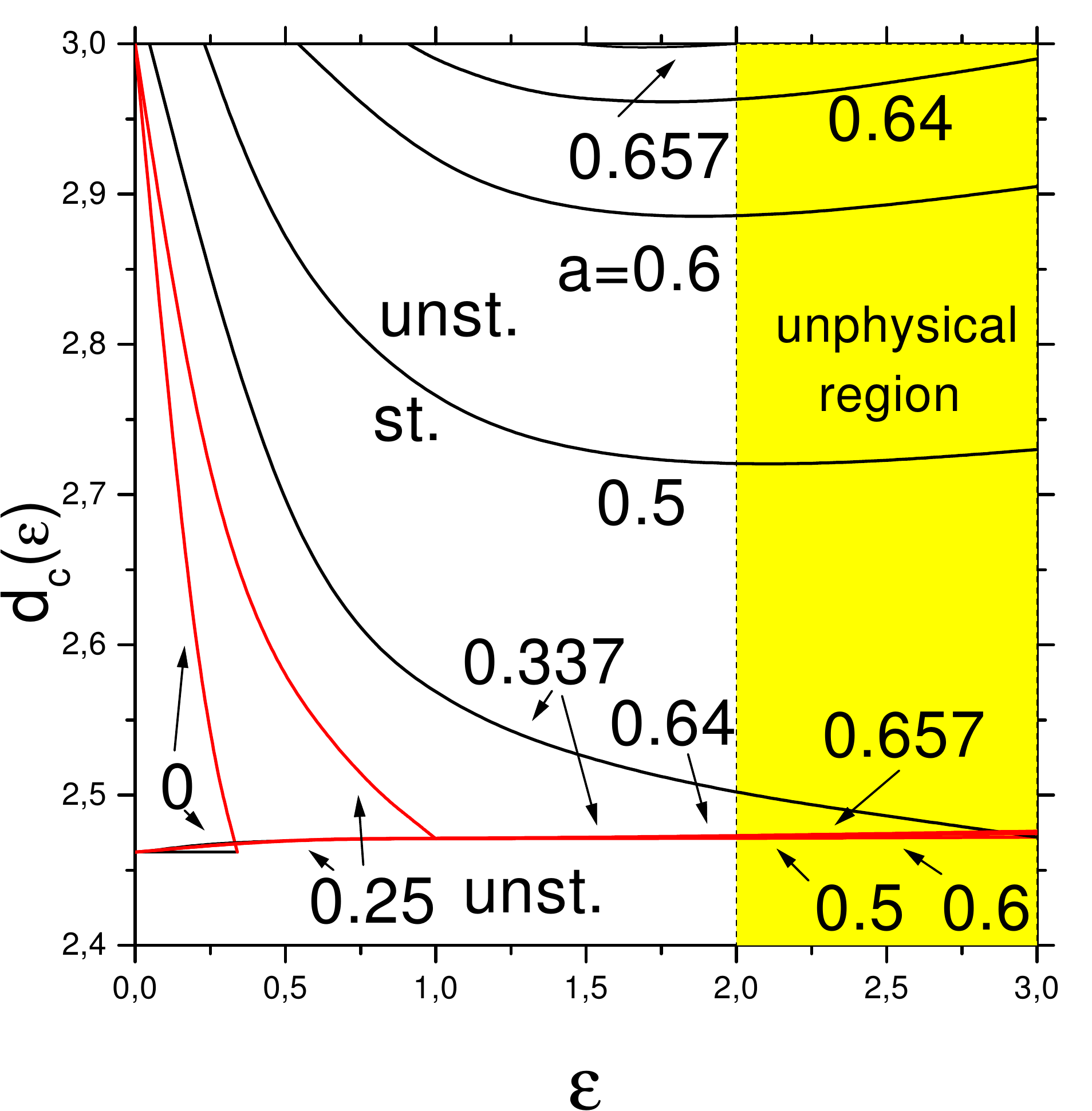}  
\caption{The borderline dimension $d_c$ between the stability regions of the magnetic fixed point
of the RG flow equations (\ref{eq:mhd2D_Zkaka1}) and (\ref{eq:mhd2D_GGGa2}) for
magnetic forcing-decay parameter $a<1$. For sufficiently small values of $a$ the magnetic
fixed point is unstable for any finite value of $\eps$, but
the region of stability grows with the growth of $a$ so that for
$a>0.658$ the magnetic point becomes stable even in three
dimensions for finite values of $\eps$. The shading shows the
region, where $\eps > 2$, in which the powerlike correlation
function (\ref{eq:double_kernel}) cannot be consistently used.}
\label{fig:mhd_fig3}
    \end{minipage}%
    \hfill
    \begin{minipage}{0.475\textwidth}
  \includegraphics[width=6cm]{\PICS 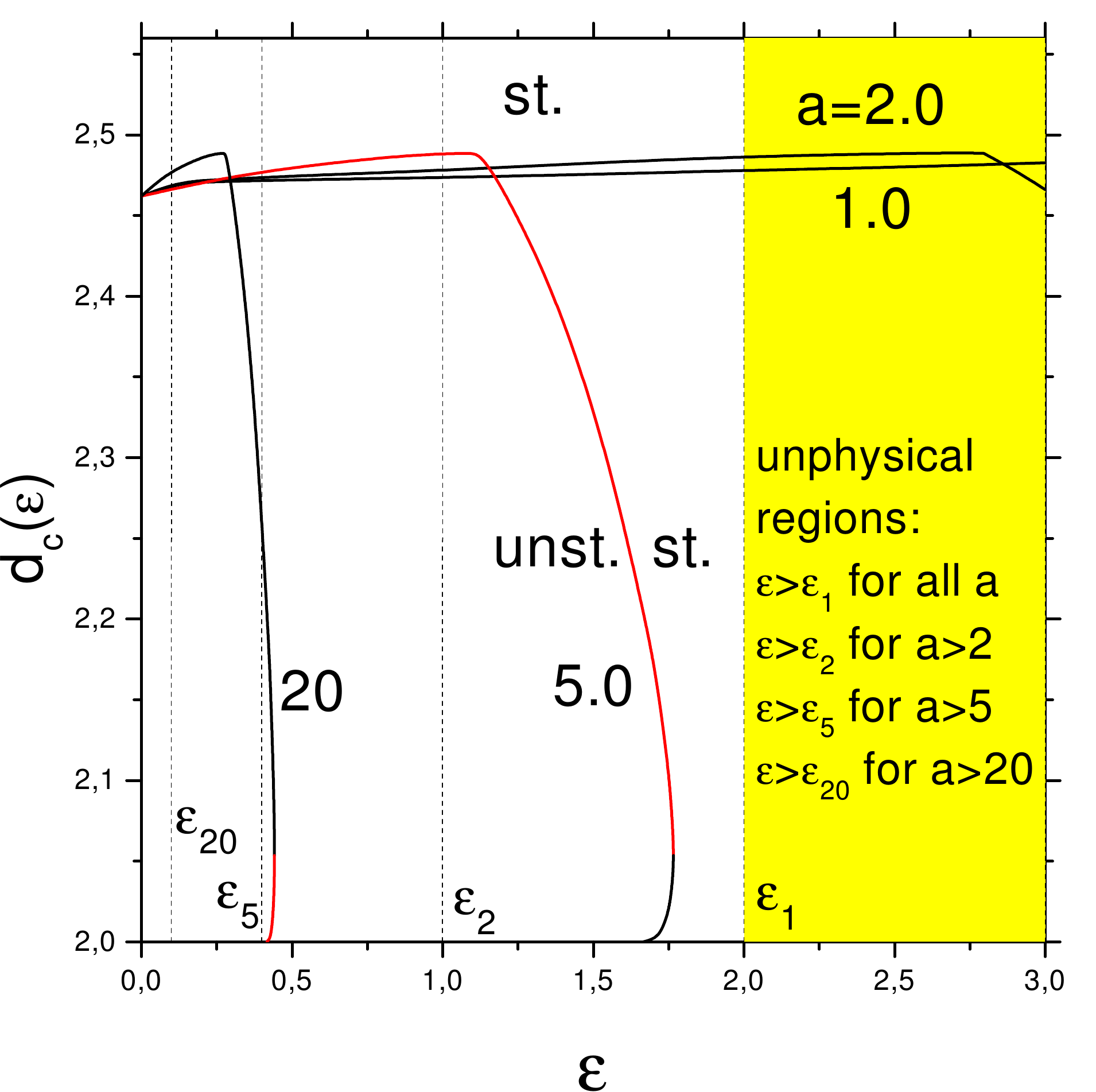}
\caption{The borderline dimension $d_c$ between the
stability regions of the magnetic fixed point of the RG equations
(\ref{eq:mhd2D_beee1}) and (\ref{eq:mhd2D_GGGa2}) for large values of the magnetic
forcing-decay parameter $a>1$. The shaded area and half-planes with vertical
dashed border lines show
regions, where $\eps > 2/a$, in which the powerlike correlation
function (\ref{eq:double_kernel}) cannot be consistently used.
}
\label{fig:mhd_fig4}
\mbox{ }\\ \mbox{ }\\
    \end{minipage}
\end{figure}

The stability of the kinetic scaling regime is
strongly affected by the behavior of magnetic fluctuations: from
Fig. \ref{fig:mhd_fig1} it is seen that the steeper falloff of the correlations of
the magnetic forcing in the wave-vector space compared with that
of the kinetic forcing the lower is the space dimension, above
which the kinetic fixed point is stable. In particular, when the
parameter $a>1.427$ the kinetic fixed point ceases to be stable
even in two dimensions. In three dimensions the kinetic scaling
regime is stable against magnetic forcing, when $a<1.15$.

The monotonic growth of the kinetic-fixed-point value of the
inverse magnetic Prandtl number $u^*$ as a function of the kinetic forcing-decay parameter
in a fixed space dimension
is depicted in Fig. \ref{fig:mhd_fig2}. The plot shows also that
$u^*$ is a monotonically decreasing function of the space
dimension at fixed $\eps$. The lowest-lying curve corresponds
to the leading order of the $\eps$ expansion~\cite{Fournier82,Adzhemyan84}
\beq
  u^*=\frac{1}{2}\left[ -1+ \sqrt{1+\frac{8\,(d+2)}{d}}\,\right].
\eeq
We are particularly interested in the stability of the magnetic
fixed point, and have carried out extensive numerical calculations
of the stability of this fixed point as a function of $\eps$
and  the space dimension $d$. The results are plotted in Figs.
\ref{fig:mhd_fig3} and  \ref{fig:mhd_fig4}.

In Fig. \ref{fig:mhd_fig3} the magnetic forcing-decay parameter $a<1$
(i.e. the kinetic-forcing correlations fall off steeper in the
wave-vector space than those the magnetic forcing) and it is seen
that for very small $a$ a slowly enough decaying kinetic forcing
renders the magnetic scaling regime unstable. In particular, this threshold is very
small in three dimensions. With the growth of
$a$ a strip of stability of the magnetic fixed point in the
$\eps$, $d$ plane appears such that the magnetic regime
remains stable in three dimensions for all allowed kinetic forcing
patterns.  It is also seen that the magnetic
fixed point is persistently unstable at $d\le 2.46$ for all
$\eps$. This borderline dimension should be compared with
that given by the $\eps$ expansion $d_c=2.85$. From the
solution it can be seen that this
significant discrepancy is due to the appearance of a stable
magnetic fixed point completely different from that found in the
$\eps$ expansion: in the latter the magnetic fixed point
is given by $g_{v1}^*=g_{v2}^*=g_{b2}^*=u^*=0$ and
$g_{b1}^*=4d(d+2)\Gamma(d/2)(4\pi)^{d/2}a\eps/(d^2-3d-32)$,
whereas at the magnetic fixed point, whose stability is plotted in
Figs. \ref{fig:mhd_fig3} and \ref{fig:mhd_fig4}, only $g_{v1}^*=u^*=0$ with
non-vanishing fixed-point values of the other couplings. Thus, the
lowering of the borderline dimension of stability of the magnetic
scaling regime is a
strong effect of the thermal fluctuations described by the
short-range terms in the forcing correlation functions.  Fig. \ref{fig:mhd_fig4}
shows the lower boundary of the stability region of the magnetic
fixed point for large values of $a$, when magnetic-forcing
correlations fall off much faster that kinetic-forcing
correlations in the wave-vector space. A remarkable feature of
both plots is the insensitivity of the lower border of the stability strip to the
value of magnetic forcing-decay parameter $a$.\\

{\section{Violated symmetries} \label{sec:violations}}
Models that we have considered so far can be in some sense regarded as ideal. 
As has been mentioned in Sec. \ref{sec:models} a traditional approach to the description of fully developed
 turbulence is based on the stochastic Navier-Stokes equation \cite{Wyld}.
The complexity of this equation leads to great difficulties
which do not allow one to solve it even in the simplest case
when one assumes the  isotropy of the system under consideration. On
the other hand, the isotropic turbulence is almost delusion and if
exists is still rather rare. Therefore, if one wants to model more
or less realistic developed turbulence, one is pushed to consider
 more general models of turbulence. This,
of course, rapidly increases complexity of the corresponding
 model which itself has to involve the part
responsible for description of the anisotropy.

In this section we give a brief overview of two possible deviation from ideality. 
First, in Sec. \ref{subsec:helicity} employing the Kraichnan model, we study violation of 
mirror-symmetry on the advection of passive quantity. Later, in Sec. \ref{subsec:str_anistropy}, the
 advection problem will be studied using velocity field with strong anisotropy taken into 
 account and in Sec. \ref{subsec:mhdstr_pamf} 
 the effect of anisotropy on stochastic magnetohydrodynamics will be studied.
In the last part, Sec. \ref{subsec:weak_wta}, the anisotropy will be introduced in the stochastic
Navier-Stokes equation itself and corresponding influence on large-scale regimes will be determined.

{\subsection{Effect of helicity} \label{subsec:helicity}}

Further topic we want to address is the effect of helicity or violation
of mirror symmetry.
Helicity is defined as the scalar product of velocity and vorticity
and its non zero value expresses mirror symmetry breaking of
turbulent flow. It plays significant role in the processes of
magnetic field generation in electrically conductive fluid
\cite{dynamo1}-\cite{Steh3} and represents one of the most important
characteristics of large-scale motions as well
\cite{Etling85}-\cite{Ponomar2003}. The presence of helicity is
observed in various natural (like large air vortices in atmosphere)
and technical flows \cite{Moffat92,Kholmyansky2001,Koprov2005}.
Despite of this fact the role of the helicity in hydrodynamical
turbulence is not completely clarified up to now.

Turbulent viscosity and diffusivity, which characterize influence of
small-scale motions on heat and momentum transport, are basic
quantities investigated in  the theoretic and applied models. The
constraint of direct energy cascade in helical turbulence  has to be
accompanied by decrease of turbulent viscosity. However, no
influence of helicity on turbulent viscosity was found in some works
\cite{Pouquet78,Zhou91}. Similar situation is observed for turbulent
diffusivity in helical turbulence. Although the  modeling
calculations demonstrate intensification of turbulent transfer in
the presence of helicity \cite{Kraich761,Drum84} direct calculation
of diffusivity  does not confirm this effect
\cite{Kraich761,Knobl77,Lipscombe91}. Helicity is the pseudoscalar
quantity hence it can be easily understood, that its influence
appears only in quadratic and higher terms of perturbation theory or
in the combination with another pseudoscalar quantities (e.g.
large-scale helicity). Really, simultaneous consideration of memory
effects and second order approximation indicate effective influence
of helicity on turbulent viscosity \cite{Belian1994,Belian1998} and
turbulent diffusivity \cite{Drum84,Dolg87,Drummond2001} already in
the limit of small or infinite correlation time.

Helicity, as we shall see below, does not affect known results in
one-loop approximation and,  therefore, it is necessary to turn to
the second order (two-loop) approximation to be able to analyze
possible consequences. It is also important to say that in the
framework of classical Kraichnan model, i.e., model of passive
advection by the Gaussian velocity field with $\delta$-like
correlations in time, it is not possible to study the influence of
the helicity because all diagrams with 
helical contribution are
identically equal to zero at all orders in the perturbation theory.
It is interesting and important to
study the helicity effects because many turbulence phenomena are
directly influenced by them (like large air vortices in atmosphere).
For example, in stochastic magnetic hydrodynamics, which studies the
turbulence in electrically conducting fluids, it leads to a
nontrivial fact of the very existence of so called "turbulent
dynamo" -- the generation of a large-scale magnetic field by the energy
of the turbulent motion
\cite{dynamo1,Moffatt,Adzhemyan87,HnJuSt01,Steh1,Steh2,Steh3}. This is
an important effect in astrophysics.

\subsubsection{The model} \label{subsubsec:hel_Model}
As we have already seen the advection of a passive scalar
field $\theta \equiv \theta(x)\equiv \theta(t, {\mx})$ by a Navier-Stokes
ensemble is governed by Eq. (\ref{eq:scalar2D_pasivo}).
Although the theoretical description of the fluid turbulence on the
basis of the "first principles", i.e., on the stochastic
Navier-Stokes (NS) equation \cite{Monin} remains an
open problem, considerable progress has been achieved in
understanding simplified model systems that share some important
properties with the real problem: shell models \cite{Dyn},
stochastic Burgers equation \cite{Burgulence} and passive advection
by random ``synthetic'' velocity fields \cite{FGV01}.

The crucial role in these studies are played by models of advected
passive scalar field \cite{Obu49}.
A simple model of a
passive scalar quantity advected by a random Gaussian velocity
field, white in time and self-similar in space (the latter property
mimics some features of a real turbulent velocity ensemble), the
so-called Kraichnan's rapid-change model \cite{Kra68}, is an
example. The interest to these models is based on two important
facts: first, as were shown by both natural and numerical
experimental investigations, the deviations from the predictions of
the classical Kolmogorov-Obukhov phenomenological theory
\cite{Monin,OrszagBook,Frisch,McComb} is even more strongly
displayed for a passively advected scalar field than for the
velocity field itself (see, e.g.,
\cite{Antonia84,Sreenivasan91,HolSig94,Pumir96,Pumir97,Pumir98,Pumir94,TonWar94,Elperin1,Elperin2,Elperin3} and
references cited therein), and second, the problem of passive
advection is much more easier to be consider from theoretical point
of view. There, for the first time, the anomalous scaling was
established on the basis of a microscopic model \cite{Kraichnan94},
and corresponding anomalous exponents was calculated within
controlled approximations 
\cite{Chertkov95b,Chertkov96,Gawedzki95,Bernard96,Shraiman96,Shraiman97,Pumir96,Pumir97,Pumir98} (see also review
\cite{FGV01} and references therein).
Within the ``zero-mode approach,''  developed in   
\cite{Chertkov95b,Chertkov96,Gawedzki95,Bernard96,Pumir96,Pumir98,Shraiman96,Shraiman97}, nontrivial anomalous   
exponents are related to the zero modes (unforced solutions)   
of the closed exact equations satisfied by the equal-time   
correlations. Within the approach based on the field theoretic   
 RG and OPE (discussed in Sec. \ref{sec:RG_theory}),   
the anomalous scaling emerges as a consequence of the existence in the   
model of composite operators with {\it negative} critical dimensions,   
which determine the anomalous exponents \cite{Eyink96,AAV98,AdzAnt98,Ant99,Ant00,KJW}.  

The standard notation for advection problem using Kraichnan model
slightly differs from the one using stochastic Navier-Stokes ensemble. Therefore
in what follows we give a brief overview of basic physical ideas
behind Kraichnan model and introduce corresponding notation.

The analog of Eq. (\ref{eq:scalar2D_pasivo}) is now given
 by the following stochastic equation
\begin{equation}
  \partial_t \theta 
  +(\mv\cdot\boldnabla)\theta-\nu_0\boldnabla^2\theta = 
  f^{\theta},
  \label{eq:hel_scalar1}
\end{equation}
where $\nu_0$ is the coefficient of molecular
diffusivity and $f^{\theta}
\equiv f^{\theta}(x)$ is a Gaussian random noise with zero mean
and correlation function
\begin{equation}
  \langle f^{\theta}(x) f^{\theta}(x^{\prime})\rangle =
  \delta(t-t^{\prime})C({\mr}/L), \,\,\, {\mr}={\mx}-{\mx^{\prime}}.
  \label{eq:hel_correlator}
\end{equation}
The noise $f^\theta$ in (\ref{eq:hel_scalar1}) maintains the
steady-state of the system but the concrete form of the correlator
is not essential. The only condition which must be fulfilled by the
function $C({\mr}/L)$ is that it must decrease rapidly for
$r\equiv |{\mr}| \gg L$, where $L$ denotes an integral scale
related to the stirring. In the case when $C$ depends not only on
the modulus of the vector ${\mr}$ but also on its direction, it
plays the role of a source of large-scale anisotropy, whereupon the
noise can be replaced by a constant gradient of scalar field.
Eq.\,(\ref{eq:hel_scalar1}) then reads
\begin{equation}
  \partial_t \theta 
  +(\mv\cdot\boldnabla)\theta - \nu_0\boldnabla^2\theta = - {\mh}\cdot{\mv}.
  \label{eq:hel_scalar2}
\end{equation}
Here, $\theta(x)$ is the fluctuation part of the total scalar field
$\Theta(x)=\theta(x)+{\mh}\cdot{\mx}$, and ${\mh}$ is a
constant vector that determines distinguished direction. The direct
formulation with a scalar gradient is even more realistic one; see,
e.g. Refs.\,\cite{HolSig94,Shraiman96,Shraiman97,Pumir96,Pumir97,Pumir98,Ant99,Ant00}.

In accordance with the generalized Kraichnan model \cite{HolSig94,Ant99} with finite
correlation time taken into account we assume
that the velocity field is driven by simple linear stochastic equation
\begin{equation}
  \partial_t v_i + R v_i=f^v_i, 
  \label{eq:hel_linearNS}
\end{equation}
where $R\equiv R(x)$ is a linear operation to be specified below
and $f^v_i\equiv f^v_i(x)$ is an external random stirring force
with zero mean and the correlator
\begin{equation}
  \langle f^v_i(x) f^v_j(x^{\prime})\rangle \equiv  D^f_{ij}(x;x^{\prime}) 
  =   
   \frac{1}{(2\pi)^{d+1}} \int \dRM\omega\int\dRM^d \mk\, P^{\rho}_{ij}({\mk}) \tilde{D}^f(\omega,\mk)
   \eRM^{-i(t-t^{\prime})+i{\mk}\cdot({\mx}-{\mx^{\prime}})}.
  \label{eq:hel_corf}
\end{equation}
It should be noted that in the SDE (\ref{eq:hel_scalar1}) the multiplicative noise due to random velocity is not
a white noise in time as in the original Kraichnan model. Therefore there is no need to specify the interpretation
of the SDE. However, in the analysis the white-noise limit will be considered and it should recalled that in this limit
the results correspond to the Stratonovich interpretation of the SDE (\ref{eq:hel_scalar1}).

 The transition to a helical fluid
corresponds to the giving up of conservation of spatial parity, and
technically, this is expressed by the fact that the correlation
function is specified in the form of mixture of a true tensor and a
pseudotensor. In our approach, it is represented by two parts of
transverse projector
\begin{equation}
  P^{\rho}_{ij}(\mk) = P_{ij}({\mk})+H_{ij}({\mk}), 
  \label{eq:hel_projectorA}
\end{equation}
which consists of non-helical standard transverse projector
$P_{ij}({\mk})$ (See Eq.(\ref{eq:double_P})) and presence of 
helicity is modeled by a term
\begin{equation}
  H_{ij}({\mk})=i \rho \, \eps_{ijl} k_l/k, 
  \label{eq:hel_H}
\end{equation}
where $k=|{\mk}|$ is the wave number and
 $\eps_{ijl}$ is Levi-Civita's
completely antisymmetric tensor of rank 3 (it is equal to $1$ or
$-1$ according to whether $(i,j,l)$ is an even or odd permutation of
$(1,2,3)$ and zero otherwise), and the real parameter of helicity,
$\rho$, characterizes the amount of helicity. Due to the requirement
of positive definiteness of the correlation function the absolute
value of $\rho$ must be in the interval $|\rho| \in \langle
0,1\rangle$ \cite{Adzhemyan87,HnJuSt01}. Physically, non-zero helical
part (proportional to $\rho$) expresses existence of non-zero
correlations $\langle{\mv}\cdot \mathrm{rot}\:{\mv}\rangle$.

Of course, due to the presence of Levi-Civita's tensor, 
the dimension of the ${\mx}$ space must be considered to be three.
 Nevertheless, in what
follows, we shall remain the $d$-dimensionality of all results which
are not related to helicity to be also able to study $d$-dependence
of non-helical case of the model.
 The correlator $\tilde{D}^f$ is chosen \cite{Ant99,Ant00,Antonov06} in the following form
\begin{equation}
  \tilde{D}^f(\omega, \mk)=g_0 \nu_0^3
  (k^2+m^2)^{2-d/2-\eps-\eta/2},
  \label{eq:hel_Df}
\end{equation}
where
\begin{equation}
  \tilde{R}(\mk)=u_0 \nu_0 (k^2+m^2)^{1-\eta/2},
  \label{eq:hel_R}
\end{equation}
the wave-number representation of $R(x)$. Positive amplitude
factors $g_0$ and $u_0$ play the role of the coupling constants of
the model. In addition, $g_0$ is a formal small
parameter of the ordinary perturbation theory. The positive
exponents $\eps$ and $\eta$ ($\eps=\mathcal{O}(\eta)$) are small
RG expansion parameters, the analogs of the parameter
$\eps=4-d$ in the $\varphi^4-$ theory. Thus, from the point of view of perturbation
theory we again have to deal with a double expansion model discussed in Sec. \ref{subsec:double_intro}. Now in the $(\eps,\eta)$-plane
around the origin $\eps=\eta=0$.

Note the presence of two scales in the problem - integral scale $L$ introduced in (\ref{eq:hel_correlator}) and
momentum scale $m$, which has appeared in (\ref{eq:hel_R}). Clearly, they have different physical origin.
 However, these two scales can be related to each other and for technical purposes  \cite{Antonov06} 
 it is reasonable to choose $L=1/m$. When not explicitly stated, this relation is always assumed.
 
 In the limit
 $k \gg m$ the functions (\ref{eq:hel_Df}) and (\ref{eq:hel_R}) take on simple
powerlike form
\begin{equation}
  \tilde{D}^f(\omega, \mk)=g_0 \nu_0^3 k^{4-d-2 \eps-\eta},\quad
  \tilde{R}(\mk)=u_0 \nu_0 k^{2-\eta},\label{eq:hel_corf1}
\end{equation}
which are used in actual calculations.
 The needed IR
regularization will be given by restrictions on the region of
integrations.

From Eqs.\,(\ref{eq:hel_linearNS}), (\ref{eq:hel_corf}), and (\ref{eq:hel_corf1}) 
 the statistics of the velocity field ${\mv}$ can be determined. It obeys
Gaussian distribution with zero mean and correlator
\begin{equation}
  \langle v_i(x) v_j(x^{\prime}) \rangle \equiv  D^v_{ij}(x;x^{\prime}) = \frac{1}{(2\pi)^{d+1}} 
  \int \dRM\omega \int\dRM^d \mk\,
  P^{\rho}_{ij}({\mk}) \tilde{D}^v(\omega,\mk)
   \eRM^{-i\omega(t-t^{\prime})+i{\mk}\cdot({\mx}  -{\mx^{\prime}})}
  \label{eq:hel_corv}
\end{equation}
with
\begin{equation}
  \tilde{D}^v(\omega, \mk) = \frac{g_0 \nu_0^3
  k^{4-d-2\eps-\eta}}{(i\omega+u_0 \nu_0  k^{2-\eta})(-i\omega+u_0 \nu_0 k^{2-\eta})}.
  \label{eq:hel_corrvelo}
\end{equation}
The correlator (\ref{eq:hel_corrvelo}) is directly related to the energy
spectrum via the frequency integral
\cite{Majda90,Majda92,Majda93,Majda94,ZhaGli92,Ant99}
\begin{equation}
  E(k)\simeq k^{d-1} \int \dRM\omega \tilde{D}^v(\omega, k) \simeq \frac{g_0 \nu_0^2}{u_0} k^{1-2\eps}.
\end{equation}
Therefore, the coupling constant $g_0$ and the exponent
$\eps$ describe the equal-time velocity correlator or,
equivalently, energy spectrum. On the other hand, the constant
$u_0$ and the second exponent $\eta$ are related to the frequency
$\omega \simeq u_0 \nu_0 k^{2-\eta}$ (or to the function
$\tilde{R}(k)$, the reciprocal of the correlation time at the wave
number $k$) which characterizes the mode $k$
\cite{Majda90,Majda92,Majda93,Majda94,ZhaGli92,Ant99,CheFalLeb96,Eyink96}. Thus, in
our notation, the value $\eps=4/3$ corresponds to the
well-known Kolmogorov "five-thirds law" for the spatial statistics
of velocity field, and $\eta=4/3$ corresponds to the Kolmogorov
frequency. Simple dimensional analysis shows that the parameters
(charges) $g_0$ and $u_0$ are related to the characteristic
ultraviolet (UV) momentum scale $\Lambda$ (of the order of inverse
Kolmogorov length) by
\begin{equation}
  g_0\simeq \Lambda^{2\eps + \eta},\quad u_0\simeq \Lambda^{\eta}.
\end{equation}

In Ref.\,\cite{HolSig94},
it was shown that the linear model (\ref{eq:hel_linearNS}) (and therefore
also the Gaussian model (\ref{eq:hel_corv}), (\ref{eq:hel_corrvelo})) is not
Galilean invariant and, as a consequence, it does not take into
account the self-advection of turbulent eddies. As a result of these
so-called "sweeping effects" the different time correlations of the
Eulerian velocity are not self-similar and depend strongly on the
integral scale; see, e.g., Ref.\,\cite{kraichnan64,kraichnan64,chen89,lvov91}. But, on the other
hand, the results presented in Ref.\,\cite{HolSig94} show that the
Gaussian model gives reasonable description of the passive advection
in the appropriate frame, where the mean velocity field vanishes.
One more argument to justify the model (\ref{eq:hel_corv}),
(\ref{eq:hel_corrvelo}) is that, in what follows, we shall be interested in
the equal-time, Galilean invariant quantities (structure functions),
which are not affected by the sweeping, and therefore, as we expect
(see, e.g., Ref.\,\cite{Ant99,Ant00,AdAnHo02}), their
absence in the Gaussian model (\ref{eq:hel_corv}), (\ref{eq:hel_corrvelo}) is not
essential.

In the end of this section, let us briefly discuss two important
limits of the considered model (\ref{eq:hel_corv}), (\ref{eq:hel_corrvelo}). 
First of them it is so-called  rapid-change model limit when
$u_0\rightarrow \infty$ and $g_0^{\prime}\equiv g_0/u_0^2=$ const,
\begin{equation}
  \tilde{D}^v(\omega, \mk)\rightarrow g_0^{\prime} \nu_0 k^{-d-2\eps + \eta},
  \label{eq:hel_rapid}
\end{equation}
and the second one is so-called quenched (time-independent or
frozen) velocity field limit which is defined by $u_0\rightarrow 0$
and $g_0^{\prime\prime}\equiv g_0/u_0=$ const,
\begin{equation}
  \tilde{D}^v(\omega, \mk)\rightarrow g_0^{\prime\prime} \nu_0^2 \pi
  \delta(\omega) k^{-d+2-2\eps},  
  \label{eq:hel_frozen}
\end{equation}
which is similar to the well-known models of the random walks in
random environment with long range correlations; see, e.g.,
Refs.\, \cite{Bouchaud1,Bouchaud2,Bouchaud3,HonKar88,HPV89,HonPis89}.
{\subsubsection{\label{subsubsec:hel_Field}Field theoretic formulation of the model}}
For completeness of our text in this and next section we shall
present and discuss  the principal moments of the RG theory of the
model defined by Eqs.\,(\ref{eq:hel_scalar2}), (\ref{eq:hel_corv}), and
(\ref{eq:hel_corrvelo}).

Using Eq. (\ref{eq:intro_DDJ-FP}) the stochastic problem
(\ref{eq:hel_scalar2})-(\ref{eq:hel_corf}) can be recast into the
equivalent field theoretic model of the doubled set of fields
$\Phi \equiv \{\theta, \theta^{\prime}, {\mv}, {\mv^{\prime}}\}$ with the action functional
\begin{equation}
  \S[\Phi] = 
  \frac{1}{2} \mv^{\prime} D^f \mv^{\prime}
   +\theta^{\prime}\left[-\partial_t \theta
  - (\mv\cdot\boldnabla) \theta+\nu_0\boldnabla^2 \theta-{\mh}\cdot{\mv}
  \right] +  \mv^{\prime} \cdot\left[-\partial_t \mv - R\mv \right] ,
  \label{eq:hel_action1}
\end{equation}
where $D_{ij}^f$ is defined in Eq.\,(\ref{eq:hel_corf}), and
 as usual $\theta^{\prime}$ and ${\mv^{\prime}}$ are auxiliary response fields.

Generating functionals of full Green functions $\G(A)$ and connected
Green functions $\W(A)$ are 
 defined by the Eq. (\ref{eq:intro_GenFunNoise2N}), where
 now  linear form $A\varphi$ is defined as
\begin{equation}
  A\Phi  =  A^{\theta}\theta+A^{\theta^{\prime}}\theta^{\prime}
  + A_i^{v} v_i+A_i^{v^{\prime}}
  v_i^{\prime}(x).
  \label{eq:hel_form}
\end{equation}
Following the arguments in \cite{Ant99}, we can put $A_i^{{\mv^{\prime}}}=0$ in Eq.\,(\ref{eq:hel_form}) and 
then perform the explicit
Gaussian integration over the auxiliary vector field ${\mv^{\prime}}$ as a consequence of the fact
that, in what follows, we shall not be interested in the Green
functions involving field ${\mv^{\prime}}$. After this integration
one is left with the field theoretic model described by the
functional action
\begin{equation}
   \S[\Phi] = 
     -\frac{1}{2} 
     \mv (D^v)^{-1} \mv +   \theta^{\prime} [-\partial_t \theta
     - (\mv\cdot\boldnabla)\theta+\nu_0\boldnabla^2\theta-{\mh}\cdot{\mv}
    ],
    \label{eq:hel_action3}
\end{equation}
where the second term
represents De Dominicis-Janssen action for the stochastic problem
(\ref{eq:hel_scalar2}) at fixed velocity field ${\mv}$, and the first
term describe the Gaussian averaging over ${\mv}$ defined
by the correlator $D^v$ in Eqs.\,(\ref{eq:hel_corv}) and
(\ref{eq:hel_corrvelo}). The latter 
explicitly reads
\begin{equation}
  \S_{\text{vel}}[\mv]  = \frac{1}{2} 
  \int \dRM t_1 \int \dRM t_2 
  \int \dRM^d \mx_1 \int \dRM^d \mx_2
  \,
  \mv_i(t_1,x_1)  D_{ij}^{-1}(t_1-t_2,\mx_1-\mx_2) \mv_j(t_2,\mx_2).
  \label{eq:scalar_vel_action}
\end{equation}

Action (\ref{eq:hel_action3}) is given in a form convenient for a
realization of the field theoretic perturbation analysis with the
standard Feynman diagrammatic technique. From the quadratic part of
the action one obtains the matrix of bare propagators. The
wave-number-frequency representation of, in what follows, important
propagators are: a) the bare propagator $\langle\theta
\theta^{\prime}\rangle_0$ defined as
\begin{equation}
  \langle\theta \theta^{\prime}\rangle_0=\langle\theta^{\prime}
  \theta\rangle^*_0=\frac{1}{-i\omega+\nu_0 k^2},
  \label{eq:scalar_vio_prop}
\end{equation}
and b) the bare propagator for the velocity field $\langle v
v\rangle_0$ given directly by Eq.\,(\ref{eq:hel_corrvelo}), namely
\begin{equation}
  \langle v_i v_j\rangle_0 = P^{\rho}_{ij}({\mk})  \tilde{D}^v(\omega, \mk),
\end{equation}
where $P^{\rho}_{ij}({\mk})$ is the transverse projector defined
in previous section by Eq.\,(\ref{eq:hel_projectorA}). Their graphical
representation is given in a similar manner as has been presented in Fig.
\ref{fig:scalar_prop}. It should be noted that the choice of the action in the rather 
standard form (\ref{eq:hel_action3}) corresponds to the choice of the diagonal value
of the propagator equal to zero: $\langle\theta
\theta^{\prime}\rangle_0(0,0)=0$.

The triple (interaction) vertex $-\theta^{\prime} v_j\partial_j
\theta = \theta^{\prime} v_j V_j \theta $ has been presented by the right picture in
Fig.\,\ref{fig:scalar_vertex}.

It is appropriate to eliminate the magnitude $h\equiv|{\mh}|$ of
the vector field ${\mh}$ from the action (\ref{eq:hel_action3}) by
rescaling of the scalar fields: $\theta\rightarrow h \theta$ and
$\theta^{\prime}\rightarrow \theta^{\prime}/h$. This representation
directly leads to the fact, which is important from the point of
view of the renormalization of the model (see next section), namely,
that the superficial divergences can be presented only in the Green
functions $\langle \theta(x_1) \cdots \theta(x_n)
\theta^{\prime}(y_1) \cdots \theta^{\prime}(y_p)\rangle$ with $n=p$,
i.e., equal number of $\theta$ and $\theta^{\prime}$ fields (see
Ref.\cite{Ant99} for details).\\

{\subsubsection{UV renormalization and RG analysis}\label{subsubsec:hel_RG}}

Detail analysis of divergences in the problem (\ref{eq:hel_action3}) was
done in Ref.\,\cite{Chkhetiani06a,Chkhetiani06b} (see also
Refs.\,\cite{Adzhemyan96,turbo,Ant99}), therefore we shall present here
only basic facts and conclusions rather than to repeat all details.
First of all, every the one-irreducible Green function with
$N_{\theta^{\prime}}<N_{\theta}$ vanish.
On the other hand, dimension analysis based on the Table I leads to
the conclusion that for any $d$, superficial divergences can be
present only in the one-irreducible Green functions $\langle
\theta^{\prime} \theta \cdots \theta \rangle$ with only one field
$\theta^{\prime}$ ($N_{\theta^{\prime}}=1$) and arbitrary number
$N_{\theta}$ of field $\theta$. Therefore, in the model under
investigation, the superficial divergences  can be found only in the
one-irreducible function $\langle \theta^{\prime} \theta \rangle$.
To remove them one needs to include into the action functional the
counterterm of the form $\theta^{\prime} \boldnabla^2 \theta$. Its
inclusion is manifested by the multiplicative renormalization of the
bare parameters $g_0, u_0$, and $\nu_0$ in action functional
(\ref{eq:hel_action3})
\begin{equation}
  \nu_0=\nu Z_{\nu}, \quad g_0=g \mu^{2\eps+\eta}Z_g, \quad u_0=u\mu^{\eta} Z_u. 
  \label{eq:hel_zetka}
\end{equation}
The standard notation is employed, where
 the dimensionless parameters $g,u$, and $\nu$ are the
renormalized counterparts of the corresponding bare ones, $\mu$ is
the scale setting parameter, and $Z_i=Z_i(g,u)$ are renormalization
constants.

\begin{table}
\centering
\begin{tabular}{| c | c | c | c | c | c | c | c | c |}
  \hline\noalign{\smallskip}
  $F$ & ${\mv}$ & $\theta$ & $\theta^{\prime}$ & $m, \Lambda, \mu$ &
  $\nu_0, \nu$ & $g_0$ & $u_0$ & $g, u, h$ 
  \\ \noalign{\smallskip}\hline\noalign{\smallskip}
  $d^k_F$ & -1 & -1 & $d+1$ & 1 & -2 & $2 \eps +\eta$ &
  $\eta$ & 0
  \\ \noalign{\smallskip}\hline\noalign{\smallskip}
  $d^{\omega}_F$ & 1 & 0 & 0 & 0 & 1 & 0 & 0 & 0 
  \\ \noalign{\smallskip}\hline\noalign{\smallskip}
  $d_F$ & 1 & -1 & $d+1$ & 1 & 0 & $2 \eps +\eta$ & $\eta$ & 0 
  \\ \noalign{\smallskip}\hline
\end{tabular}
 \caption{\label{tab:hel_table1} Canonical dimensions of the fields and
         parameters of the model under consideration.}
\end{table}

The renormalized action functional has the following form
\begin{equation}
  \S_R[\Phi] = -\frac{1}{2} 
   \mv [D^v]^{-1} \mv +
   \theta^{\prime}\left[-\partial_t \theta
  - 
  (\mv\cdot\boldnabla)\theta
  +\nu Z_1 \boldnabla^2\theta-{\mh}\cdot{\bf v}
  \right],
  \label{eq:hel_actionRen}
\end{equation}
where the correlator $D_{ij}^v$ is written in renormalized
parameters (in wave-number-frequency representation)
\begin{equation}
  \tilde{D}_{ij}^v(\omega, k) = \frac{P^{\rho}_{ij}({\mk}) g \nu^3
  \mu^{2\eps+\eta} k^{4-d-2\eps-\eta}}{(i\omega+u \nu
  \mu^{\eta} k^{2-\eta})(-i\omega+u \nu \mu^{\eta}
  k^{2-\eta})}.
  \label{eq:hel_corrveloRen}
\end{equation}
By comparison of the renormalized action (\ref{eq:hel_actionRen}) with
definitions of the renormalization constants $Z_i$, $i=g,u,\nu$
(\ref{eq:hel_zetka}) we are coming to the relations among  them:
\begin{equation}
  Z_{\nu}=Z_1,\,\,\, Z_g=Z_{\nu}^{-3},\,\,\,
  Z_u=Z_{\nu}^{-1}.
  \label{eq:hel_zetka1}
\end{equation}
The second and third relations are consequences of the absence of
the renormalization of the term with $D^v$ in renormalized action
(\ref{eq:hel_actionRen}). Renormalization of the fields, the mass parameter
$m$, and the vector ${\mh}$ is not needed, i.e., $Z_{\Phi}=1$ for
all fields, $Z_m=1$,  and also $Z_{{\mh}}=1$.

The issue of interest is, in particular, the behavior of the
equal-time structure functions
\begin{equation}
  S_{n}(r)\equiv\langle[\theta(t,{\mx})-\theta(t,{\mx '})]^{n}\rangle 
  \label{eq:hel_struc}
\end{equation}
in the inertial range, specified by the inequalities $l\sim
1/\Lambda \ll r \ll L=1/m$ ($l$ is an internal length). Here parentheses
$\langle \cdots \rangle$ mean functional average over fields
$\Phi=\{\theta, \theta', {\mv}\}$ with weight $\exp S_R[\Phi].$ In
the isotropic case, the odd functions $S_{2n+1}$ vanish, while for
$S_{2n}$ simple dimensional considerations give
\begin{equation}
  S_{2n}(r)=  \nu_0^{-n}\, r^{2n}\, R_{2n} ( r/l, r/L, g_0, u_0,  \rho), 
  \label{eq:hel_strucdim}
\end{equation}
where  $R_{2n}$ are scaling functions of dimensionless variables (see Sec. \ref{eq:RG_jednatridva}). In
principle, they can be calculated within the ordinary perturbation
theory (i.e., as series in $g_{0}$), but this is not useful for
studying inertial-range behavior: the coefficients are singular in
the limits $\ r/l \to\infty$  and/or $r/L\to 0$, which compensate
the smallness of $g_{0}$, and in order to find correct IR
behavior we have to sum the entire series. The desired summation can
be accomplished using the field theoretic RG
and OPE; see Sec. \ref{sec:RG_theory} and Refs.
\cite{Ant99,AdAnBaKaVa01,AAV98}.

The RG analysis consists
of two main stages (Sec. \ref{sec:RG_theory}). On the first stage, the multiplicative
renormalizability of the model is demonstrated and the differential
RG equations for its correlation (structure) functions  are
obtained. The asymptotic behavior of the functions like
(\ref{eq:hel_struc}) for $ r/l \gg 1$ and any fixed $r/L$ is given by IR
stable fixed points (see next section) of the RG equations and has
the form
\begin{equation}
  S_{2n}(r)= \nu_0^{-n}\, r^{2n}\, (r/l)^{-\gamma_{n}} R_{2n} (r/L,\rho),  \quad r/l \gg 1 
  \label{eq:hel_strucdim2}
\end{equation}
with yet unknown scaling functions $R_{2n} (r/L,
\rho)$. We remind the reader that the quantity $\Delta[S_{2n}]\equiv-2n +
\gamma_{n}$ is termed the critical dimension, and the exponent
$\gamma_{n}$, the difference between the critical dimension
$\Delta[S_{2n}]$ and the canonical dimension $-2n$, is called
the anomalous dimension. In the case at hand, the latter has an
extremely simple form: $\gamma_{n}=n\epsilon$. Whatever be the
functions $ R_{n} (r/L, \rho)$, the representation (\ref{eq:hel_strucdim2})
implies the existence of a scaling (scale invariance) in the IR
region ($r/l \gg 1$, $r/L$ fixed) with definite critical dimensions of
all ``IR relevant''  parameters, $\Delta[S_{2n}] =-2n+n \epsilon$,
$\Delta_{r}=-1$, $\Delta_{L^{-1}}=1$ and fixed ``irrelevant''
parameters $\nu_0$ and $l$.

On the second stage, the small $r/L$ behavior of the functions $
R_{2n} (r/L, \rho)$ is studied within the general representation
(\ref{eq:hel_strucdim2}) using the OPE technique (Sec. \ref{subsec:OPE}). It shows that, in the limit $r/L\to
0$, the functions $ R_{2n} (r/L,\rho)$ have the asymptotic forms
\begin{equation}
   R_{2n} (r/L) = \sum_{F} C_{F}(r/L)\, (r/L)^{\Delta_n},
  \label{eq:hel_ope}
\end{equation}
where $C_{F}$ are coefficients regular in $r/L$. In general, the
summation is implied over certain  renormalized composite operators
$F$  with critical dimensions $\Delta_n$. In case under
consideration the leading operators $F$ have the form $F_n=
(\partial_i \theta \partial_i \theta)^n.$ In
Sec.\,\ref{subsubsec:hel_CompOper} we shall consider them in detail where the
complete two-loop calculation \cite{Chkhetiani06b} of the critical dimensions  of the
composite operators $F_n$ will be present for arbitrary values of
$n$, $d$, $u$ and $\rho$.

The actual calculation \cite{Chkhetiani06a,Chkhetiani06b} have been performed to the 
 with two-loop approximation. The calculation of higher-order
corrections is more difficult in the models with turbulent velocity
field with finite correlation time than  in the cases with
$\delta$-correlation in time. First of all, one has to calculate
more relevant Feynman diagrams in the same order of perturbation
theory (see below). Second, and more problematic distinction is
related to the fact that the diagrams for the finite correlated case
involve two different dispersion laws, namely, $\omega \propto k^2$
for the scalar field and $\omega \propto k^{2-\eta}$ for the
velocity field. This leads to complicated expressions for
renormalization  constants even in the simplest (one-loop)
approximation \cite{Ant99,Ant00}. But, as was discussed in
\cite{Ant99,Ant00,AdAnHo02}, this difficulty can be avoided
by the calculation of all renormalization constants in an arbitrary
specific choice of the exponents $\eps$ and $\eta$ that
guarantees UV finiteness of the Feynman diagrams. From the point of
calculations the most suitable choice is to put $\eta=0$ and leave
$\eps$ arbitrary.

Thus, the knowledge of the renormalization constants  for the
special choice $\eta=0$ is sufficient to obtain all important
quantities as the $\gamma$-functions, $\beta$-functions, coordinates
of fixed points, and the critical dimensions.

Such possibility is not automatic in general. In the model under
consideration it is the consequence of an analysis which shows that
in the MS scheme all the needed anomalous dimensions are independent
of the exponents $\eps$ and $\eta$ in the two-loop
approximation. But in the three-loop approximation they can simply
appear \cite{AdAnHo02}.

In Ref.\,\cite{AdAnHo02} the two-loop corrections to the anomalous
exponents of model (\ref{eq:hel_action3}) without helicity were studied. 
Now the effect of helicity can be discussed for comparison.

Now we can  continue with renormalization of the model. The relation
$\S[\theta,\theta^{\prime},{\mv},e_0]=\S^R[\theta,\theta^{\prime},{\mv}, e, \mu]$, where $e_0$
stands for the complete set of bare parameters and $e$ stands for
renormalized one, leads to the relation $\W(A, e_0)=\W^R(A, e, \mu)$
for the generating functional of connected Green functions. By
application of the operator $\tilde{\cal{D}}_{\mu}\equiv\mu
\partial_{\mu}$ at fixed $e_0$ on both sides of the latest equation
one obtains the basic RG differential equation
\begin{equation}
  {\cal{D}}_{RG} \W^R(A,e,\mu)=0, 
  \label{eq:hel_RGE}
\end{equation}
where ${\cal{D}}_{RG}$ represents operation $\tilde{\cal{D}}_{\mu}$
written in the renormalized variables. Its explicit form is
\begin{equation}
  {\cal{D}}_{RG} = {\cal{D}}_{\mu} +
  \beta_g(g,u)\partial_g+\beta_u(g,u)\partial_u-\gamma_{\nu}(g,u){\cal{D}}_{\nu},
  \label{eq:hel_RGoper}
\end{equation}
where we denote ${\cal{D}}_x\equiv x\partial_x$ for any variable $x$
and the RG functions (the $\beta$ and $\gamma$ functions) are given
by well-known definitions and in our case, using  relations
(\ref{eq:hel_zetka1}) for renormalization constants, they have the
following form
\begin{eqnarray}
  \gamma_{\nu}&\equiv& \tilde{\cal{D}}_{\mu} \ln Z_{\nu}, 
  \label{eq:hel_gammanu}\\
  \beta_g&\equiv&\tilde{\cal{D}}_{\mu} g =g
  (-2\eps-\eta+3\gamma_{\nu}), \label{eq:hel_betag}\\
  \beta_u&\equiv&\tilde{\cal{D}}_{\mu} u =u
  (-\eta+\gamma_{\nu}).
  \label{eq:hel_betau}
\end{eqnarray}

The renormalization constant $Z_{\nu}$ is determined by the
requirement that the one-irreducible Green function $\langle
\theta^{\prime} \theta\rangle_{1-ir}$ must be UV finite when is
written in renormalized variables. In our case it means that they
have no singularities in the limit $\eps, \eta\rightarrow0$.
The one-irreducible Green function $\langle \theta^{\prime}
\theta\rangle_{1-ir}$ is related to the self-energy operator
$\Sigma_{\theta^{\prime}\theta}$ by the Dyson equation
\begin{equation}
  \langle \theta^{\prime}\theta \rangle_{1-ir}=-i\omega+\nu_0 p^2 -
  \Sigma_{\theta^{\prime}\theta}(\omega, p).
  \label{eq:hel_Dyson}
\end{equation}
Thus $Z_{\nu}$ is found from the requirement that the UV divergences
are canceled in Eq.\,(\ref{eq:hel_Dyson}) after substitution $\nu_0=\nu
Z_{\nu}$. This determines $Z_{\nu}$ up to an UV finite contribution,
which is fixed by the choice of the renormalization scheme. In the
MS scheme all the renormalization constants have the form: 1 + {\it
poles in $\eps,\eta$ and their linear combinations}. The
self-energy operator $\Sigma_{\theta^{\prime}\theta}$ is represented
by the corresponding one-irreducible diagrams. In contrast to
rapid-change model, where only one-loop diagram exists (it is
related to the fact that all higher-loop diagrams contain at least
one closed loop which is built on by only retarded propagators, thus
are automatically equal to zero), in the case with finite
correlations in time of the velocity field, higher-order corrections
are non-zero.
\begin{figure}[h]
\begin{center}
\includegraphics[width=11.5cm]{\PICS 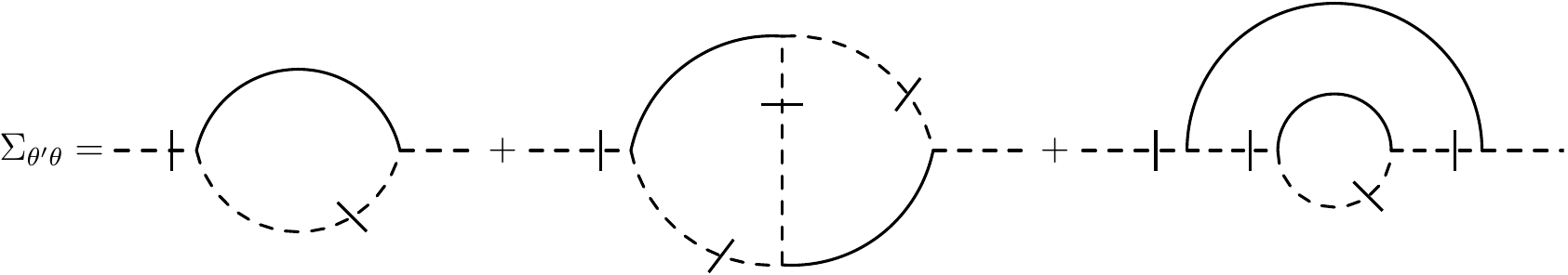}
   \caption{The one- and two-loop diagrams that contribute to their
     self-energy operator $\Sigma_{\theta'\theta}$. 
     \label{fig:hel_graphs}
   }
\end{center}
\end{figure}
In two-loop approximation the self-energy operator
$\Sigma_{\theta^{\prime}\theta}$ is defined by diagrams which are
shown in Fig.\,\ref{fig:hel_graphs}.\\

{\subsubsection{Fixed points and scaling regimes}\label{subsubsec:hel_ScalReg}}

Possible scaling regimes of a renormalizable model are directly
given by the IR stable fixed points of the corresponding system of
RG equations \cite{Zinn,Vasiliev}. In the considered model, the coordinates $g^*, u^*$ of the
fixed points are found from the system of two equations
\begin{equation}
   \beta_g(g^*,u^*)=\beta_u(g^*,u^*)=0.
\end{equation}
The beta functions $\beta_g$ and $\beta_u$ are defined in
Eqs.\,(\ref{eq:hel_betag}), and (\ref{eq:hel_betau}). To investigate the IR
stability of a fixed point the eigenvalues
of the matrix $\Omega$ 
\begin{equation}
  \Omega = \begin{pmatrix}
  \renewcommand{\arraystretch}{1.5}
  \frac{\partial \beta_g}{\partial g} &
  \frac{\partial \beta_g}{\partial u} \\ \frac{\partial \beta_u}{\partial g} & \frac{\partial \beta_u}{\partial u}
  \end{pmatrix}
  _{*}.
\end{equation}
have to be determined (see Eq. \ref{eq:RG_Omega} for general case). 

\begin{figure}[!htb]
    \begin{minipage}{0.475\textwidth}
       \mbox{ } 
        \includegraphics[width=6cm]{\PICS 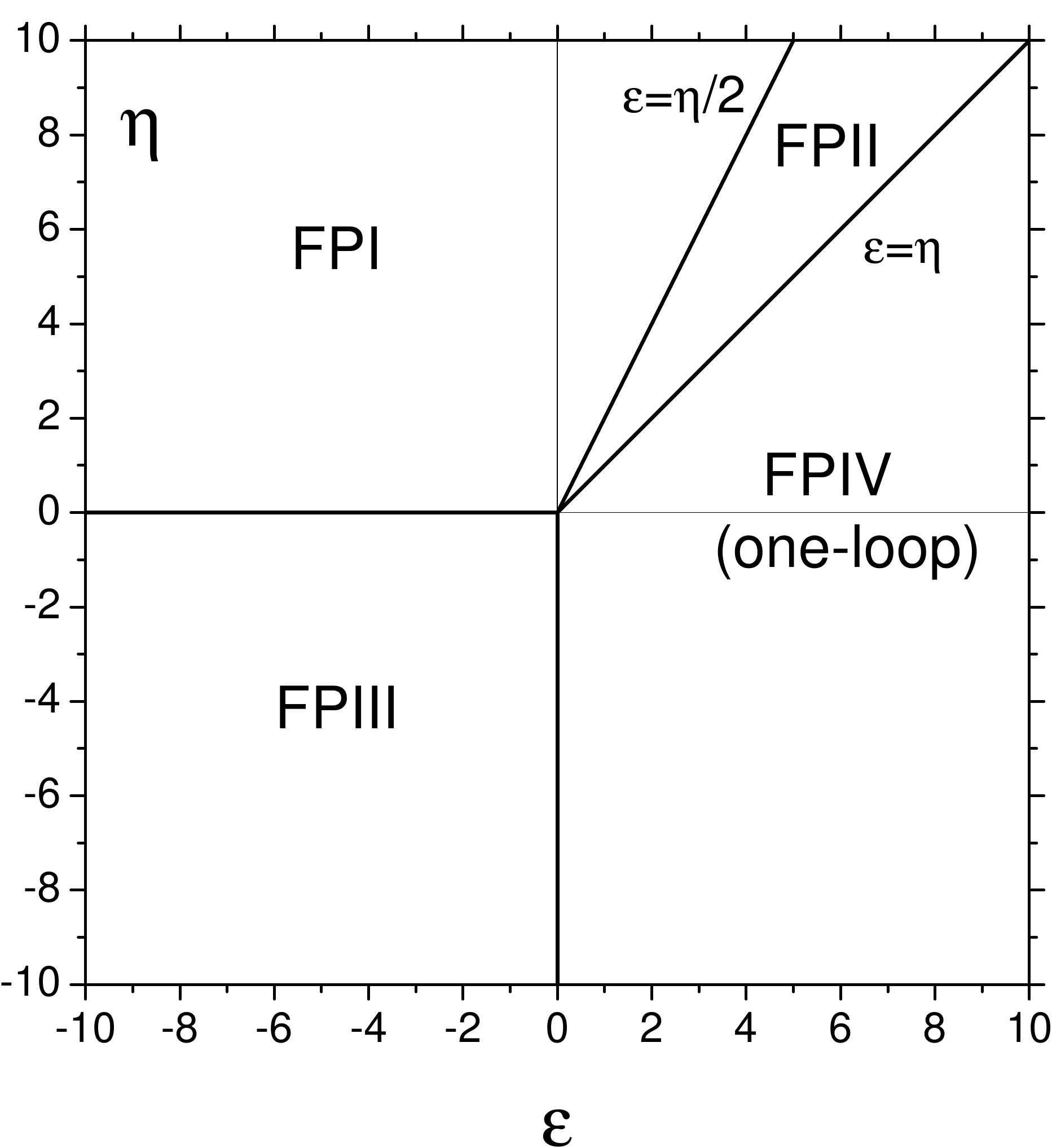}
        \caption{Regions of the stability for the fixed
       points in one-loop approximation. The regions of stability for fixed
       points FPI, FPII, and FPIII are exact, i.e., are not influenced by
       loop corrections. The fixed point FPIV is shown in one-loop
       approximation. Dependence of the FPIV on the value of $d$ in two-loop approximation
       is shown in Fig.\,\ref{fig:hel_fig5}.
      }
 \label{fig:hel_fig4}
    \end{minipage}%
    \hfill
    \begin{minipage}{0.475\textwidth}
         \mbox{ } 
        \includegraphics[width=6cm]{\PICS 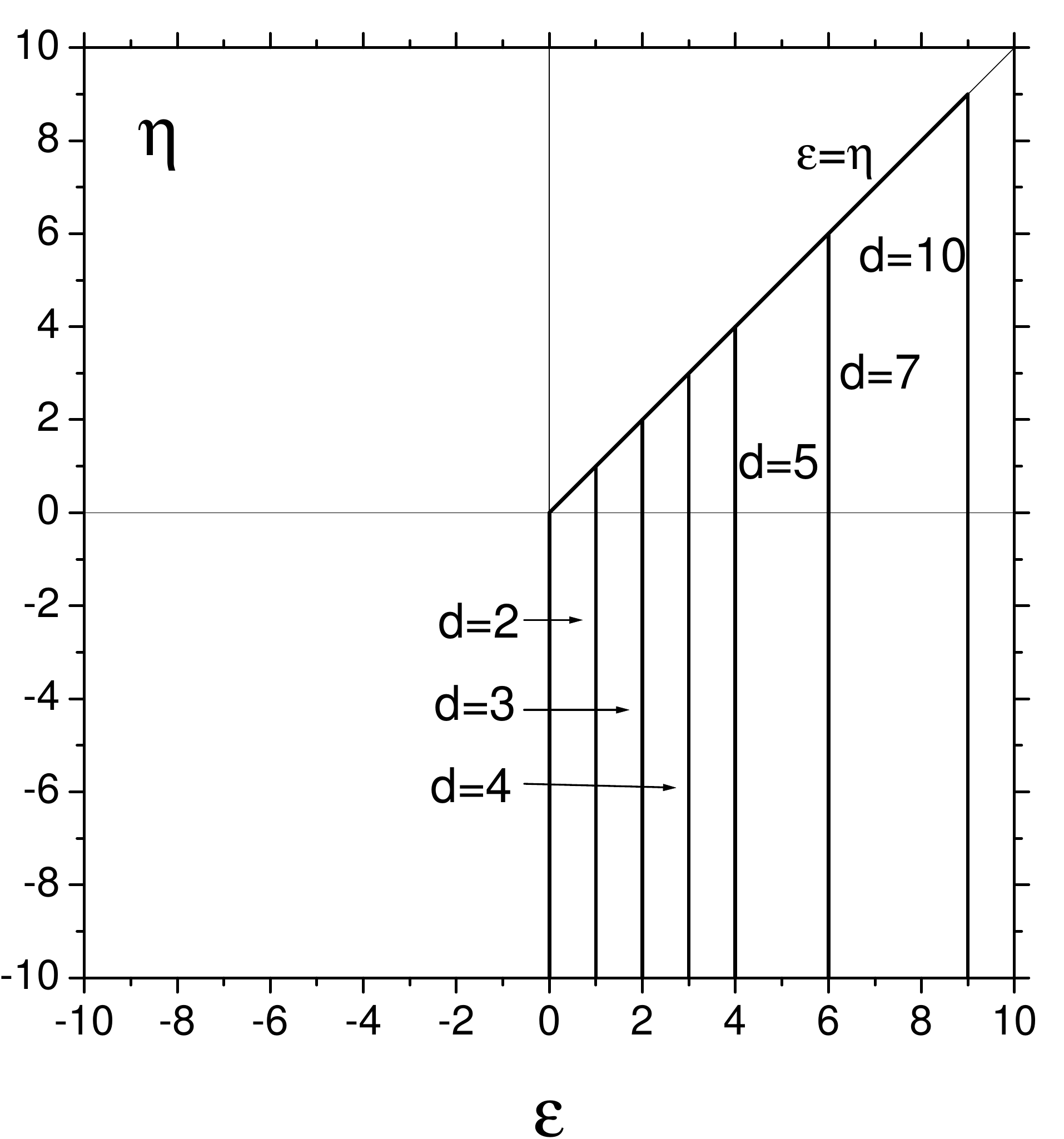}        
        \caption{Regions of the stability for the fixed
      point FPIV in two-loop approximation without helicity for different
      space dimensions $d$. The IR fixed point is stable in the region
      given by inequalities: $\eps>0, \eps>\eta$, and
      $\eps<d-1$.       
}
\label{fig:hel_fig5}
\mbox{ }\\ 
    \end{minipage}
\end{figure}

\begin{figure}[!htb]
    \begin{minipage}{0.475\textwidth}
  \includegraphics[width=6cm]{\PICS 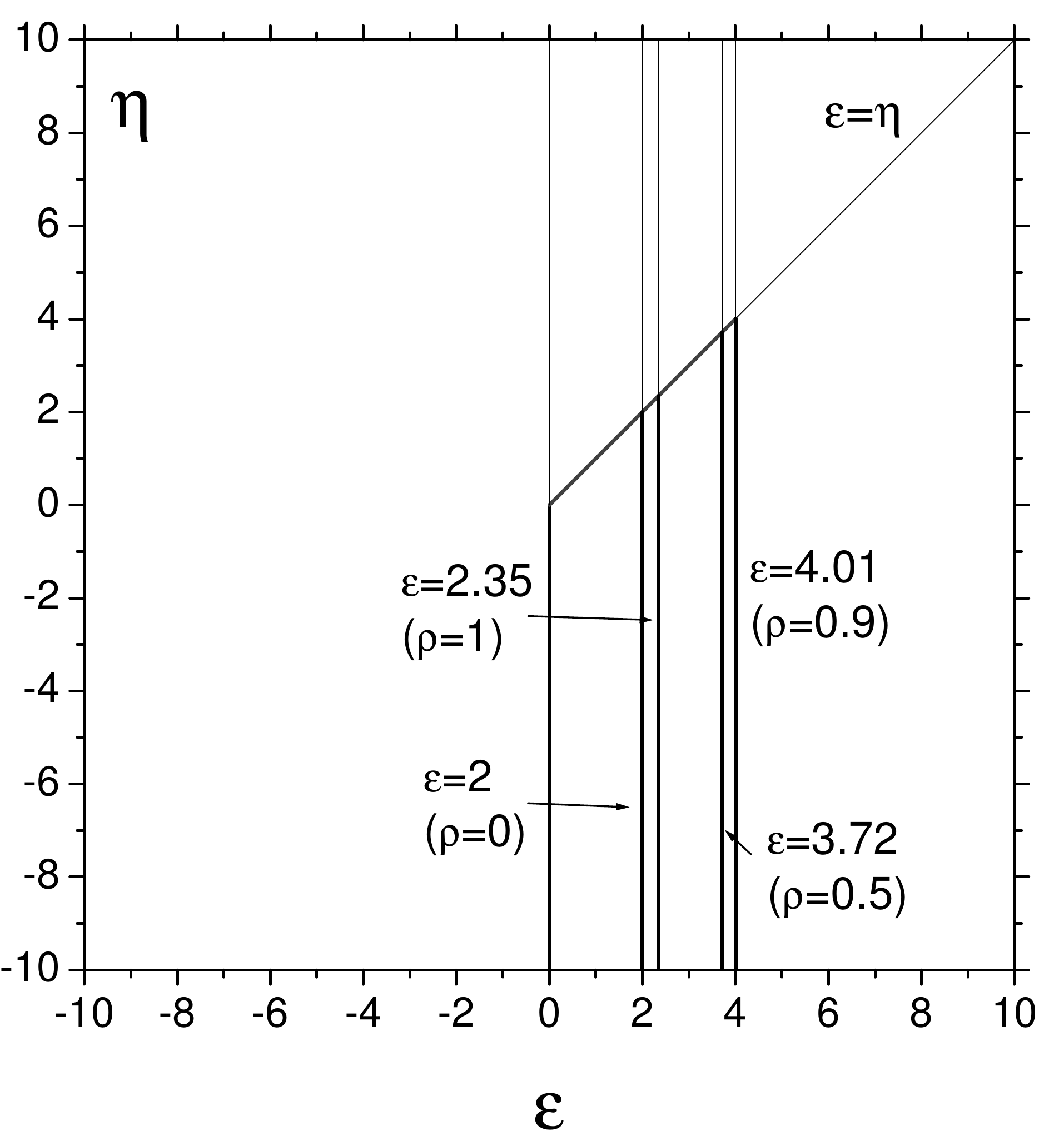}  
\caption{Regions of the stability for the fixed
    point FPIV in two-loop approximation with helicity. The IR fixed
    point is stable in the region given by inequalities: $\eps>0,
    \eps>\eta$, and $\eps<\eps_{\rho}$.
    }
   \label{fig:hel_fig6}
    \end{minipage}%
    \hfill
    \begin{minipage}{0.475\textwidth}
  \includegraphics[width=6cm]{\PICS 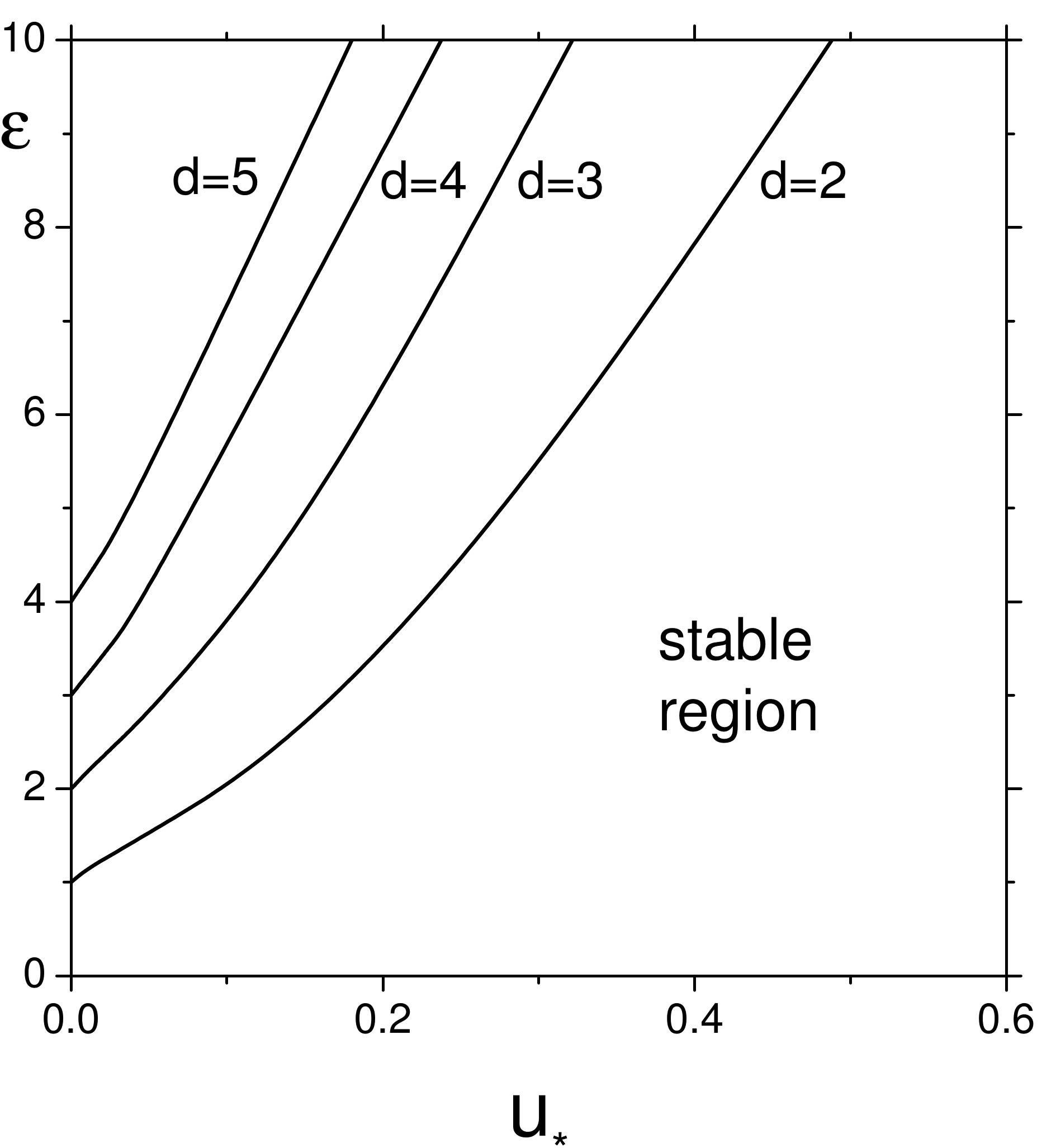}
\caption{Regions of the stability for the fixed
     point FPV in two-loop approximation without helicity. $d$ dependence
     of the stability is shown.    
}
 \label{fig:hel_fig7}
\mbox{ }\\ 
    \end{minipage}
\end{figure}

The possible scaling regimes of the model in one-loop approximation
were investigated in Ref.\,\cite{Ant99}. Our first question is
how the two-loop approximation change the picture of "phase" diagram
of scaling regimes discussed in Ref.\,\cite{Ant99}, and the
second one is what restrictions on this picture are given by
helicity (in two-loop approximation). The two-loop approximation in
the model under our consideration without helicity was studied in
Ref.\,\cite{AdAnHo02} but the question of scaling regimes from
two-loop approximation point of view was not discussed in details.

First of all, let us take a look at the rapid-change limit, which is obtained in the limit
$u\rightarrow\infty$. In this regime, it is convenient to make
transformation to new variables, namely, $w\equiv1/u$, and
$g^{\prime}\equiv g/u^2$, with the corresponding changes in the
$\beta$ functions:
\begin{equation}
  \beta_{g^{\prime}} = g^{\prime}(\eta-2\eps +\gamma_{\nu}), \quad
  \beta_w = w(\eta-\gamma_{\nu}).
\end{equation}
In the rapid-change limit $w\rightarrow0$
($u\rightarrow\infty$) the two-loop contribution to $\gamma_\nu$
is equal to zero \cite{Chkhetiani06b}. It is not surprising because in the rapid-change
model there are no higher-loop corrections to the self-energy
operator \cite{AAV98,AdAnBaKaVa01}, thus we are coming to the
one-loop result of Ref.\,\cite{Ant99} with the anomalous
dimension $\gamma_{\nu}$ of the form
\begin{equation}
  \gamma_{\nu}=\lim_{w\rightarrow 0}\frac{(d-1) g {\bar S}_d}{2
  d(1+w)} \equiv \frac{(d-1){\bar g^{\prime}}}{2 d}.
\end{equation}
In this regime we have two fixed points denoted as FPI and FPII in
Ref.\,\cite{Ant99}. The first fixed point is trivial, namely
\begin{equation}
  \mathrm{FPI}:\,\,\,\, w^*={g'}^*=0,
\end{equation}
with $\gamma_{\nu}^*=0$, and diagonal matrix $\Omega$ with
eigenvalues (diagonal elements)
\begin{equation}
  \Omega_1=\eta,\,\,\,\,\,\Omega_2=\eta-2\eps.
\end{equation}
The region of stability is shown in Fig.\,\ref{fig:hel_fig4}. The second
point is defined as
\begin{equation}
  \mathrm{FPII}:\,\,\,\,w^*=0,\,\,\,
  {\bar{g}}^{\prime *} = \frac{2d}{d-1}(2\eps-\eta),
\end{equation}
with $\gamma_{\nu}^*=2\eps-\eta$. These are exact one-loop
expressions as a result of non-existence of the higher-loop
corrections. That means that
they have no corrections of order $\mathcal{O}(\eps^2)$ and higher (we
work with assumption that $\eps \simeq \eta$, therefore it
also includes corrections of the type $\mathcal{O}(\eta^2)$ and $\mathcal{O}(\eta \eps)$). The
corresponding "stability matrix" is triangular
with diagonal elements (eigenvalues):
\begin{equation}
  \Omega_1=2(\eta-\eps),\,\,\,\,\Omega_2=2\eps-\eta.
\end{equation}
The region of stability of this fixed point is shown in
Fig.\,\ref{fig:hel_fig4}.

The frozen velocity field is mathematically obtained from the model under consideration
in the limit $u \rightarrow 0$. To study this transition it is
appropriate to change the variable $g$  to the new variable
$g^{\prime\prime} \equiv g/u$. Then the $\beta$
functions are transform to the following ones:
\begin{equation}
  \beta_{g^{\prime\prime}} = g^{\prime\prime}(-2\eps +2 \gamma_{\nu}), \quad  
  \beta_u = u(-\eta+\gamma_{\nu}).
  \label{eq:hel_betauu0}
\end{equation}
The system of $\beta$ functions (\ref{eq:hel_betauu0})
exhibits two fixed points, denoted as FPIII and FPIV in
Ref.\,\cite{Ant99}, related to the corresponding two scaling
regimes. One of them is trivial,
\begin{equation}
  \mathrm{FPIII}:\,\,\,\, u^*={g''}^*=0,
\end{equation}
with $\gamma_{\nu}^*=0$. The eigenvalues of the corresponding matrix
$\Omega$, which is diagonal in this case, are
\begin{equation}
  \Omega_1=-2\eps,\,\,\,\,\Omega_2=-\eta.
\end{equation}
Thus, this regime is IR stable only if both parameters
$\eps$, and $\eta$ are negative simultaneously as can be seen
in Fig.\,\ref{fig:hel_fig4}. The second, non-trivial, point is
\begin{equation}
  \mathrm{FPIV}:\,\,\,\, u^*=0,\,\,\,\,
  {\bar{g}}^{\prime\prime *}=-\frac{\eps}{2 {\cal
  A^{\prime\prime}}_0}-\frac{{\cal B^{\prime\prime}}_0}{2 {\cal
  A^{\prime\prime}}_0^2} \eps^2,
\end{equation}
where explicit expressions for ${\cal A^{\prime\prime}}_0$ and ${\cal B^{\prime\prime}}_0$
can be found in \cite{Chkhetiani06b}.

What is the influence of two-loop approximation on
this IR scaling regime without helicity in general $d$-dimensional
case? We denote the corresponding fixed point as FPIV$_{0}$, and its
coordinates are
\begin{equation}
  \mathrm{FPIV_0}:\,\,\,u^*=0,\,\,\,\,
  {\bar{g}}^{\prime\prime *}=\frac{2d}{d-1}\left(\eps+\frac{1}{d-1}\eps^2\right),
\end{equation}
with anomalous dimension $\gamma_{\nu}$ defined as
\begin{equation}
  \gamma_{\nu}^*=\frac{d-1}{2d}\left( {\bar{g}}^{\prime\prime *}-\frac{ ({\bar{g}}^{\prime\prime *})^2  }{2d}\right)=\eps,
\end{equation}
which is the exact one-loop result \cite{Ant99}. The eigenvalues
of the matrix $\Omega$ (taken at the fixed point) are
\begin{equation}
  \Omega_1=2\left(\eps+\frac{1}{1-d}\eps^2\right),\,\,\, \Omega_2=\eps-\eta.
\end{equation}
The eigenvalue $\Omega_2=\partial_u \beta_u|_*=-\eta+\gamma_{\nu}^*$
is also exact one-loop result. The conditions
$\bar{g}^{\prime\prime *}>0, \Omega_{1}>0$, and $\Omega_2>0$ for the
IR stable fixed point lead to the following restrictions on the
values of the parameters $\eps$ and $\eta$:
\begin{equation}
  \eps>0,\,\,\, \eps>\eta,\,\,\,\eps<d-1.
  \label{eq:hel_cond1}
\end{equation}
The region of stability is shown in Fig.\,\ref{fig:hel_fig5}. The region of
stability of this IR fixed point increases when the dimension of the
coordinate space $d$ is increasing.

For the system with helicity  the dimension of
the space is fixed for $d=3$. Thus, our starting conditions for
stable IR fixed point of this type are obtained from conditions
(\ref{eq:hel_cond1}) with explicit value $d=3$: $\eps>0,
\eps>\eta, \eps<2$. But they are valid only if
helicity is vanishing and could be changed when non-zero helicity is
present.  When helicity is present the fixed
point FPIV is given as
\begin{equation}
  u^*=0\,\,\,\,
  \bar{g}^{\prime\prime *}=3\eps+\frac32\left(1-\frac{3\pi^2\rho^2}{16}\right)\eps^2,
  \label{eq:hel_gpp}
\end{equation}
Therefore, in helical case, the situation is a little bit more
complicated as a result of a competition between non-helical and
helical term within two-loop corrections. The matrix $\Omega$ is
triangular with diagonal elements (taken already at the fixed point)
\begin{equation}
  \Omega_1 =  2\eps+
  \left(-1+\frac{3\pi^2\rho^2}{16}\right)\eps^2,\quad
  \Omega_2 = \eps-\eta,
  \label{eq:hel_lambda11}
\end{equation}
where explicit dependence of eigenvalue $\Omega_1$ on parameter
$\rho$ takes place. The requirement to have positive values for
parameter $\bar{g}^{\prime\prime *}$, and at the same time for
eigenvalues $\Omega_1, \Omega_2$ leads to the region of stable fixed
point. The results are shown in Fig.\,\ref{fig:hel_fig6}. The picture is
rather complicated due to the very existence of the "critical"
absolute value of $\rho$:
\begin{equation}
  \rho_c=\frac{4}{\sqrt{3}\pi},
\end{equation}
which is defined from the condition of vanishing of the two-loop
corrections in Eqs.\,(\ref{eq:hel_gpp}), and (\ref{eq:hel_lambda11}):
\begin{equation}
  \left(-1+\frac{3\pi^2\rho^2}{16}\right)=0.
\end{equation}
As was already discussed above, when the helicity is not present,
the system exhibits this type of fixed point (and, of course, the
corresponding scaling behavior) in the region restricted by
inequalities: $\eps>0, \eps>\eta$, and
$\eps<2$. The last condition is changing when helicity is
switched on. The important feature here is that the two-loop
contributions to $\bar{g}^{\prime\prime *}$ and $\Omega_1$ have the
same structure but opposite sign. This leads to the different
sources of conditions in the case when $|\rho|<\rho_c$ and
$|\rho|>\rho_c$, respectively. In the situation with $|\rho|<\rho_c$
the positiveness of $\Omega_1$ plays crucial role and one has the
following region of stability of IR fixed point FPIV:
\begin{equation}
  \eps>0,\,\,\, \eps>\eta,\,\,\,
  \eps<\frac{32}{16-3\pi^2\rho^2}.
  \label{eq:hel_cond3}
\end{equation}
On the other hand, in the case with $|\rho|>\rho_c$, the principal
restriction on the IR stable regime is yield by condition
$\bar{g}_*^{\prime\prime}>0$ with final IR stable region defined as
\begin{equation}
  \eps>0,\,\,\, \eps>\eta,\,\,\,
  \eps<\frac{32}{-16+3\pi^2\rho^2}.
  \label{eq:hel_cond4}
\end{equation}
Therefore, if we are continuously increasing absolute value of
helicity parameter $\rho$, the region of stability of the fixed
point defined by the last inequality in Eq.\,(\ref{eq:hel_cond3}) is
increasing too. This restriction vanishes completely when $|\rho|$
reaches the "critical" value $\rho_c$, and the picture becomes the
same as in the one-loop approximation \cite{Ant99}. In this
rather specific situation the two-loop influence on the region of
stability of fixed point is exactly zero: the helical and
non-helical two-loop contributions are canceled by each other. Then
if the absolute value of parameter $\rho$ increases further, the
last condition appears again, namely the third condition in
Eq.\,(\ref{eq:hel_cond4}), and restriction becomes stronger when $|\rho|$
tends to its maximal value, $|\rho|=1$. In this case of the maximal
breaking of mirror symmetry (maximal helicity), $|\rho|=1$, the
region of the IR stability of the fixed point is defined by
inequalities: $\eps>0, \eps=\eta$, and
$\eps<2.351$ (see Fig.\,\ref{fig:hel_fig6}). It is interesting enough
that the presence of helicity in the system leads to the enlargement
of the stability region.

\begin{figure}[!htb]
    \begin{minipage}{0.475\textwidth}
       \mbox{ } 
        \includegraphics[width=6cm]{\PICS 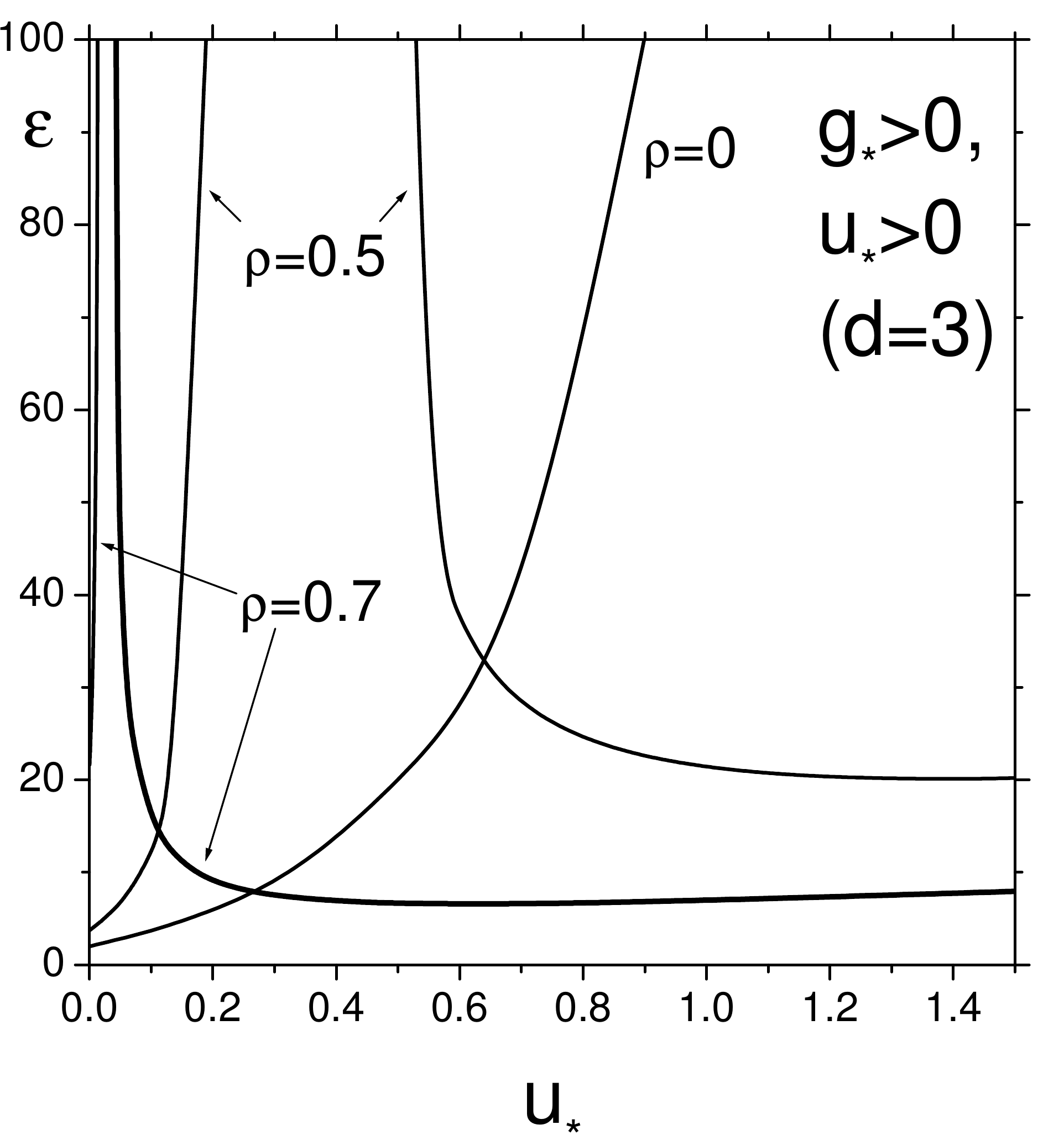}
        \caption{Regions of the stability for the fixed
           point FPV in two-loop approximation with helicity in the situation
           when $\rho<\rho_c=4/(3^{1/2}\pi)$. }
      \label{fig:hel_Fig8}
    \end{minipage}%
    \hfill
    \begin{minipage}{0.475\textwidth}
         \mbox{ } 
        \includegraphics[width=6cm]{\PICS 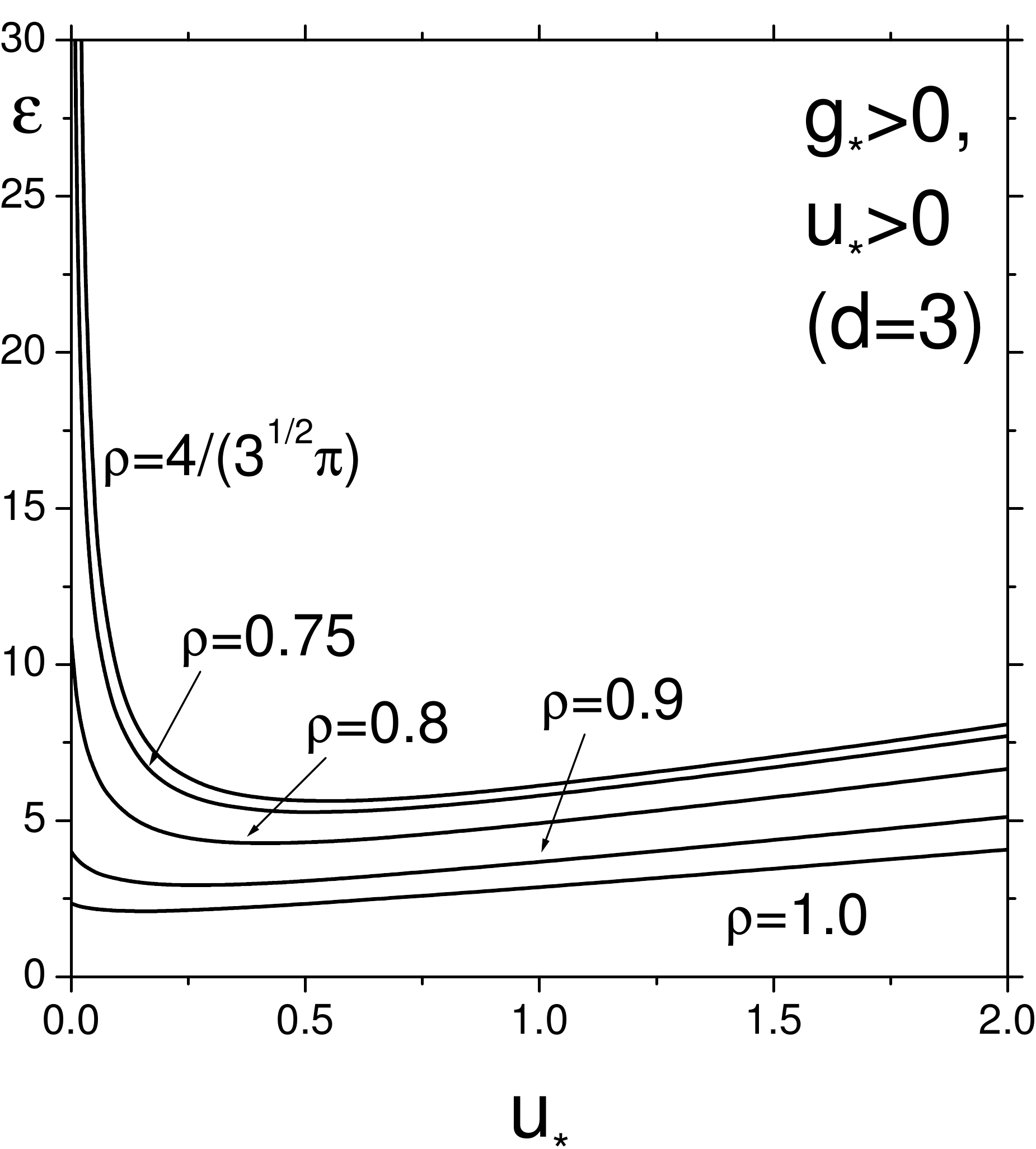}        
        \caption{ Regions of the stability for the fixed
      point FPV in two-loop approximation with helicity in the situation
      when $\rho \geq \rho_c=4/(3^{1/2}\pi)$.}
\label{fig:hel_Fig9}
    \end{minipage}
\end{figure}

The most interesting scaling regime is the one with finite
value of the fixed point for the variable $u$. But by short analysis
one immediately concludes that the system of equations (see also \cite{Ant99})
\begin{align}
  \beta_g  = g
  (-2\eps-\eta+3\gamma_{\nu})=0,\quad 
  \beta_u  = u (-\eta+\gamma_{\nu})=0
  \label{eq:hel_betau1}
\end{align}
can be fulfilled simultaneously for finite values of $g, u$ only in
the case when the parameter $\eps$ is equal to $\eta$:
$\eps=\eta$. In this case, the function $\beta_g$ is
proportional to function $\beta_u$. As a result we have not one
fixed point of this type but a curve of fixed points in the $(g,u)$- plane 
with exact one-loop result for $\gamma_{\nu}^*=\eps=\eta$
(this is already directly given by Eq.\,(\ref{eq:hel_betau1})). 
We denote the corresponding point as in Ref.\,\cite{Ant99} as FPV.
The possible values of the fixed
point for variable $u$
can be restricted (and will be restricted) as we shall discuss
below. The matrix $\Omega$ defined in Eq. (\ref{eq:RG_Omega}) has the following eigenvalues
\begin{equation}
  \Omega_1=0,\,\,\, \Omega_2=3 \bar{g}^* \left(\frac{\partial
  \gamma_{\nu}}{\partial g}\right)_*+u^*\left(\frac{\partial
  \gamma_{\nu}}{\partial u}\right)_*.
\end{equation}
The vanishing of the $\Omega_1$ is an exact result which is related
to the degeneracy of the system of Eqs. 
(\ref{eq:hel_betau1}) when nonzero solutions in respect to $g$, and $u$ are
assumed, or, equivalently, it reflects the existence of a marginal
direction in the $(g,u)$-plane along the line of the fixed points.

The analysis of the last fixed point can start with the investigation
of influence of the two-loop correction on the corresponding scaling
regime when helicity is not present in the system ($\rho=0$). In
this situation it is interesting to determine the dependence of
 scaling regime on dimension $d$. 
In Fig.\,\ref{fig:hel_fig7}, the regions of stability for the fixed point
FPV without helicity in the $\eps-u$ plane for different
space dimension $d$ are shown. It is interesting that in two-loop
case nontrivial d-dependence of IR stability appears in contrast to
one-loop approximation \cite{Ant99}.

Now the investigation of the situation with helicity follows and its
influence on the stability of the IR fixed point is analyzed. In this case we
work in three-dimension space.
The competition between helical and non-helical terms appears again
which will lead to a nontrivial restriction for the fixed point
values of variable $u$ to have positive fixed values for variable

Numerical analysis \cite{Chkhetiani06b} shows important
role is played by $\rho_c=4/(\sqrt{3}\pi)$. First, the
case $|\rho|<\rho_c$ is studied. The corresponding regions of stable IR
fixed points with $g^*>0$ is shown in Fig.\,\ref{fig:hel_Fig8}. In the case
when helicity is not present ($\rho=0$, see the corresponding curve
in Fig.\,\ref{fig:hel_Fig8}), the only restriction is given by condition
that $\Omega_2>0$, on the other hand, the condition $g^*>0$ is
fulfilled without restriction on the parameter space. When arbitrary
small helicity is present, i.e., $\rho>0$, the restriction related
to positiveness of $g^*$ arises and is stronger when $|\rho|$ is
increasing (the right curve for the concrete value of $\rho$ in
Fig.\,\ref{fig:hel_Fig8}) and becomes to play the dominant role. At the same
time, with increasing of $|\rho|$ the importance of the positiveness
of the eigenvalue $\Omega_2$ decreases (the left curve for the
concrete value of $\rho$ in Fig.\,\ref{fig:hel_Fig8}). For a given
$|\rho|<\rho_c$ there exists an interval of values of the variable
$u^*$ for which there is no restriction on the value of the
parameter $\eps$. For example, for $|\rho|=0.1$, it is
$1.128<u^*<13.502$, for $|\rho|=0.5$, $0.217<u^*<0.394$, and for
$|\rho|=0.7$, $0.019<u^*<0.029$. Now turn to the case
$|\rho|\geq\rho_c$. When $|\rho|$ acquires its "critical" value
$\rho_c$, the IR fixed point is stable for all values of $u^*>0$ and
$\eps>0$, i.e., the condition $\Omega_2>0$ becomes fulfilled
without restrictions on parameter space. On the other hand, the
condition $g^*>0$ yields strong enough restriction and it becomes
stronger when $|\rho|$ tends to its maximal value $|\rho|=1$ as it
can be seen in Fig.\,\ref{fig:hel_Fig9}).

The most important conclusion of two-loop approximation \cite{Chkhetiani06b} of the
model is the fact that the possible restrictions on the regions of
stability of IR fixed points are "pressed" to the region with rather
large values of $\eps$, namely, $\eps\geq 2$, and do not
disturb the regions with relatively small $\eps$. For
example, the Kolmogorov point ($\eps=\eta=4/3$) is not
influenced.

If $F$ denotes some multiplicatively renormalized quantity (a
parameter, a field or composite operator) then its critical
dimension is given by the expression
\begin{equation}
   \Delta[F]\equiv\Delta_F=d_F^k+\Delta_{\omega}d_F^{\omega}+\gamma^*_F\,,
   \label{eq:hel_deltaa}
\end{equation}
see, e.g., Refs.\,\cite{Adzhemyan96,turbo,Vasiliev} for details. In
Eq.\,(\ref{eq:hel_deltaa}) $d_F^{k}$ and $d_F^{\omega}$ are the canonical
dimensions of $F$, $\Delta_{\omega}=2-\gamma^*_{\nu}$ is the
critical dimension of frequency, and $\gamma^*_F$ is the value of
the anomalous dimension $\gamma_F\equiv\tilde{\cal D}_{\mu}\ln Z_F$
at the corresponding fixed point. Because the anomalous dimension
$\gamma_{\nu}$ is already exact  for all fixed points at one-loop
level, the critical dimensions of frequency $\omega$ and of fields
$\Phi\equiv\{{\mv}, \theta, \theta^{\prime}\}$  are also found
exactly at one-loop level approximation \cite{Ant99}. In our
notation they read
\begin{align}
  \Delta_{\omega} & = 2-2\eps+\eta &\mathrm{for}& \,\,\mathrm{FPII},\nonumber \\
  \Delta_{\omega} & = 2-\eps  &\mathrm{for}&\,\,\mathrm{FPIV}, \\
  \Delta_{\omega} &= 2-\eps=2-\eta &\mathrm{for}&\,\,\mathrm{FPV},\nonumber
\end{align}
and
\begin{equation}
  \Delta_{{\mv}}=1-\gamma^*_{\nu},\quad \Delta_{\theta}=-1,\quad \Delta_{\theta^{\prime}}=d+1.
\end{equation}

General equal-time two-point quantity $F(r)$ 
depends on a single distance parameter $r$ which is multiplicatively
renormalizable ($F=Z_F F^R$, where $Z_F$ is the corresponding
renormalization constant). Then the renormalized function $F^R$ must
satisfy the RG equation of the form
\begin{equation}
  ({\cal D}_{RG}+\gamma_F)F(r)=0,
\end{equation}
with operator ${\cal D}_{RG}$ given explicitly in
Eq.\,(\ref{eq:hel_RGoper}) and standardly $\gamma_F\equiv {\tilde {\cal
D}}_{\mu} \ln Z_F$. The difference between functions $F$ and $F^R$
is only in normalization, choice of parameters (bare or
renormalized), and related to this choice the form of the
perturbation theory (in $g_0$ or in $g$). The existence of a
nontrivial IR stable fixed point means that in the IR asymptotic
region $ r/l \gg 1$ and any fixed $r/L$ the function $F(r)$ takes on
the self-similar form
\begin{equation}
  F(r)\simeq\nu_0^{d_F^{\omega}} l^{-d_F}(r/l)^{-\Delta_F}
  f(r/L),
  \label{eq:hel_Fr}
\end{equation}
where the values of the critical dimensions correspond to the given
fixed point (see above in this section and Table\,\ref{tab:hel_table1}). The dependence of the scaling functions
on the argument $r/L$ in the region $r/L \ll 1$ can be studied using
the well-known OPE technique discussed in Sec. \ref{subsec:OPE}.\\

{\subsubsection{\label{subsubsec:hel_EffDiff}Effective diffusivity}}
One of the interesting object from the theoretical as well as
experimental point of view is so-called effective diffusivity $\bar
\nu$. In this section let us briefly investigate the effective
diffusivity $\bar \nu$, which replaces initial molecular diffusivity
$\nu_0$ in equation (\ref{eq:hel_scalar1}) due to the interaction of a
scalar field $\theta$ with random velocity field ${\mv}$.
Molecular diffusivity $\nu_0$ governs exponential attenuation in time of
all fluctuations in the system in the lowest approximation, which is
given by the propagator (response function)
\begin{equation}
  G(t-t',{\mk}) = \langle \theta(t,{\mk})\theta^{\prime}(t^{\prime}, {\mk}) \rangle_0 
  = \theta(t-t^{\prime})\exp(-\nu_0 k^2 (t-t^{\prime})).
\end{equation}
 Analogously,
the effective diffusivity $\bar \nu$ governs exponential attenuation of
all fluctuations described by full response function, which is
defined by Dyson equation (\ref{eq:hel_Dyson}). Its explicit expression can
be obtained by the RG approach. In accordance with general rules of
the RG (see, e.g., Ref.\,\cite{Vasiliev}) all principal parameters
of the model $g_0,u_0$ and $\nu_0$ are replaced by their effective
(running) counterparts, which satisfy Gell-Mann-Low RG equations
\begin{equation}
  s \frac{\dRM \bar g}{\dRM s}=\beta_{g}(\bar g,\bar u),\quad
  s\frac{\dRM \bar  u}{\dRM s}=\beta_{u}(\bar g, \bar u),\quad 
  s \frac{\dRM \bar \nu}{\dRM s}= -\bar \nu \gamma_{\nu}(\bar g,\bar u)\, ,
  \label{eq:hel_nu}
\end{equation}
with initial conditions $\bar g|_{s=1}=g, \bar u|_{s=1}=u, \bar
\nu|_{s=1}=\nu$. Here $s=k/\mu$,  $\beta$ and $\gamma$ functions are
defined in (\ref{eq:hel_gammanu}) - (\ref{eq:hel_betau}) and  all running
parameters clearly depend on variable $s$. Straightforward
integration (at least numerical) of  equations (\ref{eq:hel_nu}) gives way
how to find their fixed points. Instead,  one very often solves the
set of equations $\beta_{g}( g^*, u^*)=\beta_{u}( g^*, u^*)=0$
 which defines all fixed points $g^*, u^*$. Just
 this approach was used above when we classified all fixed points.
Due to special form of $\beta$-functions (\ref{eq:hel_betag}),
(\ref{eq:hel_betau}) we are able to solve equation (\ref{eq:hel_nu}) analytically.
Using Eqs.\,(\ref{eq:hel_nu}), and (\ref{eq:hel_betag}) one immediately rewrites
(\ref{eq:hel_nu}) in the form
\begin{equation}
  s \frac{\dRM\bar  \nu}{\bar \nu}= \frac{\gamma_{\nu}}{2\eps+\eta
  - 3 \gamma_{\nu}}\frac{\dRM \bar g}{\bar g}
  \label{eq:hel_nu1}
\end{equation}
which can be easily integrated. Using initial conditions the
solution acquires the form
\begin{equation}
  \bar \nu = \left(\frac{g\nu^3}{\bar g s^{2\epsilon + \eta}}\right)^{1/3}=
  \left(\frac{D_0}{\bar g k^{2\epsilon + \eta}}\right)^{1/3} ,
  \label{eq:hel_nue}
\end{equation}
where to obtain the last expression we used the equations
$g\mu^{2\epsilon+\eta} \nu^3=g_0\nu_0^3 =D_0$. We emphasize that
above solution is exact, i.e., the exponent $2\epsilon+\eta$ is
 as well. However, in infrared region $k \ll \Lambda \sim l^{-1}$,
$\bar g \rightarrow g^*,$ which can be calculated only
pertubatively. In the two-loop approximation $g^*=
{g^{(1)*}}\eps + {g^{(2)*}}\eps^2$ and after the Taylor
expansion of $(g^*)^{1/3}$ in Eq.\,(\ref{eq:hel_nue}) we obtain
\begin{equation}
  \bar \nu \approx \nu_*
  \left(\frac{D_0}{g^{(1)*}\eps}\right)^{1/3}
  k^{-\frac{2\epsilon + \eta}{3}}\,,\qquad \nu^*\equiv
  1-\frac{g^{(2)*}\eps}{3 g^{(1)*}}\,. 
  \label{eq:hel_nue1}
\end{equation}
%
Remind that for Kolmogorov values $\eps =\eta=4/3$ the
exponent in (\ref{eq:hel_nue1}) becomes equal to $-4/3.$ Let us estimate
the contribution of helicity to the effective diffusivity in
nontrivial point above denoted as FPV. 
 At this point
$\eps=\eta$ $((2\eps+\eta)/3=\eps)$ and
\begin{align}
  \nu^* & = 1- \frac{\eps} {12(1+u^*)}
  \Biggl(\frac{2(3+u^*)}{5(1+u^*)^2}{_2F_1}\left(1,1;\frac{7}{2};\frac{1}{(1+u^*)^2}\right)
  \nonumber\\
  & - \pi \rho^2
  {_2F_1}\left(\frac12,\frac12;\frac{5}{2};\frac{1}{(1+u^*)^2}\right)
  \Biggr).
  \label{eq:hel_nukon}
\end{align}

 \begin{figure}[!htb]
     \begin{minipage}{0.475\textwidth}
        \mbox{ } 
         \includegraphics[width=6cm]{\PICS 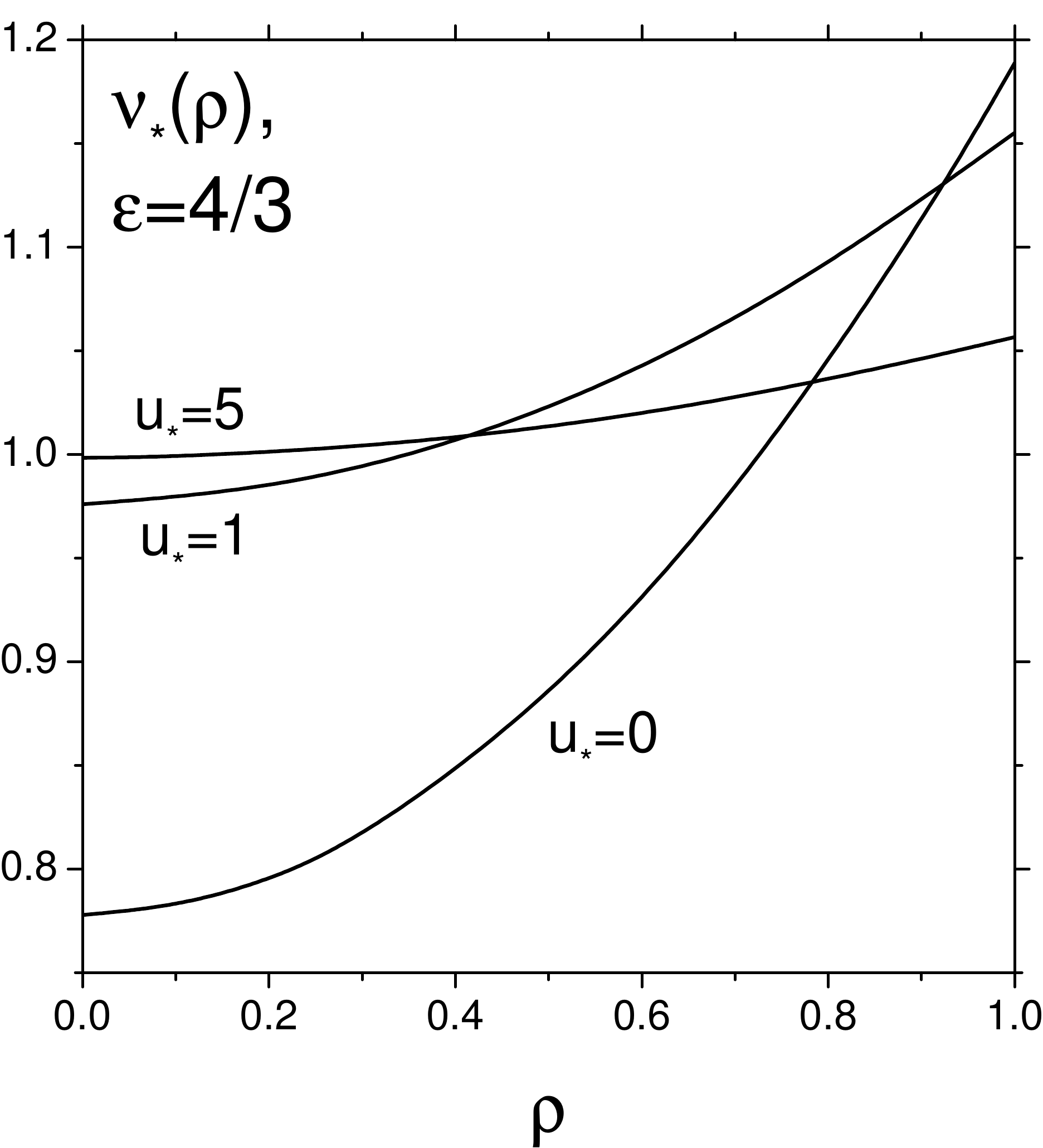}
         \caption{The dependence of $\nu^*$ on the helicity
            parameter $\rho$ for definite IR fixed point values $u^*$ of the
            parameter $u$. }
            \label{fig:hel_fig10}
     \end{minipage}%
     \hfill
     \begin{minipage}{0.475\textwidth}
          \mbox{ } 
         \includegraphics[width=6cm]{\PICS 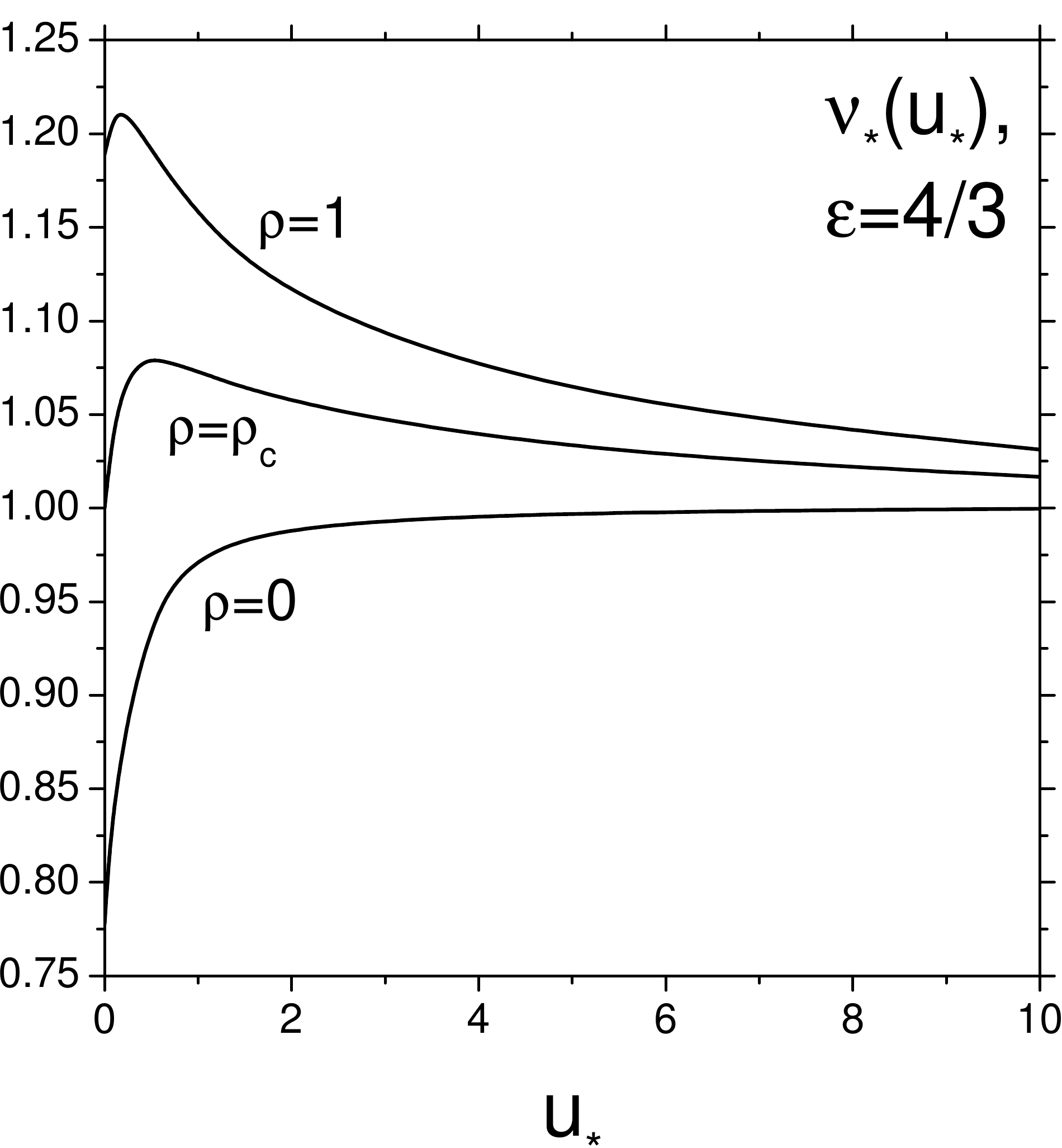}        
         \caption{
         The dependence of $\nu^*$ on the IR fixed
      point $u^*$ for the concrete values of the helicity parameter $\rho$.    }
 \label{fig:hel_fig11}
     \end{minipage}
 \end{figure}

In Figs.\,\ref{fig:hel_fig10}, and \ref{fig:hel_fig11} the dependence of the $\nu^*$
on the helicity parameter $\rho$ and the IR fixed point $u^*$ of the
parameter $u$ is shown. As one can see from these figures when
$u^*\rightarrow\infty$ (the rapid change model limit) the two-loop
corrections to $\nu^*=1$ are vanishing. Such behavior is related to
the fact, which was already stressed in the text,
that within the rapid change model there are no two and higher loop
corrections at all. On the other hand, the largest two-loop
corrections to the $\nu^*$ are given in the frozen velocity field
limit ($u^* \rightarrow 0$).
It is interesting that for all finite values of the parameter $u^*$
there exists a value of the helicity parameter $\rho$ for which the
two-loop contribution to $\nu^*$ are canceled. For example, for the
frozen velocity field limit ($u^*=0$) such situation arises when the
helicity parameter $\rho$ is equal to its "critical" value
$\rho_c=4/(\sqrt{3}\pi)$ (this situation can be seen in
Fig.\,\ref{fig:hel_fig11}). It is again the result of the competition
between the non-helical and helical parts of the the two-loop
corrections as is shown in Eq.\,(\ref{eq:hel_nukon}). Further important
feature of the expression (\ref{eq:hel_nukon}) is that it is linear in the
parameter $\eps$. Thus, when one varies the value of
$\eps$ the picture is the same as in Figs.\,\ref{fig:hel_fig10}, and
\ref{fig:hel_fig11} and only the scale of corrections is changed. In
Figs.\,\ref{fig:hel_fig10}, and \ref{fig:hel_fig11} we have shown the situation for
the most interesting case when $\eps$ is equal to its
"Kolmogorov" value, namely, $\eps=4/3$.\\

Finally, time behavior of the retarded response function $G\equiv W_{\theta \theta'} ≡ \langle\theta \theta'\rangle_\text{conn}$  in the
limit $t \rightarrow \infty $ is analyzed.
In frequency–wave vector representation $G(\omega, \mp)$ satisfies the Dyson equation (\ref{eq:hel_Dyson}) when 
one takes into account the relation $\Gamma_{\theta'\theta} \W_{\theta \theta'}=1$ that follows from Eq. (\ref{eq:RG_transform}). The
self-energy operator $\Sigma$
is expressed via multi-loop Feynman graphs and can be calculated
perturbatively. Its divergent part was found \cite{Chkhetiani06a} up to the two-loop approximation and
 its finite part with the one-loop precision was calculated.
Using the Dyson equation the response function can be written in the time–wave vector
representation
\begin{equation}
 G(t,\mp) = \int \frac{\dRM \omega}{2\pi} \eRM^{-i\omega t} G(\omega,\mp) = \int  \frac{\dRM \omega}{2\pi} \frac{\eRM^{-i\omega t}}{
 -i\omega + \nu_0 p^2 - \Sigma(\omega,\mp)}.
\end{equation}
In the lowest approximation $\Sigma(\omega,\mp) =0$; thus the integral can be easily calculated:
$G_0(t,\mp) = \theta(t) \eRM^{-i\omega_r t}$.
Here $\theta (t)$ denotes the usual step function and $\omega_r$ is a residuum
in complex plain $\omega$ in point $−i\nu_0 p^2$. Let us suppose that this situation remains the same for the full
response function $G$; i.e., the leading contribution to its asymptotic behavior for $t \rightarrow\infty $ is
determined by the residuum $\omega = \omega_r$, which corresponds to the smallest root of the dispersion
relation
\begin{equation}
 G^{-1} (\omega,\mp) = −i\omega_r + \nu_0 p^2 −\Sigma(\omega_r,\mp) = 0.
\end{equation}
It is advantageous to rewrite the last relation in the dimensionless form:
\begin{equation}
  1 - z - I (1, z) = 0,\quad 
  z \equiv \frac{i\omega_r}{\nu_0 p^2}, \quad
  I (1, z, g) \equiv \frac{\Sigma(\omega,\mp)}{\nu_0 p^2} , 
\end{equation}
which after renormalization can be rewritten in the fixed point $g^*$ () as follows:
\begin{equation}
  1 - z^∗ - I^∗ = 0,
\end{equation}
where ${\bar\nu}$ ̄is effective diffusivity (\ref{eq:hel_nue1}) and $I^*\equiv I^*(1,z^*,g^*)$ is the renormalized (finite) part of
the dimensionless self-energy operator $I$ at the fixed point $g^*$.
After some algebraic manipulations \cite{Chkhetiani06a} the decay law $G_0(t,\mp)\sim \exp(-\nu_0 p^2 t)$ is changed into
\begin{equation}
  G(t,\mp) \sim \exp(-i\omega_r t) = \exp(-z^* {\bar \nu} p^2 t),\quad t\rightarrow\infty. 
\end{equation}
Due to the existence of two complex conjugate values $z^*$ the response function $G(t,p^2)$
can be written in the asymptotic limit $t\rightarrow\infty$ in the following final form:
\begin{equation}
   G(t,p^2) \cong \sin(\nu_f p^{2-\eps}t)\exp(-\nu_\text{eff} p^{2-\eps}t) 
\end{equation}
where
\begin{align}
   \nu_\text{eff} &  = \biggl\{
   1-\eps\biggl[\frac{8}{3} + 2\ln\frac{1+u^*}{2}+\frac{3+u^*}{30(1+u^*)^3} \mbox{ }_2F_1\biggl(1,1;\frac{7}{2};\frac{1}{(1+u^*)^2}\biggl)
   \nonumber\\
   &-\frac{\pi \rho^2}{12(1+u^*)}  \mbox{ }_2F_1\biggl(\frac{1}{2},\frac{1}{2};\frac{5}{2};\frac{1}{(1+u^*)^2}\biggl)
   \biggl]\biggl\}\biggl(
   \frac{2D_0}{3(1+u^*)\eps}
   \biggl)^{1/3},\nonumber\\
   \nu_f &= \frac{\pi\eps}{2} \biggl(
   \frac{2D_0}{3(1+u^*)\eps}
   \biggl)^{1/3}.
\end{align}
It is clear that the exponential attenuation is accompanied by the oscillations.\\

{\subsubsection{Operator product expansion, critical dimensions of composite operators, and anomalous
scaling}\label{subsubsec:hel_CompOper}}
The behavior of the scaling function in
Eq.\,(\ref{eq:hel_Fr}) can be analyzed using OPE technique (Sec. \ref{subsec:OPE}). 

In what follows, we shall concentrate on the equal-time structure
functions of the scalar field defined as
\begin{equation}
  S_n(r)\equiv\langle[\theta(t,{\mx})-\theta(t,{\mx^{\prime}})]^n\rangle,\quad r=|{\mx}-{\mx^{\prime}}|,
  \label{eq:hel_Sn}
\end{equation}
which are also interesting from experimental point of view. The
representation (\ref{eq:hel_Fr}) is valid with the dimensions
$d_F^{\omega}=0$ and $d_F=\Delta_F=n\Delta_{\theta}=-n$. In general,
not only the operators which are present in the corresponding Taylor
expansion are entering into the OPE but also all possible operators
that admix to them in renormalization. In present model the leading
contribution of the Taylor expansion for the structure functions
(\ref{eq:hel_Sn}) is given by the tensor composite operators constructed
solely of the scalar gradients
\begin{equation}
  F[n,p]\equiv \partial_{i_1}\theta \cdots \partial_{i_p}\theta
  (\partial_{i}\theta
  \partial_{i} \theta)^l,
  \label{eq:hel_composite}
\end{equation}
where $n=p+2l$ is the total number of the fields $\theta$ entering
into the operator and $p$ is the number of the free vector
indices.\\

\subsubsection{ Anomalous scaling: two-loop approximation}
The influence of the helicity on the
anomalous scaling is the most interesting for the
degenerate fixed point, namely, the fixed point denoted as FPV in
Sec.\ref{subsubsec:hel_ScalReg}. In this case, the dimensions
$\Delta_{F_{np}}$ are represented in the following series in the
only independent exponent $\eps=\eta$ 
\begin{equation}
  \Delta_{F_{np}}=\eps \Delta^{(1)}_{F_{np}} + \eps^2 \Delta^{(2)}_{F_{np}}\,.
\end{equation}
The one-loop contribution has the form
\begin{equation}
  \Delta^{(1)}_{F_{np}}=\frac{2n(n-1)-(n-p)(d+n+p-2)(d+1)}{2(d+2)(d-1)}\,,
\end{equation}
which is independent of the parameter $u$. Although the fixed
point value $g_*$ 
depends on helicity
parameter $\rho$, the two-loop contribution to the critical
dimension $\Delta^{(2)}_{F_{np}}$ is independent of $\rho$. Thus,
the result is the same as that obtained in Ref.\,\cite{AdAnHo02}
(the correct formula was republished in Ref.\,\cite{AdAnHoKi05}). 

In this section, the influence of helicity on the stability of
asymptotic regimes, on the anomalous scaling, and on the effective
diffusivity was briefly reviewed in the framework of the passive scalar
advected by the turbulent flow with finite correlations in time of
the velocity field \cite{Chkhetiani06a,Chkhetiani06b}. Such investigation is important and useful for
understanding of efficiency of toy models (like Kraichnan model, and
related ones) to study the real turbulent motions by means of modern
theoretical methods including renormalization group approach. Thus,
it can be consider as the first step in investigation of the
influence of helicity in real turbulent environment.

The RG calculations \cite{Chkhetiani06b} are necessary to  two-loop
approximation in order to observe effects of helicity. It
has been mentioned that the anomalous scaling of the structure functions,
which is typical for the Kraichnan model and its numerous extensions, is not changed by the inclusion of helicity to
the incompressible fluid. It is given mathematically by the very
interesting fact that although separated two-loop Feynman diagrams
of the corresponding composite operators strongly depend on the
helicity parameter $\rho$, their sum - the critical dimension
$\Delta_{n}$ is independent of $\rho$ in the asymptotic regime
defined by IR stable fixed point. 

On the other hand, stability of possible asymptotic regimes, values
of the fixed RG points and the turbulent diffusivity strongly depend
on amount of helicity. The presence of helicity in
the system leads to the restrictions on the possible values of the
parameters of the model. The most interesting fact is the existence
of a \char`\"{}critical\char`\"{} value $\rho_{c}$ of the helicity
parameter $\rho$ which divides the interval of possible absolute
values of $\rho$ into two parts with completely different behavior.
It is related to the existence of a competition between non-helical
and helical contributions within two-loop approximation. As a result
of this competition, within of the so-called frozen limit, the
presence of helicity enlarges the region of parameter space with
stable scaling regime, and if $|\rho|=\rho_{c}$ the corresponding
two-loop restriction is vanished completely and one is coming to the
one-loop results \cite{Ant99}. Similar splitting, although more
complicated, into two nontrivial behavior of the fixed point was
also obtained in the general case with finite correlations in time
of the velocity field.

Another quantity which rather strongly depends on the helicity
parameter $\rho$ is effective diffusivity. The
value of effective diffusivity can be 50\% larger in helical case in
comparison with non-helical case.\\

{\subsection{Effect of strong anisotropy} \label{subsec:str_anistropy}}
   
Another important question addressed is the effects of   
large-scale anisotropy on inertial-range statistics of passively advected   
fields \cite{Pumir96,Pumir98,Shraiman96,Shraiman97,Ant99,Ant00,KJW,CLMV99,Lanotte99,ALM01,Arad00a,Arad00b} and the   
velocity field itself \cite{Borue,Saddoughi,Arad98,Arad99,Kurien00}. According to the classical   
Kolmogorov--Obukhov theory, the anisotropy introduced at large scales by   
the forcing (boundary conditions, geometry of an obstacle etc.) dies out   
when the energy is transferred down to the smaller scales owing to the   
cascade   
mechanism \cite{Monin,Frisch}. A number of works confirms this   
picture   
for the {\it even} correlation functions, thus giving some quantitative   
support to the aforementioned hypothesis on the restored local isotropy   
of the inertial-range turbulence for the velocity and passive fields   
\cite{Ant99,Ant00,CLMV99,Lanotte99,ALM01,Arad00a,Arad00b,Borue,Arad98,Arad99,Kurien00}. More precisely, the   
exponents describing the inertial-range scaling exhibit universality and   
hierarchy related to the degree of anisotropy, and the leading contribution   
to an even function is given by the exponent from the isotropic shell   
\cite{Ant99,Ant00,KJW,Lanotte99,ALM01,Arad00a,Arad00b,Arad98,Arad99,Kurien00}. Nevertheless, the anisotropy   
survives in the inertial range and reveals itself in {\it odd} correlation   
functions, in disagreement with what was expected on the basis of the   
cascade ideas.   
The so-called skewness factor decreases down the scales much slower than   
expected \cite{Antonia84,Sreenivasan91,Sreenivasan97,HolSig94,Pumir96,TonWar94,Pumir96,Pumir98,Shraiman96,Shraiman97}, while the higher-order odd   
dimensionless ratios (hyperskewness etc.) increase, thus signaling of   
persistent small-scale anisotropy \cite{Ant99,Ant00,CLMV99,ALM01}. The effect   
seems rather universal, being observed for the scalar \cite{Ant99,Ant00} and vector   
\cite{ALM01} fields, advected by the Gaussian rapid-change   
velocity, and for the scalar advected by the two-dimensional Navier-Stokes   
velocity field \cite{CLMV99}.   
   
Here we demonstrate the anomalous scaling behavior of a   
passive scalar advected by the time-decorrelated strongly anisotropic   
Gaussian velocity field. In contradistinction with the studies of   
\cite{HolSig94,Pumir96,TonWar94,Pumir98,Shraiman96,Shraiman97,Ant99,Ant00}, where the velocity was isotropic and   
the large-scale anisotropy was   
introduced by the imposed linear mean gradient, the uniaxial anisotropy in   
 considered model persists for all scales, leading to non-universality of   
the anomalous exponents through their dependence on the anisotropy   
parameters.   
   
The aim is twofold.   
 First, explicit inertial-range expressions for   
the structure functions  and correlation functions of the scalar   
gradients are obtained and then the corresponding anomalous exponents to   
the first order of the $\eps$ expansion are computed. The exponents   
 become non-universal through the dependence on the   
parameters describing the anisotropy of the velocity field.   
Owing to the anisotropy of the velocity statistics, the composite   
operators of different ranks mix strongly in renormalization, and   
the corresponding anomalous exponents are given by the eigenvalues   
of the matrices which are neither diagonal nor triangular   
(in contrast with the case of large-scale anisotropy). In the language of   
the zero-mode technique this means that the $SO(d)$ decompositions of the   
correlation functions (employed, e.g., in Refs. \cite{Arad00a,Arad00b})   
do not lead to the diagonalization of the   
differential operators in the corresponding exact equations.         \\

\subsubsection{Definition of the model. Anomalous scaling and ``dangerous''   
composite operators.}  \label {subsubsec:str_scenario}   
 The discussion here closely follows exposition in Sec. \ref{subsec:helicity}. 
 As has been mentioned there 
 the advection of a passive scalar field $\theta(x)\equiv\theta(t,{\mx})$   
in the rapid-change model is described by the stochastic equation (\ref{eq:hel_scalar1}).
The velocity ${\mv}(x)$ correlator now instead of Eq. (\ref{eq:hel_corv}) reads
\begin{equation}   
  \langle v_{i}(x) v_{j}(x')\rangle = D_{0}\,   
  \frac{\delta(t-t')}{(2\pi)^d}   
  \int \dRM^d {\mk}\, T_{ij}({\mk})\, (k^{2}+m^{2})^{-d/2-\eps/2}\,   
  \exp [{\rm i}{\mk}\cdot({\mx}-{\mx'})],
  \label{eq:str_3}   
\end{equation}   
where 
\begin{equation}
  \frac{D_0}{\nu_0} \equiv g_0 \equiv \Lambda^\eps
  \label{eq:str_Lambda}
\end{equation}
and as was already mentioned in Sec. \ref{subsubsec:hel_Model} the relation $m = 1/L$ holds.
In the isotropic case, the tensor quantity $T_{ij}({\mk})$ in   
(\ref{eq:str_3}) is taken to be the ordinary transverse projector    
$T_{ij}({\mk})=P_{ij}({\mk})$ (See Eq. (\ref{eq:double_P})).   
The velocity statistics is taken to be anisotropic also   
at small scales. The ordinary transverse projector is replaced   
 by the general transverse structure that possesses   
the uniaxial anisotropy:   
\begin{equation}   
  T_{ij}({\mk}) = a(\psi) P_{ij} ({\mk}) + b (\psi) \tilde n_{i}({\mk})   
  \tilde n_{j}({\mk}).   
  \label{eq:str_T}   
\end{equation}   
Here the unit vector ${\mn}$ determines the distinguished direction   
(${\mn}^{2}=1$),  
\begin{equation}
  \tilde n_{i} ({\mk})\equiv P_{ij} ({\mk})n_{j},   
  \label{eq:str_tildeN}
\end{equation}
and $\psi$ is the angle between the vectors ${\mk}$ and   
${\mn}$, so that $({\mn}\cdot{\mk})=k\cos\psi$ [note that   
$(\tilde{\mn}\cdot{\mk})=0$]. The scalar functions can be   
decomposed the Gegenbauer polynomials (the $d$-dimensional   
generalization of the Legendre polynomials, see Ref. \cite{GradRizi}):   
\begin{equation}   
  a(\psi) = \sum_{l=0}^{\infty} a_{l} P_{2l}(\cos\psi), \quad   
  b(\psi) = \sum_{l=0}^{\infty} b_{l} P_{2l}(\cos\psi).
  \label{eq:str_Legendre}   
\end{equation}   
The positivity of the correlator (\ref{eq:str_3}) leads to the conditions   
\begin{equation}   
  a(\psi) > 0, \quad a(\psi)+ b(\psi)\sin^{2}\psi >0.   
  \label{eq:str_positiv}   
\end{equation}   
In practical calculations one works with the   
special case   
\begin{equation}   
  T_{ij}({\mk}) = \biggl[1+\alpha_{1}\frac {(\mn \cdot \mk)^2} {k^2} \biggl] P_{ij} ({\mk}) + \alpha_{2}   
  \tilde n_{i}({\mk}) \tilde n_{j}({\mk}).   
  \label{eq:str_T34}   
\end{equation}   
Then the inequalities (\ref{eq:str_positiv}) reduce to $\alpha_{1,2}>-1$.   
 Later it will be shown that this choice represents nicely all the main features   
of the general model (\ref{eq:str_T}).   
   
We note that the quantities (\ref{eq:str_T}), (\ref{eq:str_T34}) possess the   
symmetry $\mn\to-\mn$. The anisotropy makes it possible to introduce   
mixed correlator $\langle{\mv}f\rangle\propto\mn   
\delta(t-t')\, C'(r/L)$ with some function $C'(r/L)$   
analogous to $C(r/L)$ from Eq. (\ref{eq:hel_correlator}). This violates the   
evenness in $\mn$ and gives rise to non-vanishing odd functions   
$S_{2n+1}$. However, this leads to no serious alterations in the   
RG analysis; this case is discussed in Sec. \ref{subsubsec:str_OPE},   
and for now we assume $\langle{\mv}f\rangle=0$. 
   
In a number of papers, e.g. \cite{HolSig94,Pumir96,TonWar94,Pumir98,Shraiman96,Shraiman97,Ant99,Ant00}, the artificial   
stirring force in Eq. (\ref{eq:hel_scalar1}) was replaced by the term $(\mh\cdot{\mv})$,   
where $\mh$ is a constant vector that determines the distinguished   
direction and therefore introduces large-scale anisotropy.   
The anisotropy gives rise to non-vanishing odd functions $S_{2n+1}$.   
The critical dimensions of all composite operators remain unchanged,   
but the irreducible tensor operators acquire nonzero mean values and   
their contributions appear on the right hand side of Eq. (\ref{eq:hel_ope}); see   
\cite{Ant99,Ant00}. This is easily understood in the language of the zero-mode   
approach: the noise $f$ and the term $(\mh\cdot{\mv})$ do not affect   
the differential operators in the equations satisfied by the equal-time   
correlations functions; the zero modes (homogeneous solutions)   
coincide in the two cases, but in the latter case the modes with   
nontrivial angular dependence should be taken into account.

The direct calculation to the order $\O(\eps)$ has shown that   
the leading exponent associated with a given rank contribution to   
Eq. (\ref{eq:hel_scalar1}) decreases monotonically with the rank \cite{Ant99,Ant00}. Hence,   
the leading term of the inertial-range behavior of an even structure   
function is determined by the same exponent as was obtained isotropic model \cite{AAHN00}, while   
the exponents related to the tensor operators determine only   
subleading corrections. A similar hierarchy was established   
 in Ref. \cite{Lanotte99} (see also \cite{ALM01})   
for the magnetic field advected passively by the rapid-change velocity   
in the presence of a constant background field, and in \cite{Arad98,Arad99,Kurien00}   
within the context of the Navier--Stokes turbulence.

In a number of papers, e.g., \cite{Barton,Carati,Denis,Kim,BHHH97},   
the RG techniques were applied to the anisotropically driven   
Navier--Stokes equation, including passive advection and magnetic   
turbulence, with the expression (\ref{eq:str_T34}) entering into the   
stirring force correlator. The detailed account can be found in   
Ref. \cite{turbo}, where some errors of the previous treatments   
 were corrected. However, these studies have up to now been   
limited to the first stage, i.e., investigation of the existence   
and stability of the fixed points and calculation of the critical   
dimensions of basic quantities. Calculation of the anomalous   
exponents in those models remains an open problem.   
   
\subsubsection{Field theoretic formulation and the Dyson--Wyld equations}   
\label {subsubsec:str_QFT}   
   
The stochastic problem is equivalent   
to the field theoretic model of the set of three fields   
$\Phi\equiv\{\theta', \theta, {\mv}\}$ with action functional   
\begin{equation}   
  S[\Phi]={1\over 2}\theta' D_{\theta}\theta' +   
  \theta' \left[ - \partial_{t} - ({\mv}\cdot\boldnabla)   
  + \nu _0\boldnabla^2 +{1\over 2}D_{vij}(0)\partial_i\partial_j\right] \theta   
  -{1\over 2}{\mv} D_{v}^{-1} {\mv}.   
  \label{eq:str_action}   
\end{equation}   
 Here $D_{\theta}$   
and $D_{v}$ are the correlators (\ref{eq:hel_correlator}) and (\ref{eq:str_3}), respectively, and
\begin{equation}
D_{vij}(0)=
{D_{0}\, }  
  \int\frac{\dRM^d{\mq}}{(2\pi)^{d}}   
  \frac{ T_{ij}(\mq) } {(q^{2}+m^{2})^{d/2+\eps/2}}
\end{equation}
is the diagonal term (in spatial variables) of the coefficient of the temporal $\delta$ function in
the velocity pair correlation function (\ref{eq:str_3}).
  
The model (\ref{eq:str_action}) corresponds to a standard Feynman   
diagrammatic technique with the triple vertex  (\ref{eq:scalar2D_vertexADV}),
propagators (\ref{eq:scalar_vio_prop}) and
\begin{align}   
  \langle \theta \theta \rangle_0=C(k)\,(\omega ^2+\nu _0^2k^4)^{-1},   
  \quad   
  \langle \theta '\theta '\rangle _0=0 ,   
  \label{eq:str_lines}   
\end{align}   
where $C(k)$ is the Fourier transform of the function $C(r/L)$ from Eq.   
(\ref{eq:hel_correlator}) and the bare propagator   
$\langle{\mv}{\mv}\rangle _0 \equiv \langle{\mv}{\mv}\rangle$   
is given by Eq. (\ref{eq:str_3}) with the transverse projector from Eqs.   
(\ref{eq:str_T}) or (\ref{eq:str_T34}).   
   
The pair correlation functions $\langle\Phi\Phi\rangle$ of the   
multicomponent field $\Phi\equiv\{\theta', \theta, {\mv}\}$ satisfy   
the standard Dyson equation, which in the component notation reduces to   
the system of two equations, cf. \cite{Monin}   

\begin{align}   
  G^{-1}(\omega, \mk ) &= -{\rm i}\omega +\nu_0 k^{2} -   
  \Sigma_{\theta'\theta} (\omega, \mk),   
  \label{eq:str_Dyson1}  \\ 
  D(\omega, \mk) &= |G(\omega, \mk)|^{2}\,   
  [C(k)+\Sigma_{\theta'\theta'} (\omega, \mk)],   
  \label{eq:str_Dyson2}   
\end{align}   
where $G(\omega, \mk)\equiv\langle\theta\theta'\rangle$ and   
$D(\omega, \mk)\equiv\langle\theta\theta\rangle$ are the exact response   
function and pair correlator, respectively, and $\Sigma_{\theta'\theta}$,   
$\Sigma_{\theta'\theta'}$ are self-energy operators represented by   
the corresponding 1-irreducible diagrams; all the other functions   
$\Sigma_{\Phi\Phi}$ in the model (\ref{eq:str_action}) vanish identically.   
   
The characteristic feature of the models like (\ref{eq:str_action}) is that   
all the skeleton multi-loop diagrams entering into the   
self-energy operators   
$\Sigma_{\theta'\theta}$, $\Sigma_{\theta'\theta'}$   
contain effectively closed circuits of retarded   
propagators $\langle\theta\theta'\rangle$   
(it is crucial here that the propagator $\langle   
{\mv}{\mv}\rangle_{0}$ in Eq. (\ref{eq:str_3}) is proportional to the   
$\delta$ function in time) and therefore vanish.   
   
Therefore the self-energy operators   
in (\ref{eq:str_Dyson1}-\ref{eq:str_Dyson2}) are given by the one-loop approximation   
exactly and have the form   
\begin{align}   
  \Sigma_{\theta'\theta} (\omega, \mk) &=   
  - \frac{D_{0}\, k_{i}k_{j}}{2}   
    \int\frac{\dRM^d{\mq}}{(2\pi)^{d}}   
    \frac{ T_{ij}(\mq) } {(q^{2}+m^{2})^{d/2+\eps/2}}.   
    \label{eq:str_sigma3} \\
  \Sigma_{\theta'\theta'} (\omega, \mk) &=   
  D_{0} \, k_{i}k_{j}   
  \int\frac{\dRM\omega'}{2\pi} \int\frac{\dRM^d{\mq}}{(2\pi)^{d}}   
  \frac{ T_{ij}(\mq) } {(q^{2}+m^{2})^{d/2+\eps/2}}\, D(\omega',\mq'),   
  \label{eq:str_sigma2}   
\end{align}   
where $\mq'\equiv{\mk}-{\mq}$.    
   
The integration over $\omega'$ on the right-hand side of   
Eq. (\ref{eq:str_sigma2}) gives the equal-time pair correlator   
\begin{equation} 
  D(\mq)=(1/{2\pi})\int{\dRM\omega'}\,D(\omega',\mq).     
  \label{eq:str_equalpair}
\end{equation}
Note that both the self-energy operators are in fact independent of   
$\omega$.
   
The integration of Eq. (\ref{eq:str_Dyson2})  over the frequency $\omega$   
gives a closed equation for the equal-time correlator. Using Eq.   
(\ref{eq:str_sigma3}) it can be written in the form   
\begin{equation}   
  2\nu_0 k^{2}D(\mk)=C(k) + D_{0}\,k_{i}k_{j}   
  \int\frac{\dRM^d{\mq}}{(2\pi)^{d}} \,   
  \frac{T_{ij}(\mq) } {(q^{2}+m^{2})^{d/2+\eps/2}}\,   
  \Bigl[D(\mq')-D(\mk)\Bigr].   
  \label{eq:str_9}   
\end{equation}   
   
Equation (\ref{eq:str_9}) can also be rewritten as a partial differential   
equation for the pair correlator in the coordinate representation,   
$D({\mr})\equiv\langle \theta   
(t,{\mx}) \theta(t,{\mx}+{\mr})\rangle$   
[we use the same notation $D$ for the coordinate   
function and its Fourier transform].   
Noting that the integral in Eq. (\ref{eq:str_9}) involves convolutions of   
the functions $D(\mk)$ and $D_{0}\,T_{ij}(\mq)/(q^{2}+m^{2})^{d/2+\eps/2}$,   
and replacing the momenta by the corresponding derivatives,   
${\rm i}k_{i} \to \partial_{i}$ and so on, the following equation  
\begin{align}   
  2\nu_0 \partial^{2}D({\mr})+C(r/L) + D_{0}\, S_{ij}({\mr}) \,   
  \partial_{i}\partial_{j} \, D({\mr}) =0   
  \label{eq:str_12}   
\end{align}   
can be obtained, where the ``effective eddy diffusivity'' is given by   
\begin{equation}   
  S_{ij}({\mr}) \equiv \int\frac{\dRM^d{\mq}}{(2\pi)^{d}} \,   
  \frac{T_{ij}(\mq)} {(q^{2}+m^{2})^{d/2+\eps/2}}   
  \Bigl[1-\exp\,({\rm i}{\mq}\cdot{\mr})\Bigr].   
  \label{eq:str_10}   
\end{equation}   
For $0<\eps <2$, equations (\ref{eq:str_9})--(\ref{eq:str_10}) allow for the limit   
$m\to0$: the possible IR divergence of the integrals at ${\mq}=0$ is   
suppressed by the vanishing of the expressions in the square brackets.   
For the isotropic case (i.e., after the substitution $T_{ij}\to P_{ij}$)   
Eq. (\ref{eq:str_12}) coincides (up to the notation) with the well-known   
equation for the equal-time pair correlator in the model \cite{Kra68,Kraichnan74,Kraichnan94,Kraichnan97}.   
   
\subsubsection{Renormalization, RG functions, and RG equations}   
\label {subsubsec:str_QFT1}   

 The analysis of the UV divergences is based on the analysis of   
 canonical dimensions introduced in Sec. \ref{sec:RG_theory}.
  \begin{table}[h!]   
 \centering
 \caption{Canonical dimensions of the fields and parameters in the   
 model (\protect\ref{eq:str_extended}).}   
 \label{tab:str_table1}   
 \begin{tabular}{| c | c | c | c | c | c | c | c | }   
 \hline\noalign{\smallskip}
 $F$ & $\theta$ & $\theta'$ & $ {\mv} $ & $\nu$, $\nu _{0}$   
 & $m$, $\mu$, $\Lambda$ & $g_{0}$ & $g$, $\alpha$, $\alpha_{0}$ 
 \\ \noalign{\smallskip}\hline\noalign{\smallskip}
 $d_{F}^{k}$ & 0 & $d$ & $-1$ & $-2$ & 1& $\eps $ & 0 
 \\ \noalign{\smallskip}\hline\noalign{\smallskip}
 $d_{F}^{\omega}$ & $-1/2$ & $1/2$ & 1 & 1 & 0 & 0 & 0 
  \\ \noalign{\smallskip}\hline\noalign{\smallskip}
 $d_{F}$ & $-1$ & $d+1$ & 1 & 0 & 1 & $\eps $ & 0 
 \\ \noalign{\smallskip}\hline\noalign{\smallskip}
 \end{tabular}   
 \end{table}   
 The dimensions for the model (\ref{eq:str_action}) are given in   
 Table \ref{tab:str_table1}, including the parameters   
 which will be introduced later on.   
 
 In the presence of anisotropy,   
it is necessary to also introduce new counterterm of the form   
$\theta'(\mn\cdot\boldnabla)^{2} \theta$, which is absent in the unrenormalized   
action functional (\ref{eq:str_action}). Therefore, the model   
(\ref{eq:str_action}) in its original formulation is not multiplicatively   
renormalizable, and in order to use the standard RG techniques it is   
necessary to extend the model by adding the new contribution to the   
unrenormalized action:   
\begin{equation}   
  \S[\Phi]=\theta' D_{\theta}\theta' /2+ \theta' \left[-\partial_{t}   
  -({\mv}\cdot\boldnabla) + \nu _0\boldnabla^2 + \chi_0\nu_0 (\mn\cdot\boldnabla)^{2} \right]   
  \theta -{\mv} D_{v}^{-1} {\mv}/2.   
  \label{eq:str_extended}   
\end{equation}   
Here $\chi_{0}$ is a new dimensionless unrenormalized parameter.   
The stability of the system implies the positivity of the total viscous   
contribution  $\nu_0 k^{2} + \chi_0\nu_0 (\mn\cdot\mk)^{2}$, which   
leads to the inequality $\chi_{0}>-1$.   
Its real (``physical'') value is zero, but this fact does   
not hinder the use of the RG techniques, in which it is first assumed   
to be arbitrary, and the equality $\chi_{0}=0$ is imposed as the   
initial condition in solving the equations for invariant variables   
(see Sec. \ref{subsubsec:str_RGE}). The zero   
value of $\chi_{0}$ corresponds to certain nonzero value of   
its renormalized analog, which can be found explicitly.   
   
For the action (\ref{eq:str_extended}), the nontrivial bare propagators   
in (\ref{eq:str_lines}) are replaced with   
\begin{align}   
  \langle \theta \theta' \rangle _0 =\langle \theta' \theta \rangle _0^* &=   
  \frac{1}{-{\rm i}\omega +\nu_0 k^2+ \chi_{0}\nu_0 (\mn\cdot \mk)^{2}} ,  \\
  \langle \theta \theta \rangle _0 &= \frac{C(k)}{ \vert -   
  {\rm i}\omega + \nu_0k^2+ \chi_{0}\nu_0 (\mn\cdot \mk)^{2} \vert ^{2}}\, .   
  \label{eq:str_lines2}   
\end{align}   
   
After the extension, the model has become multiplicatively   
renormalizable: inclusion of the counterterms is reproduced by the   
inclusion of two independent renormalization constants $Z_{1,2}$   
as coefficients in front of the counterterms.   
This leads to the renormalized action of the form   
\begin{equation}   
  \S_{R}[\Phi] = \theta' D_{\theta}\theta' /2+ \theta' \left[-\partial_{t}   
  -({\mv}\cdot\boldnabla) + \nu Z_{1}\boldnabla^2 + \chi\nu Z_{2}   
  (\mn\cdot\boldnabla)^{2} \right]   
  \theta -{\mv} D_{v}^{-1} {\mv}/2,   
  \label{eq:str_renormalized}   
\end{equation}   
or, equivalently, to the multiplicative renormalization   
of the parameters $\nu_0$, $g_{0}$ and $\chi_{0}$   
in the action functional (\ref{eq:str_extended}):   
\begin{equation}   
  \nu_0=\nu Z_{\nu},\quad g_{0}=g\mu^{\eps}Z_{g}, \quad \chi_{0}=   
  \chi Z_{\chi}.   
  \label{eq:str_18}   
\end{equation}   
The correlator (\ref{eq:str_3}) in   
(\ref{eq:str_renormalized}) is expressed in renormalized variables using Eqs.   
(\ref{eq:str_18}). The comparison of Eqs. (\ref{eq:str_extended}),   
(\ref{eq:str_renormalized}), and (\ref{eq:str_18}) leads to the relations   
\begin{equation}   
  Z_{1}=Z_{\nu}, \quad Z_{2}=Z_{\chi}Z_{\nu}, \quad Z_{g}=Z_{\nu}^{-1}.   
  \label{eq:str_18a}   
\end{equation}   
   
The beta functions are given by
\begin{equation}   
  \beta_{g}(g,\alpha)\equiv \widetilde{\cal D}_\mu g =   
  g\,(-\eps - \gamma_{g}) = g\,(-\eps + \gamma_{\nu})   
  = g\,(-\eps + \gamma_{1}),   
  \label{eq:str_RGF2}   
\end{equation}   
\begin{equation}   
  \beta_{\chi}(g,\chi)\equiv \widetilde{\cal D}_\mu \chi= -\chi  
  \gamma_{\chi}= \chi (\gamma_{1}-\gamma_{2}).   
  \label{eq:str_RGF3}   
\end{equation}   
The relation between $\beta_{g}$ and $\gamma_{\nu}$ in Eq. (\ref{eq:str_RGF2})   
results from the definitions and the last relation in (\ref{eq:str_18a}).   
   
One-loop calculation \cite{AAHN00} yields the following expressions for renormalization constants $Z_1$ and $Z_2$
\begin{align}   
  Z_{1} & = 1- \frac{g {\bar S}_d}{2d(d+2)\eps}   
  \Bigl[(d-1)(d+2)+\alpha_{1}(d+1)+\alpha_{2} \Bigr], \\
  Z_{2} & = 1- \frac{g {\bar S}_{d}}{2d(d+2)\chi\eps}   
  \Bigl[-2\alpha_{1}+\alpha_{2}(d^{2}-2) \Bigr],   
  \label{eq:str_Z}   
\end{align}   
where the geometrical factor ${\bar S}_d$ stemming from angular integration has been introduced in Eq. (\ref{eq:double_gfactors}).   
   
For the anomalous dimension $\gamma_{1}(g)$  
one obtains 
\begin{align}   
  \gamma_{1}(g)=\frac{-\eps {\cal D}_g \ln Z_{\nu}}{1-{\cal D}_g \ln Z_{\nu}}   
  =\frac{g {\bar S}_{d}}{2d(d+2)} \Bigl[(d-1)(d+2)+\alpha_{1}(d+1)+\alpha_{2} \Bigr],   
  \label{eq:str_gamma1}   
\end{align}   
and for  $\gamma_{2}(g,\alpha)$   
one has   
\begin{align}   
  \gamma_{2}(g,\alpha) & = \frac {\bigl[ (-\eps+\gamma_{1}) \D_g + \gamma_{1} 
  \D_{\chi} \Bigr] \ln Z_{2} } {1+\D_{\chi} \ln Z_{2}}=   
  \frac { -\eps \D_g \ln Z_{2} } {1+\D_{\chi} \ln Z_{2}}\\
  & =   
  \frac{g {\bar S}_{d}}{2d(d+2)\chi}   
  \bigl[-2\alpha_{1}+\alpha_{2}(d^{2}-2) \bigr]   
  \label{eq:str_gamma2}   
\end{align}   
[note that $(\D_g + \D_{\chi}) \ln Z_{2}=0$].   
The cancellation of the poles in  $\eps$ in Eqs. (\ref{eq:str_gamma1}) and   
(\ref{eq:str_gamma2}) is a consequence of the UV finiteness of the anomalous   
dimensions $\gamma_{F}$; their independence of $\eps$ is a property   
of the MS scheme. Note also that the expressions   
(\ref{eq:str_Z})--(\ref{eq:str_gamma2}) are exact, i.e., have no corrections of   
order $g^{2}$ and higher; this is a consequence of the fact that the   
one-loop approximation (\ref{eq:str_sigma3}) for the self-energy operator   
is exact.   
   
 The coordinates $g^{*},\chi^{*}$ of   
the fixed points are found from the equations  
\begin{equation}   
  \beta_{g} (g^{*},\chi^{*})=\beta_{\chi} (g^{*},\chi^{*})=0,   
  \label{eq:str_points}   
\end{equation}   
with the beta functions from Eqs. (\ref{eq:str_RGF2})-(\ref{eq:str_RGF3}).   
The type of the fixed point is determined by the eigenvalues of   
the matrix $\Omega$ defined in Eq. (\ref{eq:RG_Omega}). The IR asymptotic behavior   
is governed by the IR stable fixed points. From the explicit expressions   
(\ref{eq:str_gamma1}), (\ref{eq:str_gamma2}) it then follows that the RG equations   
of the model have the only IR stable fixed point with the coordinates   
\begin{equation}   
  g^{*} {\bar S}_{d}= \frac{2d(d+2)\eps} {(d-1)(d+2)+\alpha_{1}(d+1)   
  +\alpha_{2}}, \quad 
  \chi^{*} = \frac {-2\alpha_{1}+\alpha_{2}(d^{2}-2)}   
  {(d-1)(d+2)+\alpha_{1}(d+1)+\alpha_{2}}.   
  \label{eq:str_FP}   
\end{equation}   
For this point, both the eigenvalues of the matrix $\Omega$ equal to   
$\eps$; the values $\gamma^{*}_{1}=\gamma^{*}_{2}=\gamma^{*}_{\nu} =\eps$   
are also found exactly from Eqs. (\ref{eq:str_RGF2})-(\ref{eq:str_RGF3})   
[here and below, $\gamma^{*}_{F} \equiv \gamma_{F} (g^{*},\chi^{*})$].   
The fixed point (\ref{eq:str_FP}) is degenerate in the sense that its   
coordinates depend continuously on the anisotropy parameters   
$\alpha_{1,2}$.\footnote{ Formally, $\alpha_{1,2}$ can be treated as the   
additional coupling constants. The corresponding beta functions   
$\beta_{1,2}\equiv\widetilde{\cal D}_\mu{\alpha_{1,2}}$ vanish   
identically owing to the fact that $\alpha_{1,2}$ are not renormalized.   
Therefore the equations $\beta_{1,2}=0$ give no additional constraints   
on the values of the parameters $g,\chi$ at the fixed point.}   
   
\subsubsection{Solution of the RG equations. Invariant variables}   
 \label {subsubsec:str_RGE}   
The solution of the RG equations is discussed in Sec. \ref{sec:RGsolution}; below
we confine ourselves to only the information we need.

Consider the solution of the RG equation on the example of the   
even different-time structure functions   
\begin{equation}   
  S_{2n}(\mr,\tau)\equiv\langle[\theta(t,{\mx})   
  -\theta(t',{\mx'})]^{2n}\rangle,   
  \quad \mr\equiv\mx-\mx', \quad \tau\equiv t-t' .   
  \label{eq:str_differ}   
\end{equation}   
It satisfies the RG equation $\D_{RG} S_{2n}=0$ with the operator   
$\D_{RG} = \mathcal{D}_\mu + \beta_g\partial_g + \beta_\chi \partial_\chi - \gamma_{nu}\mathcal{D}_\nu$.
   
In renormalized variables, dimensionality considerations give   
\begin{equation}   
  S_{2n}(\mr,\tau)= \nu^{-n} r^{2n}   
  \tilde{R}_{2n}(\mu r, \tau\nu/r^{2}, r/L, g, \chi),   
  \label{eq:str_differ2}   
\end{equation}   
where $\tilde{R}_{2n}$ is a scaling function of completely dimensionless arguments (the   
dependence on $d$, $\eps$, $\alpha_{1,2}$ and the angle between the   
vectors ${\mr}$ and ${\mn}$ is also implied).   
From the RG equation the identical representation follows,   
\begin{equation}   
  S_{2n}(\mr,\tau)= (\bar\nu)^{-n} r^{2n} \tilde{R}_{2n}(1, \tau\bar\nu/r^{2}, r/L,   
  \bar g, \bar \chi),   
  \label{eq:str_differ3}   
\end{equation}   
where the invariant variables $\bar e = \bar e (\mu r, e)$   
satisfy Eq. (\ref{eq:RG_in}). The identity $\bar L \equiv L$ is a consequence of  the fact that   
$L$ is not renormalized. 
The relation between the bare and invariant variables has the form   
\begin{equation}   
  \nu_0=\bar \nu Z_{\nu}(\bar g),   
  \quad g_{0}=\bar g r^{-\eps}Z_{g}(\bar g), \quad \chi_{0}=   
  \bar\chi Z_{\chi}(\bar g, \bar \chi).   
  \label{eq:str_exo1}   
\end{equation}   
Equation (\ref{eq:str_exo1}) determines   
implicitly the invariant variables as functions of the bare parameters;   
it is valid because both sides of it satisfy the RG equation, and   
because Eq. (\ref{eq:str_exo1}) at $\mu r=1$ coincides with   
(\ref{eq:str_18}) owing to the normalization conditions.   
   
In general, the large $\mu r$ behavior of the invariant variables is   
governed by the IR stable fixed point: $\bar g\to g^{*}$,   
$\bar\chi \to\chi^{*}$ for $\mu r\to\infty$. However, in multi-charge   
problems one has to take into account that even when the IR point   
exists, not every phase trajectory (i.e., solution of Eq. (\ref{eq:str_exo1}))   
reaches it in the limit $\mu r\to\infty$. It may first pass outside   
the natural region of stability [physical region is given by the inequalities $g>0$, $\chi > -1$]   
or go to infinity within this region. Fortunately, in our case the   
constants $Z_{F}$ entering into Eq. (\ref{eq:str_exo1}) are known exactly   
from Eqs. (\ref{eq:str_Z}), and it is readily checked that the RG flow indeed   
reaches the fixed point (\ref{eq:str_FP}) for any initial conditions   
$g_{0}>0$, $\chi_{0}>-1$, including the physical case $\chi_{0}=0$.   
Furthermore, the large $\mu r$ behavior of the invariant viscosity   
$\bar\nu$ is also found explicitly from Eq. (\ref{eq:str_exo1}) and the last   
relation in (\ref{eq:str_18a}): $\bar\nu = D_{0} r^{\eps} /\bar g \to   
D_{0} r^{\eps} / g^{*}$  (we recall that $D_{0}=g_{0}\nu_0$).   
Then for $\mu r\to\infty$  and any fixed $mr$ we obtain   
\begin{equation}   
  S_{2n}(\mr,\tau)= D_{0}^{-n} r^{n(2-\eps)} {g^{*}}^{n}\,   
   R_{2n}( \tau D_{0} r^{\Delta_{t}},r/L),   
  \label{eq:str_differ4}   
\end{equation}   
where the scaling function  
\begin{equation}   
  R_{2n}( D_{0} \tau r^{\Delta_{t}},r/L)\equiv   
  \tilde{R}_{2n}(1, D_{0} \tau r^{\Delta_{t}}, r/L, g_{*},  \alpha_{*}),   
  \label{eq:str_differ5}   
\end{equation}   
has appeared
and $\Delta_{t}\equiv -2+\gamma_{\nu}^{*} =-2+\eps$   
is the critical dimension of time.  For the equal-time structure function   
(\ref{eq:hel_struc}), the first argument of $R_{2n}$ in the representation   
(\ref{eq:str_differ5}) is absent:   
\begin{equation}   
  S_{2n}(\mr)= D_{0}^{-n} r^{n(2-\eps)} {g^{*}}^{n}\,  R_{2n}(r/L),   
  \label{eq:str_100}   
\end{equation}   
where the definition of $R_{2n}$ is obvious from (\ref{eq:str_differ5}).   
It is noteworthy that Eqs. (\ref{eq:str_differ4})--(\ref{eq:str_100})   
prove the independence of the structure functions in the   
IR range (large $\mu r$  and any  $r/L$) of the viscosity coefficient or,   
equivalently of the UV scale:   
the parameters $g_{0}$ and $\nu_0$ enter into Eq. (\ref{eq:str_differ4})   
only in the form of the combination $D_{0}=g_{0}\nu_0$. A similar   
property was established in Ref. \cite{FF} for the stirred Navier--Stokes   
equation.

 In contrast to the previously mentioned models the scaling function $\tilde{R}$ in (\ref{eq:str_differ3}) contains two different scales - corresponding to
 spatial and time differences, respectively. The information about its behavior can be again using OPE method from Sec. \ref{subsec:OPE}.
 Therefore, let $F(r,\tau)$ stands for some   
multiplicatively renormalized quantity. Dimensionality considerations give   
\begin{equation}   
  F_{R}(r,\tau)= \nu^{d^{\omega}_{F}} r^{-d_{F}}   
  \tilde{R}_{F}(\mu r, \tau\nu/r^{2}, r/L, g, \alpha),   
  \label{eq:str_differ6}   
\end{equation}   
where $d^{\omega}_{F}$ and $d_{F}$ are the frequency and total canonical   
dimensions of $F$ (see Sec. \ref{subsec:UV}) and $R_{F}$ is a function of   
dimensionless arguments. The analog of Eq. (\ref{eq:str_differ3}) has the form   
\begin{equation}   
  F(r,\tau)= Z_{F} (g, \alpha) F_{R} = Z_{F} (\bar g, \bar\alpha) \,   
  (\bar\nu)^{d^{\omega}_{F}} r^{-d_{F}}   
  \tilde{R}_{F}(1, \tau\bar\nu/r^{2}, r/L, \bar g, \bar\alpha).   
  \label{eq:str_differ7}   
\end{equation}   
In the large $\mu r$ limit, one has  $Z_{F} (\bar g, \bar\alpha)   
\simeq \const (\Lambda r)^{-\gamma_{F}^{*}}$; see, e.g., \cite{AntVas97}.   
The UV scale appears in this relation from Eq. (\ref{eq:str_Lambda}). Then   
in the IR range ($\Lambda r\sim \mu r$ large, $r/L$ arbitrary) Eq.   
(\ref{eq:str_differ7}) takes on the form   
\begin{equation}   
  F(r,\tau)\simeq \const \Lambda ^{-\gamma_{F}^{*}}\,   
  D_{0}^{d^{\omega}_{F}} \, r^{-\Delta[F]}   
  R_{F}( D_{0} \tau r^{\Delta_{t}},r/L).   
  \label{eq:str_RGR}   
\end{equation}   
Here   
\begin{equation}   
  \Delta[F]\equiv\Delta_{F} = d_{F}^{k}- \Delta_{t}   
  d_{F}^{\omega}+\gamma_{F}^{*}, \quad \Delta_{t}=-2+\eps   
  \label{eq:str_32B}   
\end{equation}   
is the critical dimension of the function $F$ and the scaling   
function $R_{F}$ is related to $\tilde{R}_{F}$ as in Eq. (\ref{eq:str_differ4}).   
For nontrivial $\gamma_{F}^{*}$, the function $F$ in the IR range   
retains the dependence on $\Lambda$ or, equivalently, on $\nu_0$.   
         
\subsubsection{Renormalization and critical dimensions of   
 composite operators} \label {subsubsec:str_Operators}   
   
Operators of the form $\theta^{N}(x)$   
with the canonical dimension $d_{F}=-N$ enter into the   
structure functions (\ref{eq:hel_struc}). From Table \ref{tab:str_table1}   
in Sec. \ref{subsubsec:str_QFT1} and Eq. (\ref{eq:RG_index}) the relation   
$d_\Gamma = -N+N_{\theta}-N_{\mv} -(d+1)N_{\theta'}$ can be obtained,   
and from the analysis of the diagrams it follows that the total   
number of the fields $\theta$ entering into the function   
$\Gamma$ can never exceed the number of the fields $\theta$   
in the operator $\theta^{N}$ itself, i.e., $N_{\theta}\le N$   
(cf. item (i) in Sec. \ref{subsubsec:str_QFT1}).   
Therefore, the divergence can only exist in the functions   
with $N_{\mv}= N_{\theta'}=0$, and arbitrary value of   
$N=N_{\theta}$, for which the formal index vanishes, $d_\Gamma =0$.   
However, at least one of $N_{\theta}$ external ``tails''   
of the field $\theta$ is attached to a vertex   
$\theta'({\mv}\cdot\boldnabla)\theta$ (it is impossible to construct   
nontrivial, superficially divergent diagram of the desired   
type with all the external tails attached to the vertex $F$),   
at least one derivative $\partial$ appears as an extra   
factor in the diagram, and, consequently, the real index of   
divergence $d_\Gamma'$ is necessarily negative.   
   
This means that the operator $\theta^{N}$ requires no counterterms   
at all, i.e., it  is in fact UV finite, $\theta^{N}=Z\,[\theta^{N}]^{R}$   
with $Z=1$. It then follows that the critical dimension of   
$\theta^{N}(x)$ is simply given by the expression (\ref{eq:str_32B})   
with no correction from $\gamma_{F}^{*}$ and is therefore reduced   
to the sum of the critical dimensions of the factors:   
\begin{equation}   
  \Delta [\theta^{N}] = N\Delta[\theta] =N (-1+\eps/2).   
  \label{eq:str_2.6}   
\end{equation}   
Since the structure functions (\ref{eq:hel_struc}) or (\ref{eq:str_differ}) are linear   
combinations of pair correlators involving the operators $\theta^{N}$,   
equation (\ref{eq:str_2.6}) shows that they indeed satisfy the RG equation   
of the form (\ref{eq:RG_rgrovnica}), discussed in Sec. \ref{subsubsec:str_RGE}.   
We stress that the relation (\ref{eq:str_2.6}) was not clear {\it a priori};   
in particular, it is violated if the velocity field becomes   
non-solenoidal \cite{AdzAnt98}.

In the following, an important role will be also played by the   
tensor composite operators   
$ \partial_{i_{1}}\theta\cdots\partial_{i_{p}}\theta\,   
(\partial_{i}\theta\partial_{i}\theta)^{n}$   
constructed solely of the scalar gradients. It is convenient to deal   
with the scalar operators obtained by contracting the tensors with   
the appropriate number of the vectors $\mn$,   
\begin{equation}   
  F[N,p]\equiv [(\mn\cdot\boldnabla)\theta]^{p} (\partial_{i}\theta\partial_{i}   
  \theta)^{n}, \quad N\equiv 2n+p.   
  \label{eq:str_Fnp}   
\end{equation}   
Their canonical dimensions depend only on the total number of the fields   
$\theta$ and have the form $d_{F}=0$, $d_{F}^{\omega}=-N$.   
   
In this case, from Table \ref{tab:str_table1} and Eq. (\ref{eq:RG_index}) we obtain   
$d_\Gamma = N_{\theta}-N_{\mv} -(d+1)N_{\theta'}$,   
with the necessary condition $N_{\theta}\le N$, which follows   
from the structure of the diagrams. It is also clear from the   
analysis of the diagrams that the counterterms to these operators   
can involve the fields $\theta$, $\theta'$ only in the form of   
derivatives, $\partial\theta$, $\partial\theta'$,   
so that the real index of divergence has the form   
$d_\Gamma' = d_\Gamma -N_{\theta}-N_{\theta'}=   
-N_{\mv} -(d+2)N_{\theta'}$.   It then follows that   
superficial divergences can exist only in the Green functions with   
$N_{\mv}=N_{\theta'}=0$  and any $N_{\theta}\le N$,   
and that the corresponding operator counterterms reduce to the   
form $F[N',p']$ with $N'\le N$. Therefore, the operators   
(\ref{eq:str_Fnp}) can mix only with each other in renormalization, and   
the corresponding infinite renormalization matrix   
\begin{equation}   
  F[N,p] = \sum_{N',p'} \, Z_{[N,p]\,[N',p']} \,F^{R}[N',p']   
  \label{eq:str_Matrix}   
\end{equation}   
is in fact block-triangular, i.e., $Z_{[N,p]\,[N',p']} =0$ for $N'> N$.   
It is then obvious that the critical dimensions associated with the   
operators $F[N,p]$ are completely determined by the eigenvalues of the   
finite subblocks with $N'= N$. 
   
In the isotropic case, as well as in the presence of large-scale   
anisotropy, the elements $Z_{[N,p]\,[N,p']}$ vanish for $p<p'$,   
and the block $Z_{[N,p]\,[N,p']}$ is triangular along with the   
corresponding blocks of the matrices $U_{F}$ and $\Delta_{F}$   
from Eqs. (\ref{eq:RG_2.5}), (\ref{eq:str_32B}). In the isotropic case it can be   
diagonalized by changing to irreducible operators (scalars, vectors,   
and traceless tensors), but even for nonzero imposed gradient its   
eigenvalues are the same as in the isotropic case. Therefore, the   
inclusion of large-scale anisotropy does not affect critical   
dimensions of the operators (\ref{eq:str_Fnp}); see \cite{Ant99,Ant00}. In the   
case of small-scale anisotropy, the operators with different values   
of $p$ mix heavily in renormalization, and the matrix   
$Z_{[N,p]\,[N,p']}$ is neither diagonal nor triangular here.

The calculation of the renormalization constants   
$Z_{[N,p]\,[N,p']}$ can be illustrated within the one-loop approximation.   
Let $\Gamma(x;\theta)$ be the generating functional of the   
1-irreducible Green functions with one composite operator $F[N,p]$   
from Eq. (\ref{eq:str_Fnp}) and any number of fields $\theta$. Here 
$x$ is the argument of the operator and $\theta$ is   
the functional argument, the ``classical counterpart'' of the random   
field $\theta$. The general interest is in the $N$-th term of the   
expansion of $\Gamma(x;\theta)$ in $\theta$, which is denoted as   
$\Gamma_{N}(x;\theta)$; it has the form   
\begin{equation}   
  \Gamma_{N}(x;\theta) = \frac{1}{N!} \int \dRM x_{1} \cdots \int \dRM x_{N}   
  \, \theta(x_{1})\cdots\theta(x_{N})\,   
  \langle F[N,p](x) \theta(x_{1})\cdots\theta(x_{N})\rangle_{\rm 1-ir}.   
  \label{eq:str_Gamma1}   
\end{equation}   

The matrix of critical dimensions (\ref{eq:str_32B}) is given in the one-loop approximation by the expression
\begin{equation}   
  \Delta_{[N,p][N,p']} = N\eps/2 + \gamma^{*}_{[N,p][N,p']},   
  \label{eq:str_Dnp}   
\end{equation}   
where the asterisk implies the substitution (\ref{eq:str_FP}). The details of calculation of
$\gamma_{[N,p][N,p']}$ can be found in \cite{AAHN00}.   
   
As already said above, the critical dimensions themselves are given   
by the eigenvalues of the matrix (\ref{eq:str_Dnp}). One can check that for   
the isotropic case ($\alpha_{1,2}=0$), its elements with $p'>p$ vanish,   
the matrix becomes triangular, and its eigenvalues are simply given by   
the diagonal elements $\Delta[N,p]\equiv\Delta_{[N,p][N,p]}$. They   
are found explicitly and have the form   
\begin{equation}   
  \Delta[N,p] = N\eps/2+ \frac{2p(p-1)-(d-1)(N-p)(d+N+p)}{2(d-1)(d+2)}\,   
  \eps + \O(\eps^{2}).   
  \label{eq:str_Qnp}   
\end{equation}   
It is easily seen from Eq. (\ref{eq:str_Qnp}) that for fixed $N$ and any   
$d\ge2$, the dimension $\Delta[N,p]$ decreases monotonically with $p$   
and reaches its minimum for the minimal possible value of $p=p_{N}$,   
i.e., $p_{N}=0$ if $N$ is even and $p_{N}=1$ if $N$ is odd:   
\begin{equation}   
  \Delta[N,p] > \Delta[N,p']  \quad {\rm if} \quad  p>p' \, .   
  \label{eq:str_hier1}   
\end{equation}   
Furthermore, this minimal value $\Delta[N,p_{N}]$ decreases monotonically   
as $N$ increases for odd and even values of $N$ separately, i.e.,   
\begin{equation}   
  0\ge\Delta[2n,0]>\Delta[2n+2,0] , \quad \Delta[2n+1,1]> \Delta[2n+3,1].   
  \label{eq:str_hier2}   
\end{equation}   
A similar hierarchy is demonstrated by the critical dimensions of   
certain tensor operators in the stirred Navier--Stokes turbulence;   
see Ref. \cite{Triple} and Sec. 2.3 of \cite{turbo}. However, no   
clear hierarchy is demonstrated by neighboring even   
and odd dimensions:   
from the relations   
\begin{equation}   
  \Delta[2n+1,1]-\Delta[2n,0]=\frac{\eps(d+2-4n)}{2(d+2)}, \quad   
  \Delta[2n+2,0]- \Delta[2n+1,1] =\frac{\eps(2-d)}{2(d+2)}   
  \label{eq:str_hier3}   
\end{equation}   
it follows that the inequality $\Delta[2n+1,1]>\Delta[2n+2,0]$   
holds for any $d>2$, while the relation $\Delta[2n,0]>\Delta[2n+1,1]$   
holds only if $n$ is sufficiently large, $n>(d+2)/4$.   
\footnote{The situation is different in the presence of the linear   
mean gradient: the first term $N\eps/2$ in Eq. (\ref{eq:str_Qnp}) is then   
absent owing to the difference in canonical dimensions, and the   
complete hierarchy relations hold,   
$\Delta[2n,0]>\Delta[2n+1,1]>\Delta[2n+2,0]$; see \cite{Ant99,Ant00}. }

In what follows, we shall use the notation $\Delta[N,p]$ for the   
eigenvalue of the matrix (\ref{eq:str_Dnp}) which coincides with (\ref{eq:str_Qnp})   
for $\alpha_{1,2}=0$. Since the eigenvalues depend continuously on   
$\alpha_{1,2}$, this notation is unambiguous at least for small   
values of $\alpha_{1,2}$.

The dimension $\Delta[2,0]$ vanishes identically for any $\alpha_{1,2}$   
and to all orders in $\eps$. Like in the isotropic model, this can be   
demonstrated using the Schwinger equation of the form   
\begin{equation}   
  \int{\cal D}\Phi \frac{\delta}{\delta\theta'(x)} \left[ \theta(x) \eRM^{ \S_{R} [ \Phi]  
  + A \Phi }\right]  =0,   
  \label{eq:str_Schwi}   
\end{equation}   
(in the general sense of the word, Schwinger equations are any relations   
stating that any functional integral of a total variational derivative   
is equal to zero; see, e.g., \cite{Zinn,Vasiliev}). In (\ref{eq:str_Schwi}),   
$S_{R}$ is the renormalized action (\ref{eq:str_renormalized}) and Eq. (\ref{eq:str_Schwi}) can be   
rewritten in the form   
\begin{align}   
  &\Big\langle \theta' D_{\theta} \theta - \nabla_{t}[\theta^{2}/2] +   
  \nu Z_{1}\Delta[\theta^{2}/2] + \alpha \nu Z_{2} (\mn\cdot\boldnabla)^{2}   
  [\theta^{2}/2] -\nu Z_{1} F[2,0]  = \nonumber\\
  & - \alpha \nu Z_{2} F[2,2]   
  \Big\rangle_{A}   
-A_{\theta'} \frac{ \delta W_{R}(A)}{\delta A_{\theta}}.   
  \label{eq:str_Schwi2}   
\end{align}   
Here $D_{\theta}$ is the correlator (\ref{eq:hel_correlator}), $\langle \cdots \rangle   
_{A}$   
denotes the averaging with the weight $ \exp [\S_{R} [\Phi] +   
A \Phi]$, $\W_{R}$ is determined by Eq. (\ref{eq:intro_ActionMultiN}) with   
the replacement $\S\to \S_{R}$, and the argument $x$ common to all   
the quantities in (\ref{eq:str_Schwi2}) is omitted.   
   
The quantity $\langle F \rangle _{A}$ is the generating functional (defined in Eq. (\ref{eq:RG_xx9}))
of the correlation functions with one insertion of the operator $F$ and any number of   
the primary fields $\Phi$, therefore the UV finiteness of the   
operator $F$ is equivalent to the finiteness of the functional   
$\langle F\rangle _{A}$. The quantity in the right hand side of Eq.   
(\ref{eq:str_Schwi2}) is UV finite (a derivative of the renormalized functional   
with respect to finite argument), and so is the operator in the   
left hand side. Operators $F[2,0]$, $F[2,2]$ do not admix in   
renormalization to $\theta' D_{\theta}\theta$   
(no needed diagrams can be constructed), and to the   
operators $\nabla_{t}[\theta^{2}/2]$ and $\boldnabla^2[\theta^{2}/2]$   
(they have the form of total derivatives, and $F[N,p]$ do not   
reduce to this form). On the other hand, all the operators in   
(\ref{eq:str_Schwi2}) other than $F[N,p]$ do not admix to $F[N,p]$,   
because the counterterms of the operators (\ref{eq:str_Fnp}) can involve   
only operators of the same type; see above. Therefore,   
the operators $F[N,p]$ entering into Eq. (\ref{eq:str_Schwi2}) are independent   
of the others, and so they must be UV finite separately:   
$\nu Z_{1} F[2,0] + \alpha \nu Z_{2} F[2,2] = $ UV  finite.   
Since the operator in (\ref{eq:str_Schwi2}) is UV finite, it coincides   
with its finite part,   
\begin{equation}
  \nu Z_{1} F[2,0] + \alpha \nu Z_{2} F[2,2] =   
  \nu F^{R}[2,0] + \alpha \nu  F^{R}[2,2],   
\end{equation}
which along with the relation (\ref{eq:str_Matrix}) gives   
\begin{equation}
  Z_{1} Z_{[2,0][2,0]} + \alpha Z_{2} Z_{[2,2][2,0]} =1, \quad   
  Z_{1} Z_{[2,0][2,2]} + \alpha Z_{2} Z_{[2,2][2,2]} =\alpha,   
\end{equation}   
and therefore for the anomalous dimensions in the MS scheme one obtains   
\begin{equation}
\gamma_{1} + \gamma_{[2,0][2,0]} +\alpha \gamma_{[2,2][2,0]} =0, \quad   
\gamma_{[2,0][2,2]} + \alpha \gamma_{2} + \alpha \gamma_{[2,2][2,2]} =0.   
\end{equation}
 Bearing in mind that   
$\gamma_{1}^{*}=\gamma_{2}^{*}=\eps$ (see Sec.   
\ref{subsubsec:str_QFT1}), the conclusion can be made that among the four elements of the   
matrix $\gamma_{F}^{*}$ only two, which we take to be   
$\gamma^{*}_{[2,2][2,0]}$ and $\gamma^{*}_{[2,2][2,2]}$,   
are independent. Then the matrix of critical dimensions (\ref{eq:str_Qnp})   
takes on the form   
 \begin{align}   
   \Delta_{[2,p][2,p']} = \eps + \begin{pmatrix}
   \renewcommand{\arraystretch}{1.5}
    -\eps-\alpha_{*}   
   \gamma^{*}_{[2,2][2,0]}  & -\alpha_{*}\eps-\alpha_{*} \gamma^{*}_{[2,2][2,2]} \\   
   \gamma^{*}_{[2,2][2,0]} & \gamma^{*}_{[2,2][2,2]}    
   \end{pmatrix}
   .   
   \label{eq:str_Schwi3}   
 \end{align}   
It is then easily checked that the eigenvalue of the matrix   
(\ref{eq:str_Schwi3}), which is identified with $\Delta[2,0]$, does not   
involve unknown anomalous dimensions and vanishes identically,   
$\Delta[2,0]\equiv0$, while the second one is represented as   
\begin{equation}
   \Delta[2,2]=\eps-\alpha_{*}\gamma^{*}_{[2,2][2,0]}+\gamma^{*}_{[2,2][2,2]}.   
\end{equation}
Using the explicit $\O(\eps)$ expressions \cite{AAHN00}
one obtains to the order $\O(\eps)$:   
\begin{align}   
  \Delta[2,2]/\eps & = 2+ \Bigl\{   
  -(d-2)d(d+2)(d+4)F^{*}_{0} - (d+2)(d+4) (2+(d-2)\alpha_{1}
  \nonumber \\ 
  &+d\alpha_{2})   
    F^{*}_{1}+   
   +3(d+4)(d-2\alpha_{1}+2d\alpha_{2}) F^{*}_{2}+   
  15d (\alpha_{1}-\alpha_{2})  F^{*}_{3} \Bigr\} \Bigl/   
  \Bigl\{(d-1)
  \nonumber\\ &\times (d+4)
  [(d-1)(d+2)+(d+1)\alpha_{1}+\alpha_{2}]\Bigr\} \, ,   
  \label{eq:str_Delta22}   
\end{align}   
where $F^{*}_{n} \equiv F(1,1/2+n; d/2+n; -\alpha_{*})$ with   
$\alpha_{*}$ from Eq. (\ref{eq:str_FP}).
   
In Fig. \ref{fig:str_fig1}, we present the levels of the dimension (\ref{eq:str_Delta22})   
on the $(\alpha_{1},\alpha_{2})$-plane for $d=3$.   
We note that the dependence on $\alpha_{1,2}$   
is quite smooth, and that $\Delta[2,2]$ remains positive   
on the whole of the $(\alpha_{1},\alpha_2)$-plane, i.e., the first of the   
hierarchy relations (\ref{eq:str_hier1})-(\ref{eq:str_hier3}) remains valid also in the presence   
of anisotropy. A similar behavior takes place also for $d=2$.   
   
For $N>2$, the eigenvalues can be found analytically only within the   
expansion in $\alpha_{1,2}$. The explicit expressions can be found
in \cite{AAHN00}.
They illustrate two facts which seem to hold for all $N$:
\begin{enumerate}[(i)]
 \item The leading anisotropy correction is of order $\O(\alpha_{1,2})$   
    for $p\ne0$ and $\O(\alpha_{1,2}^{2})$ for $p=0$, so that the dimensions   
    $\gamma^{*}[N,0]$ are anisotropy independent in the {\it linear}   
    approximation, and   
 \item   
    This leading contribution depends on $\alpha_{1,2}$ only through the   
    combination $\alpha_{3}\equiv 2\alpha_{1}+d \alpha_{2}$.   
\end{enumerate}

This conjecture is confirmed by the following expressions for $N=6, 8$   
and $p=0$:   
\begin{align}   
  \gamma^{*}[6,0]/\eps= \frac{-2(d+6)}{(d+2)} - \frac{12(d-2)^{2}(d+1)   
  (d^{2}+14d+48)\alpha_{3}^{2}} {(d-1)^{2}d(d+2)^{4}(d+4)^{2}} ,   
  \label{eq:str_expansion60}   
\end{align}   
\begin{align}   
  \gamma^{*}[8,0]/\eps= \frac{-4(d+8)}{(d+2)} -   
  \frac{24(d-2)^{2}(d+1)   
  (d^{2}+18d+80)\alpha_{3}^{2}} {(d-1)^{2}d(d+2)^{4}(d+4)^{2}} .
  \label{eq:str_expansion80}   
\end{align}   
   
The eigenvalues beyond the small $\alpha_{1,2}$ expansion have been obtained   
numerically \cite{AAHN00}. Some of them are presented in Figs. \ref{fig:str_fig2}-\ref{fig:str_fig5}, namely,   
the dimensions $\Delta[n,p]$ for $n=3,4,5,6$ {\it vs} $\alpha_{1}$   
for $\alpha_{2}=0$, {\it vs} $\alpha_{1}=\alpha_{2}$, and {\it vs} $\alpha_{2}$   
for $\alpha_{1}=0$. The main conclusion that   
can be drawn from these diagrams is that the hierarchy (\ref{eq:str_hier1}-\ref{eq:str_hier3})   
demonstrated by the dimensions for the isotropic case ($\alpha_{1,2}=0$)   
holds valid for all the values of the anisotropy parameters.   
   
\subsubsection{Operator product expansion and anomalous scaling}   
  \label {subsubsec:str_OPE}   

From the operator product expansion (Sec. \ref{subsec:OPE}) we
find the following expression  for the scaling function   
$R(r/L)$ in the representation (\ref{eq:str_100}) for the correlator   
$\langle F_{1}(x)F_{2}(x') \rangle$:   
\begin{equation}   
  R(r/L)=\sum_{F} A_{F}\,\, \biggl(\frac{r}{L}\biggl)^{\Delta_{F}}, \quad r \ll L,   
  \label{eq:str_OR}   
\end{equation}   
with the coefficients $A_{F}$ regular in $(r/L)^2$.

Now let us turn to the equal-time structure functions $S_{N}$ from   
(\ref{eq:hel_struc}). From it is assumed that the mixed correlator   
$ \langle {\mv} f \rangle $ differs from zero   
(see Sec. \ref{subsubsec:str_scenario}); this does not affect the   
critical dimensions, but gives rise to non-vanishing odd structure   
functions. In general, the operators entering into the OPE are   
those which appear in the corresponding Taylor expansions, and also   
all possible operators that admix to them in renormalization   
\cite{Zinn,Vasiliev}.   
The leading term of the Taylor expansion for the function $S_{N}$   
is obviously given by the operator $F[N,N]$ from Eq. (\ref{eq:str_Fnp});   
the renormalization gives rise to all the operators $F[N',p]$ with   
$N'\le N$ and all possible values of $p$. The operators with   
$N'> N$  (whose contributions would be more important) do not   
appear in Eq. (\ref{eq:str_OR}), because they do not enter into the   
Taylor expansion for $S_{N}$ and do not admix in renormalization   
to the terms of the Taylor expansion; see Sec. \ref{subsubsec:str_Operators}.   
Therefore, combining the RG representation (\ref{eq:str_differ4}) with   
the OPE representation (\ref{eq:str_OR}) gives the desired asymptotic   
expression for the structure function in the inertial range:   
\begin{equation}   
  S_{N}(\mr)= D_{0}^{-N/2} r^{N(1-\eps/2)}\, \sum_{N'\le N} \sum_{p}   
  \Bigl\{ C_{N',p}\, (r/L)^{\Delta[N',p]}+\cdots \Bigr\}  \,.   
  \label{eq:str_struc2}   
\end{equation}   
The second summation runs over all values of $p$, allowed for a given   
$N'$; $C_{N',p}$ are numerical coefficients dependent on $\eps$, $d$,   
$\alpha_{1,2}$ and the angle $\vartheta$ between $\mr$ and $\mn$. The dots   
stand for the contributions of the operators   
other than $F[N,p]$, for example, $\partial^{2}\theta\partial^{2}\theta$;   
they give rise to the terms of order $(r/L)^{2+\O(\eps)}$ and higher   
and will be neglected in what follows.   
   
Some remarks are now in order.   
\begin{enumerate}[(i)]   
  \item If the mixed correlator $\langle{\mv}f\rangle$ is absent,   
    the odd structure functions vanish, while the contributions to   
    even functions are given only by the operators with even values   
    of $N'$. In the isotropic case ($\alpha_{1,2}=0$) only   
    the contributions with $p=0$ survive; see \cite{AAV98,AAHN00}.   
    In the presence of the anisotropy, $\alpha_{1,2}\ne0$, the   
    operators with $p\ne0$ acquire nonzero mean values, and their   
    dimensions $\Delta[N',p]$ also appear on the right hand side of   
    Eq. (\ref{eq:str_struc2}).   
  \item  
     The leading term of the small $r/L$ behavior is obviously   
     given by the contribution with the minimal possible value of   
     $\Delta[N',p]$. Now we recall the hierarchy relations   
     (\ref{eq:str_hier1}), (\ref{eq:str_hier2}), which hold for $\alpha_{1,2}=0$   
     and therefore remain valid at least for $\alpha_{1,2} \ll 1$. This means that,   
     if the anisotropy is weak enough, the leading term in Eq.   
     (\ref{eq:str_struc2}) is given by the dimension $\Delta[N,0]$ for any $S_{N}$.   
     For all the special cases studied in Sec. \ref{subsubsec:str_Operators},   
     this hierarchy persists also for finite values of the anisotropy   
     parameters, and the contribution with $\Delta[N,0]$ remains the   
     leading one for such $N$ and $\alpha_{1,2}$.   
  \item 
     Of course, it is not impossible that the inequalities   
     (\ref{eq:str_hier1}), (\ref{eq:str_hier2}) break down for some values of   
     $n$, $d$ and $\alpha_{1,2}$, and the leading contribution to Eq.   
     (\ref{eq:str_struc2}) is determined by a dimension with $N'\ne N$ and/or $p>0$.   
\end{enumerate}

Furthermore, it is not impossible that the matrix (\ref{eq:str_Dnp})   
for some $\alpha_{1,2}$ had a pair of complex conjugate eigenvalues,   
$\Delta$ and $\Delta^{*}$. Then the small $r/L$ behavior of the   
scaling function $\xi(r/L)$ entering into Eq. (\ref{eq:str_struc2})   
would involve oscillating terms of the form   
$$(r/L)^{{\rm Re}\, \Delta}   
\Bigl\{ C_{1} \cos \bigl[{\rm Im}\, \Delta\, (r/L)\bigr] +   
C_{2} \sin \bigl[{\rm Im}\, \Delta\, (r/L)\bigr] \Bigr\}, $$   
with some constants $C_{i}$.   
   
Another exotic situation emerges if the matrix (\ref{eq:str_Dnp})   
cannot be diagonalized and is only reduced to the Jordan   
form. In this case, the corresponding contribution to the scaling   
function would involve a logarithmic correction to  the powerlike   
behavior, $\, (r/L)^{\Delta}\, \bigl[C_{1}\ln (r/L)+C_{2}\bigr]$,   
where $\Delta$ is the eigenvalue related to the Jordan cell.   
However, these interesting hypothetical possibilities are not   
actually realized for the special cases studied above in   
Sec. \ref{subsubsec:str_Operators}.   
   
(iv) The inclusion of the mixed correlator   
$\langle{\mv}f\rangle\propto\mn \delta(t-t')\, C'(r/L)$ violates   
the evenness in $\mn$ and gives rise to non-vanishing odd functions   
$S_{2n+1}$ and to the contributions with odd $N'$ to the expansion   
(\ref{eq:str_struc2}) for even functions. If the hierarchy relations   
(\ref{eq:str_hier1}),   
(\ref{eq:str_hier2})  hold, the leading term for the even functions will still   
be given by the contribution with $\Delta[N,0]$. If the relations   
(\ref{eq:str_hier3}) hold, the leading term for the odd function $S_{2n+1}$   
will be given by the dimension $\Delta[2n,0]$ for $n<(d+2)/4$ and by   
$\Delta[2n+1,1]$ for $n>(d+2)/4$. Note that for the model with an imposed   
gradient, the leading terms for $S_{2n+1}$ are given by the dimensions   
$\Delta[2n+1,1]$ for all $n$; see \cite{Ant99,Ant00}. 
   
Representations similar to Eqs. (\ref{eq:str_100}), (\ref{eq:str_struc2}) can easily   
be written down for any equal-time pair correlator, provided its canonical   
and critical dimensions are known. In particular, for the operators   
$F[N,p]$ in the IR region ($\Lambda r \to\infty$, $r/L$ fixed)   
one obtains   
\begin{equation}   
\langle F[N_{1},p_{1}] F[N_{2},p_{2}] \rangle  = \nu_0 ^{-(N_{1}+N_{2})/2}   
\sum_{N,p}  \sum_{N',p'} (\Lambda r)^{-\Delta_{[N,p]}-\Delta_{[N',p']}}   
  R_{N,p;N',p'}(r/L),   
\label{eq:str_102}   
\end{equation}   
where the summation indices $N$, $N'$ satisfy the inequalities   
$N\le N_{1}$, $N'\le N_{2}$, and the indices $p$, $p'$ take on   
all possible values allowed for given $N$, $N'$. The small $r/L$   
behavior of the scaling functions $R_{N,p;N',p'}(r/L)$ has the   
form   
\begin{equation}   
\xi_{N,p; N',p'}(mr) = \sum_{N'',p''}\, C_{N'',p''}\, (r/L)^{   
\Delta[N'',p'']},   
\label{eq:str_103}   
\end{equation}   
with the restriction $N''\le N+N'$ and corresponding values of   
$p''$; $C_{N'',p''}$ are some numerical coefficients.   
   
So far, we have discussed the special case of the velocity correlator   
given by Eqs. (\ref{eq:str_3}) and (\ref{eq:str_T34}). From the explicit calculations \cite{AAHN00}
 follows that only even polynomials in the   
expansion (\ref{eq:str_Legendre}) can give contributions to the   
renormalization constants, and consequently, to the coordinates of the   
fixed point and the anomalous dimensions. For this reason,   
the odd polynomials were omitted in Eq. (\ref{eq:str_Legendre})   
from the very beginning. Moreover, it is clear from Eq.   
(\ref{eq:str_sigma3}) that only the coefficients $a_{l}$ with $l=0,1$ and   
$b_{l}$ with $l=0,1,2$ contribute to the constants $Z_{1,2}$ in Eq.   
(\ref{eq:str_renormalized}) and therefore to the basic RG functions   
(\ref{eq:str_RGF2}) and (\ref{eq:str_RGF3}) to the coordinates of the fixed point in Eq.   
(\ref{eq:str_FP}). Therefore, the fixed point in the general model (\ref{eq:str_T})   
is parametrized completely by these five coefficients; the higher   
coefficients enter only via the positivity conditions (\ref{eq:str_positiv}).   
   
Furthermore, for $\chi=0$,   
only coefficients $a_{l}$ with $l\le2$ and $b_{l}$ with $l\le3$   
can contribute to the integrals $H_{n}$ and, consequently, to the   
one-loop critical dimensions (\ref{eq:str_Dnp}). Therefore, the calculation   
of the latter essentially simplifies for the special case   
$a_{0}=1$, $a_{1}=0$ and $b_{l}=0$ for $l\le2$ in Eq. (\ref{eq:str_T}).   
Then the coordinates of the fixed point (\ref{eq:str_FP}) are the same   
as in the isotropic model, in particular, $\alpha_{*}=0$, and the   
anomalous exponents will depend on the only two parameters   
$a_{2}$ and $b_{3}$. We have performed a few sample calculations   
for this situation; the results are presented in Figs. 6--9 for   
$\Delta[n,p]$ with $n=3,4,5,6$ {\it vs} $a_{2}$ for $b_{3}=0$,   
{\it vs} $a_{2}=b_{3}$, and {\it vs} $b_{3}$ for $a_{2}=0$.   
In all cases studied, the general picture has   
appeared similar to that outlined above for the case (\ref{eq:str_T34}).   
In particular, the hierarchy of the critical dimensions, expressed   
by the inequalities (\ref{eq:str_hier1})-(\ref{eq:str_hier3}), persists also for this case.   
We may conclude that the special case (\ref{eq:str_T34}) case represents   
nicely all the main features of the general model (\ref{eq:str_T}).   
   
\begin{figure}   
 \centering
 \includegraphics[width=6cm]{\PICS 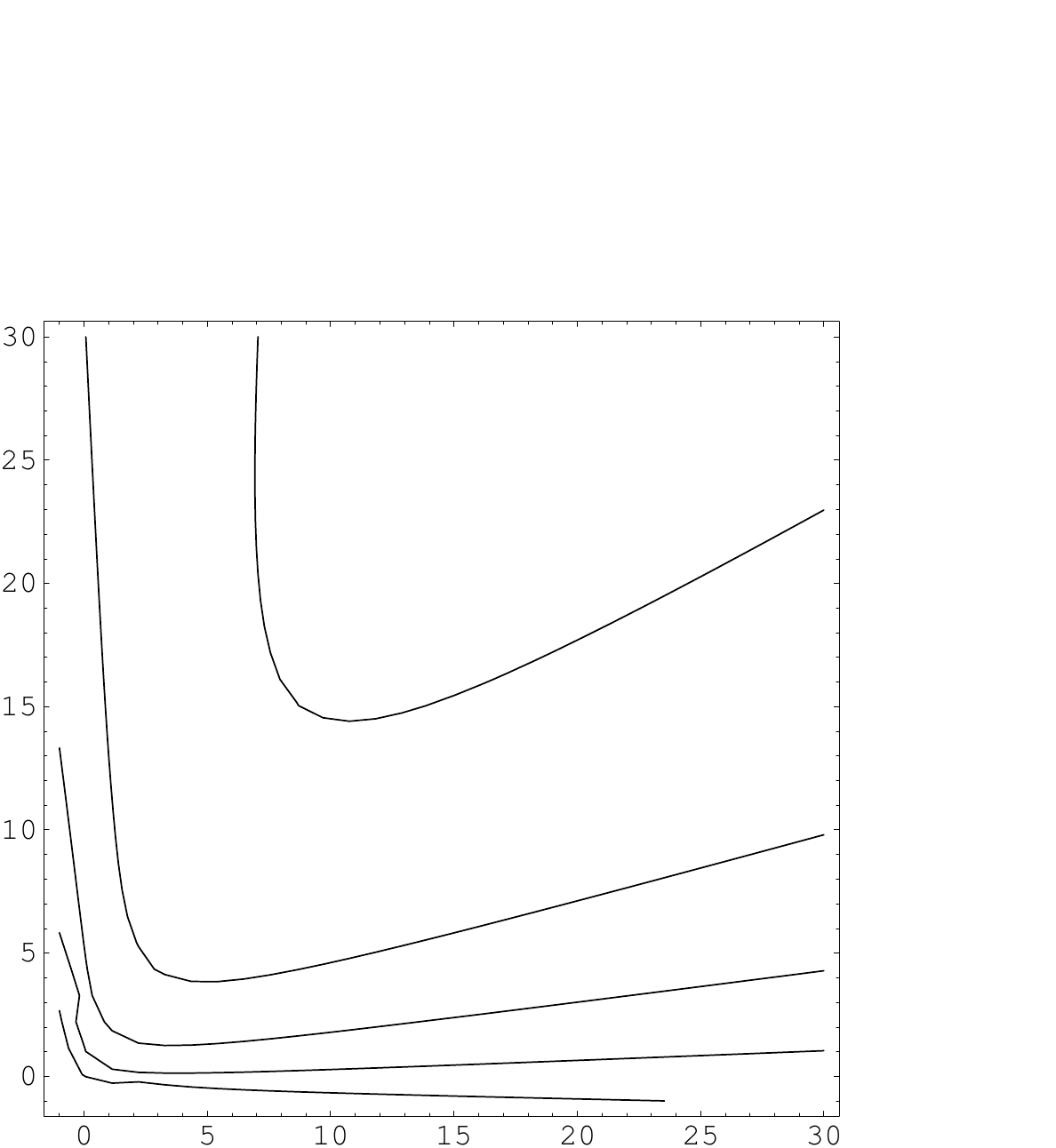} 
 \vspace{1mm}   
\caption{Levels of the dimension $\Delta[2,2]$ for $d=3$ on the plane   
$\alpha_1$--$\alpha_2$. Value changes from~$1.15$ (left-bottom)   
to~$1.4$ (right-top) with step~$0.05$.}  
\label{fig:str_fig1}
\end{figure}

\begin{figure}   
 \centering
\includegraphics[width=4.25cm]{\PICS 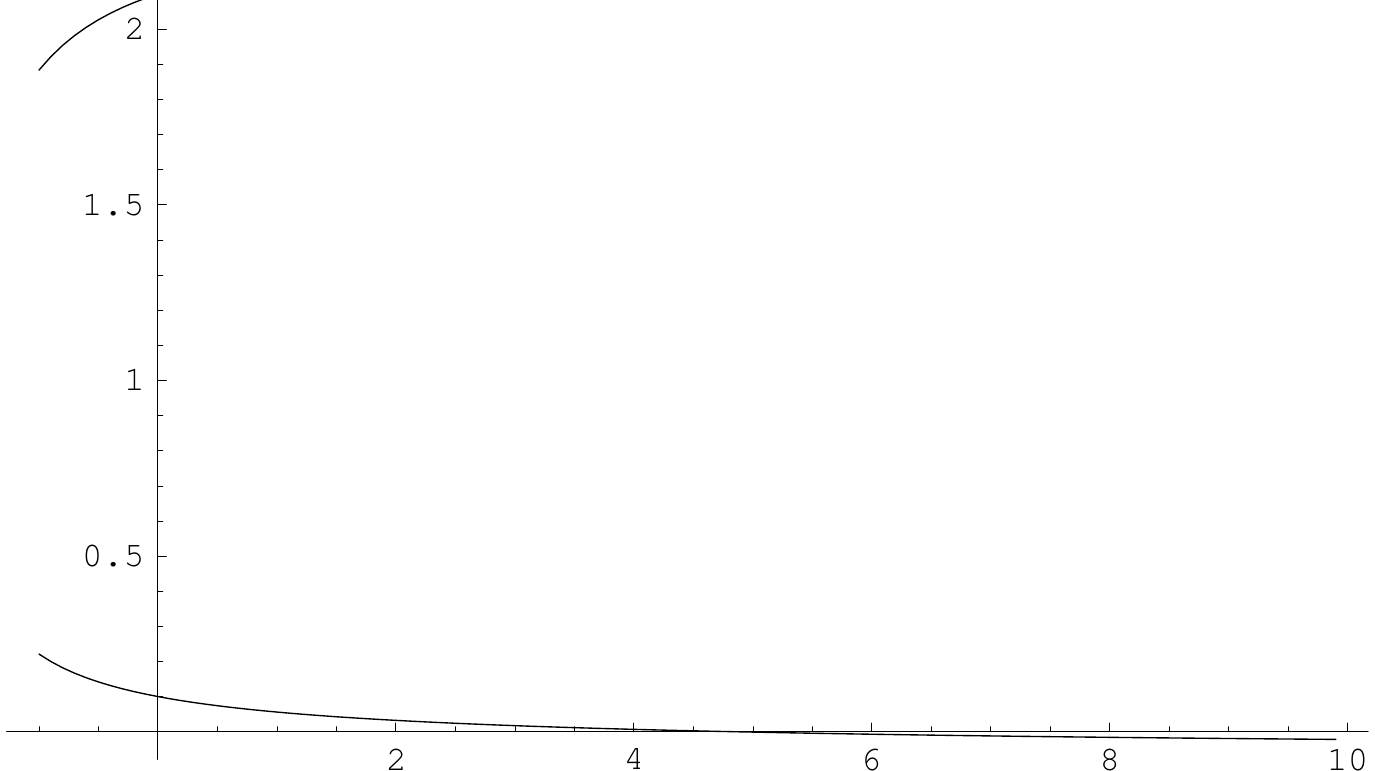}\hfill
\includegraphics[width=4.25cm]{\PICS 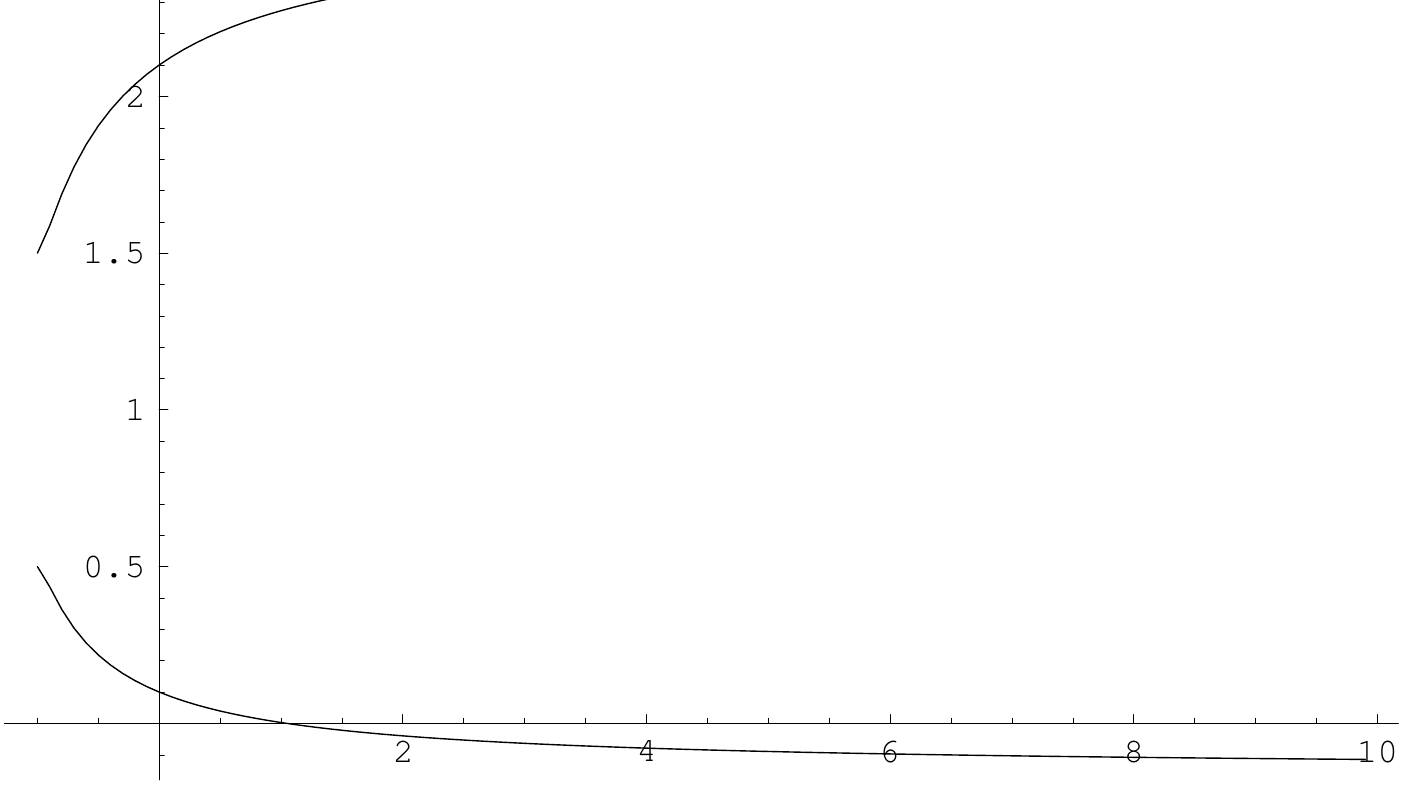}\hfill
\includegraphics[width=4.25cm]{\PICS 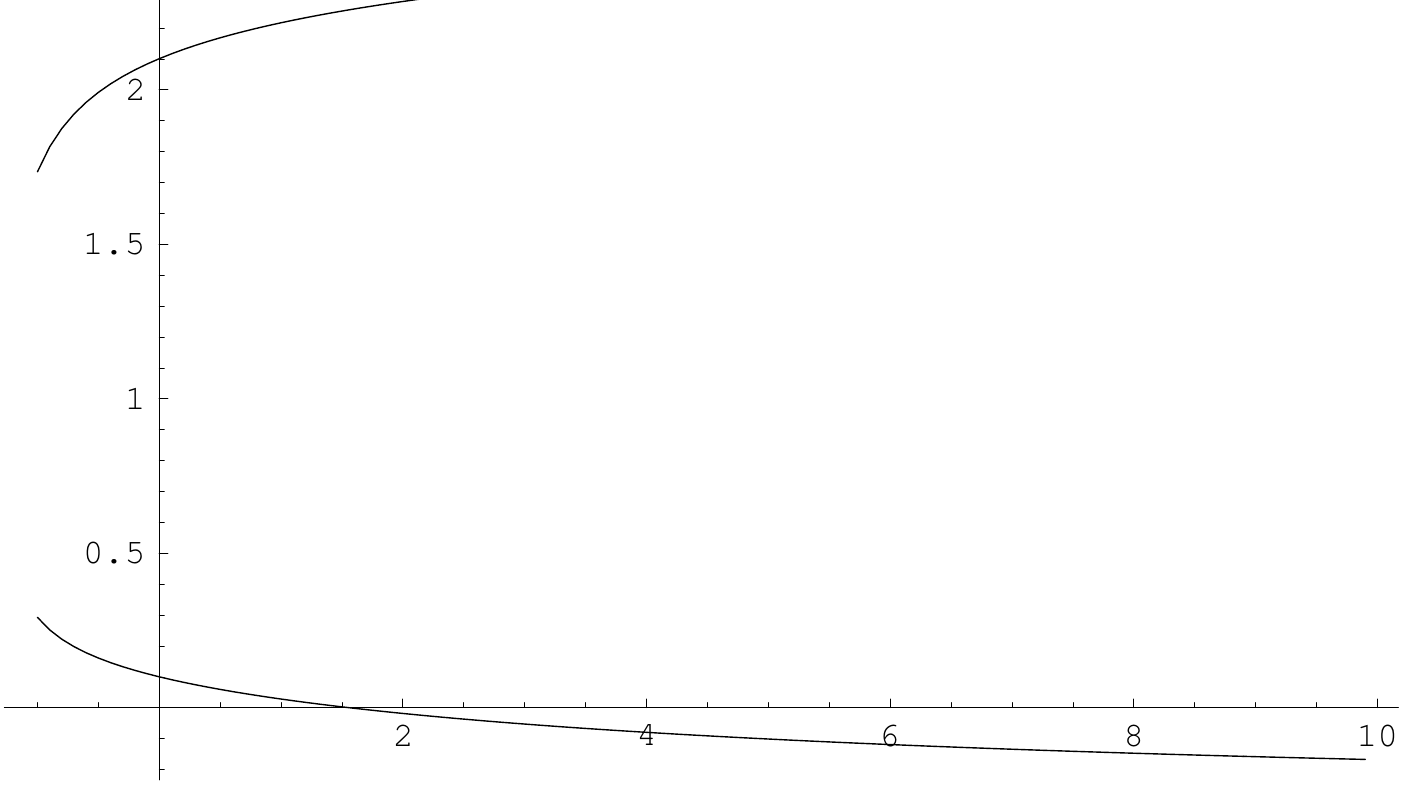}
\vspace{1mm}   
\caption{Behavior of the critical dimension   
$\Delta[3,p]$ for $d=3$ with $p=1,3$ (from below to above) {\it   
vs} $\alpha_1$ for $\alpha_2=0$---{\it left}, {\it vs}   
$\alpha\equiv\alpha_1=\alpha_2$---{\it center},{\it vs} $\alpha_2$ for   
$\alpha_1=0$---{\it right}.}   
\label{fig:str_fig2}
\end{figure}

\begin{figure}   
\centering
\includegraphics[width=4cm]{\PICS 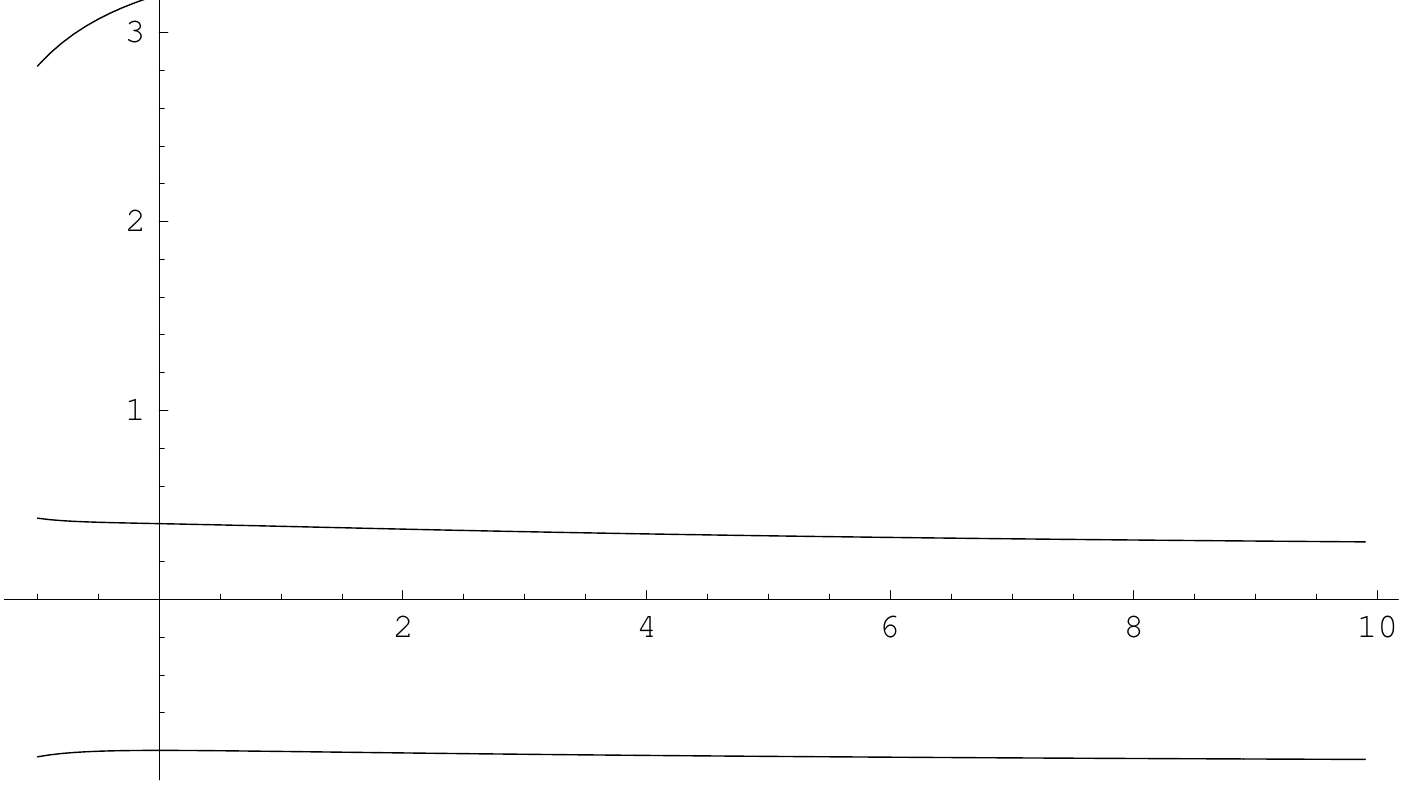}\hfill
\includegraphics[width=4cm]{\PICS 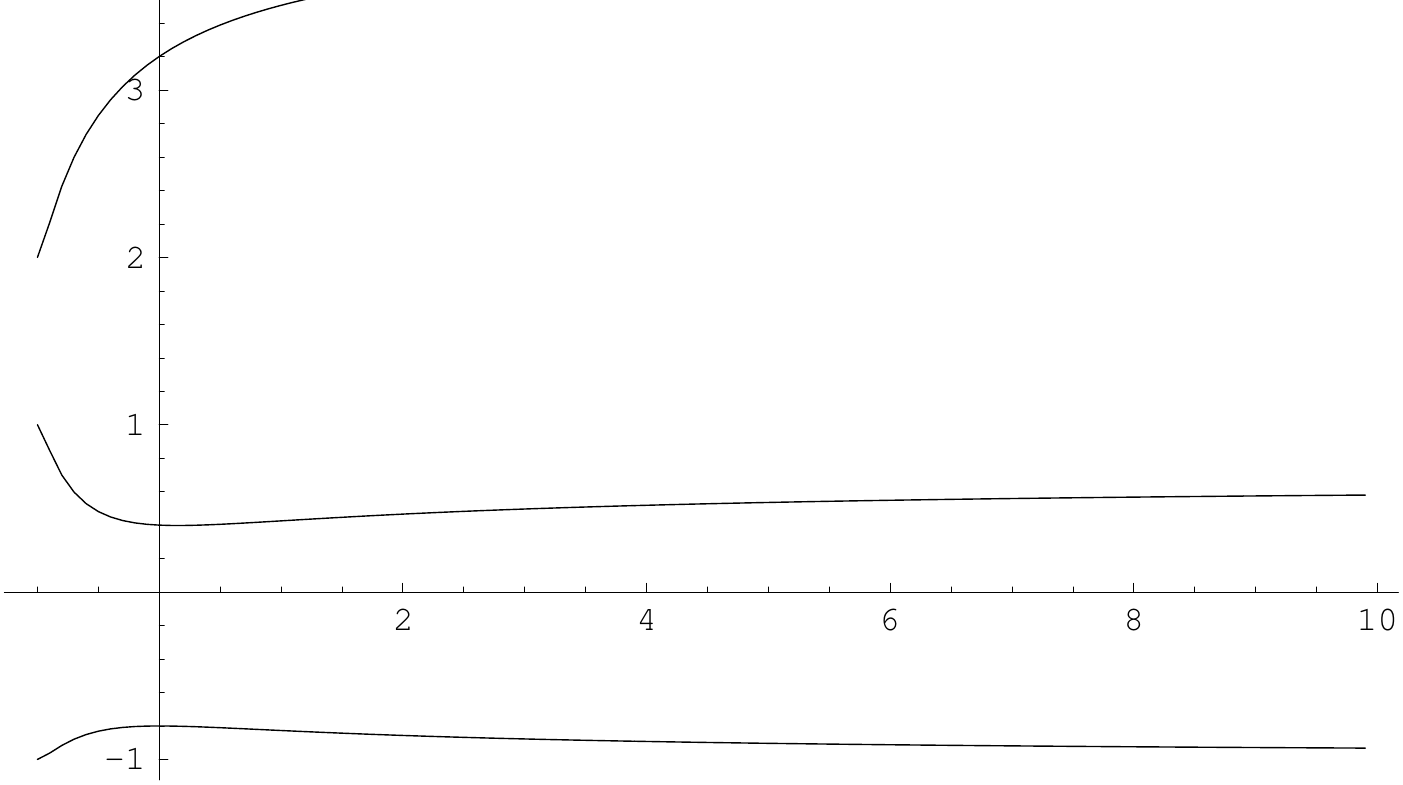}\hfill
\includegraphics[width=4cm]{\PICS 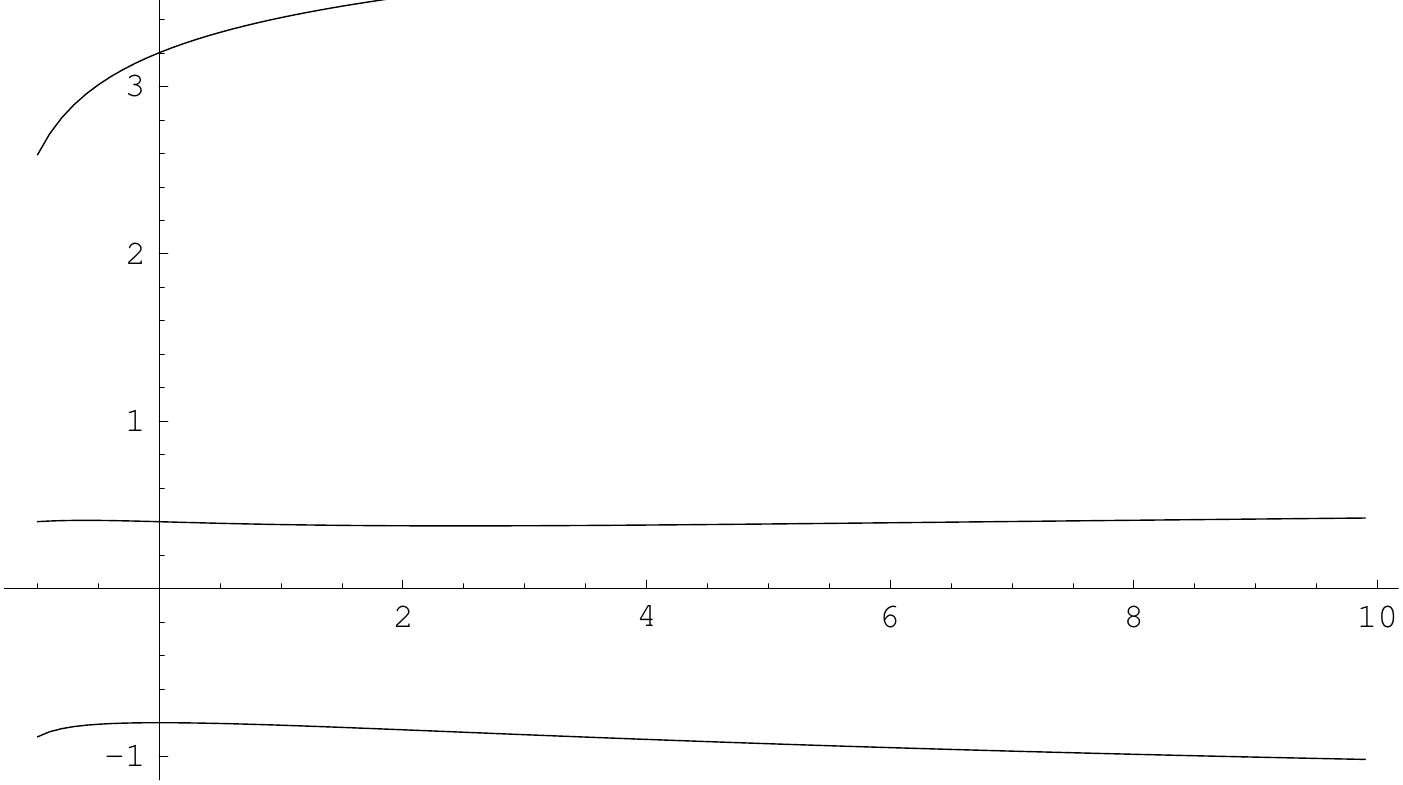}

\vspace{1mm}   
\caption{Behavior of the critical dimension   
$\Delta[4,p]$ for $d=3$ with $p=0,2,4$ (from below to above) {\it   
vs} $\alpha_1$ for $\alpha_2=0$---{\it left}, {\it vs}   
$\alpha\equiv\alpha_1=\alpha_2$---{\it center},{\it vs} $\alpha_2$ for   
$\alpha_1=0$---{\it right}.}   
\label{fig:str_fig3}
\end{figure}

\begin{figure}   
\centering
\includegraphics[width=4cm]{\PICS 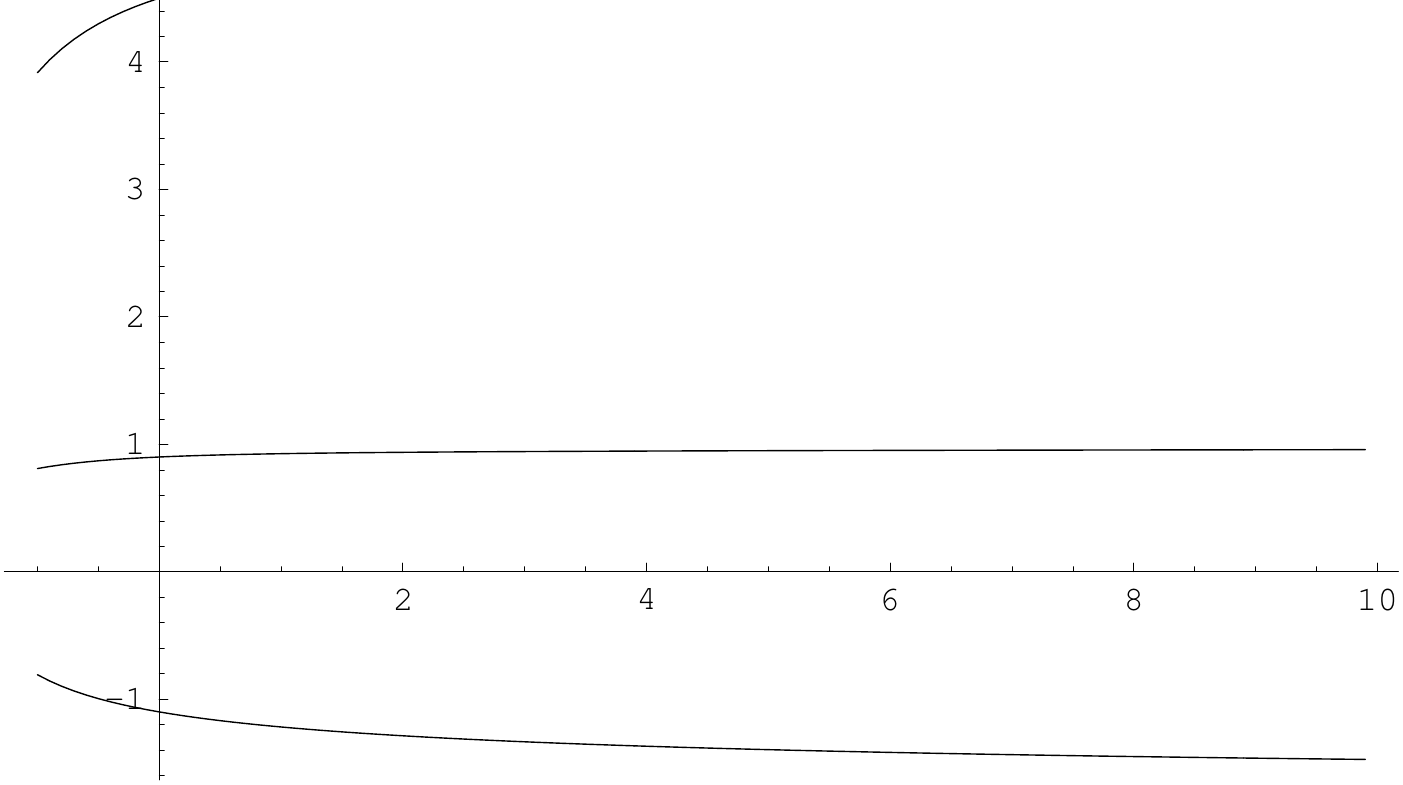}\hfill
\includegraphics[width=4cm]{\PICS 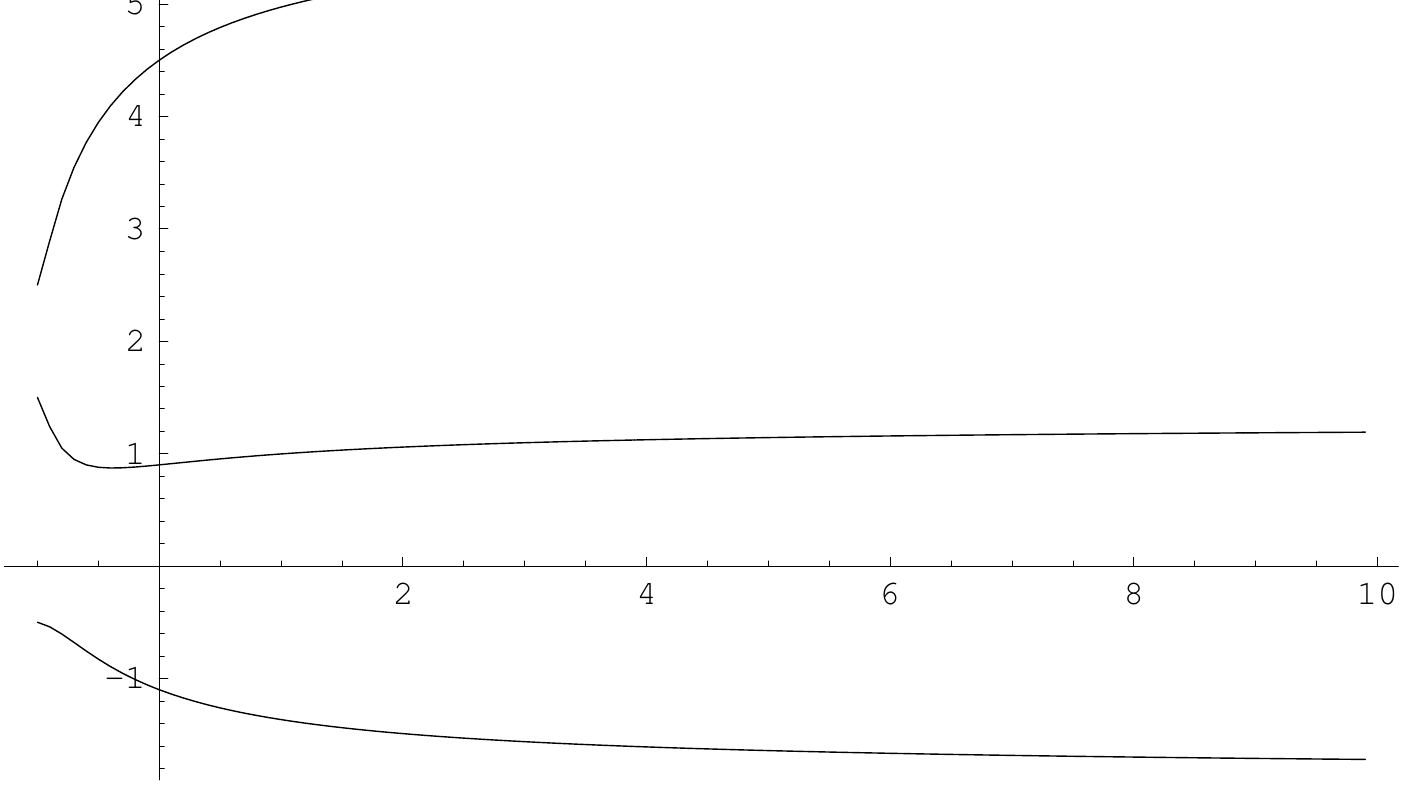}\hfill
\includegraphics[width=4cm]{\PICS 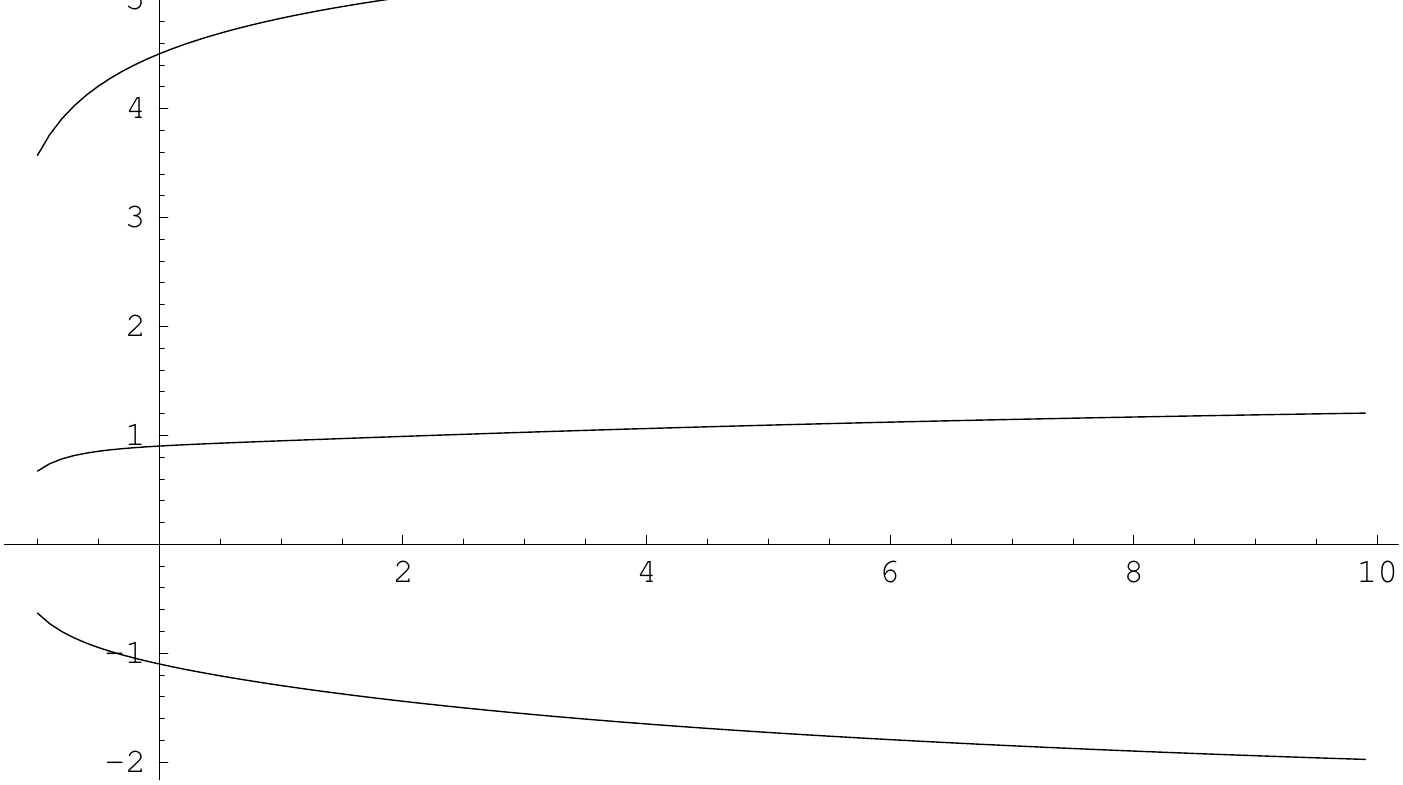}   

\vspace{1mm}   
\caption{Behavior of the critical dimension   
$\Delta[5,p]$ for $d=3$ with $p=1,3,5$ (from below to above) {\it   
vs} $\alpha_1$ for $\alpha_2=0$---{\it left}, {\it vs}   
$\alpha\equiv\alpha_1=\alpha_2$---{\it center},{\it vs} $\alpha_2$ for   
$\alpha_1=0$---{\it right}.}   
\label{fig:str_fig4}
\end{figure}

\begin{figure}   
\centering
\includegraphics[width=4cm]{\PICS 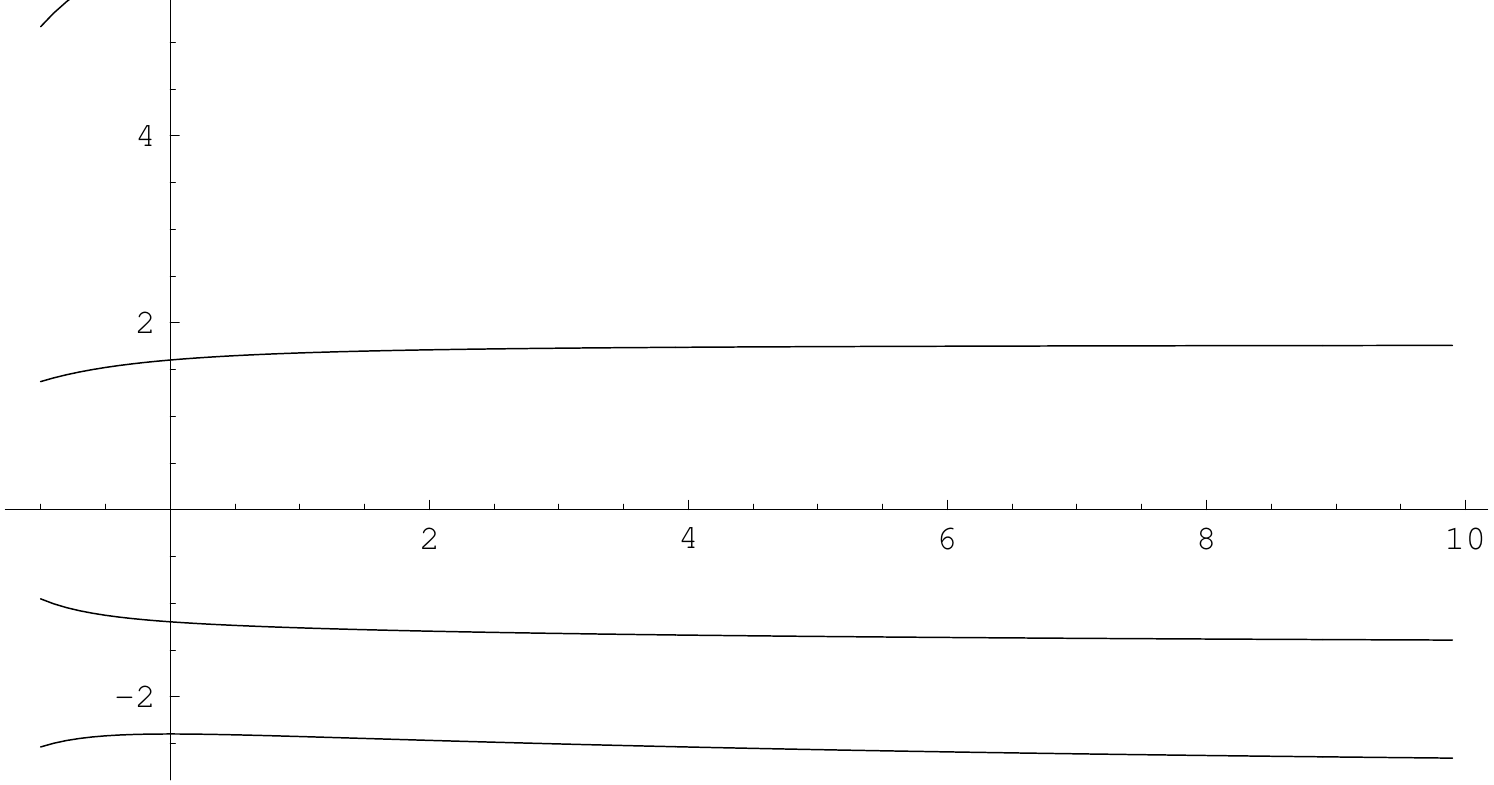}\hfill
\includegraphics[width=4cm]{\PICS 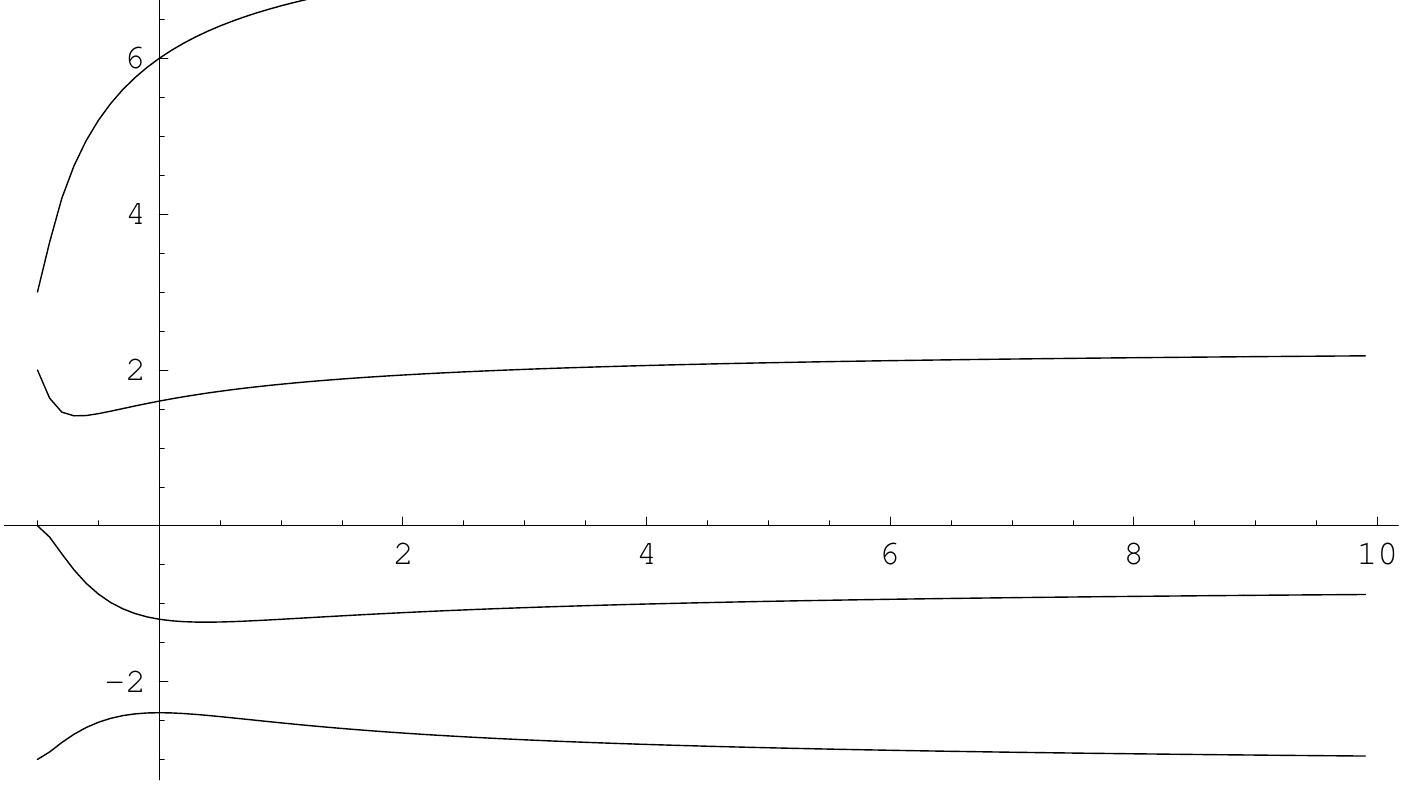}\hfill
\includegraphics[width=4cm]{\PICS 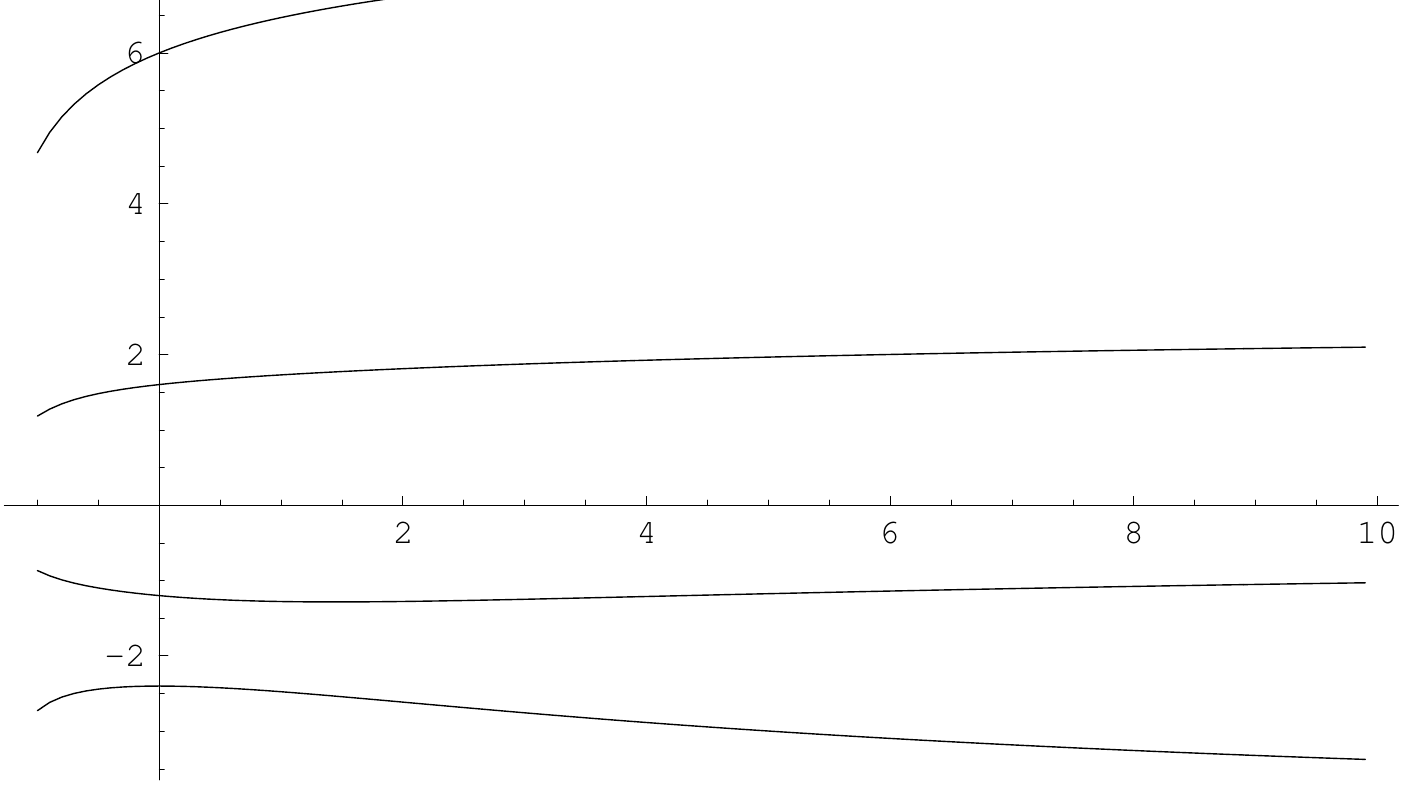}

\vspace{1mm}   
\caption{Behavior of the critical dimension   
$\Delta[6,p]$ for $d=3$ with $p=0,2,4,6$ (from below to above)   
{\it vs} $\alpha_1$ for $\alpha_2=0$---{\it left}, {\it vs}   
$\alpha\equiv\alpha_1=\alpha_2$---{\it center},{\it vs} $\alpha_2$ for   
$\alpha_1=0$---{\it right}.}   
\label{fig:str_fig5}
\end{figure}

\begin{figure}   
\centering
\includegraphics[width=4cm]{\PICS 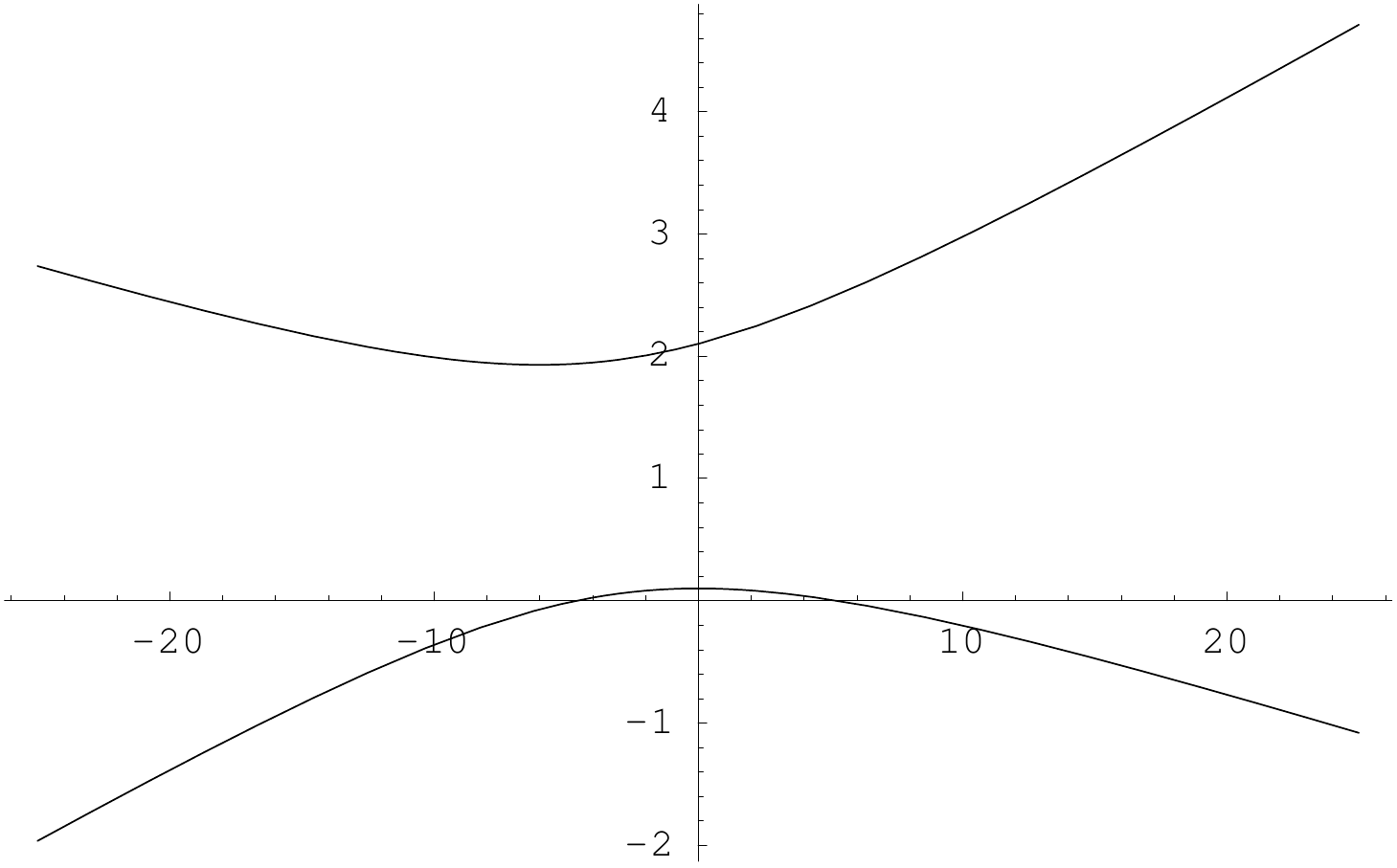}\hfill
\includegraphics[width=4cm]{\PICS 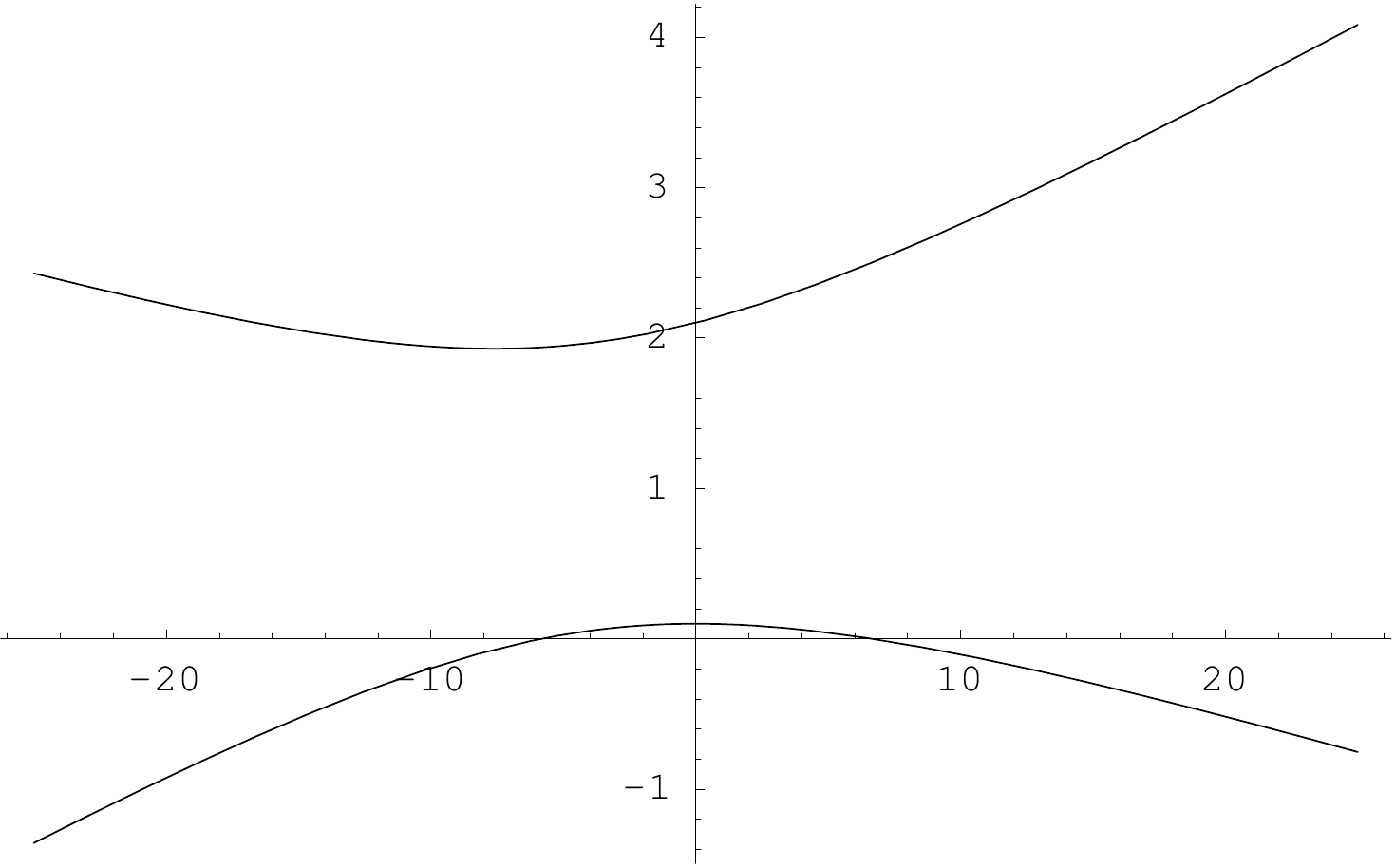}\hfill
\includegraphics[width=4cm]{\PICS 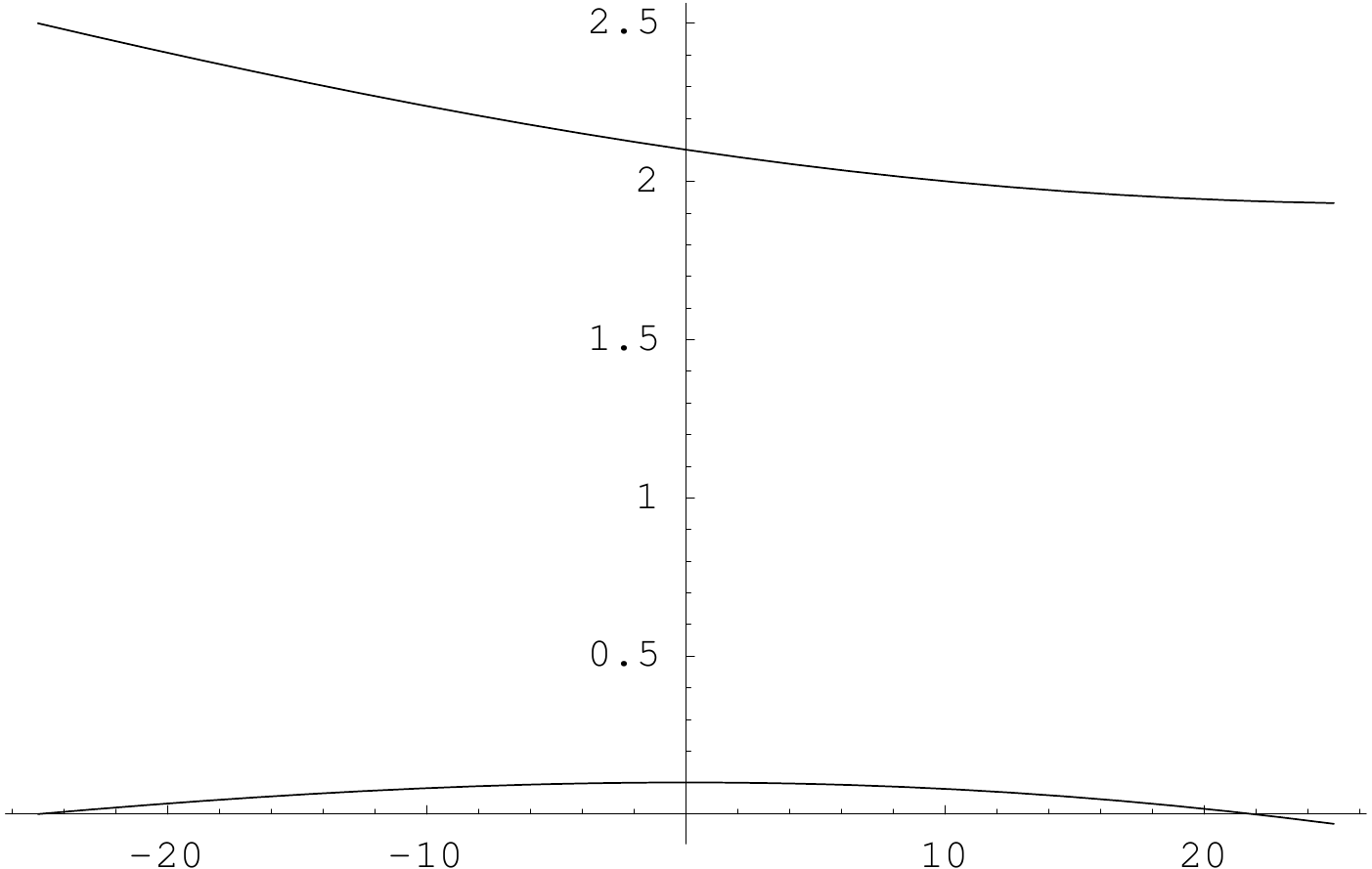}

\vspace{1mm}   
\caption{Behavior of the critical dimension   
$\Delta[3,p]$ for $d=3$ with $p=1,3$ (from below to above) {\it   
vs} $a_2$ for $b_3=0$---{\it left}, {\it vs} $a_2=b_3$---{\it   
center},{\it vs} $b_3$ for $a_2=0$---{\it right}.}   
\label{fig:str_fig6}
\end{figure}

\begin{figure}  
\centering
\includegraphics[width=4cm]{\PICS 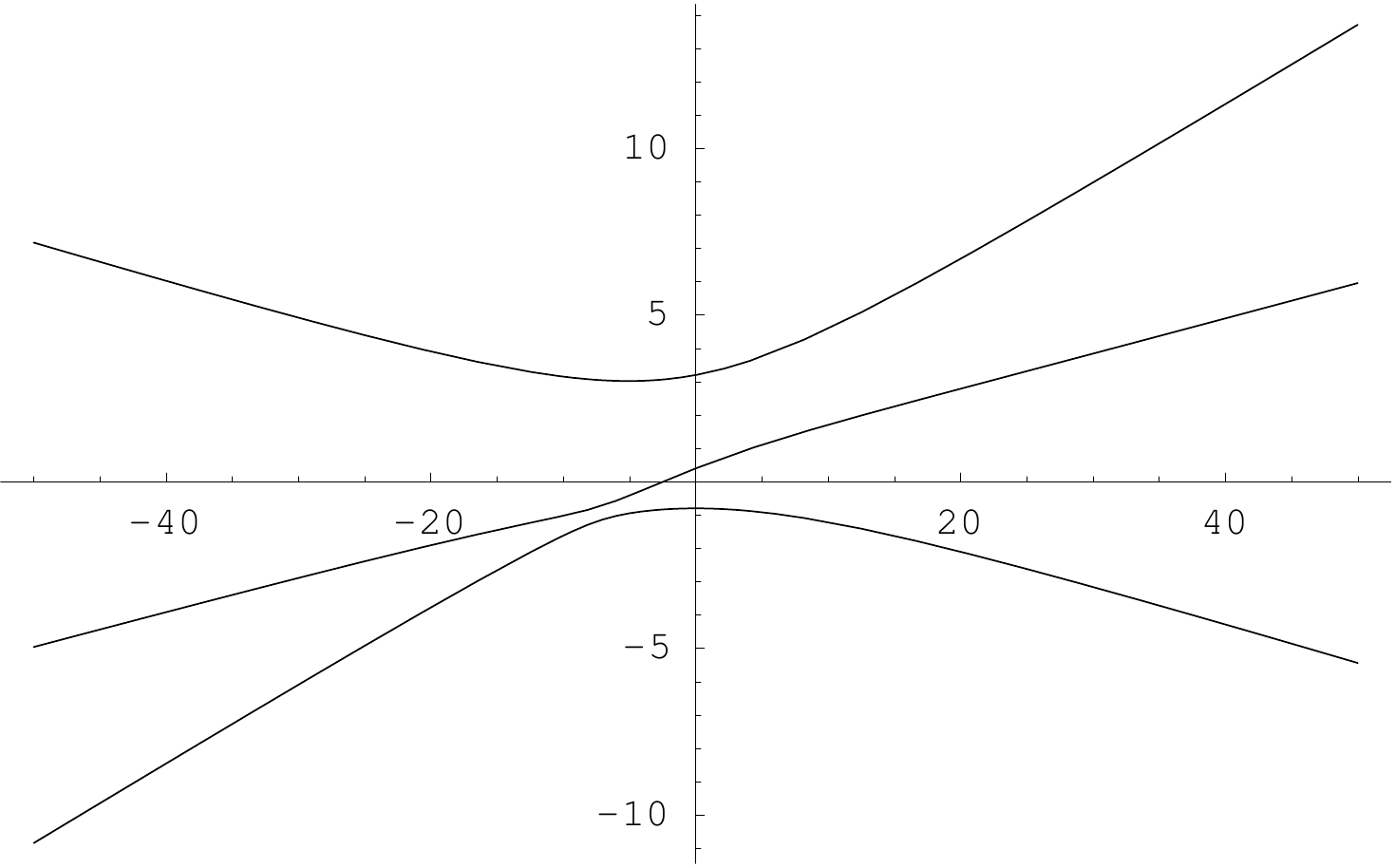}\hfill
\includegraphics[width=4cm]{\PICS 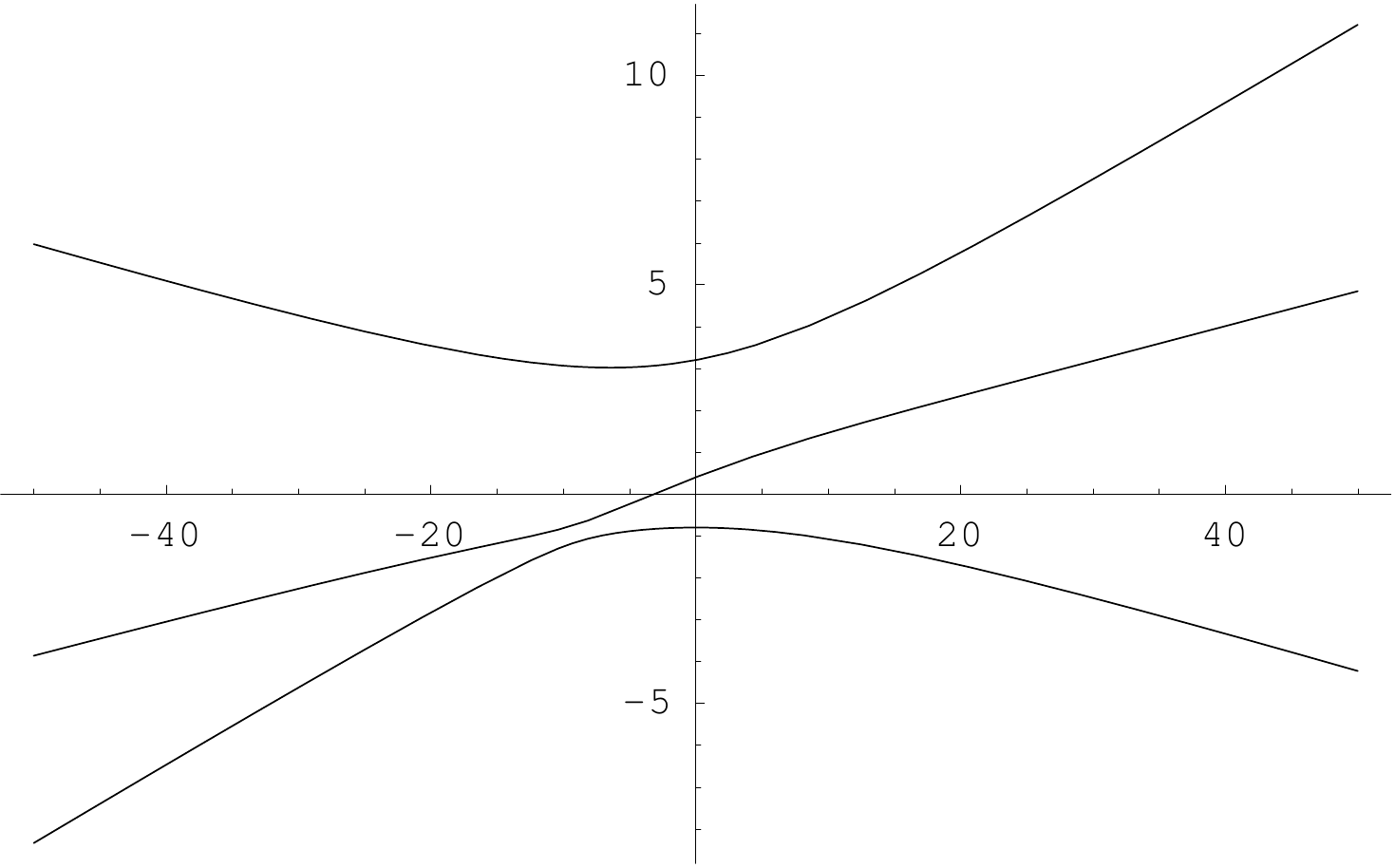}\hfill
\includegraphics[width=4cm]{\PICS 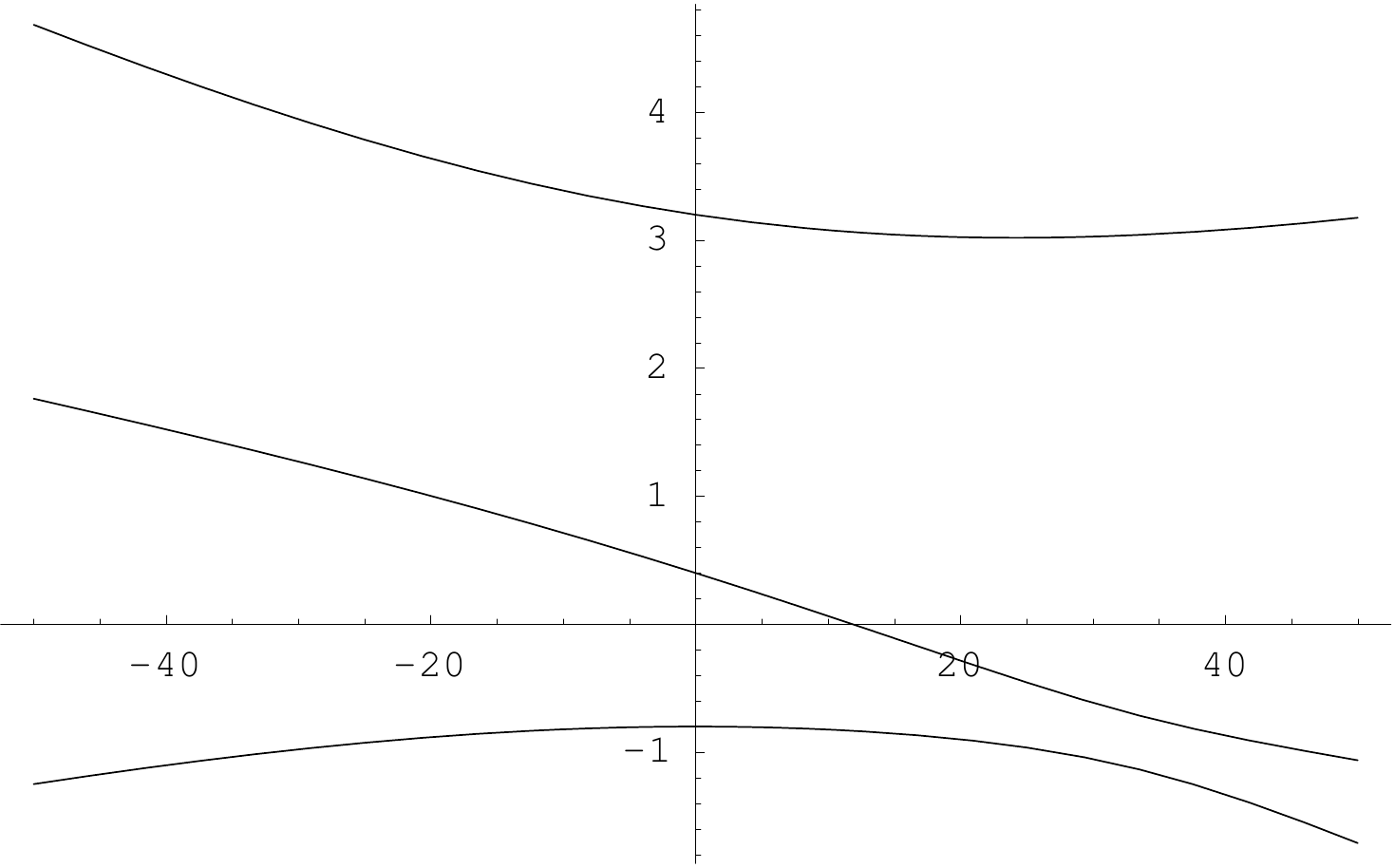}  

\vspace{1mm}   
\caption{Behavior of the critical dimension   
$\Delta[4,p]$ for $d=3$ with $p=0,2,4$ (from below to above) {\it   
vs} $a_2$ for $b_3=0$---{\it left}, {\it vs} $a_2=b_3$---{\it   
center},{\it vs} $b_3$ for $a_2=0$---{\it right}.}   
\label{fig:str_fig7}
\end{figure}

\begin{figure}   
\centering
\includegraphics[width=4cm]{\PICS 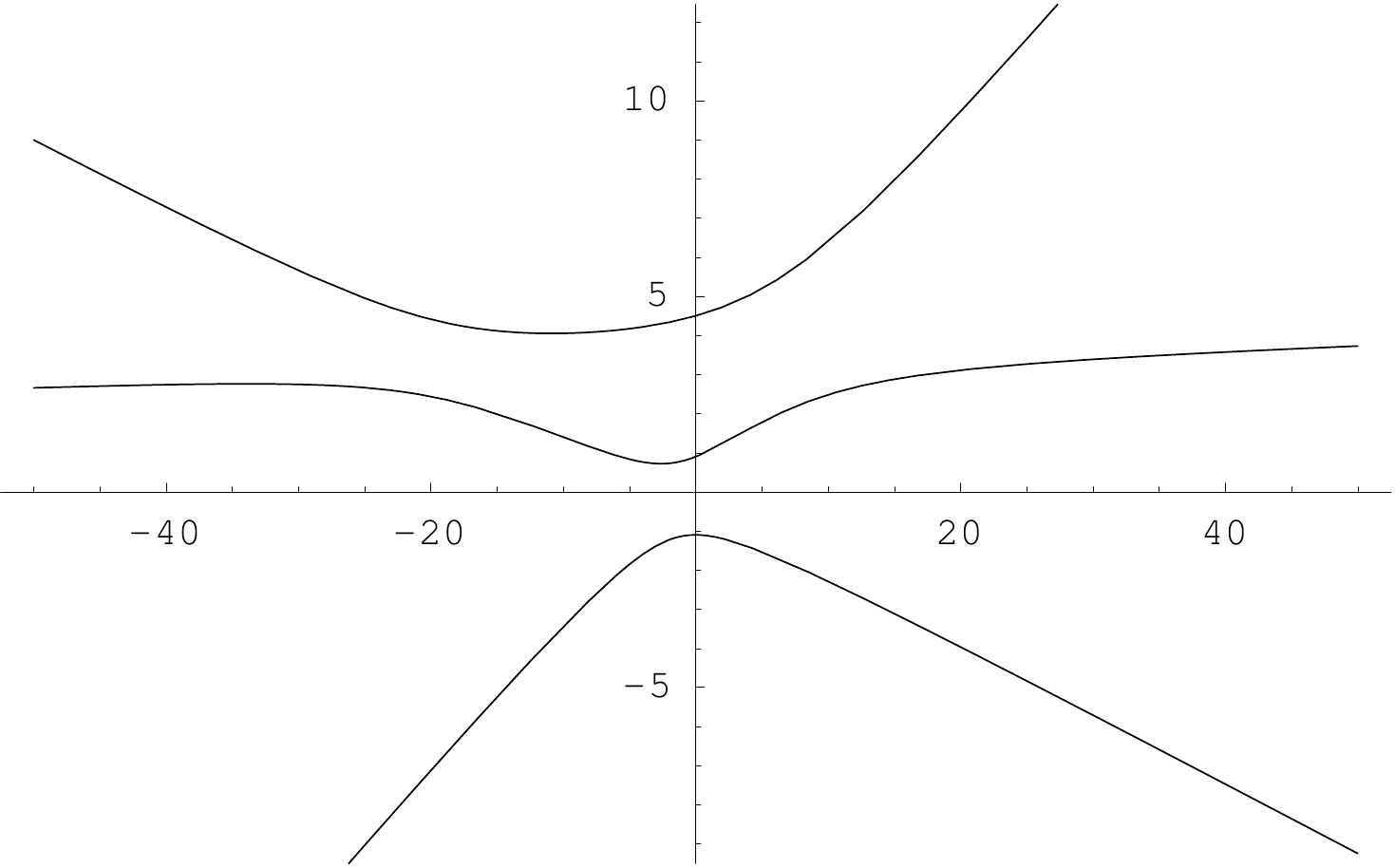}\hfill
\includegraphics[width=4cm]{\PICS 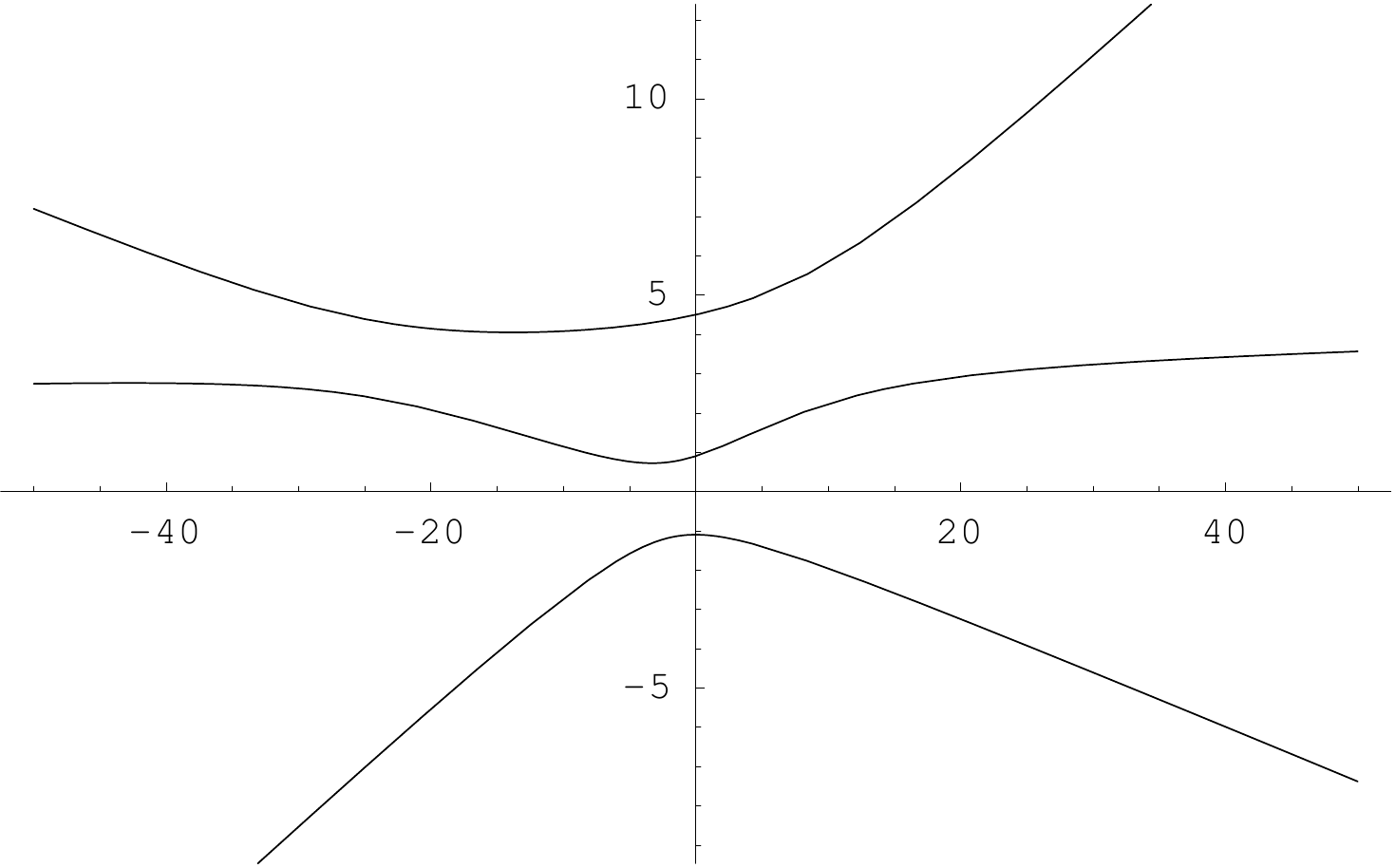}\hfill
\includegraphics[width=4cm]{\PICS 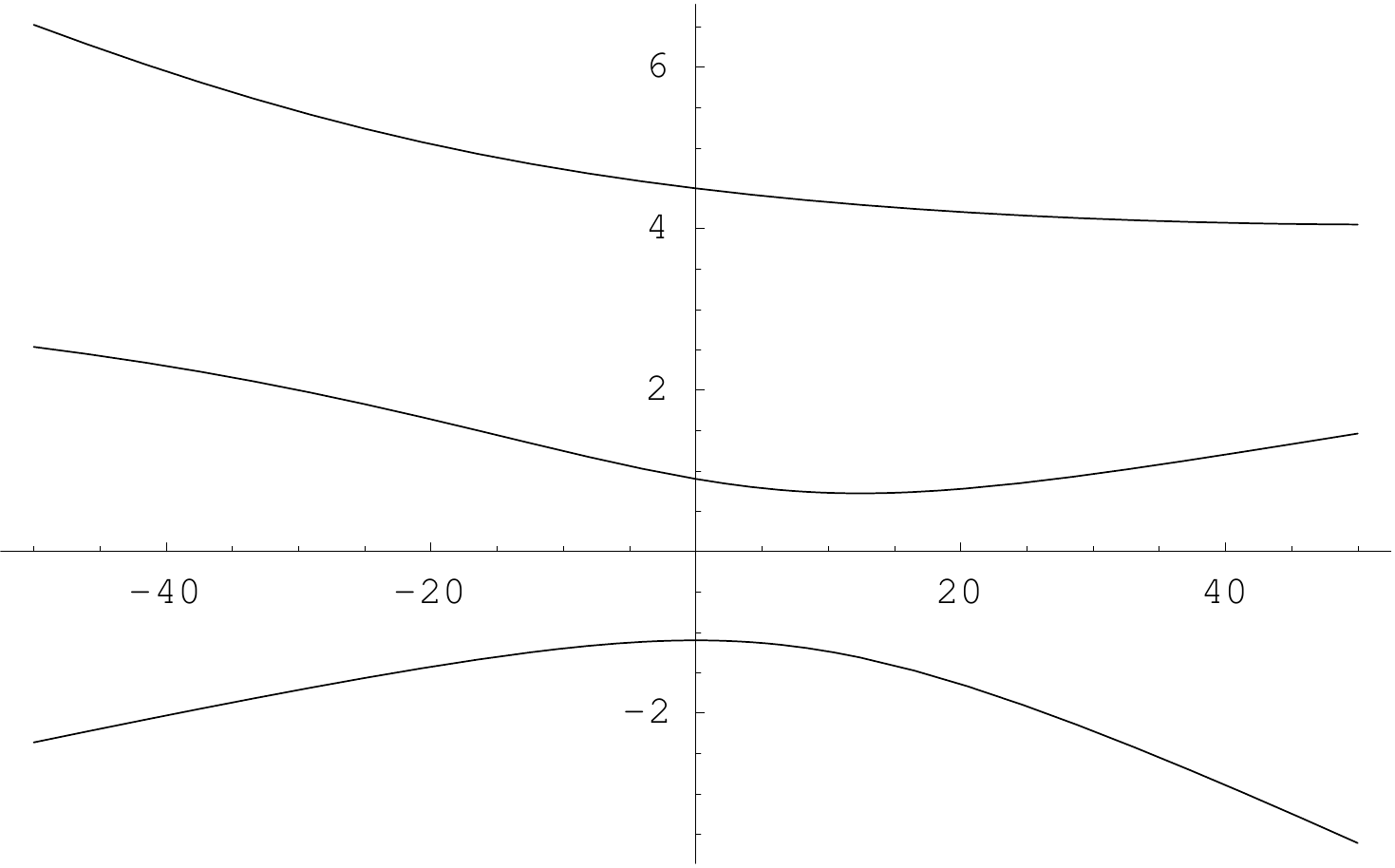}

\vspace{1mm}  
\caption{Behavior of the critical dimension  
$\Delta[5,p]$ for $d=3$ with $p=1,3,5$ (from below to above) {\it  
vs} $a_2$ for $b_3=0$---{\it left}, {\it vs} $a_2=b_3$---{\it  
center},{\it vs} $b_3$ for $a_2=0$---{\it right}.}  
\label{fig:str_fig8}
\end{figure}

\begin{figure}  
\centering
\includegraphics[width=4cm]{\PICS 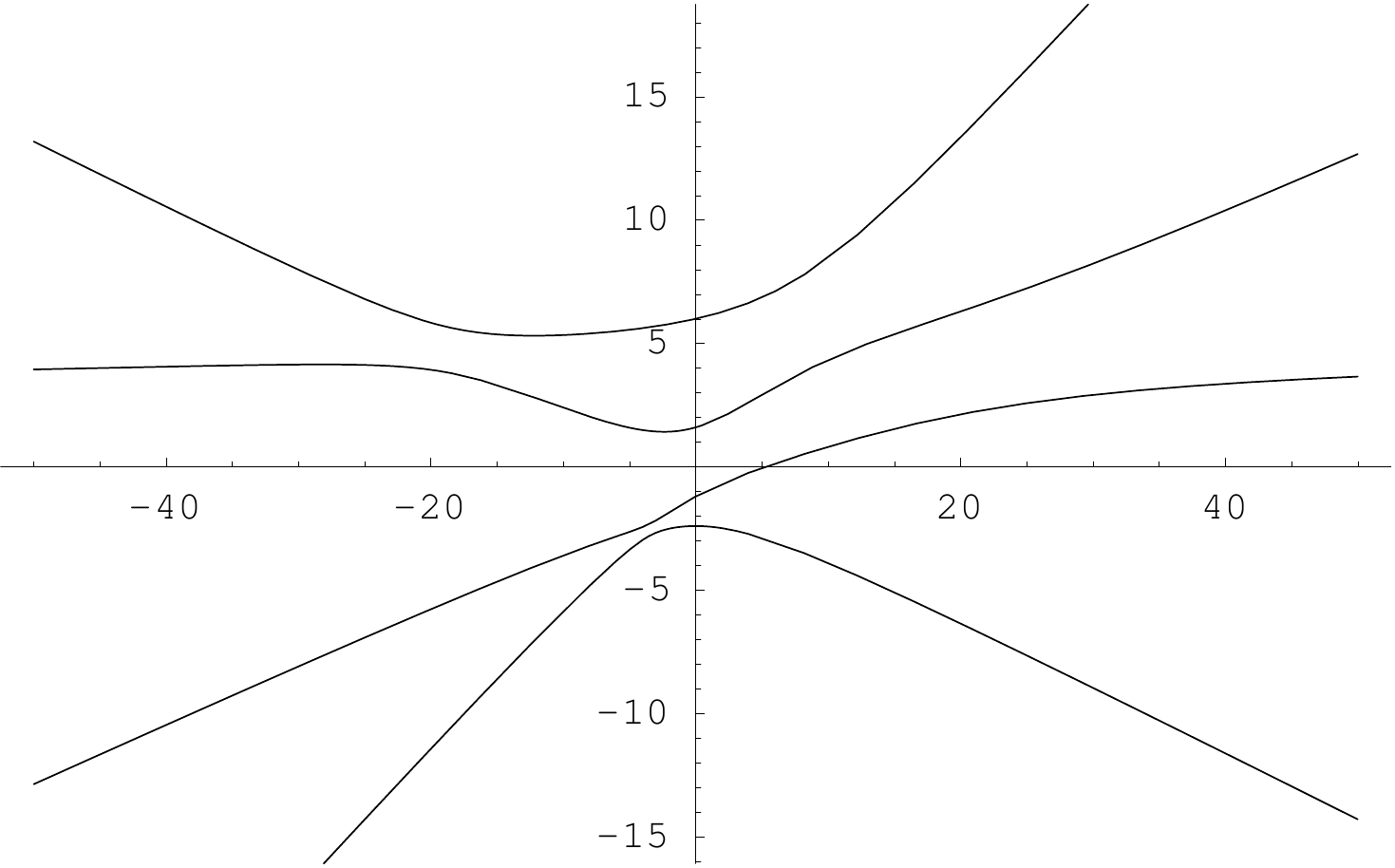}\hfill
\includegraphics[width=4cm]{\PICS 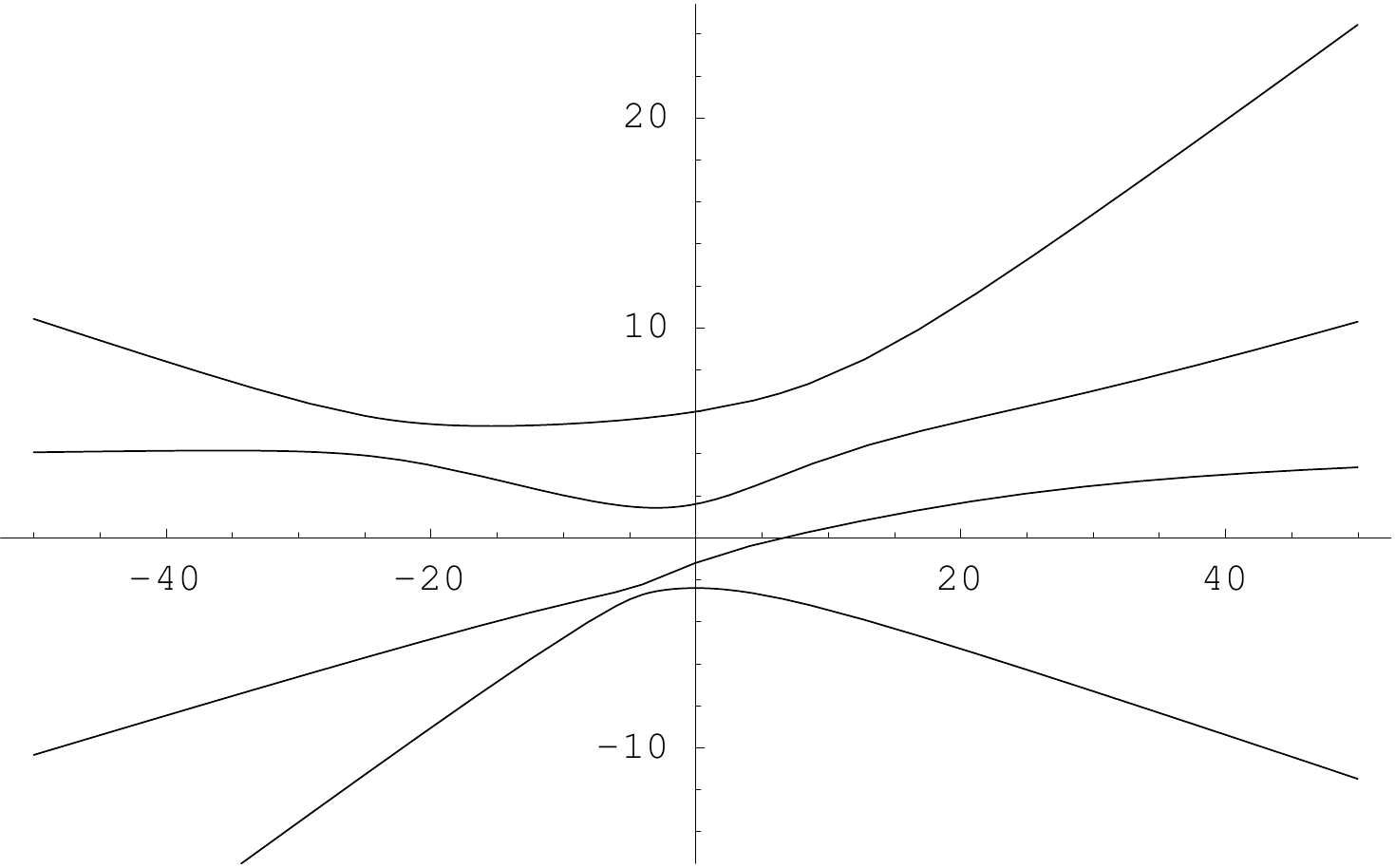}\hfill
\includegraphics[width=4cm]{\PICS 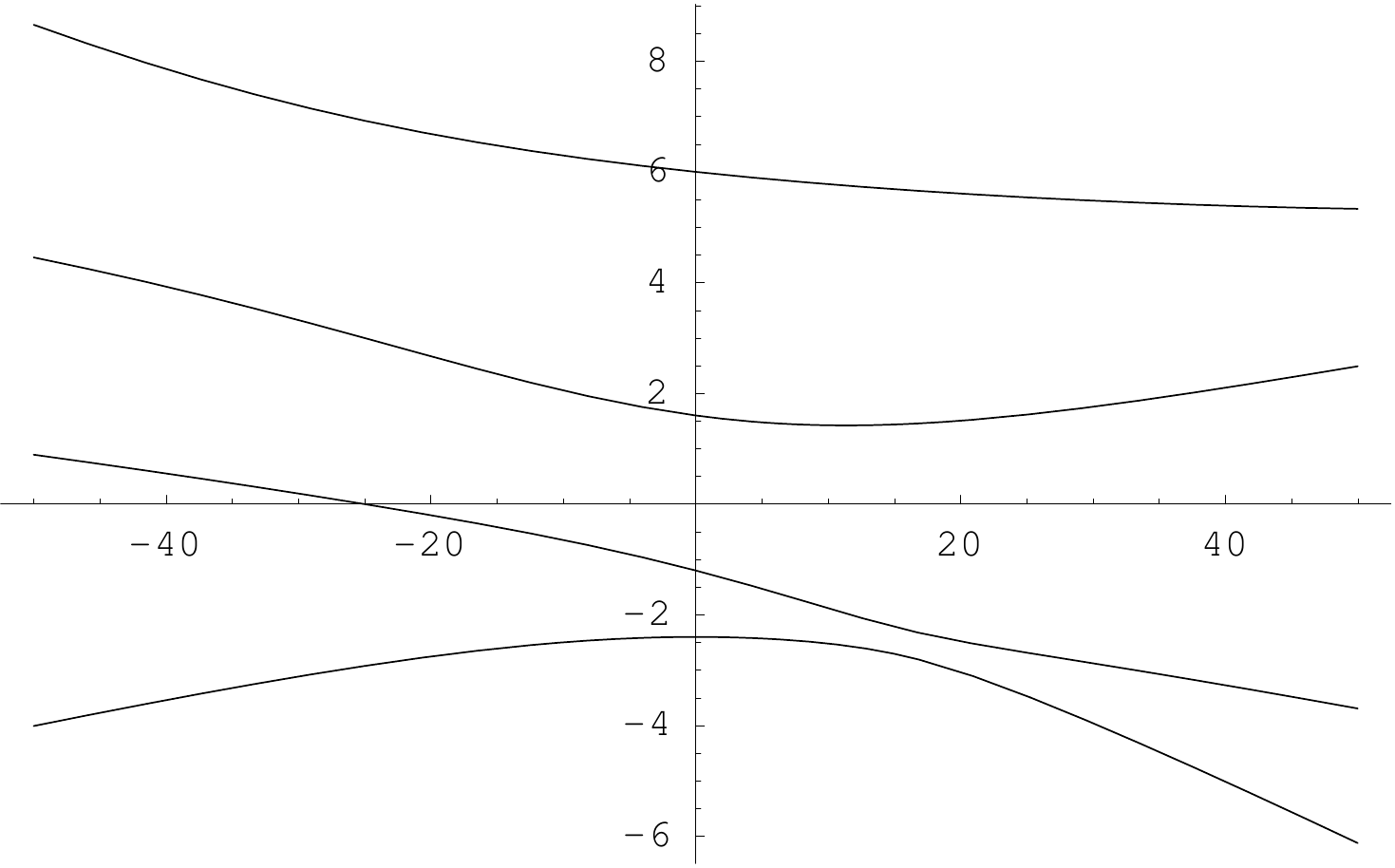}

\vspace{1mm}  
\caption{Behavior of the critical dimension  
$\Delta[6,p]$ for $d=3$ with $p=0,2,4$ (from below to above) {\it  
vs} $a_2$ for $b_3=0$---{\it left}, {\it vs} $a_2=b_3$---{\it  
center},{\it vs} $b_3$ for $a_2=0$---{\it right}.}  
\label{fig:str_fig9}
\end{figure}

The exponents are determined by the critical dimensions of composite   
operators (\ref{eq:str_Fnp}) built of the scalar gradients.   
In contrast with the isotropic velocity field, these operators in the model under consideration   
mix in renormalization such that the matrices of their critical dimensions   
are neither diagonal nor triangular. These matrices are calculated   
explicitly   
to the order $\O(\eps)$, but their eigenvalues (anomalous exponents)   
can be found explicitly only as series in $\alpha_{1,2}$   
[Eqs. (\ref{eq:str_expansion60}), (\ref{eq:str_expansion80})] or   
numerically [Figs. \ref{fig:str_fig1}--\ref{fig:str_fig9}].

In the limit of vanishing anisotropy, the exponents can be associated   
with definite tensor composite operators   
built of the scalar gradients, and exhibit a kind of hierarchy related   
to the degree of anisotropy: the less is the rank, the less is the   
dimension and, consequently, the more important is the contribution   
to the inertial-range behavior [see Eqs. (\ref{eq:str_hier1})-(\ref{eq:str_hier3})].   
   
The leading terms of the even (odd)   
structure functions are given by the scalar (vector) operators. For   
the finite anisotropy, the exponents cannot be associated with individual   
operators (which are essentially ``mixed'' in renormalization), but,   
surprising enough, the aforementioned hierarchy survives for all the   
cases studied, as is shown in Figs. \ref{fig:str_fig2}--\ref{fig:str_fig9}.   
   
The short comment about the second-order structure function $S_{2}(\mr)$ is appropriate.  
It can be studied using the RG and zero-mode techniques \cite{AAHN00}; like in the isotropic case   
\cite{Kraichnan94,Kra68,Kraichnan74,Kraichnan97}, its leading term has the form $S_{2} \propto r^{2-\eps}$,   
but the amplitude now depends on $\alpha_{1,2}$ and the angle between   
the vectors $\mr$ and $\mn$ from Eq. (\ref{eq:str_T34}). The first anisotropic   
correction has the form $(r/L)^{\Delta[2,2]}$ with the exponent   
$\Delta[2,2]=O(\eps)$ from Eq. (\ref{eq:str_Delta22}).

\begin{figure}  
\centering
 \includegraphics[width=6cm]{\PICS 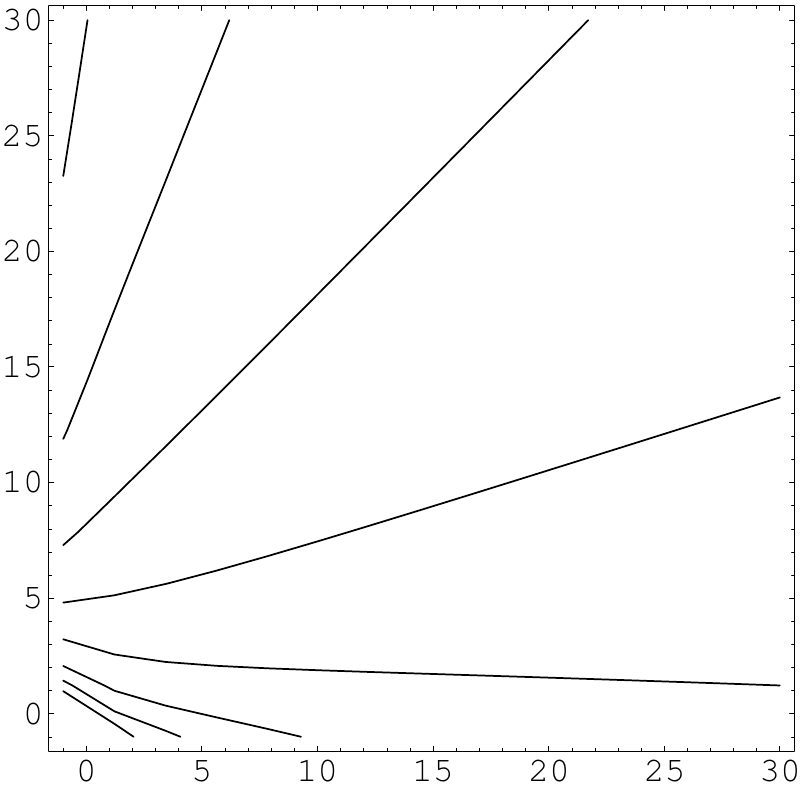}
 
 \vspace{1mm}  
\caption{Levels of the dimension $\Delta[3,1]$ for $d=3$ on the plane  
$\alpha_1$--$\alpha_2$. Value changes from~$-0.3$ (top) to~$0.1$  
(bottom) with step~$0.05$.}  
\label{fig:str_fig10}
\end{figure}

It is well known that, for the isotropic velocity field,   
the anisotropy introduced at large   
scales by the external forcing or imposed mean gradient, persists   
in the inertial range and reveals itself in {\it odd} correlation   
functions: the skewness factor $S_{3}/S_{2}^{3/2}$ decreases for   
$r/L\to 0$ but slowly 
(see Refs. \cite{Antonia84,Sreenivasan91,Sreenivasan97,HolSig94,Pumir96,TonWar94,Pumir96,Pumir98,Shraiman96,Shraiman97}),   
while the higher-order ratios $S_{2n+1}/S_{2}^{n+1/2}$ increase   
(see, e.g., \cite{Ant99,Ant00,CLMV99,ALM01}).   
   
In the case at hand, the inertial-range behavior of the skewness   
is given by $S_{3}/S_{2}^{3/2}\propto (r/L)^{\Delta[3,1]}$. For   
$\alpha_{1,2}\to 0$, the exponent $\Delta[3,1]$ is given by Eq. (\ref{eq:str_Qnp})   
with $n=3$ and $p=1$; it is positive and coincides with the result of   
Ref. \cite{Pumir96,Pumir98}. The levels of the dimension $\Delta[3,1]$ on the   
$(\alpha_{1},\alpha_{2})$-plane are shown in Fig. \ref{fig:str_fig10}. One can see that,   
if the anisotropy becomes strong enough, $\Delta[3,1]$ becomes negative   
and the skewness factor {\it increases} going down towards to the depth of   
the inertial range; the higher-order odd ratios increase already when the   
anisotropy is weak.   \\

{\subsection{Passively advected
magnetic field in the presence of strong anisotropy} \label{subsec:mhdstr_pamf}}

It is of general interest to compare passive advection of scalar and vector quantity.
Though there are some common features, also important differences are observed.
Here, the spatial structure of correlations of
fluctuations of the magnetic (vector) field ${\mb}$ in a given
turbulent fluid in the framework of the kinematic MHD
Kazantsev-Kraichnan model (KMHD) is studied. These fluctuations
are generated stochastically by a Gaussian random emf and a white
in time and anisotropic self-similar in space Gaussian drift. The
main goal is the calculation of the anomalous exponents as
functions of the anisotropy parameters of the drift.\\

\subsubsection{Kinematic MHD Kazantsev-Kraichnan model}
\label{subsubsec:mhdstr_KMHD}

Consider passive advection of a solenoidal magnetic field
${\mb} \equiv {\mb}(t,{\mx})$ in the framework of the KMHD
model described by the stochastic equation
\begin{equation}
  \partial_t {\mb}  =  \nu_0 \boldnabla^2 {\mb} - ({\mv \cdot \boldnabla}) {\mb}
  + ({\mb \cdot \boldnabla}) {\mv} + {\mf},
  \label{eq:mhdstr_K-K}
\end{equation}
where $\nu_0$ is the coefficient of the magnetic diffusivity,
and ${\mv} \equiv {\mv} (t,{\mx})$ is a random solenoidal
 velocity field. Thus, both 
${\mv}$ and ${\mb}$ are divergence-free vector fields: ${\boldnabla}
\cdot {\mv}={\boldnabla} \cdot {\mb}=0$. A transverse Gaussian
emf flux density ${\mf} \equiv {\mf} ({\mx} ,t)$ with
zero mean and the correlation function
\begin{equation}
  D_{ij}^f \equiv \langle f_i(t,{\mx}) f_j(t^{\prime},{\mx^{\prime}}) \rangle=
  \delta(t-t^{\prime})C_{ij}({\mr}/L), \,\,\,\,\ {\mr}={\mx}-{\mx^{\prime}}
  \label{eq:mhdstr_cor-b}
\end{equation}
is the source of the fluctuations of the
magnetic field ${\mb}$. The parameter $L$ represents an integral
scale related to the stirring, and $C_{ij}$ is a
function finite in the limit $L \rightarrow \infty$. In the present
treatment its precise form is irrelevant, and with no loss of
generality, we take $C_{ij}(0)=1$ in what follows. The random
velocity field ${\mv}$ obeys Gaussian statistics with zero
mean and the correlation function
\begin{equation}
  D_{ij}^v (t,{\mx})  \equiv \langle v_i (t,{\mx}) v_j (0,{\bm 0})
  \rangle = \frac {D_0 \delta (t)} {(2 \pi)^d} \int \dRM^d {\mk}\frac
  {\eRM^{i {\mk \cdot \mx}}\, T_{ij} ({\mk})} {(k^2 +
  r_l^{-2}  )^{d/2+\eps/2}},
  \label{eq:mhdstr_cor-v}
\end{equation}
where
$r_l$ is another integral scale. In general, the scale $r_l$ may be
different from the integral scale $L$, below we, however,
take $r_l \simeq L$. $D_0>0$ is an amplitude factor related to the
coupling constant $g_0$ of the model by the relation (\ref{eq:str_Lambda}). In
the isotropic case, the second-rank tensor $T_{ij}({\mk})$ in
Eq.\,(\ref{eq:mhdstr_cor-v}) has the simple form of the ordinary transverse
projector: $T_{ij}({\mk})=P_{ij}({\mk})$. The latter was defined in Eq. (\ref{eq:double_P}).

Although the structure functions $S_N(r)$  of the magnetic field
defined in analogy with (\ref{eq:hel_struc}) as
\begin{equation}
  S_{N}(r)\equiv
  \langle[b_r(t,{\mx}) - b_r(t,{\mx'})]^{N}\rangle, \qquad
  r \equiv | {\mx} - {\mx'} |, \qquad b_r = \frac{\mb \cdot \mr}{r}.
  \label{eq:mhdstr_struc}
\end{equation}
are important tools in the analysis of MHD turbulence in the
inertial range [defined by the inequalities $l \ll r \ll L$,
where $l \simeq \Lambda^{-1}$ is an internal (viscous) scale],
 here the analysis of simpler quantities (the equal-time two-point correlation functions of the composite operators)
 is invoked
$b_r^{N-m}(t,{\mx})$ and $b_r^m(t,{\mx})$)
\begin{equation}
  B_{N-m,m}(r)\equiv \langle b_r^{N-m}(t,{\mx})b_r^m(t,{\mx'})\rangle \qquad
  r \equiv | {\mx} - {\mx'} |.
  \label{eq:mhdstr_Brr}
\end{equation}
for two reasons: first, the field-theoretic approach yields the scaling behavior of these
quantities in the first place, while the scaling behavior of the structure functions
(\ref{eq:mhdstr_struc}) emerges from their representation as linear combinations of the two-point
correlation functions (\ref{eq:mhdstr_Brr}). Second, contrary to the problems of
turbulent velocity of incompressible fluid and passive scalar advected by such fluid, the
basic stochastic equation (\ref{eq:mhdstr_K-K}) is not invariant under the shift ${\mb}\to {\mb}+{\mc}$,
where ${\mc}$ is a constant vector. Thus, there is no compelling need to aim at
the analysis of more complex quantities, the structure functions, instead of their building
blocks, the correlation functions (\ref{eq:mhdstr_Brr}).

Dimensional analysis yields
\begin{equation}
  B_{N-m,m}(r)=\nu_0^{-N/2} r^{N} \tilde{R}_{N,m}(r/l,r/L),
  \label{eq:mhdstr_srtuc100}
\end{equation}
where $R_{N,m}$ are functions of dimensionless parameters.
When the random source field ${\mf}$ and the velocity field
${\mv}$ are uncorrelated, the correlation functions of odd order $B_{2n+1-m,m}$ vanish, however.
The standard perturbation expansion (series in $g_0$) is ill suited
for calculation of correlation functions (\ref{eq:mhdstr_srtuc100})
in the limit $r/l \rightarrow \infty$ and $r/L \rightarrow 0$,
due to the singular behavior of the coefficients
of the expansion. Therefore, to find the correct
IR behavior it is necessary to sum the whole series. Such
a summation can be carried out within the field-theoretic
RG and OPE given in Sec. \ref{sec:RG_theory}. 

 First, the
UV renormalization of correlation functions
(\ref{eq:mhdstr_Brr}) is carried out. As a
consequence of this the asymptotic behavior of these
functions for $r/l\gg 1$ and arbitrary
but fixed $r/L$ is given by IR stable fixed point(s) of the
corresponding RG equations and for correlation functions (\ref{eq:mhdstr_Brr})
the following asymptotic form is obtained
\begin{equation}
  \label{eq:mhdstr_Bscaling}
  B_{N-m,m}(r)\sim\nu_0^{-N/2} r_d^{N}\left({r\over l}\right)^{-\Delta[B_{N-m,m}]}
  {R}_{N,m}(r/L)\,\qquad{r\over l} \gg 1\,,
\end{equation}
where the critical dimensions $\Delta[B_{N-m,m}]$
are expressed in terms of the ''anomalous dimensions'' $\gamma^*_\nu$ and $ \gamma_{N}^*$
of the viscosity $\nu$  and the composite operators $b_r^N$, respectively, as:
\begin{equation}
  \label{eq:mhdstr_critB}
  \Delta[B_{N-m,m}]=-N\left(1-{\gamma^*_\nu\over 2}\right)+\gamma_{N-m}^*+\gamma_{m}^*\,.
\end{equation}
The scaling functions ${R}_{N,m}(r/r_l)$ in relations (\ref{eq:mhdstr_Bscaling}) remain unknown.
The critical dimensions
$\Delta[B_{N-m,m}]$ are calculated as (asymptotic) series in $\eps$
with the use of renormalized perturbation theory.

Second,
the small $r/L$ behavior of the functions ${R}_{N,m}(r/L)$
has to be estimated. This
may be done using the OPE, which leads to the following
asymptotic form in the limit $r/L \rightarrow 0$
\[
{R}_{N,m}(r/L)=\sum_{F} C_{F}(r/L) \left({r\over L}\right)^{\Delta_{F}}\,
\]
where $C_{F}(r/L)$ are coefficients regular in $r/L$. The summation
is implied over all possible renormalized scale-invariant
composite operators $F$,
and $\Delta_{F}$ are their critical dimensions.

In the limit $r/L\to 0 $ correlation function (\ref{eq:mhdstr_cor-b})
of the random source field is uniform in space, which -- as usual
in stochastic models describing passive transport \cite{AAHN00,AAV98,AdzAnt98,Ant99,Ant00}
-- brings about composite operators with negative critical dimensions
(dangerous composite operators) at the outset in the asymptotic analysis.
This takes place because the limiting behavior of the correlation function
determines the canonical scaling dimension of the magnetic field which in this case
becomes equal to $-1$. Origin of the dangerous operators is thus different from that
of the stochastic Navier-Stokes problem, where canonical field dimensions are positive and
composite operators become dangerous (i.e. acquire negative scaling dimension) only for
large enough values of the RG expansion parameter \cite{Adzhemyan96,turbo}.

The velocity fluctuation contribution to the scaling dimension in the passive transport problems
is independent of the statistical properties of the source field.
It is important to bear this in mind, because below it will be
shown that calculation of the fluctuation corrections in the present problem is very
similar to that in the case of passively advected scalar if the magnetic field
${\mb}$ is traded for the vector field $\boldnabla \theta$ -- the gradient of the scalar.
This occurs because -- as a consequence of the invariance of
the transport equation for the scalar $\theta$ with respect to the
shift $\theta \to\theta+ c$ with any constant $c$ -- in the scalar problem fluctuation
corrections to scaling behavior
are determined by composite operators constructed not
from the scalar field itself but its derivatives $\nabla \theta$.
Physically, however, these problems are different, because the usual random source for the
scalar with a correlation function $C(r/L)\to 1$ in the limit $r/L\ll 1$
corresponds to random source for the
vector field $\boldnabla \theta$ with correlations
concentrated at small separations (large wave numbers) instead of the asymptotically flat correlation
function in the coordinate space (corresponding to strong correlation at small wave numbers in
the wave-vector space).

Contributions of these dangerous operators  with negative scaling dimensions
to the OPE imply singular behavior of the scaling
functions in the limit $r/L \rightarrow 0$. The leading term is given by
the operator with the most negative critical dimension $\Delta_F$. The leading contributions to correlation functions of even order $B_{N-m,m}$
($N=2n$) are given by scalar operators
$F_{N}=({\mb}\cdot{\mb})^{N/2}$ with their critical dimensions
$\Delta_{N}=-N(1-\gamma_\nu^*/2)+\gamma_{N}^*$, which
eventually determine the nontrivial asymptotic behavior of the correlation functions $B_{N-m,m}$
of the form (the correlation function $B_{N,0}=B_{0,N}$ is a constant)
\begin{equation}
  B_{N-m,m}(r) \sim \nu_0^{-N/2}L^{N}\left({l\over L}\right)^{N\gamma_\nu^*/2}
  \left({r \over l}\right)^{-{\gamma}^*_{N-m}-{\gamma}^*_{m}}
  \left({r \over L}\right)^{{\gamma}_N^*}\sim
  r^{{\gamma}_N^*-{\gamma}^*_{N-m}-{\gamma}^*_{m}}\,.
  \label{eq:mhdstr_struc120}
\end{equation}
In the isotropic case,
the anomalous dimensions $\gamma_{N}^*$
in the one-loop approximation are
related \cite{AdzAnt98} to anomalous dimensions of composite
operators of a simpler model of passively advected scalar
field \cite{AAV98}, viz. $\gamma_{N}^*$
are given by
\begin{equation}
  \gamma_{N}^*=-{N (N+d)\eps\over 2(d+2)} + \O(\eps^2)\,, \quad N\ge 2\,, \qquad \gamma_1^*=0\,.
  \label{eq:mhdstr_critdim}
\end{equation}
From this relation it follows that
the scaling exponent in expression (\ref{eq:mhdstr_struc120}) 
${\gamma}_N^*-{\gamma}^*_{N-m}-{\gamma}^*_{m}<0$ at this order.
Below it will be shown that these relations are stable
against small-scale anisotropy.

In the anisotropic case we will assume that the statistics of the velocity field is anisotropic
at all scales and replace the ordinary transverse
projection operator in Eq.\,(\ref{eq:mhdstr_cor-v})
with the operator (\ref{eq:str_T34}), which  has been introduced in Sec. \ref{subsubsec:str_scenario}.

The strong small-scale anisotropy (\ref{eq:str_T}) affects the inertial-range asymptotic behavior
of the correlation functions (\ref{eq:mhdstr_struc120}) in two respects: the anomalous dimensions
$\gamma_N^*$ become dependent on the anisotropy parameters $\alpha_1$ and $\alpha_2$
and powerlike corrections with new anomalous dimensions appear.
Combining the results of multiplicative renormalization
and OPE in the manner sketched above for the isotropic case, we arrive at the
following expression for the inertial-range asymptotics of
the correlation functions of the passively advected vector field
$B_{N-m,m}$:
\begin{equation}
  \label{eq:mhdstr_BscalingA}
  B_{N-m,m}(r) \sim  \nu_0^{-N/2}L^{N}\left({l \over L}\right)^{{\gamma}_{\,N}^*+N\eps/2}
  \left({r \over l}\right)^{{\gamma}_{\,N}^*-\underline{\gamma}_{\,N-m}^*-\underline{\gamma}_{\,m}^*}
  \,,\qquad \underline{\gamma}_{\,1}^*=0\,,\qquad m\ge 1\,,
\end{equation}
with negative exponents
${\gamma}_{\,N}^*+N\eps/2<0$ and
${\gamma}_{\,N}^*-\underline{\gamma}_{\,N-m}^*-\underline{\gamma}_{\,m}^*<0$.
The anomalous dimensions $\gamma_{N}^*$,
$\underline{\gamma}_{\,N-m}^*$, and $\underline{\gamma}_{\,m}^*$
in (\ref{eq:mhdstr_BscalingA})
will be defined in relation (\ref{eq:mhdstr_deltaff}) and results of their numerical
calculation discussed in detail in Sec. \ref{subsubsec:mhdstr_OPE}.
The scaling behavior in expression (\ref{eq:mhdstr_BscalingA}) is similar to that of the correlation functions
of the passive scalar advected by a {\em compressible} vector field \cite{AdzAnt98}.

\subsubsection{Field-theoretic formulation, renormalization, and RG analysis}
\label{subsubsec:mhdstr_FTRG}

The stochastic problem (\ref{eq:mhdstr_K-K})--(\ref{eq:mhdstr_cor-v}) in the Stratonovich interpretation is equivalent to
the field-theoretic model of the set of the three fields
$\Phi=\{{\mb^{\prime}}, {\mb}, {\mv}\}$ with the action
functional
\begin{equation}
  \S [\Phi]  \equiv {1\over 2}\,{\mb}'D_b {\mb}' + {\mb}' \cdot [ -\partial_{t}
  -({\mv} \cdot \boldnabla) + \nu_0 \boldnabla^2 +\Sigma_{b'b}]{\mb} + {\mb}'({\mb}
  \cdot \boldnabla){\mv}
  - {1\over 2}{\mv} D_v^{-1} {\mv},
  \label{eq:mhdstr_action1}
\end{equation}
where ${\mb}'$ is an auxiliary field (all required
integrations over space-time coordinates and summations over the
vector indices are implied). To keep notation simple, we have denoted the white-noise contraction term in
dynamic action (\ref{eq:mhdstr_action1}) in terms of the self energy operator (cf. the dynamic action of the
passive scalar case (\ref{eq:str_action}) and the relation to the self-energy operator there (\ref{eq:str_sigma3}). We recall that the
self energy operator is given by a single one-loop graph corresponding to the white-noise contraction term in the limit
of white noise in the SDE).
The first five terms in
Eq.\,(\ref{eq:mhdstr_action1}) represent the De Dominicis-Janssen action
corresponding to the stochastic problem at fixed ${\mv}$ (see,
e.g., Refs.\,\cite{MSR}), the sixth term comes from the Stratonovich interpretation of the SDE, whereas the
last term represents the Gaussian averaging over ${\mv}$. The kernel functions $D_b$
and $D_v$ are the correlation functions (\ref{eq:mhdstr_cor-b}) and
(\ref{eq:mhdstr_cor-v}), respectively.

In this field-theoretic language, the correlation functions
(\ref{eq:mhdstr_Brr}) are defined as
\begin{equation}
  B_{N-m,m}(r)\equiv \int {\mathcal D} \Phi\,  b_r^{N-m}(t,{\mx})b_r^m(t,{\mx'}) \eRM^{\S[\Phi]}
  \label{eq:mhdstr_strukt}
\end{equation}
with the action $\S[\Phi]$ defined above.

From the action (\ref{eq:mhdstr_action1}) 
 the propagators for the fields
${\mb^{\prime}}$ and ${\mb}$ are directly obtained and in the
wave-vector-frequency representation they read
\begin{align}
  \langle b_i b_j^{\prime} \rangle_0 = \langle b_j^{\prime} b_i
  \rangle^*_0= \frac{P_{ij}({\mk})}{-i\omega + \nu_0 k^2},\quad
  \langle b_i b_j \rangle_0 &= \frac{C_{ij}({\mk})}{\omega^2+\nu_0^2
  k^4},\quad 
  \langle b_i^{\prime} b_j^{\prime} \rangle_0 &= 0,
  \label{eq:mhdstr_propagators} 
\end{align}
where $C_{ij}({\mk})$ is the Fourier transform of the function
$C_{ij}({\mr}/L)$ from Eq. (\ref{eq:mhdstr_cor-b}). The bare propagator of the
velocity field
$\langle {\mv} {\mv}\rangle_0 \equiv \langle
{\mv} {\mv}\rangle$ is defined by Eq.\,(\ref{eq:mhdstr_cor-v}) with the
transverse projector given by Eq. (\ref{eq:str_T34}).
The interaction in the model is  given  by the nonlinear terms
$- b^{\prime}_i ({\mv}\cdot \boldnabla) b_i + b^{\prime}_i ({\mb}\cdot \boldnabla) 
v_i \equiv b^{\prime}_i V_{ijl} v_j b_l$ with
the vertex factor which in the wave-number-frequency representation
has the following form
\[
  V_{ijl}=i(\delta_{ij} k_l -\delta_{il} k_j),
\]
where $\mk$ is momentum flowing through the corresponding prime field.
With the use of the standard power counting
\cite{Zinn,Vasiliev} 
correlation functions with superficial UV divergences
may be identified.  In the present
model superficial divergences exist only in the
 1PI Green function $\Gamma_{{\mb^{\prime}}{\mb}}$.
In the isotropic case this Green function
gives rise only to the renormalization of
the term $\nu_0{\mb}' \Delta {\mb}$
of action (\ref{eq:mhdstr_action1}) and the
corresponding
UV divergences may be fully
absorbed in the proper redefinition of the existing parameters
$g_0$, $\nu_0$.

When anisotropy is introduced, however, the situation becomes more complicated, because
the 1PI Green function $\Gamma_{{\mb^{\prime}}{\mb}}$
produces divergences corresponding to
the structure ${\mb^{\prime}} ({\mn}\cdot\boldnabla)^2 {\mb}$ in the action of the model
[due to peculiarities of the rapid-change models \cite{AAV98}
the term $({\mb^{\prime}}\cdot{\mn}) \boldnabla^2 ({\mb}\cdot{\mn})$
possible on dimensional and symmetry grounds does not appear].
The term ${\mb^{\prime}} ({\mn}\cdot\boldnabla)^2 {\mb}$
is not present in the
original unrenormalized action (\ref{eq:mhdstr_action1}), but has to be added to the renormalized action,
therefore the model is not multiplicatively renormalizable. In such a case it is customary
to extend the original action (\ref{eq:mhdstr_action1}) by including all terms needed
for the renormalization of the correlation functions and thus adding
new parameters. As a result the extended model is described \cite{HHJMS05}
by a new action of the form:
\begin{multline}
  \S [\Phi] \equiv {1\over 2}{\mb'} D_b {\mb'} + {\mb}'[ -\partial_{t}
  -({\mv} \cdot \boldnabla) + \nu_0 \boldnabla^2 + \chi_0\nu_0 ( {\mn}
  \cdot \boldnabla )^{2} +\Sigma_{b'b}]{\mb}\\
   + {\mb'}({\mb} \cdot \boldnabla){\mv}
  - {1\over 2}{\mv} D_v^{-1} {\mv},
  \label{eq:mhdstr_action2}
\end{multline}
where a new unrenormalized parameter $\chi_0$ has been introduced in the same sense as in the action (\ref{eq:str_extended}) and the self-energy
operator is calculated in terms of the extended model.

Of course, the
bare propagators (\ref{eq:mhdstr_propagators}) of the isotropic model are
modified and for the extended action (\ref{eq:mhdstr_action2}) assume the form
\begin{align}
  \langle b_i b_j^{\prime} \rangle_0 &= \langle b_j^{\prime} b_i
  \rangle^*_0= \frac{P_{ij}({\mk})}{-i\omega + \nu_0 k^2+\chi_0 \nu_0
  ({\mn}\cdot {\mk})^2}\,,
  \label{eq:mhdstr_bbprime} \\
  \langle b_i b_j \rangle_0 &= \frac{C_{ij}({\mk})}{|-i
  \omega+\nu_0
  k^2 +\chi_0 \nu_0 ({\mn}\cdot {\mk})^2|^2}\,, \label{eq:mhdstr_bb}\\
  \langle b_i^{\prime} b_j^{\prime} \rangle_0 &= 0\,.\nonumber
\end{align}
After this modification all terms needed to
remove the divergences are present in action (\ref{eq:mhdstr_action2}),
therefore the model becomes multiplicatively renormalizable allowing for the
standard RG analysis.
The corresponding renormalized action may be written down immediately:
\begin{multline}
  \S_R[\Phi]  \equiv {1\over 2}{\mb'} D_b {\mb'} + {\mb}'[
  -\partial_{t} -({\mv} \cdot \boldnabla) + \nu Z_1 \boldnabla^2 + \chi \nu
  Z_2 ( {\mn} \cdot \boldnabla )^{2} +\overline{\Sigma}_{b'b}]{\mb}\\
   + {\mb'}({\mb} \cdot
  \boldnabla){\mv}
  - {1\over 2}{\mv} D_v^{-1} {\mv}\,.
  \label{eq:mhdstr_action3}
\end{multline}
Here, $Z_1$ and $Z_2$ are the renormalization constants
in which the UV divergent parts of the 1PI response
function $\Gamma_{{\mb^{\prime}}{\mb}}$ are absorbed.
Thus, from the renormalized self-energy operator $\overline{\Sigma}_{b'b}$ in (\ref{eq:mhdstr_action3}) the divergent parts
corresponding to the chosen renormalization scheme have been subtracted.
The
renormalized action (\ref{eq:mhdstr_action3}) leads to the multiplicative
renormalization of the parameters $\nu_0, g_0$ and $\chi_0$ given by the relations (\ref{eq:str_18}).

\begin{figure}
\centering 
\includegraphics[width=4cm]{\PICS 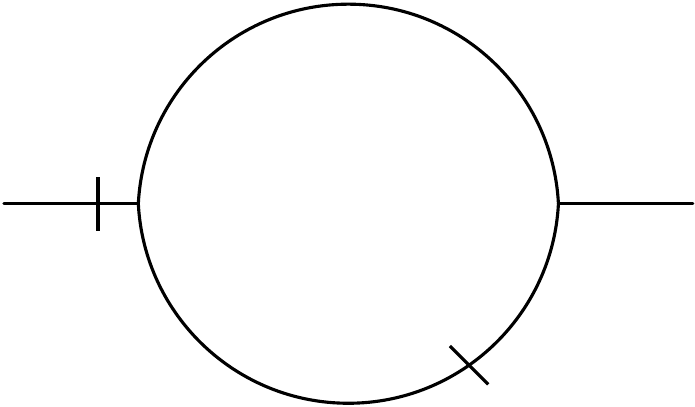}
\caption{\label{fig:mhdstr_obr1}
The (exact) graphical expression for the self-energy operator $\Sigma_{b^{\prime}b}$ of the response
function of the passive vector field.
The plain line denotes the bare
propagator (\ref{eq:mhdstr_bb}), and the line with slash
(denoting the end corresponding to the arguments of the field ${\mb^{\prime}}$) corresponds to
the bare propagator (\ref{eq:mhdstr_bbprime}).
}
\end{figure}

Identification of the unrenormalized action (\ref{eq:mhdstr_action2}) with
the renormalized one (\ref{eq:mhdstr_action3}) leads to the following
relations between the renormalization constants:
\begin{equation}
  \label{eq:mhdstr_Zrelation}
  Z_1=Z_{\nu},\,\,\,\,Z_2=Z_{\chi}Z_{\nu},\,\,\,\,Z_g=Z_{\nu}^{-1}\,.  
\end{equation}
As has been discussed after Eq. (\ref{eq:str_Dyson2}) the rapid-change-like models (\ref{eq:mhdstr_action2})
have a distinguished property that in all multi-loop diagrams of the
self-energy operator $\Sigma_{b^{\prime} b} $ closed circuits of the retarded
bare propagators $\langle {\mb} {\mb^{\prime}}\rangle_0$ are produced,
because the propagator $\langle{\mv} {\mv}\rangle_0$ is proportional
to the $\delta$ function in time.
As a result, also the one-loop self energy operator $\Sigma_{b^{\prime} b} $
with the graphical notation of Fig. \ref{fig:mhdstr_obr1} is exact.

The divergent part of the graph in Fig. \ref{fig:mhdstr_obr1} is
 \begin{align}
   \Sigma_{b^{\prime} b}(p) &= - \frac{g \nu {\bar S}_d}{2
   d (d+2)\eps}
   \biggl\{\biggl[(d-1)(d+2)+ \alpha_1(d+1)+\alpha_2\biggl]
   p^2-(2\alpha_1-(d^2-2)\alpha_2) \nonumber\\
   & \times ({\mn}\cdot {\mp})^2 \biggl\}\, .
   \label{eq:mhdstr_sigmaa}
 \end{align}
The expression (\ref{eq:mhdstr_sigmaa}) leads to a straightforward determination of the
renormalization constants $Z_{1}$ and $Z_{2}$:
\begin{align}
  Z_1 &= 1-\frac{g C_d}{2 d (d+2) \eps} \left[(d-1)(d+2)+
  \alpha_1(d+1)+\alpha_2 \right]
  \,,\nonumber \\
  Z_2 &= 1-\frac{g C_d}{2 d (d+2) \chi \eps}
  \left[-2\alpha_1+(d^2-2)\alpha_2\right] \,.
  \label{eq:mhdstr_Z12}
\end{align}
The RG differential equations for
the renormalized correlation functions of the fields read
\begin{equation}
  ({\mathcal D}_{\mu} + \beta_g \partial_g + \beta_{\chi}
  \partial_{\chi} - \gamma_{\nu} {\mathcal D}_{\nu})
  \langle {\mb}(t,{\mx}){\mb}(t,{\mx}')\rangle_R =0\,
  \label{eq:mhdstr_rgeq1}
\end{equation}
with the definition standard definition from Sec. \ref{subsec:UV}. 
The $\beta$ functions can be written as follows
\begin{equation}
  \label{eq:mhdstr_betadef}
  \beta_g\equiv \widetilde {\mathcal D}_{\mu}
  g=g(-\eps+\gamma_1)\,,\,\,\,\,\, \beta_{\chi}\equiv \widetilde
  {\mathcal D}_{\mu} \chi=\chi(\gamma_1-\gamma_2)\, .
\end{equation}

The anomalous dimensions $\gamma_{1}$ and $\gamma_{2}$ can be computed using definition (\ref{rg_gamma}) and Eqs. (\ref{eq:mhdstr_Z12})
\begin{align}
  \gamma_1 &=\frac{g C_d}{2 d (d+2)} \left[(d-1)(d+2)+
  \alpha_1(d+1)+\alpha_2 \right] \,,
  \label{eq:mhdstr_anomal1}\\
  \gamma_2 &= 1-\frac{g C_d}{2 d (d+2) \chi}
  \left[-2\alpha_1+(d^2-2)\alpha_2\right] \,. 
  \label{eq:mhdstr_anomal2}
\end{align}
It should be emphasized that both the renormalization
constants (\ref{eq:mhdstr_Z12}) and the corresponding anomalous dimensions
(\ref{eq:mhdstr_anomal1}) and (\ref{eq:mhdstr_anomal2}) in the present model are exact, i.e., they have
no corrections of order $g^2$ or higher.

The fixed points ($g^*$, $\chi^*$) of the RG equations are defined by the system of two
equations
\begin{equation}
  \label{eq:mhdstr_fixedpoint}
  \beta_g(g^*,\chi^*)=0\,,\qquad \beta_\chi(g^*,\chi^*)=0\,.
\end{equation}
It should be noted that as a consequence of relations
(\ref{eq:mhdstr_Zrelation}), (\ref{eq:mhdstr_betadef}) and (\ref{eq:mhdstr_fixedpoint})
at any fixed point with $g^*\ne 0$ the anomalous dimension of viscosity
assumes the exact value $\gamma_\nu^*=\eps$.

The IR stability of a fixed point is
determined by the eigenvalues of the
matrix
\[
  \Omega=
  \renewcommand{\arraystretch}{1.5}
  \begin{pmatrix}
    \frac{\partial\beta_g}{\partial g} & \frac{\partial\beta_g}{\partial \chi}\\
    \frac{\partial \beta_\chi}{\partial g} & \frac{\partial \beta_\chi}{\partial \chi}
  \end{pmatrix}
  _{*}
\]
are positive.
Calculation shows that the RG equations have
only one non-trivial IR stable fixed point defined by expressions
\begin{align}
  g_*&=\frac{2 d (d+2) \eps}{C_d\left[(d-1)(d+2)+
  \alpha_1(d+1)+\alpha_2\right]}\,,
  \label{eq:mhdstr_fixg} \\
  \chi_*&=\frac{-2\alpha_1+(d^2-2)\alpha_2}{(d-1)(d+2)+
  \alpha_1(d+1)+\alpha_2}. 
  \label{eq:mhdstr_fixchi}
\end{align}
Both eigenvalues of the stability matrix
$\Omega$
are equal to $\eps$ at this fixed point, therefore, the IR
fixed point (\ref{eq:mhdstr_fixg}), (\ref{eq:mhdstr_fixchi})
is stable for $\eps>0$ and all values of the anisotropy
parameters $\alpha_1$ and $\alpha_2$.

Rather unexpectedly, the $\beta$ functions and, consequently, the fixed points
of the present model of
passively advected vector field are exactly the same as
in the model of passively advected scalar field
\cite{AAHN00}. In Sec. \ref{subsubsec:mhdstr_OPE} it will be shown that
this similarity is extended to the anomalous scaling dimensions of the composite operators
in the OPE representation of the correlation functions as well.

The fixed point (\ref{eq:mhdstr_fixg}), (\ref{eq:mhdstr_fixchi}) governs the behavior of solutions of
Eqs. (\ref{eq:mhdstr_rgeq1})
and the like, and at large scales far from viscous length
$r\gg l$ at any fixed ratio $r/L$ yields the scaling
form
\begin{equation}
  \label{eq:mhdstr_bbscaling} \langle {\mb}(t,{\mx}){\mb}(t,{\mx}-{\mr})\rangle
  =D_0^{-1} r^{2-\eps}  R_2(r/L)\,,
\end{equation}
for the {\em unrenormalized} correlation function (we remind that due to the absence of
field renormalization renormalized and unrenormalized correlation functions are equal but expressed
in terms of different variables). It should be noted, however, that
the scaling function $R_2(r/L)$ in Eq. (\ref{eq:mhdstr_bbscaling}) is not determined by the RG Eqs.
(\ref{eq:mhdstr_rgeq1}).

The correlation functions (\ref{eq:mhdstr_Brr})
contain UV divergences additional to those included in the
renormalization constants (\ref{eq:mhdstr_Z12}).
These additional divergences due to
composite operators
(products of fields and their derivatives with coinciding space and time
arguments)
can be dealt with in a manner similar to that applied to
the divergences in
the usual correlation functions \cite{Vasiliev}.

\subsubsection{Renormalization and critical dimensions of composite operators}
\label{subsubsec:mhdstr_OPE}
Two-point correlation functions (\ref{eq:mhdstr_strukt}) are averages of products of composite
operators at two separate space points. These composite operators are
integer powers of the field $b_r$ and contain additional UV
divergences, which also may be removed by a suitable renormalization
procedure discussed in Sec. \ref{subsec:OPE}.

The two-point correlation functions are, however, quantities
with insertions of two composite operators. This is the crucial difference with the previous
 models (and also Sec.  \ref{subsec:OPE}). Nevertheless, field theoretic RG technique is capable of 
dealing with this problem.
 It would seem that
we would have to consider renormalization of products of two
composite operators as well, the aim being then to render
UV finite all 1PI correlation functions with two insertions of
composite operators. Superficially divergent correlation functions
with operator insertions are identified by power counting similar to
that of the basic renormalization. In the present
model such a power counting shows that insertion of products of composite operators of the structure
${\mb}^m(t,{\mx}){\mb}^n(t,{\mx}')$ does not bring about any new superficial divergences
and it is thus sufficient to renormalize the composite operators
themselves only in order to make the two-point correlation functions UV finite.
Therefore, from the RG analysis of composite operators it follows
-- by virtue of relations (\ref{eq:RG_2.2}) and (\ref{eq:RG_2.5}) -- that
the two-point correlation function $B_{N-m,m}$ given in Eq. (\ref{eq:mhdstr_BscalingA}) may be expressed as a functional average
of a quadratic form 
\begin{equation}
  \label{eq:mhdstr_SNF}
  B_{N-m,m}(r)=
  \sum\limits_{\alpha,\beta}B_{\alpha\beta}
  \left\langle {{\bar F}_\alpha^R}\left(t,{\mx}+{{ 1\over 2}}{\mr}\right) {{\bar F}_\beta^R }
  \left(t,{\mx}-{1\over 2}{\mr}\right)\right\rangle_{\!\!R}
\end{equation}
with coefficients $B_{\alpha\beta}$ independent of spatial coordinates and basis operators have been introduced in Eq. (\ref{eq:RG_2.5}).
Each term in expression (\ref{eq:mhdstr_SNF}) obeys the following
asymptotic form in the limit $l \ll r$, $r \lesssim L$
\begin{equation}
  \label{eq:mhdstr_basicF}
  \left\langle  {\bar F}_\alpha^R\left(t,{\mx}+{{ 1\over 2}}{\mr}\right) {\bar F}_\beta^R
  \left(t,{\mx}-{1\over 2}{\mr}\right)\right\rangle_{\!\!R}\sim
  D_0^{d^\omega_{\alpha}+d^\omega_{\beta}}
  r^{-\Delta_{\alpha}-\Delta_{\beta}}
  r_d^{\gamma^*_{\alpha}+\gamma^*_{\beta}}
  \Xi_{\alpha\beta}\left({r\over L}\right)
\end{equation}
with the scaling functions $\Xi_{\alpha\beta}$ still to be
determined.

The physically interesting range of scales, however, is the inertial range,
specified by the inequalities $l \ll r\ll L$. The limit $r\ll L$
may be explored with the use of the OPE \cite{Zinn,Vasiliev} as was
already discussed in Sec. \ref{subsec:OPE}.
The basic statement of the OPE theory is summarized in the relation (\ref{eq:RG_2.44}), which we know
write as follows
\begin{equation}
  F^R_\alpha\left(t,{\mx}+{{ 1\over 2}}{\mr}\right){{F}^R_\beta}\left(t,{\mx}
  -{{ 1\over 2}}{\mr}\right)
  =\sum_{\gamma} C_{\alpha\beta\gamma} ({\mr}) {F}^R_\gamma(t,{\mx}),
  \label{eq:mhdstr_OPE}
\end{equation}
where the functions $C_{\alpha\beta\gamma}$  are the Wilson coefficients regular
in $1/L$, and ${F}^R_\gamma$ are renormalized local
composite operators which appear in the
formal Taylor expansion with respect to ${\mr}$ together with all operators
that mix with them in renormalization. If these operators have
additional vector indices, they are contracted with the
corresponding indices of the coefficients $C_{\alpha\beta\gamma}$.

Without loss of generality we may take the expansion
on the right-hand side of Eq. (\ref{eq:mhdstr_OPE}) in terms of the basis operators with
definite critical dimensions $\Delta_{\mathcal{F}}$. The renormalized
correlation function $\langle F^R_\alpha{{F}^R_\beta} \rangle_R$
is obtained by averaging Eq. (\ref{eq:mhdstr_OPE}) with the weight
generated by the renormalized action, the quantities
$\langle \mathcal{F} \rangle_R$
appear now only on the right-hand side. Their asymptotic behavior
for $r/L\to 0$ is found from the corresponding RG equations
and is of the form $\langle \mathcal{F} \rangle \propto  L^{-\Delta_\mathcal{F}}$.
Comparison of the expression for a given function $\langle F^R_\alpha{F}^R_\beta \rangle_R$ in terms of
the IR scaling representation of correlation functions of the basis operators (\ref{eq:mhdstr_basicF})
on one hand and the OPE representation brought about by relation (\ref{eq:mhdstr_OPE}) on the other in the
limit $L\to \infty$ allows to find the asymptotic form of the scaling functions
$\Xi_{\alpha\beta}(r/L)$ in relation (\ref{eq:mhdstr_basicF}).

The two-point correlation functions are products of
integer powers of the field $b_r$ of the form
$b_r^{N-m}(t,{\mx})b_r^{m}(t,{\mx}')$.
Thus, at the leading order in ${\mr}$ their OPE contains operators
of the closed set generated by the operator $b_r^{N}(t,{\mx})$.
Power counting and analysis of the structure of graphs shows
that this set of composite operators contains only operators consisting
of exactly $N$ components of the vector field ${\mb}$, viz. the tensor operators constructed solely of
the fields ${\mb}$ without derivatives: $b_{i_1} ... b_{i_p}
(b_i b_i)^l$ with $p+2l=N$. It is convenient to deal with the scalar operators
obtained by contracting the tensor with the appropriate number of
the anisotropy vectors ${\mn}$:
\begin{equation}
  F[N,p](t,{\mx}) \equiv [{\mn} \cdot {\mb}(t,{\mx})]^{p} [{\mb}^2(t,{\mx})]^{l}  
  \label{eq:mhdstr_CO}
\end{equation}
with $N \equiv 2l+p$. Analysis of graphs
shows that composite operators (\ref{eq:mhdstr_CO}) with different
$N$ do not mix in renormalization, and
therefore the corresponding  renormalization matrix $Z_{[N,p][N',p']}$
is in fact block-diagonal, i.e., $Z_{[N,p][N',p']}=0$ for
$N'\ne N$ and
\[
  F[N,p]=\sum_{l=0}^{[N/2]} Z_{[N,p][N,N-2l]} F^R[N,N-2l]\,,
\]
where $[N/2]$ stands for the integer part of the rational number $N/2$ for odd $N$
(although the odd-order correlation functions $B_{2n+1-m,m}$ vanish,
renormalization of the even-order correlation functions $B_{2n-m,m}$ involves odd-order composite operators).
Each block with fixed $N$ gives rise to a $(N+1)\times(N+1)$ matrix of critical dimensions
whose eigenvalues at the IR stable fixed point are the critical dimensions $\Delta[N,p]$ of the set
of operators $F[N,p]$.

Taking into account that renormalization of the composite operators
$b_r^{N-m}$ and $b_r^m$ in
the correlation function $B_{N-m,m}$ involves operators of the sets
$F[N-m,p]$ and $F[m,q]$, respectively, whereas the leading contribution to the OPE involves the
set $F[N,s]$, the basis-operator decomposition of the correlation function may be written as
\begin{align}  
  B_{N-m,m}(r) & \sim
  \frac{r_d^{N}}{\nu_0^{N/2}}  
  \sum\limits_{l_1=0}^{[(N-m)/2]}\sum\limits_{l_2=0}^{[m/2]}\sum\limits_{l_3=0}^{N/2}
  A_{l_1l_2l_3}^{Nm}({r/L}) \left({r\over L}\right)^{\Delta[N,N-2l_3]}
  \nonumber\\
  & \times \left({r\over l}\right)^{-\Delta[N-m,N-m-2l_1]-\Delta[m,m-2l_2]} 
  \label{eq:mhdstr_Sbb}
\end{align}
where the coefficients $A_{l_1l_2l_3}^{Nm}({r/L})$ are regular in $(r/L)^{2}$.

\begin{figure}
\centering
       \includegraphics[width=4.25cm]{\PICS 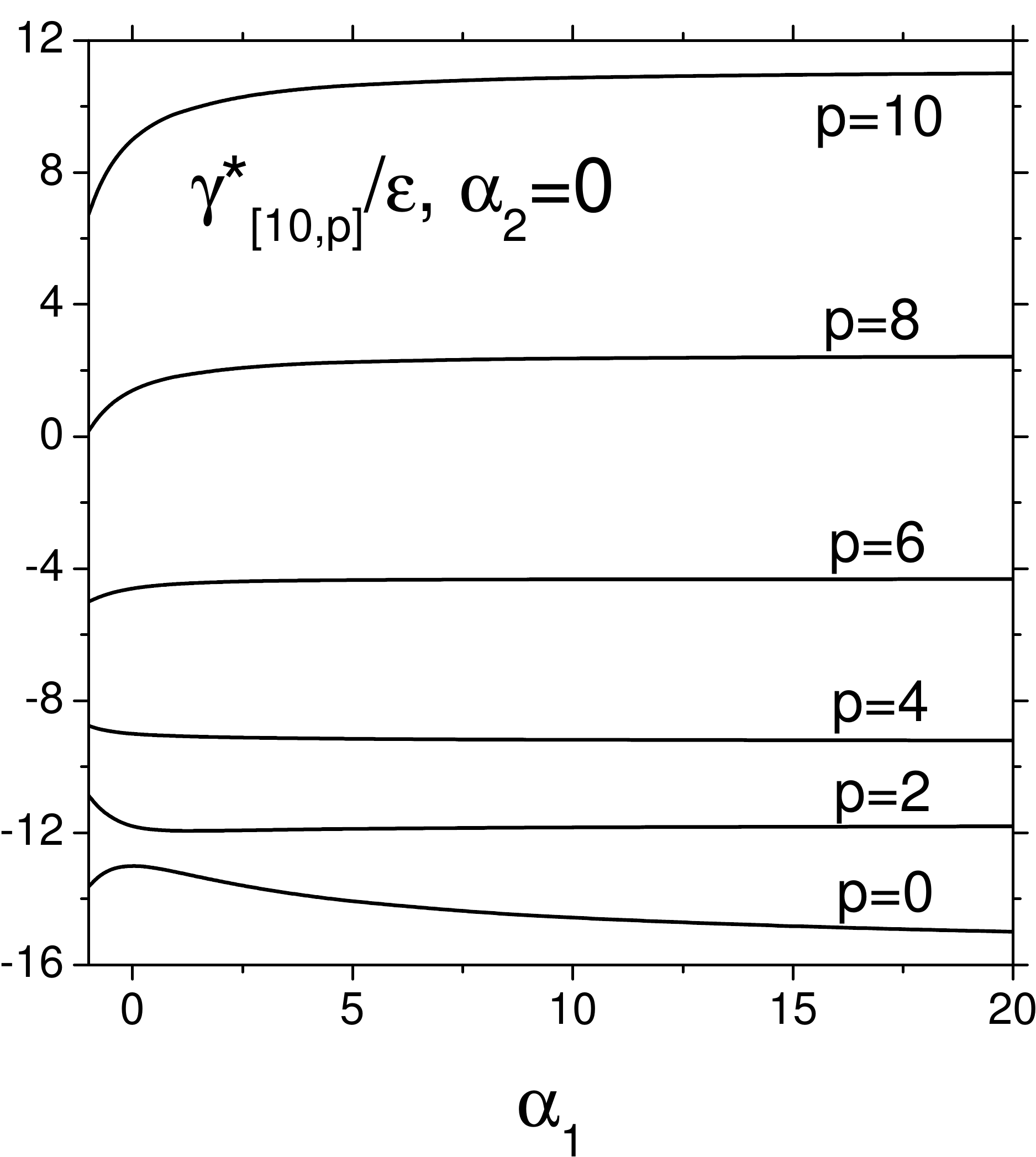}\hfill
       \includegraphics[width=4.25cm]{\PICS 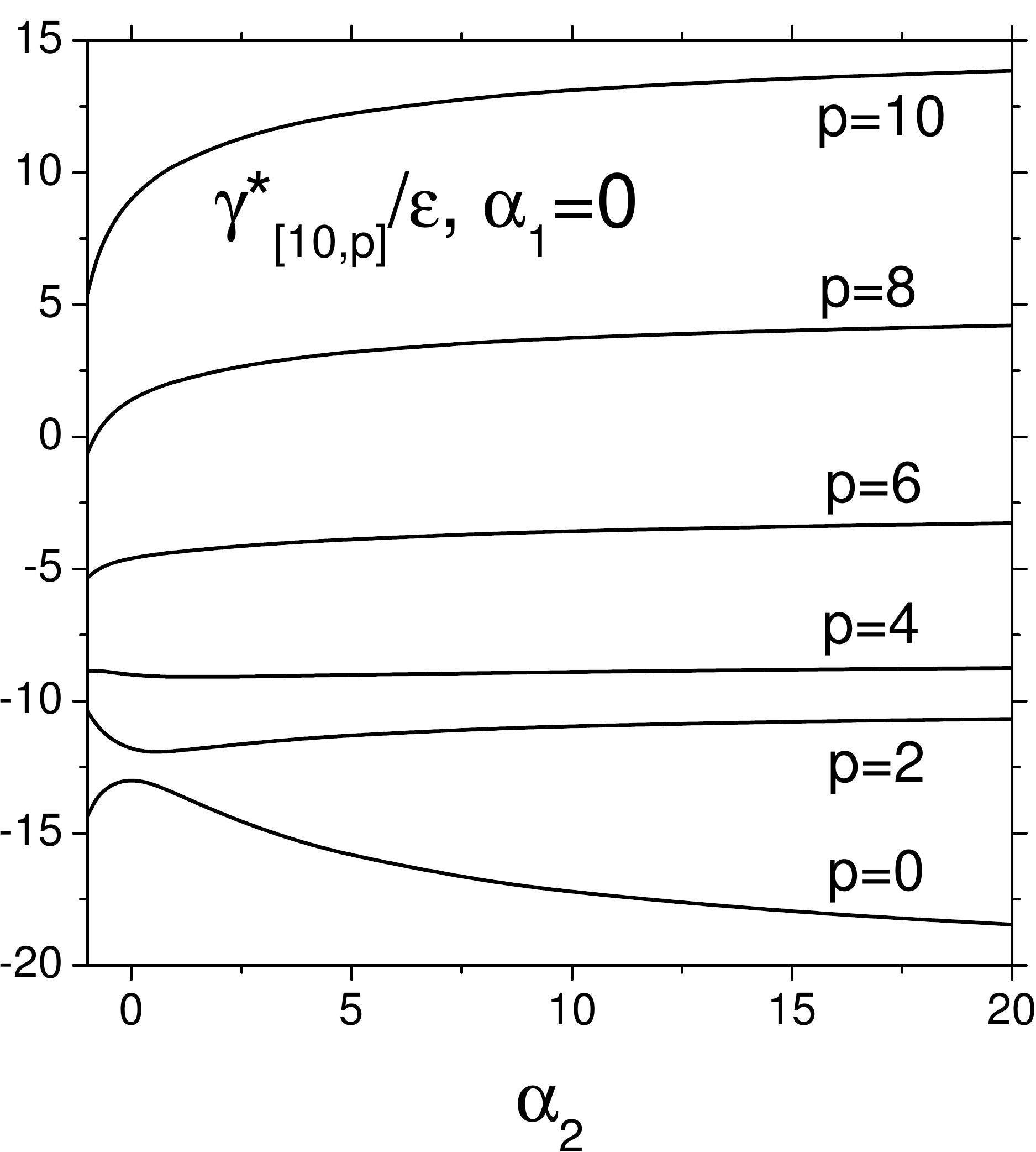}\hfill
       \includegraphics[width=4.25cm]{\PICS 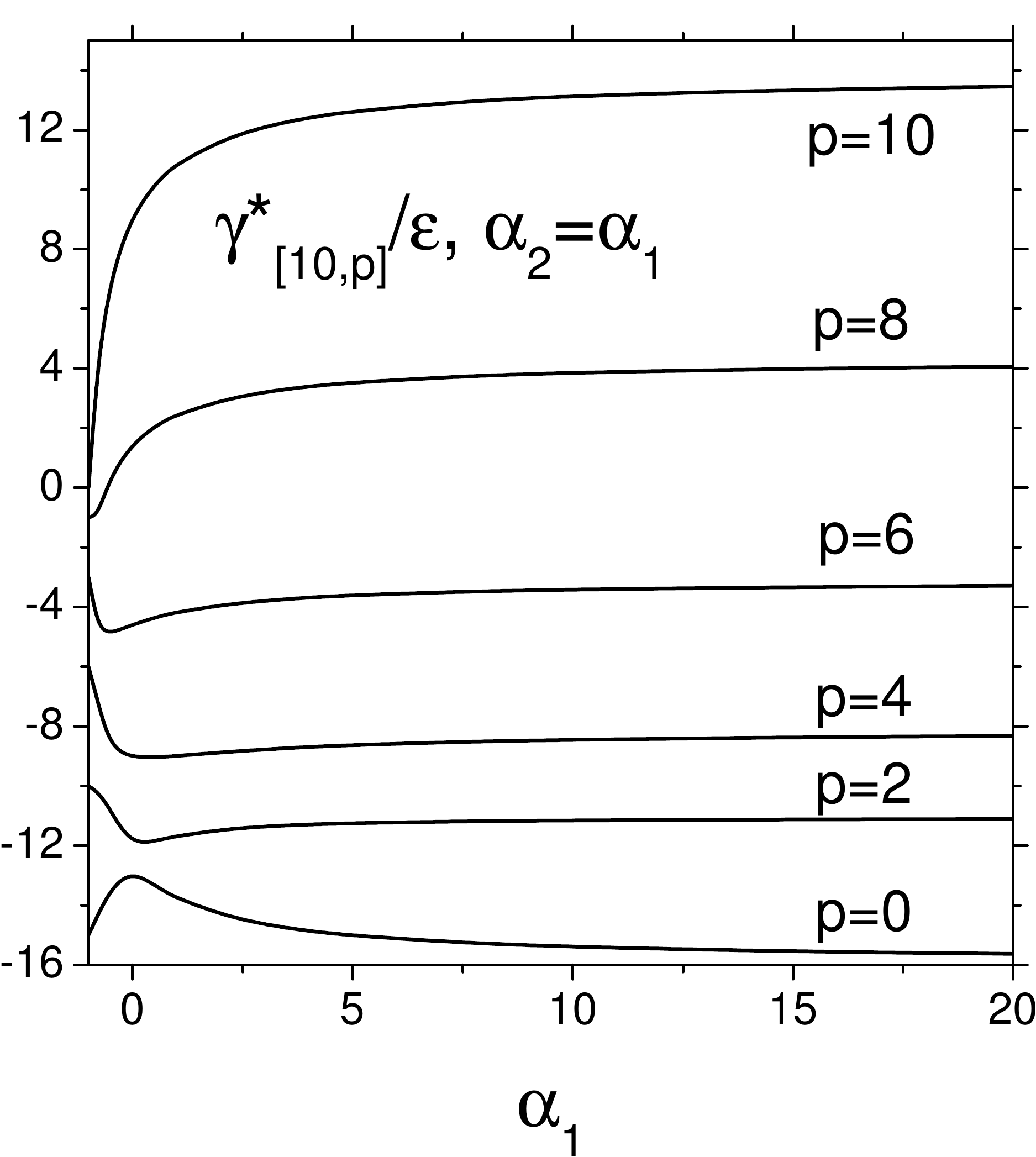}
\caption{Behavior of the anomalous dimension
$\gamma^*_{[10,p]}/\eps$ in space dimension $d=3$ and for
representative values of $p$ as functions of anisotropy parameters
$\alpha_1$ and $\alpha_2$.  \label{fig:mhdstr_fig1}}
\end{figure}

\begin{figure}
\centering
       \includegraphics[width=4.25cm]{\PICS 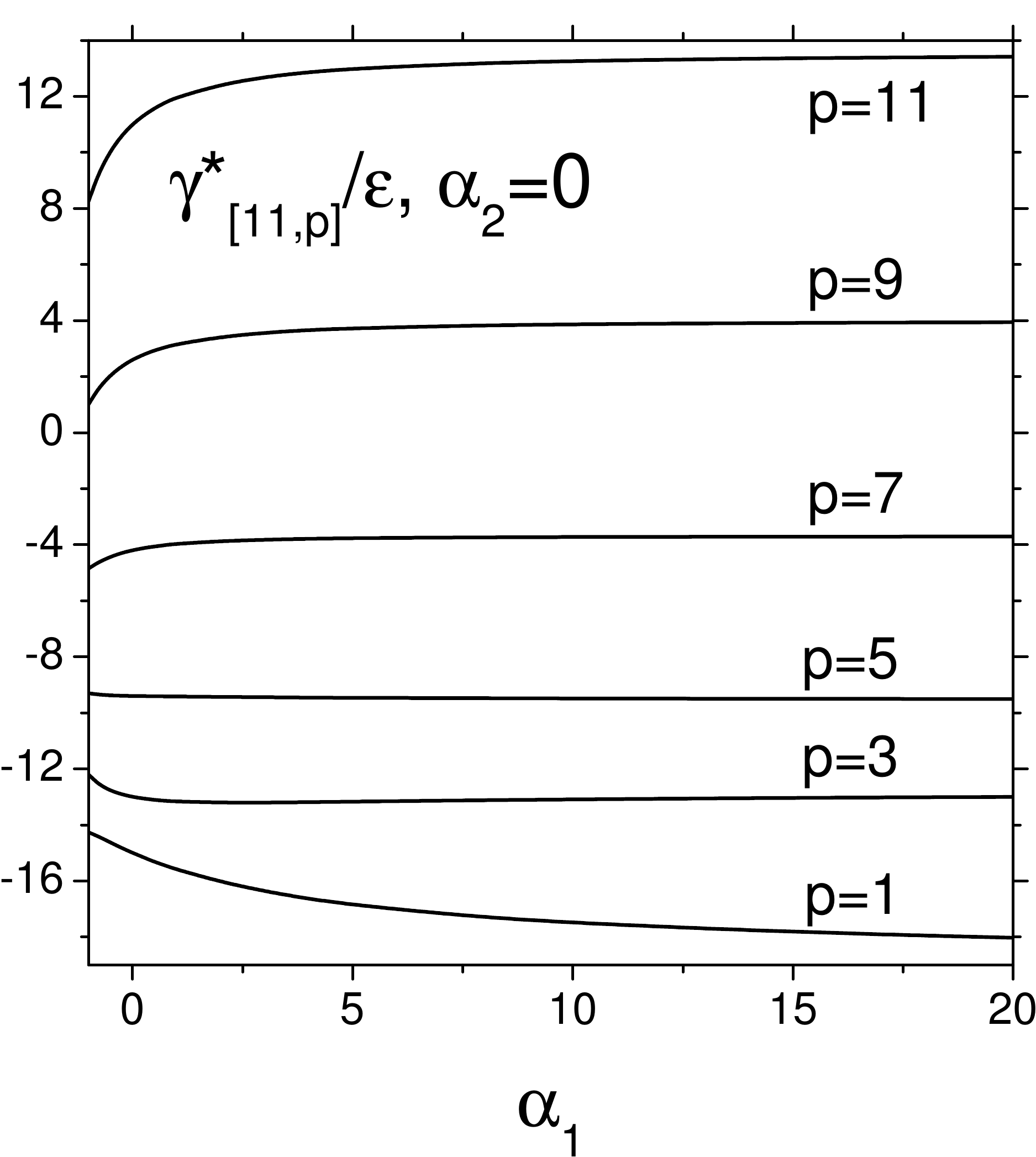}\hfill
       \includegraphics[width=4.25cm]{\PICS 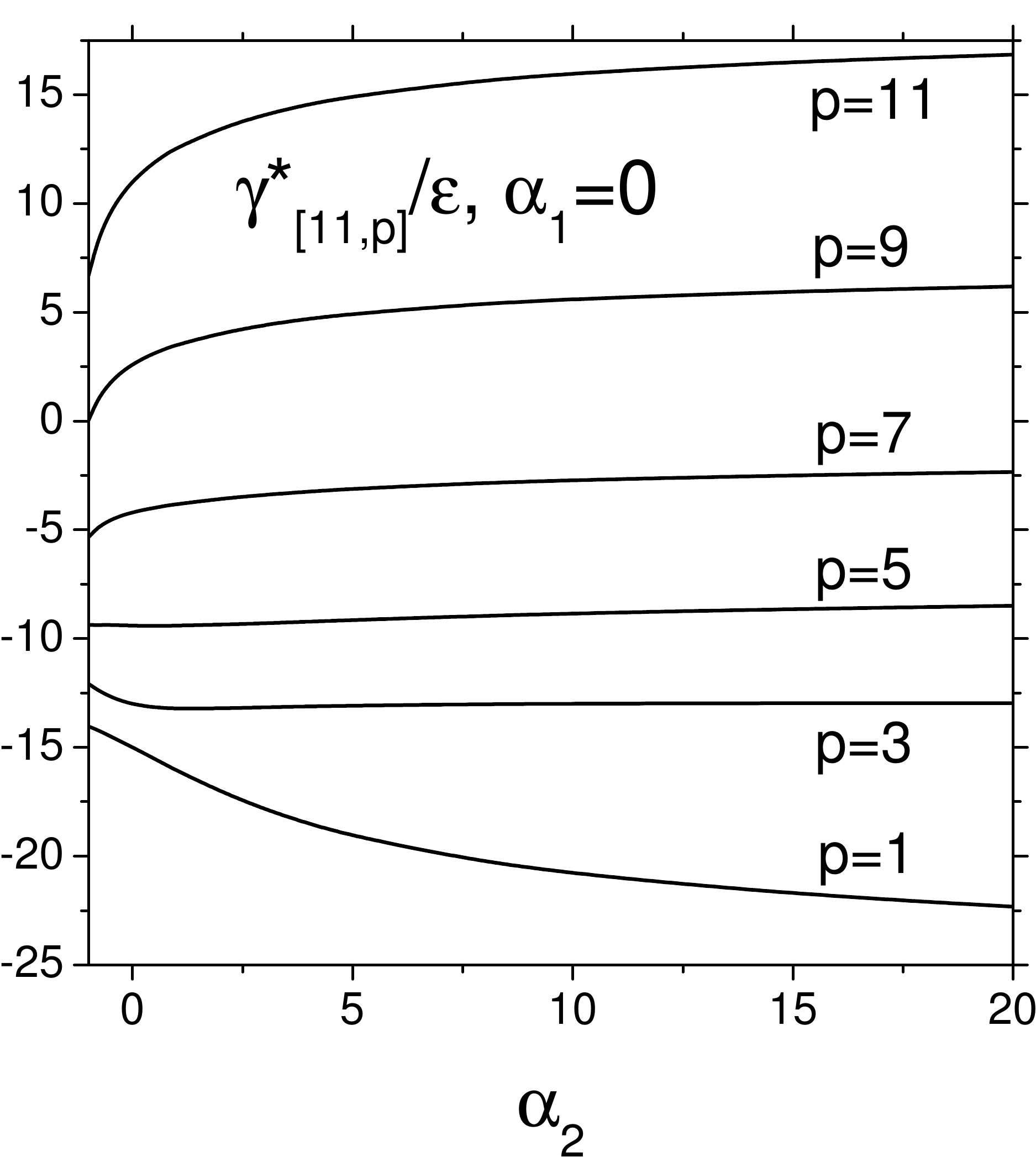}\hfill
       \includegraphics[width=4.25cm]{\PICS 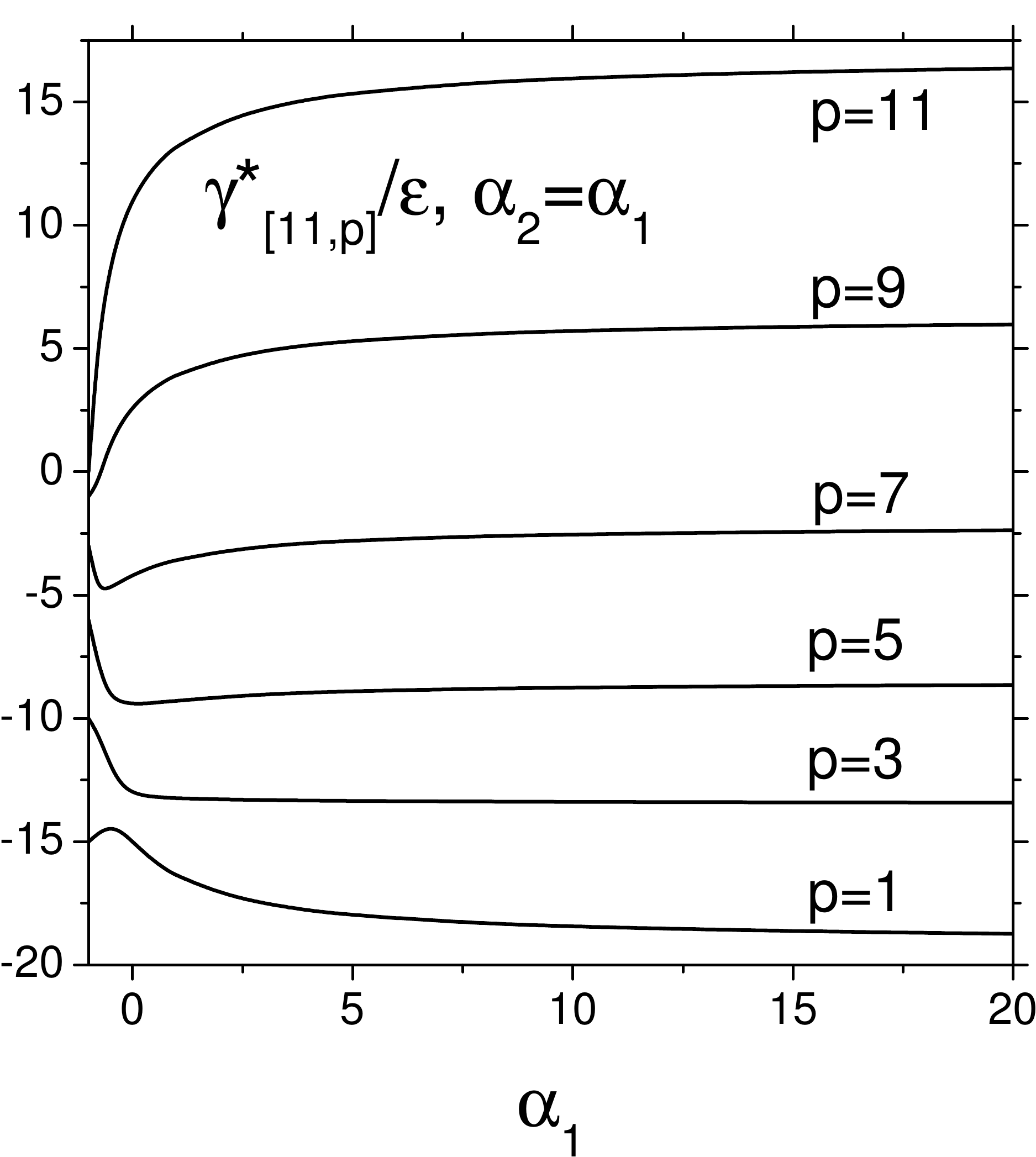}
\caption{Behavior of the anomalous dimension
$\gamma^*_{[11,p]}/\eps$ in space dimension $d=3$ and for
representative values of $p$ as functions of anisotropy parameters
$\alpha_1$ and $\alpha_2$.  \label{fig:mhdstr_fig2}}
\end{figure}

\begin{figure}
\centering
       \includegraphics[width=4.25cm]{\PICS 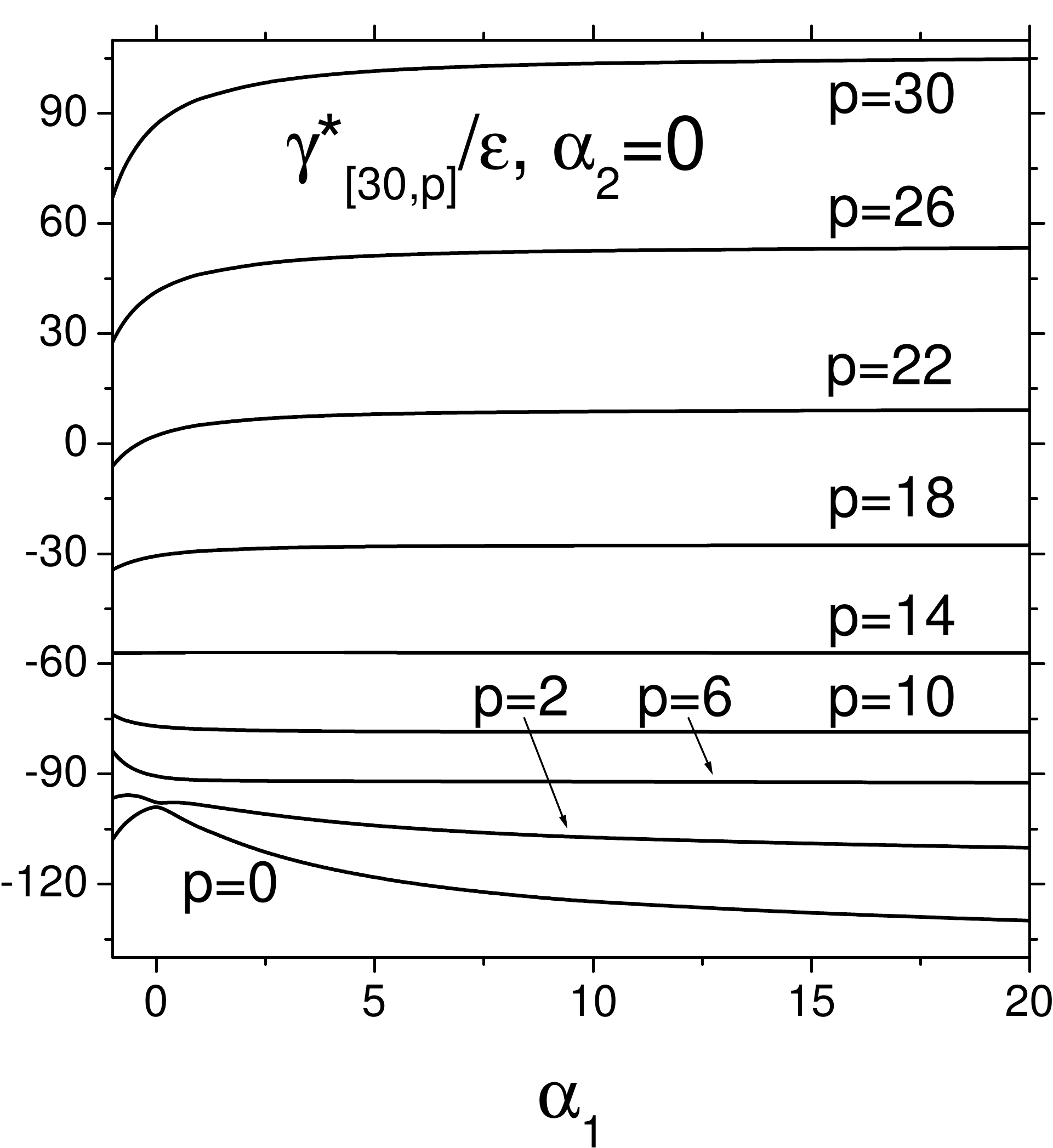}\hfill
       \includegraphics[width=4.25cm]{\PICS 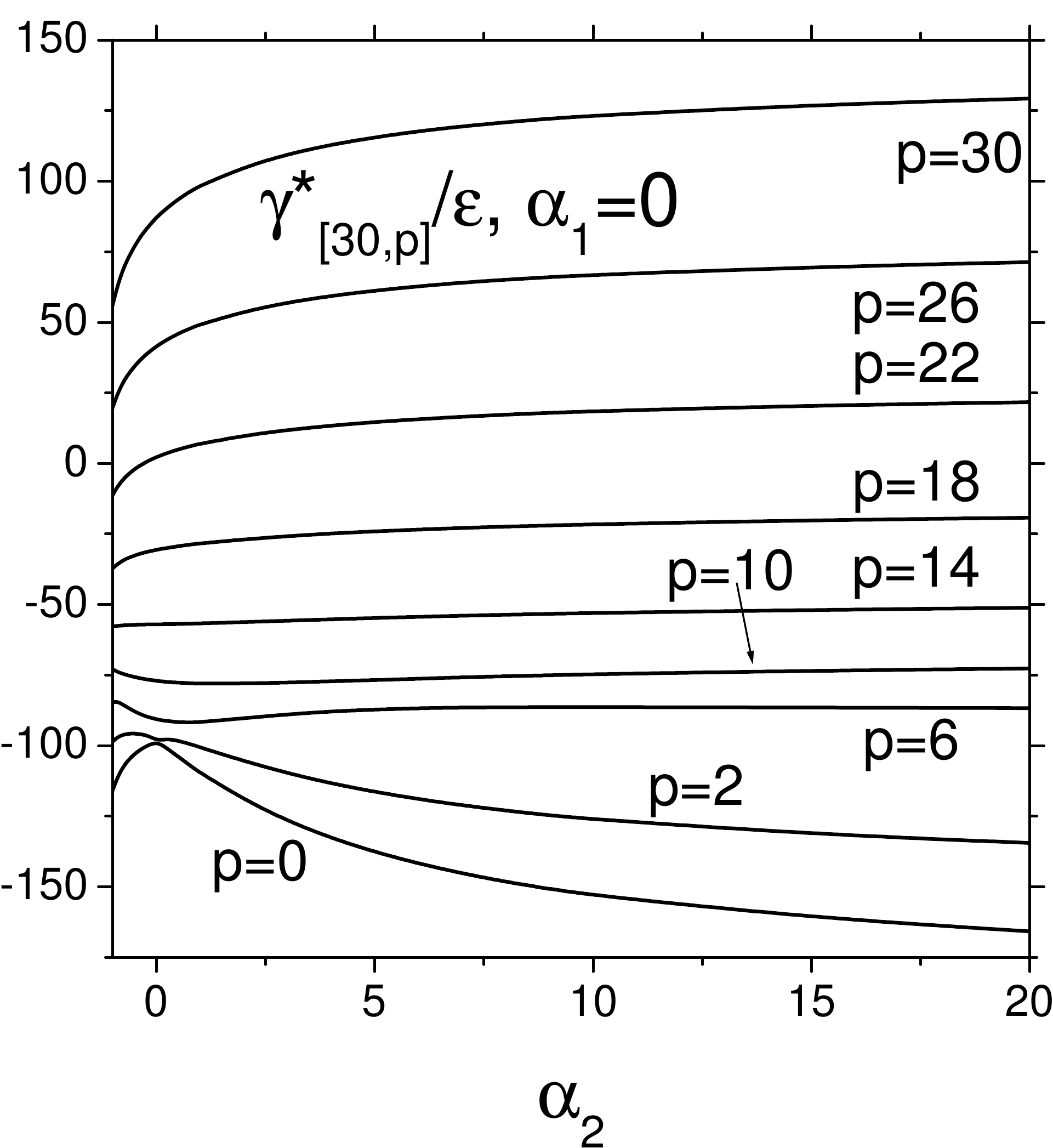}\hfill
       \includegraphics[width=4.25cm]{\PICS 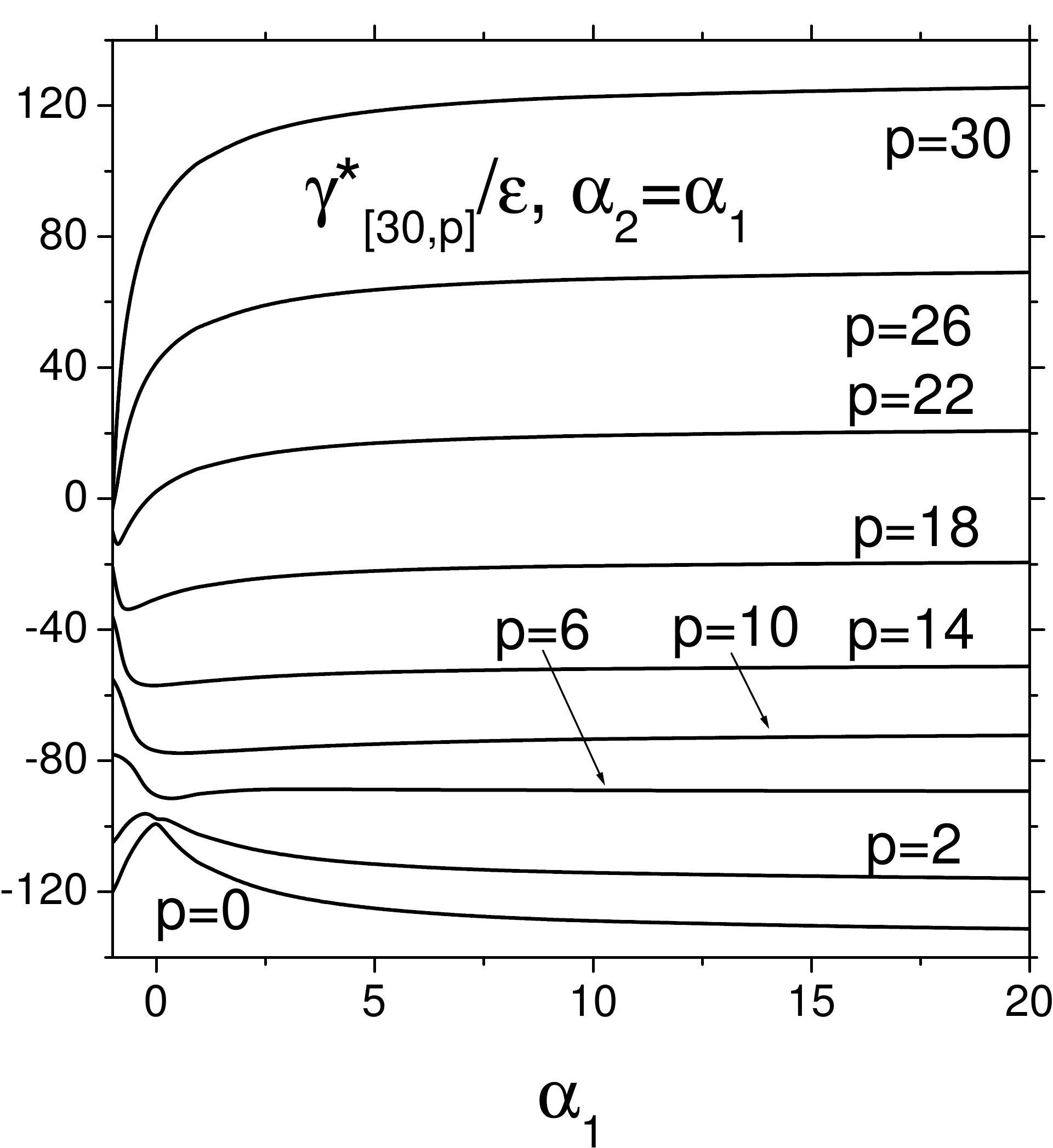}
\caption{Behavior of the anomalous dimension
$\gamma^*_{[30,p]}/\eps$ in space dimension $d=3$ and for
representative values of $p$ as functions of anisotropy parameters
$\alpha_1$ and $\alpha_2$.  \label{fig:mhdstr_fig3}}
\end{figure}

\begin{figure}
\centering
       \includegraphics[width=4.25cm]{\PICS 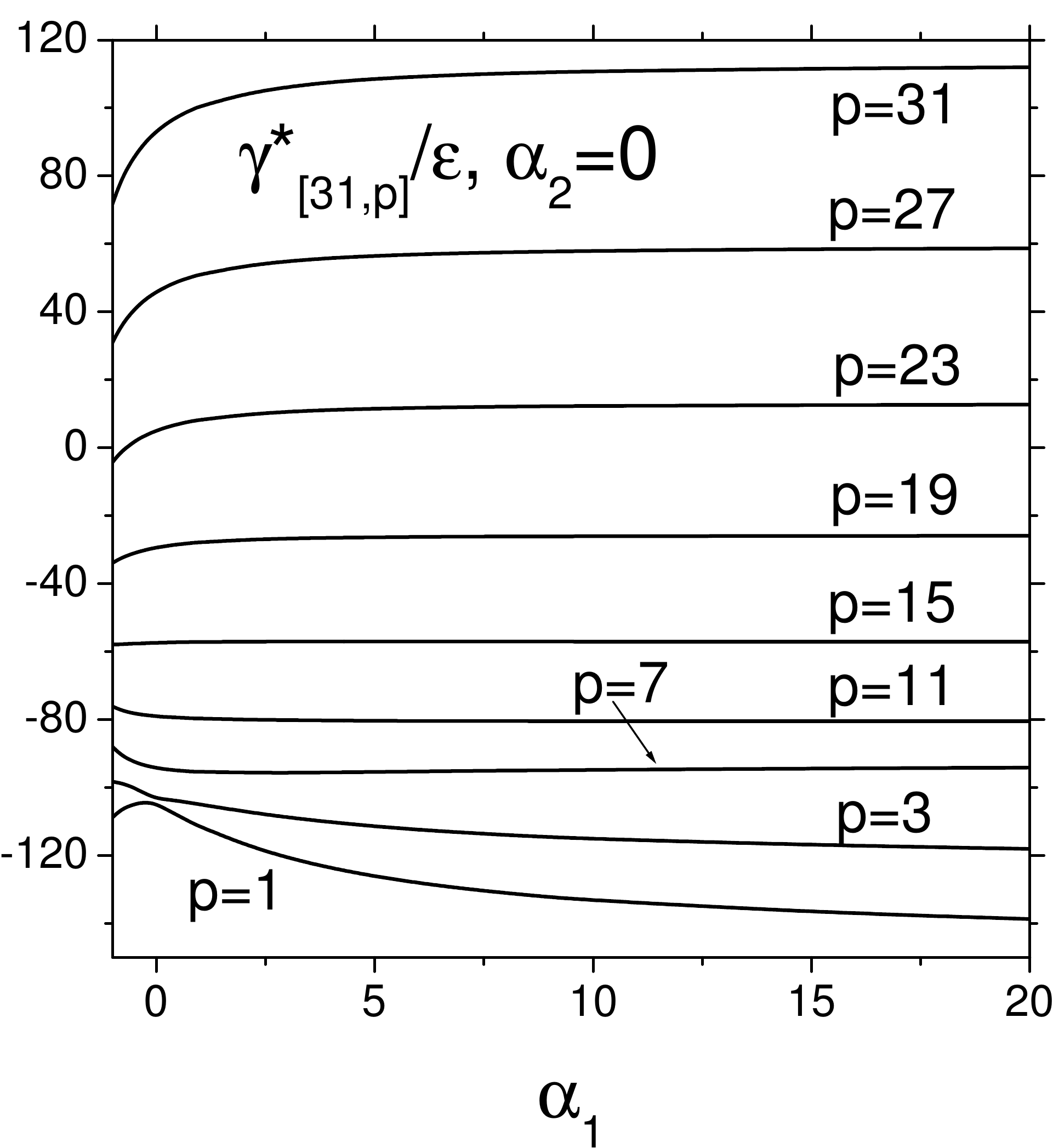}\hfill
       \includegraphics[width=4.25cm]{\PICS 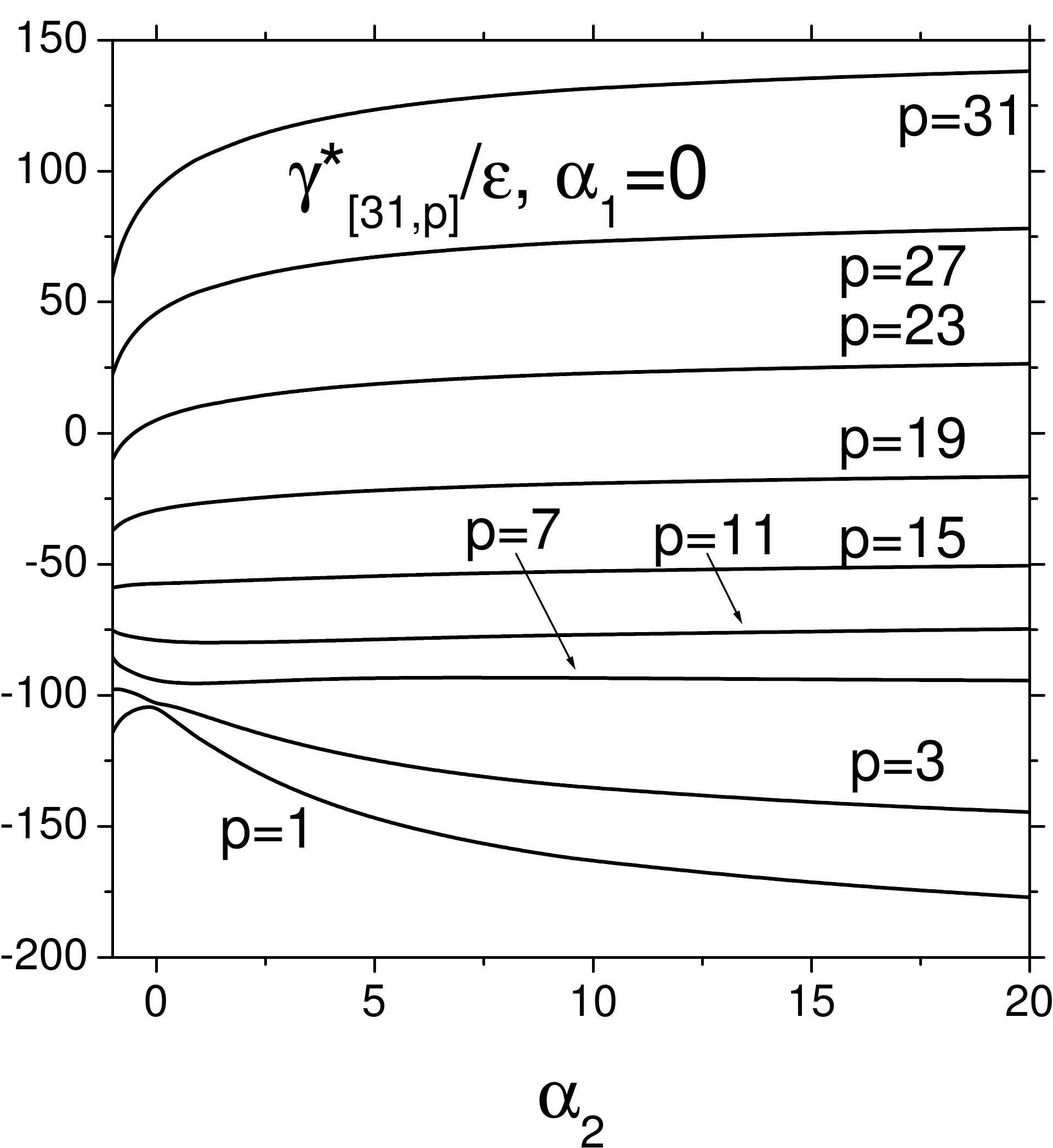}\hfill
       \includegraphics[width=4.25cm]{\PICS 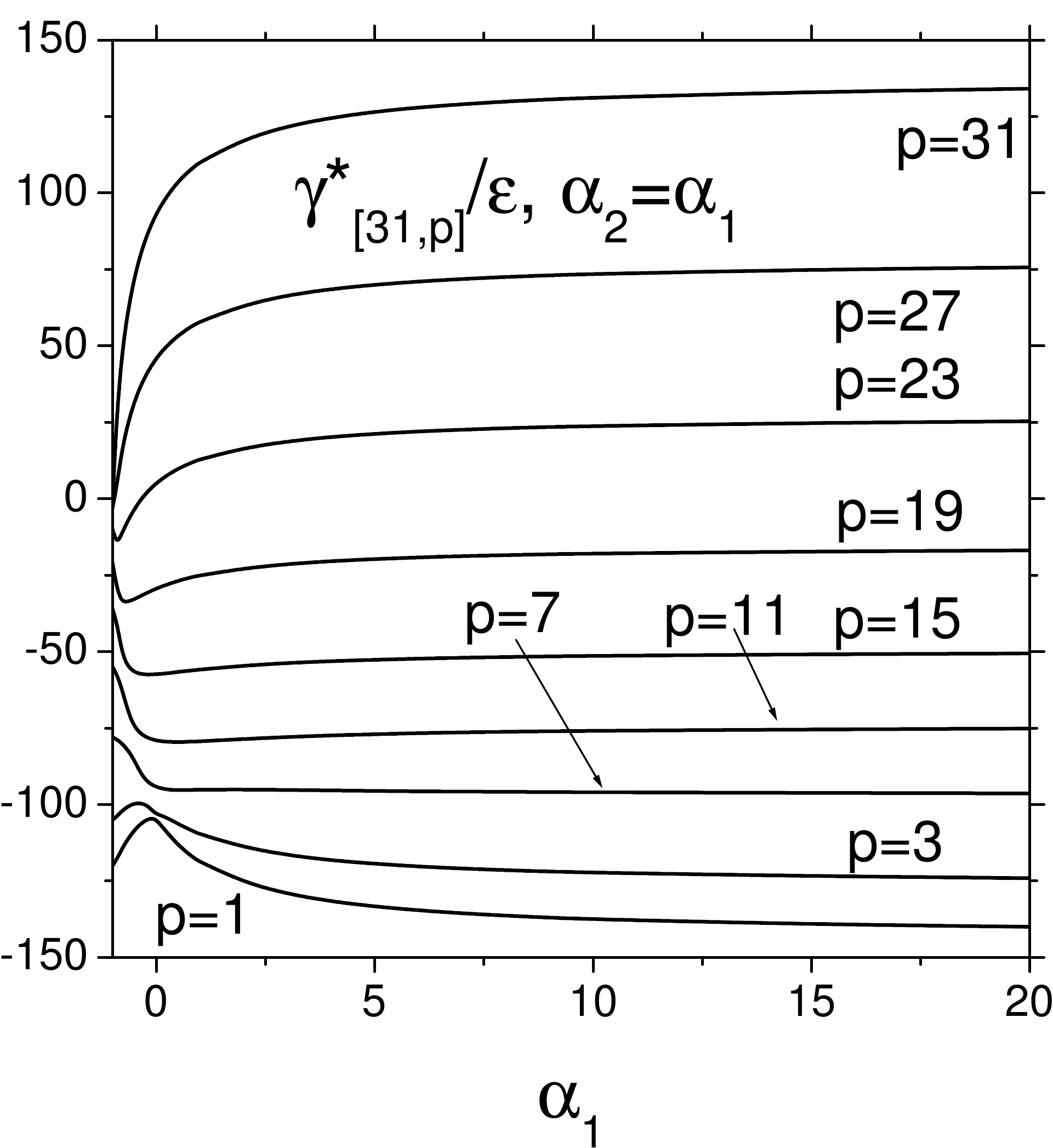}
\caption{Behavior of the anomalous dimension
$\gamma^*_{[31,p]}/\eps$ in space dimension $d=3$ and for
representative values of $p$ as functions of anisotropy parameters
$\alpha_1$ and $\alpha_2$.  \label{fig:mhdstr_fig4}}
\end{figure}

\begin{figure}
\centering
       \includegraphics[width=4.25cm]{\PICS 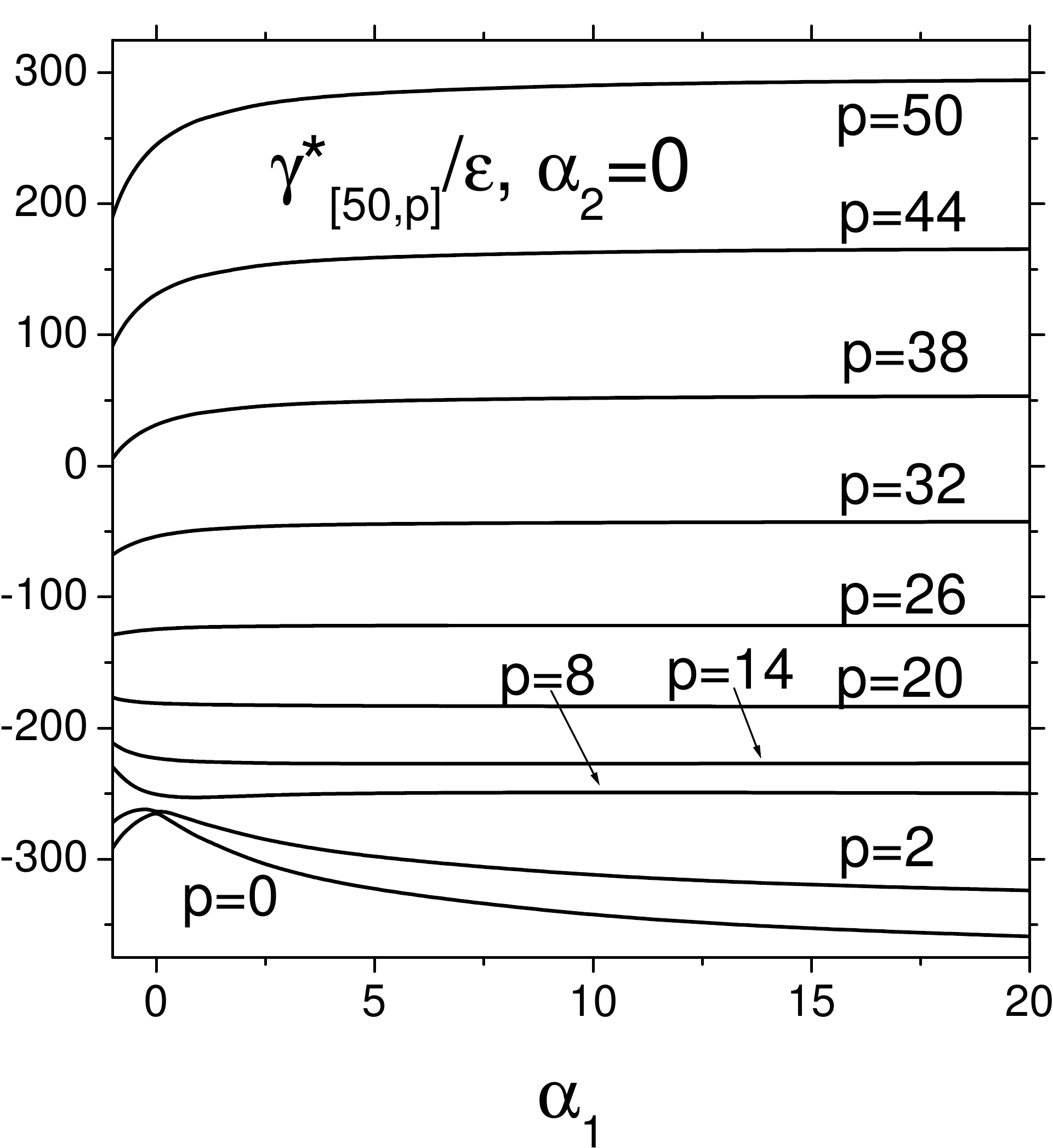}\hfill
       \includegraphics[width=4.25cm]{\PICS 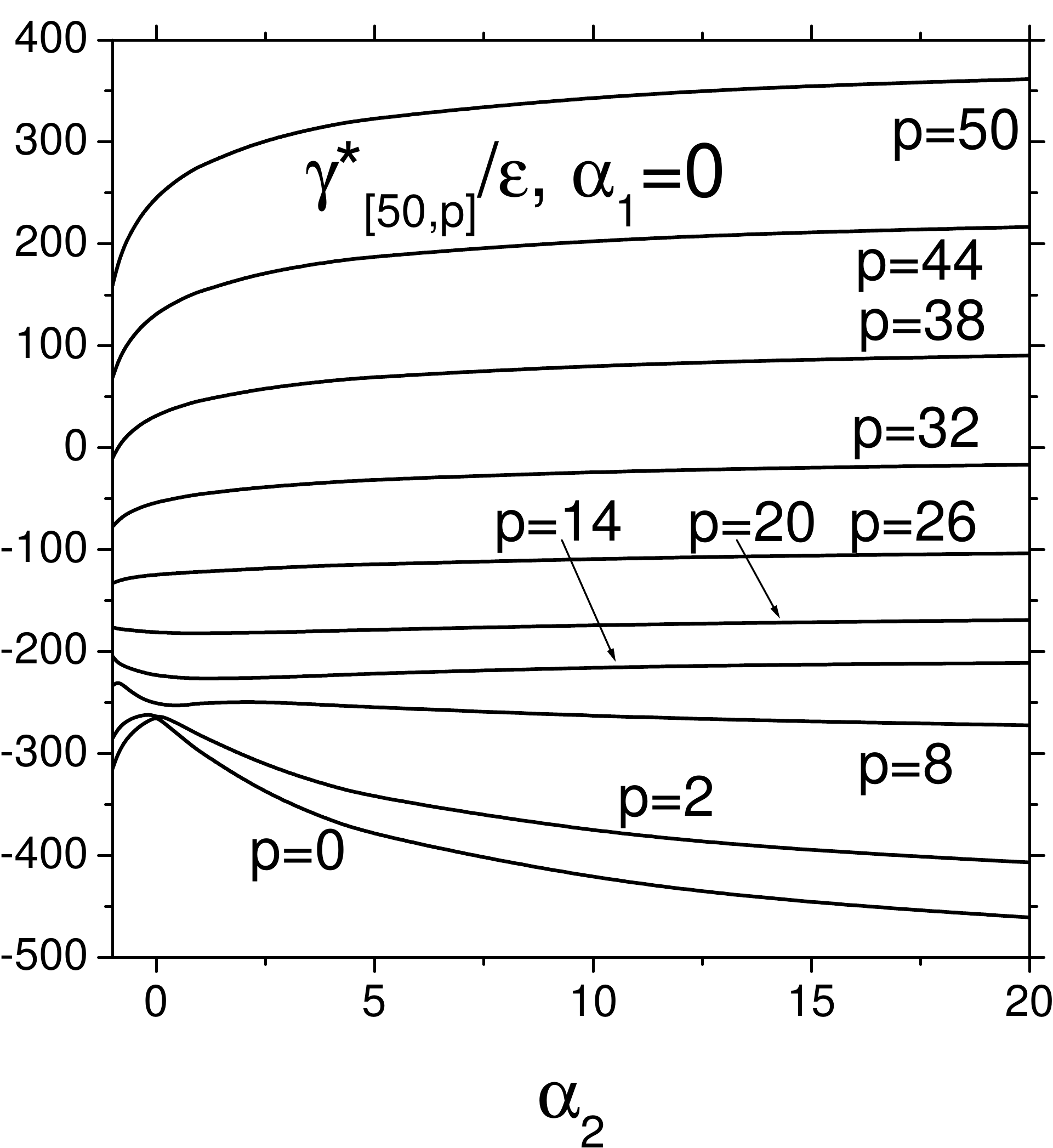}\hfill
       \includegraphics[width=4.25cm]{\PICS 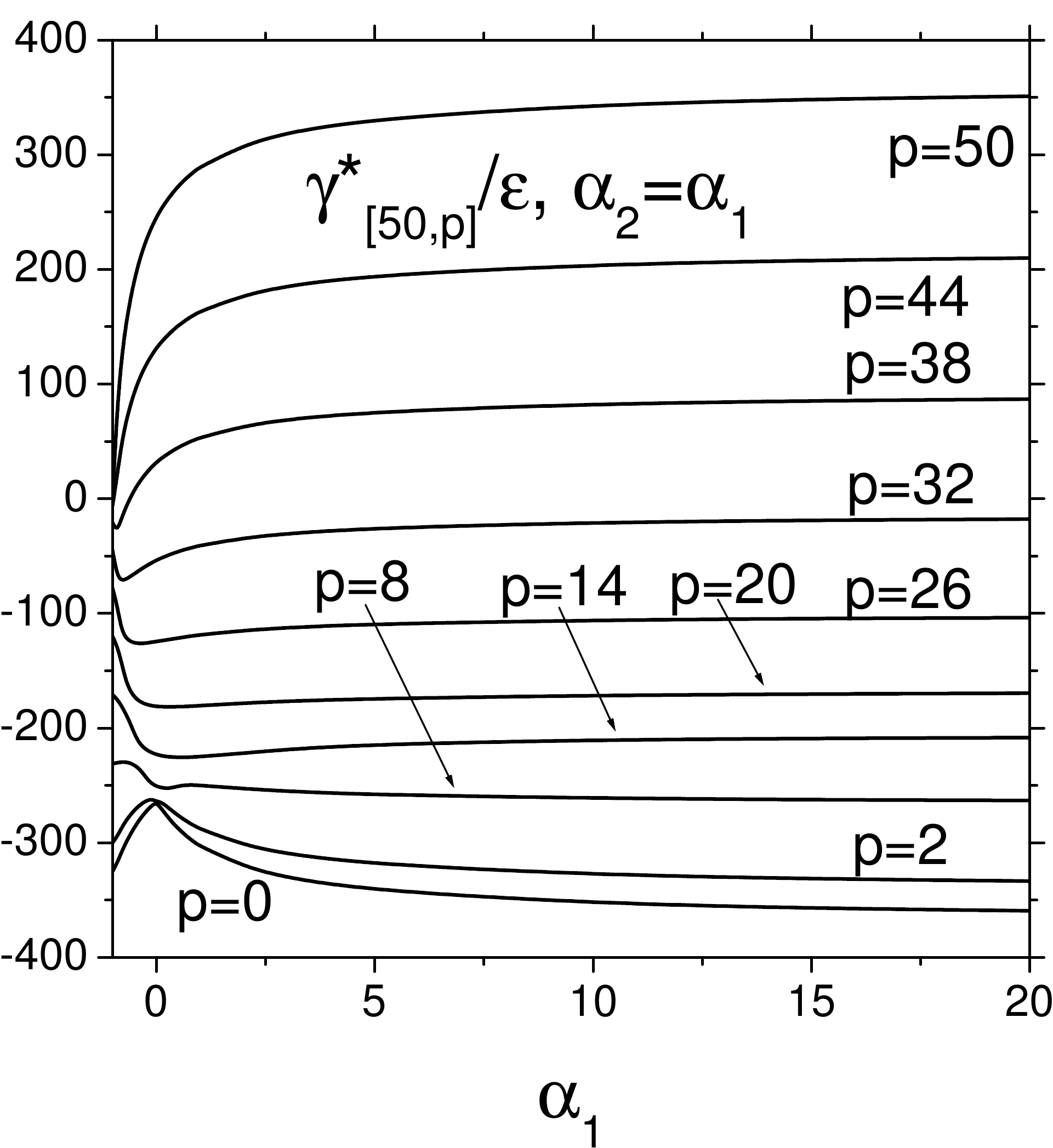}
\caption{Behavior of the anomalous dimension
$\gamma^*_{[50,p]}/\eps$ in space dimension $d=3$ and for
representative values of $p$ as functions of anisotropy parameters
$\alpha_1$ and $\alpha_2$.  \label{fig:mhdstr_fig5}}
\end{figure}

\begin{figure}
\centering
       \includegraphics[width=4.25cm]{\PICS 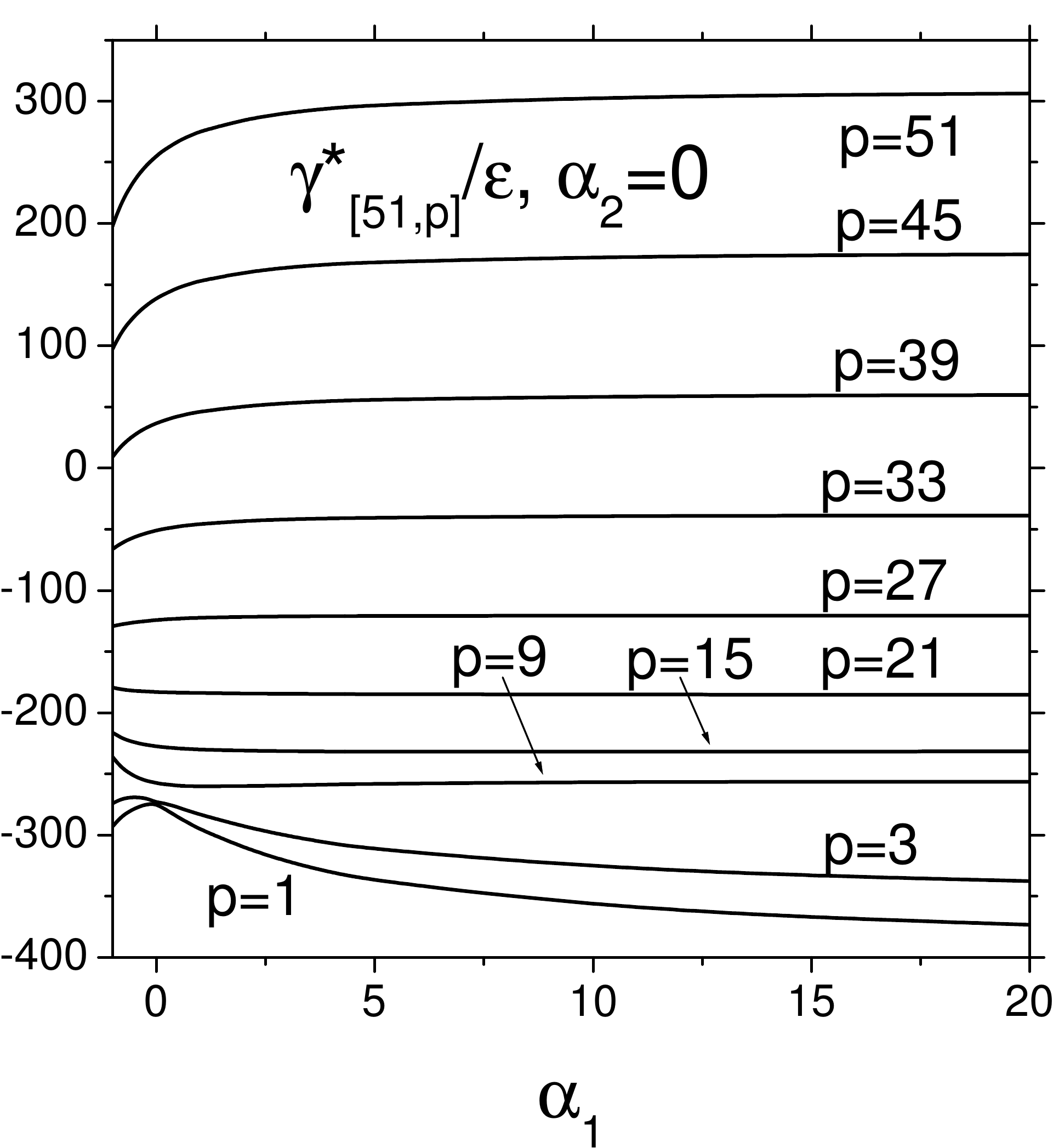}\hfill
       \includegraphics[width=4.25cm]{\PICS 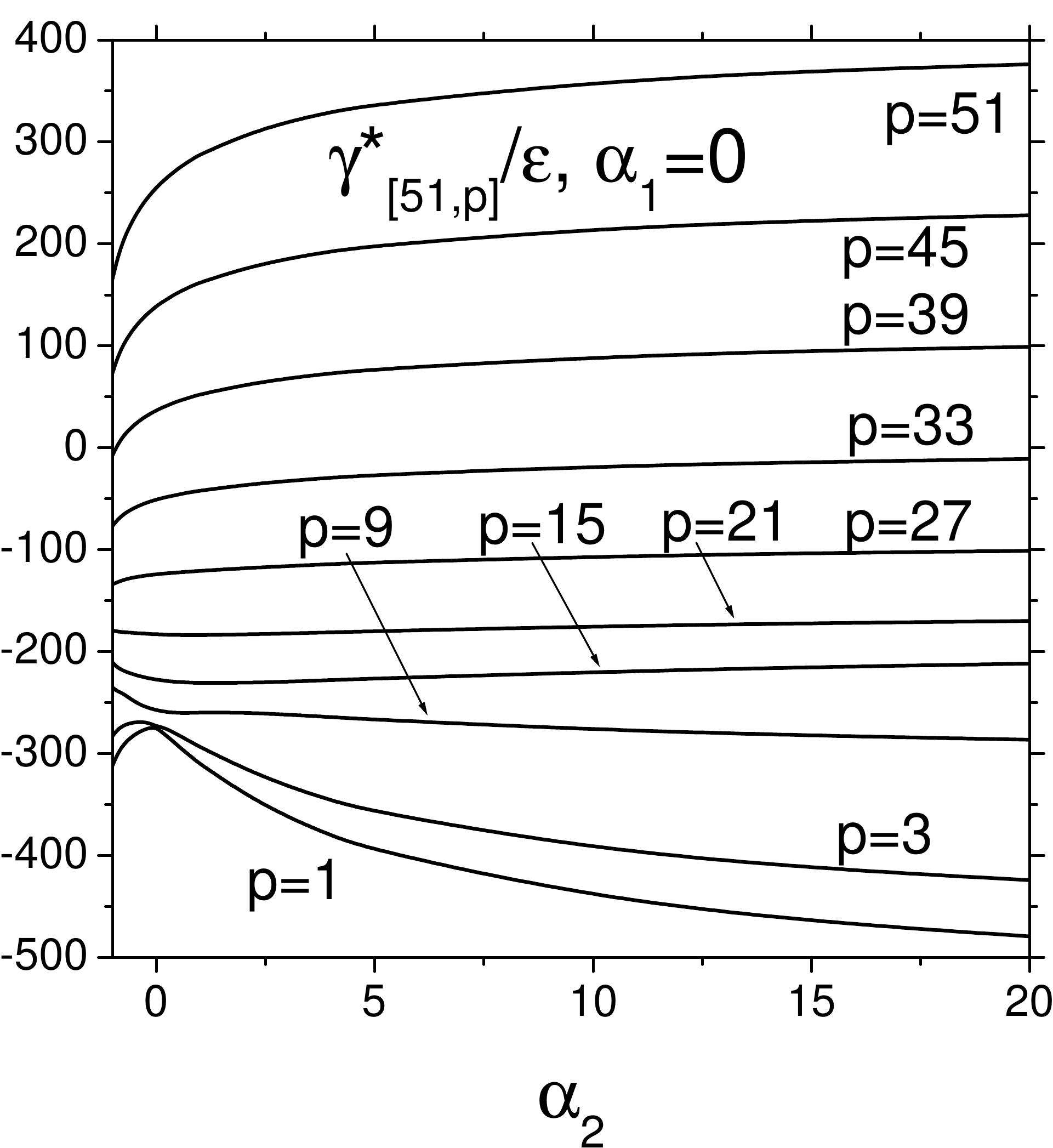}\hfill
       \includegraphics[width=4.25cm]{\PICS 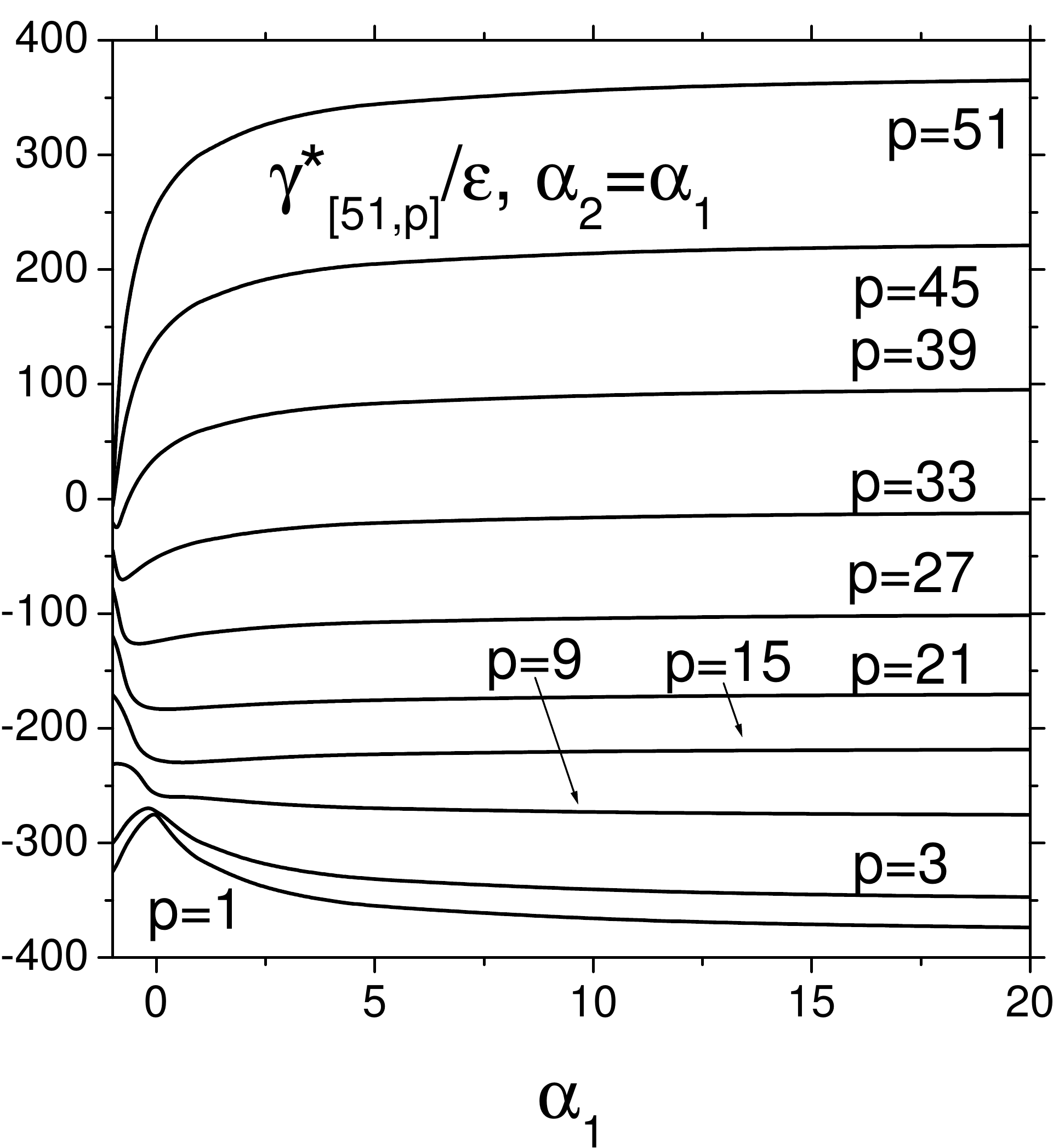}
\caption{Behavior of the anomalous dimension
$\gamma^*_{[51,p]}/\eps$ in space dimension $d=3$ and for
representative values of $p$ as functions of anisotropy parameters
$\alpha_1$ and $\alpha_2$.  \label{fig:mhdstr_fig6}}
\end{figure}

The decomposition (\ref{eq:mhdstr_Sbb}) reveals the
inertial-range scaling form of the correlation functions.
The leading singular contribution in the limit $L\to \infty$, $l\to 0$ is given by
the basis operator $\mathcal{F}[N,s]$ with the minimal critical dimension $\Delta[N,s]$
and operators $\mathcal{F}[N-m,p]$ and $\mathcal{F}[m,q]$ with the minimal sum of critical dimensions
$\Delta[N-m,p]+\Delta[m,q]$. We also remind that the critical dimensions of the basis operators have
the structure $\Delta[N,p]=-N\left(1-{\displaystyle\eps\over\displaystyle 2}\right)+\gamma_{[N,p]}^*$,
where $\gamma_{[N,p]}^*$ is the anomalous dimension. Therefore, in expression (\ref{eq:mhdstr_Sbb})
other contributions than the anomalous dimensions cancel in the power of separation distance $r$. As a result,
the correlation functions have asymptotic powerlike behavior as $r /L \to~0$
with the minimal anomalous dimension in the basis set generated by the composite
operator $b_r^{N}(t,{\mx})$ and the sum of minimal anomalous dimensions
in the basis set brought about by operators $b_r^{N-m}(t,{\mx})$ and $b_r^{m}(t,{\mx})$.
Calculation shows that these anomalous dimensions grow with the number of
field components in the anisotropy direction
and thus the minimal anomalous dimension $\underline{\gamma}_{\,N}^*$ in any set with fixed $N$ is
$\gamma_{[N,0]}^*=\gamma_{N}^*$ for
$N$ even and $\gamma_{[N,1]}^*$ for $N$ odd. Therefore, the leading asymptotic term of the correlation
functions in the inertial range is of the form (we remind that $N$ is an even integer here)
\begin{equation}
  \label{eq:mhdstr_BNmasy}
  B_{N-m,m}(r) \sim  \nu_0^{-N/2}L^{N}\left({l \over L}\right)^{{\gamma}_{\,N}^*+N\eps/2}
  \left({r \over l}\right)^{{\gamma}_{\,N}^*-\underline{\gamma}_{\,N-m}^*-\underline{\gamma}_{\,m}^*}
  \,,\qquad \underline{\gamma}_{\,1}^*=0\,,\qquad m\ge 1\,.
\end{equation}
Numerical results at one-loop order yield negative exponents
${\gamma}_{\,N}^*+N\eps/2<0$ and
${\gamma}_{\,N}^*-\underline{\gamma}_{\,N-m}^*-\underline{\gamma}_{\,m}^*<0$,
see Figs. \ref{fig:mhdstr_fig1} -- \ref{fig:mhdstr_fig6}.

A few words about the structure functions (\ref{eq:mhdstr_struc}) are in order. These structure functions
may be expressed as linear combinations of two-point correlation functions, in which the constant
term corresponding to $B_{N,0}$ is always present in even-order structure functions. The present
asymptotic analysis does not allow for direct comparison
of this constant and the powerlike
asymptotics predicted in relation (\ref{eq:mhdstr_BNmasy}) for the ''genuine'' two-point correlation functions
in the inertial range. If a powerlike behavior is to be detected, however, this constant must be small
at least compared to the leading powerlike term. One-loop results show that the leading powerlike term
in the inertial range corresponds to the two-point correlation function $B_{N-1,1}$ with the asymptotic
behavior ($\gamma_{\,1}^*=0$)
\begin{equation}
  \label{eq:mhdstr_BN-11}
  B_{N-1,1}(r) \sim  \nu_0^{-N/2} L^{N}\left({l \over L}\right)^{{\gamma}_{\,N}^*+N\eps/2}
  \left({r \over l}\right)^{{\gamma}_{\,N}^*-\underline{\gamma}_{\,N-1}^*}.
\end{equation}
A detailed account
of calculation
of the matrix of the renormalization
constants $Z_{[N,p][N,p']}$
(which may be readily extended
to investigation of all related problems)
has been given
in Ref. \cite{AAHN00} for the advection of a passive scalar and in Ref. \cite{HHJMS05} for passive vector admixture,
therefore we will not describe all details of the
determination of renormalization constants in the present vector model, rather we
will discuss its specific features.

It turned out that not only the $\beta$ functions in
the vector and scalar models coincide, but
the one-loop renormalization matrices as well.
This nontrivial fact stems from
the similarities of the mathematical structure of both models.
In the model of scalar advection
\cite{AAHN00} the composite operators
$\partial_{i_1}\theta ... \partial_{i_p}\theta (\partial_i\theta
\partial_i \theta)^l$ constructed solely of the scalar gradients
of the scalar admixture $\theta$
are needed for calculation of the asymptotic behavior of the
correlation functions, whereas
in vector case \cite{HHJMS05} the main contribution is given by
composite operators constructed solely of the fields ${\mb}$
without derivatives. As direct inspection of the
relevant diagrams shows, the tensor structures arising upon
functional averaging in both cases
are in fact identical, which yields the same renormalization matrix
$Z_{[N,p][N,p']}$ in both models. 

The only nonzero elements of the matrix $Z_{[N,p][N,p']}$ are
\begin{align*}
  Z_{[N,p][N,p-2]}&=\frac{g {\bar S}_d}{d^2-1}\frac{1}{4\eps}Q_1\,,
  &Z_{[N,p][N,p]}&=1+\frac{g {\bar S}_d}{d^2-1}\frac{1}{4\eps}Q_2\,,  \\
  Z_{[N,p][N,p+2]}&=\frac{g {\bar S}_d}{d^2-1}\frac{1}{4\eps}Q_3\,,
 &Z_{[N,p][N,p+4]}&=\frac{g {\bar S}_d}{d^2-1}\frac{1}{4\eps}Q_4\,,
\end{align*}
with the coefficients $Q_i$ given in \cite{HHJMS05} and ${\bar S}_d$ was given in Eq. (\ref{eq:double_gfactors}).
The nontrivial elements of the matrix of anomalous dimensions $\gamma_{[N,p][N,p']}$ are
\begin{align}
  \gamma_{[N,p][N,p-2]}&=-\frac{g {\bar S}_d}{4(d^2-1)}Q_1\,,
  &\gamma_{[N,p][N,p]}&=-\frac{g {\bar S}_d}{4(d^2-1)}Q_2\,, \nonumber \\
  \gamma_{[N,p][N,p+2]}&=-\frac{g {\bar S}_d}{4(d^2-1)}Q_3\,,
  &\gamma_{[N,p][N,p+4]}&=-\frac{g {\bar S}_d}{4(d^2-1)}Q_4\,, 
  \label{eq:mhdstr_andim}
\end{align}
and the matrix of critical dimensions (\ref{eq:RG_32B}) is thus
\begin{equation}
  \Delta_{[N,p][N,p']}=-N\left(1-{\eps\over 2}\right)\delta_{pp'} +
  \gamma^*_{[N,p][N,p']}\,,
  \label{eq:mhdstr_deltaff}
\end{equation}
where the asterisk stands for the value at the fixed point
(\ref{eq:mhdstr_fixg}), (\ref{eq:mhdstr_fixchi}). This represents the critical dimensions
of the composite operators (\ref{eq:mhdstr_CO}) at the first order in
$\eps$. It should to be stressed that in contrast to the value of
the fixed point (\ref{eq:mhdstr_fixg}), (\ref{eq:mhdstr_fixchi}), which has no higher
order corrections, the expressions for anomalous dimensions
(\ref{eq:mhdstr_andim}) have non-vanishing corrections of order $g^2$ and higher.

The critical dimensions $\Delta[N,p]=-N\left(1-{\eps/ 2}\right)+
\gamma_{[N,p]}^*$ are given by the eigenvalues of the matrix
(\ref{eq:mhdstr_deltaff}). 

Since the result for the anomalous dimensions is the same as in
Ref. \cite{AAHN00} for the admixture of a passive scalar, all
conclusions about the hierarchical behavior of the critical
dimensions of the composite operators are also valid in the
analysis of the vector model. Nevertheless, the inertial-range asymptotic behavior
of the correlation functions
in these two problems is completely different, because, first, in the scalar problem
single-point products of the scalar are not renormalized, while in the vector problem
they are, and, second, the leading contribution to the OPE is given by the products of derivatives
of the scalar, whereas in the vector problem products of the field components themselves
yield the leading contribution.

In Ref. \cite{AAHN00}
the behavior of the critical dimensions
$\Delta[N,p]$ for $N=2,3,4,5,$ and $6$ was
numerically studied.  The main conclusion is
that the dimensions $\Delta_N$ remain negative in anisotropic case
and decrease monotonically as $N$ increases for odd and even
values of $N$ separately.

Here the main attention is paid to  the investigation
of the composite operators (\ref{eq:mhdstr_CO}) for relatively large values
of $N$, namely we will analyze cases with
$N=10,11,20,21,30,31,40,41,50$, and $51$. The aim has been to
find out whether hierarchies which hold for small values of $N$
remain valid for significantly larger values of $N$, and the answer turned out to be
in the affirmative.

In Figs. \ref{fig:mhdstr_fig1}-\ref{fig:mhdstr_fig6}  behavior of the
eigenvalues of the matrix of anomalous dimensions $\gamma^*_{[N,p]}$ for
relatively large values of the $N$ are shown. It can be seen that only real
eigenvalues exist in all cases, and also their hierarchical
behavior discussed in Ref. \cite{AAHN00} is conserved. At
first sight the curves for $p=0$ and $p=2$ in the even case and
the curves for $p=1$ and $p=3$ in the odd case in
Figs. \ref{fig:mhdstr_fig3}-\ref{fig:mhdstr_fig6} appear to be crossing at the point $\alpha_1 =
\alpha_2=0$ but in fact the curves are only visually
running very near together at that point which is a mathematical
consequence of the formulas for critical dimensions in the
infinitesimal limit $\alpha_1 \rightarrow 0$ and $\alpha_2
\rightarrow 0$.

\begin{figure}
   \centering
       \includegraphics[width=5.cm]{\PICS 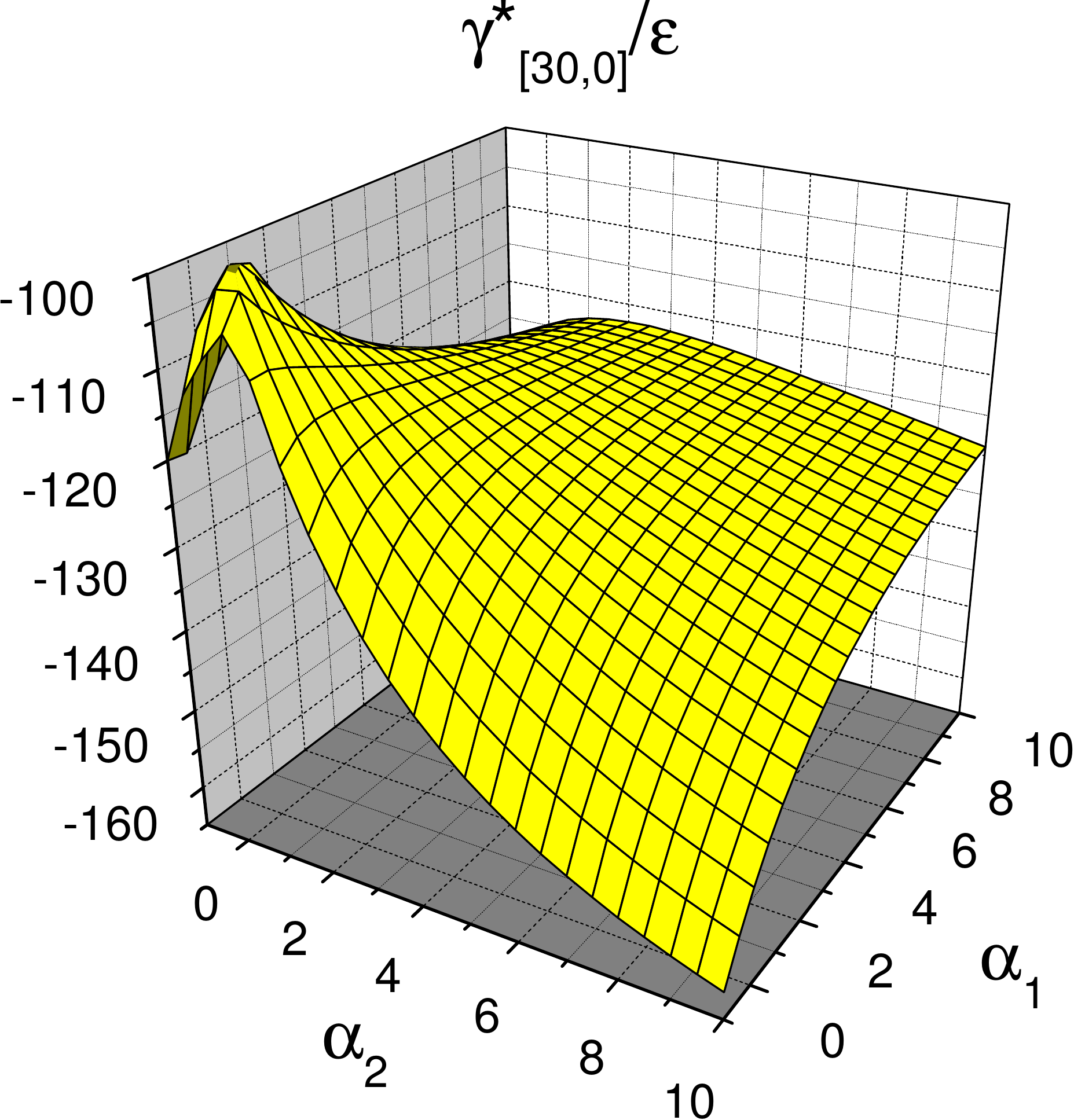}\hfill
       \includegraphics[width=5.cm]{\PICS 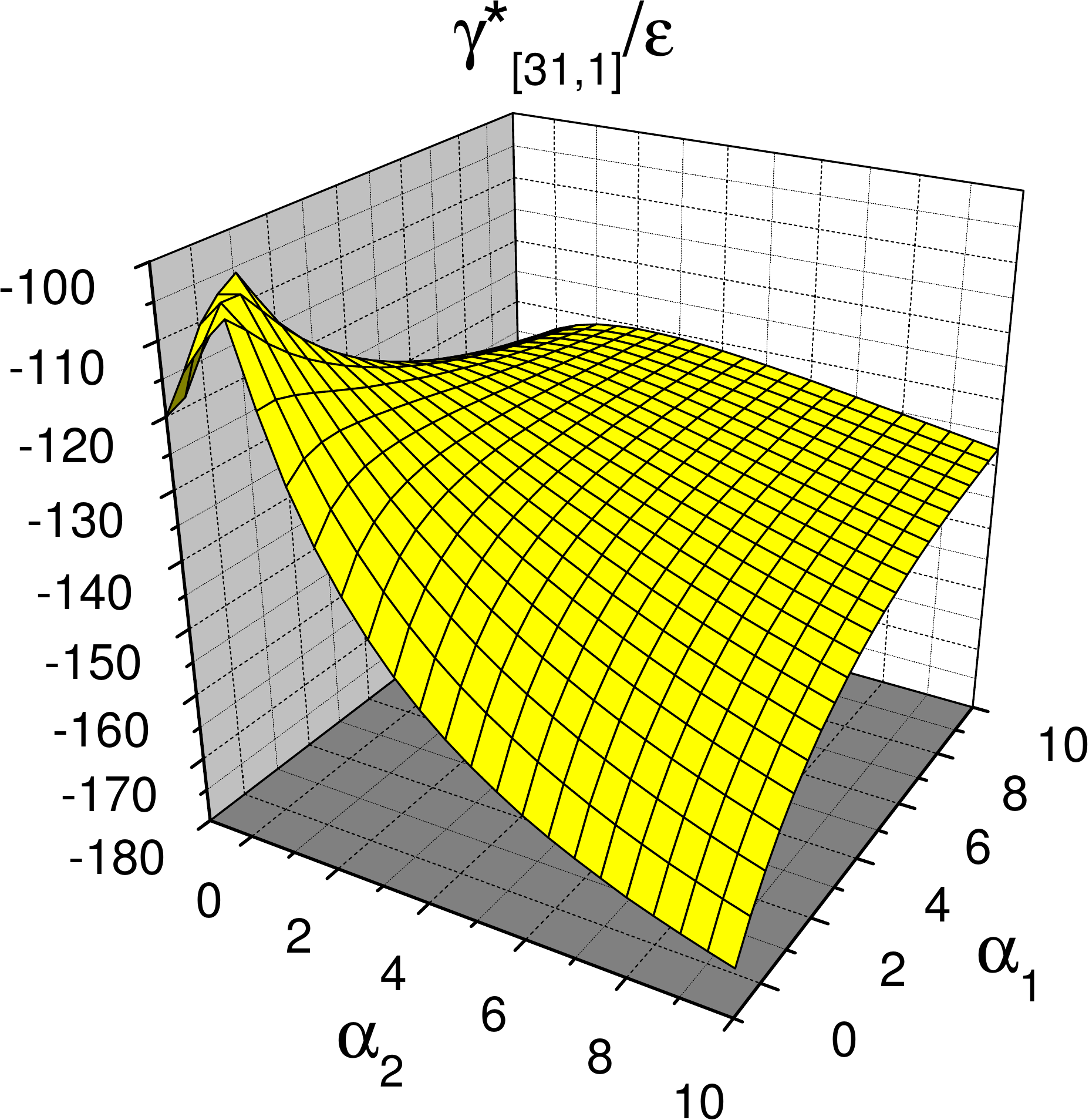}
\caption{Behavior of the anomalous dimensions
$\gamma^*_{[30,0]}/\eps$ and $\gamma^*_{[31,1]}/\eps$ in space
dimension $d=3$ as functions of anisotropy parameters $\alpha_1$
and $\alpha_2$.  \label{fig:mhdstr_fig7}}
\end{figure}

\begin{figure}
\centering
       \includegraphics[width=5.cm]{\PICS 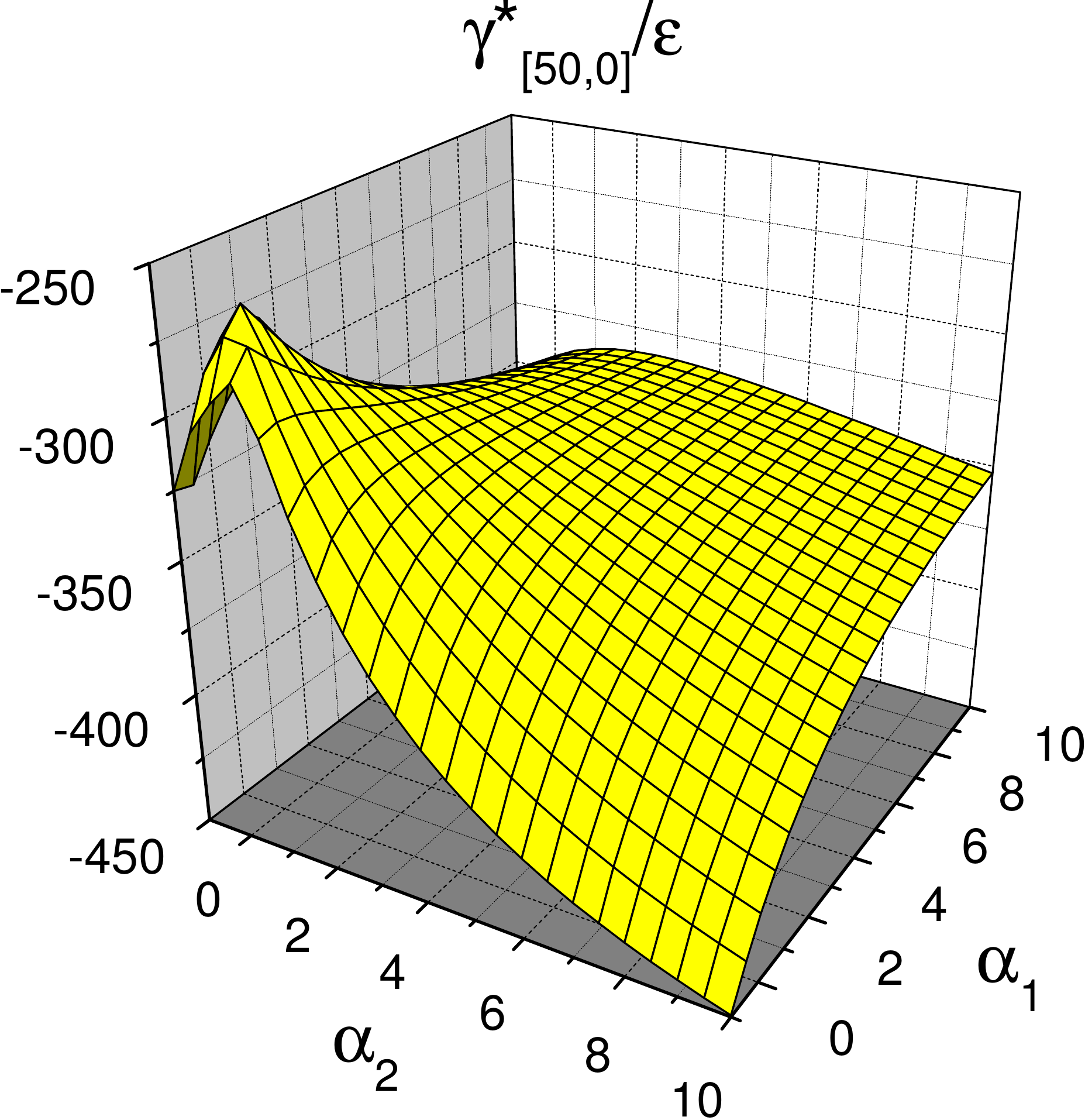}\hfill
       \includegraphics[width=5.cm]{\PICS 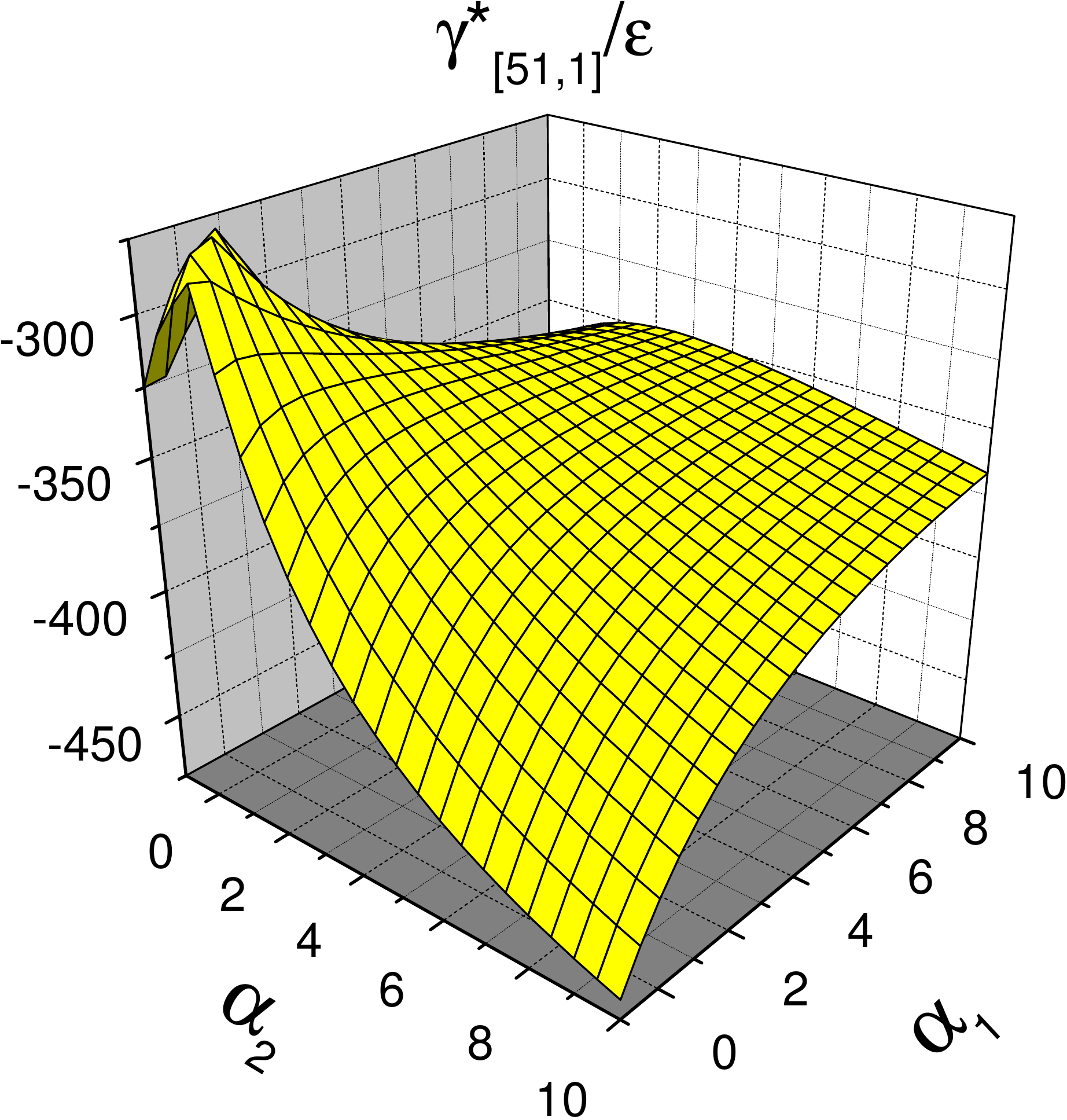}
\caption{Behavior of the anomalous dimensions
$\gamma^*_{[50,0]}/\eps$ and $\gamma^*_{[51,1]}/\eps$ in space
dimension $d=3$ as functions of anisotropy parameters $\alpha_1$
and $\alpha_2$.  \label{fig:mhdstr_fig8}}
\end{figure}

In Figs. \ref{fig:mhdstr_fig7}, \ref{fig:mhdstr_fig8} the eigenvalues are
presented as the functions of two variables $\alpha_1$ and
$\alpha_2$ for $N=30,31,50,51$ for the first the most singular
modes $p=0$ for even $N$ and $p=1$ for odd ones. It can clearly
be seen
that as values of the parameters $\alpha_{1}$ and $\alpha_{2}$ increase the anomalous dimensions
become more negative approaching some saturated values, therefore the
anisotropy amplifies the anomalous scaling.

From the
mathematical point of view the present model is found to be
similar to the model of a passive scalar quantity advected by a
Gaussian strongly anisotropic velocity field \cite{AAHN00} in
that the fluctuation contributions to the critical dimensions
$\Delta_F$ of the OPE representation of the
structure functions coincide in both cases and thus the
hierarchical dependence on the degree of anisotropy is also the
same.
Here, numerical calculation of the critical dimensions
$\Delta_F$ in the one-loop approximation has been extended to
dimensions related to correlation functions of order $N=51$ to
explore possible departures from powerlike asymptotic behavior.
However, contrary to the scalar case, in the inertial range the
leading terms of the structure functions of the magnetic field
themselves are shown to be coordinate independent with powerlike
corrections whose exponents are generated by the calculated
critical dimensions.

It is shown that in the inertial range the leading-order powerlike
asymptotic behavior of the correlation functions
is determined by the critical dimensions brought about already in the isotropic case,
which, however, acquire rather strong dependence on the parameters of anisotropy.
Powerlike corrections also appear with critical dimensions generated entirely by anisotropic
velocity fluctuations.
We have calculated numerically the anomalous
correction exponents up to order $N=51$ to explore possible oscillatory
modulation
or logarithmic corrections to the leading powerlike asymptotics, but have
found no sign of this kind of behavior: all calculated corrections have
had purely powerlike behavior.
Our results show that the exponents of the powerlike corrections tend
to decrease with increasing relative impact of the anisotropy.

From the renormalization-group point of view the present model of
passively advected vector field in the presence of
strong anisotropy is technically similar to
that of passively advected scalar field \cite{AAHN00}. In
particular, the $\beta$ functions and the one-loop
contributions to renormalization matrices of relevant
composite operators are the same. Since
in the published analysis of the scalar problem \cite{AAHN00} there were some misprints,
we have also presented corrected complete results of the calculation of the
renormalization matrices.

However, physically the two models differ significantly: instead of the
anomalous powerlike growth of the correlation and structure functions of the scalar problem
in the inertial range, in the present
vector case a powerlike falloff is predicted for the correlation and structure functions.
Moreover, in the scalar problem velocity fluctuation contributions to the inertial-range
scaling exponents are (small in $\eps$) corrections to the exponents determined by canonical dimensions
of the fields, whereas in the vector problem the inertial-range scaling exponents are solely determined by the
fluctuation contributions.

{\subsection{Developed turbulence with weak anisotropy} \label{subsec:weak_wta}}
Usually, the RG approach is applied to the
isotropic models of developed turbulence. However, the field theoretical approach can 
be used (with some modifications) in the theory of anisotropic
developed turbulence.
A crucial question immediately arises, whether the principal
properties of the isotropic case and anisotropic one are the same at
least at the qualitative level. If they are, then it is possible to
consider the isotropic case as a first step in the investigation of
real systems.
On this way of transition from the isotropic developed turbulence
into the anisotropic one we have to learn whether the scaling regime
does remain stable under this transition.
That means, whether the stable fixed points of
the RG equations remain stable under the influence of anisotropy.

During the last thirty years a few papers have appeared in which the above
question has been considered. In some cases it has been found out
that stability really takes place (see,e.g.\cite{Barton,Kim}). On the
other hand, existence of systems without such a stability has
been proved too.
As has been shown in Ref.\,\cite{Denis}, in the anisotropic
magnetohydrodynamic developed turbulence a stable regime generally
does not exist. In \cite{Kim,BHHH97}, $d$-dimensional models
with $d>2$ were investigated for two cases: weak anisotropy
\cite{Kim} and strong one \cite{BHHH97}, and it has been shown that the
stability of the isotropic fixed point is lost for dimensions
$d<d_c=2.68$. It has also been shown that stability of the fixed
point even for dimension $d=3$ takes place only for sufficiently
weak anisotropy.
The only
problem in these investigations is that it is impossible to use
them in the case $d=2$ because new UV divergences
appear in the Green functions when one considers $d=2$,
and they were not taken into account in \cite{Kim,BHHH97}.

In paper \cite{HonNal96}, a correct treatment of the two-dimensional
isotropic turbulence  has been given. The correctness in the
renormalization procedure has been reached by introduction into
the model a new local term (with a new coupling constant) which
allows one to remove additional UV-divergences. From this point of
view, the results obtained earlier for anisotropical developed
turbulence presented in \cite{Olla94} and based on the paper
\cite{Ronis87} (the results of the last paper are in conflict with
\cite{HonNal96}) cannot be considered as correct because they are
inconsistent with the basic requirement of the UV-renormalization,
namely with the requirement of the localness of the counterterms
\cite{Collins,Zinn}.

The authors of the paper \cite{Runov97} have used the
double-expansion procedure introduced in Sec. \ref{subsec:double_intro} for investigation of
developed turbulence with weak anisotropy for $d=2$. In such a
perturbative approach the deviation of the spatial dimension from
$d=2$, $\Delta=(d-2)/2$, and the deviation of the exponent of the
powerlike correlation function of random forcing from their
critical values, $\eps$, play the role of the expansion
parameters.

The main result of the paper \cite{Runov97} was the conclusion that the
two-dimensional fixed point is not stable under weak anisotropy.
It means that two-dimensional turbulence is very sensitive to the anisotropy
and no stable scaling regimes exist in this case.
In the case $d=3$, for both the isotropic turbulence and anisotropic one,
as it has been mentioned above,
existence of the stable fixed point, which  governs  the Kolmogorov
asymptotic regime, has been established by means of the RG approach
by using the analytical regularization procedure \cite{Barton,Kim,BHHH97}.
One can make analytical continuation from $d=2$ to the
three-dimensional turbulence (in the same sense as in the theory
of critical phenomena) and verify
whether the stability of the fixed point (or, equivalently,
stability of the Kolmogorov scaling regime) is restored.
From the analysis made in Ref.\,\cite{Runov97} it follows that it is
impossible to restore the   stable regime by transition from
dimension $d=2$ to $d=3$. We suppose  that the main reason for the above
described discrepancy is related to the straightforward
application of the standard MS scheme. In the standard MS scheme
one works  with the purely divergent part of the Green functions only,
and in concrete calculations its dependence on the space dimension
$d$ resulting from the tensor nature of these Green functions is
neglected (see Sect.3).
In the case of isotropic models, the stability of the fixed points
is independent of dimension $d.$
However, in anisotropic models the stability of fixed points
depends on the dimension $d$, and consideration of the tensor
structure of Feynman graphs in the analysis of their divergences
becomes important.

Here we describe modified MS scheme \cite{HJJS01} in which
we keep the $d-$dependence of UV-di\-ver\-gen\-ces of graphs.
There it was confirmed that after such a
modification the $d$-dependence is correctly taken into account and
can be used in investigation of whether it is possible to restore the
stability of the anisotropic developed turbulence for some
dimension $d_c$ when going from a two-dimensional system to
a three-dimensional one.
In the limit of infinitesimally weak anisotropy for the physically
most reasonable value of $\eps=2$, the value of the borderline
dimension is $d_c=2.44.$
Below  the borderline dimension, the stable regime of the fixed point of
the isotropic developed turbulence is lost by influence of weak
anisotropy.\\

\subsubsection{Description of the model}

As has already
been discussed above, the aim is now to study fully developed
turbulence with assumed weak anisotropy. It means that
the parameters that describe deviations from the fully isotropic case
are sufficiently small and allow one to forget about corrections
of higher degrees (than linear) which are made by them.

In the statistical theory of anisotropically developed turbulence the
turbulent flow can be described by a random velocity field
$\mv (t,\mx)$
and its evolution is given by the stochastically forced Navier-Stokes equation
\begin{equation}
  {\partial_t \mv} +
  ({\mv} \cdot \boldnabla)\mv
  -\nu_0 \boldnabla^2 \mv - \mf^A = \mf, 
  \label{eq:weak_NS}
\end{equation}
which is slightly modified with respect to the isotropic case (\ref{eq:double_NS}).
The incompressibility of the fluid is again assumed. The term $\mf^A$ is related
to anisotropy and will be specified later.  The large-scale random
force per  unit mass $\mf$ is assumed to have Gaussian
statistics defined by the averages
\begin{align}
  \langle f_i \rangle =0,\,\, \langle f_i(\mx_1 ,t)f_j(\mx_2 ,t)
  \rangle=D_{ij}(\mx_1 - \mx_2 ,t_1-t_2).
\end{align}
The two-point correlation matrix
\begin{align}
  D_{ij}(\vec x ,t)=\delta (t) \int \frac{\dRM^d \mk}{(2 \pi)^{d}}
  \tilde{D}_{ij}(\mk) \exp (i \mk \cdot \mx)
  \label{eq:weak_COR}
\end{align}
is convenient to be parameterized in the following way \cite{Barton,Denis}:
\begin{align}
  \tilde{D}_{ij}(\mk)=g_0 \nu_0^3 k^{4-d-2\eps}
  \biggl\{ \biggl[1+\alpha_{10} \frac{(\mn\cdot\mk)^2}{k^2} \biggl] P_{ij} (\mk) + \alpha_{20} \tilde{n}_i(\mk)\tilde{n}_j(\mk)\biggl\},
  \label{eq:weak_D}
\end{align}
where $\tilde{\mn}_i$ has been introduced in Eq. (\ref{eq:str_tildeN}).
 In what it is assumed that parameters $\alpha_{10}$ and $\alpha_{20}$ are small
enough and generate only small deviations from the isotropy case.

The action of the fields $\mv$ and $\mv^{,}$
is given in the compact form
\begin{align}
  \S & = \frac{1}{2} 
  v^{,}_{i}D_{ij} v^{,}_{j}  +   \mv^{,}
  \cdot\biggl[-\partial_t \mv -(\mv \cdot \boldnabla)\mv +
  \nu_0 {\boldnabla}^2 \mv +\mf^A \biggl].
  \label{eq:weak_action1}
\end{align}

Now we can return back to give an explicit form of the anisotropic
dissipative term $\mf^{A}$. When $d>2$ the UV-divergences  are
only present  in the one-particle-irreducible Green function
$\langle \mv^{,} \mv \rangle$. To remove them, one needs to introduce
into the action in addition to the counterterm 
$\mv^{,} {\boldnabla}^2 \mv$ (the only counterterm needed in the
isotropic model) the following ones $\mv^{,} (\mn \cdot
\boldnabla)^2 \mv$, $(\mn \cdot \mv^{,}) {\boldnabla}^2 (\mn 
\cdot \mv)$ and $(\mn \cdot \mv^{,}) (\mn \cdot {\boldnabla})^2 (\mn \cdot \mv)$. 
These additional terms are
needed to remove divergences related to anisotropic structures. In
this case ($d>2$), one can use the above action (\ref{eq:weak_action1})
with (\ref{eq:weak_D}) to solve the anisotropic turbulent problem.
Therefore, in order to arrive at the multiplicative renormalizable
model, it is necessary to take the term $\mf^A$ in the form
\begin{align}
  \mf^A=\nu_0 \biggl[\chi_{10}(\mn \cdot \boldnabla)^2 \mv +
  \chi_{20} \mn {\nabla}^2 (\mn \cdot \mv) +
  \chi_{30} \mn (\mn \cdot \boldnabla)^2 (\mn \cdot
  \mv) \biggl]\,.
  \label{eq:weak_sila}
\end{align}
Bare parameters $\chi_{10}$, $\chi_{20}$ and $\chi_{30}$
characterize the weight of the individual structures in
(\ref{eq:weak_sila}).

As was discussed in Sec. \ref{subsec:double_intro} more complicated situation arises in the specific  case $d=2$
where new divergences appear. They are related to the
1-irreducible Green function $\langle\mv^{,} \mv^{,} \rangle$ which is
finite when $d>2$. Here one comes to a problem how to remove these
divergences because the term in our action, which contains
a structure of this type is nonlocal, namely $\mv^{,} k^{4-d-2
\eps} \mv^{,}$. The only correct way of solving the above
problem is to introduce into the action a new local term of the form
$\mv^{,} {\nabla}^2 \mv^{,}$ (isotropic case)
\cite{HonNal96}. In the anisotropic case, we have to introduce
additional counterterms $\mv^{,} (\mn \cdot \boldnabla)^2 \mv^{,}$,
$(\mn \cdot \mv^{,}) {\boldnabla}^2 (\mn \cdot
\mv^{,})$ and $(\mn \cdot \mv^{,}) (\mn \cdot
{\boldnabla})^2 (\mn \cdot \mv^{,})$. 

In this case, the kernel (\ref{eq:weak_D}) corresponding to the correlation matrix
$D_{ij}(\mx_1 - \mx_2, t_2-t_1)$ in action (\ref{eq:weak_action1}) is
replaced by the expression
\begin{align}
  \tilde{{ D}}_{ij}(\mk) & = g_{10} \nu_0^3 k^{2-2\Delta
  -2\eps} \biggl\{ \biggl[ 1+\alpha_{10} \frac{(\mn\cdot\mk)^2}{k^2} \biggl] P_{ij}(\vec k) + \alpha_{20}
  \tilde{\mn}_i(\mk)\tilde{\mn}_j(\mk) \biggl\} \nonumber \\ 
  & +  g_{20} \nu_0^3 k^2
  \biggl\{ \biggl[1+\alpha_{30} \frac{(\mn\cdot\mk)^2}{k^2} \biggl] P_{ij}(\mk)+ \biggl[\alpha_{40} +
  \alpha_{50} \frac{(\mn\cdot\mk)^2}{k^2} \biggl] \tilde{\mn}_i(\mk)\tilde{\mn}_j(\mk) \biggl\} \,.
  \label{eq:weak_D1}
\end{align}
Here 
$g_{20}$, $\alpha_{30}$, $\alpha_{40}$ and $\alpha_{50}$ are new
parameters of the model, and the parameter $g_0$ in Eq.\,(\ref{eq:weak_D})
is now renamed as $g_{10}$. One can see that in such a formulation
the counterterm $\mv^{,} {\boldnabla}^2 \mv^{,}$ and all
anisotropic terms can be taken into account by renormalization of
the coupling constant $g_{20}$, and the parameters $\alpha_{30}$,
$\alpha_{40}$ and $\alpha_{50}$.

It has to be stressed that the last term of the $\vec f^A$ in Eq.
(\ref{eq:weak_sila}) which is characterized by the parameter $\chi_{30}$
and the term of the correlation matrix (\ref{eq:weak_D1}) related to the
parameter $\alpha_{50}$ are of the order $\O(n^4)$ in contrast to
the others which are either $\O(n^0)$ (the isotropic terms) or
$\O(n^2)$. Because we work in the limit of weak anisotropy this
fact has its consequence in the vanishing of their values at the
fixed point. On the other hand, the eigenvalues of the stability
matrix which correspond to the parameters $\chi_{30}$ and
$\alpha_{50}$ are of the same order, $\O(\varepsilon)$, as the
eigenvalues which correspond to the other parameters and they play
important role in determination of stability of the regime.

The action (\ref{eq:weak_action1}) with the kernel $\tilde{D}_{ij}(\mk)$
(\ref{eq:weak_D1})
is given in the form convenient for
realization of the quantum field perturbation analysis with
the standard Feynman diagram technique. From the quadratic  part of the
action one obtains the matrix of bare propagators. Their wave-number -
frequency representation reads
\begin{align}
  \Delta_{ij}^{vv}(\omega_k,\mk) & =  -\frac{K_3}{K_1 K_2} P_{ij}
  +  \frac{1}{K_1 (K_2+\tilde K (1-\xi_k^2))} \nonumber\\
  & \times\bigg[
  \frac{\tilde K K_3}{K_2}+
  \frac{\tilde K (K_3 + K_4 (1-\xi_k^2))}{(K_1+\tilde K (1-\xi_k^2))}-K_4
  \bigg] \tilde{\mn}_i(\mk) \tilde{\mn}_j(\mk) \nonumber \\
  \Delta_{ij}^{v v^{,}}(\omega_k,\mk) & =  \frac{1}{K_2} P_{ij} -
  \frac{\tilde K}{K_2 (K_2+\tilde K (1-\xi_k^2))} \tilde{\mn}_i(\mk) \tilde{\mn}_j(\mk)\,,
\end{align}
where for brevity following notation has been used
\begin{align}
  \xi_k & = \frac{\mn\cdot\mk}{k},\nonumber\\
  K_1 & = i \omega_k + \nu_{0}k^2 + \nu_0 \chi_{10} (\mn \cdot
  \mk)^2\,, \nonumber \\ 
  K_2 & =  - i \omega_k + \nu_{0}k^2 + \nu_0
  \chi_{10} (\mn \cdot \mk)^2\,, \nonumber \\
  K_3 & = -g_{10}\nu_0^3 k^{2-2\Delta -2\eps} (1+\alpha_{10}\xi_k^2) -
  g_{20}\nu_0^3 k^2 (1+\alpha_{30}\xi_k^2)\,, \nonumber \\ 
  K_4 & = 
  -g_{10}\nu_0^3 k^{2-2\Delta -2\eps} \alpha_{20} - g_{20}\nu_0^3
  k^2 (\alpha_{40}+\alpha_{50}\xi_k^2)\,, \nonumber \\
  \tilde K & =
  \nu_0 \chi_{20} k^2 + \nu_0 \chi_{30} (\mn \cdot \mk)^2\,.
  \label{eq:weak_parts}
\end{align}
The propagators are written in the form suitable also for strong
anisotropy when the parameters $\alpha_{i0}$ are not small.
In the case of weak anisotropy, it is possible to make the expansion and
work only with linear terms with respect to all parameters which
characterize anisotropy.  

\subsubsection{RG-analysis and stability of the fixed point}

Using the standard analysis of quantum field theory (see e.g.
\cite{Adzhemyan96,Vasiliev,Collins,Zinn}), one can find out that the UV divergences
of one-particle-irreducible Green functions $\langle \mv' \mv \rangle_\text{1-ir}$ and
$\langle \mv' \mv' \rangle_\text{1-ir}$
are quadratic in the wave vector.
The last one takes place only in the case when
dimension of the space is two. All terms needed for removing
the divergences are included in the action (\ref{eq:weak_action1}) with
(\ref{eq:weak_sila}) and kernel (\ref{eq:weak_D1}). This leads to the fact that
 the model is multiplicatively renormalizable. Thus, one can
immediately write down the renormalized action in wave-number -
frequency representation with ${\boldnabla} \rightarrow i\mk,
\partial_t \rightarrow -i\omega_k$ (all needed integrations and
summations are assumed)
\begin{align}
  \S^{R} [\mv, \mv '] & =  \frac{1}{2} v_i^{,}\biggl\{ g_1 \nu^3 \mu^{2 \eps}
  k^{2-2\Delta -2\eps} \biggl(1+\alpha_1 \xi_k^2\biggl)P_{ij} +
  \alpha_2 \tilde{\mn}_i(\mk) \tilde{\mn}_j(\mk) \nonumber \\ 
  & +  g_2 \nu^3 \mu^{-2\Delta} k^2
  \biggl[\biggl( Z_5 + Z_6 \alpha_3 \xi_k^2\biggl) P_{ij} + \biggl(Z_7 \alpha_4
  +Z_8 \alpha_5 \xi_k^2 \biggl) \tilde{\mn}_i(\mk) \tilde{\mn}_j(\mk) \biggl] \biggl\} v_{j}^{,} 
  \nonumber \\
  & +  v_i^{,} \biggl\{(i \omega_k -Z_1 \nu k^2)P_{ij}
  -\nu k^2 \biggl[Z_2 \chi_1 \xi_k^2 P_{ij} +
  \biggl(Z_3 \chi_2 + Z_4 \chi_3 \xi_k^2 \biggl)\nonumber\\
  &\times \tilde{\mn}_i(\mk) \tilde{\mn}_j(\mk) \biggl] \biggl\}v_{j}
   +  \frac{1}{2}v_i^{,} v_j v_l V_{ijl}.
  \label{eq:weak_action3}
\end{align}
Quantities $g_i$, $\chi_i$,
$\alpha_3$, $\alpha_4$, $\alpha_5$ and $\nu$ are the renormalized
counterparts of bare ones and $Z_i$ are renormalization constants
which are expressed via the UV divergent parts of the functions
$\langle \mv' \mv \rangle_\text{1-ir}$ and $\langle \mv' \mv'\rangle_\text{1-ir}$. Their general form in one
loop approximation is
\begin{align}
  Z_i=1-F_i\,  \mbox{Poles}_i^{\Delta , \eps}\,.
  \label{eq:weak_zet}
\end{align}

In the standard MS scheme the amplitudes $F_i$ are only some functions of
 $g_i$, $\chi_i$, $\alpha_3$, $\alpha_4,$ $\alpha_5$  and are independent
of $d$ and $ \eps.$ The terms  $\mbox{Poles}_i^{\Delta ,
\eps}$ are given by linear combinations of the poles
$\frac{1}{2\eps}$, $\frac{1}{2\Delta}$ and $\frac{1}{4\eps
+2\Delta}$ (for $\Delta \rightarrow 0, \eps \rightarrow 0 $).
The amplitudes $F_i=F_i^{(1)} F_i^{(2)}$ are a product of two multipliers
$F_i^{(1)}, F^{(2)}_i.$ One of them, say, $F_i^{(1)}$ is a multiplier
originating from the divergent part of the Feynman diagrams, and
the second one $F_i^{(2)}$ is connected only with the tensor nature of
the diagrams. For illustration purposes let us consider the following simple example (UV-divergent diagram)
\begin{equation}
  I({\mk}, {\mn}) \equiv
  n_i n_j k_l k_m\int \dRM^d {\mq} \frac{1}{(q^2+m^2)^{1+2\Delta}}
  \biggl(\frac{q_iq_jq_lq_m}{q^4}
  -\frac{\delta_{ij}q_lq_m+\delta_{il}q_jq_m+\delta_{jl}q_iq_m}{3q^2}
  \biggl),
\end{equation}
where summations over repeated indices are implied. It can be simplified in the following way: 
\begin{equation}
  I({\mk}, {\mn}) \equiv n_i n_j k_l k_m S_{ijlm}\int_0^{\infty}
  \dRM q^2 \frac{q^{2\Delta}}{2(q^2+m^2)^{1+2\Delta}},
\end{equation}
 where
\begin{equation}
  S_{ijlm}=\frac{S_d}{d(d+2)}\biggl(\delta_{ij}\delta_{lm}+\delta_{il}\delta_{jm}+
  \delta_{im}\delta_{jl}-\frac{(d+2)}{3}(\delta_{ij}\delta_{lm}+
  \delta_{il}\delta_{jm}+\delta_{im}\delta_{jl})\biggl),
\end{equation}

\begin{equation}
  \int_0^{\infty} \dRM q^2
  \frac{q^{2\Delta}}{2(q^2+m^2)^{1+2\Delta}}=\frac{\Gamma{(\Delta
  +1)}\Gamma{(\Delta)}}{2 m^{2\Delta}\Gamma(2\Delta+1)}.
\end{equation}  
 The purely UV divergent part manifests
itself as the pole in $\Delta$; therefore, one finds 
$$\mbox{UV
div. part of}\,\,\,\, I= \frac{1}{2\delta}(F^{(2)}_1 k^2+F^{(2)}_2
({\mn\cdot\mk})^2),
$$
where $F^{(2)}_1=F^{(2)}_2/2 = (1-d)S_d/3d(d+2)$ ($F_1^{(1)}=F_2^{(1)}=1$).

In the standard MS scheme one puts $d=2$ in $F^{(2)}_1,F^{(2)}_2$;
therefore
the $d$-dependence of these multipliers is ignored. For the theories
with vector fields and, consequently, with tensor diagrams, where
the sign of values of fixed points and/or their stability depend
on the dimension $d$, the procedure, which eliminates the
dependence of multipliers of the type $F^{(2)}_1,F^{(2)}_2$ on $d,$
is not completely correct because one is not able to control the
stability of the fixed point when drives to $d=3.$ In the analysis
of Feynman diagrams slightly modified MS scheme was proposed \cite{HJJS01} in
such a way that we keep the $d$-dependence of $F$ in (\ref{eq:weak_zet}).
The following calculations of RG functions ($\beta$ functions and
anomalous dimensions) allow one to arrive at the results which are in
qualitative agreement with the results obtained  in
the framework of the simple analytical regularization scheme
\cite{BHHH97}, i.e., the fixed point was obtained \cite{HJJS01} which is not stable for
$d=2,$ but whose stability is restored for a borderline
dimension $2<d_c<3.$

The transition from the action (\ref{eq:weak_action1}) to the
renormalized one (\ref{eq:weak_action3}) is given by the introduction of
the following renormalization constants  $Z$:
\begin{align}
  \nu_0=\nu Z_{\nu}\,,\,\,
  g_{10}=g_1 \mu^{2\eps}Z_{g_1}\,,\,\,
  g_{20}=g_2 \mu^{-2\Delta}Z_{g_2}\,,\,\,
  \chi_{i0}=\chi_i Z_{\chi_i}\,,\,\,
  \alpha_{(i+2)0}=\alpha_{i+2}Z_{\alpha_{i+2}},
  \label{eq:weak_ren}
\end{align}
where $i=1,2,3$. By comparison of the corresponding terms in the
action (\ref{eq:weak_action3}) with definitions of the renormalization
constants $Z$ for the parameters (\ref{eq:weak_ren}), one can immediately
write down relations between them
\begin{align}
  Z_{\nu} &  = Z_1,\quad
  Z_{g_1} & = Z_{1}^{-3},\quad
  Z_{g_2} & = Z_{5}Z_1^{-3},\quad
  Z_{\chi_{i}} & = Z_{1+i}Z_{1}^{-1},\quad
  Z_{\alpha_{i+2}} & = Z_{i+5}Z_{5}^{-1}, 
  \label{eq:weak_rel}
\end{align}
where again $i=1,2,3$.

In the one-loop approximation, divergent one-irreducible Green
functions are  
$\langle \mv' \mv\rangle_\text{1-ir}$ and $\langle \mv' \mv'\rangle_\text{1-ir}$
Corresponding divergent parts of these diagrams
$\Gamma^{v^{,}v^{,}},$ $ \Gamma^{v^{,}v}$ have the structure
\begin{align}
  \Gamma_{v^{,}v^{,}}& = \frac{1}{2}\nu^3 A 
  \Bigg[\frac{g_1^2}{4\eps +2\Delta} \left(a_1 \Delta_{ij} k^2+
  a_2 \delta_{ij} (\mn \cdot \mk)^2+
  a_3 n_i n_j k^2 + a_4 n_i n_j (\mn \cdot \mk)^2\right) \nonumber \\
  & + \frac{g_1 g_2}{2\eps}
  \left(b_1 \delta_{ij} k^2+
  b_2 \delta_{ij} (\mn \cdot \mk)^2+
  b_3 n_i n_j k^2 + b_4 n_i n_j (\mn \cdot \mk)^2\right) \nonumber \\
  & + \frac{g_2^2}{-2\Delta}
  \left(c_1 \delta_{ij} k^2+
  c_2 \delta_{ij} (\mn \cdot \mk)^2+
  c_3 n_i n_j k^2 + c_4 n_i n_j (\mn \cdot \mk)^2\right)
  \Bigg]\,,\nonumber \\
  \Gamma_{v^{,}v} & = -\nu A  \Bigg[
  \frac{g_1}{2\eps} \left(d_1 \delta_{ij} k^2+
  d_2 \delta_{ij} (\mn \cdot \mk)^2+
  d_3 n_i n_j k^2 + d_4 n_i n_j (\mn \cdot \mk)^2\right) \nonumber \\
  & + \frac{g_2}{-2\Delta} \left(e_1 \delta_{ij} k^2+
  e_2 \delta_{ij} (\mn \cdot \mk)^2+
  e_3 n_i n_j k^2 + e_4 n_i n_j (\mn \cdot \mk)^2\right) \Bigg]\,,
\end{align}
where parameter $A$ and functions $a_i$, $b_i$, $c_i$, $d_i$ and
$e_i$ are not not needed for a discussion of basic idea. Their explicit
expressions can be found in \cite{HJJS01}.
The counterterms are built up from these divergent
parts  which lead to the following equations for
renormalization constants:
\begin{align}
  Z_1 & =  1- A \left( \frac{g_1}{2\eps} d_1 - \frac{g_2}{2\Delta}
  e_1\right)\,, \nonumber \\ 
  Z_{1+i} & =  1- \frac{A}{\chi_i}
  \left(\frac{g_1}{2\eps} d_{1+i} - \frac{g_2}{2\Delta}
  e_{1+i}\right)\,, \nonumber \\
  Z_5 & =  1+
  \frac{A}{2}\left(\frac{g_1^2}{g_2}\frac{a_1}{4\eps +2\Delta} +
  \frac{g_1}{2\eps} b_1 - \frac{g_2}{2\Delta} c_1\right)\,,
  \nonumber \\ 
  Z_{5+i} & =  1+ \frac{A}{2 \alpha_{i+2}} \left(
  \frac{g_1^2}{g_2}\frac{a_{i+1}}{4\eps +2\Delta} +
  \frac{g_1}{2\eps} b_{i+1} - \frac{g_2}{2\Delta}
  c_{i+1}\right)\,,\nonumber \\ 
  i & = 1,2,3\,.
\end{align}

By substitution of the anomalous dimensions $\gamma_i$
(\ref{gamma}) into the expressions for the $\beta$-functions one
obtains
\begin{align}
  \beta_{g_1} & = g_1 \bigg[-2\eps + 3 A (g_1 d_1 + g_2 e_1) \bigg]\,, \nonumber \\
  \beta_{g_2} & = g_2 \left[2\Delta + 3 A (g_1 d_1 + g_2 e_1)+
   \frac{A}{2}\left(\frac{g_1^2}{g_2} a_1 +
  g_1 b_1 + g_2 c_1\right) \right]\,,
  \nonumber \\
  \beta_{\chi_i} & = - A \left[\left(g_1 d_{i+1} + g_2 e_{i+1}\right)-
  \chi_i (g_1 d_1 + g_2 e_1)\right]\,, \nonumber \\
  \beta_{\alpha_{i+2}} & = 
  -\frac{A}{2} \left[-\left(\frac{g_1^2}{g_2} a_{i+1}+
  g_1 b_{i+1} + g_2 c_{i+1}\right) +
  \alpha_{i+2}\left(\frac{g_1^2}{g_2} a_1 + g_1 b_1 + g_2 c_1\right)
  \right]\,,
  \nonumber \\
  i & = 1,2,3\,.
\end{align}

Now we have all necessary tools at hand to investigate the fixed
points and their stability. In {\it the isotropic case} all
parameters which are connected with the anisotropy are equal to zero,
and one can immediately find the Kolmogorov fixed point, namely:
\begin{align}
  g_{1*} & = \frac{1}{A}\,\,
  \frac{8 (2 + d) \eps (2 \eps - 3 d (\Delta + \eps) +
  d^2 (3 \Delta + 2 \eps))}{9(-1 + d)^3 d (1 + d)(\Delta +
  \eps)}\,,\nonumber \\
  g_{2*} & = \frac{1}{A}\,\,
  \frac{8 (-4-2 d+ 2 d^2 +d^3)\eps^2}{9(-1 + d)^3 d (1 + d)(\Delta +
  \eps)}\,,
  \label{eq:weak_IZO}
\end{align}
where corresponding $\Omega$ matrix
has the following eigenvalues:
\begin{align}
  \lambda_{1,2} & = \frac{1}{6 d (d-1)}\Bigg\{6 d \Delta (d-1)+4\eps
  (2- 3 d + 2 d^2)
  \nonumber \\
  &\pm  \Bigg[ (6 d \Delta (1-d) -4\eps (2-3 d + 2 d^2))^2 \nonumber \\
  & -  12 d (d-1) \eps (12 d \Delta (d-1) +4 \eps (2-3 d + 2 d^2))
  \Bigg]^{\frac{1}{2}}
  \Bigg\}\,.
\end{align}
From detailed analysis of these eigenvalues it follows \cite{HJJS01} that in the
interesting region of parameters, namely $\eps>0$ and $\Delta
\geq0$ (it corresponds to $d \geq 2$) the above computed fixed
point is stable. In the limit $d=2$, this fixed point is in
agreement with that given in \cite{HonNal96,Runov97}.

When one considers {\it the weak anisotropy case} the situation
becomes more complicated because of necessity to use all system of
$\beta$-functions if one wants to analyze the stability of the
fixed point. It is also possible to find analytical expressions
for the fixed point in this more complicated case because in the weak
anisotropy limit it is enough to calculate linear corrections of
$\alpha_1$ and $\alpha_2$ to all the quantities.

To investigate the stability of the fixed point it is necessary to
apply it in the  matrix (\ref{eq:RG_Omega}). Analysis of this matrix shows
us that it can be written in the block-diagonal form: $(6\times
6)(2\times 2)$. The $(2\times 2)$ part is given by the
$\beta$-functions of the parameters $\alpha_5$ and $\chi_3$ and,
namely, this block is responsible for the existence of the borderline
dimension $d_c$ because one of its eigenvalues, say
$\lambda_1(\eps,d,\alpha_1,\alpha_2)$, has a solution $d_c \in
\langle 2, 3\rangle$ of the equation
$\lambda_1(\eps,d_c,\alpha_1,\alpha_2)=0$  for the defined values
of $\eps, \alpha_1, \alpha_2$. Te details of determination of fixed points' structure
and the corresponding eigenvalues of the stability matrix
responsible for instability can be found in \cite{HJJS01}.

\begin{figure}[!htb]
    \begin{minipage}{0.475\textwidth}
       \mbox{ } 
        \includegraphics[width=6cm]{\PICS 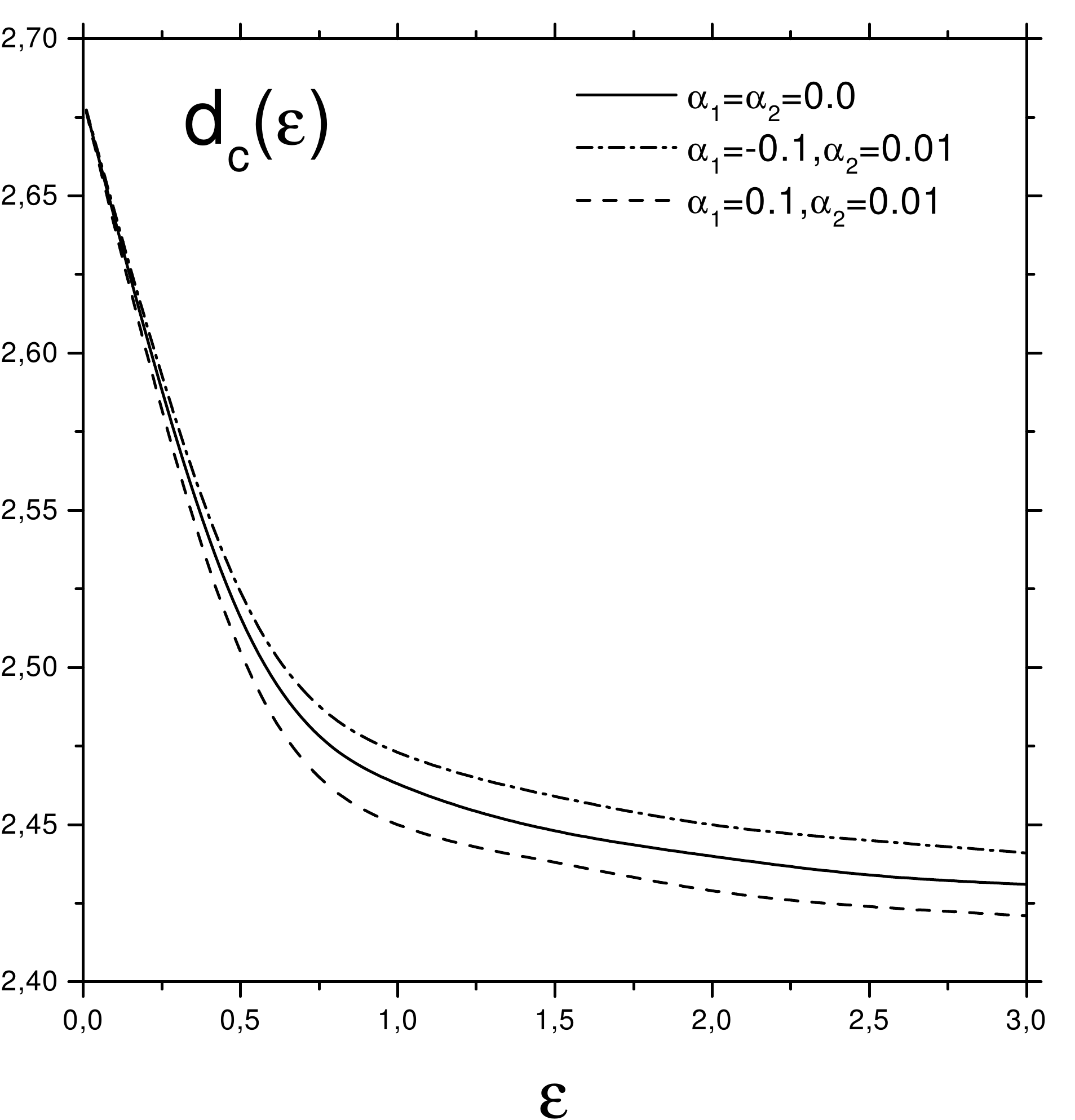}
        \caption{Dependence of the borderline dimension
$d_c$ on the parameter $\epsilon$ for concrete values of
$\alpha_1$ and $\alpha_2$.
      }
 \label{fig:weak_fig4}
    \end{minipage}%
    \hfill
    \begin{minipage}{0.475\textwidth}
         \mbox{ } 
        \includegraphics[width=6cm]{\PICS 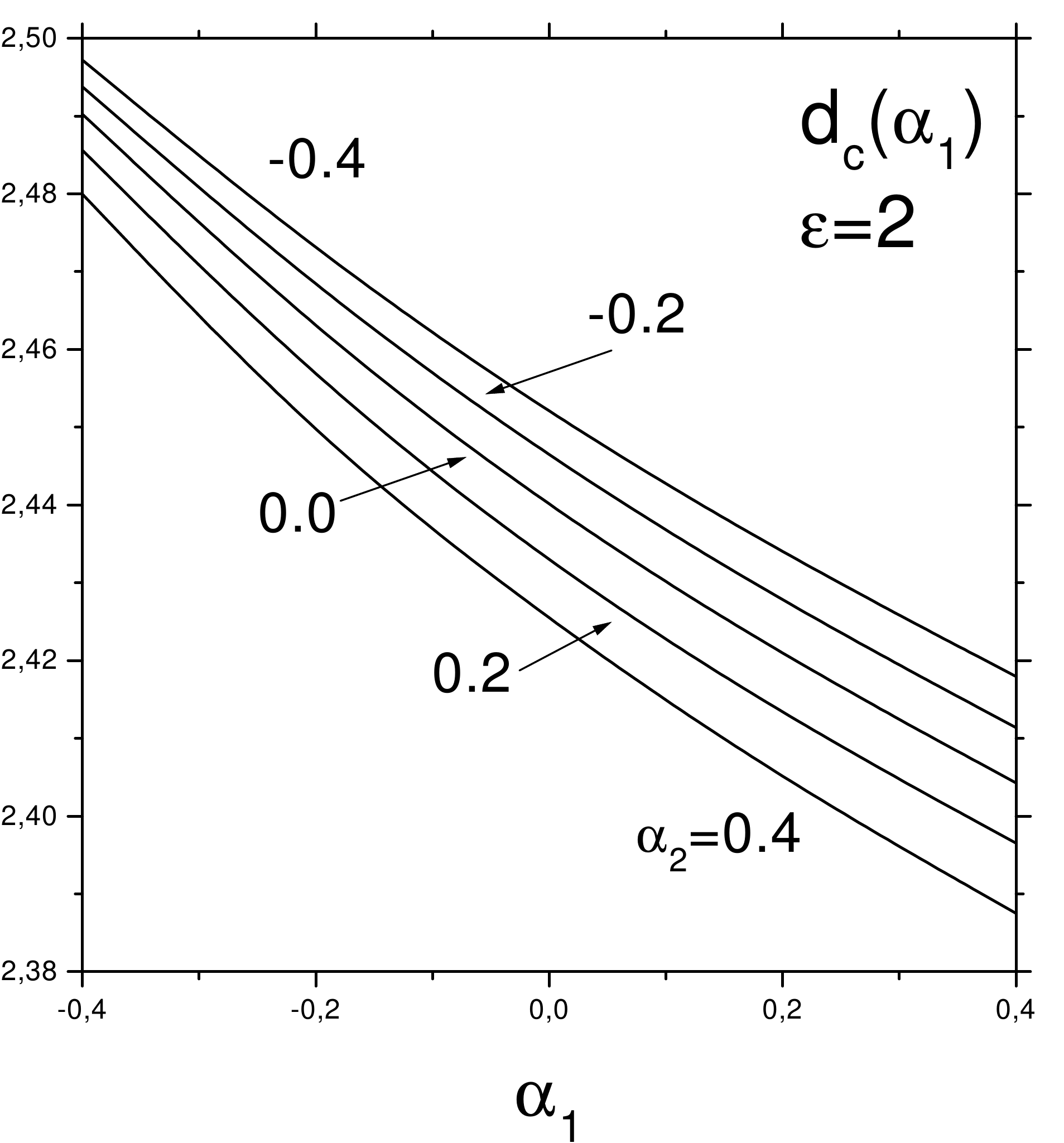}        
        \caption{ Dependence of the borderline dimension
$d_c$ on the parameters $\alpha_1$ and $\alpha_2$ for physical
value $\epsilon=2$.
}
\label{fig:weak_fig5}
    \end{minipage}
\end{figure}

For the energy pumping regime ($\eps=2$) and $\alpha_1=\alpha_2=0$
 the critical dimension $d_c=2.44$ is found.
This is the case when one supposes only the
fact of anisotropy. Using nonzero values of $\alpha_1$ and $\alpha_2$
one can also estimate the influence of these parameters on the
borderline dimension $d_c$. It is interesting to calculate the
dependence of  $d_c$ on the parameter $\eps$ too. In
Fig.\,\ref{fig:weak_fig4}, this dependence and the dependence on small values
of $\alpha_1$ and $\alpha_2$ are presented. As one can see from this
figure $d_c$ increases when $\eps \rightarrow 0$ and also the
parameters $\alpha_1$ and $\alpha_2$ give small corrections to $d_c$.
In Fig.\,\ref{fig:weak_fig5}, the dependence of $d_c$ on $\alpha_1$ and
$\alpha_2$ for $\eps=2$ is presented.

{\section{Reaction-diffusion problems} \label{sec:reactions}}

Reaction-diffusion models introduced in Sec. \ref{subsubsec:diff_limit} play an important
role in non-equilibrium physics. In contrast to the models of critical dynamics they are often
described by non-hermitian operators. This leads to intrigued and interesting phenomena \cite{Krapivsky,HHL08}.
This will serve us as an illustration of possible use of technique introduced in Secs. \ref{subsec:langevin} and \ref{subsec:master}.
For an analysis of resulting field-theoretic actions again theoretical approaches from Sec.  \ref{sec:RG_theory} will
be used. The main aim here is to analyze scaling behavior of two important models:
\begin{enumerate}
  \item annihilation process $A+A \rightarrow \varnothing$.
  \item directed bond percolation process.
\end{enumerate}

Velocity fluctuations are hardly avoidable
in any of experiments. For example,  a vast majority of chemical reactions occurs
at finite temperature, which is inevitably encompassed with the presence of a thermal noise. 
Furthermore, disease spreading and chemical reactions could be affected by the turbulent
advection to a great extent.
Fluid dynamics is in general described by the Navier-Stokes equations \cite{Landau_fluid}.
A general solution of these equations remains an open question
 \cite{Frisch,Monin}. 
 For both of these problems very important role is played by the properties of the environment in which
 these processes take place.
 It can be expected that
  impurities and defects,  which are not taken into account
 in the original formulation, could cause a change in the universal properties of the model. 
 This is believed to be one of the reasons  why there are
not so many direct experimental realizations \cite{RRR03,TKCS07} of the percolation process itself.
 A study of deviations from the ideal situation could proceed in different routes and this 
 still constitutes a topic of an ongoing debate \cite{HHL08}.
A substantial effort has been made  in studying  a long-range interaction using
L{\'e}vy-flight jumps \cite{Jan99,Hin06,Hin07}, effects
 of immunization \cite{Hin01,JanTau04}, mutations \cite{Sarkar15}, feedback of the environment on
 the percolating density \cite{SarkarBasu14}, or in the presence of spatially quenched
 disorder \cite{Janssen97}.
 In general, the novel behavior is observed  with a  possibility that critical
 behavior is lost or profoundly changed.
 For example, the latter is observed in annihilation process \cite{Hnatich00}.
 There was shown that thermal fluctuations makes otherwise IR stable regime unstable.
  On the other hand,
 the presence of a quenched disorder in percolation process causes a shift
 of the critical fixed point to the unphysical region.
 This leads to such interesting phenomena as an activated
 dynamical scaling or Griffiths singularities \cite{MorDic96,CGM98,Vojta05,Vojta06}.
{\subsection{Effect of velocity fluctuations on $A+A\rightarrow\varnothing$} \label{subsec:AA}}
The irreversible annihilation reaction $\textit{A} +\textit{A} \rightarrow
\varnothing$ is a fundamental model of non-equilibrium physics.
The reacting particles perform chaotic motion due to diffusion or some
external advection field such as atmospheric eddy and may react after the mutual collision with
constant microscopic probability $K_0$ per unit time. Implicitly it is assumed that resulting molecule is
 inert, i.e. chemically inactive, and has no influence on the
movement of the reacting $A$ particles. Many reactions of this type are observed in diverse
chemical, biological or physical systems \cite{Derrida95,Kroon93}.
The usual approach to this kind of problems is based on the use of the kinetic rate equation.
It leads to a self-consistent description analogous to the mean-field approximation in the
theory of critical phenomena. The basic assumption of the rate-equation approach is that the particle
density is spatially homogeneous. This homogeneity can be thought as a consequence of
either an infinite mobility of the reactants or of a very small probability
that a chemical reaction actually occurs when reacting entities meet each other.
On the other hand, if the particle mobility becomes sufficiently small, or
equivalently, if the microscopic reaction probability becomes
large enough there is a possible transition to a new regime where is is more
probable that the given particle reacts with local neighbors than with distant
particles. This behavior is known as the diffusion-controlled regime \cite{Kampen,Kang84}.
For the annihilation reaction limited assumption of the density homogeneity leads to the following
equation for the mean particle number
\begin{equation}
   \partial_t n(t) = -K_0 n^2(t).
  \label{eq:react_mpn}
\end{equation}
This equation predicts a long-time asymptotic decay as $n(t)\sim t^{-1}$ and the decay exponent does not depend on
the space dimension. This is a common situation observed in the mean field theory.
 However, it turns out \cite{Lee94,Cardy96} that in lower space dimensions $d\leq 2$ the assumption of
spatially uniform density, or equivalently of negligible density fluctuations of reacting particles, is not appropriate.
The upper critical dimension for this reaction $d_c=2$ \cite{Lee94}, above which mean field
approximation is valid. Re-entrancy of the diffusing particles \cite{Itzykson} in low space dimensions leads to
 effective decrease of the decay exponent.\\
\mbox{ }\quad  A typical reaction occurs in liquid or gaseous environment.
Thermal fluctuations of this underlying environment cause
additional advection of the reacting particles. Therefore, it is
interesting to study the influence of the advection field
on the annihilation process.\\
\mbox{ }\quad Most of the
renormalization-group analyses of the effect of random drift
on the annihilation reaction  $\textit{A} +\textit{A} \rightarrow \varnothing$ in the framework
of the Doi approach
have been carried out for the case of a quenched random drift field. Potential random drift with 
long-range \cite{Park,Chung}
and short-range correlations \cite{Richardson99} have been studied as well as ''turbulent'' flow 
(i.e. quenched solenoidal random field) with potential
disorder \cite{Deem2,Tran}. For a more realistic description of a turbulent flow time-dependent 
velocity field would be more appropriate.
In Ref. \cite{Tran} dynamic disorder with a given Gaussian distribution has been considered, whereas
the most ambitious approach
on the basis of a velocity field generated by the stochastic Navier-Stokes equation has been 
introduced here by two of the present authors  \cite{Hnatich00}.
From the point of the Navier-Stokes equation the situation near the critical dimension $d_c=2$ of 
the pure reaction model is even more intriguing
due to the properties of the Navier-Stokes equation (see Sec. \ref{subsec:NS-double}).
 The aim of this part is to examine the IR behavior of the
annihilation process under the influence of advecting velocity fluctuations and to determine its
stability in the second order of the perturbation theory.
Using mapping procedure based on the Doi formalism
 (see Sec. \ref{subsec:master}) an effective field-theoretic model for the annihilation process is constructed.
The RG method is applied on the model
in the field-theoretic formulation, which is the most efficient in calculations beyond
the one-loop order,
and within the two-parameter expansion the renormalization
constants and fixed points of the renormalization group are determined in the two-loop approximation.
The non-linear integro-differential equation, which includes first non-trivial
corrections to the (\ref{eq:react_mpn}), is obtained for the mean particle number and it is
shown how the information about IR asymptotics can be extracted from it
in the case of a homogeneous system. This equation
allows to investigate heterogeneous systems as well with the account of the effect
density and advecting velocity fluctuations. The solution of the equation
in the heterogeneous case requires numerical calculations and was considered in \cite{Busa2016}. 
\subsubsection{Field-theoretic model of the annihilation reaction}
The goal is to analyze anomalous kinetics of the irreversible single-species
annihilation reaction
\begin{equation}
  A+A \xrightarrow{K_0} \varnothing,
  \label{eq:react_rea}
\end{equation}
with the unrenormalized (mean field) rate constant $K_0$. This can be considered
as a special case of the Verhulst model \ref{eq:intro_Verhulst}.
The corresponding field-theoretic action \ref{eq:intro_Peliti-Verhulst} with velocity fluctuations
taken into account reads
\begin{align}
   \S_{1} & =  - \int^{\infty}_{0} \dRM t \int \dRM^d{\mx}\, \{ \psi' \partial_{t} \psi +
  \psi'\boldnabla\cdot({\mv}\psi) -D_{0}\psi'\boldnabla^{2}\psi
   \nonumber\\*
  &+  \lambda_{0} D_{0}[2\psi'+(\psi')^{2}]\psi^{2} +
  n_{0}\int \dRM^d{\mx}\, \psi'({\mx},0)  \}.
  \label{eq:react_S_1}
\end{align}
In the usual fashion the diffusion constant $D_0$ has been extracted from the rate constant $K_0=\lambda_0 D_0$.

 The most realistic description of the velocity field ${\mv}(x)$ is
based on the use of the stochastic Navier-Stokes equation (\ref{eq:double_NS}).
  Averaging over the random velocity field ${\mv}$  is done with the
 "weight" functional $\mathcal{W} = \rm{e}^{\S_\text{NS}}$,
 where $S_\text{NS}$ is the effective action for the advecting velocity field (\ref{eq:double_NS_action}).

 It is
 easily seen that the studied model
 contains three different types of propagators $\Delta^{vv'}, \Delta^{vv}$
 and $\Delta^{\psi\psi'}$.
 The former two were given in (\ref{eq:scalar2D_PropY}) and the latter one reads
\begin{align}  
  \Delta^{\psi\psi'}(\omega_k,{\mk}) & = \frac{1}{-i\omega_k+D_0k^2}.
  \label{eq:react_prop}
\end{align}
\begin{figure}[h!]
  \centering
  \resizebox{0.4\columnwidth}{!}{%
  \includegraphics{\PICS 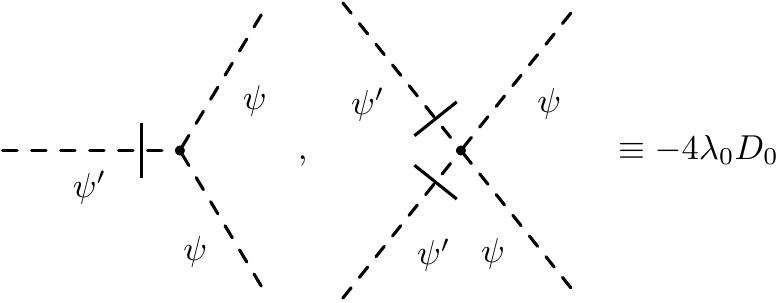}  
  }
  \caption{Interaction vertices responsible for density fluctuations and their corresponding vertex factor}
  \label{fig:ver_react}       
\end{figure}
The two reaction vertices derived from the functional (\ref{eq:react_S_1}) are depicted on Fig. \ref{fig:ver_react}
and physically describe the density fluctuations of the reactant particles.
  It should be stressed that in this model no backward influence of the reactants on the velocity field is assumed.
Therefore, the model given by actions (\ref{eq:react_S_1}) and (\ref{eq:double_NS}) may be
characterized as a model for the advection of the passive chemically active admixture.

\begin{figure}[h!]
  \centering
  \resizebox{0.45\columnwidth}{!}{%
  \includegraphics{\PICS 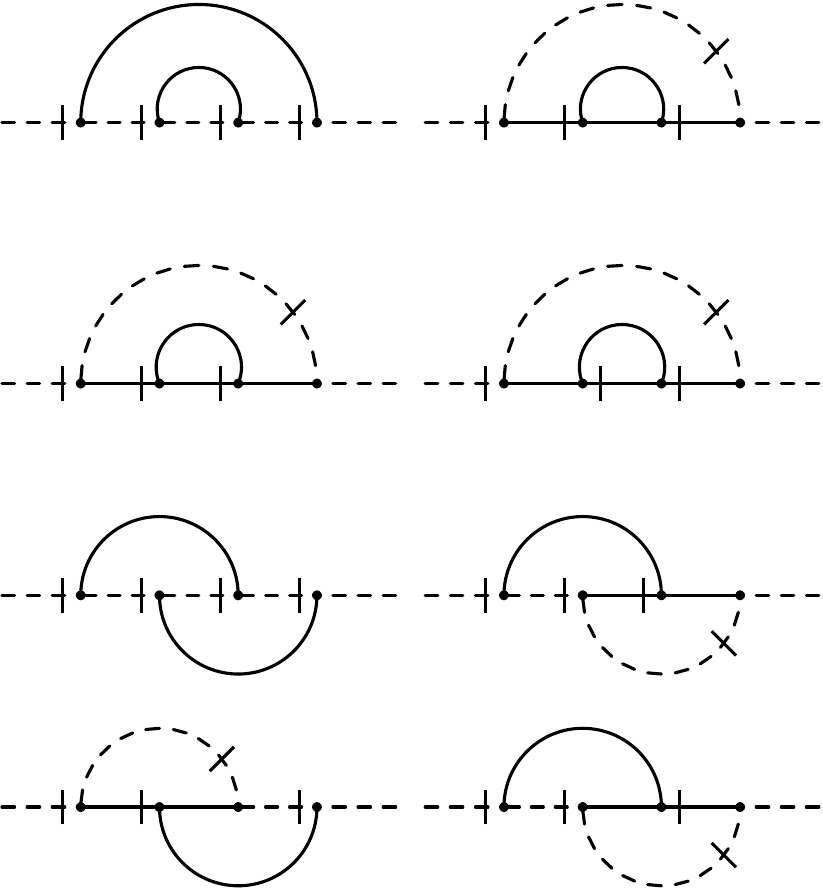}  
  }
  \hfill
  \resizebox{0.45\columnwidth}{!}{%
  \includegraphics{\PICS 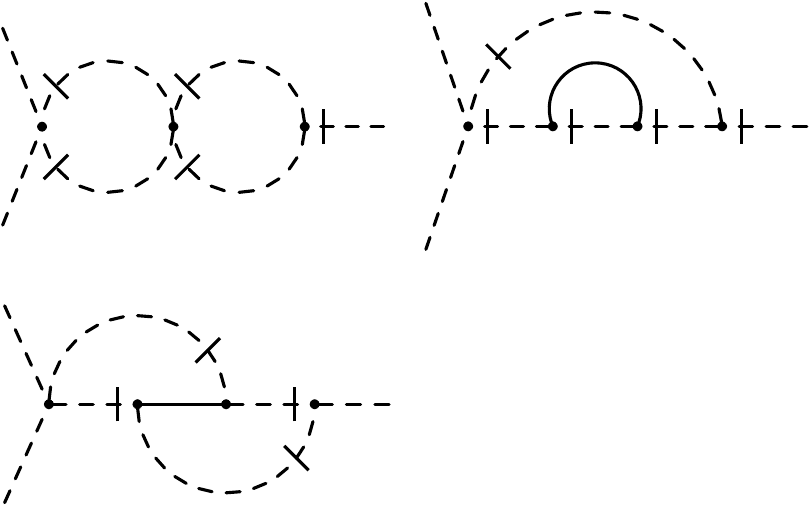}  
  }
  \caption{Two-loop graphs needed for UV renormalization of $\Gamma_{\psi'\psi}$(left) and
  $\Gamma_{\psi'\psi\psi}$(left) 1PI functions.}
  \label{fig:react_2loop}       
\end{figure}

In the one-loop approximation velocity fluctuations have no effect on the renormalization of 
the interaction vertex \cite{Hnatich00}, whereas in the two-loop approximation situation becomes more
complicated and all graphs in Fig. \ref{fig:react_2loop} must be considered \cite{HHL13}.
\subsubsection{UV renormalization of the model}
 In what follows we will employ the
modified minimal subtraction  scheme. 
What we mean here, is the ray scheme \cite{AHKV05} introduced in Sec. \ref{subsec:double_intro}, in
which the two regularizing parameters $\varepsilon$, $\Delta$ ($\varepsilon$ has been introduced
in (\ref{eq:double_kernel}) and $\Delta$ in (\ref{dz})) are taken proportional
to each other: $\Delta=\xi \varepsilon$, where the coefficient $\xi$ is arbitrary but fixed. In this case,
only one independent small parameter, say, $\varepsilon$ remains and the notion of minimal subtraction
becomes meaningful. In the modified scheme, as usual, certain geometric factors are not expanded in $\varepsilon$ (for details see \cite{AHKV05}).

In order to apply the dimensional regularization for the evaluation 
of renormalization constants, an analysis of possible superficial
divergences has to be performed. 
For the power counting in the actions (\ref{eq:react_S_1}) and
(\ref{eq:double_NS_action}) the scheme from Sec. \ref{subsec:UV} is employed, in which to each
quantity $Q$ two canonical dimensions are assigned, one with respect
to the wave number $d_Q^k$ and the other to the frequency
$d_Q^\omega$. The normalization for these dimensions is given by Eq. (\ref{eq:RG_normal}).
 The canonical dimensions
for fields and parameters of the model are derived from the condition for action
to be a scale-invariant quantity, i.e. to have a zero canonical dimension.

The quadratic part of the action (\ref{eq:react_S_1}) determines only the
canonical dimension of the quadratic product $\psi^\dagger\psi$. In
order to keep both terms in the nonlinear part of the action
\begin{equation}
  \lambda_0 D_0 \int
  \dRM t \int\dRM^d\mx\,[2\psi^\dagger+(\psi^\dagger)^2]\psi^2,
  \label{eq:react_quad_part}
\end{equation}
the field $\psi^\dagger$ must be dimensionless. If
the field $\psi^\dagger$ has a positive canonical dimension, which
is the case for $d>2$, then
the quartic term should be discarded as irrelevant by the power
counting. The action with the cubic term only, however, does not
generate any loop integrals
corresponding to the density fluctuations and thus is uninteresting for the analysis of fluctuation effects in the
space dimension $d=2$. One-loop calculation has been performed in \cite{Hnatich00} and
the two-loop approximation can be found in the work \cite{HHL13}, where all computational details can found. Similar
approach was used using Kraichnan model for generating velocity fluctuations \cite{Hnatic11b}. There the RG analysis
was somewhat simpler due to the absence of interaction between velocity fluctuations.

Using the normalization choice (\ref{eq:RG_normal}),
we are able to obtain the canonical dimensions for all the fields and parameters in the
$d$-dimensional space. The results
are summarized in Table \ref{tab:canon_react}. The dimensions of quantities related
to velocity can be found in Tab. \ref{tab:canon}.
\begin{table}
  \centering
  \begin{tabular}{| c | c | c | c | c |}
     \hline\noalign{\smallskip}
    $Q$ & $\psi$ & $\psi'$ &  $\nu_0,D_0$ & $\lambda_0$  
    \\  \noalign{\smallskip}\hline\noalign{\smallskip}
    $d_Q^\omega $& $0 $ & $0$  & $1$ & $0$ 
    \\  \noalign{\smallskip}\hline\noalign{\smallskip}
    $d_Q^k$ & $d$ & $0$ &  $-2$ & $-2\Delta$  
    \\  \noalign{\smallskip}\hline\noalign{\smallskip}
    $d_Q $& $d$ & $0$  &  $0$  & $-2\Delta$  
    \\ \noalign{\smallskip}\hline
  \end{tabular}
 \caption{Canonical dimensions for the parameters and the fields of the model with action (\ref{eq:react_S_1}).}
  \label{tab:canon_react}
\end{table}
Here, $d_Q=d^k_Q+2d_Q^\omega$ is the total canonical dimension and it is determined from the condition that the parabolic
differential operator of the diffusion and Navier-Stokes equation scales uniformly under the simultaneous momentum and frequency dilatation
$k\rightarrow \mu k, \omega\rightarrow \mu^2\omega$.

The total canonical dimension of an arbitrary one-particle irreducible Green 1PI
 is given by the relation (\ref{eq:RG_diver}).
Superficial UV divergences may be present
only in those $\Gamma$ functions for which $d_\Gamma$ is a non-negative integer.
Using the dimensions of the fields from Table \ref{tab:canon_react}
we see that the superficial degree of divergence for a 1PI function $\Gamma$ is given by
the expression
\begin{equation}
d_\Gamma=4-N_v-N_{\tilde{v}}-2N_\psi.
\end{equation}
However, the real degree of divergence $\delta_\Gamma$ is smaller, because of the
structure of the interaction vertex (\ref{eq:scalar2D_vertexADV}), which allows for factoring out
the operator $\boldnabla$ to each external line ${v'}$.
Thus
the real divergence exponent $\delta_\Gamma$
may be expressed as
\begin{equation}
  \delta_\Gamma\equiv d_\Gamma-N_{{v'}} = 4-N_v-2N_{v'}-2N_\psi
  \label{eq:real_div_exp}
\end{equation}
Although the canonical dimension for
the field $\psi'$ is zero, there is no proliferation of superficial divergent
graphs with arbitrary number of external $\psi'$ legs. This is due to the
condition
 $n_{\psi'}\leq n_\psi$, which
may be established by a straightforward analysis of the
Feynman
graphs \cite{Lee94}.
As has already been shown \cite{AdAnHo02} the divergences in 1PI Green functions containing at least one velocity field ${\mv}$
may be removed by a single counterterm of the form $\psi' \boldnabla^2 \psi$.

Brief analysis shows that the UV divergences are expected only for the 1PI Green functions listed in
Table \ref{tab:canon_green}.
\begin{table}
  \centering
  \begin{tabular}{|c|c|c|c|c|c|c|c|c|c|}
    \hline\noalign{\smallskip}
     $\Gamma_{1-ir}$ & $\langle\psi' \psi\rangle $& $\langle\psi'\psi v\rangle$
                  & $\langle {v'} v\rangle $ & $ \langle {v'} v v\rangle $
                  & $\langle {v'}{v'}\rangle$
                  & $\langle \psi' \psi^2\rangle $ 
   \\
    \noalign{\smallskip}\hline\noalign{\smallskip}
    $d_\Gamma$ & $2$ & $1$
         & $2$ & $1$ & $2$
             & $0$ 
               \\
    $\delta_\Gamma$ & $2$ & $1$
         & $1$ & $0$ & $0$
             & $0$ 
              \\
    \noalign{\smallskip}\hline
  \end{tabular}
  \caption{Canonical dimensions for the 1PI divergent Green functions of the model}
  \label{tab:canon_green}
\end{table}
This theoretical analysis leads to the following renormalization of the parameters $g_0, D_0$ and $u_0$:
\begin{align}
  &  g_1 = g_{10}\mu^{-2\epsilon} Z_1^3, &g_2&=g_{20}\mu^{2\Delta}Z_1^3 Z_3^{-1},  &\nu&=\nu_0 Z_1^{-1} \nonumber \\
  & u=u_0 Z_1 Z_2^{-1}, &\lambda& = \lambda_0 \mu^{2\Delta} Z_2 Z_4^{-1},  &D&=D_0Z_2^{-1},
  \label{eq:ren_rel}
\end{align}
where  the inverse Prandtl number 
has been introduced in the same way as was done for advection of passive scalar in action \ref{eq:scalar2D_pasivo}.
From Table \ref{tab:canon_react} it follows that $u$ is purely dimensionless quantity ($d^k_u=d^\omega_u=d_u=0$). 
In terms of the introduced renormalized parameters the total renormalized action for the annihilation reaction in a fluctuating
velocity field is
\begin{align}
 \S_{1R} &= \int \dRM^d{\mr} \int_0^\infty \dRM t \biggl\{
  \psi'\partial_t\psi +\psi'\boldnabla\cdot({\mv}\psi)-u\nu Z_2\boldnabla^2\psi
   +  \lambda u \nu \mu^{-2\Delta} Z_4[2\psi^\dagger  \nonumber\\
   & +(\psi')^2]\psi^2\biggl\}
  +n_0\int \dRM^d{\mr}    \psi'({\mr},0).
  \label{eq:total_act}
\end{align}
The renormalization constants $Z_i,i=1,2,3,4$ are to be calculated perturbatively through
the calculation of the UV divergent parts of the 1PI functions
$\Gamma_{{v'}v}$, $\Gamma_{{v'}{v'}}$,
$\Gamma_{\psi'\psi}$, $ \Gamma_{\psi'\psi^2}$ and $\Gamma_{(\psi')^2\psi^2}$. Interaction terms
corresponding to these functions have to be
added to the original action $\S=\S_1+\S_2$ with the aim to ensure UV finiteness of all Green
functions generated by the renormalized action $\S_R$. At this stage the main goal is to
calculate the renormalization constants $Z_i,i=1,2,3,4$.
The explicit expressions for them could be found in \cite{HHL13} and are not needed here.
\subsubsection{IR stable fixed points and scaling regimes}
RG analysis reveals an existence of four IR stable fixed points and
 one IR unstable fixed point. In this section they are presented together with their regions of stability.
For brevity, in the following the abbreviation $\overline{g}\equiv g\overline{S_d}$ for
the parameters $\{g_{10},g_{20},\lambda_0 \}$ or their renormalized counterparts is assumed. 
 
({\it i}$\,$) The trivial (Gaussian) fixed point
\begin{equation}
  \label{eq:react_Gaussian}
  \overline{g_1}^*=\overline{g_2}^*=\overline{\lambda}^*=0\,,
\end{equation}
with no restrictions on the inverse Prandtl number $u$. The Gaussian fixed point is stable, when
\begin{equation}
  \label{eq:react_Gstable}
  \varepsilon<0\,,\qquad \Delta >0\,.
\end{equation}
and physically corresponds to the case, when the mean-field solution is valid and fluctuation
effects negligible.

({\it ii}$\,$)
The short-range (thermal) fixed point
\begin{align}
  \label{eq:react_SR}
   \overline{g_1}^* &=0,\quad &\overline{g_2}^*&=-16\Delta+8(1+2R)\Delta^2,\nonumber\\
   u^* & = \frac{\sqrt{17}-1}{2}-1.12146\Delta,\quad
  &\overline{\lambda}^*&=-\Delta+\frac{\Delta^2}{2}\,(\xi-2.64375),
\end{align}
where $R=-0.168$ is a numerical constant. At this point
 local correlations of the random force dominate over the long-range
correlations. This fixed point
has the following basin of attraction
\begin{align}
  \label{eq:react_SRstable}
   &\Delta-\frac{2R-1}{2}\Delta^2 < 0, &2\varepsilon&+3\Delta-\frac{3\Delta^2}{2} <0,\nonumber\\
   &\Delta+\frac{1}{2}\Delta^2 < 0, &\Delta&+0.4529\Delta\varepsilon<0
\end{align}
and corresponds to anomalous decay faster than that due to density fluctuations only, but
slower than the mean-field decay.

({\it iii}$\,$)
The kinetic fixed point (corresponds to regime (\ref{eq:scalar2D_fix11})) with finite rate coefficient:
 \begin{align}
  &\overline{g_{1}}^{\ast}=
 \frac{32}{9}\,\frac{\varepsilon\,(2\varepsilon+3\Delta)}
 {\varepsilon+\Delta}+g_{12}^*(\xi)\varepsilon^2,
  &\overline{g_{2}}^{\ast}&=\frac{32}{9}\,
 \frac{\varepsilon^2}{\Delta+\varepsilon}+g_{22}^*(\xi)\varepsilon^2,\nonumber\\
 &u^*=\frac{\sqrt{17}-1}{2}+u_1^*(\xi)\varepsilon,
 &\overline{\lambda}^*&=-\frac{2}{3}(\varepsilon+3\Delta)+\frac{1}{9\pi}(3\Delta+\varepsilon)(Q\varepsilon-\Delta).
\label{eq:react_K1}
 \end{align}
 Here $Q=1.64375$. The fixed point (\ref{eq:react_K1}) is stable, when inequalities
\begin{equation}
  \label{eq:react_K1stable}
  \text{\rm{Re }}
  \Omega_{\pm}>0\,,\qquad\varepsilon>0\,,\qquad -\frac{2}{3}\varepsilon<\Delta<-\frac{1}{3}\varepsilon,
\end{equation}
are fulfilled, where
\begin{align}
  \Omega_{\pm} & = \Delta+\frac{4}{3}\varepsilon\pm\frac{\sqrt{9\Delta^2-12\varepsilon\Delta-8\varepsilon^2}}{3} +
  \frac{2\varepsilon}{9}\biggl(
    \frac{4\varepsilon(\varepsilon+3\Delta)R-6\varepsilon^2-12\varepsilon\Delta-9\Delta^2}
  {\sqrt{9\Delta^2-12\Delta-8\varepsilon^2}} \nonumber\\
  &-(3+2R)\varepsilon
  - 3\Delta\biggl).
\end{align}
The decay rate controlled by this fixed point
of the average number density is faster than the decay rate induced by dominant local
force correlations, but still slower than the mean-field decay rate.

({\it iv}$\,$)
The kinetic fixed point with vanishing rate coefficient:
\begin{align}
  \overline{g_{1}}^{\ast} &=
  \frac{32}{9}\,\frac{\varepsilon\,(2\varepsilon+3\Delta)}
  {\varepsilon+\Delta}+g_{12}^*(\xi)\varepsilon^2,
  &\overline{g_{2}}^{\ast}& = \frac{32}{9}\,
  \frac{\varepsilon^2}{\Delta+\varepsilon}+g_{22}^*(\xi)\varepsilon^2, \nonumber\\*
  u^* & = \frac{\sqrt{17}-1}{2}+u_1^*(\xi)\varepsilon,  
  &\overline{\lambda}^*& = 0\,.
  \label{eq:react_K2}
 \end{align}
 The expressions for $g_{12}^*(\xi), g_{22}^*(\xi)$ and $u_1^*(\xi)$ are rather cumbersome. They can
 be found in \cite{HHL13}, therefore we do not repeat them here.
This fixed point is stable, when the long-range correlations of the random
force are dominant
\begin{equation}
  \label{eq:react_K2stable}
  \mathrm{Re }
  \Omega_{\pm}>0,\quad\varepsilon>0,\quad \Delta>-\frac{1}{3}\varepsilon\,,
\end{equation}
and corresponds to reaction kinetics with the normal (mean-field like) decay rate.\\
({\it v}$\,$)
Driftless fixed point given by
\begin{equation}
   \overline{g_1}^*=\overline{g_2}^*=0,\quad u^* \mbox{ not fixed},\quad \overline{\lambda}^*=-2\Delta,
\end{equation}
with the following eigenvalues of stability matrix (\ref{eq:RG_Omega})
\begin{equation}
  \Omega_1=-2\varepsilon,\quad \Omega_2=-\Omega_4=2\Delta,\quad \Omega_3=0.
\end{equation}
An analysis of
the structure of the fixed points and the basins of attraction leads to the following
physical picture of the effect of the random stirring on the
reaction kinetics. Anomalous behavior always emerges below two dimensions, when the
{\it local} correlations are dominant in the spectrum of the random forcing
[the short-range fixed point ({\it ii}$\,$)]. However, the random stirring
gives rise to an effective reaction rate faster than the density-fluctuation
induced reaction rate even in this case. The anomaly is present
(but with still faster decay, see the next Section) also, when the long-range part
of the forcing spectrum is effective, but the powerlike falloff of the correlations is
fast [this regime is governed by the kinetic fixed point ({\it iii}$\,$)].
Note that this is different from the case in which the solenoidal
random velocity field is time-independent, in which case there is no fixed
point with $\lambda^*\ne 0$\cite{Deem2}.
At slower spatial falloff of correlations, however, the anomalous reaction kinetics
is replaced by a mean-field-like behavior
[this corresponds to the kinetic fixed point ({\it iv}$\,$)]. In particular,
in dimensions $d>1$ this is the situation
for the value $\varepsilon=2$ which corresponds to the Kolmogorov spectrum of the velocity field in
fully developed turbulence.
Thus, long-range correlated forcing gives rise to a random velocity field, which
tends to suppress the effect of density fluctuations on the reaction kinetics below two dimensions.\\
\begin{figure}
  \centering
  \resizebox{0.45\columnwidth}{!}{%
  \includegraphics{\PICS 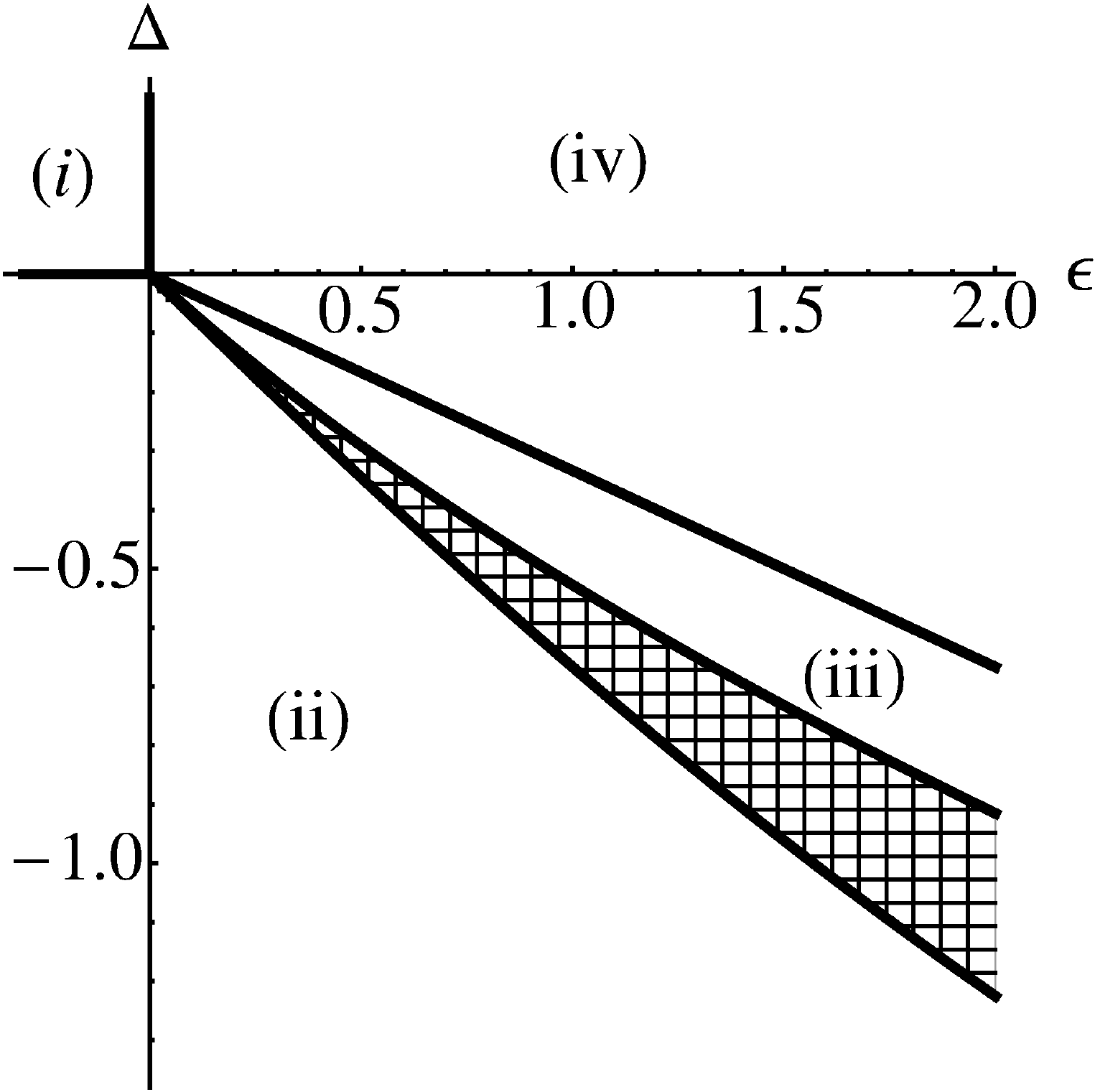}
  }
  \caption{Regions of stability for the IR stable fixed points $({\it i})-({\it iv})$ in $(\eps,\Delta)$-plane.}
  \label{fig:react_regions_of_stability}
\end{figure}

For better illustration, regions of stability for fixed points $({\it i})-({\it iv})$ are
depicted in Fig.\ref{fig:react_regions_of_stability}. Wee see
that in contrast to the one-loop approximation \cite{Hnatich00}, overlap (dashed region)
between regions of stability of fixed points $({\it ii})$ and $({\it iii})$ is observed.
It is a common situation in the perturbative RG approach that higher order terms
lead to either gap or overlap between neighboring stability regions. The physical realization
of the large-scale behavior then depends on the initial state of the system.
\subsubsection{Long-time asymptotics of number density}
Because the renormalization and calculation of the fixed points of the RG are
carried out at two-loop level,it is possible to find the first two terms of the
$\varepsilon$, $\Delta$ expansion of the average number density, which corresponds to
solving the stationarity equations at the one-loop level.
The simplest way to find the average number density is to calculate it from the stationarity
condition of the functional Legendre transform~\cite{Vasiliev}
(which is often called the effective action) of the generating functional
 obtained by replacing the unrenormalized action by the renormalized
one in the weight functional. This is a convenient way to avoid any
summing procedures used~\cite{Lee94} to take into account the
higher-order terms in the initial number density $n_0$.
 The standard aim in reaction-diffusion problems is the solution for the number density. Therefore 
the expectation values of the fields ${\mv}$ and $\tilde{ {\mv} }$ can be set to zero
at the outset (but retain, of course, the propagator and the correlation function).
Therefore,
at the second-order approximation 
the effective renormalized action for this model is
\begin{equation}
  \label{eq:react_GammaAA1loop}
  \Gamma_R  =  \S_1+\frac{1}{4}
  \raisebox{-4.4ex}{ \includegraphics[width=3.cm]{\PICS 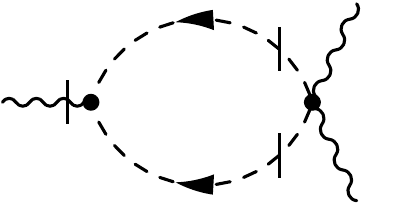}
    }
  +\frac{1}{8}
  \raisebox{-4.4ex}{ \includegraphics[width=3.cm]{\PICS 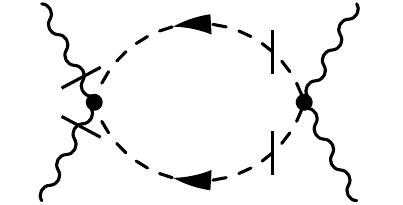}
  }+
  \raisebox{-4.ex}{ \includegraphics[width=3.cm]{\PICS 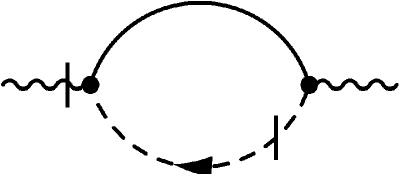}
  } +\ldots\,,
\end{equation}
where $\S_1$ is the action (\ref{eq:react_S_1}) (within a convention $\S_\text{NS}=0$ in the effective action)
and graphs are shown together with their symmetry coefficients.
The slashed wavy line corresponds to the field $\psi^\dagger$ and the single wavy line
to the field $\psi$.
 The stationarity equations for the variational functional
\begin{equation}
  \frac{\delta \Gamma_R}{\delta \psi'} = \frac{\delta \Gamma_R}{\delta \psi} = 0
  \label{eq:react_var_func}
\end{equation}
  give rise to the equations
\begin{align}
  \label{eq:react_EqAA1}
  \partial_t \psi & = u\nu Z_2 \nabla^2 \psi-2\lambda u\nu \mu^{-2\Delta}Z_4 \left(1+\psi' \right)\psi^2  
   + 4 u^2\nu^2\lambda^2\mu^{-4\Delta}
   \nonumber \\
   &\times \int\limits_0^\infty\dRM t' \int \dRM^d \my 
   \,( \Delta^{\psi\psi'})^2(t-t',\mx-\my) \psi^2(t',\my) 
   +4u^2\nu^2\lambda^2\mu^{-4\Delta}\psi'(t,\mx)
   \nonumber \\
   & \times\int\limits_0^\infty\! \dRM t'\int\!\dRM^d \my\,( \Delta^{\psi\psi'})^2(t-t',\mx-\my)\psi^2(t',\my)
    \nonumber\\
   & +\doo{}{x_i}\int\limits_0^\infty\! \dRM t' \int \! \dRM^d \my \, \Delta_{ij}^{vv}(t-t',\mx-\my)
  \doo{}{x_j}\Delta^{\psi\psi'}(t-t',\mx-\my) \psi(t',\my)
   +\ldots,
\end{align}
and
\begin{align}
  -\partial_t \psi' &= u\nu Z_2\nabla^2\psi' - 2\lambda u\nu\mu^{-2\Delta}Z_4
   \left[2\psi' + \left(\psi' \right)^2\right]\psi 
  +8u^2\nu^2\lambda^2\mu^{-4\Delta}
  \nonumber\\
  &\times\int\limits_0^\infty\!\dRM t'\int\!\dRM^d {\my}(
  \Delta^{\psi \psi' })^2(t'-t,\my-\mx)
  \psi' (t',\my)
  \psi(t,\vx)+4u^2\nu^2\lambda^2\mu^{-4\Delta}
  \nonumber\\
  &\times\int\limits_0^\infty\!\dRM t'\int\!\dRM^d {\my}( 
  \Delta^{\psi \psi' })^2(t'-t,\my-\mx)
  \left[\psi' (t',\my)\right]^2 \psi(t,\mx)
  \nonumber\\
  &+\int\limits_0^\infty\!\dRM t' \int \! \dRM^d \my \, \Delta_{ji}^{vv}(t'-t,\my-\mx)
  \doo{}{x_i}\Delta^{\psi\psi'}(t'-t,\my-\mx) \doo{}{y_j}\psi' (t',\my)
  +\ldots
  \label{eq:react_EqAA2}
\end{align}
In (\ref{eq:react_EqAA1}) and (\ref{eq:react_EqAA2}), in the integral terms it is sufficient to put all renormalization constants equal to unity.
Substituting the solution $\psi'=0$ of (\ref{eq:react_EqAA2}) into (\ref{eq:react_EqAA1}) one arrives at the
fluctuation-amended rate equation in the form
\begin{align}
  \label{eq:react_RateEqAA1loop}
  \partial_t \psi & =\! u\nu Z_2 \nabla^2 \psi\!-\!2\lambda u\nu \mu^{-2\Delta}Z_4 \psi^2 \!+
  4u^2\nu^2\mu^{-4\Delta}\lambda^2 \!\int\limits_0^\infty\! \dRM t'\!\!\! \int\!  \dRM^d \my (
  \Delta^{\psi\psi' })^2(t-t',\mx-\my)
  \nonumber \\
  &\times \psi^2(t',\my) 
  +\doo{}{x_i}\int\limits_0^\infty\! \dRM t' \int \!\dRM^d \my \, \Delta_{ij}^{vv}(t-t',\mx-\my)  
  \doo{}{x_j}\Delta^{\psi  \psi' }(t-t',\mx-\my) \psi(t',\my)
  \nonumber\\
  &+\ldots,
\end{align}
This is a nonlinear partial integro-differential equation, whose explicit solution is not known.
It is readily seen that for a homogeneous solution the term resulting from the third graph in
 (\ref{eq:react_GammaAA1loop}) vanishes and
hence the influence of the velocity field on the homogeneous annihilation process would be only through the
renormalization of the coefficients $\lambda$ and $D$. However, in case of a nonuniform density field $\psi$
the effect of velocity fluctuations is explicit in (\ref{eq:react_RateEqAA1loop}). Such a solution can be most probably
found only numerically.

To arrive at an analytic solution, we restrict ourselves to the homogeneous density $n(t)=\langle\psi(t)\rangle$.
In this case the last term in (\ref{eq:react_RateEqAA1loop}) vanishes together with the Laplace operator term and
the remaining coordinate integral
may be calculated explicitly .
The propagator is the diffusion kernel of the renormalized model (the system is considered in the general 
space dimension $d$)
\begin{equation}
   \Delta^{\psi\psi'}(t-t',\mx)
   =\frac{\theta(t-t')}{\left[4\pi u\nu(t-t')\right]^{d/2}}\exp\left[-\frac{x^2}{4u\nu(t-t')}\right].
  \label{eq:react_diffkernel}
\end{equation}
As noted above, for calculation of the one-loop contribution it is sufficient to put the 
renormalization constant $Z_2=1$ in the propagator
$\Delta^{\psi\psi'}$.
Therefore,
evaluation of the Gaussian coordinate integral in (\ref{eq:react_RateEqAA1loop}) yields
\begin{equation}
  \label{eq:react_integral}
  \int\! \dRM^d{\my}\,( \Delta^{\psi\psi'})^2(t-t',\mx-\my)=
  \frac{\theta(t-t')}{\left[8\pi u\nu(t-t')\right]^{d/2}}
\end{equation}
and the ordinary integro-differential equation
\begin{equation}
  \label{eq:react_RateEqAA1loopUniform}
  \dee{ n(t)}{t}= -2\lambda u\nu \mu^{-2\Delta}Z_4  n^2(t)
  +
  4\lambda^2u^2\nu^2\mu^{-4\Delta}\int\limits_0^t\!\dRM t'\,\frac{n^2(t')}{\left[8\pi u\nu(t-t')\right]^{d/2}}\,.
\end{equation}
is obtained. 
Naively one can assume that concentrating on homogeneous solution $n(t)$ the information about spatial
fluctuations has been lost. However, the integral term in (\ref{eq:react_RateEqAA1loopUniform}) corresponds these fluctuations and cause rather heavy effect
 even on the homogeneous solution. In particular, the integral in (\ref{eq:react_RateEqAA1loopUniform}) diverges at
the upper limit in space dimensions $d\ge 2$. This is a consequence of the UV divergences in the model
above the critical dimension $d_c=2$ and near the critical dimension is remedied by the UV renormalization of the
model. To see this, subtract and add the term $n^2(t)$ in the integrand to obtain
\begin{align}
  \label{eq:react_RateEqAA1loopUniform2}
  \dee{ n(t)}{t} &= -2\lambda u\nu \mu^{-2\Delta}Z_4  n^2(t)
  +
  4\lambda^2u^2\nu^2\mu^{-4\Delta}n^2(t)\int\limits_0^t\!\frac{\dRM t'}{\left[8\pi u\nu(t-t')\right]^{d/2}}
  \nonumber\\
  &+ 4\lambda^2u^2\nu^2\mu^{-4\Delta}\int\limits_0^t\!\dRM t'\,\frac{n^2(t')-n^2(t)}{\left[8\pi u\nu(t-t')\right]^{d/2}}\,.
\end{align}
The last integral here is now convergent at least near two dimensions, provided the solution $n(t)$ is
a continuous function. This is definitely the case for the iterative solution constructed below.
The divergence in the first integral in (\ref{eq:react_RateEqAA1loopUniform2}) may be explicitly calculated
below two dimensions and
is canceled -- in the leading order in the parameter $\Delta$ -- by the one-loop term of the 
renormalization constant $Z_4$ 
\cite{HHL13}.
Expanding the right-hand side of (\ref{eq:react_RateEqAA1loopUniform2}) in the
parameter $\Delta$ to the next-to-leading
order the equation
\begin{align}
  \label{eq:react_RateEqAA1loopUniform3}
  \dee{ n(t)}{t} & = -2\lambda  u\nu\mu^{-2\Delta}   n^2(t)
  +2\lambda u\nu\mu^{-2\Delta}    n^2(t)\left\{\frac{\lambda}{4\pi}\,
  \left[\gamma+\ln\left(2u\nu\mu^2 t\right)\right]\right\}
  \nonumber\\
  & +\frac{\lambda^2u\nu\mu^{-2\Delta}}{2\pi}\int\limits_0^t\!\dRM t'\,\frac{n^2(t')-n^2(t)}{t-t'}
\end{align}
can be derived
without divergences near two dimensions. Here, the factor $\mu^{-2\Delta}$ has been retained intact in order not
to spoil the consistency of scaling dimensions in different terms of the equation. In 
(\ref{eq:react_RateEqAA1loopUniform3}), $\gamma=0.57721$ is Euler's constant and
 the coupling constant $\lambda$ and the parameter
$\Delta$ have been considered to be small parameters of the same order taking into account the magnitudes of the parameters
in the basins of attraction
of the fixed points of the RG. The leading-order approximation for $n(t)$ is given by the first term on the
right-hand side of (\ref{eq:react_RateEqAA1loopUniform3}) and it is readily seen that after substitution of this
expression the integral
term in (\ref{eq:react_RateEqAA1loopUniform3}) is of the order of $\lambda^3$ and thus negligible in the
present next-to-leading-order calculation. In this
approximation, Eq. (\ref{eq:react_RateEqAA1loopUniform3}) yields
\begin{equation}
  \label{eq:react_NLsolution}
  n(t)=
  \frac{n_0}{1+2{\lambda} u\nu t\left\{1+ \frac{\displaystyle\lambda}{\displaystyle 4\pi}\left[1-\gamma
  -\ln\left(2u\nu\mu^2 t\right)\right]\right\}\mu^{-2\Delta}n_0}\,,
\end{equation}
where $n_0$ is the initial number density.

Green functions $\W_R$ differ from the unrenormalized $\W=\langle\Phi\ldots\Phi\rangle$ \cite{Vasiliev} only by the
choice of parameters and thus one may write
\begin{equation}
  \W_R( g,\nu,\mu,\ldots) = \W(g_0,\nu_0,\ldots),
  \label{eq:react_diff_GF}
\end{equation}
where $g_0=\{g_{10},g_{20},u_0,\lambda_0\}$ is the full set of the bare parameters and dots denotes
all variables unaffected by the renormalization procedure. The independence of renormalization mass
parameter $\mu$ is expressed by the equation $\mu\partial_\mu \W_R= 0$.
Using this equation the RG equation for the mean particle number $n(t)$ is readily obtained:
\begin{equation}
  \left(\mu\frac{\partial}{\partial\mu}+\sum\limits_g\beta_g\frac{\partial}{\partial g}
  -\gamma_1\nu\frac{\partial}{\partial\nu}\right)n(t,\mu,\nu,n_0,g)=0,
  \label{eq:react_RG}
\end{equation}
where in the second term the sum runs over all charges $g_1,g_2,u$ and $\lambda$ of the model.
From macroscopic point of view the most interesting is
 long-time behavior of the system ($t\rightarrow\infty$), therefore
 the scale setting parameter $\mu$ should be traded for the time variable. Canonical scale invariance (Sec. \ref{eq:RG_jednatridva} and \cite{turbo})
 yields relations
\begin{equation}
  \left(\mu\frac{\partial}{\partial \mu}-2\nu\frac{\partial}{\partial \nu}+
   dn_0\frac{\partial}{\partial n_0}-d \right) n(t,\mu,\nu,n_0,g)=0,
   \label{eq:react_scale_mom}
\end{equation}
\begin{equation}
  \left( -t\frac{\partial }{\partial t}+\nu\frac{\partial}{\partial \nu}\right)   n(t,\mu,\nu,n_0,g)=0,
   \label{eq:react_scale_freq}
\end{equation}
where the first equation expresses scale invariance with respect to wave number and the second equation
with respect to time.
Eliminating partial derivatives with respect to parameter $\mu$ and viscosity $\nu$ we
obtain the Callan-Symanzik equation for the mean particle number:
\begin{equation}
   \biggl[(2-\gamma_1)t\frac{\partial}{\partial t}+\sum_{g} \beta_g \frac{\partial}{\partial g}-
   dn_0 \frac{\partial}{\partial n_0}+d \biggr] n\left(t,\mu,\nu,n_0,g\right)=0
   \label{eq:react_Callan}
\end{equation}
To separate information given by the RG, consider the dimensionless normalized mean particle number
\begin{equation}
  \label{eq:react_normaln}
  \frac{n}{n_0}=\Phi\left(\nu\mu^2 t,\lambda u\,\frac{n_0}{\mu^d},g\right)\,.
\end{equation}
For the asymptotic analysis, it is convenient to express the particle density in the combination used here.
Solution of (\ref{eq:react_Callan}) by the method of characteristics yields
\begin{equation}
  \Phi\left(\nu\mu^2 t,\lambda u\,{n_0\over \mu^d},g\right)
  =\Phi\left({\nu}\mu^2 \tau,\overline{\lambda}\overline{u}\,{\overline{n}_0\over \mu^d},\overline{g}\right)
   \label{eq:react_solNLRG}
\end{equation}
where $\tau$ is the convenient time scale.
In Eq. (\ref{eq:react_solNLRG}), $\overline{g}$ and $\overline{n}_0$
are the first integrals (discussed in Sec. \ref{sec:RGsolution}) of the system of differential equations
\begin{equation}
  t\frac{\dRM }{\dRM t}\overline{g}=-\frac{\beta_g (\overline{g})}{2-\gamma_1(\overline{g})}\,,\quad
  t\frac{\dRM}{\dRM t}\overline{n}_0=d\frac{\overline{n}_0}{2-\gamma_1(\overline{g})}.
  \label{eq:react_first_int}
\end{equation}
Here $\overline{g}=\{\overline{g_1},\overline{g_2},\overline{u},\overline{\lambda} \}$ with
initial conditions $\overline{g}|_{t=\tau}=g$ and $\overline{n}_0|_{t=\tau}=n_0$.
In particular,
\begin{equation}
  \label{eq:react_initialdensity}
  \overline{\lambda}\overline{u}\,\overline{n}_0=\lambda u\,n_0\,\left({t\over \tau}\right)
  \exp\left[{\displaystyle  \int_\tau^t{\displaystyle \gamma_4 \dRM s\over\displaystyle  (2-\gamma_1)s}}\right]\,.
\end{equation}
The asymptotic expression of the integral on the right-hand side of (\ref{eq:react_initialdensity}) in the
vicinity of the IR-stable fixed point $g^*$ is of the form
\begin{equation}
  \label{eq:react_asyamplitude}
  { \int_\tau^t{\displaystyle \gamma_4 \dRM s\over\displaystyle  (2-\gamma_1)s}}
  \operatornamewithlimits{\sim}_{t\to\infty}{\gamma_4^*\over 2-\gamma_1^*}\ln\left({t\over\tau}\right)+
  {2\over 2-\gamma_1^*}\int\limits_\tau^\infty\!{ (\gamma_4-\gamma_4^*)\dRM s\over  (2-\gamma_1)s}
  ={\gamma_4^*\over 2-\gamma_1^*}\,\ln\left({t\over \tau}\right) + {c}_n(\tau)\,,
\end{equation}
corrections to which vanish in the limit $t\to\infty$. Quantities with asterisk always refer to corresponding fixed point value.
In (\ref{eq:react_asyamplitude}) and henceforth, the
notation $\gamma_1^*=\gamma_1\left(g^*\right)$ has been used.
From the point of view of the long-time asymptotic
behavior the next-to-leading term in (\ref{eq:react_asyamplitude}) is an inessential constant.
In the vicinity of the fixed point
\begin{align}
  \label{eq:react_initialdensityasy}
  \overline{\lambda}\overline{u}\,{\overline{n}_0\over \mu^d}
  \sim \lambda u\,{n_0\over \mu^d} \left({t\over \tau}\right)^{\displaystyle 1+{\displaystyle
  \gamma_4^*\over \displaystyle 2-\gamma_1^*}} {C}_n
  \equiv
  \lambda u\,{n_0\over \mu^d} \left({t\over \tau}\right)^\alpha {C}_n
  \equiv \overline{y}\,{C}_n\,,
\end{align}
where a shorthand notation $\overline{y}$ has been introduced for the long-time scaling of the normalized number density as well as
the dimensional normalization constant
\[
 {C}_n=\eRM^{d\tilde{c}_n(\tau)}\,.
\]
and the decay exponent
\begin{equation}
  \label{eq:react_defalpha}
  \alpha=\displaystyle 1+{\displaystyle \gamma_4^*\over \displaystyle 2-\gamma_1^*}
\end{equation}
The asymptotic behavior of the normalized particle density is described by the scaling function $f(x,y)$:
\begin{align}
  \Phi\left(\nu\mu^2 t,\lambda u\,{n_0\over \mu^d},g\right)
  \sim\Phi\left({\nu}\mu^2 \tau, {C}_n\overline{y},{g}^*\right)\equiv
  f\left({\nu}\mu^2 \tau, {C}_n\overline{y}\right)\,.
  \label{eq:react_scalingf}
\end{align}
The scaling function $f(x,y)$ describing the asymptotic behavior of the normalized number density
$\Phi=n/n_0$ is a function of two dimensionless argument only,
whereas the generic $\Phi$ has six dimensionless arguments (all four coupling constants on top of the
scaling arguments of $f(x,y)$).
We recall that the generic solution of the Callan-Symanzik equation (\ref{eq:react_Callan}) does not give the explicit
functional form of the function $n=n_0\Phi$, which may to determined from the solution (\ref{eq:react_NLsolution}) of the
stationarity equation of the variational problem for the effective potential.
The free parameters in the variables of the scaling function $f(x,y)$ correspond to the choice of units of
these variables, whereas the
objective information is contained in the form of the scaling function~\cite{Vasiliev,turbo}. Here, it
is convenient to use the explicit solution (\ref{eq:react_NLsolution})
to obtain the $\varepsilon$, $\Delta$ expansion for the inverse $h(x,y)={1/ f(x,y)}$ of the scaling 
function. From solution (\ref{eq:react_NLsolution}) the generic expression
\begin{align}
  \label{eq:react_ExpansionOfScaling}
  h(x,y)={1\over f(x,y)}
  =1+2xy\left\{1+{\displaystyle{{\lambda}^* }\over \displaystyle 4\pi}\left[1-\gamma
  -\ln\left(2u^*x\right)\right]\right\}
\end{align}
is obtained,
the substitution in which of the various fixed-point values $\lambda^* $ 
(at the leading order $\lambda^*\approx 2\pi\overline{\lambda}^*$)
and $u^*$ in the leading approximation yields the corresponding $\varepsilon$, $\Delta$ expansions.

Below, we list the scaling functions  $h(x,y)$ and the dynamic exponents $\alpha$ at the stable fixed points
in the next-to-leading-order approximation.\\
({\it i}$\,$) At the trivial (Gaussian) fixed point (\ref{eq:react_Gaussian})
the mean-field behavior takes place with
\begin{align}
  \label{eq:react_hGaussian}
  h(x,y)&=1+2xy\,,\nonumber\\*
  \alpha&=1\,.
\end{align}

({\it ii}$\,$)
The thermal (short-range) fixed point
(\ref{eq:react_SR}) leads to scaling function and decay exponent
\begin{align}
\label{eq:react_hSR}
  h(x,y)=1+2xy\left\{1-{\displaystyle{\Delta }\over \displaystyle 2}\left[1-\gamma
  -\ln \left(\sqrt{17}-1\right)\,x\right]\right\}
  ,\quad
  \alpha=1+{\Delta\over 2}+{\Delta^2\over 2}\,.
\end{align}
Here, the last coefficient is actually a result of numerical calculation, which in the standard accuracy 
of Mathematica \cite{mathematica}
is equal to $0.5$. The authors of \cite{HHL13} have not been able to sort out this result analytically, but think that most probably the coefficient of
the $\Delta^2$ term in the decay exponent $\alpha$ in (\ref{eq:react_hSR}) really is ${1\over 2}$.

({\it iii}$\,$)
The kinetic fixed point with an anomalous reaction rate
(\ref{eq:react_K1}) corresponds to
\begin{align}
  \label{eq:react_hK1}
  h(x,y)& = 1+2xy \biggl\{ 1-{\displaystyle{\varepsilon+3\Delta }\over \displaystyle 3}
  \left[1-\gamma-\ln \left(\sqrt{17}-1\right)\,x\right]\biggl\}
  ,\quad
  \alpha = 1+{3\Delta+\varepsilon\over 3-\varepsilon}\,,
\end{align}
with an exact value of the decay exponent. 

({\it iv}$\,$)
At the kinetic fixed point with mean-field-like reaction rate
(\ref{eq:react_K2}) we obtain
\begin{align}
\label{eq:react_hK2}
 h(x,y)& = 1+2xy,\quad 
\alpha=1\,.
\end{align}
In the actual asymptotic expression corresponding to (\ref{eq:react_scalingf}) the argument $y\to \widetilde{C}_n\overline{y}$
is different from that of the Gaussian fixed point.

To complete the picture, we recapitulate -- with a little bit more detail --
the asymptotic behavior of the number density in the physical space dimension $d=2$ predicted
within the present approach \cite{Hnatich00} (it turns out that for these conclusions the one-loop
calculation is sufficient).
On the ray $\varepsilon\le 0$, $\Delta=0$ logarithmic corrections to the mean-field decay take place. The integral
determining the asymptotic behavior of the variable (\ref{eq:react_initialdensity}) yields in this case
\begin{equation}
  \label{eq:react_asyamplitudelog}
  { \int_\tau^t{\displaystyle \gamma_4 \dRM s\over\displaystyle  (2-\gamma_1)s}}
  \operatornamewithlimits{\sim}_{t\to\infty}-{1\over 2}\,\ln\ln\left({t\over \tau}\right) + {c}_n(\tau)\,,
\end{equation}
with corrections vanishing in the limit $t\to\infty$. Therefore, in the vicinity of the fixed point
\begin{equation}
  \label{eq:react_initialdensityasylog}
  \overline{\lambda}\overline{u}\,{\overline{n}_0\over \mu^d}
  \sim \lambda u\,{n_0\over \mu^d} \left({t\over \tau}\right)\ln^{-1/2}\left({t\over \tau}\right) {C}_n
  \equiv \overline{y}\,  {C}_n\,.
\end{equation}
The scaling function $h$ is of the simple form
\begin{equation}
  \label{eq:react_hlog}
  h(x,y)=1+2xy\nonumber
\end{equation}
and gives rise to asymptotic decay slower than in the mean-field case by a logarithmic factor:
\[
  {n}\sim {\ln^{1/2}\left(t/ \tau\right)\over 2\nu\lambda u {C}_n t}\,.
\]
It is worth noting that this logarithmic slowing down is weaker than that brought about 
the density fluctuations only \cite{Peliti86}
and this change is produced even by the ubiquitous thermal fluctuations of the fluid, when the reaction is taking
place in gaseous or liquid media.

On the open ray $\varepsilon> 0$, $\Delta=0$ the kinetic fixed point with mean-field-like 
reaction rate (\ref{eq:react_K2}) is
stable and the asymptotic behavior is given by (\ref{eq:react_hK2}) regardless of the value of the falloff exponent of the
random forcing in the Navier-Stokes equation. In particular, only the amplitude factor in
the asymptotic decay rate in two dimensions is affected by the developed turbulent flow with Kolmogorov scaling, which
corresponds to the value $\varepsilon=2$.
This is in accordance with the results obtained in the case of quenched solenoidal flow with long-range correlations 
\cite{Deem2,Tran} as
well as with the usual picture of having the maximal reaction rate in a well-mixed system.

To conclude, the effect of density and velocity
fluctuations on the reaction kinetics of
the single-species decay $A+A\to \emptyset$ universality class has been analyzed within the
framework of field-theoretic renormalization group and calculated the
scaling function and the decay exponent of the mean particle density for the four asymptotic patterns
predicted by the RG.

{\subsection{Effect of velocity fluctuations on directed bond percolation} \label{subsec:percolation}}
Another important model in non-equilibrium physics is directed percolation (DP)  
 bond process \cite{Stauffer,HHL08}. Its aim is to describe phase transition
 between absorbing and active phase. In contrast to the annihilation process (\ref{eq:react_rea})
 also spreading process $A \rightarrow 2A$, and death process
 $A \rightarrow \varnothing$ are allowed \cite{JanTau04}.
 As pointed out by  Janssen and Grassberger \cite{Janssen81,Grassberger82}, necessary
conditions for corresponding universality class are: i) a unique absorbing state, ii) short-ranged interactions,
 iii) a positive order parameter and iv) no extra symmetry or additional slow variables.
 Among a few models described within this framework we name population dynamics,
 reaction-diffusion problems \cite{Odor04}, percolation processes \cite{JanTau04}, hadron
 interactions \cite{Cardy80}, etc. 

We focus on the directed bond percolation process in the presence
of advective velocity fluctuations. 
However, to provide more insight we restrict ourselves
 to a more decent problem.
Namely, we assume that the velocity field is given by the Gaussian 
velocity ensemble with prescribed  statistical properties \cite{Kra68,Ant99}.
 Although this assumption appears  as
 oversimplified,  compared to the realistic flows at  the first sight, it nevertheless captures essential
 physical information about advection processes \cite{Kra68,FGV01,turbo}.
 
Recently, there has been  increased interest in  different advection problems in 
compressible turbulent flows \cite{Benzi09,Pig12,Volk14,depietro15}. These studies show that
compressibility plays a decisive role for population dynamics or chaotic mixing of colloids.
 The aim here is to analyze an influence of 
 compressibility \cite{AdzAnt98,Ant00}
 on the critical properties of the directed bond percolation process \cite{HHL08}. 
  To this end, 
  the advective field is described by the Kraichnan model with finite correlation time, in which
  not only a solenoidal (incompressible) but also
 a potential (compressible) part of the velocity statistics is involved.
 Similarly as the annihilation process from the previous part 
 the model under consideration corresponds to
 the passive advection of the reacting scalar.

 The details of the calculations can be found in the work \cite{Antonov16}. Here, only main steps are reviewed.
\subsubsection{The model \label{sec:model}}

A continuum description of DP in terms of a density 
$\psi = \psi(t,\mx)$ of infected individuals typically arises from
a coarse-graining procedure in which a large number of
fast microscopic degrees of freedom are averaged out. A loss of the physical
information is supplemented by a Gaussian noise in a resulting Langevin equation.
Obviously, a correct mathematical description has to be in conformity regarding
the absorbing state condition: $\psi = 0 $ is always a stationary state
and no microscopic fluctuation can change that. 
The coarse grained stochastic equation then reads \cite{JanTau04}
\begin{equation}
  \partial_t {\psi}  = D_0 (\boldnabla^2 - \tau_0)\psi  - 
   \frac{g_0 D_0}{2}\psi^2
  + \xi,
  \label{eq:react_basic}
\end{equation}
where $\xi$ denotes the noise term, $D_0$ 
is the diffusion constant, $g_0$ is the coupling constant and $\tau_0$ measures
 a deviation from the threshold value for injected probability. It can be thought
 as an analog to the temperature (mass) variable in the standard $\varphi^4-$theory 
 \cite{JanTau04,Zinn}.
 Due to  dimensional reasons,  we have extracted the 
 dimensional part from the interaction term.

 It can be shown \cite{JanTau04} that
the Langevin equation (\ref{eq:react_basic}) captures the gross properties
 of the percolation process and contains essential physical information about the
 large-scale behavior of the non-equilibrium phase 
 transition between the active $(\psi > 0)$ and the absorbing state $(\psi = 0)$.    
The Gaussian noise term $\xi$ with zero mean
 has to satisfy the absorbing state condition. Its
 correlation function can be chosen in the following form
\begin{equation}
   \langle \xi(t_1,\mx_1) \xi(t_2,\mx_2) \rangle = g_0 D_0 \psi(t_1,\mx_1) 
   \delta(t_1-t_2) \delta^{(d)}(\mx_1 - \mx_2)
   \label{eq:react_noise_correl}
\end{equation}
up to irrelevant contributions \cite{Tauber2014}. It should be noted that due to the dependence of
the noise correlations on the density, in the stochastic problem (\ref{eq:react_basic}), (\ref{eq:react_noise_correl})
the noise is multiplicative. It is customary to use the It\^{o} interpretation in this case \cite{JanTau04} and we stick
to this convention.

A further step consists in  incorporating  of the velocity fluctuations into
the model (\ref{eq:react_basic}). The 
 standard route \cite{Landau_fluid} based on 
   the replacement (\ref{eq:intro_convective})   
   is not sufficient due to the assumed compressibility. As shown in \cite{AntKap10},  the following
   replacement is then adequate
\begin{equation}
  \partial_t \rightarrow \partial_t +({\bm v}\cdot\boldnabla)+a_0 ({\bm \nabla}\cdot{\bm v}),
  \label{eq:react_subs}
\end{equation}
where $a_0$ is an additional positive parameter, whose significance will be discussed later.

Note that the last term in (\ref{eq:react_subs}) contains
a divergence of the velocity field $\bm v$ and thus ${\bm \nabla}$ operator does not act on
what could possibly follow.

Following  \cite{Ant00},  we
 consider the velocity field to be a random Gaussian variable with zero mean and 
 a translationally invariant correlator given as follows:
\begin{equation}
  \langle v_i(t,{\bm x}) v_j (0,{\bm 0}) \rangle =
  \int \frac{{\mathrm d} \omega}{2\pi}
  \int \frac{{\mathrm d}^d {\bm k}}{(2\pi)^d} 
  [P_{ij}^{k} + \alpha Q_{ij}^{k}] D_v (\omega,\mk) {\mathrm e}^{-i\omega  t  +{\bm k}\cdot {\bm x}},
  \label{eq:react_vel_correl}
\end{equation}
where in contrast to the Sec. \ref{subsubsec:hel_RG} we consider the kernel function $D_v(\omega,\mk)$ in the form
\begin{equation}
  D_v (\omega,\mk) = 
  \frac{g_{10} u_{10} D_0^3 k^{4-d-y-\eta}}{\omega^2 + u_{10}^2 D_0^2 (k^{2-\eta})^2}.
  \label{eq:react_kernelD}
\end{equation}
The reason is that the model (\ref{eq:react_basic}) has an upper critical dimension $d_c=4$ and
in perturbation theory poles in $d-d_c$ are expected. In accordance with the standard notation
\cite{Zinn,Vasiliev} we therefore retain symbol $\varepsilon$ for the expression $4-d$ and
$y$ for a scaling exponent of velocity field. Just for completeness we note that parameter $\eps$
from Secs. \ref{sec:models}-\ref{subsec:AA} is related to $y$ as follows $y=2\eps.$
The incompressible version of the considered model, $\alpha=0$, has been studied  in 
works \cite{AntKap08,AntKap10,Ant11,SarkarBasu12,DP13}. The velocity field here is a coloured
noise field and does not require any specification of interpretation. It should be noted, however, the
rapid change limit discussed below corresponds to the white-noise problem in the Stratonovich interpretation.

The stochastic problem (\ref{eq:react_basic}-\ref{eq:react_kernelD})
 can be cast into a field-theoretic form and the resulting dynamic functional \cite{Antonov16} reads
\begin{equation}
   \S[\Phi] = \S_{ \text{diff}}[\psi',\psi]
   + \S_{\text{vel}}[\mv]
   + \S_{\text{int}}[\Phi], 
   \label{eq:react_bare_act}
\end{equation}
where $\Phi=\{{\psi'},\psi,\mv \}$ stands for the complete set of fields
and ${\psi'}$ is the auxiliary  response field \cite{MSR}. 
The first term represents a free part of the equation (\ref{eq:react_basic})
and is given by the following expression:
\begin{equation}
  \S_{ \text{diff}}[\psi',\psi] =  
  \int \dRM t \int \dRM^{d} \mx \biggl\{
  {\psi'}[
  -\partial_t + D_0\boldnabla^2 - D_0\tau_0
  ]\psi \biggl\}.
  \label{eq:react_act_diffuse}
\end{equation}
Averaging over the velocity fluctuations is performed with the action (\ref{eq:scalar_vel_action})
 and fhe final interaction part can be written as
\begin{equation}
  \S_{\text{int}}[\Phi]  = 
  \int \dRM t \int \dRM^{d} \mx
  \biggl\{  
  \frac{D_0\lambda_0}{2} [{\psi'}-{\psi}
  ]{\psi'}\psi
  +\frac{u_{20}}{2D_0} {\psi'} \psi
  \mv^2 -  {\psi'} (\mv\cdot\boldnabla) \psi 
  -a_0 {\psi'} (\boldnabla\cdot\mv)\psi
  \biggl\}.
  \label{eq:react_inter_act}
\end{equation}

All but the third term in (\ref{eq:react_inter_act}) directly stem from the nonlinear
terms in (\ref{eq:react_basic}) and (\ref{eq:react_subs}).
The third term proportional to $\propto {\psi'}\psi\mv^2$ deserves a special consideration. 
The presence of this term is prohibited in the original Kraichnan model due
to the underlying Galilean invariance. However, in present case the general form of
the velocity kernel function does not lead to such restriction. Moreover, by direct
inspection of the perturbative expansion, one can show that this kind of  term is indeed generated
under RG transformation.
This term was considered
for the first time in \cite{DP13}, where the incompressible case
is analyzed.

Let us also note that for the linear advection-diffusion equation \cite{Ant00,Landau_fluid}, the
choice $a_0=1$ corresponds to the conserved quantity $\psi$ (advection of a
density field), whereas for the choice $a_0=0$ the conserved
quantity is ${\psi'}$ (advection of a tracer field).
 From the point of view of the renormalization 
group, the introduction of $a_0$ is necessary,
because it ensures multiplicative renormalizability of the model \cite{AntKap10}.

As was mentioned in Sec. \ref{subsubsec:functional_repre}, basic ingredients of any stochastic theory, correlation and 
response functions of the concentration 
field $\psi(t,\mx)$, can  be
computed as functional averages with respect to the weight functional $\exp \S$
with action (\ref{eq:react_bare_act}).
Further, the field-theoretic formulation summarized in (\ref{eq:react_act_diffuse})-(\ref{eq:react_inter_act})
 has an additional advantage to be amenable to the full machinery of (quantum) field theory
 reviewed in Sec. \ref{sec:RG_theory}.
Next,  the RG perturbative technique is applied that allows 
to study the model in the vicinity of its upper critical dimension $d_c=4$.
 By direct inspection of the Feynman diagrams one can 
 observe that the real expansion parameter is 
 rather $\lambda_0^2$ than $\lambda_0$. This is a direct consequence of the duality symmetry 
 \cite{JanTau04}
 of the action for the pure percolation problem with respect to time inversion
 \begin{equation}
   \psi(t,\mx) \rightarrow -{\psi'}(-t,\mx),\quad
   {\psi'}(t,\mx) \rightarrow -\psi(-t,\mx).
   \label{eq:react_time_sym}
 \end{equation}
 Therefore, it is convenient to consider a
 new charge $g_{20}$ 
 \begin{equation}
   g_{20} = \lambda_0^2
   \label{eq:react_new_g2}
 \end{equation}
 and express the perturbation calculation in terms of this parameter.

\subsubsection{Fixed points and scaling regimes \label{sec:regimes}}

From the technical point of view the model is an example of multicharge problem with five charges  $\{g_1,g_2,u_1,u_2,a\}$.
The $\beta$-function, are now given by the expressions
\begin{align}
   \beta_{g_1} &= g_1 (-y + 2\gamma_D-2\gamma_v), 
      &\beta_{g_2}& = g_2 (-\eps -\gamma_{g_2}), 
      &\beta_{a} &= - a \gamma_{a}.
      \nonumber \\
      \beta_{u_1} &= u_1(-\eta +\gamma_D), 
      &\beta_{u_2}& = - u_2 \gamma_{u_2}.
  \label{eq:react_beta_functions}
\end{align}
It turns out \cite{Antonov16} that for some fixed points the computation of the eigenvalues
of the matrix (\ref{eq:RG_Omega}) is cumbersome and rather unpractical. In those
cases it is possible to obtain information about the stability from
analyzing  RG flow equations (\ref{eq:RG_gellmann}). 
 Using approach from Sec. (\ref{subsec:UV}) the following relations
\begin{align}
   & \Delta_{\tilde{\psi}} = \frac{d}{2} + \gamma_{{\psi}^{\prime *}},\quad
   \Delta_{\psi} = \frac{d}{2} + \gamma_{\psi^*},\quad
   \Delta_\tau = 2 + \gamma_\tau^*.
\end{align}
are derived in straightforward manner.
Important information about the  physical system can be read out from the behavior
of correlation functions, which can be expressed
in terms of the  cumulant Green functions. In the percolation problems one is typically
interested \cite{HHL08,JanTau04} in the behavior of the following functions
\begin{enumerate}[a)]
  \item The number $N(t,\tau)$ of active particles generated by a seed at the origin
        \begin{equation}
           N(t) = \int \dRM^d \mr\mbox{ } \langle \psi (t,\mr) {\psi^{\prime }} (0,{\bm 0}) \rangle_\text{conn}  ,
           \label{eq:react_scale_N}
        \end{equation}
        where the notation from Eq. (\ref{eq:1PI_writing}) has been employed.
  \item The mean square radius $R^2(t)$ of percolating particles, which 
        started from the origin at time $t=0$
        \begin{equation}
           R^2(t) = \frac{\int \dRM^d\mr\mbox{ } \mr^2 \langle \psi (t,\mr) {\psi^{\prime }} (0,{\bm 0}) \rangle_\text{conn}  }
           {2d\int \dRM^d\mr\mbox{ }  \langle \psi (t,\mr) {\psi^{\prime }} (0,{\bm 0}) \rangle_\text{conn}  }.
           \label{eq:react_scale_R}
        \end{equation}
  \item Survival probability $P(t)$ of an active cluster originating
        from a seed at the origin (see \cite{Janssen05} for derivation)
        \begin{equation}
           P(t) = - \lim_{k\rightarrow\infty} \langle 
           {\psi'}(-t,{\bm 0}) \eRM^{-k\int \dRM^d\mr\mbox{ } \psi(0,\mr)}
           \rangle.
           \label{eq:react_scale_P}
        \end{equation}
\end{enumerate}

By straightforward analysis \cite{JanTau04} it can be shown that the scaling behavior of
these functions is given by the asymptotic relations
\begin{align}
   R^2(t) \sim 
   t^{2/\Delta_\omega}, \quad   
   N(t) \sim t^{-(\gamma_{\psi^*} +  \gamma_{{\psi}^{\prime *}})/\Delta_\omega},\quad
   P(t) \sim t^{-(d+\gamma_{\psi^*} +  \gamma_{{\psi}^{\prime *}})/2\Delta_\omega} .
   \label{eq:react_scalingGF}
\end{align}

Although
to some extent it is possible to obtain coordinates of the fixed points, the
eigenvalues of the matrix (\ref{eq:RG_Omega}) in this case pose a more severe technical problem.
Hence, in order to gain some physical insight into the 
structure of the model, the overall analysis is divided into special cases and
analyzed separately.

The fixed points for rapid change model are listed in Tab. \ref{tab:perkol_rchm}, for frozen velocity field
in Tab. \ref{tab:perkol_fvf} and for an illustration purposes nontrivial point in Tab. \ref{tab:perkol_nontrivial1}.
For  convenience a new parameter $a'$ has been introduced via the relation
$a'=(1-2a)^2$ and NF stands for Not Fixed, i.e., for the given FP the corresponding value
of a charge coordinate could not be unambiguously determined. 
\begin{table*}
  \begin{center}
    \begin{tabular}{| c | c | c | c | c |}
      \hline\noalign{\smallskip}
      \fp{I}{} & $g_1'^{*}$ & $g_2^{*}$ & $u_2^{*}$ & $a'^{*}$ 
       \\  \noalign{\smallskip}\hline\noalign{\smallskip}
      \fp{I}{1} & $0$ & $0$ &  NF & NF 
       \\  \noalign{\smallskip}\hline\noalign{\smallskip}
      \fp{I}{2} & $0$ & $\frac{2\eps}{3}$&  $0$ & $0$ 
       \\  \noalign{\smallskip}\hline\noalign{\smallskip}
      \fp{I}{3} & $\frac{4\xi}{3+\alpha}$ &  $0$ & $0$ & NF 
       \\  \noalign{\smallskip}\hline\noalign{\smallskip}
      \fp{I}{4} & $-\frac{4\xi}{3+\alpha}$ &  $0$ & $\frac{1}{2}$ & $0$ 
       \\  \noalign{\smallskip}\hline\noalign{\smallskip}
      \fp{I}{5} & $\frac{24\xi-2\eps}{3(5+2\alpha)}$ & 
      $ \frac{4\eps(3+\alpha) -24\xi}{3(5+2\alpha)}$&  $0$ & $0$
       \\  \noalign{\smallskip}\hline\noalign{\smallskip}
      \fp{I}{6} & $\frac{2[\eps -4\xi]}{9+2\alpha}$ &
      $ \frac{4\eps(3+\alpha) +24\xi}{3(9+2\alpha)}$  & 
      $\frac{(3+\alpha)\eps -3\xi(7+2\alpha)}{3(3+\alpha)[\eps - 4\xi]}$ & 
      $0 $
       \\  \noalign{\smallskip}\hline\noalign{\smallskip}
      \fp{I}{7} & $-\frac{\xi}{3+\alpha}$ & $2\xi$  & $1$ &
      $-\frac{3(5+2\alpha)}{\alpha}+\frac{2(3+\alpha)\eps}{\alpha\xi}$
      \\ \hline\noalign{\smallskip}
    \end{tabular}
      \caption{List of all fixed points obtained in the rapid-change limit.  The coordinate
  $w^*$ is equal to $0$ for all points.}
       \label{tab:perkol_rchm}
  \end{center}
\end{table*}

\begin{table*}
  \begin{center}
    \begin{tabular}{| c | c | c | c | c | c |}
       \hline\noalign{\smallskip}
      \fp{II}{} & $g_1^{*}$ & $g_2^{*}$ & $u_2^{*}$ & ${a'}^{*}$ 
       \\  \noalign{\smallskip}\hline\noalign{\smallskip}
      \fp{II}{1} & $0$ & $0$ & NF & NF
       \\  \noalign{\smallskip}\hline\noalign{\smallskip}
      \fp{II}{2} & $0$ & $\frac{2\eps}{3}$ & $0$ & $0$ 
       \\  \noalign{\smallskip}\hline\noalign{\smallskip}
      \fp{II}{3} & $\frac{2y}{9}(3-\alpha)$ & $0$ & $\frac{\alpha}{2(\alpha-3)}$ & 
      $0$ 
       \\  \noalign{\smallskip}\hline\noalign{\smallskip}
      \fp{II}{4} & $\frac{2 (\eps-y)}{2 \alpha-9}$ & $\frac{4 [3 \eps + 2y ( \alpha- 6)]}
      {2 \alpha-9}$ & 
      $1$ & $\frac{\eps(12-\alpha)+5y(\alpha-6)}{\alpha(\eps-y)}$ 
       \\  \noalign{\smallskip}\hline\noalign{\smallskip}
      \fp{II}{5} & $-\frac{2 [6 \eps+5 y( \alpha -3)]}{3 (9+\alpha)}$ & $0$ & $
      \frac{3 [\eps+y(\alpha-1)]}{6 \eps+5y  (\alpha -3)}$ & 
           $\frac{18\eps-(\alpha-6)(\alpha-3)y}{\alpha[6\eps+5(\alpha-3)y]}$
       \\  \noalign{\smallskip}\hline\noalign{\smallskip}      
      \fp{II}{6} & NF & $0$ & NF & NF
       \\  \noalign{\smallskip}\hline\noalign{\smallskip}
      \fp{II}{7} & $g_1^{*}$ & $g_2^{*}$ & $u_2^{*}$ & $0$ 
       \\  \noalign{\smallskip}\hline\noalign{\smallskip}
      \fp{II}{8} & $g_1^{*}$ & $g_2^{*}$ & $u_2^{*}$ & $0$ 
      \\ \hline\noalign{\smallskip}
    \end{tabular}
      \caption{List of all fixed points obtained in the frozen velocity limit. The value
  of the charge $u_1^*$ is equal to  $0$ for all points. }
  \label{tab:perkol_fvf}
  \end{center}
\end{table*}

\begin{table*}
  \begin{center}
    \begin{tabular}{| c | c | c | c | c | c | c |}       
       \hline\noalign{\smallskip}
      \fp{}{} & $g_1^{*}$ & $g_2^{*}$ & $u_1^*$ & $u_2^{*}$ & ${a'}^{*}$ 
       \\  \noalign{\smallskip}\hline\noalign{\smallskip}
      \fp{II}{7} & $0.532193$ & $9.89135$ & $0$ & $0.37859$ & $0$ 
       \\  \noalign{\smallskip}\hline\noalign{\smallskip}
      \fp{III}{1} & $0.365039$ & $6.38225$ & $0.24709$ & $0.352422$ & $0$ 
       \\  \noalign{\smallskip}\hline\noalign{\smallskip}
      \fp{III}{2} & $0.399062$ & $7.29847$ & $0.148951$ & $0.35954$ & $0$        
       \\ \hline\noalign{\smallskip}
    \end{tabular}     
     \caption{Coordinates of the IR stable fixed points obtained by numerical
  integration of flow equations (\ref{eq:RG_gellmann}) the model (\ref{eq:react_bare_act}). The relevant
  parameters were fixed as follows $\alpha=110,\eps = 1$ and
     $y=2\eta=8/3$ (Kolmogorov regime).
   }
  \label{tab:perkol_nontrivial1}
  \end{center}
\end{table*}

\subsubsection{Rapid change \label{subsec:rapid}}
First,  an analysis of the rapid-change limit is performed.
It is convenient \cite{Ant99,Ant00}
to introduce the new variables $g'_1$ and $w $ given by
\begin{equation}
   g_1' = \frac{g_1}{u_1}, \quad w=\frac{1}{u_1}.
   \label{eq:react_def_rapid}
\end{equation}
The rapid change limit then corresponds to fixed points with a coordinate $w^* = 0$.
The beta-functions for the charges (\ref{eq:react_def_rapid})
are easily obtained 
\begin{align}
   & \beta_{g_1'} = g_1' (\eta - y +\gamma_D - 2\gamma_v) , \quad
   & &\beta_{w} = w(\eta - \gamma_D ).   
\end{align}
Analyzing the resulting system of equations  seven possible regimes
 can be found. 
Due to the cumbersome form of the matrix
(\ref{eq:RG_Omega}),  the determination of all the corresponding eigenvalues in 
 an explicit form is impossible. In particular, for nontrivial fixed points (with non-zero coordinates 
 of $g_1',g_2$ and $u_2$) the resulting expressions
are of a quite unpleasant form. Nevertheless, using numerical software \cite{mathematica}
it is possible to obtain
all the necessary information about the fixed points' structure and in this way
the boundaries between the corresponding regimes have been obtained.
In the analysis it is advantageous to exploit additional constraints
following from the physical interpretation of the charges. For example,  
 $g_1'$ describes the density of kinetic energy of the velocity fluctuations,
 $g_2$ is equal to $\lambda^2$ and $a'$ will be later
 on introduced 
 as $(1-2a)^2$. Hence, it is clear that these
 parameters have to be non-negative real numbers. Fixed points that 
 violate this condition can be immediately discarded as non-physical.
 
 Out of seven possible fixed points, only 
four are IR stable: \fp{I}{1}, \fp{I}{2}, \fp{I}{5} and \fp{I}{6}. 
Thus,  only regimes which correspond to those points
could be in principle realized in real physical systems. As
expected \cite{Ant00},  the coordinates of these fixed points 
and the scaling behavior 
of the Green functions 
depend only on the parameter $\xi=y-\eta$.
In what follows, we restrict our discussion
only to them. 

\begin{figure}[h!]
  \centering
  \includegraphics[width=7cm]{\PICS 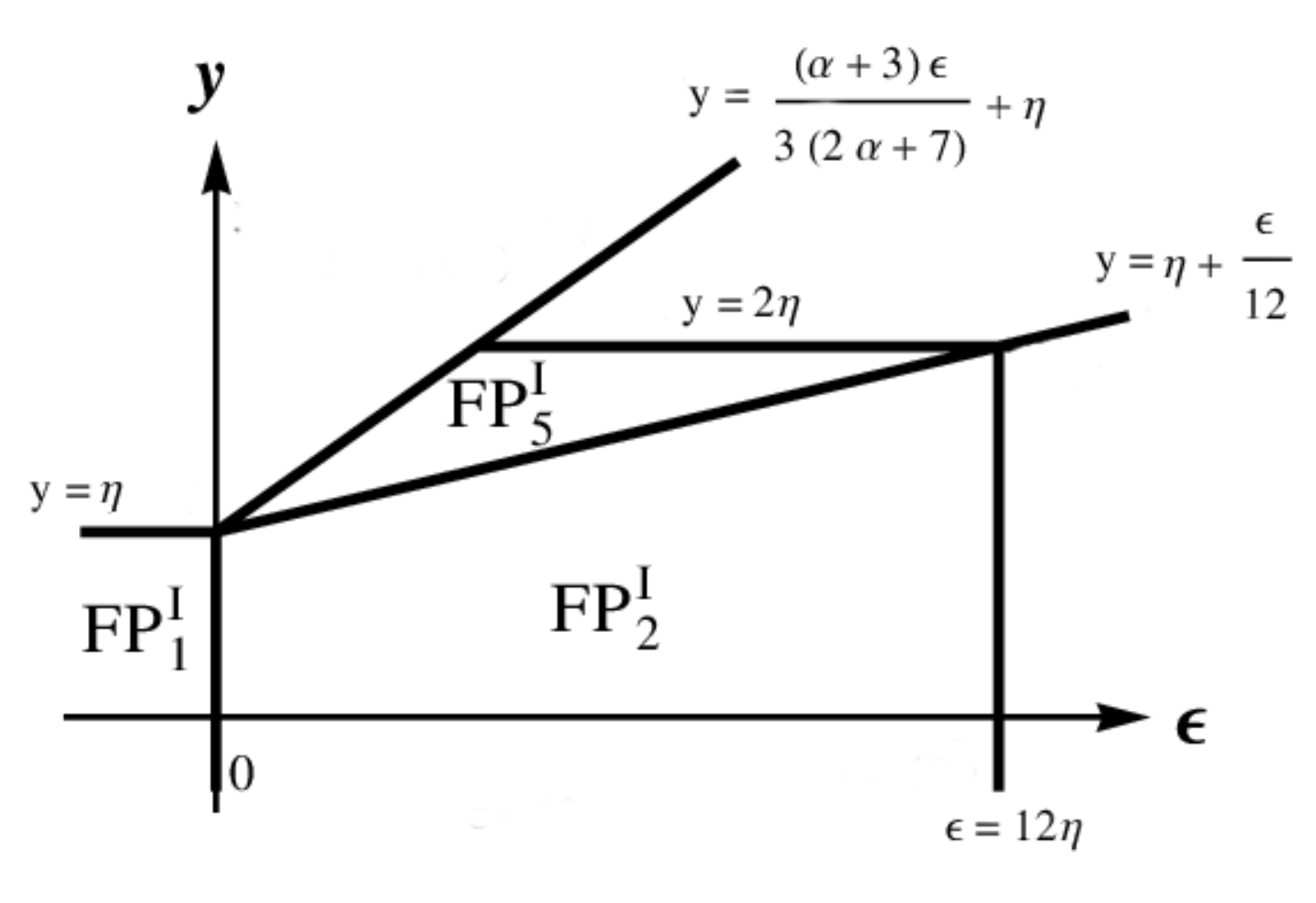}
  \caption{A qualitative sketch of the 
  regions of stability for the fixed points in the limit of the rapid-change model.
The borders between the regions are depicted with the bold lines.
  }
  \label{fig:stab_rch}
\end{figure}

The \fp{I}{1} represents the free (Gaussian) FP for which all interactions are
irrelevant and ordinary perturbation theory is applicable.
As expected, this regime is IR stable  in the region
\begin{equation}
   y < \eta,\quad \eta>0,\quad \eps < 0.
   \label{eq:react_FPI1_region}
\end{equation}
  The latter condition ensures that the system is considered above the upper
 critical dimension $d_c=4$.
For  \fp{I}{2} the correlator of the velocity field is irrelevant and 
this point describes standard the DP universality class \cite{JanTau04} and is IR stable 
in the region
\begin{equation}
  \eps > 0,\quad \eps/12 + \eta > y,\quad \eps < 12\eta. 
  \label{eq:react_FPI2_region}
\end{equation} 
The remaining two fixed points constitute nontrivial regimes for which velocity
fluctuations as well as percolation interaction become relevant.
The \fp{I}{5} is IR stable in the region given by
\begin{equation}
 (\alpha+3)\eps > 3(2\alpha+7)(y-\eta),\quad 12(y-\eta)>\eps,\quad 2\eta>y.
 \label{eq:react_FPI5_region}
\end{equation}
The boundaries for \fp{I}{6} can be only computed by numerical calculations. 

\begin{figure}[h!]
  \includegraphics[width=4.25cm]{\PICS 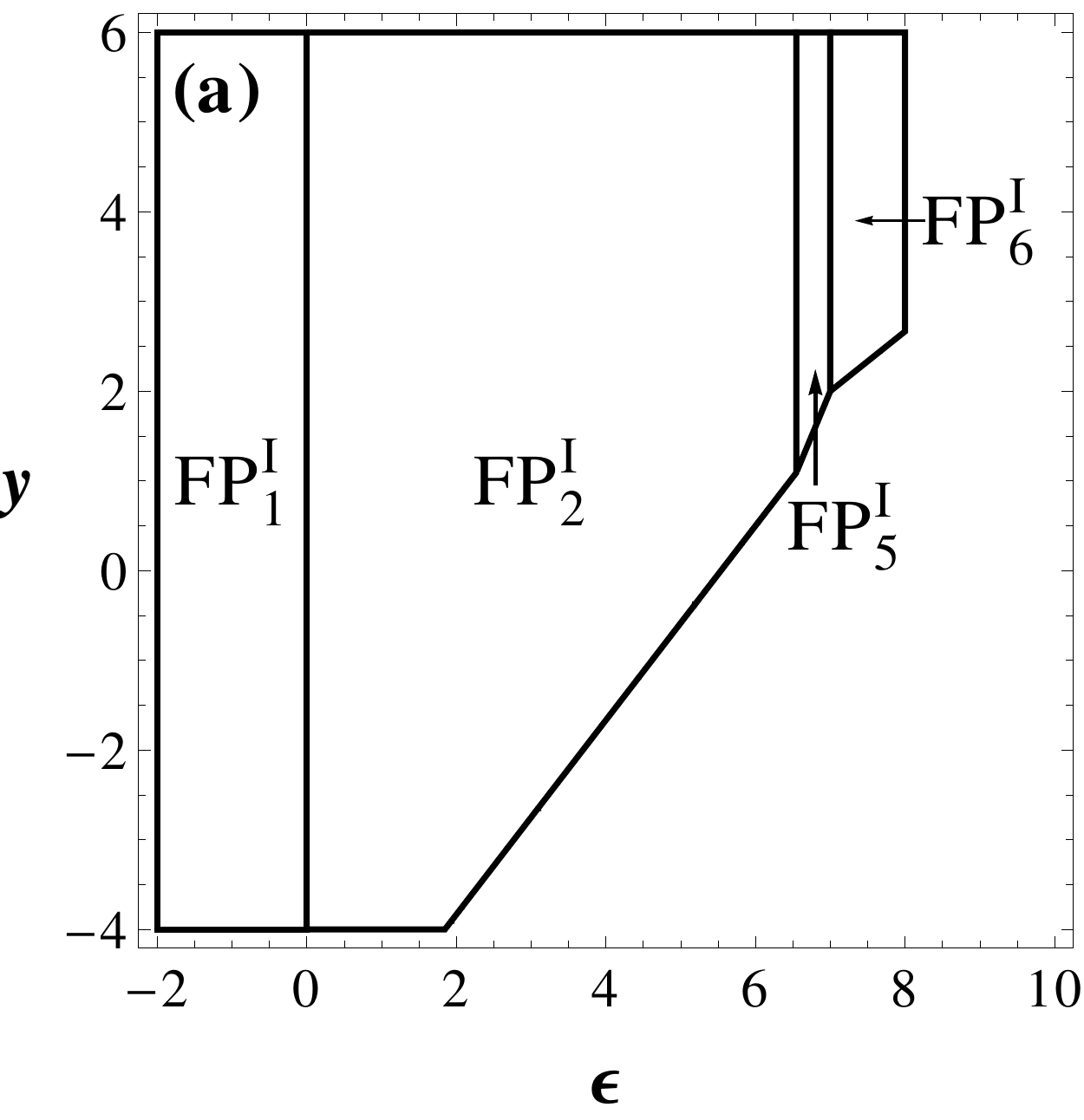}
  \includegraphics[width=4.25cm]{\PICS 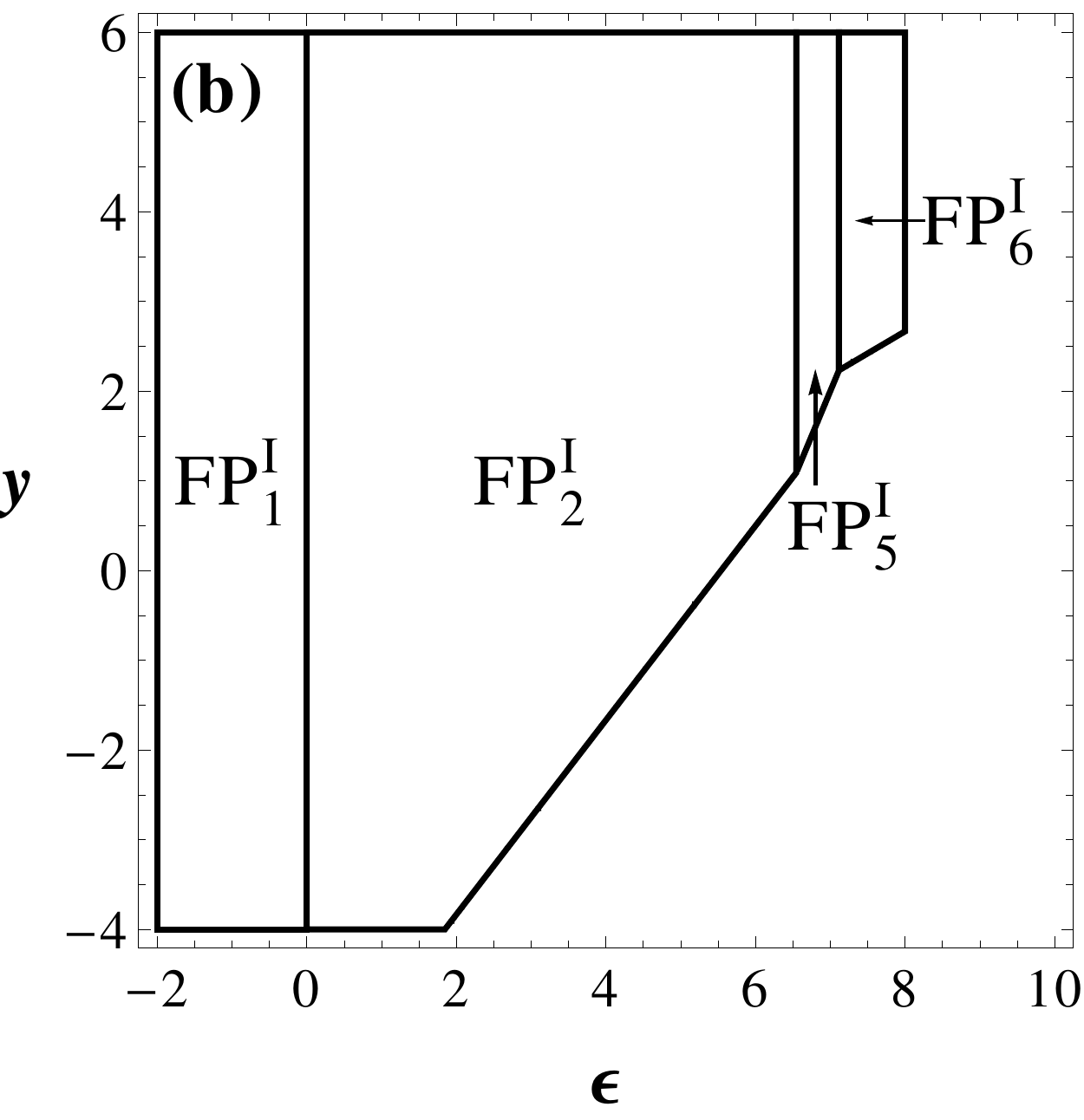}
  \includegraphics[width=4.25cm]{\PICS 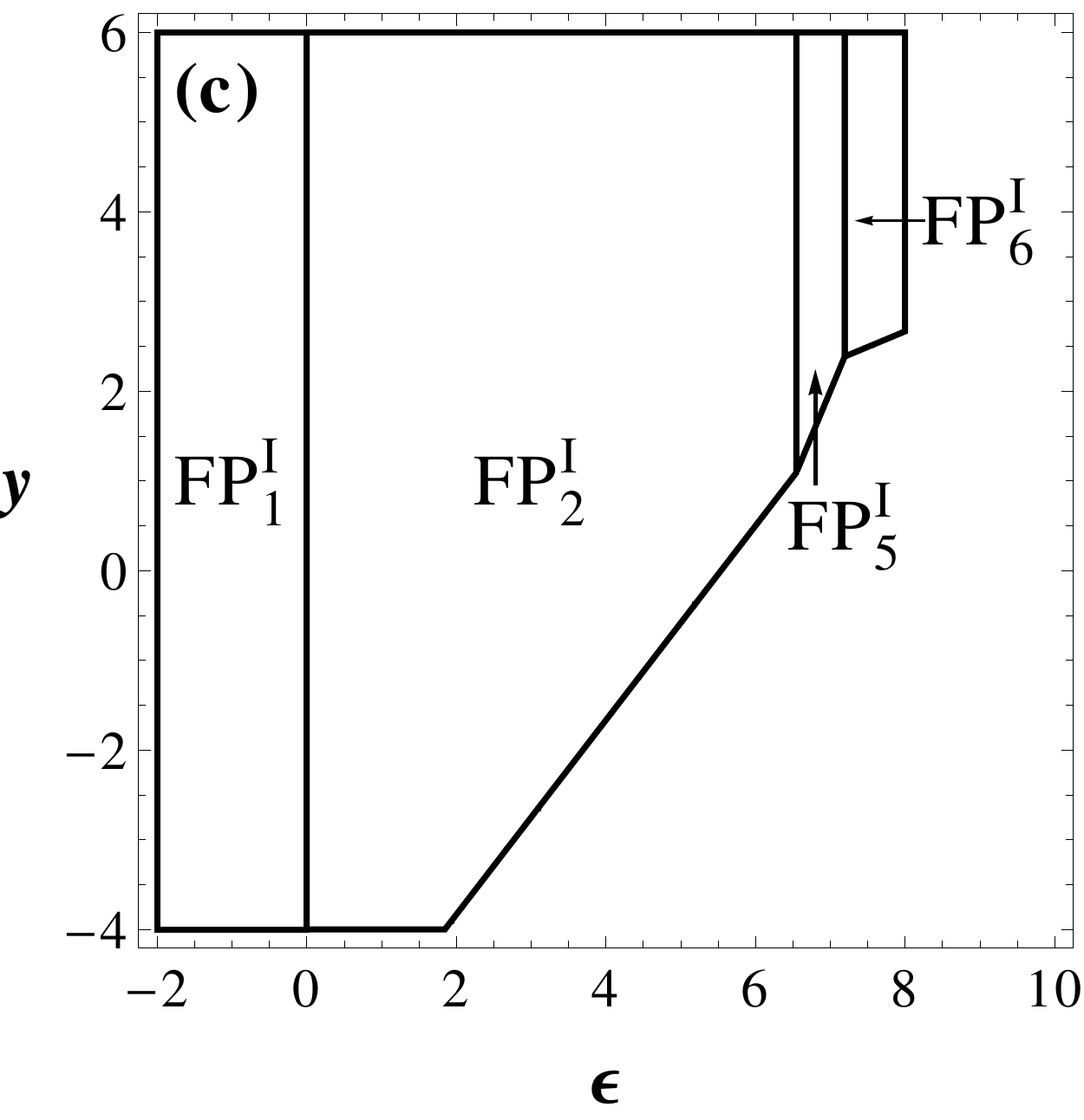}
  \caption{Fixed points' structure for the thermal noise situation 
	(\ref{eq:react_cond_thermal}). 
	From above to bottom the compressibility parameter $\alpha$ attains
	consecutively the values: (a) $\alpha=0$, (b) $\alpha=5$ and (c) 
	$\alpha=100$.	}
  \label{fig:thermal}	
\end{figure}

Using the information about the phase boundaries, a qualitative picture of the phase diagram
can be constructed. In Fig. \ref{fig:stab_rch} the situation in the plane $(\eps,y)$ is depicted.
The compressibility affects 
 only the outer boundary of \fp{I}{5}. The larger value of $\alpha$ the larger area
of stability. Also the realizability of the regime \fp{I}{5} 
crucially depends on the nonzero value of $\eta$.

The important subclass of the rapid-change  limit constitutes thermal velocity
fluctuations, which
are characterized by the quadratic dispersion law \cite{FNS77}. In formulation (\ref{eq:react_kernelD}) this is achieved
by considering the following relation:
\begin{equation}
  \eta = 6 +y - \eps   
  \label{eq:react_cond_thermal}
\end{equation}
which follows directly from  expression (\ref{eq:hel_rapid}).
The situation for increasing values of the parameter $\alpha$ is depicted in
Fig. \ref{fig:thermal}. We see that for physical space dimensions 
$d=3\mbox{ }(\eps=1)$ and $d=2\mbox{ }(\eps=2)$ the only stable regime is that of pure DP.
The nontrivial regimes \fp{I}{5} and \fp{I}{6} are realized only in the nonphysical
region for large values of $\eps$. This numerical result confirms our
previous expectations \cite{AntKap08,AntKap10}. 
It was pointed out \cite{Hnatich00,HHL13} that
genuine thermal fluctuations could change IR stability of the given universality class. 
However, this is not realized for the percolation process.
\subsubsection{Regime of frozen velocity field \label{subsec:frozen}}

The regime of  the frozen
velocity field
corresponds to the charge constraint $u_{1}^*=0$.
 As analysis show eight possible fixed points are obtained.
However, only three
of them (\fp{II}{1}, \fp{II}{2} and \fp{II}{7}) could be physically realized (IR stable).
 
\begin{figure}[h!]
  \includegraphics[width=4.25cm]{\PICS 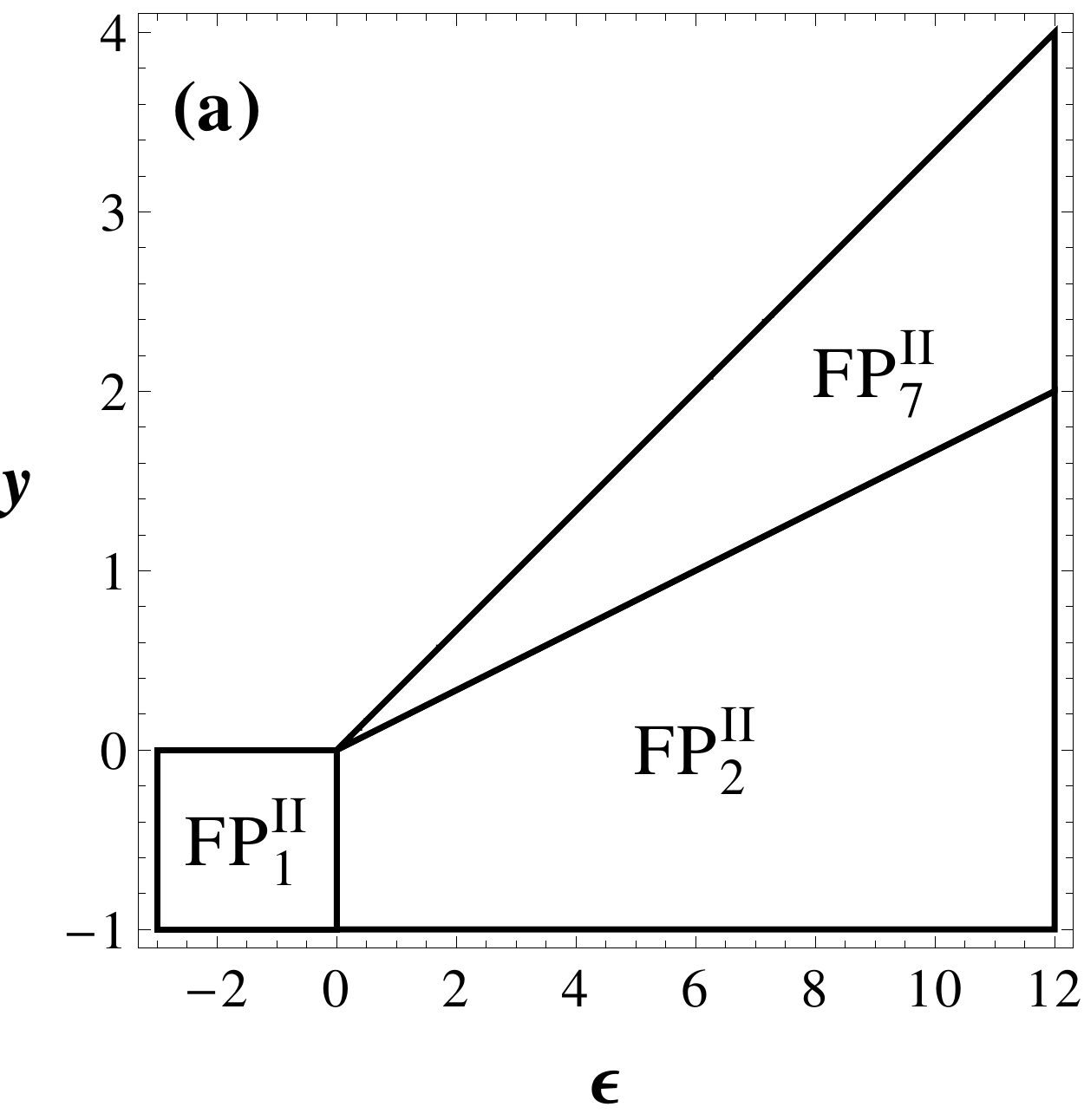}
  \includegraphics[width=4.25cm]{\PICS 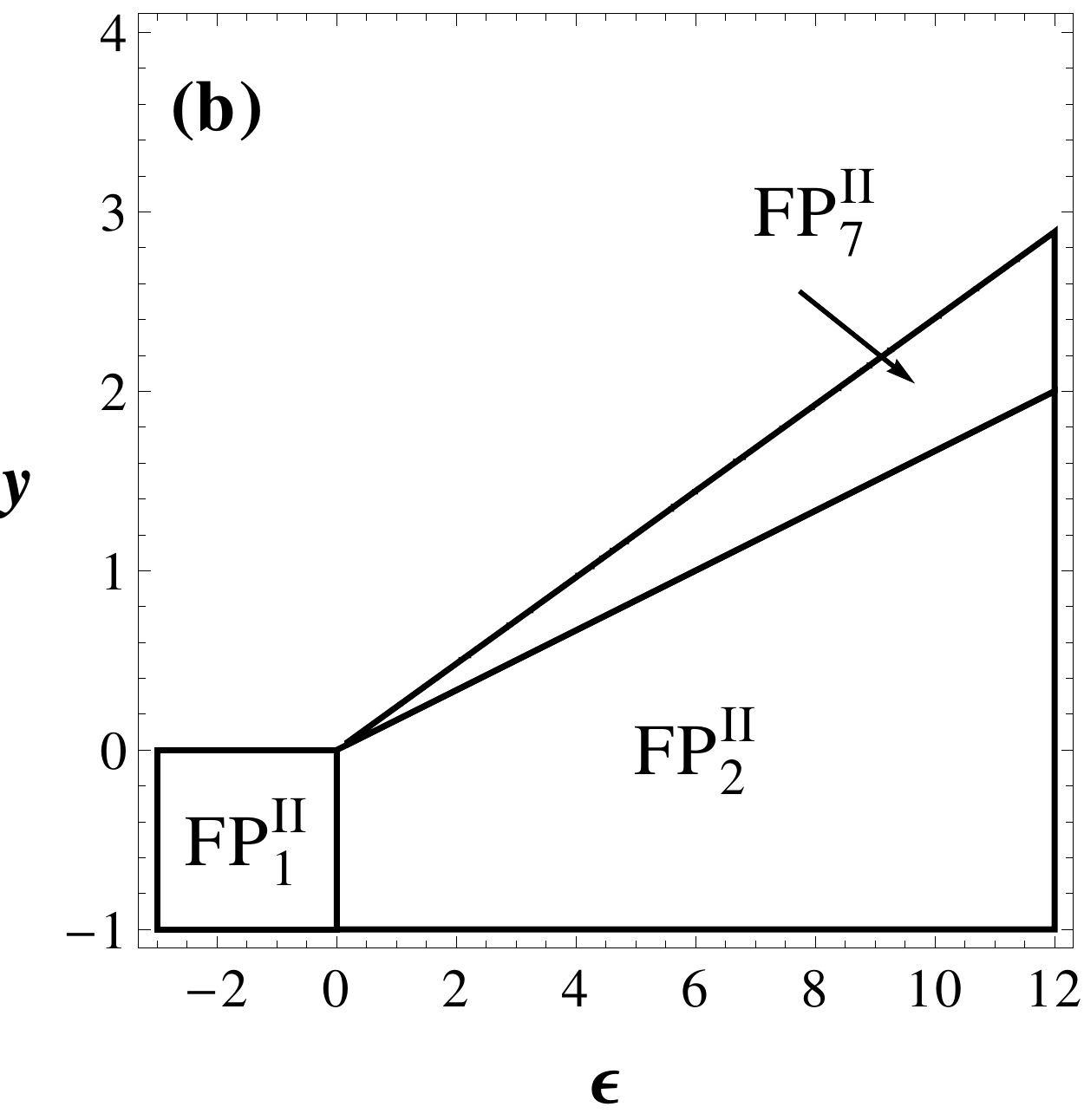}
  \includegraphics[width=4.25cm]{\PICS 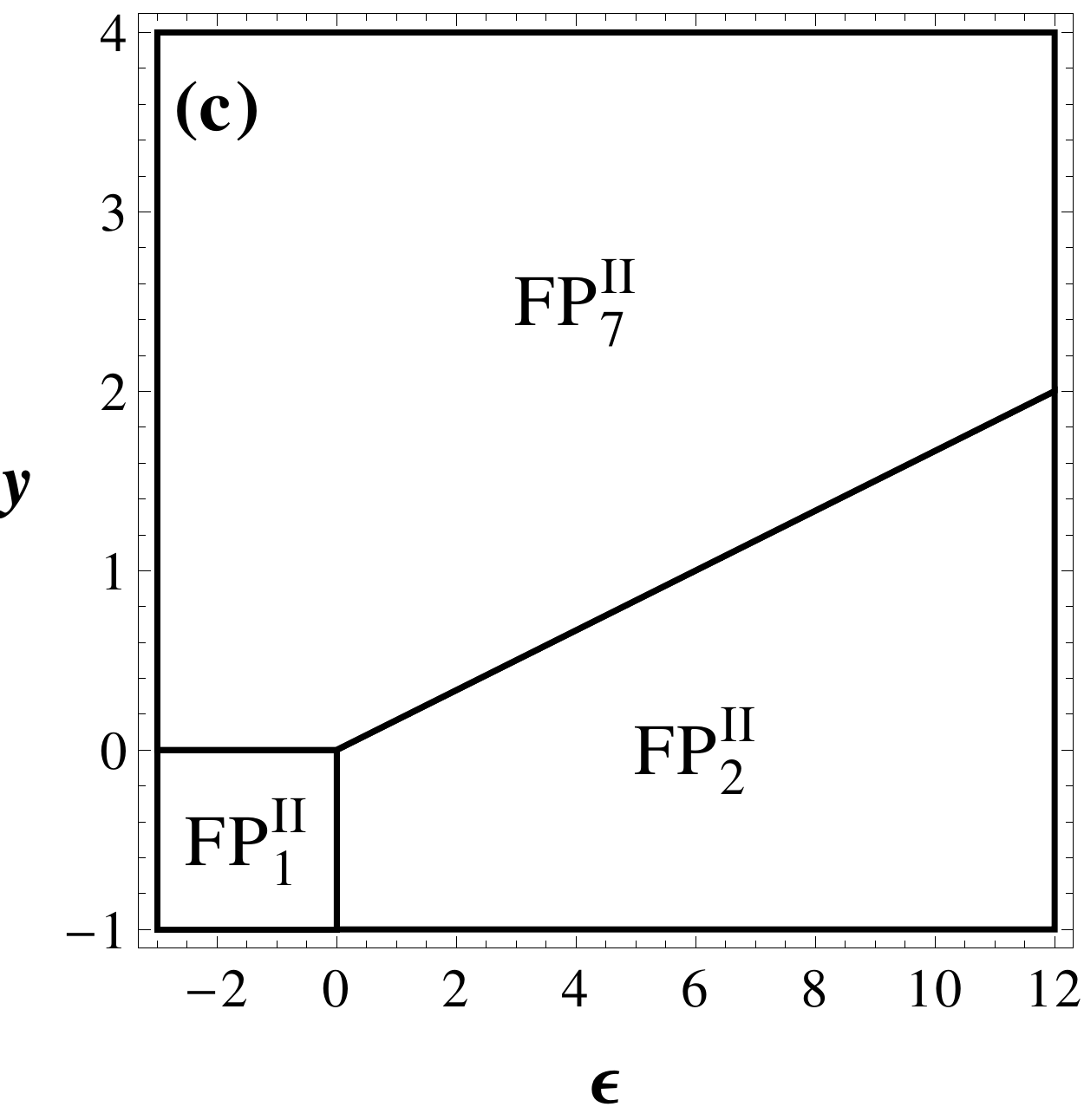}
  \caption{Fixed points' structure for frozen velocity case
        with $\eta=0$.
	From above to bottom the compressibility parameter $\alpha$ attains
	consecutively the values: (a) $\alpha=0$, |(b) $\alpha=3.5$ and (c) $\alpha=8$.}
	\label{fig:FVF_struct}	
\end{figure}
The fixed point \fp{II}{1} describes the free (Gaussian) theory. It is stable in the region
\begin{equation}
   y<0,\quad \eps < 0, \quad \eta < 0.
   \label{eq:react_free_frozen} 
\end{equation}

For  \fp{II}{2}
the velocity field is asymptotically irrelevant and the only 
relevant interaction is due to the percolation process itself. This regime is stable in
 the region
\begin{equation}
   \eps > 6y, \quad \eps > 0,\quad \eps > 12\eta.
   \label{eq:react_DP_frozen}
\end{equation}

On the other hand,  \fp{II}{7} represents a truly nontrivial regime for which both velocity and
percolation are relevant. 
Since for the points \fp{II}{1} and \fp{II}{2} 
the velocity field could be effectively neglected, the trivial observation is that
these boundaries do not depend on the value of the parameter $\alpha$. The stability region
of \fp{II}{7} can be computed only numerically. 

In order to illustrate the  influence of compressibility on the stability in the nontrivial regime
\fp{II}{7},  let us take a look at situation for $\eta=0$. For other values of $\eta$ the situation
remains qualitatively the same. The situation for increasing values of $\alpha$ is depicted
 in Fig. \ref{fig:FVF_struct}. For $\alpha=0$ there is a region
of stability for \fp{II}{7}, which shrinks for the immediate value $\alpha=3.5$ to a smaller
area. Numerical
analysis \cite{Antonov16} shows that this shrinking continues well down to the value $\alpha=6$. A further increase
 of $\alpha$ leads to a substantially larger region of stability for the given
FP. Already for $\alpha=8$ this region covers all the rest of the $(y,\eps)$ plane.
The compressibility thus changes profoundly a simple picture expected from an incompressible case.
 Altogether the advection process becomes
more efficient due to the combined effects of  compressibility and the nonlinear terms.
\subsubsection{Turbulent advection \label{subsec:turbulent}}
In the last part the focus is on a special case of the turbulent advection.
Main aim is to determine whether Kolmogorov regime \cite{Frisch}, which
corresponds to the choice $y=2\eta=8/3$, could  lead to  a new
nontrivial regime for the percolation process.
In this section,  the parameter $\eta$ is always considered to attain its Kolmogorov value, $4/3$.
For a better visualization we present two-dimensional regions of stability
in the plane $(\eps,y)$ for different values of the parameter $\alpha$.
\begin{figure}[h!]
  \includegraphics[width=4.25cm]{\PICS 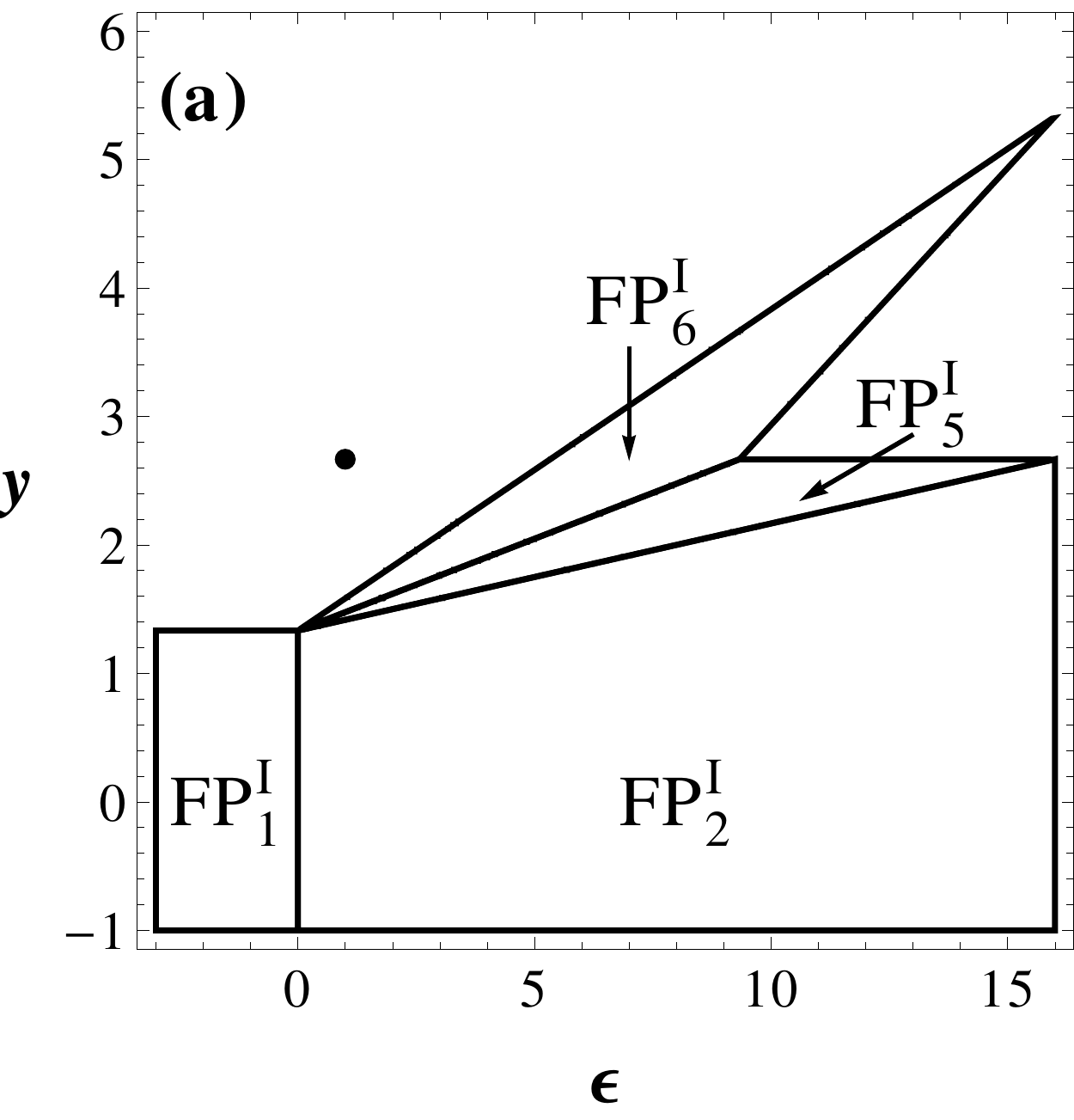} 
  \includegraphics[width=4.25cm]{\PICS 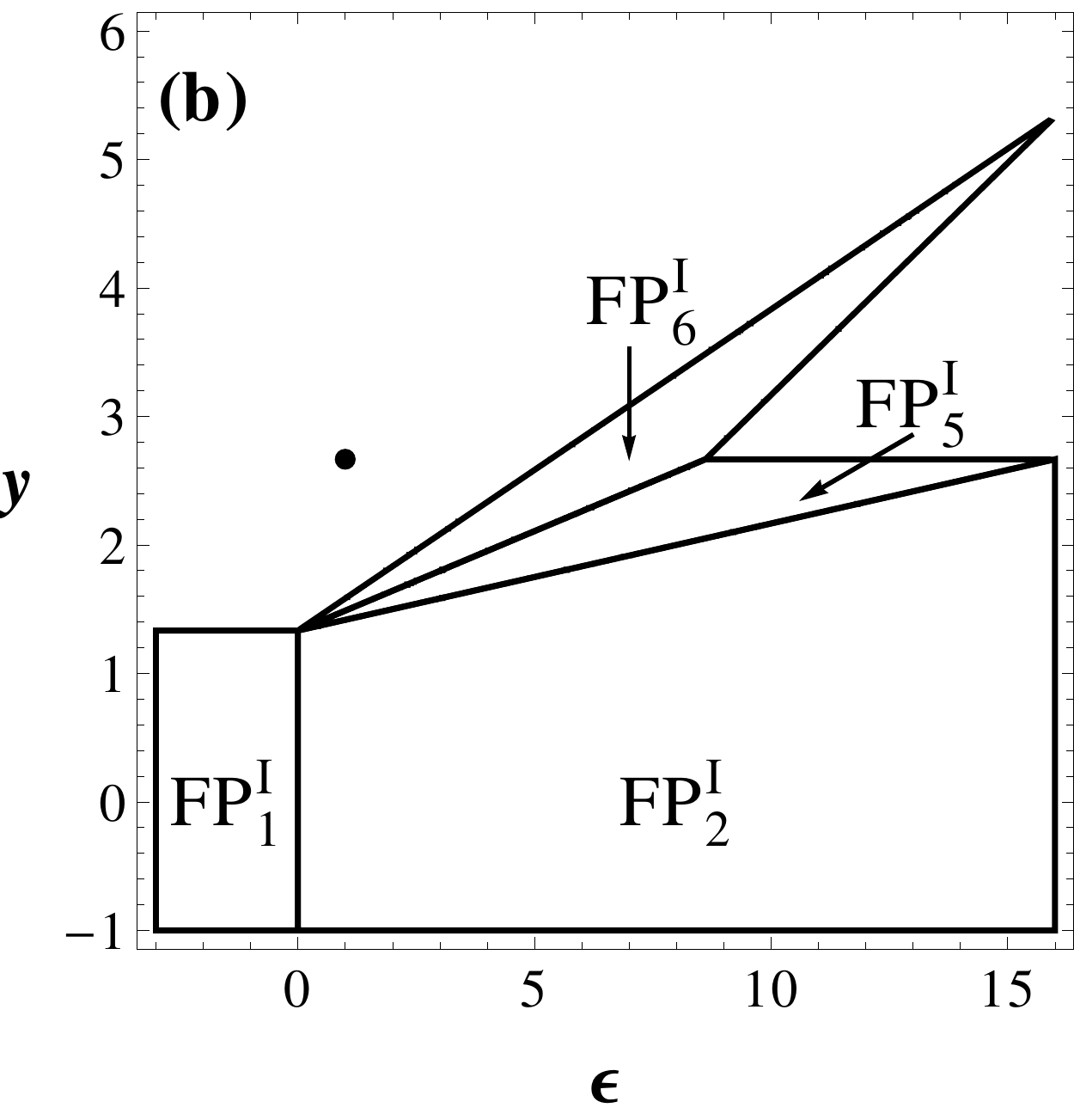}
  \includegraphics[width=4.25cm]{\PICS 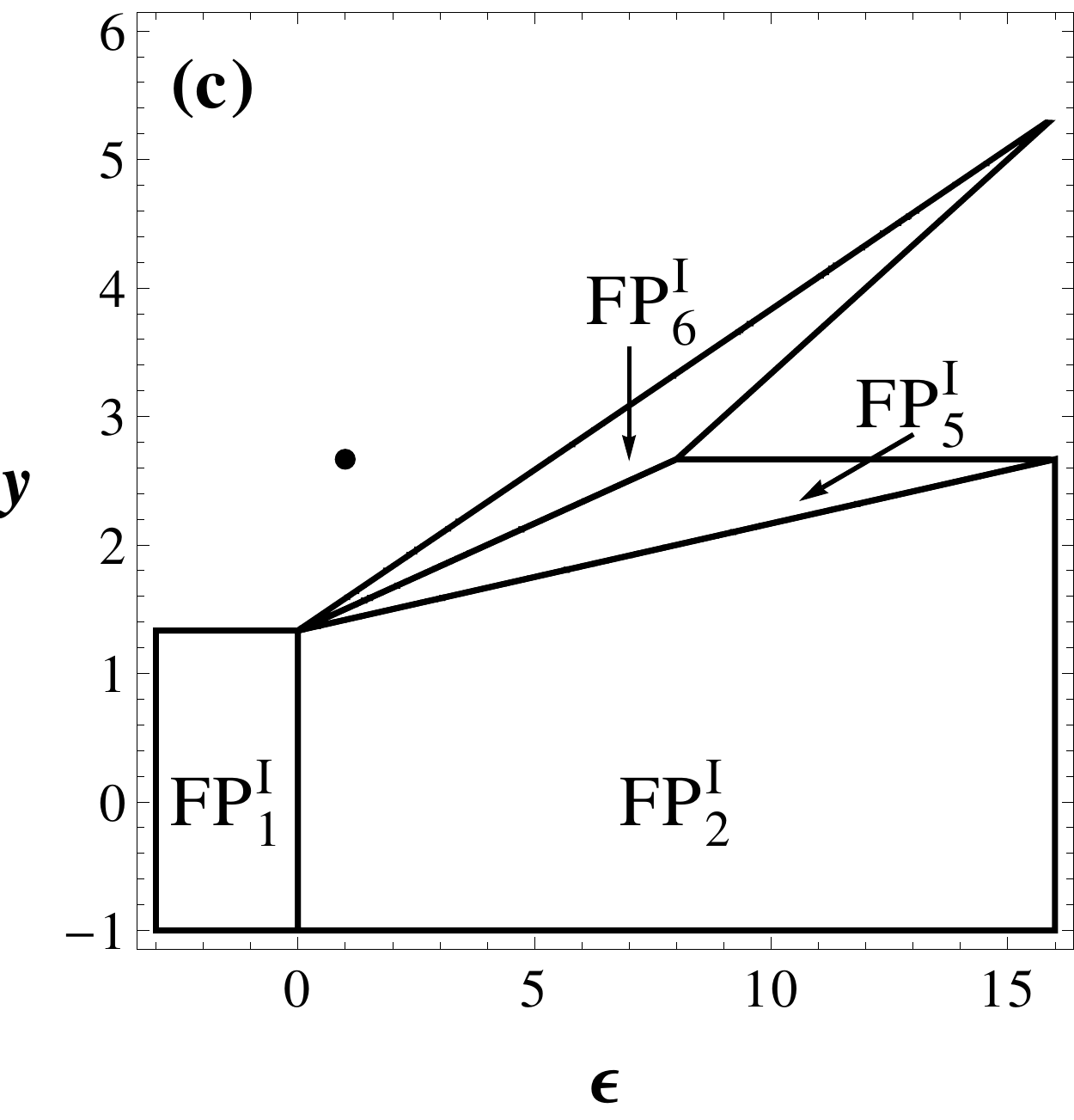}  
  \caption{  
  Fixed points' structure for rapid change model with $\eta=4/3$. 
	From above to bottom the compressibility parameter $\alpha$ attains
	consecutively the values: (a) $\alpha=0$, (b) $\alpha=5$ and (c) $\alpha=\infty$. 
	The dot denotes the coordinates of the
	three-dimensional Kolmogorov regime.}
  \label{fig:rapid}	
\end{figure}

First, let us reanalyze the situation for the rapid-change model. The result
is depicted in Fig.~\ref{fig:rapid}. It is clearly visible that for this case
 a realistic turbulent scenario ($\varepsilon=1$ or $\varepsilon=2$) falls out of the 
possible stable regions. This result is expected because the rapid-change model with
vanishing time-correlations could not properly describe well-known
turbulent properties \cite{Frisch,Monin}. We also observe that compressibility
mainly affects the boundaries between the regions \fp{I}{5} and \fp{I}{6}. However, this
happens mainly in the nonphysical region.

\begin{figure}[h!]
  \includegraphics[width=4.25cm]{\PICS 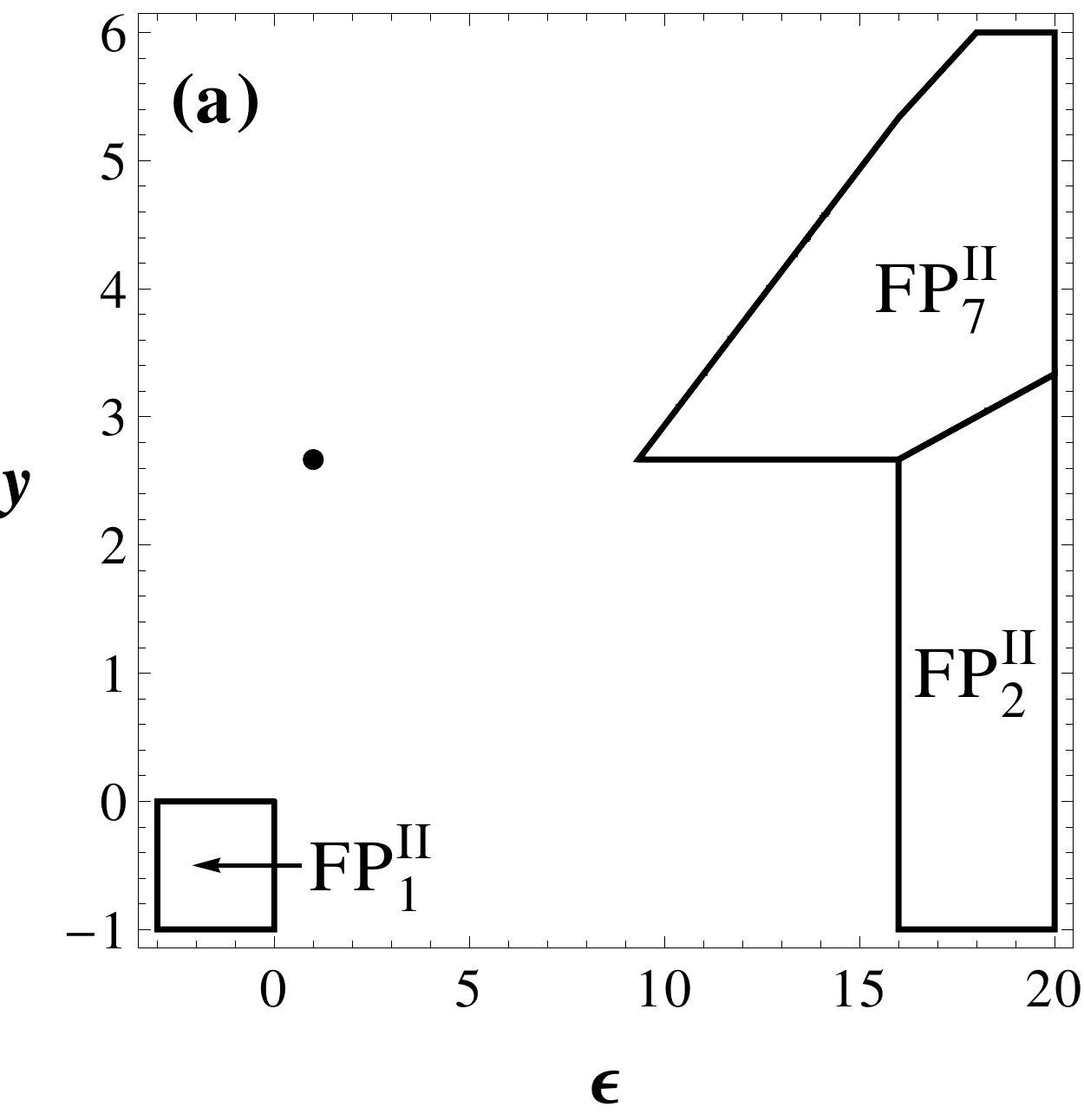}
  \includegraphics[width=4.25cm]{\PICS 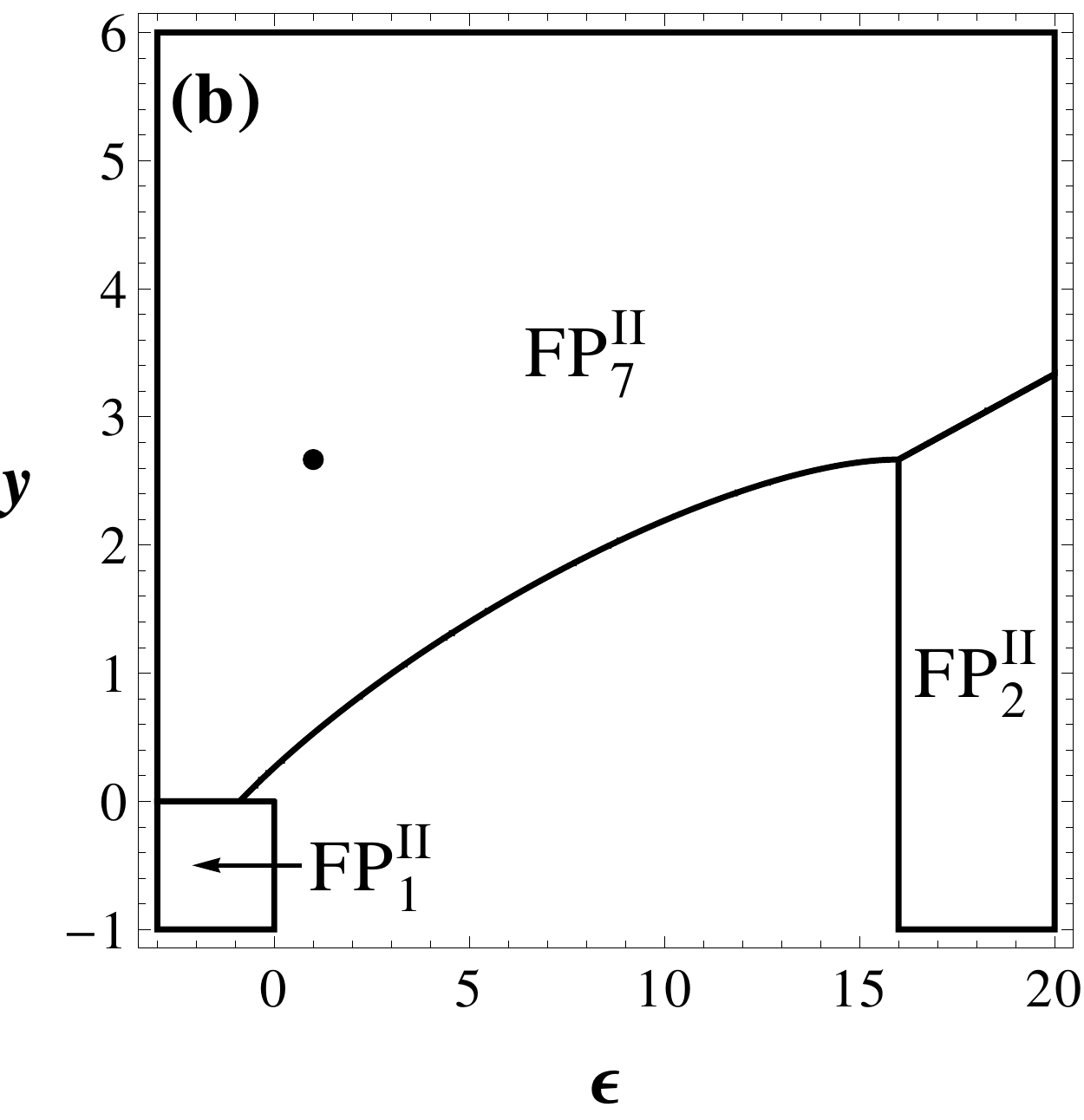}
  \includegraphics[width=4.25cm]{\PICS 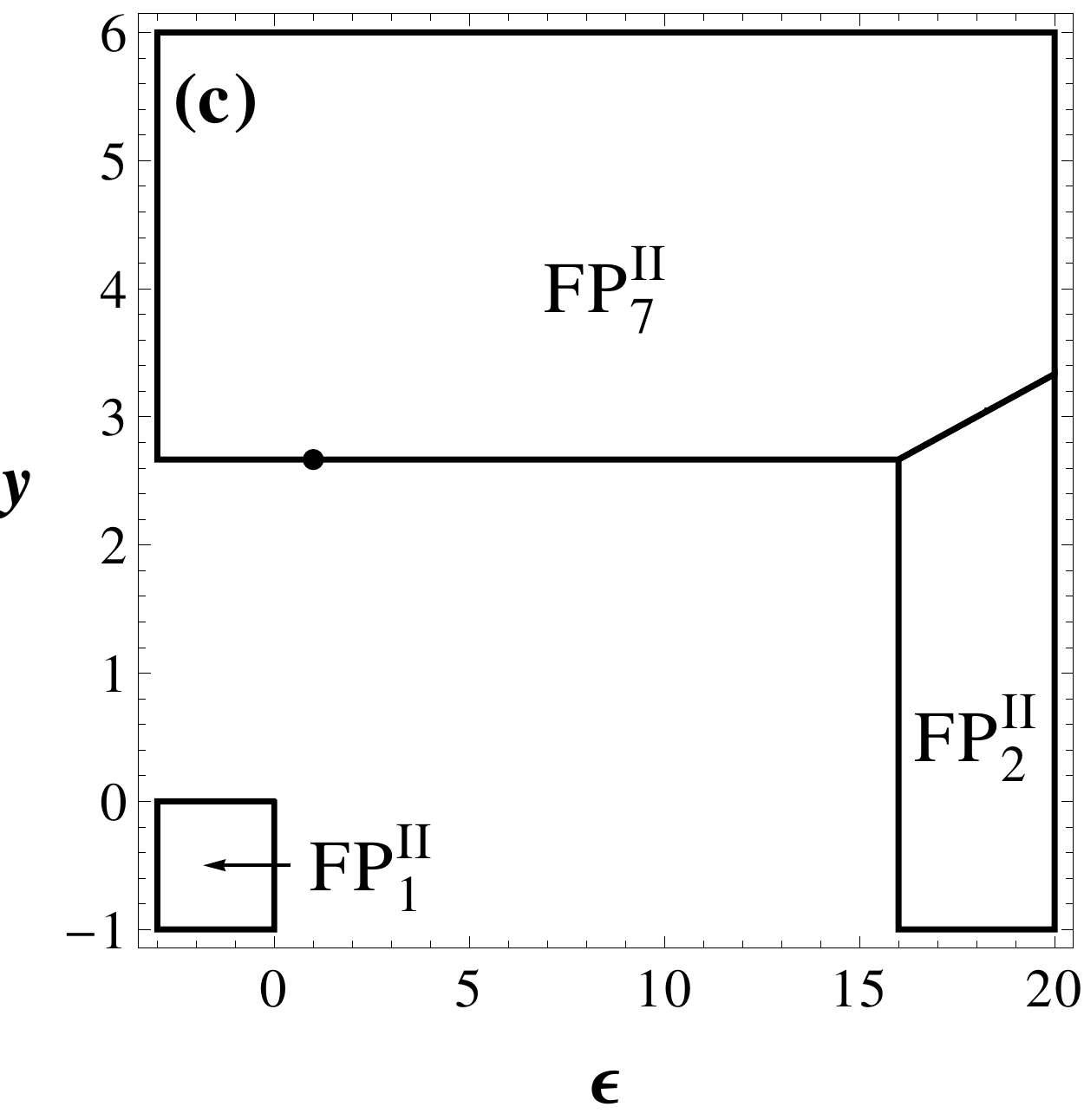}
  \caption{
  Fixed points' structure for  the frozen velocity case
        with $\eta=4/3$. From above to bottom the compressibility parameter $\alpha$ attains
	consecutively the values: (a) $\alpha=0$, (b) $\alpha=8$ and (c) $\alpha=\infty$.
	The dot denotes the coordinates of  the
	three-dimensional Kolmogorov regime.
  }
  \label{fig:frozen}  
\end{figure}

Next, let us make a
similar analysis for the frozen velocity field. The corresponding stability
regions are depicted in Fig. \ref{fig:frozen}.
Here it can be seen that the situation is more complex. 
The regime \fp{II}{2} is situated in the non-physical region and could not be realized. 
For small values of  the parameter $\alpha$
the Kolmogorov regime (depicted by a point) does not belong to the frozen velocity limit. 
However,
from a special value $\alpha = 6$ up to $\alpha\rightarrow\infty$ the Kolmogorov
regime belongs to the frozen velocity limit. Note that the bottom line for the region of
stability of \fp{II}{7} is exactly given by $y=4/3$. We observe that compressibility affects
mainly the boundary of the nontrivial region. We conclude that the presence of 
compressibility has  a stabilizing effect on the regimes where nonlinearities are relevant.

Finally, let us take  look at the nontrivial regime, which means that no
special requirements were laid upon the parameter $u_1$.  The corresponding differential
equations for the RG flow (\ref{eq:RG_gellmann}) have been analyzed numerically \cite{Antonov16}. 
The behavior of the  RG flows has been found as follows.
There exists a borderline in the plane $(\varepsilon,\alpha_c)$ given approximately by the expression
\begin{equation}
  \alpha_c = -12.131\varepsilon + 117.165.
  \label{eq:react_alpha_crit}
\end{equation}
Below $\alpha_c$,
only the frozen velocity regime corresponding to \fp{II}{7} is stable. 
Above $\alpha_c$, three fixed points \fp{II}{7}, \fp{III}{1} and \fp{III}{2} are observed. Whereas
two of them (\fp{II}{7} and \fp{III}{1}) 
are IR stable, the remaining one \fp{III}{2} is unstable in the IR regime. Again one of the stable FPs
 corresponds to \fp{II}{7}, but the new FP is a regime with finite correlation time. 
 Since all free parameters $(\eps,\eta,y,\alpha)$ are the same for both points, 
 which of the two points
 will be realized depends on the initial values of the bare parameters.
A similar situation is observed for the stochastic magnetohydrodynamic
turbulence \cite{HHJ01}, where the crucial role is played by a forcing decay-parameter $a$ (See Eq. (\ref{eq:mhd2D_corel2}).
\begin{figure}[h!]
  \centering
  \includegraphics[width=6cm]{\PICS 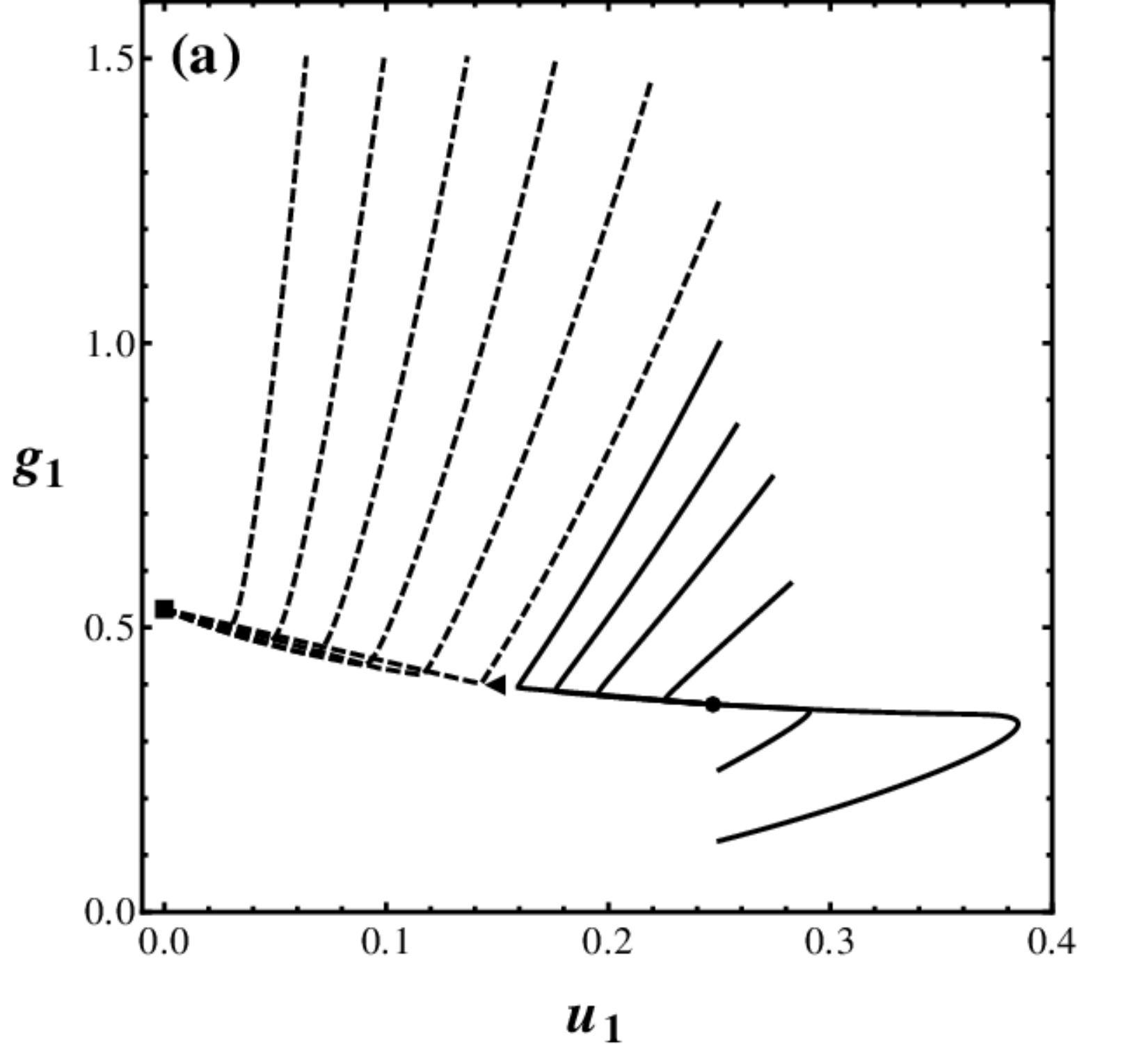} 
   \includegraphics[width=6cm]{\PICS 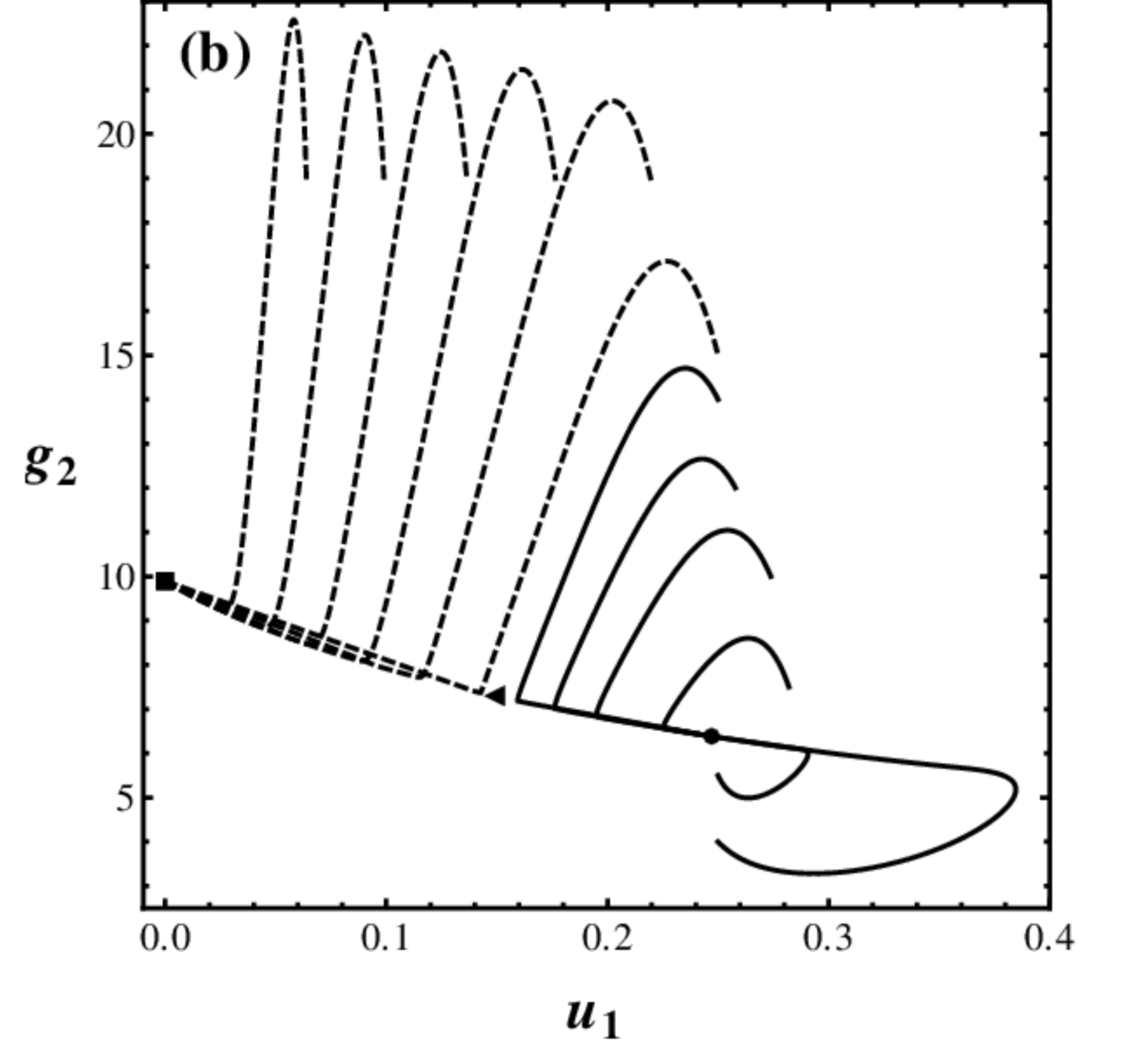}  
  \caption{
    Demonstration of the
    RG trajectories flows' in:  (a) the plane $(g_1,u_1)$ and (b) the plane $(g_2,u_1)$ for three dimensional
    ($\eps=1$) turbulent advection with
    $\alpha=110$. The square $\blacksquare$  denotes frozen velocity regime \fp{II}{7}, 
    triangle $\blacktriangleleft$ corresponds to the unstable regime \fp{III}{2} and
    circle $\bullet$ to the nontrivial
    regime \fp{III}{1} for which the time correlations are relevant. Dashed lines
    corresponds to the chosen flows to the point \fp{II}{7}, whereas
    the full lines to the flows to the other stable point \fp{III}{1}.
    \label{fig:flow2D}  
  }
\end{figure}  

For illustration purposes the projections of the RG flow onto the planes $(g_1,u_1)$
and $(g_2,u_1)$ are
depicted in Fig. \ref{fig:flow2D}. The two stable points are clearly separated
by the unstable one.

\section{Conclusion}
Methods of quantum field theory including functional differential equations, functional integrals, renormalization group and operation
product expansion have been successfully applied for a description of second order phase transitions \cite{Zinn,Vasiliev,Amit}.
They cover a broad set of problems, such as critical behavior in ferromagnets, transition to the superfluid phase in $\mbox{}^4$He, multicriticality
and many others.
Nowadays, these methods and the related theoretical framework form a cornerstone of the research area of critical phenomena. 

Classical dynamic systems, such as developed turbulence, turbulent transport phenomenon, magnetohydrodynamics and reaction-diffusion
models, could not be described within the standard equilibrium statistical physics. All these problems
are examples of systems far from equilibrium. They are fundamentally different from systems near the thermal equilibrium
in two crucial points. First, they do not possess equivalent to the Gibbs thermal state. Second, their underlying dynamics cannot -- as a rule --
be described by a single hermitian evolution operator.

Nevertheless, these problems and problems in quantum field theory share several common properties. 
Namely, the fluctuation-dissipation condition is broken in systems far from equilibrium and strong correlations are present over large spatial scales.
This often causes diverging correlation length, which allows a description in terms of continuous fields.
The fact that classical systems can be considered as euclidean versions of quantum models
induced a great stimulus in application of field-theoretic methods in the aforementioned classical problems.

The second half of 70s was a crucial period when quantum field theory and renormalization group were used for the first time
in theory of developed isotropic turbulence, stochastic magnetohydrodynamics and for analysis of transport phenomena in random environments.
The effort paid off and led, e.g., to a proof of the second Kolmogorov hypothesis about the independence of statistical correlations in velocity fluctuations
of viscosity in the inertial interval. This fact is directly related to the scaling with the celebrated universal
Kolmogorov exponents.
Furthermore, an endeavor has been undertaken with a goal of a controllable calculation of non-universal quantities such as the Kolmogorov
constant, Prandtl number and explanation of large-scale generation of magnetic fields due to spontaneous symmetry breaking.
The critical exponents of composite operators were computed allowing to analyze behavior of experimentally measurable
quantities, e.g., rate of energy dissipation on large scales. It is known that strong fluctuations of this quantity 
could violate Kolmogorov scaling,  especially for higher order structure functions of the velocity field. 
Generally speaking, the research in this area based on field-theoretical methods was and still is very fruitful and
has led to a number of important results summarized in many review articles.

Consequently it became clear that it was important to study not only ideal systems but also more realistic systems with anisotropy, non-homogeneities, broken parity symmetry and compressible fluid. These effects could profoundly change the macroscopic behavior: not only stability of universal regimes but also values of critical exponents. 

Due to the fact that most chemical reactions occur in fluids, it was clear that
the effect of hydrodynamic fluctuations stemming e.g. from thermal fluctuations must be properly taken into account for reaction-diffusion systems, phase transitions and percolation problems. 

During last two decades the anomalous scaling in the developed turbulence (known as intermittency) has been intensively studied. 
The results have theoretically confirmed the existence of anomalous scaling in open non-equilibrium systems. Study of such systems
has also brought about a further research activity in quantum-field methods. Mainly, the paradigmatic $\eps$ expansion has been improved and
specific algorithms have been created for more effective calculation of composite operators and multi-loop Feynman diagrams.

In this article it has been our aim summarize and elucidate theoretical approaches and results found in last twenty years in
classical stochastic problems far from equilibrium. We have tried to discuss a rather broad set of problems to demonstrate the robustness and
effectiveness of functional methods and  the renormalization-group technique.

The first part of the article, which is of methodological nature, reveals the importance of functional
methods suitable for systems with multiplicative noise. Functional methods in studying systems
with intrinsic noise and the double expansion scheme have been presented in detail. The latter
is an invaluable tool in problems involving hydrodynamic fluctuations.
  
The second part has been devoted to the study of systems with some violation from ideality.
The effect of helicity and anisotropy has been investigated in the Kraichnan model as well as in the Navier-Stokes equation.
A general conclusion is that both violations have a substantial effect on the stability of universal regimes and
values of critical exponents. Positions of non-universal fixed points as well as their regions of stability change
profoundly with the parameters characterizing quantitatively the violation from ideality. They influence
the crossover behavior between different universality classes as well.

An analysis of paradigmatic models that describe spreading of a passive scalar admixture and a vector quantity (magnetic field)
has revealed similarities but also differences between these two cases. The r\^ole played by tensor structures
of the fields and effects thereof on the phenomenon of intermittency and the values of critical dimensions of composite operators have been inferred.
The results clearly show that critical dimensions for given harmonics describing anisotropy are identical. However, the
behavior of scaling functions is completely different due to the leading term.
In the last part it is shown that in realistic models, in which, e.g., compressibility of the environment is allowed, new universality 
classes appear. Their existence can be traced back to new interactions, which in ideal systems are usually prohibited
by symmetry reasons.

All the mentioned results and observations generate new questions and challenges. In the future, it is desirable to
 investigate new universal patterns, crossovers between them and to develop effective techniques for higher loop calculations. 
 From a qualitative point of view, a study of more complicated reaction schemes is called for. In an analysis of intermittent behavior,
development of new models with velocity fluctuations -- as close to real fluids as possible -- is needed.

\subsection*{Acknowledgments}
The authors thank Nikolai Antonov, Loran Adzhemyan, and Mikhail Nalimov
for many illuminating and valuable discussions.
The work was supported by VEGA grant No. $1/0222/13$ 
 of the Ministry of Education, Science, Research and Sport of the Slovak Republic. 
 

\bibliography{mybib}{}
\bibliographystyle{amsplain}

\end{document}